# Catalog of Unconventional Magnons in Collinear Magnets

Xiaobing Chen[1,2], Yuntian Liu[1], Pengfei Liu[1], Yutong Yu[1], Jun Ren[1], Jiayu Li[1], Ao Zhang[1] and Qihang Liu[1,2,3*]

[1]*Department of Physics and Guangdong Basic Research Center of Excellence for Quantum Science, Southern University of Science and Technology, Shenzhen 518055, China*

[2]*Quantum Science Center of Guangdong–Hong Kong–Macao Greater Bay Area (Guangdong), Shenzhen 518045, China*

[3]*Guangdong Provincial Key Laboratory of Computational Science and Material Design, Southern University of Science and Technology, Shenzhen 518055, China*

[*]Email: liuqh@sustech.edu.cn

## Abstract

**Topological magnons have garnered significant interest for their potential in both fundamental research and device applications, owing to their exotic, uncharged, yet topologically protected boundary modes. However, their comprehension has been hindered by the absence of fundamental symmetry descriptions of magnetic materials, which are primarily governed by isotropic Heisenberg interactions in spin Hamiltonians. The ensuing magnon dispersions enable gapless magnon band nodes that go beyond the scenario of representation theory of the magnetic space groups (MSGs), thus referred to as unconventional magnons. Here we developed spin space group (SSG) theory to elucidate collinear magnetic configurations, classifying the 1421 collinear SSGs into four types, constructing their band representations, and providing a comprehensive tabulation of unconventional magnons, such as duodecuple points, octuple nodal lines, and charge-4 octuple points. Based on the MAGNDATA database, we identified 498 collinear magnets with unconventional magnons, among which over 200 magnon band structures were obtained by using first-principles calculations and linear spin wave theory. Additionally, we evaluated the influence of the spin-orbit coupling-induced exchange interaction in these magnets and found that more than 80% are predominantly governed by the Heisenberg interactions, indicating that SSG serves as an ideal framework for describing magnon band nodes in most**

**3*d*, 4*d* and half-filled 4*f* collinear magnets. Our work offers new pathways for exploring uncharged transports in magnonic systems, holding promise for advancements in next-generation spintronic devices.**

Symmetry plays a crucial role in comprehending various phenomena in condensed matter, particularly for the dispersion and topology of diverse quasiparticles. The electronic topological diagnosis has enjoyed recent success in identifying thousands of materials with topological electronic bands[1-6], and there is increasing interest in extending these principles to bosonic quasiparticles like magnons. Magnonic systems provide a fertile playground for bosonic topology, e.g., Dirac[7-9], Weyl[10-12] and triply-degenerate[13] magnons, leading to plenty of exotic phenomena[14] such as charge-free topologically protected boundary modes[7,10,15], magnon thermal Hall effect[13,16] and magnon spin Nernst effect[17,18]. Thus, finding magnetic materials that contain such topological magnons is an area of high interest and demand, with potential applications in next-generation ultrafast spintronic devices and quantum computing[19,20].

However, extending the success of topological quantum chemistry theory and material diagnosis from electrons to magnons faces a significant challenge in the fundamental symmetry description of magnetic materials. The conventional framework for describing the symmetry of magnetic materials[21] and designing magnon band nodes[22-24] enforced by symmetry is based on magnetic space groups (MSGs). These groups completely lock the rotational operations in spin space and real space. Nevertheless, the magnon dispersions, especially in collinear magnets with small spin-orbital coupling (SOC) effects [e.g., Dzyaloshinskii-Moriya (DM) interaction[25,26]], predominantly depend on the isotropic Heisenberg exchange interaction[27]. The corresponding spin Hamiltonian permits more symmetry operations with separated spin and spatial operations than MSGs. As a result, even though the material candidates for experimental measurements are rare, the existing experimentally observed magnon spectra still cannot be entirely explained by MSGs[27-30]. Examples include the observed Dirac and sextuple points in collinear antiferromagnet (AFM) $Cu_3TeO_6$[9,31], the two-fold nodal plane in collinear ferromagnet (FM) Gd[32,33], and doubly degenerate magnon modes in collinear AFMs[17].

To address these challenges, enhanced symmetry groups that describe the magnon band nodes beyond the conventional MSGs, termed unconventional magnons, are required. Spin space groups (SSGs), which were first proposed in the 1960s (yet overlooked until the recent development of AFM spintronics[29,30,34-54]), provide a framework that allows for decoupling of spatial and spin operations, and an examination of the symmetries associated with magnon dispersions[27,28,30]. However, the absence of magnonic material databases and fundamental SSG theories has impeded the understanding of emergent quasiparticles and the systematic search for unconventional magnonic systems.

In this work, we develop the SSG theory describing collinear magnetic configurations, which are typically dictated by Heisenberg exchange interaction, and apply it to the comprehensive search of unconventional magnons in real materials. We categorize the 1421 collinear SSGs into four types, corresponding to all the type-I, III, and IV MSGs yet with more spatial-free spin symmetries and thus different band representations. By employing the representation theory, we not only capture the band degeneracies to reconcile the results of previous experiments, but also tabulate all possible unconventional magnons such as duodecuple (12-fold) nodal point, charge-4 octuple (8-fold) point, octuple nodal line, quadruple nodal plane, and altermagnetic-splitting chiral magnon. Thanks to the $SO(2)$ spin symmetry, the nature of unconventional magnons explored here can be straightforwardly applied to electronic structures without considering spin-1/2 double-group representations. We further conduct high-throughput *ab initio* calculations on the magnon band dispersions and representations for 348 computable collinear magnets within MAGNDATA material database[55], which adopts experimentally confirmed magnetic structures. Based on an efficient and systematic diagnosis, we identify more than 200 unconventional magnonic materials by the calculated bilinear magnetic exchange coefficients and magnon dispersions. By comparing the results with and without SOC, we evaluate the magnitude of anisotropic interactions and thus the validity of unconventional magnons within the SSG scenario. Finally, we provide the full array of data of collinear SSG theory, including general positions, spin Wyckoff positions, spin Brillouin zones, little co-groups of wavevectors, band representations, and the magnon dispersions of calculated magnets in our homemade online program and database FINDSPINGROUP (findspingroup.com)[56].

***Categories of 1421 collinear spin space groups***

Beginning with 90 spin point groups describing collinear magnetic orders[29], 1421 SSGs for collinear magnets can be constructed by group extension[52]. A typical SSG operation is written as $\{O_s||O_r\}$, where $O_s$ and $O_r$ denote the operation in spin space and real space, respectively. In general, SSGs can be expressed as $G_S = G_{NSS} \times G_{SO}$, where $G_{SO}$ stands for the spin-only group that only contains spin operations $\{O_s||E\}$; $G_{NSS}$ stands for the nontrivial spin space group that contains no pure spin operations[57]. For collinear magnetic structures, $G_{SO} = Z_2^{\mathrm{K}} \ltimes SO(2)$, where $SO(2) = \{U_z(\phi), \phi \in [0,2\pi)\}$ contains full spin rotations along the spin axis $z$, $Z_2^K = \{E, TU_{\boldsymbol{n}}(\pi)\}$ contains the product of time-reversal $T$ and a two-fold spin rotation about any axis $\boldsymbol{n}$ perpendicular to the $z$-axis[29].

For collinear FM materials with only one type of magnetic sublattice, spatial operations cannot couple any spin operations that change the spin direction. Therefore, the nontrivial spin space group is given by $G_{NSS} = \{E||G\}$, where $G$ is the space group (230 in total). These SSGs are also sufficient to describe the symmetry of collinear ferrimagnet (FIM) that contains multiple spin sublattices yet without any symmetry connecting them. For collinear AFM with two sublattices carrying opposite spins, $G_{NSS}$ is obtained by group extension $\{E||G_\uparrow\} + \{U_{\boldsymbol{n}}(\pi)||AG_\uparrow\}$, where sublattice space group $G_\uparrow$ contains all spatial symmetries that do not exchange atoms from different sublattices; $\{U_{\boldsymbol{n}}(\pi)||A\}$ is the symmetry operation that simultaneously reverses the spins and exchanges the two sublattices.

In analogy with the construction of type-I, III and IV MSGs, 1421 collinear SSGs are constructed, including 230 type-I SSGs describing collinear FM/FIM and 1191 SSGs describing collinear AFM. Compared with 674 type-III MSGs, we further categorize collinear SSGs using $\{U_{\boldsymbol{n}}(\pi)||A\}$, including 252 type-II SSGs with $A = P$ (space inversion), and 422 type-III SSGs with $A = C_n, PC_n$ $(n = 2,4)$ (space rotations and rotoinversions). Such classification naturally leads to the separation of conventional PT-symmetric antiferromagnets with two-fold spin degenerate bands (type-II SSGs) and the recently emerged altermagnets with AFM-induced spin splitting (type-III SSGs)[37,40,41]. Besides, there are 517 type-IV SSGs describing AFMs with $A = \tau$ (fractional lattice translation). All collinear SSGs have been tabulated in Supplementary Information S2 and are summarized in Fig. 1**a**. While exhibiting one-to-one correspondence with 1421 MSGs, 1421 collinear SSGs manifest more spin symmetries, including spin $SO(2)$, $TU_{\boldsymbol{n}}(\pi)$ and $\{U_{\boldsymbol{n}}(\pi)||A\}$, which are crucial for more band degeneracies beyond the regime of MSGs (see a detailed comparison in Supplementary Information S1). We list the SSGs of all collinear magnets in MAGNDATA in our online database FINDSPINGROUP[56].

***Workflow***

Figure 2**a** outlines the procedure to obtain all types of unconventional magnons and their realization in collinear magnets documented in MAGNDATA. Starting from 1421 collinear SSGs, we establish the position-space description of collinear SSGs, where the spin site groups for all spin Wyckoff positions can be defined and characterized by 90 collinear spin point groups. Among these groups, only 32 FM spin point groups support nonzero magnetic moments, resulting in a reduction of the total number of spin Wyckoff positions from 12481 to 6368. Meanwhile, we introduce a momentum-space description for collinear SSGs, exhausting the spin Brillouin zones and high-symmetry $k$-points using 24 mapped centrosymmetric symmorphic space groups. Subsequently, by employing Zak's method[58], we derive the irreducible little co-representations of magnons in collinear SSGs and enumerate all band representations hosting unconventional magnons. Finally, we construct $k \cdot p$ effective models around the degenerate points and evaluate the node topology characterized by chiral charge or monopole charge, identifying all unconventional magnons beyond the scope of MSGs. The general positions, Wyckoff positions, spin site groups, spin Brillouin zones, $k$-little groups, and magnon band representations for all collinear SSGs are available in our online database FINDSPINGROUP.

For material realization, we adopt MAGNDATA database that contains more than 2000 commensurate magnetic structures determined by neutron scattering measurements. By spin group symmetry identification, we filter out 1223 collinear magnets (shown in Fig 1**b**) with their SSGs and spin Wyckoff positions of magnetic ions. According to the corresponding band representations, we identify 498 materials hosting unconventional magnons, including 5 FMs, 17 FiMs and 476 AFMs, which can be further classified into 56 PT AFMs, 203 altermagnets and 217 Uτ AFMs. After excluding magnetic structures with fractional occupancy or more than 80 atoms in the spin primitive cell, we perform density functional theory (DFT) and Wannier projection calculations on 348 materials. After filtering out the cases failing of convergence and Wannierization, we obtain the bilinear magnetic exchange coefficients of 223 collinear magnets by using TB2J code. The magnon band structures of the Heisenberg spin Hamiltonians are calculated by the linear spin wave theory. Specifically, Holstein-Primakoff transformation $S^+ \approx \sqrt{2S}a, S^- \approx \sqrt{2S}a^\dagger, S^z = S - a^\dagger a$ connects the magnon creation $a^\dagger$ (annihilation $a$) operator with the electron annihilation $S^-$ (creation $S^+$) operator[59]. Finally, we obtain the unconventional magnons and corresponding irreducible co-representations (co-irreps) in 203 collinear magnets, while the magnon Hamiltonians of the other 20 systems are negative definite, leading to the magnetic instability.

***Catalog of unconventional magnons***

The bases of magnon bands $S^\pm$ transform as the irreducible representations (irreps) of the spin site groups of the magnetic ions. For collinear magnets, the spin site group always have $SO(2)$ spin rotation symmetry, $S^\pm$ thus transform as $m_s = \pm 1$ irreps of $SO(2)$ in spin space, while the real-space part only provides the identity irrep for $S^\pm$. As a result, there are three main types of symmetry resulting in extra degeneracies of magnon bands: (i) Unitary spin-space symmetry (USS): the combination of $SO(2)$ with unitary $\{U_{\boldsymbol{n}}(\pi)||A\}$ symmetry pairs two conjugated one-dimensional (1D) irreps for $S^\pm$ into a two-dimensional (2D) irrep in spin space. (ii) Antiunitary spin-space symmetry (ASS): combination of $SO(2)$ with antiunitary $\{T||A\}$ pairs two conjugated 1D irreps for $S^\pm$ into a 2D co-irrep in spin space. (iii) Antiunitary real-space symmetry (ARS): antiunitary $\{TU_{\boldsymbol{n}}(\pi)||E\}$ symmetry pairs two conjugated irreps of real-space operations (see Methods).

We next note several features of the magnon band degeneracies based on the co-irreps of collinear SSG: i) Despite the absence of $T$, type-I SSGs have the same band degeneracy and topology with the corresponding space group $G$ combining $T$ due to $\{TU_{\boldsymbol{n}}(\pi)||E\}$ symmetry. This is also applicable in searching electronic orbital multiplets of collinear FM[60]. ii) For collinear SSGs describing AFM, magnon band structures are doubly degenerate throughout the Brillouin zone for all type-II and type-IV SSGs, because of ASS $\{T||P\}$ and USS $\{U_{\boldsymbol{n}}(\pi)||\tau\}$ at an arbitrary $k$ point, respectively. iii) The topological charges at nodal points or lines formed by two opposite-spin branches are opposite for type-II SSGs, but identical for type-IV SSGs. iv) Magnon band structures with type-III SSGs show AFM-induced chirality splitting, the magnonic analog of spin splitting in altermagnets[39,61]. Therefore, the unconventional magnons can be classified into five categories: (1) topological charge-neutral ($C$ = 0) quasiparticles, including duodecuple, octuple, sextuple and triple magnons; (2) chiral quasiparticles ($C \neq 0$), including C–4 octuple, C–4 sextuple, C–8 Dirac and C–2 triple magnons; (3) octuple nodal line magnon; (4) quadruple nodal plane magnon; and (5) altermagnetic-splitting chiral magnon. The correspondence between co-irreps and unconventional magnons in all collinear SSGs is provided in Supplementary Information S3.

Figure 2**b** summarizes the statistics of material candidates hosting the five categories of unconventional magnons. Notably, about 40% of the collinear magnets possess at least one type of unconventional magnons. The two major types are quadruple nodal plane magnon (22.34%) and altermagnetic chiral magnon (16.45%). This is consistent with that the proportion of altermagnets and Uτ AFMs is relatively large in collinear magnets with unconventional magnons in Fig 1**b**. In Table I we list some representative materials with their SSGs, co-irreps, and magnon node topology, where the diagnosis results of all 498 materials with unconventional magnon are provided in Supplementary Information S4. Specifically, we note that even the total topological charges of the Dirac and sextuple points in some PT AFMs are zero[22], their node topology can be characterized by a nonzero $C_2$ monopole charge because the composite Weyl or triple points of the two spin channels host opposite topological charges (see Methods). Therefore, breaking $\{T||P\}$ symmetry, e.g., by an external magnetic field, could be feasible for the detection of spatially resolved thermal conductivity, i.e., “hidden” magnon thermal Hall effect.

***Representative materials with unconventional magnons***

Now we present calculated materials manifesting various unconventional magnons in each type of SSGs, including their degeneracy and node topology. The calculated magnon band structures of all the 203 candidates are provided in Supplementary Information S5. We begin with sextuple and triple points in type-I collinear SSG $I^1\bar{4}^13^1d^{\infty m}1$ in FM $Gd_4Sb_3$ (Figs. 3**a**-3**c**). Within the regime of MSG, since that collinear magnets cannot be cubic, triple or sextuple magnons are thus absent. However, we find triple magnons at the Γ point, indicating that the SSG maintains the cubic nature of the lattice. In addition, due to the ARS $\{TU_{\boldsymbol{n}}(\pi)\|C_{2z}|\tau_{y/2}\}$, two triple magnons stick together forming a sextuple magnon at the boundary of Brillouin zone (H point). This is a typical case where the decoupled spin and space rotations in type-I SSGs provide more fertile platform for quasiparticles than MSGs. We note that similar ARS was found to protect the two-fold nodal plane in collinear FM Gd[32].

Compared with the collinear FM, spin space can lead to additional degeneracy by USS and/or ASS in collinear AFM. The presented example is PT AFM $Cu_3TeO_6$ (Figs. 3**d**-3**f**), whose magnon dispersion were measured by inelastic neutron scattering experiments[9,31]. Specifically, the magnon bands are doubly degenerate throughout the whole Brillouin zone, with several sextuple and Dirac magnons identified at the Γ, P and H points. These features are all beyond the regime of MSG, where the corresponding MSG $R\bar{3}'$ predicts non-degenerate band at an arbitrary $k$ point and at most two-fold band crossings at any high-symmetry points (see Supplementary Information S7.3.2). In sharp contrast, these band degeneracies are perfectly described using co-irreps of the type-II cubic SSG $I^{\bar{1}}a^{\bar{1}}\bar{3}^{\infty m}1$. As discussed above, the ASS $\{T\|P\}$ at arbitrary $k$ points protects doubly degenerate band throughout the Brillouin zone[22]. Furthermore, the little group ${}^{\bar{1}}m^{\bar{1}}\bar{3}^{\infty m}1$ allows sextuple magnons at Γ and H points with $C_2$–2 monopole charge and thus the hidden magnon thermal Hall effect (Table I). By symmetry diagnosis, we predict that such hidden effects could also be detected in $K_yFe_{2-x}Se_2$, a widely-explored iron-based superconductor[62,63]. We further note that the highest dimension of the symmetry-protected magnon bands in type-II SSGs is twelve. A representative example is the R point in experimentally synthesized $Pr_5Mo_3O_{16}$, where a collinear AFM order is implemented (Supplementary Information S7.3.5).

Another interesting case is the octuple magnon, which is found at the A point in type-III SSG $P^{\bar{1}}\bar{6}^{\bar{1}}2^{1}c^{\infty m}1$, in FeS (Figs. 3**g**-3**i**). Such an octuple magnon is also the synergistic effect of the ASS ($\{U_{\boldsymbol{n}}(\pi)\|m_z|\tau_{z/2}\}$) and ARS ($\{T\|C_{2x}|0\}$). Moreover, the bands off the high-symmetry point are nondegenerate, leading to AFM-induced magnon chirality splitting along, e.g., Γ-A and A-K directions. We expect that magnon chirality splitting in collinear AFM can lead to the appearance of non-zero Berry curvature, magnon orbital angular momentum, and magnon nonlinear thermal Hall in the absence of SOC, which will further expand the scope of relevant field of spintronics[64-67]. We list all the 203 diagnosed altermagnets and the momentum-position of magnon chirality splitting (also electronic spin splitting) from MAGNDATA in Supplementary Information S4.6.

The magnon quasiparticles carry various topological charges, some of which are absent in all encyclopedias of MSGs[68-70] but protected by SSG symmetries. One striking case is the charge–4 (C–4) octuple magnons (Figs. 3**j**-3**l**) in collinear $Fe_{0.35}NbS_2$ with a type-IV SSG $P_a{}^{1}2_1{}^{1}2_1{}^{1}2_1{}^{\infty m}1$. Due to the $\{U_{\boldsymbol{n}}(\pi)||E|\tau\}$ and ARS $\{T||C_{2z}|\tau_{x/2}\}$ of the little group of $k_z = \pi$ plane, the magnon bands form quadruple Z-U-R-T nodal plane and octuple quasiparticle at the R point. Different from type-II SSGs, the sign of Berry curvature from the two pairs of four-fold degenerate bands connected by $\{U_{\boldsymbol{n}}(\pi)||E|\tau\}$ are identical for type-IV SSGs, resulting in C–4 octuple magnons as the superposition of two C–2 Dirac points at R. In addition, a pair of Dirac points with topological charge C = –2 are observed along the high-symmetry line G-±Y. As a consequence, the total topological charge in the whole Brillouin zone is zero, obeying the Nielsen-Ninomiya no-go theorem[71]. Furthermore, three perpendicular k-planes ($k_x$, $k_y$, $k_z = \pi$) intersecting at the R point form a quadruple nodal plane network. Detailed discussions on topological charges and the quadruple nodal plane network in $Fe_{0.35}NbS_2$ are provided in Supplementary Information S7.3.4. We predict cases of magnon band topology such as C–4 sextuple and C–8 Dirac magnons as summarized in Supplementary Information S4.

***SOC effects***

To evaluate the applicability of collinear SSGs in diagnosing unconventional magnons, we fully consider SOC effects by calculating the bilinear magnetic exchange interactions, including antisymmetric DM interaction ($D$), exchange anisotropy ($J^{ani}$), and off-diagonal symmetric anisotropy ($\Gamma$), and compare them with the isotropic Heisenberg exchange interaction ($J$) in 223 candidate materials. Comparing with SOC-free calculations, the high-throughput SOC calculations yield over 10 times of computational costs. Notably, single-ion anisotropy is excluded in our analysis because in collinear magnets it does not affect the symmetry of the magnon Hamiltonian, the band representations and the nodal topology (see Supplementary Information S7). Table II shows that in 82% of the calculated collinear magnets, DM interaction, exchange anisotropy (e.g., Kitaev interaction), and off-diagonal symmetric anisotropy are at least an order of magnitude smaller than the Heisenberg interaction, indicating the validity of SSG for these systems. Specifically, 87% of the $3d$ and $4d$ magnets are dominated by Heisenberg interactions, while the remaining ones possess strong DM interactions, such as the reported altermagnet $Fe_2Mo_3O_8$ with DM interaction-induced magnon polaron[72]. In addition, the half-filled $4f$ magnets are also largely compatible with the framework of SSGs, such as $Gd^{3+}$ in $GdAlO_3$ and GdAgSn. However, the $5d$ and $5f$ magnets failed due to the strong DM interaction or exchange anisotropy. These results indicates that SSG serves as an ideal framework for describing magnon band nodes in most $3d$, $4d$ and $4f$ collinear magnets. Detailed information regarding the calculated exchange interactions for 223 candidates is available in Supplementary Information S6.

For collinear magnets, there exists an effective time-reversal symmetry $TU_n(\pi)$ that squares to one. Consequently, magnons in collinear magnets fall into the class AI in the tenfold way, which do not support strong topological insulating phase in 3D[73,74]. Recent studies show that SOC-free weak topological insulators could exist in altermagnets[75]. On the other hand, under nonlinear spin wave theory, the introduction of magnon-magnon interaction does not break the $SO(2)$ symmetry in collinear magnets. Thus, it does not change the band degeneracy, while it may cause band renormalization[76]. However, adding SOC term to collinear spin Hamiltonian typically open a small gap, sometimes rendering the emergence of class AII and $\mathbb{Z}_2$ gapped topological phase. For example, the introduction of DM interaction in pyrochlore, honeycomb and kagome FMs can transform Dirac magnons into topological magnon bands[77-79]. In this case, the description of SSGs is still useful in that it can be used to search the SOC-induced small gap, which is often the perquisite of magnon topological materials, such as Chern insulators and axion insulators. Furthermore, the incorporation of DM interaction and magnon-magnon interactions can cause multiple topological phase transitions.[80,81] Overall, understanding the evolution from node topology to band topology with the introduction of SOC effect and its impact on magnon transport will be valuable for the development of magnon-based spintronic devices, and is left for future studies.

## *Discussion*

Importantly, we emphasize that while our focus is on magnon systems, the main results of unconventional quasiparticles originated from band degeneracies can be straightforwardly applied to electronic systems. In the framework of (magnetic) space group, the double-valued representation of spin-1/2 fermions requires the so-called double group to describe an additional -1 phase of $2\pi$ rotation. In sharp contrast, the particularity of collinear SSGs renders that the dimension of the band representations for both fermions and bosons remains invariant within $4\pi$ rotation. This is because the infinite spin-only group $SO(2)$ has the double-covering group of itself, and can thus be regarded as either a single group or a double group. Therefore, $SO(2)$ always provides 1D irreps labeled by spin angular momentum $m_s = \pm 1/2$ for electrons and $m_s = \pm 1$ for magnons. The only difference between fermions and bosons in collinear SSGs occurs solely within the phase in irrep matrices. The comparison between band representations for electrons and magnons in collinear SSGs is provided in Supplementary Information S1.6.4.

The main outcome of this work is twofold. First, we present a comprehensive representation theory of 1421 collinear SSGs, which includes the spin Wyckoff positions, spin Brillouin zones, and band representations associated with each wavevector. Such framework allows for the full tabulation of unconventional magnons beyond the MSGs. Second, we provide a dictionary of over 200 candidate materials that host unconventional magnons, identified through high-throughput calculations. The integration of SSGs and the collinear magnets listed in the MAGNDATA database facilitates efficient and systematic characterization of magnons exhibiting nontrivial nodal topology. By virtue of this, the unconventional magnons have the potential to host topologically protected, uncharged surface states, leaving fruitful transport and neutron scattering signatures for experiments.

## Methods

**Band representation theory for SSGs.** Band representations are constructed from atomic limit as introduced by Zak[58]. Firstly, we construct the spin site group $G_S^{\boldsymbol{q}}$ of the spin space group $G_S$, the corresponding character table and the irreps of orbit basis $\rho_{S\pm}^{\boldsymbol{q}}$, where $\boldsymbol{q}$ is any point in the unit cell of magnetic lattice. Among these spin site groups, only 32 FM spin point groups can support nonzero magnetic moments and carry magnon. Secondly, to induce full band representations $\rho_{S^{\pm}} = \rho_{S^{\pm}}^{\boldsymbol{q}} \uparrow G_S$, we seek a coset decomposition of $G_S^{q}$. All orbits of the Wyckoff position $\{\boldsymbol{q}_\alpha = g_\alpha \boldsymbol{q}_1 | \boldsymbol{q}_\alpha \in G_S\}, \alpha = 1, 2, \ldots, n$ with multiplicity $n$ of the Wyckoff position are derived. The SSG element $g_\alpha$, combined with the translation $\mathbb{T}$, generate the decomposition of $G_S$ with respect to the $G_S^{\boldsymbol{q}}$:

$$G_S = \bigcup_{\alpha} g_\alpha \left( G_S^{\boldsymbol{q}} \ltimes \mathbb{T} \right) \tag{1}$$

The full band representation $\rho_{S^{\pm}} = \left( \rho_{S^{\pm}}^{\boldsymbol{q}} \uparrow G_S \right)$ is induced from orbital representation $\rho_{S^{\pm}}^{\boldsymbol{q}}$, then we restrict it to band representations of k-little groups $\rho_{S^{\pm}}^{\boldsymbol{k}} = \rho_{S^{\pm}} \downarrow G_S^{\boldsymbol{k}}$ with ingredients:

$$\chi_{\rho_{S^{\pm}}^{\boldsymbol{k}}}(g) = \begin{cases} \sum_{\alpha} e^{-i[R(g)k \cdot t_{\alpha\alpha}]} \chi_{\rho_{S^{\pm}}^{\boldsymbol{q}}} \left( g_\alpha^{-1} \{E \| E | -t_{\alpha\alpha}\} g g_\alpha \right) & g \in G_{ss}^{q} \\ 0 & g \notin G_{ss}^{q} \end{cases} \tag{2}$$

where $t_{\alpha\alpha} = g\boldsymbol{q}_\alpha - \boldsymbol{q}_\alpha$. In this step, the character table of the unitary part of $G_S^{\boldsymbol{k}}$ is also constructed. At last, we perform sum rules[21] to account for the introduction of anti-unitary operations and the band representation $\rho_{S^{\pm}}^{\boldsymbol{k}}$ in Spin Brillouin zone is determined. The construction of band representation for SSG $I^{\bar{1}}a^{\bar{1}}\bar{3}^{\infty m}1$ ($Cu_3TeO_6$ case) can be found in Supplementary Information S1.6.

**Two mechanisms of extra twofold degeneracy provided by spin space.** For collinear AFM, $G_S$ have the form of $(\{E \| G_{\uparrow}\} + \{U_{\boldsymbol{n}}(\pi) \| A G_{\uparrow}\}) \times Z_2^{\mathrm{K}} \ltimes SO(2)$. Here we briefly show how the combination of $SO(2)$ spin symmetry with $\{U_{\boldsymbol{n}}(\pi) \| A\}$ or $\{T \| A\}$ will pair two 1D representations into 2D irreps for $S^{\pm}$ in spin space.

Despite the pure space rotations ($G_{\uparrow}^{k}$) and pure spin rotations from $SO(2)$ symmetry in the little group $G_S^{k}$, three symmetry operations account for the twofold degeneracy in spin space, including $\{U_{\boldsymbol{n}}(\pi) \| A\}$, $\{T \| A\}$ and $\{T U_{\boldsymbol{n}}(\pi) \| E\}$. Here we show how these operations acting with the $U_{\boldsymbol{z}}(\phi)$ in the $\{S^{+}, S^{-}\}$ basis.

The matrix representations of spin rotations and the time reversal are written as:

$$D\left(U_z(\phi)\right) = \begin{pmatrix} e^{-i\phi} & 0 \\ 0 & e^{i\phi} \end{pmatrix},\ D\left(U_x(\pi)\right) = \begin{pmatrix} 0 & 1 \\ 1 & 0 \end{pmatrix},\ D(T) = \begin{pmatrix} 0 & -1 \\ -1 & 0 \end{pmatrix} K,$$

where $K$ is the complex conjugation operator, and we use $\boldsymbol{n} = x$.

For $\{U_x(\pi)\|A\}$:

$$\{U_{\boldsymbol{x}}(\pi)\|A\}\{U_{\boldsymbol{z}}(\phi)\|E\}(\{U_{\boldsymbol{x}}(\pi)\|A\})^{-1} = \{U_{\boldsymbol{z}}(-\phi)\|E\}$$

$$\Rightarrow \left(\{E\|G_{\uparrow}^{k}\} + \{U_{\boldsymbol{n}}(\pi)\|AG_{\uparrow}^{k}\}\right) \ltimes SO(2) \cong \{E\|G_{\uparrow}^{k}\} \times D_{\infty} \tag{3}$$

where $D_{\infty}$ provide 2D irreps for $S^{\pm}$.

For $\{T\|A\}$:

$$(\{T\|A\}\{U_{\boldsymbol{z}}(\phi)\|E\})^2 \begin{pmatrix} \psi_{i,S^+}(k) \\ \psi_{i,S^-}(k) \end{pmatrix} = \begin{pmatrix} e^{2i\phi}\psi_{i,S^+}(k) \\ e^{-2i\phi}\psi_{i,S^-}(k) \end{pmatrix} \tag{4}$$

where the time reversal binds two conjugated 1D irreps in spin space into one 2D co-irrep.

For $\{TU_{\boldsymbol{x}}(\pi)\|E\}$:

$$(\{TU_{\boldsymbol{x}}(\pi)\|E\}\{U_{\boldsymbol{z}}(\phi)\|E\})^2 \begin{pmatrix} \psi_{i,S^+}(k) \\ \psi_{i,S^-}(k) \end{pmatrix} = \begin{pmatrix} \psi_{i,S^+}(k) \\ \psi_{i,S^-}(k) \end{pmatrix} \tag{5}$$

$\{TU_{\boldsymbol{x}}(\pi)\|E\}$ cannot contribute to the new degeneracy.

Therefore, we can conclude that both $\{U_{\boldsymbol{n}}(\pi)\|A\}$ and $\{T\|A\}$ can lead to the emergence of 2D irreps in spin space. More details on band representations of collinear SSGs can be found in Supplementary Information S1.6.

**Density functional theory (DFT) calculations.** All DFT calculations herein are performed using projector augmented wave method, implemented in Vienna ab initio simulation package (VASP)[82,83]. The generalized gradient approximation of the Perdew-Burke-Ernzerhof-type exchange-correlation potential[84] is adopted. For all candidate materials, we employ a cutoff energy of 500eV, which typically leads to numerical convergence. We use Γ-centered Monkhorst-Pack meshes[85], with the standard for each direction being the product of the number of k-points and the lattice length greater than 45 Å. For the *d*- and *f*-electron magnetic atoms, the initial magnetic moments are set to 5 and 7 $\mu_B$, respectively. To include the effect of electron correlation, the DFT+U approach within the rotationally invariant formalism[86] were performing with the $U_{eff}$ values based on the reported value from literature, which are provided in Supplementary Information S5 for each material. To get an accurate determination of the exchange interactions, a self-consistency convergence within $10^{-7}$ eV has been achieved in our calculations. Tight-binding models are constructed from DFT bands using the WANNIER90 package[87,88], and then the TB2J code[89] is used to extract the magnetic exchange parameters. The spin-exchange cutoff distance is set to truncate when the absolute value of the remaining Heisenberg exchange coefficients J is one-thousandth of the largest J value or numerically less than 0.001 meV. Detailed parameters of Heisenberg exchange interactions for calculating the magnon band structure can be found in the Supplementary Information S6.

**Magnon band structure calculations.** The magnon band structure are all calculated using the linear spin wave theory and the Heisenberg spin Hamiltonian. The ground state spin Hamiltonian can be changed into quadratic Hamiltonian by performing the Holstein-Primakoff transformation and Fourier transformation:

$$H = \sum_{k} \psi^{\dagger}(k) H(k) \psi(k) \tag{6}$$

where $\psi^{\dagger}(k) = \left(a_{k1}^{\dagger} \dots a_{km}^{\dagger} a_{-k1} \dots a_{-km}\right)^{T}, H(k) = \begin{pmatrix} h(k) & g(k) \\ g(k)^{\dagger} & h(-k)^{T} \end{pmatrix}$

$$h(k)_{ab} = S\left[\sum_{R_{ij}} \left(\alpha_{ab} J_{\tau_a,\tau_b+R_{ij}}\right) \cdot e^{ikR_{ij}} - \delta_{ab} \sum_{R_{ij},\ c} \left(\gamma_{ac} J_{\tau_a,\tau_c+R_{ij}}\right)\right] \tag{7}$$

$$g(k)_{ab} = S \sum_{R_{ij}} \left(\lambda_{ab} J_{\tau_a,\tau_b+R_{ij}}\right) \cdot e^{ikR_{ij}} \tag{8}$$

where $\delta_{ab}$ is the Kronecker delta, $R$ and $\tau$ represent the lattice translation vector and the position of magnetic ions in the lattice basis, respectively. When $\vec{S}_a$ is parallel with $\vec{S}_b$, $\alpha_{ab} = 1, \gamma_{ab} = 1, \lambda_{ab} = 0$; when $\vec{S}_a$ is antiparallel with $\vec{S}_b$, $\alpha_{ab} = 0, \gamma_{ab} = -1, \lambda_{ab} = -1$.

Based on above, the eigenvalues and eigenvectors of magnon Hamiltonian can be calculated by diagonalizing $H(k) \cdot I_{-}$, where $I_{-} = \begin{pmatrix} I_m & 0 \\ 0 & -I_m \end{pmatrix}$, and $I_m$ is the *m*-directional identity matrix.

**Topological charges.** We characterize the topology of the degenerate point by calculating the topological charge. For nodal point, we calculate the Wilson loops on a sphere enclosing the nodal point[90,91]:

$$W(\theta) = \oint A(k) dk \tag{9}$$

where $\theta$ is the polar angle of the sphere, $A(k) = i\langle\psi(k)|\nabla|\psi(k)\rangle$ is the Berry connection.

Now we show the different symmetry operations on $A(k)$:

$$PA(k) = -A(-k), TA(k) = A(-k), UA(k) = A(k) \tag{10}$$

As a result, the topological charge is zero at $P$-symmetric k points since $PW(\theta) = W(\pi - \theta)$, where Wilson loop is symmetric about the $\theta = \pi/2$ plane. This conclusion can be generalized to $PC_n$-invarinat $k$ point, where $C_n$ is any proper space rotation. However, the time reversal $T$ and spin rotation $U$ does not give any constrains on the topological charge.

For collinear FMs, when the $k$-little group is chiral, it can host a nonzero topological charge. For collinear AFMs, the magnon Hamiltonian can be separated into two spin channels with spin angular momentum $S = \pm 1$, the two degenerate spin channels should be connected by $TA$ or $UA$. If the $k$-little sublattice group $G_{\uparrow}^{k}$ is chiral, it can host a nonzero topological charge $C_{\uparrow}$ in the spin up channel. The two spin channels can possess identical or opposite topological charges when the two sublattice are connected by proper or improper $A$. In the former, the topological charge will be doubled as $C = 2C_{\uparrow}$, while $C = 0$ in the latter. For the compensated charge with improper $A$, we can define a monopole charge $C_2 = (C_{\uparrow} - C_{\downarrow})/2$.

Therefore, the nodal points of two degenerate branches possess opposite (identical) topological charges for type-II (IV) SSGs due to the $PT$ ($U\tau$) symmetry in the whole spin Brillouin zone. For type-III SSGs, if $G_{\uparrow}^{k}$ is chiral and the $k$-little group $G^{k}$ contain $TA$ and $UA$ operation, the two magnon branches will degenerate with doubled or compensated topological charge when $A$ is proper or improper. However, when the $k$-little group does not contain $TA$ and $UA$ operation, the two magnon branches will split and the sign of topological charge in two channels is irrelevant.

Detailed information about the nonzero topological charges of magnonic irreps in collinear SSGs at symmetry-protected degeneracies is provided on our online program FINDSPINGROUP[56].

**Acknowledgements**

This work was supported by National Key R&D Program of China under Grant No.

2020YFA0308900, the National Natural Science Foundation of China under Grant No. 12274194, Guangdong Provincial Key Laboratory for Computational Science and Material Design under Grant No. 2019B030301001, Guangdong Provincial Quantum Science Strategic Initiative (Grant No. GDZX2401002), Shenzhen Science and Technology Program (Grants No. RCJC20221008092722009 and No. 20231117091158001), Innovative Team of General Higher Educational Institutes in Guangdong Province (No. 2020KCXTD001), the Science, Technology and Innovation Commission of Shenzhen Municipality (No. ZDSYS20190902092905285), the Open Fund of the State Key Laboratory of Spintronics Devices and Technologies (Grant No. SPL-2407) and Center for Computational Science and Engineering of Southern University of Science and Technology.

**References**

1. Bradlyn, B. et al. Topological quantum chemistry. *Nature* **547**, 298 (2017).
2. Kruthoff, J. et al. Topological Classification of Crystalline Insulators through Band Structure Combinatorics. *Phys. Rev. X* **7**, 041069 (2017).
3. Elcoro, L. et al. Magnetic topological quantum chemistry. *Nat. Commun.* **12**, 5965 (2021).
4. Zhang, T. et al. Catalogue of Topological Electronic Materials. *Nature* **566**, 475 (2019).
5. Vergniory, M. G. et al. A complete catalogue of high-quality topological materials. *Nature* **566**, 480 (2019).
6. Tang, F., Po, H. C., Vishwanath, A. & Wan, X. Comprehensive search for topological materials using symmetry indicators. *Nature* **566**, 486 (2019).
7. Yuan, B. et al. Dirac Magnons in a Honeycomb Lattice Quantum XY Magnet $CoTiO_3$. *Phys. Rev. X* **10**, 011062 (2020).
8. Chen, L. et al. Topological Spin Excitations in Honeycomb Ferromagnet $CrI_3$. *Phys. Rev. X* **8**, 041028 (2018).

9. Bao, S. et al. Discovery of coexisting Dirac and triply degenerate magnons in a three-dimensional antiferromagnet. *Nat. Commun.* **9**, 2591 (2018).

10. Li, F.-Y. et al. Weyl magnons in breathing pyrochlore antiferromagnets. *Nat. Commun.* **7**, 12691 (2016).

11. Mook, A., Henk, J. & Mertig, I. Tunable Magnon Weyl Points in Ferromagnetic Pyrochlores. *Phys. Rev. Lett.* **117**, 157204 (2016).

12. Su, Y., Wang, X. S. & Wang, X. R. Magnonic Weyl semimetal and chiral anomaly in pyrochlore ferromagnets. *Phys. Rev. B* **95**, 224403 (2017).

13. Hwang, K., Trivedi, N. & Randeria, M. Topological Magnons with Nodal-Line and Triple-Point Degeneracies: Implications for Thermal Hall Effect in Pyrochlore Iridates. *Phys. Rev. Lett.* **125**, 047203 (2020).

14. McClarty, P. A. Topological Magnons: A Review. *Annu. Rev. Condens. Matter Phys.* **13**, 171 (2022).

15. Kondo, H. & Akagi, Y. Dirac Surface States in Magnonic Analogs of Topological Crystalline Insulators. *Phys. Rev. Lett.* **127**, 177201 (2021).

16. Li, S. & Nevidomskyy, A. H. Topological Weyl magnons and thermal Hall effect in layered honeycomb ferromagnets. *Phys. Rev. B* **104**, 104419 (2021).

17. Cheng, R., Okamoto, S. & Xiao, D. Spin Nernst Effect of Magnons in Collinear Antiferromagnets. *Phys. Rev. Lett.* **117**, 217202 (2016).

18. Zyuzin, V. A. & Kovalev, A. A. Spin Hall and Nernst effects of Weyl magnons. *Phys. Rev. B* **97**, 174407 (2018).

19. Hetényi, B., Mook, A., Klinovaja, J. & Loss, D. Long-distance coupling of spin qubits via topological magnons. *Phys. Rev. B* **106**, 235409 (2022).

20. Baltz, V. et al. Antiferromagnetic spintronics. *Rev. Mod. Phys.* **90**, 015005 (2018).

21. Dimmock, J. O. & Wheeler, R. G. Symmetry Properties of Wave Functions in

Magnetic Crystals. *Phys. Rev.* **127**, 391 (1962).

22. Li, K. et al. Dirac and Nodal Line Magnons in Three-Dimensional Antiferromagnets. *Phys. Rev. Lett.* **119**, 247202 (2017).
23. Karaki, M. J. et al. An efficient material search for room-temperature topological magnons. *Sci. Adv.* **9**, eade7731 (2023).
24. Corticelli, A., Moessner, R. & McClarty, P. A. Identifying and Constructing Complex Magnon Band Topology. *Phys. Rev. Lett.* **130**, 206702 (2023).
25. Dzyaloshinsky, I. A thermodynamic theory of "weak" ferromagnetism of antiferromagnetics. *J. Phys. Chem. Solids* **4**, 241 (1958).
26. Moriya, T. Anisotropic superexchange interaction and weak ferromagnetism. *Phys. Rev.* **120**, 91 (1960).
27. Brinkman, W. Magnetic Symmetry and Spin Waves. *J. Appl. Phys.* **38**, 939 (1967).
28. Brinkman, W. F. & Elliott, R. J. Theory of spin-space groups. *Proc. R. Soc. A* **294**, 343 (1966).
29. Liu, P. et al. Spin-Group Symmetry in Magnetic Materials with Negligible Spin-Orbit Coupling. *Phys. Rev. X* **12**, 021016 (2022).
30. Corticelli, A., Moessner, R. & McClarty, P. A. Spin-space groups and magnon band topology. *Phys. Rev. B* **105**, 064430 (2022).
31. Yao, W. et al. Topological spin excitations in a three-dimensional antiferromagnet. *Nat. Phys.* **14**, 1011 (2018).
32. Scheie, A. et al. Dirac Magnons, Nodal Lines, and Nodal Plane in Elemental Gadolinium. *Phys. Rev. Lett.* **128**, 097201 (2022).
33. Scheie, A. et al. Spin-exchange Hamiltonian and topological degeneracies in elemental gadolinium. *Phys. Rev. B* **105**, 104402 (2022).
34. Zelezny, J., Zhang, Y., Felser, C. & Yan, B. Spin-Polarized Current in Noncollinear

Antiferromagnets. *Phys. Rev. Lett.* **119**, 187204 (2017).

35. Liu, P., Zhang, A., Han, J. & Liu, Q. Chiral Dirac-like fermion in spin-orbit-free antiferromagnetic semimetals. *The Innovation* **3**, 100343 (2022).
36. Yang, J., Liu, Z.-X. & Fang, C. Symmetry invariants in magnetically ordered systems having weak spin-orbit. Preprint at *arXiv* https://arxiv.org/abs/2105.12738 (2022).
37. Šmejkal, L., Sinova, J. & Jungwirth, T. Beyond Conventional Ferromagnetism and Antiferromagnetism: A Phase with Nonrelativistic Spin and Crystal Rotation Symmetry. *Phys. Rev. X* **12**, 031042 (2022).
38. Ma, H.-Y. et al. Multifunctional antiferromagnetic materials with giant piezomagnetism and noncollinear spin current. *Nat. Commun.* **12**, 2846 (2021).
39. Šmejkal, L., Sinova, J. & Jungwirth, T. Emerging Research Landscape of Altermagnetism. *Phys. Rev. X* **12**, 040501 (2022).
40. Hayami, S., Yanagi, Y. & Kusunose, H. Momentum-Dependent Spin Splitting by Collinear Antiferromagnetic Ordering. *J. Phys. Soc. Jpn.* **88**, 123702 (2019).
41. Yuan, L.-D. et al. Giant momentum-dependent spin splitting in centrosymmetric low-*Z* antiferromagnets. *Phys. Rev. B* **102**, 014422 (2020).
42. Guo, P. J. et al. Eightfold Degenerate Fermions in Two Dimensions. *Phys. Rev. Lett.* **127**, 176401 (2021).
43. Shao, D. F. et al. Spin-neutral currents for spintronics. *Nat. Commun.* **12**, 7061 (2021).
44. Bai, H. et al. Observation of Spin Splitting Torque in a Collinear Antiferromagnet $RuO_2$. *Phys. Rev. Lett.* **128**, 197202 (2022).
45. Bai, H. et al. Efficient Spin-to-Charge Conversion via Altermagnetic Spin Splitting Effect in Antiferromagnet $RuO_2$. *Phys. Rev. Lett.* **130**, 216701 (2023).
46. Bose, A. et al. Tilted spin current generated by the collinear antiferromagnet

ruthenium dioxide. *Nat. Electron.* **5**, 267 (2022).

47. Feng, Z. et al. An anomalous Hall effect in altermagnetic ruthenium dioxide. *Nat. Electron.* **5**, 735 (2022).

48. Shao, D. F. et al. Neel Spin Currents in Antiferromagnets. *Phys. Rev. Lett.* **130**, 216702 (2023).

49. Zhu, Y.-P. et al. Observation of plaid-like spin splitting in a noncoplanar antiferromagnet. *Nature* **626**, 523-528 (2024).

50. Krempaský, J. et al. Altermagnetic lifting of Kramers spin degeneracy. *Nature* **626**, 517-522 (2024).

51. Xiao, Z. et al. Spin space groups: full classification and applications. *Phys. Rev. X* **14**, 031037 (2024).

52. Chen, X. et al. Enumeration and representation theory of spin space groups. *Phys. Rev. X* **14**, 031038 (2024).

53. Jiang, Y. et al. Enumeration of spin-space groups: Toward a complete description of symmetries of magnetic orders. *Phys. Rev. X* **14**, 031039 (2024).

54. Ghorashi, S. A. A., Hughes, T. L. & Cano, J. Altermagnetic routes to Majorana modes in zero net magnetization. *Phys. Rev. Lett.* **133**, 106601 (2024).

55. Gallego, S. V. et al. MAGNDATA: towards a database of magnetic structures. I. The commensurate case. *J. Appl. Crystallogr.* **49**, 1750 (2016).

56. https://findspingroup.com/.

57. Litvin, D. B. Spin point groups. *Acta Crystallogr. A* **33**, 279 (1977).

58. Zak, J. Band representations of space groups. *Phys. Rev. B* **26**, 3010 (1982).

59. Holstein, T. & Primakoff, H. Field Dependence of the Intrinsic Domain Magnetization of a Ferromagnet. *Phys. Rev.* **58**, 1098 (1940).

60. Li, J. et al. Designing light-element materials with large effective spin-orbit

coupling. *Nat. Commun.* **13**, 919 (2022).

61. Šmejkal, L. et al. Chiral magnons in altermagnetic $RuO_2$. *Phys. Rev. Lett.* **131**, 256703 (2023).
62. Pomjakushin, V. Y. et al. Room temperature antiferromagnetic order in superconducting $X_yFe_{2-x}Se_2$ (X = Rb, K): a neutron powder diffraction study. *J. Phys.: Condens. Matter* **23**, 156003 (2011).
63. Yu, W. et al. $^{77}$Se NMR study of the pairing symmetry and the spin dynamics in $K_yFe_{2-x}Se_2$. *Phys. Rev. Lett.* **106**, 197001 (2011).
64. Liu, Y. et al. Switching magnon chirality in artificial ferrimagnet. *Nat. Commun.* **13**, 1264 (2022).
65. Neumann, R. R., Mook, A., Henk, J. & Mertig, I. Orbital Magnetic Moment of Magnons. *Phys. Rev. Lett.* **125**, 117209 (2020).
66. Fishman, R. S., Gardner, J. S. & Okamoto, S. Orbital Angular Momentum of Magnons in Collinear Magnets. *Phys. Rev. Lett.* **129**, 167202 (2022).
67. Go, G., An, D., Lee, H.-W. & Kim, S. K. Magnon Orbital Nernst Effect in Honeycomb Antiferromagnets without Spin–Orbit Coupling. *Nano Lett.* **24**, 5968 (2024).
68. Yu, Z. M. et al. Encyclopedia of emergent particles in three-dimensional crystals. *Sci. Bull.* **67**, 375 (2022).
69. Liu, G.-B. et al. Systematic investigation of emergent particles in type-III magnetic space groups. *Phys. Rev. B* **105**, 085117 (2022).
70. Zhang, Z. et al. Encyclopedia of emergent particles in type-IV magnetic space groups. *Phys. Rev. B* **105**, 104426 (2022).
71. Nielsen, H. B. & Ninomiya, M. Absence of neutrinos on a lattice:(II). Intuitive topological proof. *Nuclear Physics B* **193**, 173-194 (1981).
72. Bao, S. et al. Direct observation of topological magnon polarons in a multiferroic

material. *Nat. Commun.* **14**, 6093 (2023).

73. Altland, A. & Zirnbauer, M. R. Nonstandard symmetry classes in mesoscopic normal-superconducting hybrid structures. *Phys. Rev. B* **55**, 1142 (1997).
74. Xu, Q.-R. et al. Squaring the fermion: The threefold way and the fate of zero modes. *Phys. Rev. B* **102**, 125127 (2020).
75. Zhang, M.-H., Xiao, L. & Yao, D.-X. Topological magnons in a collinear altermagnet. Preprint at *arXiv* https://arxiv.org/abs/2407.18379 (2024).
76. Pershoguba, S. S. et al. Dirac magnons in honeycomb ferromagnets. *Phys. Rev. X* **8**, 011010 (2018).
77. Zhang, L., Ren, J., Wang, J.-S. & Li, B. Topological magnon insulator in insulating ferromagnet. *Phys. Rev. B* **87**, 144101 (2013).
78. Mook, A., Henk, J. & Mertig, I. Edge states in topological magnon insulators. *Phys. Rev. B* **90**, 024412 (2014).
79. Owerre, S. A first theoretical realization of honeycomb topological magnon insulator. *J. Phys.: Condens. Matter* **28**, 386001 (2016).
80. Mook, A., Plekhanov, K., Klinovaja, J. & Loss, D. Interaction-stabilized topological magnon insulator in ferromagnets. *Phys. Rev. X* **11**, 021061 (2021).
81. Lu, Y.-S., Li, J.-L. & Wu, C.-T. Topological phase transitions of Dirac magnons in honeycomb ferromagnets. *Phys. Rev. Lett.* **127**, 217202 (2021).
82. Kresse, G. & Furthmüller, J. Efficient iterative schemes for *ab initio* total-energy calculations using a plane-wave basis set. *Phys. Rev. B* **54**, 11169 (1996).
83. Kresse, G. & Joubert, D. From ultrasoft pseudopotentials to the projector augmented-wave method. *Phys. Rev. B* **59**, 1758 (1999).
84. Perdew, J. P., Burke, K. & Ernzerhof, M. Generalized Gradient Approximation Made Simple. *Phys. Rev. Lett.* **77**, 3865 (1996).

85. Monkhorst, H. J. & Pack, J. D. Special points for Brillouin-zone integrations. *Phys. Rev. B* **13**, 5188 (1976).

86. Anisimov, V. I., Zaanen, J. & Andersen, O. K. Band theory and Mott insulators: Hubbard U instead of Stoner I. *Phys. Rev. B* **44**, 943 (1991).

87. Mostofi, A. A. et al. wannier90: A tool for obtaining maximally-localised Wannier functions. *Comput. Phys. Commun.* **178**, 685 (2008).

88. Marzari, N. et al. Maximally localized Wannier functions: Theory and applications. *Rev. Mod. Phys.* **84**, 1419 (2012).

89. He, X., Helbig, N., Verstraete, M. J. & Bousquet, E. TB2J: A python package for computing magnetic interaction parameters. *Comput. Phys. Commun.* **264**, 107938 (2021).

90. Yu, R. et al. Equivalent expression of $Z_2$ topological invariant for band insulators using the non-Abelian Berry connection. *Phys. Rev. B* **84**, 075119 (2011).

91. Berry, M. V. Quantal phase factors accompanying adiabatic changes. *Proc. R. Soc. A* **392**, 45 (1997).

**Table I. | Summary of unconventional magnons identified in prototypical candidate materials**. The superscript "†" represents that the nodal feature has been experimentally observed. The superscript "*" represents that the material is synthesized in experiment, but the magnetic structure is artificially imposed. "$C$–$n$" means that the topological charge $C$ is $|C| = n$. "$C_2$–$m$" means that the monopole charge $C_2 = (C_\uparrow - C_\downarrow)/2$ is $|C_2| = m$. DCP: duodecuple point; ONL: octuple nodal line; OP: octuple point; SP: sextuple point; QNPL: Quadruple nodal plane; DP: Dirac point; TP: triple point.

| SSG | $k$ | BRs at $k$ | Type | Material |
|---|---|---|---|---|
| 220.220.1.1<br>$(I^{1}\bar{4}^{1}3^{1}d^{\infty m}1)$ | H (1, 1, 1)<br>Γ (0, 0, 0) | $H_4^S H_5^S(6)$<br>$\Gamma_4^S(3)$,$\Gamma_5^S(3)$ | SP<br>TP | $Gd_4Sb_3$ (FM) |
| 227.227.1.1<br>$(F^{1}d^{1}\bar{3}^{1}m^{\infty m}1$ | Γ (0, 0, 0) | $\Gamma_5^{S,+}(3)$ | TP | $Lu_2V_2O_7^{\dagger}$ (FM) |
| 212.212.1.1<br>$(P^{1}4_3{}^{1}3^{1}2^{\infty m}1)$ | Γ (0, 0, 0) | $\Gamma_4^S(3)$,$\Gamma_5^S(3)$ | $C$–2 TP | $LiFe_5O_8$ (FIM) |
| 218.222.1.1<br>$(P^{\bar{1}}n^{\bar{1}}\bar{3}^{1}n^{\infty m}1)$ | R (1/2, 1/2, 1/2) | $R_4^S R_5^S(12)$ | DCP | $Pr_5Mo_3O_{16}^{*}$<br>(PT AFM) |
| 199.206.1.1<br>$(I^{\bar{1}}a^{\bar{1}}\bar{3}^{\infty m}1)$ | Γ (0, 0, 0)<br>H (1, 1, 1) | $\Gamma_4^S(6)$<br>$H_4^S(6)$ | $C_2$–2 SP | $Cu_3TeO_6^{\dagger}$<br>(PT AFM) |
| 159.190.1.1<br>$(P^{\bar{1}}\bar{6}^{\bar{1}}2^{1}c^{\infty m}1)$ | A (0, 0, 1/2) | $A_3^S A_3^S(8)$ | OP | FeS<br>(Altermagnet) |
| 62.63.2.1<br>$(P_B{}^{1}n^{1}m^{1}a^{\infty m}1)$ | Q (1/2, 1/2, $w$)<br>W ($u$, $v$, 1/2) | $Q_1^S Q_1^S(8)$<br>$W_1^S W_2^S(4)$ | ONL<br>QNPL | $Mn_5Si_3$<br>(Uτ AFM) |
| 19.18.2.1<br>$(P_c{}^{1}2_1{}^{1}2_1{}^{1}2_1{}^{\infty m}1)$ | R (1/2, 1/2, 1/2)<br>L (1/2, $v$, $w$)<br>N ($u$, 1/2, $w$)<br>W ($u$, $v$, 1/2) | $R_1^S R_1^S(8)$<br>$L_1 L_1(4)$<br>$N_1 N_1(4)$<br>$W_1 W_1(4)$ | $C$–4 OP<br>QNPL<br>net | $Fe_{0.35}NbS_2$<br>(Uτ AFM) |
| 195.197.2.1<br>$(P_I{}^{1}2^{1}3^{\infty m}1)$ | Γ (0,0,0)<br>Γ (0,0,0)<br>R (1/2,1/2,1/2) | $\Gamma_2^S \Gamma_3^S(4)$<br>$\Gamma_4^S(6)$<br>$R_4^S(6)$ | $C$–8 DP<br>$C$–4 SP<br>$C$–4 SP | $LaMn_3Cr_4O_{12}$<br>(Uτ AFM) |

**Table II. | Statistics of bilinear exchange interactions with SOC**. The second columns present the number of collinear magnets classified by the outermost shell of magnetic ions in the first column. The third to fifth columns evaluate the number of materials in which the magnitude of DM ($D$), exchange anisotropy ($J^{ani}$), and anisotropic symmetric (Γ) terms does not exceed 10% of that of Heisenberg term ($J$), respectively. The last column gives the number of $J$ dominated magnets, in which none of $|D|$, $|J^{ani}|$ and $|\Gamma|$ exceeds 10% $|J|$. Detailed statistical methods and exchange parameters can be found in Supplementary Information S6.

| Magnetic ions | Materials | < 10% $\|J\|$ | | | $J$ dominated |
|---|---|---|---|---|---|
| | | $\|D\|$ | $\|J^{ani}\|$ | $\|\Gamma\|$ | |
| 3d | 194 | 175 | 182 | 189 | 169 |
| 4d | 7 | 5 | 5 | 6 | 5 |
| 4f | 12 | 11 | 10 | 11 | 9 |
| 5d | 6 | 1 | 0 | 3 | 0 |
| 5f | 4 | 1 | 0 | 3 | 0 |
| Total | 223 | 193 | 197 | 212 | 183 |

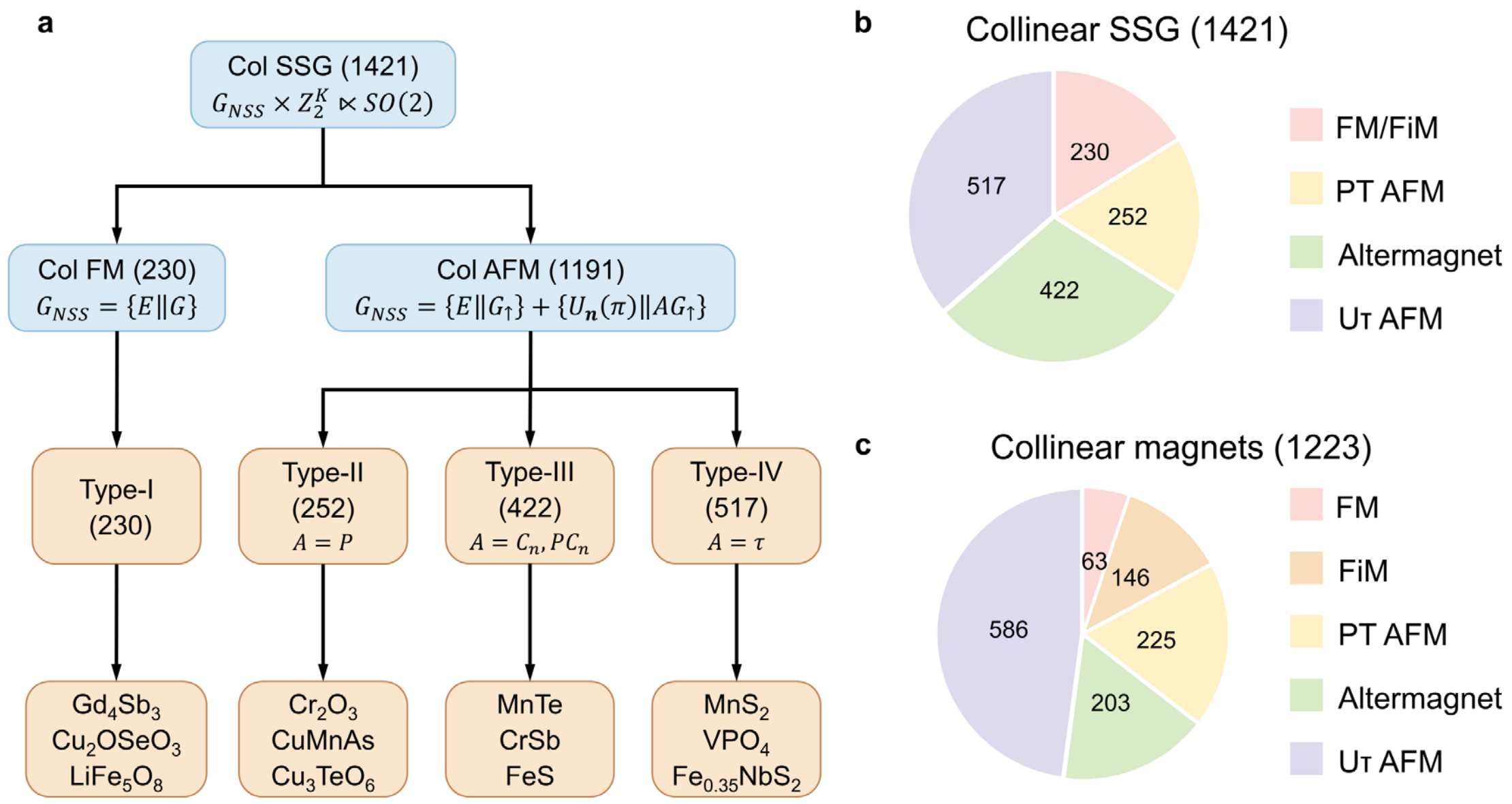

**Fig. 1 | Classification of 1421 collinear SSGs for collinear magnets. a,** For collinear FM, the spin space group $G_{NSS} = \{E\|G\}$ can be constructed from the space group $G$. For collinear AFM, the sublattice space group $G_\uparrow$, the space group $G$ of magnetic cell and the space operation $A$ that connects different magnetic sublattice are necessary for constructing $G_{NSS} = \{E\|G_\uparrow\} + \{U_{\boldsymbol{n}}(\pi)\|AG_\uparrow\}$. The last column lists the material candidates for each category. Col: collinear; $G_{NSS}$: nontrivial SSG; $\tau$: fractional lattice translation; $P$: spatial inversion; $C_n$: spatial rotations. **b**, Statistics of four types of collinear magnets in collinear SSGs. **c**, Statistics of four types of collinear magnets in the MAGNDATA database.

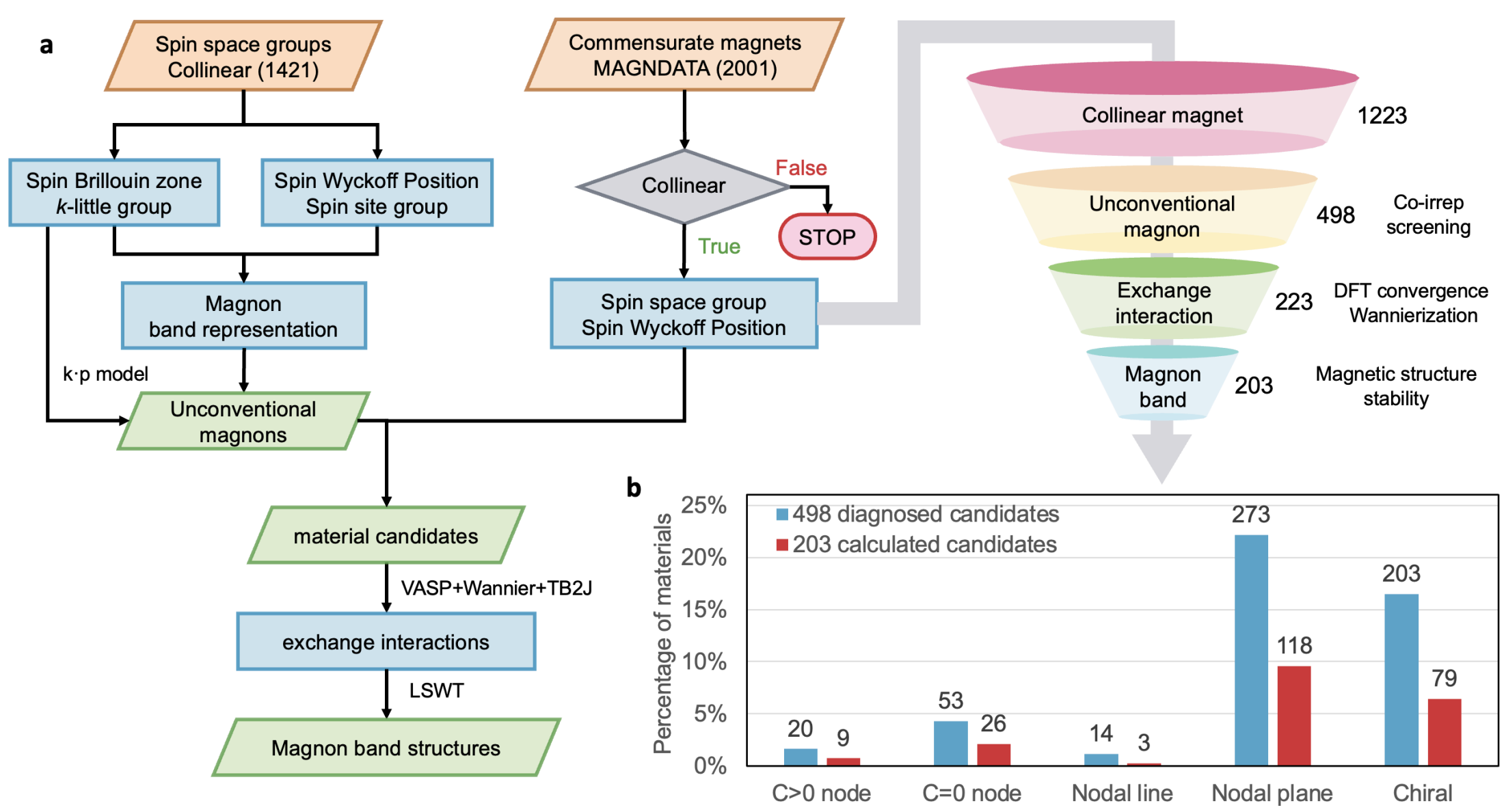

**Fig. 2 | Workflow of the construction of the unconventional magnon database. a,** First, 1223 collinear magnets are identified from the MAGNDATA database. Second, after the identification of the SSG, the co-irrep screenings are performed to obtain the material candidates hosting unconventional magnons. Third, the high-throughput calculations are performed to obtain all bilinear exchange interactions. Finally, unconventional magnon database with 203 magnon band structures and identified band representations are constructed. **b,** Statistics of different types of unconventional magnons in the collinear magnets from the MAGNDATA database.

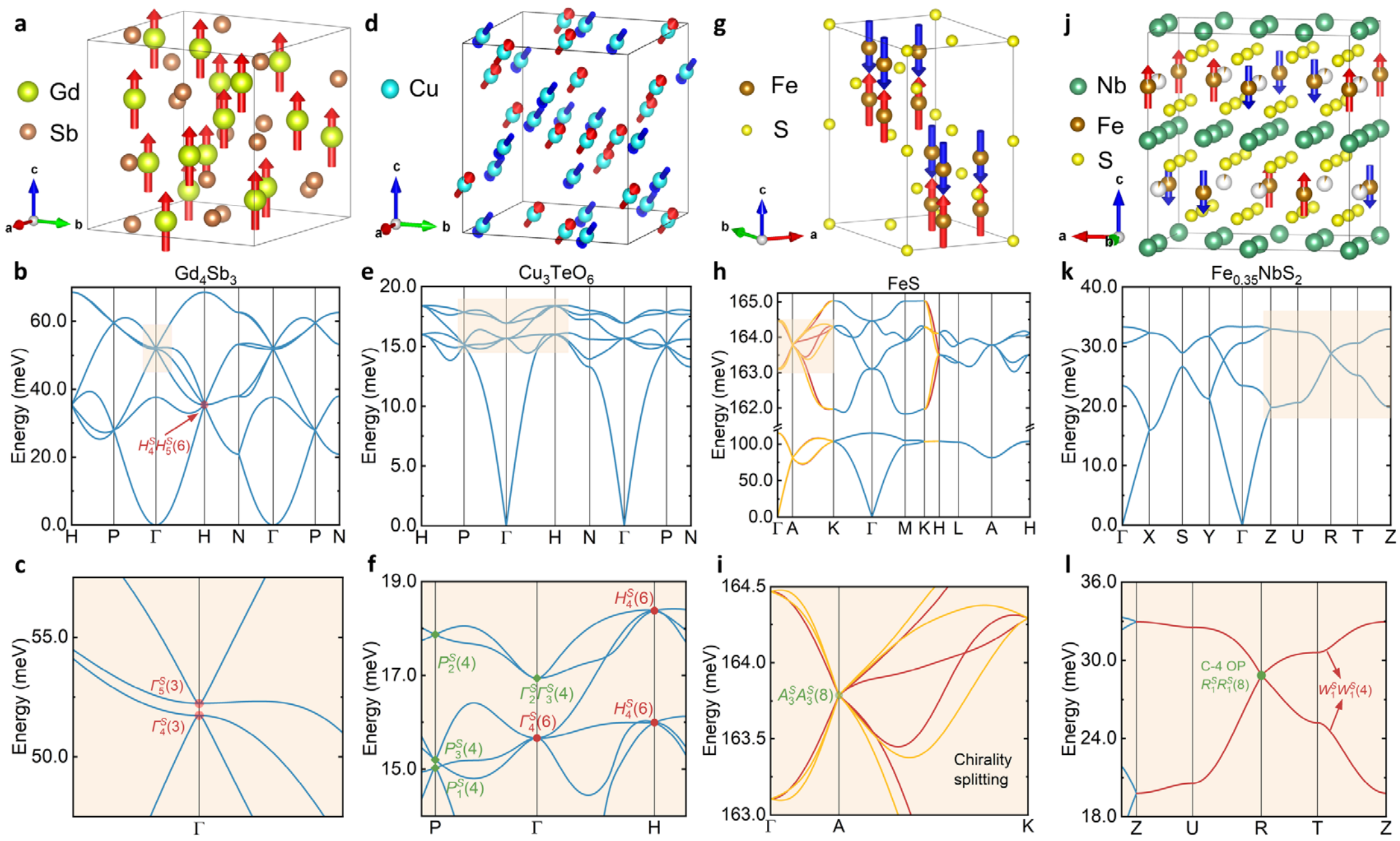

**Fig. 3 | Four candidates hosting unconventional magnons. a-c,** The collinear FM $Gd_4Sb_3$, which has six-, four- and three-fold band crossings. **d-f,** The collinear AFM $CuTe_3O_6$ with twelve-fold band crossing at R point. **g-i,** The altermagnet FeS, which has octuple degeneracy at A point and magnon chirality splitting along A-K line. The red and yellow lines represent the magnon chirality splitting bands with magnon spin S=1 and S=-1, respectively, while the blue lines represent spin degenerate bands. **j-l,** The collinear AFM $Fe_{0.35}NbS_2$ with C-4 octuple magnons at R point, and red lines stand for the quadruple Z-U-R-T nodal plane. Schematic of Brillouin zone for each candidate can be found in Supplementary Information S7.

# Supplementary Information for

# "Catalog of Unconventional Magnons in Collinear Magnets"

Xiaobing Chen[1,2], Yuntian Liu[1], Pengfei Liu[1], Yutong Yu[1], Jun Ren[1], Jiayu Li[1], Ao Zhang[1] and Qihang Liu[1,2,3*]

[1]*Department of Physics and Guangdong Basic Research Center of Excellence for Quantum Science, Southern University of Science and Technology, Shenzhen 518055, China*

[2]*Quantum Science Center of Guangdong–Hong Kong–Macao Greater Bay Area (Guangdong), Shenzhen 518045, China*

[3]*Guangdong Provincial Key Laboratory of Computational Science and Material Design, Southern University of Science and Technology, Shenzhen 518055, China*

[*]Email: liuqh@sustech.edu.cn

## S1. Introduction to collinear SSGs

## S1.1. Relation between MSGs and collinear SSGs

Magnetic space groups (MSGs) are constructed from space groups (SGs) by introducing the anti-symmetry operation $1'$ that reverses the spin. In MSGs, the magnetic moment is treated as an axial vector associated with the current loop symmetry[1]. Such treatment naturally corresponds to completely locked lattice space and spin space, corresponding to the presenece of SOC[2]. Thus, any operations in MSG can be interpreted as

$$g_M = f(r) \cdot f(S) \cdot A(S) \tag{1}$$

where $A$ is restricted to the identity or time reversal $T = 1'$. Take $g_M = 2'_{001}$ as an example, its effect upon magnetic ions is as follows: the atomic positions are transformed with the operation $f(r) = 2_{001}$, while the atomic magnetic moment transforms under $f(S) \cdot A(S) = 2_{001}T = m_{001}$.

In comparison, any operation in collinear SSG can be interpreted as:

$$g_S = f(r) \cdot A(S) \tag{2}$$

where $A$ is also restricted to the identity or time reversal $T = \bar{1}$. Note that we take the convention that denotes T by $1'$ in MSGs and by $\bar{1}$ in SSGs, respectively. Take $g_S = {}^{\bar{1}}2_{001} = \{T\|2_{001}\}$ as an example, its effect upon magnetic atoms is as follows: the atomic positions are transferred with the operation $f(r) = 2_{001}$, while the atomic magnetic moments change under the influence of $A(S) = T$. In summary, $h$, $h'$ in MSGs and ${}^{1}h$, ${}^{\bar{1}}h$ in SSGs can be expressed in the notations of SSGs by

$$h = \{f(h)\|h\}, \qquad h' = \{Tf(h)\|h\}, \qquad {}^{1}h = \{E\|h\}, \qquad {}^{\bar{1}}h = \{T\|h\} \tag{3}$$

where $f(h)$ stands for the proper rotation part of $h$.

Supplementary Table I. Four types of MSGs, where $H$ is an index-2 normal subgroup of space group $G$.

| MSG | form | number |
|---|---|---|
| Type-I | $M_I = G$ | 230 |
| Type-II | $M_{II} = G + TG$ | 230 |
| Type-III | $M_{III} = H + TAH$ | 674 |
| Type-IV | $M_{IV} = H + T\tau H$ | 517 |

Typically, the 1651 MSGs are often categorized into four types, i.e., 230 colorless groups (type-I), 230 grey groups (type-II), 674 black-white groups (type-III) with ordinary Bravais lattices and 517 black-white groups (type-IV) with black-white Bravais lattices, as listed in Supplementary Table I. Type-I MSGs are identical to ordinary SGs $G$. Type-II MSGs describing nonmagnetic solids can be constructed by the product of all symmetry operations in ordinary SGs with the time reversal $T$. Type-III and IV MSGs are constructed using a group $H$, which is an index-2 normal subgroups of $G$. The difference between these two types lies in the representatives of the coset decompositions of group $G$ using $H$. In type-III, it is an arbitrary rotation operation $A$, while in type-IV, it is a pure translation operation $\tau$.

Collinear magnets exhibit one-dimensional spin geometry, featured by only positive and negative directions. Consequently, for nontrivial SSGs describing collinear magnets, the symmetry on spin is restricted to identity operation (1) or time reversal ($\bar{1}$), and the subgroup $H$ represents the SG symmetry of magnetic sublattice. In this vein, $G_{\uparrow}$ is termed as sublattice group in collinear SSGs, which corresponds to the subgroup $H$ in MSGs. When describing a collinear ferromagnet (FM) or ferrimagnet (FiM) that possesses only a single magnetic sublattice (or symmetry-unrelated multiple magnetic sublattices), the nontrivial part of collinear SSGs can be written as $G_S^I = \{E\|G\}$, corresponding to 230 type-I collinear SSGs and exhibiting one-to-one correspondence with the 230 type-I MSGs. The full type-I SSGs describing collinear FMs/FiMs obey the form:

$$G_S^{FM} = (\{E\|G\}) \times G_{SO} \tag{4}$$

where the spin only group $G_{SO} = Z_2^K \ltimes SO(2)$.

For collinear antiferromagnet (AFM) with two symmetry-related magnetic sublattices with opposite magnetic moments, the nontrivial part of collinear SSGs can be constructed using the sublattice group $G_{\uparrow}$, which is an index-2 normal subgroup of $G$. As a result, the collinear nontrivial SSGs describing AFMs possess the form of $\{E\|G_{\uparrow}\} + \{T\|AG_{\uparrow}\}$, and thus exhibit a one-to-one correspondence to 1191 black-and-white MSGs as shown in Supplementary Table II. The collinear AFM SSGs possess the form:

$$G_S^{AFM} = (\{E\|G_{\uparrow}\} + \{T\|AG_{\uparrow}\}) \times G_{SO} \tag{5}$$

Thanks to the spin only group $G_{SO} = Z_2^K \ltimes SO(2)$ describing collinear magnets, where $Z_2^K = \{\{E\|E\}, \{TU_{\boldsymbol{n}}(\pi)\|E\}\}$, the $\{T\|AG_{\uparrow}\}$ part is equivalent to $\{U_{\boldsymbol{n}}(\pi)\|AG_{\uparrow}\}$ and $\{TU_{\boldsymbol{n}}(\pi)\|AG_{\uparrow}\}$. Since the construction of the representation theory requires a starting point of maximal unitary subgroups, we use $\{U_{\boldsymbol{n}}(\pi)\|AG_{\uparrow}\}$ in the main text.

To distinguish collinear SSGs with AFM-induced spin splitting, we further categorize the 674 collinear SSGs into 252 type-II collinear SSGs that exhibit PT symmetry, while the remaining 422 type-III collinear SSGs manifest altermagnetic spin splitting. The 517 type-IV collinear SSGs are in one-to-one correspondence with type-IV SSGs. We note that type-II MSGs do not have any corresponding SSGs since they do not describe magnetically ordered systems.

Supplementary Table II. Types of collinear SSGs. Here $A$ represents the real-space operation that connects the opposite-spin sublattice, while $\{E\|G_{\uparrow}\}$ represents the sublattice SG, $P, C_n, \tau$ represent inversion,

rotation and fractional translation in real space, respectively.

| Type | SSG form | $A$ | number | Correspondence |
|---|---|---|---|---|
| I | $G_S^I = \{E \Vert G\} \times G_{SO}$ | | 230 | Type-I MSGs |
| II | $G_S^{II} = (\{E \Vert G_\uparrow\} + \{T \Vert PG_\uparrow\}) \times G_{SO}$ | $P$ | 252 | Type-III MSGs |
| III | $G_S^{III} = (\{E \Vert G_\uparrow\} + \{T \Vert AG_\uparrow\}) \times G_{SO}$ | $C_n, PC_n$ | 422 | |
| IV | $G_S^{IV} = (\{E \Vert G_\uparrow\} + \{T \Vert \tau G_\uparrow\}) \times G_{SO}$ | $\tau$ | 517 | Type-IV MSGs |

Although the 1421 collinear SSGs and the 1421 magnetically ordered MSGs are in a one-to-one correspondence, due to the existence of the spin-only group, the quotient groups of collinear SSGs with respect to the translation group are still infinite groups, resulting in different irreducible co-representations between collinear SSGs and MSGs. IN contrast, the quotient groups of MSGs with respect to the translation group are finite groups. In addition, there are strike differences in the general position, Wyckoff position (WP), site symmetry group, and Brillouin zone. The main differences between them are outlined in Supplementary Table III, while a comprehensive explanation of the construction methods for these components can be found in the subsequent sections.

Supplementary Table III. The similarities and differences between the 1421 one-to-one corresponding collinear SSGs and MSGs in terms of general position, Wyckoff position, site symmetry group, Brillouin zone, band representation, and topological charge at high-symmetry points. SPG: spin point group; MPG: magnetic point group.

| | | Collinear SSGs | Type-I, III, IV MSGs |
|---|---|---|---|
| Translational quotient group | | infinite group | finite group |
| General position | atomic position | same | |
| | magnetic moment | only one component $\pm m_z$ | transforms as axial vector |
| Site symmetry group of magnetic ions | | 32 collinear SPGs | 31 MPGs |
| Wyckoff position of magnetic ions | | 6368 WPs | 9212 WPs |
| Brillouin zone | arithmetic crystal class | 24 (centrosymmetric) | 73 |
| Band representation (dimension) | nodal point | 12, 8, 6, 4, 3, 2, 1 | 6, 4, 3, 2, 1 (*) |
| | nodal line | 8, 4, 2 | 4, 2 |
| | nodal plane | 4, 2 | 2 |

(*) Sextuple and triple nodal point does not exist in MSGs describing collinear magnets.

For example, in collinear SSGs, the magnetic moment at any general position is constrained by $SO(2)$ to have at most only one degree of freedom, either up or down along the spin axis $z$, while in the MSGs, the magnetic moment is constrained by the proper part of spatial operations. Since any operations in spin space do not affect the atomic position in lattice space, there are a total of 12481 WPs in both the 1421 collinear SSGs and the 1421 type-I, III and IV MSGs. However, they differ in the site symmetry groups and WPs that support nonzero localized magnetic moments and magnon basis, which correspond to 32 collinear spin point groups and 31 magnetic point groups, respectively. Consequently, a total of 6368 and 9212 WPs for nonzero localized magnetic moments are obtained in collinear SSGs and MSGs, respectively. In terms of the Brillouin zone, the presence of $\{TU_{\boldsymbol{n}}(\pi)\|E\}$ is equivalent to inversion symmetry for k-points, so the Brillouin zone of collinear SSGs always belongs to 24 centrosymmetric arithmetic crystal classes, while the 1421 magnetic groups correspond to all 73 arithmetic crystal classes.

## S1.2. General positions

The general positions of a certain collinear SSG can be constructed using the international symbol. All general positions of 1421 collinear SSGs are listed on our online database FINDSPINGROUP[3].

For example, the type-II collinear SSG $I^{\bar{1}}a^{\bar{1}}\bar{3}^{\infty m}1$ (199.206.1.1) can be generated by $\{\bar{1}\|m_{100}|1/2,1/2,0\}$, $\{\bar{1}\|\bar{3}^{+}_{111}|0\}$ and $\{1\|1|1/2,1/2,1/2\}$. All Seitz symbols and their corresponding coordinations of $G_{NSS}$ are listed in Supplementary Table IV. Note that due to the presence of spin-only group ${}^{\infty m}1$, the magnetic moments are constrained as 0, 0, ±*z*, where the spin axis *z* is independent with the atomic coordinates *a*, *b*, *c*.

Supplementary Table IV. General Positions of collinear SSG $I^{\bar{1}}a^{\bar{1}}\bar{3}^{\infty m}1$ (199.206.1.1).

| Seitz | Coordination |
|---|---|
| $\{1\|1|0\}$ | *a*, *b*, *c*, 0, 0, *z* |
| $\{1\|2_{100}|0,0,1/2\}$ | *a*, −*b*, −*c*+1/2, 0, 0, *z* |
| $\{1\|2_{010}|1/2,0,0\}$ | −*a*+1/2, *b*, −*c*, 0, 0, *z* |
| $\{1\|2_{001}|0,1/2,0\}$ | −*a*, −*b*+1/2, *c*, 0, 0, *z* |
| $\{1\|3^{+}_{111}|0\}$ | *c*, *a*, *b*, 0, 0, *z* |
| $\{1\|3^{-}_{111}|0\}$ | *b*, *c*, *a*, 0, 0, *z* |
| $\{1\|3^{-}_{1\bar{1}\bar{1}}|1/2,0,0\}$ | −*b*+1/2, *c*, −*a*, 0, 0, *z* |
| $\{1\|3^{+}_{1\bar{1}\bar{1}}|0,1/2,0\}$ | −*c*, −*a*+1/2, *b*, 0, 0, *z* |
| $\{1\|3^{-}_{\bar{1}1\bar{1}}|0,1/2,0\}$ | −*b*, −*c*+1/2, *a*, 0, 0, *z* |
| $\{1\|3^{+}_{\bar{1}1\bar{1}}|0,0,1/2\}$ | *c*, −*a*, −*b*+1/2, 0, 0, *z* |
| $\{1\|3^{-}_{\bar{1}\bar{1}1}|0,0,1/2\}$ | *b*, −*c*, −*a*+1/2, 0, 0, *z* |

| $\{1\|3^{+}_{\bar{1}\bar{1}1}\|1/2,0,0\}$ | $-c+1/2, a, -b, 0, 0, z$ |
|---|---|
| $\{1\|1\|1/2,1/2,1/2\}$ | $a+1/2, b+1/2, c+1/2, 0, 0, z$ |
| $\{1\|2_{100}\|1/2,1/2,0\}$ | $a+1/2, -b+1/2, -c, 0, 0, z$ |
| $\{1\|2_{010}\|0,1/2,1/2\}$ | $-a, b+1/2, -c+1/2, 0, 0, z$ |
| $\{1\|2_{001}\|1/2,0,1/2\}$ | $-a+1/2, -b, c+1/2, 0, 0, z$ |
| $\{1\|3^{+}_{111}\|1/2,1/2,1/2\}$ | $c+1/2, a+1/2, b+1/2, 0, 0, z$ |
| $\{1\|3^{-}_{111}\|1/2,1/2,1/2\}$ | $b+1/2, c+1/2, a+1/2, 0, 0, z$ |
| $\{1\|3^{-}_{1\bar{1}\bar{1}}\|0,1/2,1/2\}$ | $-b, c+1/2, -a+1/2, 0, 0, z$ |
| $\{1\|3^{+}_{1\bar{1}\bar{1}}\|1/2,0,1/2\}$ | $-c+1/2, -a, b+1/2, 0, 0, z$ |
| $\{1\|3^{-}_{\bar{1}1\bar{1}}\|1/2,0,1/2\}$ | $-b+1/2, -c, a+1/2, 0, 0, z$ |
| $\{1\|3^{+}_{\bar{1}1\bar{1}}\|1/2,1/2,0\}$ | $c+1/2, -a+1/2, -b, 0, 0, z$ |
| $\{1\|3^{-}_{\bar{1}\bar{1}1}\|1/2,1/2,0\}$ | $b+1/2, -c+1/2, -a, 0, 0, z$ |
| $\{1\|3^{+}_{\bar{1}\bar{1}1}\|0,1/2,1/2\}$ | $-c, a+1/2, -b+1/2, 0, 0, z$ |
| $\{\bar{1}\|\bar{1}\|0\}$ | $-a, -b, -c, 0, 0, -z$ |
| $\{\bar{1}\|m_{100}\|0,0,1/2\}$ | $-a, b, c+1/2, 0, 0, -z$ |
| $\{\bar{1}\|m_{010}\|1/2,0,0\}$ | $a+1/2, -b, c, 0, 0, -z$ |
| $\{\bar{1}\|m_{001}\|0,1/2,0\}$ | $a, b+1/2, -c, 0, 0, -z$ |
| $\{\bar{1}\|\bar{3}^{+}_{111}\|0\}$ | $-c, -a, -b, 0, 0, -z$ |
| $\{\bar{1}\|\bar{3}^{-}_{111}\|0\}$ | $-b, -c, -a, 0, 0, -z$ |
| $\{\bar{1}\|\bar{3}^{-}_{1\bar{1}\bar{1}}\|1/2,0,0\}$ | $b+1/2, -c, a, 0, 0, -z$ |
| $\{\bar{1}\|\bar{3}^{+}_{1\bar{1}\bar{1}}\|0,1/2,0\}$ | $c, a+1/2, -b, 0, 0, -z$ |
| $\{\bar{1}\|\bar{3}^{-}_{\bar{1}1\bar{1}}\|0,1/2,0\}$ | $b, c+1/2, -a, 0, 0, -z$ |
| $\{\bar{1}\|\bar{3}^{+}_{\bar{1}1\bar{1}}\|0,0,1/2\}$ | $-c, a, b+1/2, 0, 0, -z$ |
| $\{\bar{1}\|\bar{3}^{-}_{\bar{1}\bar{1}1}\|0,0,1/2\}$ | $-b, c, a+1/2, 0, 0, -z$ |
| $\{\bar{1}\|\bar{3}^{+}_{\bar{1}\bar{1}1}\|1/2,0,0\}$ | $c+1/2, -a, b, 0, 0, -z$ |
| $\{\bar{1}\|\bar{1}\|1/2,1/2,1/2\}$ | $-a+1/2, -b+1/2, -c+1/2, 0, 0, -z$ |
| $\{\bar{1}\|m_{100}\|1/2,1/2,0\}$ | $-a+1/2, b+1/2, c, 0, 0, -z$ |
| $\{\bar{1}\|m_{010}\|0,1/2,1/2\}$ | $a, -b+1/2, c+1/2, 0, 0, -z$ |
| $\{\bar{1}\|m_{001}\|1/2,0,1/2\}$ | $a+1/2, b, -c+1/2, 0, 0, -z$ |
| $\{\bar{1}\|\bar{3}^{+}_{111}\|1/2,1/2,1/2\}$ | $-c+1/2, -a+1/2, -b+1/2, 0, 0, -z$ |

| $\{\bar{1}\|\bar{3}^{-}_{\bar{1}11}\|1/2,1/2,1/2\}$ | $-b+1/2$, $-c+1/2$, $-a+1/2$, 0, 0, $-z$ |
|---|---|
| $\{\bar{1}\|\bar{3}^{-}_{1\bar{1}\bar{1}}\|0,1/2,1/2\}$ | $b$, $-c+1/2$, $a+1/2$, 0, 0, $-z$ |
| $\{\bar{1}\|\bar{3}^{+}_{1\bar{1}\bar{1}}\|1/2,0,1/2\}$ | $c+1/2$, $a$, $-b+1/2$, 0, 0, $-z$ |
| $\{\bar{1}\|\bar{3}^{-}_{\bar{1}1\bar{1}}\|1/2,0,1/2\}$ | $b+1/2$, $c$, $-a+1/2$, 0, 0, $-z$ |
| $\{\bar{1}\|\bar{3}^{+}_{\bar{1}1\bar{1}}\|1/2,1/2,0\}$ | $-c+1/2$, $a+1/2$, $b$, 0, 0, $-z$ |
| $\{\bar{1}\|\bar{3}^{-}_{\bar{1}\bar{1}1}\|1/2,1/2,0\}$ | $-b+1/2$, $c+1/2$, $a$, 0, 0, $-z$ |
| $\{\bar{1}\|\bar{3}^{+}_{\bar{1}\bar{1}1}\|0,1/2,1/2\}$ | $c$, $-a+1/2$, $b+1/2$, 0, 0, $-z$ |

## S1.3. Spin Wyckoff positions and Spin site groups

In this section, we will discuss the position-space action of the symmetries of the collinear SSGs. First, we define the spin site-symmetry group (abbreviated as spin site group) $G_S^{\boldsymbol{q}}$ in a crystal that leaves the point **q** in a crystal invariant under a collinear SSG, which comprises operations $g_i$:

$$g_i = \{g_{s_i} \| R_i | \tau_i\}, \tag{6}$$

where $g_{s_i}$ represents identity $E = 1$ or time reversal $T = \bar{1}$ in nontrivial SSG, while $g_{s_i}$ represents $TU_n(\pi)$, $U_z(\phi)$ or the combination of the aforementioned two operations in spin-only group of collinear SSG; $R_i$ denotes a real-space symmetry operation that is either the identity, inversion, a rotation or a rotation-inversion; and $\tau_i$ is a real-space pure translation.

The action of $g_i$ on $\boldsymbol{q}$ can be expressed as:

$$g_i \boldsymbol{q} = R_i \boldsymbol{q} + \tau_i, \tag{7}$$

where only $R_i$ and $\tau_i$ act on $\boldsymbol{q}$ due to the invariance of spatial coordinates under any spin-space operations (including $\bar{1}$ and any proper rotation symmetries in spin space).

Those operations $g_i^q$ that leave the point invariant form $G_S^{\boldsymbol{q}}$, i.e., $g_i^q \boldsymbol{q} = \boldsymbol{q}$. In additional, the elements $g_\alpha \in G_S$ but $g_\alpha \notin G_S^{\boldsymbol{q}}$ in magnetic primitive cell form the cosets $G_S/G_S^{\boldsymbol{q}}$, which is the set of points $\{\boldsymbol{q}_\alpha = g_\alpha \boldsymbol{q} | g_\alpha \notin G_S^{\boldsymbol{q}}\}$, where $\alpha = 1,2,\ldots,n$. These points $\boldsymbol{q}_\alpha$ is also called as Wyckoff positions, and $n$ represents the multiplicity. Therefore, the spin site group $G_S^{\boldsymbol{q}}$ is a finite subgroup of the SSG $S$, i.e., $G_S^{\boldsymbol{q}} \subset G_S$. For collinear SSGs, $G_S^{\boldsymbol{q}}$ is necessarily isomorphic to one of the 90 collinear spin point groups. The collinear spin point groups divide into 32 FM type-I and 58 AFM type-II collinear spin point groups, which also corresponds to 32 colorless type-I and 58 black-and-white type-III magnetic point groups.

Note that the nontrivial group of collinear SSG is isomorphic to a certain MSG and the spin-only group does not influence the multiplicities and the atomic positions of Wyckoff positions, while the constrains on magnetic moments are changed at Wyckoff positions. However, since the collinear SSG and its corresponding MSG varies in coupled operation in spin space, the site symmetry groups and the magnetic moment coordinates are different.

For type-I collinear SSGs describing FMs/FiMs, magnetic moments of all Wyckoff positions are constrained as (0, 0, *z*). For type-II, III and IV collinear SSGs describing AFMs with the form of $G_S^{AFM} = (\{E\|G_\uparrow\} + \{T\|AG_\uparrow\}) \times G_{SO}$:

1. If $G_S^{\boldsymbol{q}}$ contains any operation in $\{T\|AG_\uparrow\}$, magnetic moments of this set of points $\boldsymbol{q}_\alpha$ are also zero. These sites do not support the onsite local magnetic moment and the magnon excitation.
2. If $G_S^{\boldsymbol{q}}$ only contains operation in $\{E\|G_\uparrow\}$, set the magnetic moment of $\boldsymbol{q}$ as (0, 0, *z*). The magnetic moment of the set of points $\{\boldsymbol{q}_\alpha = g_\alpha \boldsymbol{q} | g_\alpha \notin G_S^{\boldsymbol{q}}\}$ can be easily determined as (0, 0, *z*) for $g_\alpha \in \{E\|G_\uparrow\}$, while (0, 0, -*z*) for $g_\alpha \in \{T\|AG_\uparrow\}$.

The spin Wyckoff positions of the type-II collinear SSG $G_S = I^{\bar{1}}a^{\bar{1}}\bar{3}^{\infty m}1$ (199.206.1.1) are shown in Supplementary Table V.

Supplementary Table V. Spin Wyckoff positions of $G_S = I^{\bar{1}}a^{\bar{1}}\bar{3}^{\infty m}1$ (199.206.1.1).

| WP | Spin site group $G_S^{\boldsymbol{q}}$ | Coordinates |
|---|---|---|
| | | (0, 0, 0) + (1/2, 1/2, 1/2) |
| 48e | ${}^{1}1^{\infty m}1$ | (*a*,*b*,*c* \| 0,0,*z*), (*a*,-*b*,-*c*+1/2 \| 0,0,*z*),<br>(-*a*+1/2,*b*,-*c* \| 0,0,*z*), (-*a*,-*b*+1/2,*c* \| 0,0,*z*),<br>(*c*,*a*,*b* \| 0,0,*z*),(*b*,*c*,*a* \| 0,0,*z*),<br>(-*b*+1/2,*c*,-*a* \| 0,0,*z*), (-*c*,-*a*+1/2,*b* \| 0,0,*z*),<br>(-*b*,-*c*+1/2,*a* \| 0,0,*z*), (*c*,-*a*,-*b*+1/2 \| 0,0,*z*),<br>(*b*,-*c*,-*a*+1/2 \| 0,0,*z*), (-*c*+1/2,*a*,-*b* \| 0,0,*z*),<br>(-*a*,-*b*,-*c* \| 0,0,-*z*), (-*a*,*b*,*c*+1/2 \| 0,0,-*z*),<br>(*a*+1/2,-*b*,*c* \| 0,0,-*z*), (*a*,*b*+1/2,-*c* \| 0,0,-*z*), |
| | | (-*c*,-*a*,-*b* \| 0,0,-*z*), (-*b*,-*c*,-*a* \| 0,0,-*z*),<br>(*b*+1/2,-*c*,*a* \| 0,0,-*z*), (*c*,*a*+1/2,-*b* \| 0,0,-*z*),<br>(*b*,*c*+1/2,-*a* \| 0,0,-*z*), (-*c*,*a*,*b*+1/2 \| 0,0,-*z*),<br>(-*b*,*c*,*a*+1/2 \| 0,0,-*z*), (*c*+1/2,-*a*,*b* \| 0,0,-*z*) |
| 24d | ${}^{1}2^{\infty m}1$ | (*a*,0,1/4 \| 0,0,*z*), (-*a*+1/2,0,3/4 \| 0,0,*z*),<br>(1/4,*a*,0 \| 0,0,*z*), (0,1/4,*a* \| 0,0,*z*),<br>(0,3/4,-*a*+1/2 \| 0,0,*z*), (1/4,-*a*,1/2 \| 0,0,*z*),<br>(-*a*,0,3/4 \| 0,0,-*z*), (*a*+1/2,0,1/4 \| 0,0,-*z*),<br>(3/4,-*a*,0 \| 0,0,-*z*), (0,3/4,-*a* \| 0,0,-*z*),<br>(0,1/4,*a*+1/2 \| 0,0,-*z*), (3/4,*a*,1/2 \| 0,0,-*z*) |
| 16c | ${}^{1}3^{\infty m}1$ | (*a*,*a*,*a* \| 0,0,*z*), (*a*,-*a*,-*a*+1/2 \| 0,0,*z*),<br>(-*a*+1/2,*a*,-*a* \| 0,0,*z*), (-*a*,-*a*+1/2,*a* \| 0,0,*z*),<br>(-*a*,-*a*,-*a* \| 0,0,-*z*), (-*a*,*a*,*a*+1/2 \| 0,0,-*z*),<br>(*a*+1/2,-*a*,*a* \| 0,0,-*z*), (*a*,*a*+1/2,-*a* \| 0,0,-*z*) |
| 8b | ${}^{\bar{1}}\bar{3}^{\infty m}1$ | (1/4,1/4,1/4 \| 0,0,0), (1/4,3/4,1/4 \| 0,0,0),<br>(1/4,1/4,3/4 \| 0,0,0), (3/4,1/4,1/4 \| 0,0,0) |
| 8a | ${}^{\bar{1}}\bar{3}^{\infty m}1$ | (0,0,0 \| 0,0,0), (0,0,1/2 \| 0,0,0),<br>(1/2,0,0 \| 0,0,0), (0,1/2,0 \| 0,0,0) |

To construct magnon band representations, the first step is to identify the possible spin site groups of all collinear SSGs, where the on-site magnetization $S_z$ transforms as the total symmetric irrep of the spin site groups. This constraint reduces the number of possibile spin site groups to 32 collinear FM spin point groups out of 90 collinear spin point groups. In comparison, 31 of 122 magnetic point groups are compatible with magnetic ordered systems. As a result, in type-I to IV collinear SSGs, a total of 6368 out of 12481 spin Wyckoff positions with local magnetic moment are obtained, specially 1731 out of 1731, 981 out of 2641, 1458 out of 3314 and 2198 out of 4795 spin Wyckoff positions, respectively. On the other hand, in type-I, III and IV MSGs, a total of 9212 out of 12481 magnetic Wyckoff positions with local magnetic moment are obtained, specifically 1210 out of 1731, 4337 out of 5955, 3665 out of 4795 magnetic Wyckoff positions, respectively. The coordinations, multiplicities, and site symmetry groups of the Wyckoff positions of 1421 collinear SSGs are listed on our online database FINDSPINGROUP[3].

## S1.4. Spin Brillouin zone and momentum stars

In this section, we will establish the momentum-space description of the collinear SSGs. In reciprocal space, coordinates are indexed by crystal momentum **k**, and the shapes of the reciprocal cells (i.e., Brillouin zones) are determined by the reciprocal lattice vectors. To begin, we review the momentum-space description method that used in literature for space groups and magnetic space groups.

Supplementary Table VI. Auxiliary space group $G^A$ for 1651 magnetic space groups in identifying the symmetry points in the magnetic Brillouin zone. Here the space group $H$ is either a *translationengleiche* or a *klassengleiche* subgroup of $G = H + AH$ for type-III or IV magnetic space groups.

| MSG | form | auxiliary space group $G^A$ |
|---|---|---|
| Type-I | $G_M^I = G$ | $G$ |
| Type-II | $G_M^{II} = G + TG$ | $\begin{cases} G + PG & P \notin G \\ G & P \in G \end{cases}$ |
| Type-III | $G_M^{III} = H + TAH$ | $\begin{cases} H + PAH & P \notin H \text{ and } P \notin AH \\ H & P \notin H \text{ but } P \in AH \\ G & P \in H \text{ and } P \in AH \end{cases}$ |
| Type-IV | $G_M^{IV} = H + T\tau H$ | $\begin{cases} H + P(G - H) & P \notin H \\ H & P \in H \end{cases}$ |

The classification and enumeration of the translation strata of the irreducible representations of 230 space groups was initially solved by Wintgen[4], where the auxiliary space group was adopted to identify all symmetry points in the Brillouin zone of 230 space groups. The reciprocal space group always belongs to one of 73 types of symmorphic space groups, or 73 arithmetic crystal classes. Subsequently, Angelova generalized the Wintgen's method for magnetic space groups with the assistance of an auxiliary non-magnetic space group[5]. This approach enables the identification of symmetry points in the Brillouin zone of magnetic crystals (magnetic Brillouin zone). First, we review the auxiliary non-magnetic space group that proposed by Angelova. According to Angelova's method, time reversal $T$ is analogous to space inversion $P$ in the origin due to $T\mathbf{k} = P\mathbf{k} = -\mathbf{k}$. As a consequence, the symmetry points in the magnetic Brillouin zone correspond to those in a non-magnetic auxiliary space group $G^A$, which can be generateed for any magnetic space group $M$ by substituting time reversal with space inversion in the origin as shown in Supplementary Table VI.

Supplementary Table VII. Auxiliary space group $G^A$ for 1421 collinear SSGs in identifying the symmetry points in the spin Brillouin zone.

| Type | SSG form | $A$ | auxiliary space group $G^A$ |
|---|---|---|---|
| I | $G_S^I = \{E\|G\} \times G_{SO}$ | | $\begin{cases} G+PG & P \notin G \\ G & P \in G \end{cases}$ |
| II | $G_S^{II} = (\{E\|G_\uparrow\} + \{T\|PG_\uparrow\}) \times G_{SO}$ | $P$ | $G$ |
| III | $G_S^{II} = (\{E\|G_\uparrow\} + \{T\|AG_\uparrow\}) \times G_{SO}$ | $C_n, PC_n$ | $\begin{cases} G+PG & P \notin G_\uparrow \\ G & P \in G_\uparrow \end{cases}$ |
| IV | $G_S^{IV} = (\{E\|G_\uparrow\} + \{T\|\tau G_\uparrow\}) \times G_{SO}$ | $\tau$ | $\begin{cases} H+PH & P \notin G_\uparrow \\ H & P \in G_\uparrow \end{cases}$ |

Following Angelova's method, a non-magnetic auxiliary space group $G^A$ can be generated for any collinear SSG by substituting the improper spin rotation with space inversion $P$ and the proper spin rotation with the identity $E$, respectively (shown in Supplementary Table VII). It is important to note that the $Z_2^K$ in collinear SSG also affect the choose of auxiliary space group $G^A$ because $Z_2^K$ falls into $P$ in $G^A$. As a consequence, one can map each of the 1421 collinear SSGs into one of the 24 centrosymmetric symmorphic space groups, rather than 73 symmorphic space groups for MSGs. In Supplementary Table VIII, we list all 24 types of spin Brillouin zones used in collinear SSGs and all maximal wavevectors. For this work, we have specifically implemented the spin Brillouin zone and k-little cogroups on the online database FINDSPINGROUP[3], through which users can access the momentum stars and the spin Brillouin zones of the collinear SSGs. This method can be also generalized to construct the spin Brillouin zone of any SSGs.

Supplementary Table VIII. Spin Brillouin zones and the k-vectors for 24 centrosymmetric symmorphic space groups $G^A$ (its arithmetic crystal class[6]) used in determining the spin Brillouin zone of collinear SSGs.

| $G^A$ | Symbol | Brillouin zone | k-vector |
|---|---|---|---|
| $P\bar{1}$<br>(2)<br>$[\bar{1}P]$ | a1 | | Γ:(0,0,0), R:(1/2,1/2,1/2), T:(0,1/2,1/2),<br>U:(1/2,0,1/2), V:(1/2,1/2,0), X:(1/2,0,0),<br>Y:(0,1/2,0), Z:(0,0,1/2), GP:(*u*,*v*,*w*) |

| $P2/m$<br>(10)<br>$[2/mP]$ | m1 | | A:(1/2,0,1/2), B:(0,0,1/2), C:(1/2,1/2,0), D:(0,1/2,1/2), E:(1/2,1/2,1/2), Γ:(0,0,0), Y:(1/2,0,0), Z:(0,1/2,0), F:($u$,0,$w$), G:($u$,1/2,$w$), Λ:(0,$v$,0), U:(1/2,$v$,1/2), V:(0,$v$,1/2), W:(1/2,$v$,0), GP:($u$,$v$,$w$) |
|---|---|---|---|
| $C2/m$<br>(12)<br>$[2/mC]$ | m2 | | Γ:(0,0,0), M:(0,1,1/2), Y:(0,1,0), L:(1/2,1/2,1/2), A:(0, 0, 1/2), V:(1/2,1/2,0), B:($u$,0,$w$), Λ:(0,$v$,0), U:(0,$v$,1/2), GP:($u$,$v$,$w$) |
| $Pmmm$<br>(47)<br>$[mmmP]$ | o1 | | Γ:(0,0,0), R:(1/2,1/2,1/2), S:(1/2,1/2,0), T:(0,1/2,1/2), U:(1/2,0,1/2), X:(1/2,0,0), Y:(0,1/2,0), Z:(0,0,1/2), A:($u$,0,1/2), B:(0,$v$,1/2), C:($u$,1/2,0), D:(1/2,$v$,0), Δ:(0,$v$,0), E:($u$,1/2,1/2), G:(1/2,0,$w$), H:(0,1/2,$w$), Λ:(0,0,$w$), P:(1/2,$v$,1/2), Q:(1/2,1/2,$w$), Σ:($u$,0,0), K:(0,$v$,$w$), L:(1/2,$v$,$w$), M:($u$,0,$w$), N:($u$,1/2,$w$), V:($u$,$v$,0), W:($u$,$v$,1/2), GP:($u$,$v$,$w$) |
| $Cmmm$<br>(65)<br>$[mmmC]$ | o2C | | Γ:(0,0,0), T:(1,0,1/2), Y:(1,0,0), Z:(0,0,1/2), R:(1/2,1/2,1/2), S:(1/2,1/2,0), A:($u$,0,1/2), B:(0,$v$,1/2), Δ:(0,$v$,0), H:(1,0,$w$), Λ:(0,0,$w$), Σ:($u$,0,0), D:(1/2,1/2,$w$), K:(0,$v$,$w$), M:($u$,0,$w$), P:($u$,$v$,0), Q:($u$,$v$,1/2), GP:($u$,$v$,$w$) |
| $Ammm$<br>(65)<br>$[mmmA]$ | o2A | | Γ:(0,0,0), T:(1/2,0,1), Y:(0,0,1), Z:(1/2,0,0), R:(1/2,1/2,1/2), S:(0,1/2,1/2), A:(1/2,0,$w$), B:(1/2,$v$,0), Δ:(0,$v$,0), H:($u$,0,1), Λ:($u$,0,0), Σ:(0,0,$w$), D:($u$,1/2,1/2), K:($u$,$v$,0), M:($u$,0,$w$), P:(0,$v$,$w$), Q:(1/2,$v$,$w$), GP:($u$,$v$,$w$) |
| $Fmmm$<br>(69)<br>[$mmmF$] | o3 | | Γ:(0,0,0), T:(1,0,0), Y:(0,1,0), Z:(0,0,1), L:(1/2,1/2,1/2), A:$(u,0,1)$, B:$(0,v,1)$, Δ:$(0,v,0)$, H:$(0,1,w)$, Λ:$(0,0,w)$, Σ:$(u,0,0)$, E:$(0,v,w)$, J:$(u,0,w)$, M:$(u,v,0)$, GP:$(u,v,w)$ |
| $Immm$<br>(71)<br>[$mmmI$] | o4 | | Γ:(0,0,0), X:(1,1,1), R:(1/2,0,1/2), S:(0,1/2,1/2), T:(1/2,1/2,0), W:(1/2,1/2,1/2), Δ:$(0,v,0)$, Λ:$(0,0,w)$, Σ:$(u,0,0)$, A:$(0,v,w)$, B:$(u,0,w)$, C:$(u,v,0)$, D:$(u,1/2,1/2)$, P:$(1/2,1/2,w)$, Q:$(1/2,v,1/2)$, GP:$(u,v,w)$ |
| $P4/m$<br>(83)<br>[$4/mP$] | t1 | | A:(1/2,1/2,1/2), Γ:(0,0,0), M:(1/2,1/2,0), Z:(0,0,1/2), R:(0,1/2,1/2), X:(0,1/2,0), Λ:$(0,0,w)$, V:$(1/2,1/2,w)$, D:$(u,v,0)$, Δ:$(0,v,0)$, E:$(u,v,1/2)$, S:$(u,u,1/2)$, Σ:$(u,u,0)$, T:$(u,1/2,1/2)$, U:$(0,v,1/2)$, W:$(0,1/2,w)$, Y:$(u,1/2,0)$, B:$(0,v,w)$, C:$(u,u,w)$, F:$(u,1/2,w)$, GP:$(u,v,w)$ |
| $I4/m$<br>(87)<br>[$4/mI$] | t2 | | Γ:(0,0,0), M:(1,1,1), P:(1/2,1/2,1/2), X:(1/2,1/2,0), N:(1/2,0,1/2), Λ:$(0,0,w)$, W:$(1/2,1/2,w)$, Q:$(1/2,v,1/2)$, Σ:$(u,0,0)$, Y:$(u,1-u,0)$, Δ:$(u,u,0)$, A:$(u,u,w)$, B:$(u,0,w)$, C:$(u,v,0)$, GP:$(u,v,w)$ |
| $P4/mmm$ (123) [$4/mmmP$] | t3 | | A:(1/2,1/2,1/2), Γ:(0,0,0), M:(1/2,1/2,0), Z:(0,0,1/2), R:(0,1/2,1/2), X:(0,1/2,0), Λ:(0,0,w), V:(1/2,1/2,w), Δ:(0,v,0), S:(u,u,1/2), Σ:(u,u,0), T:(u,1/2,1/2), U:(0,v,1/2), W:(0,1/2,w), Y:(u,1/2,0), B:(0,v,w), C:(u,u,w), D:(u,v,0), E:(u,v,1/2), F:(u,1/2,w), GP:(u,v,w) |
| $I4/mmm$ (139) [$4/mmmI$] | t4 | | Γ:(0,0,0), M:(1,1,1), P:(1/2,1/2,1/2), X:(1/2,1/2,0), N:(1/2,0,1/2), Λ:(0,0,w), Δ:(u,u,0), Q:(1/2,v,1/2), Σ:(u,0,0), W:(1/2,1/2,w), Y:(u,1-u,0), A:(u,u,w), B:(u,0,w), C:(u,v,0), GP:(u,v,w) |
| $P\bar{3}$ (147) [$\bar{3}P$] | hP1 | | A:(0,0,1/2), Γ:(0,0,0), H:(1/3,1/3,1/2), K:(1/3,1/3,0), L:(1/2,0,1/2), M:(1/2,0,0), Δ:(0,0,w), P:(1/3,1/3,w), B:(u,v,0), C:(u,u,w), D:(u,0,w), E:(u,v,1/2), Λ:(u,u,0), Q:(u,u,1/2), R:(u,0,1/2), Σ:(u,0,0), U:(1/2,0,w), GP:(u,v,w) |
| $R\bar{3}$ (148) [$\bar{3}R$] | hR1 | | Γ:(0,0,0), T:(0,0,3/2), F:(0,1/2,1), L:(-1/2,1/2,1/2), Λ:(0,0,w), Σ:(u,-2*u,0), Y:(u,u,3/2), C:(u,-u,w), GP:(u,v,w) |
| $P\bar{3}1m$ (162) [$\bar{3}1mP$] | hP2 | | A:(0,0,1/2), Γ:(0,0,0), H:(1/3,1/3,1/2), K:(1/3,1/3,0), L:(1/2,0,1/2), M:(1/2,0,0), Δ:(0,0,w), P:(1/3,1/3,w), Λ:(u,u,0), Σ:(u,0,0), Q:(u,u,1/2), R:(u,0,1/2), U:(1/2,0,w), B:(u,v,0), C:(u,u,w), D:(u,0,w), E:(u,v,1/2), GP:(u,v,w) |
| $P\bar{3}m1$<br>(164)<br>[$\bar{3}m1P$] | hP3 | | A:(0,0,1/2), Γ:(0,0,0), H:(1/3,1/3,1/2), K:(1/3,1/3,0), L:(1/2,0,1/2), M:(1/2,0,0), Δ:(0,0,$w$), P:(1/3,1/3,$w$), Λ:($u$,$u$,0), Σ:($u$,0,0), Q:($u$,$u$,1/2), R:($u$,0,1/2), U:(1/2,0,$w$), B:($u$,$v$,0), C:($u$,$u$,$w$), D:($u$,0,$w$), E:($u$,$v$,1/2), GP:($u$,$v$,$w$) |
| $R\bar{3}m$<br>(166)<br>[$\bar{3}mR$] | hR2 | | Γ:(0,0,0), T:(0,0,3/2), F:(0,1/2,1), L:(-1/2,1/2,1/2), Λ:(0,0,$w$), Σ:($u$,-2*$u$,0), Y:($u$,$u$,3/2), C:($u$,-$u$,$w$), GP:($u$,$v$,$w$) |
| $P6/m$<br>(175)<br>[$6/mP$] | hP4 | | A:(0,0,1/2), Γ:(0,0,0), H:(1/3,1/3,1/2), K:(1/3,1/3,0), L:(1/2,0,1/2), M:(1/2,0,0), Δ:(0,0,$w$), P:(1/3,1/3,$w$), Λ:($u$,$u$,0), Σ:($u$,0,0), Q:($u$,$u$,1/2), R:($u$,0,1/2), U:(1/2,0,$w$), B:($u$,$v$,0), C:($u$,$u$,$w$), D:($u$,0,$w$), E:($u$,$v$,1/2), GP:($u$,$v$,$w$) |
| $P6/mmm$<br>(191)<br>[$6/mmmP$] | hP5 | | A:(0,0,1/2), Γ:(0,0,0), H:(1/3,1/3,1/2), K:(1/3,1/3,0), L:(1/2,0,1/2), M:(1/2,0,0), Δ:(0,0,$w$), P:(1/3,1/3,$w$), Λ:($u$,$u$,0), Σ:($u$,0,0), Q:($u$,$u$,1/2), R:($u$,0,1/2), U:(1/2,0,$w$), B:($u$,$v$,0), C:($u$,$u$,$w$), D:($u$,0,$w$), E:($u$,$v$,1/2), GP:($u$,$v$,$w$) |
| $Pm\bar{3}$<br>(200)<br>[$m\bar{3}P$] | c1 | | Γ:(0,0,0), R:(1/2,1/2,1/2), M:(1/2,1/2,0), X:(0,1/2,0), Δ:(0,$v$,0), T:(1/2,1/2,$w$), Z:($u$,1/2,0), ZA:(1/2,$u$,0), Λ:($u$,$u$,$u$), A:($u$,$v$,0), B:($u$,1/2,$w$), S:($u$,1/2,$u$), Σ:($u$,$u$,0), C:($u$,$u$,$w$), GP:($u$,$v$,$w$) |
| $Fm\bar{3}$<br>(202)<br>$[m\bar{3}F]$ | c2 | | Γ:(0,0,0), X:(0,1,0), L:(1/2,1/2,1/2),<br>W:(1/2,1,0), Δ:(0,*v*,0), Σ:(*u*,*u*,0), Λ:(*u*,*u*,*u*),<br>V:(*u*,1,0), Q:(1/2,*v*,1-*v*), A:(*u*,*v*,0), C:(*u*,*u*,*w*),<br>GP:(*u*,*v*,*w*) |
| $Im\bar{3}$<br>(204)<br>$[m\bar{3}I]$ | c3 | | Γ:(0,0,0), H:(1,1,1), P:(1/2,1/2,1/2),<br>N:(1/2,1/2,0), Δ:(0,*v*,0), Λ:(*u*,*u*,*u*), A:(*u*,*v*,0),<br>D:(1/2,1/2,*w*), G:(1+*u*,1-*u*,1), Σ:(*u*,*u*,0),<br>C:(*u*,*u*,*w*), GP:(*u*,*v*,*w*) |
| $Pm\bar{3}m$<br>(221)<br>$[m\bar{3}mP]$ | c4 | | Γ:(0,0,0), R:(1/2,1/2,1/2), M:(1/2,1/2,0),<br>X:(0,1/2,0), Δ:(0,*v*,0), T:(1/2,1/2,*w*), Λ:(*u*,*u*,*u*),<br>S:(*u*,1/2,*u*), Σ:(*u*,*u*,0), Z:(*u*,1/2,0), A:(*u*,*v*,0),<br>B:(*u*,1/2,*w*), C:(*u*,*u*,*w*), GP:(*u*,*v*,*w*) |
| $Fm\bar{3}m$<br>(225)<br>$[m\bar{3}mF]$ | c5 | | Γ:(0,0,0), X:(0,1,0), L:(1/2,1/2,1/2),<br>W:(1/2,1,0), Δ:(0,*v*,0), Λ:(*u*,*u*,*u*), Q:(1/2,*v*,1-*v*),<br>Σ:(*u*,*u*,0), V:(*u*,1,0), A:(*u*,*v*,0), C:(*u*,*u*,*w*),<br>GP:(*u*,*v*,*w*) |
| $Im\bar{3}m$<br>(229)<br>$[m\bar{3}mI]$ | c6 | | Γ:(0,0,0), H:(1,1,1), P:(1/2,1/2,1/2),<br>N:(1/2,1/2,0), Δ:(0,*v*,0), Λ:(*u*,*u*,*u*),<br>D:(1/2,1/2,*w*), G:(1+*u*,1-*u*,1), Σ:(*u*,*u*,0),<br>A:(*u*,*v*,0), C:(*u*,*u*,*w*), GP:(*u*,*v*,*w*) |

## S1.5. Little cogroups

In this section, we will introduce the concepts of little groups, little cogroups in SSGs. For spin-space symmetry operations $g_i = \{S_i \| R_i | \tau_i\}$, given a wave vector **k** in a crystal, the action of $g_i$ on **k** can be expressed as:

$$g_i\mathbf{k} = h(S_i)R_i\mathbf{k}, \tag{8}$$

where $h(S_i) = \begin{cases} 1 & det(S_i) = 1 \\ -1 & det(S_i) = -1 \end{cases}$.

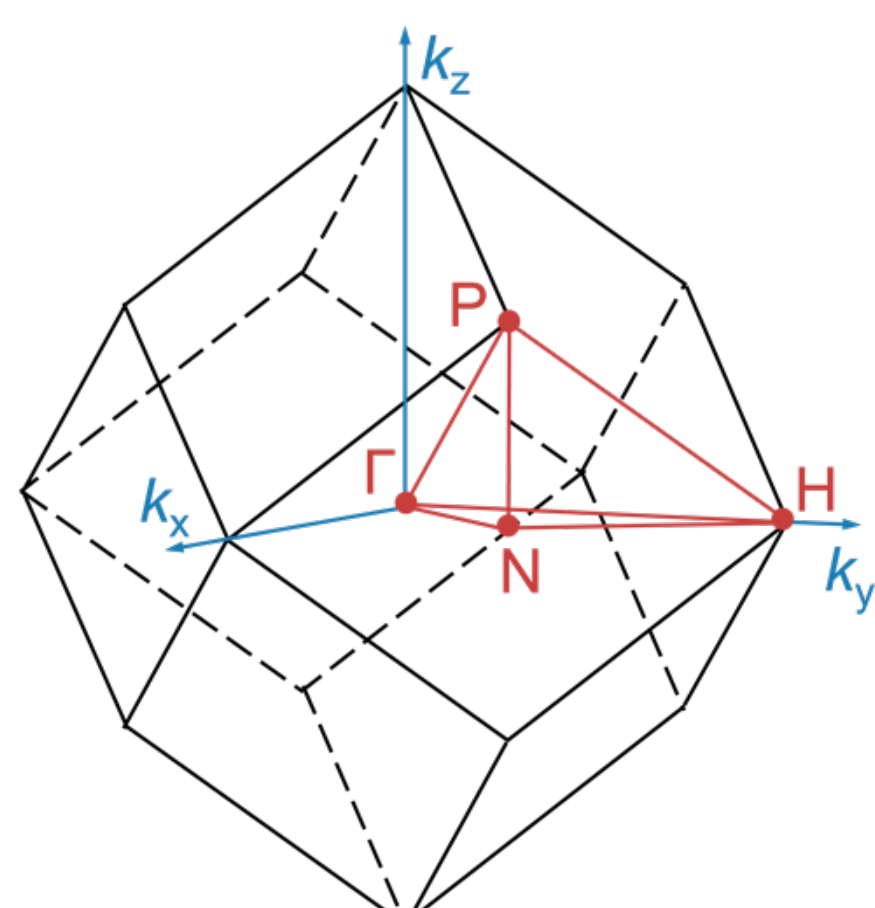

Supplementary Fig. 1. Spin Brillouin zone of the collinear SSG $I^{\bar{1}}a^{\bar{1}}\bar{3}^{\infty m}1$ (199.206.1.1).

Time reversal in spin space and $R_i$ in real space change the wave vector **k**, while **k** is invariant under proper rotation in spin space and lattice translation in real space. Those operations $g_i^k$ that leave the wave vector **k** invariant form the site-symmetry group $G_S^{\boldsymbol{k}}$, i.e., $g_i^k\mathbf{k} = \mathbf{k} + \boldsymbol{K}$, where $\boldsymbol{K}$ is an integer-valued linear combination of the reciprocal lattice vectors $\boldsymbol{K}_{a,b,c}$. As lattice translations in real space leave **k** points invariant, then $G_S^{\boldsymbol{k}}$ contains the group of lattice translation at any point. We can also extend this definition to a little cogroup $\tilde{G}_S^{\boldsymbol{k}}$, which is given by the point group of $G_S^{\boldsymbol{k}}$. However, the mometum-space co-irreps of $G_S^{\boldsymbol{k}}$ can be different from those of $\tilde{G}_S^{\boldsymbol{k}}$ due to nonsymmorphic operations. Below we show the spin Brillouin zone, the momentum stars and the little cogroups of the type-II collinear SSG $I^{\bar{1}}a^{\bar{1}}\bar{3}^{\infty m}1$ (199.206.1.1) in Supplementary Fig. 1 and Supplementary Table IX.

Supplementary Table IX. List of momentum stars and the corresponding little cogroups of the collinear SSG $I^{\bar{1}}a^{\bar{1}}\bar{3}^{\infty m}1$ (199.206.1.1).

| wavevector **k** | little cogroup $\tilde{G}_S^{\boldsymbol{k}}$ |
| --- | --- |
| Γ:(0,0,0) | ${}^{\bar{1}}m^{\bar{1}}\bar{3}^{\infty m}1$ |
| H:(1,1,1) | ${}^{\bar{1}}m^{\bar{1}}\bar{3}^{\infty m}1$ |

| P:(1/2,1/2,1/2) | ${}^{\bar{1}}m{}^{\bar{1}}\bar{3}{}^{\infty}1$ |
|---|---|
| N:(1/2,1/2,0) | ${}^{1}2/{}^{2}m{}^{\infty m}1$ |
| Δ:(0,$v$,0) | ${}^{2}m{}^{\bar{1}}m{}^{2}m{}^{\infty}1$ |
| Λ:($u$,$u$,$u$) | ${}^{\bar{1}}\bar{3}{}^{\infty}1$ |
| A:($u$,$v$,0) | ${}^{m}2/{}^{2}m{}^{\infty}1$ |
| D:(1/2,1/2,$w$) | ${}^{1}2/{}^{\bar{1}}m{}^{\infty}1$ |
| G:(1+$u$,1-$u$,1) | ${}^{m}2/{}^{2}m{}^{\infty}1$ |
| Σ:($u$,$u$,0) | ${}^{m}2/{}^{2}m{}^{\infty}1$ |
| C:($u$,$u$,$w$) | ${}^{\bar{1}}\bar{1}{}^{\infty}1$ |
| GP:($u$,$v$,$w$) | ${}^{\bar{1}}\bar{1}{}^{\infty}1$ |

# S1.6. Band representations

This section mainly introduces the representation matrices of different symmetry operations in spin space for magnons, which gives rise to the two mechanisms of extra double degeneracy in spin space as stated in Methods. At the same time, taking $Cu_3TeO_6$ as an example, the detailed process of constructing band representations based on the spin site group is shown.

### S1.6.1. Matrix representations in the magnon basis

In this section, as we mainly focus on the degeneracy of spin space, we simplify the operations in collinear SSGs. The spin space symmetry operations $\{T\|E\}$, $\{TU_{\boldsymbol{n}}(\pi)\|E\}$, $\{U_{\boldsymbol{n}}(\pi)\|E\}$ and $\{U_z(\phi)\|E\}$ are denoted as $T$, $TU_{\boldsymbol{n}}(\pi)$, $U_{\boldsymbol{n}}(\pi)$ and $U_z(\phi)$, repsectively.

Here we start from the electronic basis using $\{|\uparrow\rangle, |\downarrow\rangle\}$: (**$\boldsymbol{n}$=($n_x$, $n_y$, $n_z$)**)

$$D\big(U_{\boldsymbol{n}}(\theta)\big) = exp\left(-\frac{i\sigma\cdot\boldsymbol{n}\theta}{2}\right) = \begin{pmatrix} \cos\frac{\theta}{2} - in_z\sin\frac{\theta}{2} & \left(-in_x - n_y\right)\sin\frac{\theta}{2} \\ \left(-in_x + n_y\right)\sin\frac{\theta}{2} & \cos\frac{\theta}{2} + in_z\sin\frac{\theta}{2} \end{pmatrix} \tag{9}$$

$$D\big(U_z(\phi)\big) = \begin{pmatrix} e^{-i\phi/2} & 0 \\ 0 & e^{-i\phi/2} \end{pmatrix} \tag{10}$$

$$D\big(U_x(\pi)\big) = \begin{pmatrix} 0 & -i \\ -i & 0 \end{pmatrix} \tag{11}$$

$$D\left(U_y(\pi)\right) = \begin{pmatrix} 0 & -1 \\ 1 & 0 \end{pmatrix} \tag{12}$$

The spin operators can be expressed in basis of $\{|\uparrow\rangle, |\downarrow\rangle\}$:

$$S^Z = |\uparrow\rangle\langle\uparrow| - |\downarrow\rangle\langle\downarrow| \quad (13)$$

$$S^+ = |\uparrow\rangle\langle\downarrow| \quad (14)$$

$$S^- = |\downarrow\rangle\langle\uparrow| \quad (15)$$

In the $\{S^+, S^-\}$ basis, matrix representations of the time reversal and spin rotations are

$$D\big(U_x(\pi)\big) = \begin{pmatrix} 0 & 1 \\ 1 & 0 \end{pmatrix} \quad (16)$$

$$D\left(U_y(\pi)\right) = \begin{pmatrix} 0 & -1 \\ -1 & 0 \end{pmatrix} \quad (17)$$

$$D\big(U_z(\phi)\big) = \begin{pmatrix} e^{-i\phi} & 0 \\ 0 & e^{-i\phi} \end{pmatrix} \quad (18)$$

$$D(T) = \begin{pmatrix} 0 & -1 \\ -1 & 0 \end{pmatrix} K \quad (19)$$

$$D\big(U_z(\phi)\big) = \begin{pmatrix} e^{-i\phi} & 0 \\ 0 & e^{i\phi} \end{pmatrix} \quad (20)$$

Then we have

$$\begin{pmatrix} S^+ \\ S^- \end{pmatrix} \xrightarrow{U_x(\pi)} \begin{pmatrix} S^- \\ S^+ \end{pmatrix} \xrightarrow{U_x(\pi)} \begin{pmatrix} S^+ \\ S^- \end{pmatrix} \Rightarrow D\left(\big(U_x(\pi)\big)^2\right) = D(E) \quad (21)$$

$$\begin{pmatrix} S^+ \\ S^- \end{pmatrix} \xrightarrow{T} \begin{pmatrix} -S^+ \\ -S^- \end{pmatrix} \xrightarrow{T} \begin{pmatrix} S^+ \\ S^- \end{pmatrix} \Rightarrow D(T^2) = D(E) \quad (22)$$

$$\begin{pmatrix} S^+ \\ S^- \end{pmatrix} \xrightarrow{TU_x(\pi)} \begin{pmatrix} -S^- \\ -S^+ \end{pmatrix} \xrightarrow{TU_x(\pi)} \begin{pmatrix} S^+ \\ S^- \end{pmatrix} \Rightarrow D\left(\big(TU_x(\pi)\big)^2\right) = D(E) \quad (23)$$

$$\begin{pmatrix} S^+ \\ S^- \end{pmatrix} \xrightarrow{TU_x(\pi)U_z(\phi)} \begin{pmatrix} -e^{i\phi}S^- \\ -e^{-i\phi}S^+ \end{pmatrix} \xrightarrow{TU_x(\pi)U_z(\phi)} \begin{pmatrix} S^+ \\ S^- \end{pmatrix} \Rightarrow D\left(\big(TU_x(\pi)U_z(\phi)\big)^2\right) = D(E) \quad (24)$$

Based on Eq. (24), we can conclude that the combined effect of $TU_{\boldsymbol{n}}(\pi)$ and $U_z(\phi)$ does not contribute to the formation of new degeneracy in spin space. Additionally, this is also the origin of the overall spin splitting in the whole Brillouin zone under the spin only group $G_{SO} = Z_2^K \ltimes SO(2)$ in collinear ferromagnets. Here we take $\boldsymbol{n} = x$ in $TU_{\boldsymbol{n}}(\pi)$ for brief. This is also valid for any direction perpendicular to the spin axis *z*.

Similarly,

$$\begin{pmatrix} S^+ \\ S^- \end{pmatrix} \xrightarrow{TU_z(\phi)} \begin{pmatrix} -e^{-i\phi}S^+ \\ -e^{i\phi}S^- \end{pmatrix} \xrightarrow{TU_z(\phi)} \begin{pmatrix} e^{-2i\phi}S^+ \\ e^{2i\phi}S^- \end{pmatrix} \Rightarrow \sum_{\beta \in TU_z(\phi)} \chi(\beta^2) = \int_0^{2\pi} 2\cos(2\phi) = 0 \quad (25)$$

According to the Dimmock and Wheeler character sum rule[7]:

$$\sum_{\beta \in AG} \chi(\beta^2) = \begin{cases} g & Type\ (a) \quad no \\ -g & Type\ (b) \quad double \\ 0 & Type\ (c) \quad double \end{cases} \quad (26)$$

Eq. (25) belongs to type (c), indicating the double degeneracy in spin space when combining $T$ and $U_z(\phi)$. Since there is no pure time reversal $T$ in collinear SSGs, the degeneracy in collinear magnets actually originates from the combined effect of $\{T\|A\}$ and $U_z(\phi)$, where $A$ represents pure lattice space operations that do not affect the degeneracy in spin space. Since the lattice space only provides identity representation for magnon, the orbital basis will not be influenced.

Finally,

$$\begin{pmatrix} S^+ \\ S^- \end{pmatrix} \xrightarrow{(U_x(\pi))^{-1}} \begin{pmatrix} S^- \\ S^+ \end{pmatrix} \xrightarrow{U_z(\phi)} \begin{pmatrix} e^{i\phi}S^- \\ e^{-i\phi}S^+ \end{pmatrix} \xrightarrow{U_x(\pi)} \begin{pmatrix} e^{-i(-\phi)}S^+ \\ e^{i(-\phi)}S^- \end{pmatrix} \tag{27}$$

$$\begin{pmatrix} S^+ \\ S^- \end{pmatrix} \xrightarrow{U_z(\phi)^{-1}} \begin{pmatrix} e^{i\phi}S^+ \\ e^{-i\phi}S^- \end{pmatrix} \xrightarrow{U_x(\pi)} \begin{pmatrix} e^{i\phi}S^- \\ e^{-i\phi}S^+ \end{pmatrix} \xrightarrow{U_z(\phi)} \begin{pmatrix} e^{i(2\phi)}S^- \\ e^{-i(2\phi)}S^+ \end{pmatrix} \tag{28}$$

From Eqs. (27) and (28), we can conclude that $U_x(\pi)U_z(\phi)\left(U_x(\pi)\right)^{-1} = U_z(-\phi)$ and $U_z(\phi)U_x(\pi)U_z(\phi)^{-1} = U_z(2\phi)U_x(\pi)$. Therefore, the combination of $U_{\boldsymbol{n}}(\pi)$ and $U_z(\phi)$ contribute to the emergence of infinite group $D_\infty = \infty 2$, which provides two-dimensional irreducible representations for magnon spin $S = \pm 1$. Actually, there is also no pure in-plane spin rotation $U_{\boldsymbol{n}}(\pi)$ in collinear SSGs, the degenracy in collinear magnets originates from the combined effect of $\{U_{\boldsymbol{n}}(\pi)\|A\}$ and $U_z(\phi)$, where $A$ represents pure lattice space operations that do not affect the degeneracy in spin space.

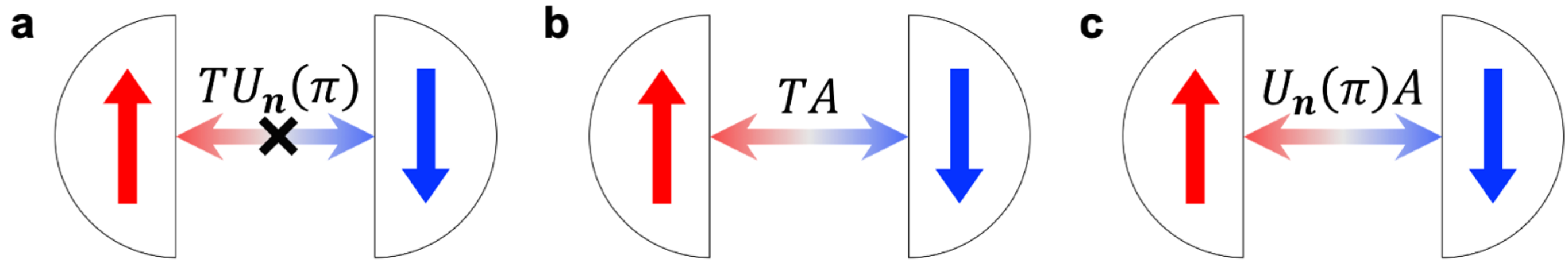

Supplementary Fig. 2 **a**. Spin splitting when combining $TU_{\boldsymbol{n}}(\pi)$ and $U_z(\phi)$. **b**. Spin degenerate when combining $TA$ ($\{T\|A\}$) and $U_z(\phi)$. **c**. Spin degenerate when combining $U_{\boldsymbol{n}}(\pi)A$ ($\{U_{\boldsymbol{n}}(\pi)\|A\}$) and $U_z(\phi)$.

From the above demonstration, we can conclude that in the collinear SSGs, $SO(2)$ actually provides two spin channels for magnons with S = 1 and -1. Whether these two spin channels are degenerate can be determined by the existence of $\{U_{\boldsymbol{n}}(\pi)\|A\}$ and $\{T\|A\}$ symmetry operations in the k-little groups. These rules can be simply depicted using Supplementary Fig. 2. For further discussions on the band representations of k-little groups in collinear SSGs and their relations with type-II MSGs, please refer to Supplementary Information S1.6.3.

### S1.6.2. Band representations in type-II SSGs 199.206.1.1 for magnon

Below we take the $Cu_3TeO_6$ case as an example, whose SSG is $G_S = I^1a^{\bar{1}}\bar{3}^{\infty m}1$ (199.206.1.1), and the spin Wyckoff position for the magnetic ion Cu is 24*d*. The spin space group of $I^1a^{\bar{1}}\bar{3}^{\infty m}1$ can be constructed from the $G_\uparrow = I2_13$, the connecting spin symmetry $\{U_n(\pi)\|\bar{1}|0\}$, and the spin only part $G_{SO}$ semidirect-producted from $Z_2^K = \{E, TU_n(\pi)\}$ and the $SO(2)$ group. Below we take $n = x$ and denote $U_n(\pi)$ and $U_z(\phi)$ as $\{2_{100}\|1|0\}$ and $\{\infty_{001}\|1|0\}$ using international notations, respectively.

The spin site group of Cu (24*d*) is $G_S^{24d} = (\{1\|1|0\} + \{1\|2_{100}|1/2,1/2,0\}) \times G_{SO}$, and the maximal unitary group is $H_S^{24d} = \{\infty_{001}\|1|0\} \times (\{1\|1|0\} + \{1\|2_{100}|1/2,1/2,0\})$. Following the band representation theory for SSGs as provided in Methods. We first construct the character table of is $H_S^{24d}$ (shown in Supplementary Table X), where the "orbital" representation for $S^{\pm}$ carrying magnon is $\rho_{S^{\pm}}^{24d} = \Gamma_1^{\mathrm{S}}(1) + \Gamma_{-1}^{\mathrm{S}}(1)$.

Supplementary Table X. The character table for $H_S^{24d} = {}^{1}2^{\infty}1$.

| $H_S^{24d}$ | $\{1\|1|0\}$ | $\{\infty_{001}\|1|0\}$ | $\{1\|2_{100}|1/2,1/2,0\}$ | $\{\infty_{001}\|2_{100}|1/2,1/2,0\}$ |
|---|---|---|---|---|
| $\Gamma_1^{\mathrm{S}}$ | 1 | $e^{-i\phi}$ | 1 | $e^{-i\phi}$ |
| $\Gamma_{-1}^{\mathrm{S}}$ | 1 | $e^{i\phi}$ | 1 | $e^{i\phi}$ |

Now we build the full band representation of space group induced from the orbital representation of a spin site group. Here we seek a coset decomposition of $G_S^{\boldsymbol{q}}$ in $G_S$. All orbits of the Wyckoff position $\{\boldsymbol{q}_\alpha = g_\alpha \boldsymbol{q}_1 | \boldsymbol{q}_\alpha \in G_S\}, \alpha = 1, 2, \ldots, n$ with multiplicity $n$ of the Wyckoff position are derived. The SSG element $g_\alpha$, combined with the translation $\mathbb{T}$, generate the decomposition of $G_S$ with respect to the $G_S^{24d}$:

$$G_S = \bigcup_{\alpha} g_\alpha \left( G_S^{24d} \ltimes \mathbb{T} \right) \tag{29}$$

In this case, $g_\alpha = \left( \{1\|1|0\} + \{1\|2_{001}|1/2,0,1/2\} + \left\{1\middle\|3_{11\bar{1}}^{+}|1/2,1/2,0\right\} + \{1\|3_{111}^{+}|0\} + \{1\|3_{\bar{1}11}^{-}|0\} + \left\{1\middle\|3_{\bar{1}\bar{1}\bar{1}}^{-}|0,1/2,1/2\right\} \right) \times (\{1\|1|0\} + \{2_{100}\|\bar{1}|0\})$

The full band representation $\rho_{S^{\pm}} = \left( \rho_{S^{\pm}}^{\boldsymbol{q}} \uparrow G_S \right)$ is induced from orbital representation $\rho_{S^{\pm}}^{\boldsymbol{q}}$, then we restrict it to band representations of k-little groups $\rho_{S^{\pm}}^{\boldsymbol{k}} = \rho_{S^{\pm}} \downarrow G_S^{\boldsymbol{k}}$ with ingredients:

$$\chi_{\rho_{S^{\pm}}^{\boldsymbol{k}}}(g) = \begin{cases} \sum_{\alpha} e^{-i[R(g)k \cdot t_{\alpha\alpha}]} \chi_{\rho_{S^{\pm}}^{\boldsymbol{q}}} \left( g_\alpha^{-1} \{E\|E|-t_{\alpha\alpha}\} g g_\alpha \right) & g \in G_S^{24d} \\ 0 & g \notin G_S^{24d} \end{cases} \tag{30}$$

where

$$t_{\alpha\alpha} = g\boldsymbol{q}_\alpha - \boldsymbol{q}_\alpha \tag{31}$$

$$\chi_{\rho_{S^{\pm}}^{\boldsymbol{k}}}(\{\infty_{001}\|1|0\}) = 6\left[ \chi_{\rho_{S^{\pm}}^{24d}}(\{\infty_{001}\|1|0\}) + \chi_{\rho_{S^{\pm}}^{24d}}(\{\infty_{001}^{-1}\|1|0\}) \right] \tag{32}$$

$$\chi_{\rho_{S^\pm}^{\boldsymbol{k}}}(\{\infty_{001}\|2_{001}|1/2,0,1/2\}) = e^{i\pi(k_x-k_z)}\left[\left(e^{-i\pi k_y}+e^{-i3\pi k_y}\right)m+\left(e^{-i\pi k_y}+e^{i\pi k_y}\right)n\right] \tag{33}$$

$$\chi_{\rho_{S^\pm}^{\boldsymbol{k}}}(\{\infty_{001}\|2_{010}|0,1/2,1/2\}) = e^{i\pi(k_z-k_y)}\left[\left(e^{-i\pi k_x}+e^{-i3\pi k_x}\right)m+\left(e^{-i\pi k_x}+e^{i\pi k_x}\right)n\right] \tag{34}$$

$$\chi_{\rho_{S^\pm}^{\boldsymbol{k}}}(\{\infty_{001}\|2_{100}|1/2,1/2,0\}) = e^{i\pi(k_y-k_x)}\left[\left(e^{-i\pi k_z}+e^{-i3\pi k_z}\right)m+\left(e^{-i\pi k_z}+e^{i\pi k_z}\right)n\right] \tag{35}$$

where $m = \chi_{\rho_{S^\pm}^{24d}}(\{\infty_{001}\|2_{100}|0,0,1/2\}), n = \chi_{\rho_{S^\pm}^{24d}}(\{\infty_{001}^{-1}\|2_{100}|0,0,1/2\})$. For other elements $g \in G_{SS}^{k}$, $\chi\left(\rho_{S^\pm}^{k}(g)\right) = 0$. At last, we perform sum rules in Eq. (26) to account for the introduction of anti-unitary operations and the band representation $\rho_{S^\pm}^{\boldsymbol{k}}$ in Spin Brillouin zone is determined.

Here we take Γ, N point as examples:

1. Γ (0, 0, 0) point:

$$G_S^{\Gamma}/\mathbb{T} \cong (G_\uparrow + \{2_{100}\|\bar{1}|0\}G_\uparrow) \times G_{SO} \cong D_z(\infty) \times G_\uparrow \times Z_2^K \tag{36}$$

The character table for Γ point can be constructed from $23$ and $D_z(\infty) = \infty 2$ as shown in Supplementary Table XI, and we denote the four classes of $23$ as $C_1 = E$, $C_2 = 3C_2$, $C_3 = 4C_3^{-1}$, $C_4 = 4C_3$. For brief, we denote $a_i = \{E\|C_i\}, b_i = \{U_z(\pm\phi)\|C_i\}, c_i = \{U_x(\pi)\|-C_i\}$ (i = 1~4), $\omega = -(1+\sqrt{3}i)/2$ and $t = \cos\phi$. (The last column is the result of sum rule.)

Supplementary Table XI. The character table for $^{\bar{1}}m^{\bar{1}}\bar{3}^{\infty}1$ at Γ point.

| | $a_1$ | $a_2$ | $a_3$ | $a_4$ | $b_1$ | $b_2$ | $b_3$ | $b_4$ | $b_1$ | $b_2$ | $b_3$ | $b_4$ | Type |
|---|---|---|---|---|---|---|---|---|---|---|---|---|---|
| $\Gamma_1^S$ | 2 | 2 | 2 | 2 | $2t$ | $2t$ | $2t$ | $2t$ | 0 | 0 | 0 | 0 | (a) |
| $\Gamma_2^S$ | 2 | 2 | $2\omega$ | $2\omega^*$ | $2t$ | $2t$ | $2\omega t$ | $2\omega^* t$ | 0 | 0 | 0 | 0 | (c) |
| $\Gamma_3^S$ | 2 | 2 | $2\omega^*$ | $2\omega$ | $2t$ | $2t$ | $2\omega^* t$ | $2\omega t$ | 0 | 0 | 0 | 0 | (c) |
| $\Gamma_4^S$ | 6 | -2 | 0 | 0 | $6t$ | $-2t$ | 0 | 0 | 0 | 0 | 0 | 0 | (a) |

For Γ point, Eqs. (32)~(35) can be calculated as:

$$\chi_{\rho_{S^\pm}^{\Gamma}}(\{U_z(\phi)\|1|0\}) = 12\cos\phi \tag{37}$$

$$\chi_{\rho_{S^\pm}^{\Gamma}}(\{U_z(\phi)\|2_{001}|1/2,0,1/2\}) = 4\cos\phi \tag{38}$$

$$\chi_{\rho_{S^\pm}^{\Gamma}}(\{U_z(\phi)\|2_{010}|0,1/2,1/2\}) = 4\cos\phi \tag{39}$$

$$\chi_{\rho_{S^\pm}^{\Gamma}}(\{U_z(\phi)\|2_{100}|1/2,1/2,0\}) = 4\cos\phi \tag{40}$$

Then we can get $\rho_{S^\pm}^{\Gamma} = \Gamma_1^S(2) + \Gamma_2^S(2) + \Gamma_3^S(2) + \Gamma_4^S(6)$

Now we take the antiunitary element into consideration using sum rule in Eq. (26) , and the result is $\rho_{S^\pm}^{\Gamma} = \Gamma_1^S(2) + \Gamma_2^S\Gamma_3^S(4) + \Gamma_4^S(6)$.

2. N (1/2, 1/2, 0)

The little group of N point is

$$G_S^{\mathrm{N}}/\mathbb{T} \cong \left(G_\uparrow^N + \{U_x(\pi)\|\bar{1}|0\}G_\uparrow^N\right) \times G_{SO} \cong D_z(\infty) \times G_\uparrow^N \times Z_2^K \tag{41}$$

where $G_\uparrow^N = \{\{1\|1|0\},\{1\|2_{001}|1/2,0,1/2\}\}$.

Since N point is at the boundary of spin Brillouin zone, the central extension should be taken into consideration.

(1) The number of kernel subgroup $g$

$$\mu(h_i, h_j) = e^{-2\pi i \vec{g}_i \cdot \vec{w}_j} \tag{42}$$

where $h_i, h_j \in P_N = \{\{1\|1|0\},\{1\|2_{001}|1/2,0,1/2\}\}$, $\vec{w}_j$ is the translation part of $h_j$

$$\vec{g}_i = h_i^{-1}k - k \tag{43}$$

$$\vec{g}_1 = (0,\ 0,\ 0), \vec{g}_{2_{001}} = (-1,\ -1,\ 0) \Rightarrow \mu(h_i, h_j) = \pm 1, g = 2 \tag{44}$$

(2) kernel subgroup: $t = ma + nb + oc.\ k_N = (1/2,\ 1/2,\ 0)$

$$\mathbb{T}_N = \{\{1\|1|ma + nb\},\{1\|1|oc\}\}; m + n = even,\ \forall\ intergers\ o \tag{45}$$

(3) Central extension:

$$\bar{H}_S^{N^*} \cong D_z(\infty) \times (\{1\|1|0\} + \{1\|1|0,1,0\}) \times (\{1\|1|0\} + \{1\|2_{001}|1/2,0,1/2\}) \cong D_z(\infty) \times D_2 \tag{46}$$

(4) Find $\Gamma^{(\mu)}(\{1\|1|0,1,0\}\mathbb{T}_N) = -\mathbb{I}$

(5) Decomposition: $\rho_{S^\pm}^{\mathrm{N}} = 3\mathrm{N}_1^S(2) + 3\mathrm{N}_2^S(2)$

Following these procedures, we can get all band representations at different high symmetry points, which is shown in Supplementary Table XII.

Supplementary Table XII. Band representation of the SSG $I^{\bar{1}}a^{\bar{1}}\bar{3}^{\infty m}1$ (199.206.1.1) and Wyckoff position 24d ($x$, 0, 1/4).

| **Band representation** | | $I^{\bar{1}}a^{\bar{1}}\bar{3}^{\infty m}1$ (199.206.1.1) |
|---|---|---|
| k-vector of $G$ | k-vector of $G_\uparrow$ | $\boldsymbol{A^S \uparrow G\ (12)}$ |
| Γ:(0,0,0) | Γ:(0,0,0) | $\Gamma_1^S(2) + \Gamma_2^S\Gamma_3^S(4)+\Gamma_4^S(6)$ |
| H:(1,1,1) | H:(1,1,1) | $2H_4^S(6)$ |
| P:(1/2,1/2,1/2) | P:(1/2,1/2,1/2) | $P_1^S(4) + P_2^S(4) + P_3^S(4)$ |

| N:(1/2,1/2,0) | N:(1/2,1/2,0) | $3N_1^S(2)+3N_2^S(2)$ |
|---|---|---|
| Δ:(0,*v*,0) | Δ:(0,*v*,0) | $4\Delta_1^S(2)+2\Delta_2^S(2)$ |
| Λ:(*u*,*u*,*u*) | Λ:(*u*,*u*,*u*) | $2\Lambda_1^S(2)+2\Lambda_2^S(2)+2\Lambda_3^S(2)$ |
| A:(*u*,*v*,0) | A:(*u*,*v*,0) | $6A_1^S(2)$ |
| D:(1/2,1/2,*w*) | D:(1/2,1/2,*w*) | $3D_1^S(2)+3D_2^S(2)$ |
| G:(1+*u*,1-*u*,1) | G:(1+*u*,1-*u*,1) | $6G_1^S(2)$ |
| Σ:(*u*,*u*,0) | Σ:(*u*,*u*,0) | $6\Sigma_1^S(2)$ |
| C:(*u*,*u*,*w*) | C:(*u*,*u*,*w*) | $6C_1^S(2)$ |
| GP:(*u*,*v*,*w*) | GP:(*u*,*v*,*w*) | $6GP_1^S(2)$ |

### S1.6.3. Relation with type-II MSGs

Now let us return to the construction of collinear FM and AFM SSGs in Eqs. (4) and (5). It can be observed that collinear SSGs involve two new types of operations that are absence in space group $G$ or $G_{\uparrow}$. The first type is the spin-only operations, namely $TU_n(\pi)$ and $U_z(\phi)$. In collinear SSGs describing AFM, there exists another new type of operations, namely $U_n(\pi)A$ and $TA$, which connect the two opposite-spin sublattices. Considering the effects of $TU_n(\pi)$, $U_n(\pi)A$ and $TA$ on $SO(2)$ symmetry, the spin degeneracy mechanism can be achieved, which have been been demonstrated in the Methods. This is because the $SO(2)$ symmetry provides the two well-defined spin channels. From this, we can conclude that the band representation of collinear SSGs can be achieved by using the band representations of space group $G$ or $G_{\uparrow}$ plus $Z_2^{\mathrm{K}}$, and the spin degeneracy mechanisms. In other word, the band representation of collinear SSGs can be derived from the band representations of type-II MSGs and the spin degeneracy mechanisms. A detailed demonstration will be provided later.

Next, starting from the specific form of the little co-group at the k-point, we will prove that the band representation of the collinear SSGs can also be obtained with the help of the type-II MSG. The little group $G_S^{\boldsymbol{k}}$ of collinear SSGs can have one of these forms:

$$G_S^{\boldsymbol{k}}=\{E\|L^{\boldsymbol{k}}\}\times SO(2) \tag{47}$$

$$G_S^{\boldsymbol{k}}=\{E\|L^{\boldsymbol{k}}\}\times\left(E+TU_{\boldsymbol{n}}(\pi)\right)\times SO(2) \tag{48}$$

$$G_S^{\boldsymbol{k}}=\{E\|L^{\boldsymbol{k}}\}\times(E+TU_{\boldsymbol{n}}(\pi)A)\times SO(2) \tag{49}$$

$$G_S^{\boldsymbol{k}} = \{E\|L^{\boldsymbol{k}}\} \times (E + TA) \times SO(2) \tag{50}$$

$$G_S^{\boldsymbol{k}} = \{E\|L^{\boldsymbol{k}}\} \times (E + U_{\boldsymbol{x}}(\pi)A) \ltimes SO(2) \tag{51}$$

$$G_S^{\boldsymbol{k}} = \{E\|L^{\boldsymbol{k}}\} \times (E + TA) \times (E + U_{\boldsymbol{x}}(\pi)A) \ltimes SO(2) \tag{52}$$

where $L^{\boldsymbol{k}}$ is the little group of $L = G$ or $G_{\uparrow}$ for collinear FM or AFM SSGs, $A$ is a coset representative with $A \notin L^{\boldsymbol{k}}$ but $A^2 \in L^{\boldsymbol{k}}$. Next, we will analyze these six cases one by one:

(1) When the little group $G_S^{\boldsymbol{k}}$ has the form of Eq. (47), the irreps can be constructed from the single-valued irreps $\boldsymbol{K}_i^R(n_i)$ of space group $L$ and the conjugated pair of irreps of $SO(2)$, $\boldsymbol{K}_{+1}^S(1)$ and $\boldsymbol{K}_{-1}^S(1)$. Here $\boldsymbol{K}_i^R(n_i)$ stands for the $i$-th, $n_i$-dimensional irrep at momentum $\boldsymbol{K}$ in real space $R$, while $\boldsymbol{K}_{\pm1}^S(1)$ represents the two conjugated one-dimensional irreps of $m = \pm1$ in spin space $S$ for magnon. As a result, the co-irrep of $G_S^{\boldsymbol{k}}$ is $\boldsymbol{K}_i^R(n_i)\boldsymbol{K}_{+1}^S(1) + \boldsymbol{K}_i^R(n_i)\boldsymbol{K}_{-1}^S(1)$. Since we focus on the band degeneracy, or the dimension of irrep, we abbreviated them as $2\boldsymbol{K}_i(n_i)$.

(2) When the little group $G_S^{\boldsymbol{k}}$ has the form of Eq. (48), the combination of $TU_{\boldsymbol{x}}(\pi)A$ and $SO(2)$ cannot combine two conjugated irreps in spin space according to Eq. (5) in the Methods. As a result, the co-irreps of $G_S^{\boldsymbol{k}}$ is obtained from the single-valued co-irreps $\boldsymbol{K}_i^R(n_i)$ of space group $L + TL$ and $\boldsymbol{K}_{+1}^S(1)$ and $\boldsymbol{K}_{-1}^S(1)$ from $SO(2)$ symmetry in spin space. As a result, the co-irreps of $G_S^{\boldsymbol{k}}$ can also be abbreviated as $2\boldsymbol{K}_i^R(n_i)$.

(3) When the little group $G_S^{\boldsymbol{k}}$ has the form of Eq. (49), the combination of $TU_{\boldsymbol{n}}(\pi)A$ and $SO(2)$ cannot combine two conjugated irreps in spin space according to Eq. (24). As a result, the co-irrep of $G_S^{\boldsymbol{k}}$ is $\boldsymbol{K}_i^R(n_i)\boldsymbol{K}_{+1}^S(1) + \boldsymbol{K}_i^R(n_i)\boldsymbol{K}_{-1}^S(1)$, which is abbreviated as $2\boldsymbol{K}_i^R(n_i)$.

(4) When the little group $G_S^{\boldsymbol{k}}$ has the form of Eq. (50), the combination of $TA$ and $SO(2)$ contributes to the emergence of two-dimensional co-irreps $\boldsymbol{K}_1^S(2)$ according to Eq. (25). Then the co-irreps of the little group $G_S^{\boldsymbol{k}}$ can be written as $\boldsymbol{K}_i^R(n_i)\boldsymbol{K}_1^S(2)$, which is abbreviated as $\boldsymbol{K}_i(2n_i)$.

(5) When the little group $G_S^{\boldsymbol{k}}$ has the form of Eq. (51), $(E + U_{\boldsymbol{n}}(\pi)A) \ltimes SO(2) \cong \infty 2$ provides two-dimensional irrep $\boldsymbol{K}_1^S(2)$ in spin space as shown in Eqs. (27) and (28). Then the co-irreps of the little group $G_S^{\boldsymbol{k}}$ can be written as $\boldsymbol{K}_i^R(n_i)\boldsymbol{K}_1^S(2)$, which is abbreviated as $\boldsymbol{K}_i(2n_i)$.

(6) When the little group $G_S^{\boldsymbol{k}}$ has the form of Eq. (52). $(E + U_{\boldsymbol{n}}(\pi)A) \ltimes SO(2) \cong \infty 2$ provides a two-dimensional irrep $\boldsymbol{K}_1^S(2)$ in spin space. Since $\boldsymbol{K}_1^S(2)$ in spin space is real irrep now, the introduction of $TA$ cannot contribute to further extra degeneracy. In this way, the co-irreps of the little group $G_S^{\boldsymbol{k}}$ can be written as $\boldsymbol{K}_i^R(n_i)\boldsymbol{K}_1^S(2)$, which is abbreviated as $\boldsymbol{K}_i(2n_i)$.

In summary, the co-irreps of little group $G_S^{\boldsymbol{k}}$ can be obtained from the single-valued irreps of type-II MSG with the form of $L + TL$ and the degeneracy in spin space arised from the combination of $U_{\boldsymbol{n}}(\pi)A$ or $TA$ and $SO(2)$. In Supplementary Table XIII, the relation between type-I~IV collinear SSGs and the form of the k-little group $G_S^{\boldsymbol{k}}$ is given. We have observed that type-I collinear SSG encompasses only the first three cases, resulting in spin-splitting in the whole Brillouin zone. Type-II collinear SSG is restricted to the fourth and sixth cases due to PT symmetry at any k-point, while type-IV SSG is limited to the fifth and sixth cases by Uτ symmetry. Type-III SSG can potentially match the first to sixth cases. When type-III SSG possesses inversion symmetry in $G_{\uparrow}$, it leads to $PTU_n(\pi)$ symmetry at any k-point, corresponding to the second and sixth cases.

By conducting a comprehensive analysis of the band representation, we establish the relationships between the four types of collinear SSGs and collinear magnets, including collinear FMs/FiMs, PT AFMs, altermagnets, and Uτ AFMs, as illustrated in Supplementary Fig. 1 of the main text.

For case (4)~(6), when $L^{\boldsymbol{k}}$ hosts nonzero topological charge in one spin channel, the topological charge at this spin-degenerate point will be doubled or compensated, dependent on whether $A$ is a proper or improper operation in lattice space. As a result, the nonzero topological charge in one channel will be compensated in type-II collinear SSGs thanks to the $PT$ symmetry. However, it will be doubled in type-IV collinear SSGs thanks to the $U_{\boldsymbol{n}}(\pi)\tau$ symmetry. In type-III SSGs, it depends on the $A$ of $U_{\boldsymbol{n}}(\pi)A$ and $TA$ that connects two spin channels.

Supplementary Table XIII. Existence of six types of the little group $G_S^{\boldsymbol{k}}$ in type-I~IV SSGs.

| $G_S^{\boldsymbol{k}}$ | spin space | I | II | III | IV |
| --- | --- | --- | --- | --- | --- |
| $\{E\|\|L^{\boldsymbol{k}}\} \times SO(2)$ | No | yes | no | yes | no |
| $\{E\|\|L^{\boldsymbol{k}}\} \times \big(E + TU_n(\pi)\big) \ltimes SO(2)$ | No | yes | no | yes | no |
| $\{E\|\|L^{\boldsymbol{k}}\} \times (E + TU_n(\pi)A) \ltimes SO(2)$ | No | yes | no | yes | no |
| $\{E\|\|L^{\boldsymbol{k}}\} \times (E + TA) \ltimes SO(2)$ | Doubled | no | yes | yes | no |
| $\{E\|\|L^{\boldsymbol{k}}\} \times (E + U_n(\pi)A) \ltimes SO(2)$ | Doubled | no | no | yes | yes |

| $\{E\|L^k\}\times\left(E+TU_n(\pi)\right)\times\left(E+U_n(\pi)A\right)\ltimes SO(2)$ | Doubled | no | yes | yes | yes |
| --- | --- | --- | --- | --- | --- |

### S1.6.4. Relation with band representations for electronic band structures

Now we turn to the electronic band structures of collinear SSGs with the help of Eqs. (9)-(12) in Supplementary Information S1.6.1.

The matrix representation of time reversal can be expressed in basis of $\{|\uparrow\rangle,|\downarrow\rangle\}$ as:

$$D(T)=\begin{pmatrix}0 & 1\\ -1 & 0\end{pmatrix}K \tag{53}$$

Then we have

$$\begin{pmatrix}|\uparrow\rangle\\ |\downarrow\rangle\end{pmatrix}\xrightarrow{U_x(\pi)}\begin{pmatrix}-i|\downarrow\rangle\\ -i|\uparrow\rangle\end{pmatrix}\xrightarrow{U_x(\pi)}\begin{pmatrix}-|\uparrow\rangle\\ -|\downarrow\rangle\end{pmatrix}\Rightarrow D\left(\left(U_x(\pi)\right)^2\right)=-D(E) \tag{54}$$

$$\begin{pmatrix}|\uparrow\rangle\\ |\downarrow\rangle\end{pmatrix}\xrightarrow{T}\begin{pmatrix}|\downarrow\rangle\\ -|\uparrow\rangle\end{pmatrix}\xrightarrow{T}\begin{pmatrix}-|\uparrow\rangle\\ -|\downarrow\rangle\end{pmatrix}\Rightarrow D(T^2)=-D(E) \tag{55}$$

$$\begin{pmatrix}|\uparrow\rangle\\ |\downarrow\rangle\end{pmatrix}\xrightarrow{TU_x(\pi)}\begin{pmatrix}-i|\uparrow\rangle\\ -i|\downarrow\rangle\end{pmatrix}\xrightarrow{TU_x(\pi)}\begin{pmatrix}|\uparrow\rangle\\ |\downarrow\rangle\end{pmatrix}\Rightarrow D\left(\left(TU_x(\pi)\right)^2\right)=D(E) \tag{56}$$

$$\begin{pmatrix}|\uparrow\rangle\\ |\downarrow\rangle\end{pmatrix}\xrightarrow{TU_x(\pi)U_z(\phi)}\begin{pmatrix}ie^{i\phi/2}|\uparrow\rangle\\ -ie^{-i\phi/2}|\downarrow\rangle\end{pmatrix}\xrightarrow{TU_x(\pi)U_z(\phi)}\begin{pmatrix}|\uparrow\rangle\\ |\downarrow\rangle\end{pmatrix}\Rightarrow D\left(\left(TU_x(\pi)U_z(\phi)\right)^2\right)=D(E) \tag{57}$$

Based on Eq. (57), we can conclude that the combined effect of $TU_{\boldsymbol{n}}(\pi)$ and $U_z(\phi)$ does not contribute to the formation of new degeneracy in spin space. Additionally, this is also the origin of the overall electronic spin splitting in the whole Brillouin zone under the spin only group $G_{SO}=Z_2^K\ltimes SO(2)$ in collinear ferromagnets. Here we take $\boldsymbol{n}=x$ in $TU_{\boldsymbol{n}}(\pi)$ for brief. This is also valid for any direction perpendicular to the spin axis *z*.

Similar to magnon system, in electronic band structures, the combination of $TA$ and $U_z(\phi)$ brings the spin degenearte band as given in Eq. (58), where $A$ represents pure lattice space operations that do not affect the degeneracy in spin space.

$$\begin{pmatrix}|\uparrow\rangle\\ |\downarrow\rangle\end{pmatrix}\xrightarrow{TU_z(\phi)}\begin{pmatrix}e^{-i\phi/2}|\uparrow\rangle\\ -e^{i\phi/2}|\downarrow\rangle\end{pmatrix}\xrightarrow{TU_z(\phi)}\begin{pmatrix}-e^{-i\phi}|\uparrow\rangle\\ -e^{i\phi}|\downarrow\rangle\end{pmatrix}\Rightarrow\sum_{\beta\in TU_z(\phi)}\chi(\beta^2)=\int_0^{2\pi}2\cos(2\phi)=0 \tag{58}$$

However, unlike the magnon system where the lattice space only provides the identity representation, different orbital representations in lattice space correspond to different orbital bases in electronic band structures (e.g., *s*, *p*, *d*, *f* ...). The specific form of $A$ in lattice space determines the linear combination of orbital bases that contribute to the spin degenerate band. This is also where the electronic band structures is more complex compared to the magnon dispersion in terms of collinear SSGs.

$$\begin{pmatrix}|\uparrow\rangle\\|\downarrow\rangle\end{pmatrix}\xrightarrow{\left(U_x(\pi)\right)^{-1}}\begin{pmatrix}i|\downarrow\rangle\\i|\uparrow\rangle\end{pmatrix}\xrightarrow{U_z(\phi)}\begin{pmatrix}ie^{-i\phi/2}|\downarrow\rangle\\ie^{i\phi/2}|\uparrow\rangle\end{pmatrix}\xrightarrow{U_x(\pi)}\begin{pmatrix}e^{i\phi/2}S^+\\e^{-i\phi/2}S^-\end{pmatrix} \tag{59}$$

$$\begin{pmatrix}|\uparrow\rangle\\|\downarrow\rangle\end{pmatrix}\xrightarrow{U_z(\phi)^{-1}}\begin{pmatrix}e^{i\phi/2}|\uparrow\rangle\\e^{-i\phi/2}|\downarrow\rangle\end{pmatrix}\xrightarrow{U_x(\pi)}\begin{pmatrix}-ie^{-i\phi/2}|\downarrow\rangle\\-ie^{i\phi/2}|\uparrow\rangle\end{pmatrix}\xrightarrow{U_z(\phi)}\begin{pmatrix}-ie^{-i\phi}|\downarrow\rangle\\-ie^{i\phi}|\uparrow\rangle\end{pmatrix} \tag{60}$$

From Eqs. (59) and (60), we can conclude that $U_x(\pi)U_z(\phi)\left(U_x(\pi)\right)^{-1} = U_z(-\phi)$ and $U_z(\phi)U_x(\pi)U_z(\phi)^{-1} = U_z(2\phi)U_x(\pi)$ . Note that $U_z(-\phi) = U_z(\phi)^{-1} = U_z(4\pi - \phi)$ for electrons. Therefore, the combination of $U_{\boldsymbol{n}}(\pi)A$ and $U_z(\phi)$ contribute to the emergence of infinite double group $D_\infty = \infty 2$, which provides two-dimensional irreducible representations for electronic spin $m_s = \pm 1/2$.

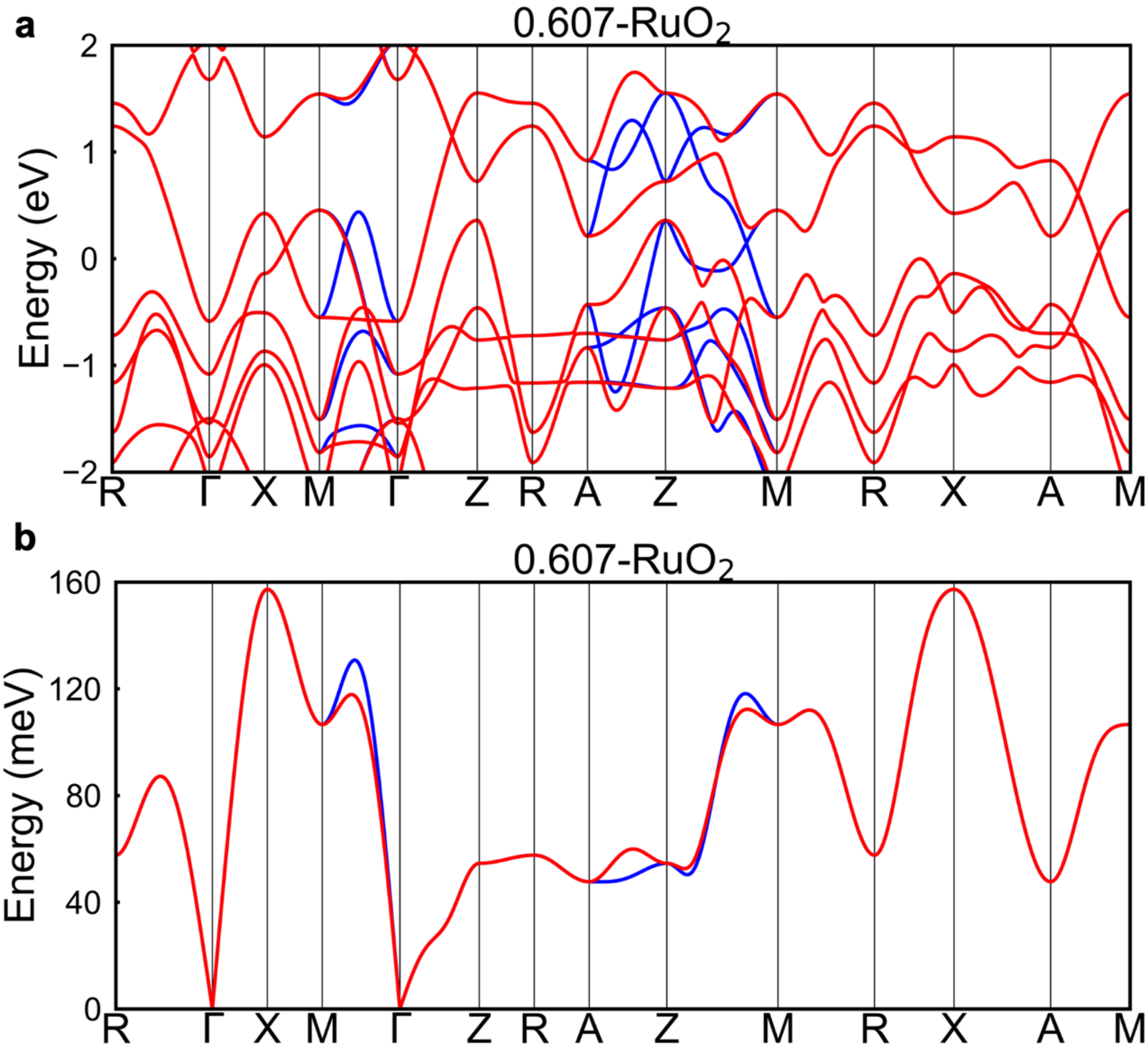

Supplementary Fig. 3. SOC-free **a,** electronic and **b**, magnon band structure of $RuO_2$ with projection of spin components. The red and blue line represent the spin up and down of electronic (magnon) spin channel, respectively.

We can conclude that in collinear SSGs, the band degeneracies of the magnon and electronic band representations are exactly the same. Although there are differences in the representation matrices due to the $2\pi$ and $4\pi$ properties for spinless (magnon) and spinful (electron) particles, the dimension of the band representations are still the same. The only difference lies in that the magnon in lattice space can only give the identity representation, while the electron in lattice space can carry different representations, thereby inducing more band representations under the same Wyckoff Positions. The overall construction of the band representations for magnonic and electronic band structures can be achieved through the type-II MSGs of the sublattice in one spin channel and the three rules we conclude to determine the spin degeneracy.

We also provide the electronic and magnon band structure in the whole Brillouin zone for $RuO_2$ in Supplementary Fig. 3, where the momentum position of altermagnetic band splitting and magnon chirality splitting are the same.

# S2. Table of collinear SSGs

## S2.1. Group and material statistics

The full list of type-I, II, III and IV collinear SSGs are given in Supplementary Table XIV~XVII, respectively. In Supplementary Fig. 4, we show the distribution of different types of SSGs or materials in seven lattice systems.

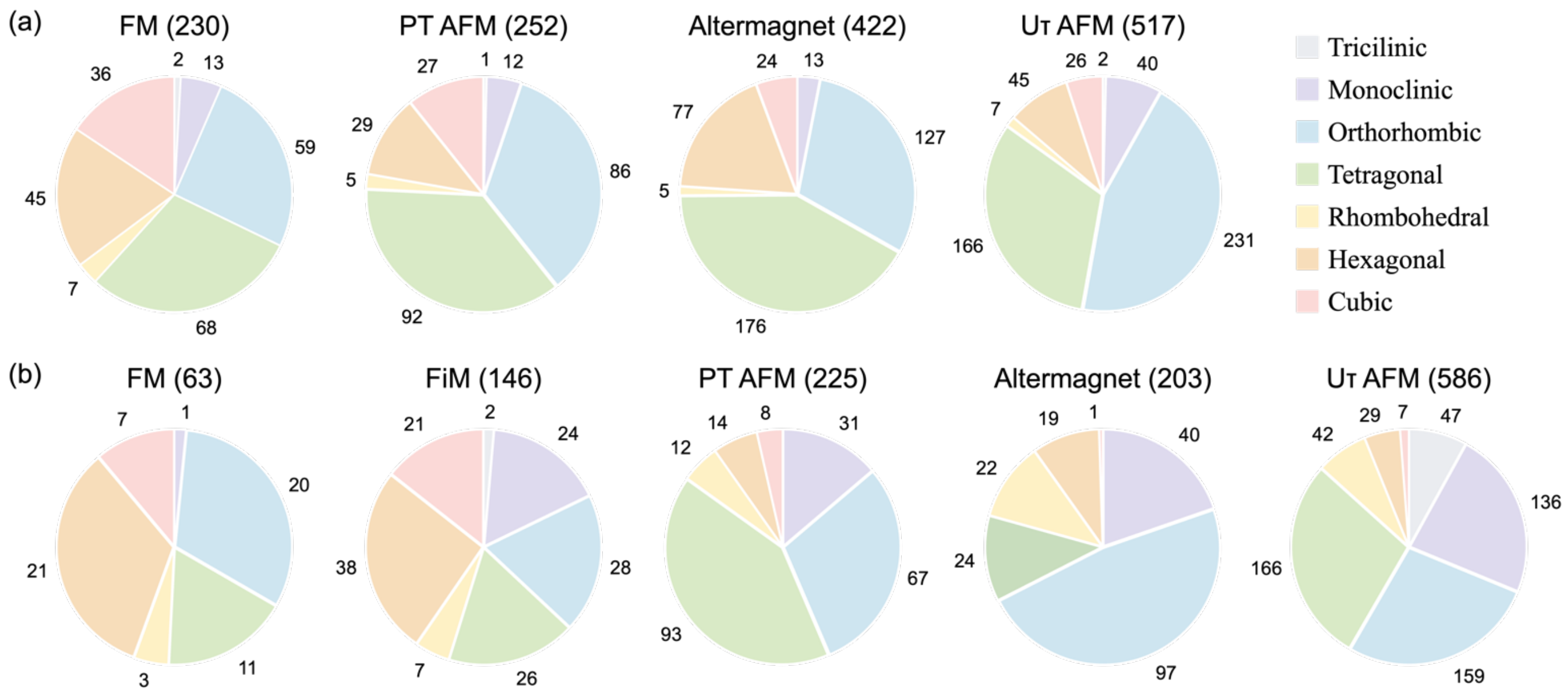

Supplementary Fig. 4. Statistics on the lattice systems of different magnetic structures in the (a) collinear SSGs and (b) 1223 collinear magnets in the MAGNDATA database.

## S2.2. Type-I collinear SSGs

Now we give the full list of 230 type-I SSGs that describing collinear FMs or FIMs in Supplementary Table XIV. The type-I SSGs have the form of $G_S = \{E\|G\} \times Z_2^{\mathrm{K}} \ltimes SO(2)$.

Supplementary Table XIV. Full tabulation of 230 type-I collinear SSGs describing collinear FMs and FiMs. The four columns represent the group number using a four-index nomenclature[8], the one-to-one correspondence with MSGs, the international notation of collinear SSGs and the spin Brillouin zone, respectively. For detailed information about the notation used in the spin Brillouin zone, please refer to Supplementary Table VIII in Supplementary Information S1.

| $G_{NSS}$ | MSG | $G_{S}$ | SBZ |
|---|---|---|---|
| 1.1.1.1 | 1.1 | $P^{1}1^{\infty m}1$ | a1 |
| 2.2.1.1 | 2.4 | $P^{1}\bar{1}^{\infty m}1$ | a1 |
| 3.3.1.1 | 3.1 | $P^{1}2^{\infty m}1$ | m1 |
| 4.4.1.1 | 4.7 | $P^{1}2_{1}{}^{\infty m}1$ | m1 |
| 5.5.1.1 | 5.13 | $C^{1}2^{\infty m}1$ | m2 |
| 6.6.1.1 | 6.18 | $P^{1}m^{\infty m}1$ | m1 |
| 7.7.1.1 | 7.24 | $P^{1}c^{\infty m}1$ | m1 |
| 8.8.1.1 | 8.32 | $C^{1}m^{\infty m}1$ | m2 |
| 9.9.1.1 | 9.37 | $C^{1}c^{\infty m}1$ | m2 |
| 10.10.1.1 | 10.42 | $P^{1}2/^{1}m^{\infty m}1$ | m1 |
| 11.11.1.1 | 11.50 | $P^{1}2_{1}/^{1}m^{\infty m}1$ | m1 |
| 12.12.1.1 | 12.58 | $C^{1}2/^{1}m^{\infty m}1$ | m2 |
| 13.13.1.1 | 13.65 | $P^{1}2/^{1}c^{\infty m}1$ | m1 |
| 14.14.1.1 | 14.75 | $P^{1}2_{1}/^{1}c^{\infty m}1$ | m1 |
| 15.15.1.1 | 15.85 | $C^{1}2/^{1}c^{\infty m}1$ | m2 |
| 16.16.1.1 | 16.1 | $P^{1}2^{1}2^{1}2^{\infty m}1$ | o1 |
| 17.17.1.1 | 17.7 | $P^{1}2^{1}2^{1}2_{1}{}^{\infty m}1$ | o1 |
| 18.18.1.1 | 18.16 | $P^{1}2_{1}{}^{1}2_{1}{}^{1}2^{\infty m}1$ | o1 |
| 19.19.1.1 | 19.25 | $P^{1}2_{1}{}^{1}2_{1}{}^{1}2_{1}{}^{\infty m}1$ | o1 |
| 20.20.1.1 | 20.31 | $C^{1}2^{1}2^{1}2_{1}{}^{\infty m}1$ | o2C |
| 21.21.1.1 | 21.38 | $C^{1}2^{1}2^{1}2^{\infty m}1$ | o2C |
| 22.22.1.1 | 22.45 | $F^{1}2^{1}2^{1}2^{\infty m}1$ | o3 |
| 23.23.1.1 | 23.49 | $I^{1}2^{1}2^{1}2^{\infty m}1$ | o4 |
| 24.24.1.1 | 24.53 | $I^{1}2_{1}{}^{1}2_{1}{}^{1}2_{1}{}^{\infty m}1$ | o4 |
| 25.25.1.1 | 25.57 | $P^{1}m^{1}m^{1}2^{\infty m}1$ | o1 |
| 26.26.1.1 | 26.66 | $P^{1}m^{1}c^{1}2_{1}{}^{\infty m}1$ | o1 |
| 27.27.1.1 | 27.78 | $P^{1}c^{1}c^{1}2^{\infty m}1$ | o1 |
| 28.28.1.1 | 28.87 | $P^{1}m^{1}a^{1}2^{\infty m}1$ | o1 |
| 29.29.1.1 | 29.99 | $P^{1}c^{1}a^{1}2_{1}{}^{\infty m}1$ | o1 |
| 30.30.1.1 | 30.111 | $P^{1}n^{1}c^{1}2^{\infty m}1$ | o1 |
| 31.31.1.1 | 31.123 | $P^{1}m^{1}n^{1}2_{1}{}^{\infty m}1$ | o1 |
| 32.32.1.1 | 32.135 | $P^{1}b^{1}a^{1}2^{\infty m}1$ | o1 |
| 33.33.1.1 | 33.144 | $P^{1}n^{1}a^{1}2_{1}{}^{\infty m}1$ | o1 |
| 34.34.1.1 | 34.156 | $P^{1}n^{1}n^{1}2^{\infty m}1$ | o1 |
| 35.35.1.1 | 35.165 | $C^{1}m^{1}m^{1}2^{\infty m}1$ | o2C |
| 36.36.1.1 | 36.172 | $C^{1}m^{1}c^{1}2_{1}{}^{\infty m}1$ | o2C |
| 37.37.1.1 | 37.180 | $C^{1}c^{1}c^{1}2^{\infty m}1$ | o2C |
| 38.38.1.1 | 38.187 | $A^{1}m^{1}m^{1}2^{\infty m}1$ | o2A |
| 39.39.1.1 | 39.195 | $A^{1}e^{1}m^{1}2^{\infty m}1$ | o2A |
| 40.40.1.1 | 40.203 | $A^{1}m^{1}a^{1}2^{\infty m}1$ | o2A |
| 41.41.1.1 | 41.211 | $A^{1}e^{1}a^{1}2^{\infty m}1$ | o2A |
| 42.42.1.1 | 42.219 | $F^{1}m^{1}m^{1}2^{\infty m}1$ | o3 |

| 43.43.1.1 | 43.224 | $F^1d^1d^12^{\infty m}1$ | o3 |
|---|---|---|---|
| 44.44.1.1 | 44.229 | $I^1m^1m^12^{\infty m}1$ | o4 |
| 45.45.1.1 | 45.235 | $I^1b^1a^12^{\infty m}1$ | o4 |
| 46.46.1.1 | 46.241 | $I^1m^1a^12^{\infty m}1$ | o4 |
| 47.47.1.1 | 47.249 | $P^1m^1m^1m^{\infty m}1$ | o1 |
| 48.48.1.1 | 48.257 | $P^1n^1n^1n^{\infty m}1$ | o1 |
| 49.49.1.1 | 49.265 | $P^1c^1c^1m^{\infty m}1$ | o1 |
| 50.50.1.1 | 50.277 | $P^1b^1a^1n^{\infty m}1$ | o1 |
| 51.51.1.1 | 51.289 | $P^1m^1m^1a^{\infty m}1$ | o1 |
| 52.52.1.1 | 52.305 | $P^1n^1n^1a^{\infty m}1$ | o1 |
| 53.53.1.1 | 53.321 | $P^1m^1n^1a^{\infty m}1$ | o1 |
| 54.54.1.1 | 54.337 | $P^1c^1c^1a^{\infty m}1$ | o1 |
| 55.55.1.1 | 55.353 | $P^1b^1a^1m^{\infty m}1$ | o1 |
| 56.56.1.1 | 56.365 | $P^1c^1c^1n^{\infty m}1$ | o1 |
| 57.57.1.1 | 57.377 | $P^1b^1c^1m^{\infty m}1$ | o1 |
| 58.58.1.1 | 58.393 | $P^1n^1n^1m^{\infty m}1$ | o1 |
| 59.59.1.1 | 59.405 | $P^1m^1m^1n^{\infty m}1$ | o1 |
| 60.60.1.1 | 60.417 | $P^1b^1c^1n^{\infty m}1$ | o1 |
| 61.61.1.1 | 61.433 | $P^1b^1c^1a^{\infty m}1$ | o1 |
| 62.62.1.1 | 62.441 | $P^1n^1m^1a^{\infty m}1$ | o1 |
| 63.63.1.1 | 63.457 | $C^1m^1c^1m^{\infty m}1$ | o2C |
| 64.64.1.1 | 64.469 | $C^1m^1c^1e^{\infty m}1$ | o2C |
| 65.65.1.1 | 65.481 | $C^1m^1m^1m^{\infty m}1$ | o2C |
| 66.66.1.1 | 66.491 | $C^1c^1c^1m^{\infty m}1$ | o2C |
| 67.67.1.1 | 67.501 | $C^1m^1m^1e^{\infty m}1$ | o2C |
| 68.68.1.1 | 68.511 | $C^1c^1c^1e^{\infty m}1$ | o2C |
| 69.69.1.1 | 69.521 | $F^1m^1m^1m^{\infty m}1$ | o3 |
| 70.70.1.1 | 70.527 | $F^1d^1d^1d^{\infty m}1$ | o3 |
| 71.71.1.1 | 71.533 | $I^1m^1m^1m^{\infty m}1$ | o4 |
| 72.72.1.1 | 72.539 | $I^1b^1a^1m^{\infty m}1$ | o4 |
| 73.73.1.1 | 73.548 | $I^1b^1c^1a^{\infty m}1$ | o4 |
| 74.74.1.1 | 74.554 | $I^1m^1m^1a^{\infty m}1$ | o4 |
| 75.75.1.1 | 75.1 | $P^14^{\infty m}1$ | t1 |
| 76.76.1.1 | 76.7 | $P^14_1{}^{\infty m}1$ | t1 |
| 77.77.1.1 | 77.13 | $P^14_2{}^{\infty m}1$ | t1 |
| 78.78.1.1 | 78.19 | $P^14_3{}^{\infty m}1$ | t1 |
| 79.79.1.1 | 79.25 | $I^14^{\infty m}1$ | t2 |
| 80.80.1.1 | 80.29 | $I^14_1{}^{\infty m}1$ | t2 |
| 81.81.1.1 | 81.33 | $P^1\bar{4}^{\infty m}1$ | t1 |
| 82.82.1.1 | 82.39 | $I^1\bar{4}^{\infty m}1$ | t2 |
| 83.83.1.1 | 83.43 | $P^14/^1m^{\infty m}1$ | t1 |
| 84.84.1.1 | 84.51 | $P^14_2/^1m^{\infty m}1$ | t1 |
| 85.85.1.1 | 85.59 | $P^14/^1n^{\infty m}1$ | t1 |
| 86.86.1.1 | 86.67 | $P^{1}4_{2}/^{1}n^{\infty m}1$ | t1 |
| 87.87.1.1 | 87.75 | $I^{1}4/^{1}m^{\infty m}1$ | t2 |
| 88.88.1.1 | 88.81 | $I^{1}4_{1}/^{1}a^{\infty m}1$ | t2 |
| 89.89.1.1 | 89.87 | $P^{1}4^{1}2^{1}2^{\infty m}1$ | t3 |
| 90.90.1.1 | 90.95 | $P^{1}4^{1}2_{1}{}^{1}2^{\infty m}1$ | t3 |
| 91.91.1.1 | 91.103 | $P^{1}4_{1}{}^{1}2^{1}2^{\infty m}1$ | t3 |
| 92.92.1.1 | 92.111 | $P^{1}4_{1}{}^{1}2_{1}{}^{1}2^{\infty m}1$ | t3 |
| 93.93.1.1 | 93.119 | $P^{1}4_{2}{}^{1}2^{1}2^{\infty m}1$ | t3 |
| 94.94.1.1 | 94.127 | $P^{1}4_{2}{}^{1}2_{1}{}^{1}2^{\infty m}1$ | t3 |
| 95.95.1.1 | 95.135 | $P^{1}4_{3}{}^{1}2^{1}2^{\infty m}1$ | t3 |
| 96.96.1.1 | 96.143 | $P^{1}4_{3}{}^{1}2_{1}{}^{1}2^{\infty m}1$ | t3 |
| 97.97.1.1 | 97.151 | $I^{1}4^{1}2^{1}2^{\infty m}1$ | t4 |
| 98.98.1.1 | 98.157 | $I^{1}4_{1}{}^{1}2^{1}2^{\infty m}1$ | t4 |
| 99.99.1.1 | 99.163 | $P^{1}4^{1}m^{1}m^{\infty m}1$ | t3 |
| 100.100.1.1 | 100.171 | $P^{1}4^{1}b^{1}m^{\infty m}1$ | t3 |
| 101.101.1.1 | 101.179 | $P^{1}4_{2}{}^{1}c^{1}m^{\infty m}1$ | t3 |
| 102.102.1.1 | 102.187 | $P^{1}4_{2}{}^{1}n^{1}m^{\infty m}1$ | t3 |
| 103.103.1.1 | 103.195 | $P^{1}4^{1}c^{1}c^{\infty m}1$ | t3 |
| 104.104.1.1 | 104.203 | $P^{1}4^{1}n^{1}c^{\infty m}1$ | t3 |
| 105.105.1.1 | 105.211 | $P^{1}4_{2}{}^{1}m^{1}c^{\infty m}1$ | t3 |
| 106.106.1.1 | 106.219 | $P^{1}4_{2}{}^{1}b^{1}c^{\infty m}1$ | t3 |
| 107.107.1.1 | 107.227 | $I^{1}4^{1}m^{1}m^{\infty m}1$ | t4 |
| 108.108.1.1 | 108.233 | $I^{1}4^{1}c^{1}m^{\infty m}1$ | t4 |
| 109.109.1.1 | 109.239 | $I^{1}4_{1}{}^{1}m^{1}d^{\infty m}1$ | t4 |
| 110.110.1.1 | 110.245 | $I^{1}4_{1}{}^{1}c^{1}d^{\infty m}1$ | t4 |
| 111.111.1.1 | 111.251 | $P^{1}\bar{4}^{1}2^{1}m^{\infty m}1$ | t3 |
| 112.112.1.1 | 112.259 | $P^{1}\bar{4}^{1}2^{1}c^{\infty m}1$ | t3 |
| 113.113.1.1 | 113.267 | $P^{1}\bar{4}^{1}2_{1}{}^{1}m^{\infty m}1$ | t3 |
| 114.114.1.1 | 114.275 | $P^{1}\bar{4}^{1}2_{1}{}^{1}c^{\infty m}1$ | t3 |
| 115.115.1.1 | 115.283 | $P^{1}\bar{4}^{1}m^{1}2^{\infty m}1$ | t3 |
| 116.116.1.1 | 116.291 | $P^{1}\bar{4}^{1}c^{1}2^{\infty m}1$ | t3 |
| 117.117.1.1 | 117.299 | $P^{1}\bar{4}^{1}b^{1}2^{\infty m}1$ | t3 |
| 118.118.1.1 | 118.307 | $P^{1}\bar{4}^{1}n^{1}2^{\infty m}1$ | t3 |
| 119.119.1.1 | 119.315 | $I^{1}\bar{4}^{1}m^{1}2^{\infty m}1$ | t4 |
| 120.120.1.1 | 120.321 | $I^{1}\bar{4}^{1}c^{1}2^{\infty m}1$ | t4 |
| 121.121.1.1 | 121.327 | $I^{1}\bar{4}^{1}2^{1}m^{\infty m}1$ | t4 |
| 122.122.1.1 | 122.333 | $I^{1}\bar{4}^{1}2^{1}d^{\infty m}1$ | t4 |
| 123.123.1.1 | 123.339 | $P^{1}4/^{1}m^{1}m^{1}m^{\infty m}1$ | t3 |
| 124.124.1.1 | 124.351 | $P^{1}4/^{1}m^{1}c^{1}c^{\infty m}1$ | t3 |
| 125.125.1.1 | 125.363 | $P^{1}4/^{1}n^{1}b^{1}m^{\infty m}1$ | t3 |
| 126.126.1.1 | 126.375 | $P^{1}4/^{1}n^{1}n^{1}c^{\infty m}1$ | t3 |
| 127.127.1.1 | 127.387 | $P^{1}4/^{1}m^{1}b^{1}m^{\infty m}1$ | t3 |
| 128.128.1.1 | 128.399 | $P^{1}4/^{1}m^{1}n^{1}c^{\infty m}1$ | t3 |

| 129.129.1.1 | 129.411 | $P^{1}4/^{1}n^{1}m^{1}m^{\infty m}1$ | t3 |
|---|---|---|---|
| 130.130.1.1 | 130.423 | $P^{1}4/^{1}n^{1}c^{1}c^{\infty m}1$ | t3 |
| 131.131.1.1 | 131.435 | $P^{1}4_{2}/^{1}m^{1}m^{1}c^{\infty m}1$ | t3 |
| 132.132.1.1 | 132.447 | $P^{1}4_{2}/^{1}m^{1}c^{1}m^{\infty m}1$ | t3 |
| 133.133.1.1 | 133.459 | $P^{1}4_{2}/^{1}n^{1}b^{1}c^{\infty m}1$ | t3 |
| 134.134.1.1 | 134.471 | $P^{1}4_{2}/^{1}n^{1}n^{1}m^{\infty m}1$ | t3 |
| 135.135.1.1 | 135.483 | $P^{1}4_{2}/^{1}m^{1}b^{1}c^{\infty m}1$ | t3 |
| 136.136.1.1 | 136.495 | $P^{1}4_{2}/^{1}m^{1}n^{1}m^{\infty m}1$ | t3 |
| 137.137.1.1 | 137.507 | $P^{1}4_{2}/^{1}n^{1}m^{1}c^{\infty m}1$ | t3 |
| 138.138.1.1 | 138.519 | $P^{1}4_{2}/^{1}n^{1}c^{1}m^{\infty m}1$ | t3 |
| 139.139.1.1 | 139.531 | $I^{1}4/^{1}m^{1}m^{1}m^{\infty m}1$ | t4 |
| 140.140.1.1 | 140.541 | $I^{1}4/^{1}m^{1}\mathrm{c}^{1}m^{\infty m}1$ | t4 |
| 141.141.1.1 | 141.551 | $I^{1}4_{1}/^{1}a^{1}m^{1}d^{\infty m}1$ | t4 |
| 142.142.1.1 | 142.561 | $I^{1}4_{1}/^{1}a^{1}c^{1}d^{\infty m}1$ | t4 |
| 143.143.1.1 | 143.1 | $P^{1}3^{\infty m}1$ | hP1 |
| 144.144.1.1 | 144.4 | $P^{1}3_{1}{}^{\infty m}1$ | hP1 |
| 145.145.1.1 | 145.7 | $P^{1}3_{2}{}^{\infty m}1$ | hP1 |
| 146.146.1.1 | 146.10 | $R^{1}3^{\infty m}1$ | hR1 |
| 147.147.1.1 | 147.13 | $P^{1}\bar{3}^{\infty m}1$ | hP1 |
| 148.148.1.1 | 148.17 | $R^{1}\bar{3}^{\infty m}1$ | hR1 |
| 149.149.1.1 | 149.21 | $P^{1}3^{1}1^{1}2^{\infty m}1$ | hP2 |
| 150.150.1.1 | 150.25 | $P^{1}3^{1}2^{1}1^{\infty m}1$ | hP3 |
| 151.151.1.1 | 151.29 | $P^{1}3_{1}{}^{1}1^{1}2^{\infty m}1$ | hP2 |
| 152.152.1.1 | 152.33 | $P^{1}3_{1}{}^{1}2^{1}1^{\infty m}1$ | hP3 |
| 153.153.1.1 | 153.37 | $P^{1}3_{2}{}^{1}1^{1}2^{\infty m}1$ | hP2 |
| 154.154.1.1 | 154.41 | $P^{1}3_{2}{}^{1}2^{1}1^{\infty m}1$ | hP3 |
| 155.155.1.1 | 155.45 | $R^{1}3^{1}2^{\infty m}1$ | hR2 |
| 156.156.1.1 | 156.49 | $P^{1}3^{1}m^{1}1^{\infty m}1$ | hP3 |
| 157.157.1.1 | 157.53 | $P^{1}3^{1}1^{1}m^{\infty m}1$ | hP2 |
| 158.158.1.1 | 158.57 | $P^{1}3^{1}c^{1}1^{\infty m}1$ | hP3 |
| 159.159.1.1 | 159.61 | $P^{1}3^{1}1^{1}c^{\infty m}1$ | hP2 |
| 160.160.1.1 | 160.65 | $R^{1}3^{1}m^{\infty m}1$ | hR2 |
| 161.161.1.1 | 161.69 | $R^{1}3^{1}c^{\infty m}1$ | hR2 |
| 162.162.1.1 | 162.73 | $P^{1}\bar{3}^{1}1^{1}m^{\infty m}1$ | hP2 |
| 163.163.1.1 | 163.79 | $P^{1}\bar{3}^{1}1^{1}c^{\infty m}1$ | hP2 |
| 164.164.1.1 | 164.85 | $P^{1}\bar{3}^{1}m^{1}1^{\infty m}1$ | hP3 |
| 165.165.1.1 | 165.91 | $P^{1}\bar{3}^{1}c^{1}1^{\infty m}1$ | hP3 |
| 166.166.1.1 | 166.97 | $R^{1}\bar{3}^{1}m^{\infty m}1$ | hR2 |
| 167.167.1.1 | 167.103 | $R^{1}\bar{3}^{1}c^{\infty m}1$ | hR2 |
| 168.168.1.1 | 168.109 | $P^{1}6^{\infty m}1$ | hP4 |
| 169.169.1.1 | 169.113 | $P^{1}6_{1}{}^{\infty m}1$ | hP4 |
| 170.170.1.1 | 170.117 | $P^{1}6_{5}{}^{\infty m}1$ | hP4 |
| 171.171.1.1 | 171.121 | $P^{1}6_{2}{}^{\infty m}1$ | hP4 |
| 172.172.1.1 | 172.125 | $P^{1}6_{4}{}^{\infty m}1$ | hP4 |
| 173.173.1.1 | 173.129 | $P^{1}6_{3}{}^{\infty m}1$ | hP4 |
| 174.174.1.1 | 174.133 | $P^{1}\bar{6}^{\infty m}1$ | hP4 |
| 175.175.1.1 | 175.137 | $P^{1}6/^{1}m^{\infty m}1$ | hP4 |
| 176.176.1.1 | 176.143 | $P^{1}6_{3}/^{1}m^{\infty m}1$ | hP4 |
| 177.177.1.1 | 177.149 | $P^{1}6^{1}2^{1}2^{\infty m}1$ | hP5 |
| 178.178.1.1 | 178.155 | $P^{1}6_{1}{}^{1}2^{1}2^{\infty m}1$ | hP5 |
| 179.179.1.1 | 179.161 | $P^{1}6_{5}{}^{1}2^{1}2^{\infty m}1$ | hP5 |
| 180.180.1.1 | 180.167 | $P^{1}6_{2}{}^{1}2^{1}2^{\infty m}1$ | hP5 |
| 181.181.1.1 | 181.173 | $P^{1}6_{4}{}^{1}2^{1}2^{\infty m}1$ | hP5 |
| 182.182.1.1 | 182.179 | $P^{1}6_{3}{}^{1}2^{1}2^{\infty m}1$ | hP5 |
| 183.183.1.1 | 183.185 | $P^{1}6^{1}m^{1}m^{\infty m}1$ | hP5 |
| 184.184.1.1 | 184.191 | $P^{1}6^{1}c^{1}c^{\infty m}1$ | hP5 |
| 185.185.1.1 | 185.197 | $P^{1}6_{3}{}^{1}c^{1}m^{\infty m}1$ | hP5 |
| 186.186.1.1 | 186.203 | $P^{1}6_{3}{}^{1}m^{1}c^{\infty m}1$ | hP5 |
| 187.187.1.1 | 187.209 | $P^{1}\bar{6}^{1}m^{1}2^{\infty m}1$ | hP5 |
| 188.188.1.1 | 188.215 | $P^{1}\bar{6}^{1}c^{1}2^{\infty m}1$ | hP5 |
| 189.189.1.1 | 189.221 | $P^{1}\bar{6}^{1}2^{1}m^{\infty m}1$ | hP5 |
| 190.190.1.1 | 190.227 | $P^{1}\bar{6}^{1}2^{1}c^{\infty m}1$ | hP5 |
| 191.191.1.1 | 191.233 | $P^{1}6/^{1}m^{1}m^{1}m^{\infty m}1$ | hP5 |
| 192.192.1.1 | 192.243 | $P^{1}6/^{1}m^{1}c^{1}c^{\infty m}1$ | hP5 |
| 193.193.1.1 | 193.253 | $P^{1}6_{3}/^{1}m^{1}c^{1}m^{\infty m}1$ | hP5 |
| 194.194.1.1 | 194.263 | $P^{1}6_{3}/^{1}m^{1}m^{1}c^{\infty m}1$ | hP5 |
| 195.195.1.1 | 195.1 | $P^{1}2^{1}3^{\infty m}1$ | c1 |
| 196.196.1.1 | 196.4 | $F^{1}2^{1}3^{\infty m}1$ | c2 |
| 197.197.1.1 | 197.7 | $I^{1}2^{1}3^{\infty m}1$ | c3 |
| 198.198.1.1 | 198.9 | $P^{1}2_{1}{}^{1}3^{\infty m}1$ | c1 |
| 199.199.1.1 | 199.12 | $I^{1}2_{1}{}^{1}3^{\infty m}1$ | c3 |
| 200.200.1.1 | 200.14 | $P^{1}m^{1}\bar{3}^{\infty m}1$ | c1 |
| 201.201.1.1 | 201.18 | $P^{1}n^{1}\bar{3}^{\infty m}1$ | c1 |
| 202.202.1.1 | 202.22 | $F^{1}m^{1}\bar{3}^{\infty m}1$ | c2 |
| 203.203.1.1 | 203.26 | $F^{1}d^{1}\bar{3}^{\infty m}1$ | c2 |
| 204.204.1.1 | 204.30 | $I^{1}m^{1}\bar{3}^{\infty m}1$ | c3 |
| 205.205.1.1 | 205.33 | $P^{1}a^{1}\bar{3}^{\infty m}1$ | c1 |
| 206.206.1.1 | 206.37 | $I^{1}a^{1}\bar{3}^{\infty m}1$ | c3 |
| 207.207.1.1 | 207.40 | $P^{1}4^{1}3^{1}2^{\infty m}1$ | c4 |
| 208.208.1.1 | 208.44 | $P^{1}4_{2}{}^{1}3^{1}2^{\infty m}1$ | c4 |
| 209.209.1.1 | 209.48 | $F^{1}4^{1}3^{1}2^{\infty m}1$ | c5 |
| 210.210.1.1 | 210.52 | $F^{1}4_{1}{}^{1}3^{1}2^{\infty m}1$ | c5 |
| 211.211.1.1 | 211.56 | $I^{1}4^{1}3^{1}2^{\infty m}1$ | c6 |
| 212.212.1.1 | 212.59 | $P^{1}4_{3}{}^{1}3^{1}2^{\infty m}1$ | c4 |
| 213.213.1.1 | 213.63 | $P^{1}4_{1}{}^{1}3^{1}2^{\infty m}1$ | c4 |
| 214.214.1.1 | 214.67 | $I^{1}4_{1}{}^{1}3^{1}2^{\infty m}1$ | c6 |

| 215.215.1.1 | 215.70 | $P^{1}\bar{4}^{1}3^{1}m^{\infty m}1$ | c4 |
|---|---|---|---|
| 216.216.1.1 | 216.74 | $F^{1}\bar{4}^{1}3^{1}m^{\infty m}1$ | c5 |
| 217.217.1.1 | 217.78 | $I^{1}\bar{4}^{1}3^{1}m^{\infty m}1$ | c6 |
| 218.218.1.1 | 218.81 | $P^{1}\bar{4}^{1}3^{1}n^{\infty m}1$ | c4 |
| 219.219.1.1 | 219.85 | $F^{1}\bar{4}^{1}3^{1}c^{\infty m}1$ | c5 |
| 220.220.1.1 | 220.89 | $I^{1}\bar{4}^{1}3^{1}d^{\infty m}1$ | c6 |
| 221.221.1.1 | 221.92 | $P^{1}m^{1}\bar{3}^{1}m^{\infty m}1$ | c4 |
| 222.222.1.1 | 222.98 | $P^{1}n^{1}\bar{3}^{1}n^{\infty m}1$ | c4 |
| 223.223.1.1 | 223.104 | $P^{1}m^{1}\bar{3}^{1}n^{\infty m}1$ | c4 |
| 224.224.1.1 | 224.110 | $P^{1}n^{1}\bar{3}^{1}m^{\infty m}1$ | c4 |
| 225.225.1.1 | 225.116 | $F^{1}m^{1}\bar{3}^{1}m^{\infty m}1$ | c5 |
| 226.226.1.1 | 226.122 | $F^{1}m^{1}\bar{3}^{1}c^{\infty m}1$ | c5 |
| 227.227.1.1 | 227.128 | $F^{1}d^{1}\bar{3}^{1}m^{\infty m}1$ | c5 |
| 228.228.1.1 | 228.134 | $F^{1}d^{1}\bar{3}^{1}c^{\infty m}1$ | c5 |
| 229.229.1.1 | 229.140 | $I^{1}m^{1}\bar{3}^{1}m^{\infty m}1$ | c6 |
| 230.230.1.1 | 230.145 | $I^{1}a^{1}\bar{3}^{1}d^{\infty m}1$ | c6 |

## S2.3. Type-II collinear SSGs

The 1191 type-II, III and IV SSGs that describe collinear AFMs are tabulated in Supplementary Table XV~XVII, respectively. These SSGs have the form of $G_S = (\{E\|G_{\uparrow}\} + \{T\|AG_{\uparrow}\}) \times Z_2^{\mathrm{K}} \ltimes SO(2)$.

Supplementary Table XV. Full tabulation of 252 type-II collinear SSGs describing collinear PT AFMs. The five columns represent the group number using a four-index nomenclature[8], the one-to-one correspondence with MSGs, the representative lattice-space element $A$ that connects the two opposite-spin sublattices, the international notation of collinear SSGs and the spin Brillouin zone, respectively. For detailed information about the notation used in the spin Brillouin zone, please refer to Supplementary Table VIII in Supplementary Information S1.

| $G_{NSS}$ | MSG | $A$ | $G_S$ | SBZ |
|---|---|---|---|---|
| 1.2.1.1 | 2.6 | $\{-1\|0\}$ | $P^{\bar{1}}\bar{1}^{\infty m}1$ | a1 |
| 6.10.1.1 | 10.44 | $\{-1\|0\}$ | $P^{\bar{1}}2/{}^{1}m^{\infty m}1$ | m1 |
| 3.10.1.1 | 10.45 | $\{-1\|0\}$ | $P^{1}2/{}^{\bar{1}}m^{\infty m}1$ | m1 |
| 6.11.1.1 | 11.52 | $\{-1\|0\}$ | $P^{\bar{1}}2_1/{}^{1}m^{\infty m}1$ | m1 |
| 4.11.1.1 | 11.53 | $\{-1\|0\}$ | $P^{1}2_1/{}^{\bar{1}}m^{\infty m}1$ | m1 |
| 8.12.1.1 | 12.60 | $\{-1\|0\}$ | $C^{\bar{1}}2/{}^{1}m^{\infty m}1$ | m2 |
| 5.12.1.1 | 12.61 | $\{-1\|0\}$ | $C^{1}2/{}^{\bar{1}}m^{\infty m}1$ | m2 |
| 7.13.1.1 | 13.67 | $\{-1\|0\}$ | $P^{\bar{1}}2/{}^{1}c^{\infty m}1$ | m1 |
| 3.13.1.1 | 13.68 | $\{-1\|0\}$ | $P^{1}2/{}^{\bar{1}}c^{\infty m}1$ | m1 |
| 7.14.1.1 | 14.77 | $\{-1\|0\}$ | $P^{\bar{1}}2_1/{}^{1}c^{\infty m}1$ | m1 |
| 4.14.1.1 | 14.78 | $\{-1\|0\}$ | $P^{1}2_1/{}^{\bar{1}}c^{\infty m}1$ | m1 |
| 9.15.1.1 | 15.87 | $\{-1\|0\}$ | $C^{\bar{1}}2/{}^{1}c^{\infty m}1$ | m2 |
| 5.15.1.1 | 15.88 | $\{-1\|0\}$ | $C^{1}2/{}^{\bar{1}}c^{\infty m}1$ | m2 |
| 25.47.1.1 | 47.251 | $\{-1\|0\}$ | $P^{\bar{1}}m^{1}m^{1}m^{\infty m}1$ | o1 |
| 16.47.1.1 | 47.253 | $\{-1\|0\}$ | $P^{\bar{1}}m^{\bar{1}}m^{\bar{1}}m^{\infty m}1$ | o1 |
| 34.48.1.1 | 48.259 | $\{-1\|0\}$ | $P^{\bar{1}}n^{1}n^{1}n^{\infty m}1$ | o1 |
| 16.48.1.1 | 48.261 | $\{-1\|0\}$ | $P^{\bar{1}}n^{\bar{1}}n^{\bar{1}}n^{\infty m}1$ | o1 |
| 28.49.1.1 | 49.267 | $\{-1\|0\}$ | $P^{\bar{1}}c^{1}c^{1}m^{\infty m}1$ | o1 |
| 27.49.1.1 | 49.268 | $\{-1\|0\}$ | $P^{1}c^{1}c^{\bar{1}}m^{\infty m}1$ | o1 |
| 16.49.1.1 | 49.271 | $\{-1\|0\}$ | $P^{\bar{1}}c^{\bar{1}}c^{\bar{1}}m^{\infty m}1$ | o1 |
| 30.50.1.1 | 50.279 | $\{-1\|0\}$ | $P^{\bar{1}}b^{1}a^{1}n^{\infty m}1$ | o1 |
| 32.50.1.1 | 50.280 | $\{-1\|0\}$ | $P^{1}b^{1}a^{\bar{1}}n^{\infty m}1$ | o1 |
| 16.50.1.1 | 50.283 | $\{-1\|0\}$ | $P^{\bar{1}}b^{\bar{1}}a^{\bar{1}}n^{\infty m}1$ | o1 |
| 26.51.1.1 | 51.291 | $\{-1\|0\}$ | $P^{\bar{1}}m^{1}m^{1}a^{\infty m}1$ | o1 |
| 28.51.1.1 | 51.292 | $\{-1\|0\}$ | $P^{1}m^{\bar{1}}m^{1}a^{\infty m}1$ | o1 |
| 25.51.1.1 | 51.293 | $\{-1\|0\}$ | $P^{1}m^{1}m^{\bar{1}}a^{\infty m}1$ | o1 |
| 17.51.1.1 | 51.297 | $\{-1\|0\}$ | $P^{\bar{1}}m^{\bar{1}}m^{\bar{1}}a^{\infty m}1$ | o1 |
| 30.52.1.1 | 52.307 | $\{-1\|0\}$ | $P^{\bar{1}}n^{1}n^{1}a^{\infty m}1$ | o1 |
| 33.52.1.1 | 52.308 | $\{-1\|0\}$ | $P^{1}n^{\bar{1}}n^{1}a^{\infty m}1$ | o1 |
| 34.52.1.1 | 52.309 | $\{-1\|0\}$ | $P^{1}n^{1}n^{\bar{1}}a^{\infty m}1$ | o1 |
| 17.52.1.1 | 52.313 | $\{-1\|0\}$ | $P^{\bar{1}}n^{\bar{1}}n^{\bar{1}}a^{\infty m}1$ | o1 |
| 30.53.1.1 | 53.323 | $\{-1\|0\}$ | $P^{\bar{1}}m^{1}n^{1}a^{\infty m}1$ | o1 |
| 28.53.1.1 | 53.324 | $\{-1\|0\}$ | $P^{1}m^{\bar{1}}n^{1}a^{\infty m}1$ | o1 |
| 31.53.1.1 | 53.325 | $\{-1\|0\}$ | $P^{1}m^{1}n^{\bar{1}}a^{\infty m}1$ | o1 |
| 17.53.1.1 | 53.329 | $\{-1\|0\}$ | $P^{\bar{1}}m^{\bar{1}}n^{\bar{1}}a^{\infty m}1$ | o1 |
| 29.54.1.1 | 54.339 | $\{-1\|0\}$ | $P^{\bar{1}}c^{1}c^{1}a^{\infty m}1$ | o1 |
| 32.54.1.1 | 54.340 | $\{-1\|0\}$ | $P^{1}c^{\bar{1}}c^{1}a^{\infty m}1$ | o1 |
| 27.54.1.1 | 54.341 | $\{-1\|0\}$ | $P^{1}c^{1}c^{\bar{1}}a^{\infty m}1$ | o1 |
| 17.54.1.1 | 54.345 | $\{-1\|0\}$ | $P^{\bar{1}}c^{\bar{1}}c^{\bar{1}}a^{\infty m}1$ | o1 |
| 26.55.1.1 | 55.355 | $\{-1\|0\}$ | $P^{\bar{1}}b^{1}a^{1}m^{\infty m}1$ | o1 |
| 32.55.1.1 | 55.356 | $\{-1\|0\}$ | $P^{1}b^{1}a^{\bar{1}}m^{\infty m}1$ | o1 |
| 18.55.1.1 | 55.359 | $\{-1\|0\}$ | $P^{\bar{1}}b^{\bar{1}}a^{\bar{1}}m^{\infty m}1$ | o1 |
| 33.56.1.1 | 56.367 | $\{-1\|0\}$ | $P^{\bar{1}}c^{1}c^{1}n^{\infty m}1$ | o1 |
| 27.56.1.1 | 56.368 | $\{-1\|0\}$ | $P^{1}c^{1}c^{\bar{1}}n^{\infty m}1$ | o1 |
| 18.56.1.1 | 56.371 | $\{-1\|0\}$ | $P^{\bar{1}}c^{\bar{1}}c^{\bar{1}}n^{\infty m}1$ | o1 |
| 28.57.1.1 | 57.379 | $\{-1\|0\}$ | $P^{\bar{1}}b^{1}c^{1}m^{\infty m}1$ | o1 |
| 26.57.1.1 | 57.380 | $\{-1\|0\}$ | $P^{1}b^{\bar{1}}c^{1}m^{\infty m}1$ | o1 |
| 29.57.1.1 | 57.381 | $\{-1\|0\}$ | $P^{1}b^{1}c^{\bar{1}}m^{\infty m}1$ | o1 |
| 18.57.1.1 | 57.385 | $\{-1\|0\}$ | $P^{\bar{1}}b^{\bar{1}}c^{\bar{1}}m^{\infty m}1$ | o1 |
| 31.58.1.1 | 58.395 | $\{-1\|0\}$ | $P^{\bar{1}}n^{1}n^{1}m^{\infty m}1$ | o1 |
| 34.58.1.1 | 58.396 | $\{-1\|0\}$ | $P^{1}n^{1}n^{\bar{1}}m^{\infty m}1$ | o1 |
| 18.58.1.1 | 58.399 | $\{-1\|0\}$ | $P^{\bar{1}}n^{\bar{1}}n^{\bar{1}}m^{\infty m}1$ | o1 |
| 31.59.1.1 | 59.407 | $\{-1\|0\}$ | $P^{\bar{1}}m^{1}m^{1}n^{\infty m}1$ | o1 |
| 25.59.1.1 | 59.408 | $\{-1\|0\}$ | $P^{1}m^{1}m^{\bar{1}}n^{\infty m}1$ | o1 |
| 18.59.1.1 | 59.411 | $\{-1\|0\}$ | $P^{\bar{1}}m^{\bar{1}}m^{\bar{1}}n^{\infty m}1$ | o1 |
| 33.60.1.1 | 60.419 | $\{-1\|0\}$ | $P^{\bar{1}}b^{1}c^{1}n^{\infty m}1$ | o1 |
| 30.60.1.1 | 60.420 | $\{-1\|0\}$ | $P^{1}b^{\bar{1}}c^{1}n^{\infty m}1$ | o1 |
| 29.60.1.1 | 60.421 | $\{-1\|0\}$ | $P^{1}b^{1}c^{\bar{1}}n^{\infty m}1$ | o1 |
| 18.60.1.1 | 60.425 | $\{-1\|0\}$ | $P^{\bar{1}}b^{\bar{1}}c^{\bar{1}}n^{\infty m}1$ | o1 |
| 29.61.1.1 | 61.435 | $\{-1\|0\}$ | $P^{\bar{1}}b^{1}c^{1}a^{\infty m}1$ | o1 |
| 19.61.1.1 | 61.437 | $\{-1\|0\}$ | $P^{\bar{1}}b^{\bar{1}}c^{\bar{1}}a^{\infty m}1$ | o1 |
| 26.62.1.1 | 62.443 | $\{-1\|0\}$ | $P^{\bar{1}}n^{1}m^{1}a^{\infty m}1$ | o1 |
| 33.62.1.1 | 62.444 | $\{-1\|0\}$ | $P^{1}n^{\bar{1}}m^{1}a^{\infty m}1$ | o1 |
| 31.62.1.1 | 62.445 | $\{-1\|0\}$ | $P^{1}n^{1}m^{\bar{1}}a^{\infty m}1$ | o1 |
| 19.62.1.1 | 62.449 | $\{-1\|0\}$ | $P^{\bar{1}}n^{\bar{1}}m^{\bar{1}}a^{\infty m}1$ | o1 |
| 40.63.1.1 | 63.459 | $\{-1\|0\}$ | $C^{\bar{1}}m^{1}c^{1}m^{\infty m}1$ | o2C |
| 38.63.1.1 | 63.460 | $\{-1\|0\}$ | $C^{1}m^{\bar{1}}c^{1}m^{\infty m}1$ | o2C |
| 36.63.1.1 | 63.461 | $\{-1\|0\}$ | $C^{1}m^{1}c^{\bar{1}}m^{\infty m}1$ | o2C |
| 20.63.1.1 | 63.465 | $\{-1\|0\}$ | $C^{\bar{1}}m^{\bar{1}}c^{\bar{1}}m^{\infty m}1$ | o2C |
| 41.64.1.1 | 64.471 | $\{-1\|0\}$ | $C^{\bar{1}}m^{1}c^{1}e^{\infty m}1$ | o2C |

| 39.64.1.1 | 64.472 | $\{-1\|0\}$ | $C^{1}m^{\bar{1}}c^{1}e^{\infty m}1$ | o2C |
|---|---|---|---|---|
| 36.64.1.1 | 64.473 | $\{-1\|0\}$ | $C^{1}m^{1}c^{\bar{1}}e^{\infty m}1$ | o2C |
| 20.64.1.1 | 64.477 | $\{-1\|0\}$ | $C^{\bar{1}}m^{\bar{1}}c^{\bar{1}}e^{\infty m}1$ | o2C |
| 38.65.1.1 | 65.483 | $\{-1\|0\}$ | $C^{\bar{1}}m^{1}m^{1}m^{\infty m}1$ | o2C |
| 35.65.1.1 | 65.484 | $\{-1\|0\}$ | $C^{1}m^{1}m^{\bar{1}}m^{\infty m}1$ | o2C |
| 21.65.1.1 | 65.487 | $\{-1\|0\}$ | $C^{\bar{1}}m^{\bar{1}}m^{\bar{1}}m^{\infty m}1$ | o2C |
| 40.66.1.1 | 66.493 | $\{-1\|0\}$ | $C^{\bar{1}}c^{1}c^{1}m^{\infty m}1$ | o2C |
| 37.66.1.1 | 66.494 | $\{-1\|0\}$ | $C^{1}c^{1}c^{\bar{1}}m^{\infty m}1$ | o2C |
| 21.66.1.1 | 66.497 | $\{-1\|0\}$ | $C^{\bar{1}}c^{\bar{1}}c^{\bar{1}}m^{\infty m}1$ | o2C |
| 39.67.1.1 | 67.503 | $\{-1\|0\}$ | $C^{\bar{1}}m^{1}m^{1}e^{\infty m}1$ | o2C |
| 35.67.1.1 | 67.504 | $\{-1\|0\}$ | $C^{1}m^{1}m^{\bar{1}}e^{\infty m}1$ | o2C |
| 21.67.1.1 | 67.507 | $\{-1\|0\}$ | $C^{\bar{1}}m^{\bar{1}}m^{\bar{1}}e^{\infty m}1$ | o2C |
| 41.68.1.1 | 68.513 | $\{-1\|0\}$ | $C^{\bar{1}}c^{1}c^{1}e^{\infty m}1$ | o2C |
| 37.68.1.1 | 68.514 | $\{-1\|0\}$ | $C^{1}c^{1}c^{\bar{1}}e^{\infty m}1$ | o2C |
| 21.68.1.1 | 68.517 | $\{-1\|0\}$ | $C^{\bar{1}}c^{\bar{1}}c^{\bar{1}}e^{\infty m}1$ | o2C |
| 42.69.1.1 | 69.523 | $\{-1\|0\}$ | $F^{\bar{1}}m^{1}m^{1}m^{\infty m}1$ | o3 |
| 22.69.1.1 | 69.525 | $\{-1\|0\}$ | $F^{\bar{1}}m^{\bar{1}}m^{\bar{1}}m^{\infty m}1$ | o3 |
| 43.70.1.1 | 70.529 | $\{-1\|0\}$ | $F^{\bar{1}}d^{1}d^{1}d^{\infty m}1$ | o3 |
| 22.70.1.1 | 70.531 | $\{-1\|0\}$ | $F^{\bar{1}}d^{\bar{1}}d^{\bar{1}}d^{\infty m}1$ | o3 |
| 44.71.1.1 | 71.535 | $\{-1\|0\}$ | $I^{\bar{1}}m^{1}m^{1}m^{\infty m}1$ | o4 |
| 23.71.1.1 | 71.537 | $\{-1\|0\}$ | $I^{\bar{1}}m^{\bar{1}}m^{\bar{1}}m^{\infty m}1$ | o4 |
| 46.72.1.1 | 72.541 | $\{-1\|0\}$ | $I^{\bar{1}}b^{1}a^{1}m^{\infty m}1$ | o4 |
| 45.72.1.1 | 72.542 | $\{-1\|0\}$ | $I^{1}b^{1}a^{\bar{1}}m^{\infty m}1$ | o4 |
| 23.72.1.1 | 72.545 | $\{-1\|0\}$ | $I^{\bar{1}}b^{\bar{1}}a^{\bar{1}}m^{\infty m}1$ | o4 |
| 45.73.1.1 | 73.550 | $\{-1\|0\}$ | $I^{\bar{1}}b^{1}c^{1}a^{\infty m}1$ | o4 |
| 24.73.1.1 | 73.552 | $\{-1\|0\}$ | $I^{\bar{1}}b^{\bar{1}}c^{\bar{1}}a^{\infty m}1$ | o4 |
| 46.74.1.1 | 74.556 | $\{-1\|0\}$ | $I^{\bar{1}}m^{1}m^{1}a^{\infty m}1$ | o4 |
| 44.74.1.1 | 74.557 | $\{-1\|0\}$ | $I^{1}m^{1}m^{\bar{1}}a^{\infty m}1$ | o4 |
| 24.74.1.1 | 74.560 | $\{-1\|0\}$ | $I^{\bar{1}}m^{\bar{1}}m^{\bar{1}}a^{\infty m}1$ | o4 |
| 75.83.1.1 | 83.46 | $\{-1\|0\}$ | $P^{1}4/^{\bar{1}}m^{\infty m}1$ | t1 |
| 81.83.1.1 | 83.47 | $\{-1\|0\}$ | $P^{\bar{1}}4/^{\bar{1}}m^{\infty m}1$ | t1 |
| 77.84.1.1 | 84.54 | $\{-1\|0\}$ | $P^{1}4_2/^{\bar{1}}m^{\infty m}1$ | t1 |
| 81.84.1.1 | 84.55 | $\{-1\|0\}$ | $P^{\bar{1}}4_2/^{\bar{1}}m^{\infty m}1$ | t1 |
| 75.85.1.1 | 85.62 | $\{-1\|0\}$ | $P^{1}4/^{\bar{1}}n^{\infty m}1$ | t1 |
| 81.85.1.1 | 85.63 | $\{-1\|0\}$ | $P^{\bar{1}}4/^{\bar{1}}n^{\infty m}1$ | t1 |
| 77.86.1.1 | 86.70 | $\{-1\|0\}$ | $P^{1}4_2/^{\bar{1}}n^{\infty m}1$ | t1 |
| 81.86.1.1 | 86.71 | $\{-1\|0\}$ | $P^{\bar{1}}4_2/^{\bar{1}}n^{\infty m}1$ | t1 |
| 79.87.1.1 | 87.78 | $\{-1\|0\}$ | $I^{1}4/^{\bar{1}}m^{\infty m}1$ | t2 |
| 82.87.1.1 | 87.79 | $\{-1\|0\}$ | $I^{\bar{1}}4/^{\bar{1}}m^{\infty m}1$ | t2 |
| 80.88.1.1 | 88.84 | $\{-1\|0\}$ | $I^{1}4_1/^{\bar{1}}a^{\infty m}1$ | t2 |
| 82.88.1.1 | 88.85 | $\{-1\|0\}$ | $I^{\bar{1}}4_1/^{\bar{1}}a^{\infty m}1$ | t2 |
| 99.123.1.1 | 123.341 | $\{-1\|0\}$ | $P^{1}4/^{\bar{1}}m^{1}m^{1}m^{\infty m}1$ | t3 |
| 111.123.1.1 | 123.344 | $\{-1\|0\}$ | $P^{\bar{1}}4/^{\bar{1}}m^{\bar{1}}m^{1}m^{\infty m}1$ | t3 |

| 115.123.1.1 | 123.346 | $\{-1\vert 0\}$ | $P^{\bar{1}}4/^{\bar{1}}m^{1}m^{\bar{1}}m^{\infty m}1$ | t3 |
|---|---|---|---|---|
| 89.123.1.1 | 123.347 | $\{-1\vert 0\}$ | $P^{1}4/^{\bar{1}}m^{\bar{1}}m^{\bar{1}}m^{\infty m}1$ | t3 |
| 103.124.1.1 | 124.353 | $\{-1\vert 0\}$ | $P^{1}4/^{\bar{1}}m^{1}c^{1}c^{\infty m}1$ | t3 |
| 112.124.1.1 | 124.356 | $\{-1\vert 0\}$ | $P^{\bar{1}}4/^{\bar{1}}m^{\bar{1}}c^{1}c^{\infty m}1$ | t3 |
| 116.124.1.1 | 124.358 | $\{-1\vert 0\}$ | $P^{\bar{1}}4/^{\bar{1}}m^{1}c^{\bar{1}}c^{\infty m}1$ | t3 |
| 89.124.1.1 | 124.359 | $\{-1\vert 0\}$ | $P^{1}4/^{\bar{1}}m^{\bar{1}}c^{\bar{1}}c^{\infty m}1$ | t3 |
| 100.125.1.1 | 125.365 | $\{-1\vert 0\}$ | $P^{1}4/^{\bar{1}}n^{1}b^{1}m^{\infty m}1$ | t3 |
| 111.125.1.1 | 125.368 | $\{-1\vert 0\}$ | $P^{\bar{1}}4/^{\bar{1}}n^{\bar{1}}b^{1}m^{\infty m}1$ | t3 |
| 117.125.1.1 | 125.370 | $\{-1\vert 0\}$ | $P^{\bar{1}}4/^{\bar{1}}n^{1}b^{\bar{1}}m^{\infty m}1$ | t3 |
| 89.125.1.1 | 125.371 | $\{-1\vert 0\}$ | $P^{1}4/^{\bar{1}}n^{\bar{1}}b^{\bar{1}}m^{\infty m}1$ | t3 |
| 104.126.1.1 | 126.377 | $\{-1\vert 0\}$ | $P^{1}4/^{\bar{1}}n^{1}n^{1}c^{\infty m}1$ | t3 |
| 112.126.1.1 | 126.380 | $\{-1\vert 0\}$ | $P^{\bar{1}}4/^{\bar{1}}n^{\bar{1}}n^{1}c^{\infty m}1$ | t3 |
| 118.126.1.1 | 126.382 | $\{-1\vert 0\}$ | $P^{\bar{1}}4/^{\bar{1}}n^{1}n^{\bar{1}}c^{\infty m}1$ | t3 |
| 89.126.1.1 | 126.383 | $\{-1\vert 0\}$ | $P^{1}4/^{\bar{1}}n^{\bar{1}}n^{\bar{1}}c^{\infty m}1$ | t3 |
| 100.127.1.1 | 127.389 | $\{-1\vert 0\}$ | $P^{1}4/^{\bar{1}}m^{1}b^{1}m^{\infty m}1$ | t3 |
| 113.127.1.1 | 127.392 | $\{-1\vert 0\}$ | $P^{\bar{1}}4/^{\bar{1}}m^{\bar{1}}b^{1}m^{\infty m}1$ | t3 |
| 117.127.1.1 | 127.394 | $\{-1\vert 0\}$ | $P^{\bar{1}}4/^{\bar{1}}m^{1}b^{\bar{1}}m^{\infty m}1$ | t3 |
| 90.127.1.1 | 127.395 | $\{-1\vert 0\}$ | $P^{1}4/^{\bar{1}}m^{\bar{1}}b^{\bar{1}}m^{\infty m}1$ | t3 |
| 104.128.1.1 | 128.401 | $\{-1\vert 0\}$ | $P^{1}4/^{\bar{1}}m^{1}n^{1}c^{\infty m}1$ | t3 |
| 114.128.1.1 | 128.404 | $\{-1\vert 0\}$ | $P^{\bar{1}}4/^{\bar{1}}m^{\bar{1}}n^{1}c^{\infty m}1$ | t3 |
| 118.128.1.1 | 128.406 | $\{-1\vert 0\}$ | $P^{\bar{1}}4/^{\bar{1}}m^{1}n^{\bar{1}}c^{\infty m}1$ | t3 |
| 90.128.1.1 | 128.407 | $\{-1\vert 0\}$ | $P^{1}4/^{\bar{1}}m^{\bar{1}}n^{\bar{1}}c^{\infty m}1$ | t3 |
| 99.129.1.1 | 129.413 | $\{-1\vert 0\}$ | $P^{1}4/^{\bar{1}}n^{1}m^{1}m^{\infty m}1$ | t3 |
| 113.129.1.1 | 129.416 | $\{-1\vert 0\}$ | $P^{\bar{1}}4/^{\bar{1}}n^{\bar{1}}m^{1}m^{\infty m}1$ | t3 |
| 115.129.1.1 | 129.418 | $\{-1\vert 0\}$ | $P^{\bar{1}}4/^{\bar{1}}n^{1}m^{\bar{1}}m^{\infty m}1$ | t3 |
| 90.129.1.1 | 129.419 | $\{-1\vert 0\}$ | $P^{1}4/^{\bar{1}}n^{\bar{1}}m^{\bar{1}}m^{\infty m}1$ | t3 |
| 103.130.1.1 | 130.425 | $\{-1\vert 0\}$ | $P^{1}4/^{\bar{1}}n^{1}c^{1}c^{\infty m}1$ | t3 |
| 114.130.1.1 | 130.428 | $\{-1\vert 0\}$ | $P^{\bar{1}}4/^{\bar{1}}n^{\bar{1}}c^{1}c^{\infty m}1$ | t3 |
| 116.130.1.1 | 130.430 | $\{-1\vert 0\}$ | $P^{\bar{1}}4/^{\bar{1}}n^{1}c^{\bar{1}}c^{\infty m}1$ | t3 |
| 90.130.1.1 | 130.431 | $\{-1\vert 0\}$ | $P^{1}4/^{\bar{1}}n^{\bar{1}}c^{\bar{1}}c^{\infty m}1$ | t3 |
| 105.131.1.1 | 131.437 | $\{-1\vert 0\}$ | $P^{1}4_{2}/^{\bar{1}}m^{1}m^{1}c^{\infty m}1$ | t3 |
| 112.131.1.1 | 131.440 | $\{-1\vert 0\}$ | $P^{\bar{1}}4_{2}/^{\bar{1}}m^{\bar{1}}m^{1}c^{\infty m}1$ | t3 |
| 115.131.1.1 | 131.442 | $\{-1\vert 0\}$ | $P^{\bar{1}}4_{2}/^{\bar{1}}m^{1}m^{\bar{1}}c^{\infty m}1$ | t3 |
| 93.131.1.1 | 131.443 | $\{-1\vert 0\}$ | $P^{1}4_{2}/^{\bar{1}}m^{\bar{1}}m^{\bar{1}}c^{\infty m}1$ | t3 |
| 101.132.1.1 | 132.449 | $\{-1\vert 0\}$ | $P^{1}4_{2}/^{\bar{1}}m^{1}c^{1}m^{\infty m}1$ | t3 |
| 111.132.1.1 | 132.452 | $\{-1\vert 0\}$ | $P^{\bar{1}}4_{2}/^{\bar{1}}m^{\bar{1}}c^{1}m^{\infty m}1$ | t3 |
| 116.132.1.1 | 132.454 | $\{-1\vert 0\}$ | $P^{\bar{1}}4_{2}/^{\bar{1}}m^{1}c^{\bar{1}}m^{\infty m}1$ | t3 |
| 93.132.1.1 | 132.455 | $\{-1\vert 0\}$ | $P^{1}4_{2}/^{\bar{1}}m^{\bar{1}}c^{\bar{1}}m^{\infty m}1$ | t3 |
| 106.133.1.1 | 133.461 | $\{-1\vert 0\}$ | $P^{1}4_{2}/^{\bar{1}}n^{1}b^{1}c^{\infty m}1$ | t3 |
| 112.133.1.1 | 133.464 | $\{-1\vert 0\}$ | $P^{\bar{1}}4_{2}/^{\bar{1}}n^{\bar{1}}b^{1}c^{\infty m}1$ | t3 |
| 117.133.1.1 | 133.466 | $\{-1\vert 0\}$ | $P^{\bar{1}}4_{2}/^{\bar{1}}n^{1}b^{\bar{1}}c^{\infty m}1$ | t3 |
| 93.133.1.1 | 133.467 | $\{-1\vert 0\}$ | $P^{1}4_{2}/^{\bar{1}}n^{\bar{1}}b^{\bar{1}}c^{\infty m}1$ | t3 |
| 102.134.1.1 | 134.473 | $\{-1\vert 0\}$ | $P^{1}4_{2}/^{\bar{1}}n^{1}n^{1}m^{\infty m}1$ | t3 |

| 111.134.1.1 | 134.476 | $\{-1\|0\}$ | $P^{\bar{1}}4_2/^{\bar{1}}n^{\bar{1}}n^{1}m^{\infty m}1$ | t3 |
|---|---|---|---|---|
| 118.134.1.1 | 134.478 | $\{-1\|0\}$ | $P^{\bar{1}}4_2/^{\bar{1}}n^{1}n^{\bar{1}}m^{\infty m}1$ | t3 |
| 93.134.1.1 | 134.479 | $\{-1\|0\}$ | $P^{1}4_2/^{\bar{1}}n^{\bar{1}}n^{\bar{1}}m^{\infty m}1$ | t3 |
| 106.135.1.1 | 135.485 | $\{-1\|0\}$ | $P^{1}4_2/^{\bar{1}}n^{1}b^{1}c^{\infty m}1$ | t3 |
| 114.135.1.1 | 135.488 | $\{-1\|0\}$ | $P^{\bar{1}}4_2/^{\bar{1}}m^{\bar{1}}b^{1}c^{\infty m}1$ | t3 |
| 117.135.1.1 | 135.490 | $\{-1\|0\}$ | $P^{\bar{1}}4_2/^{\bar{1}}m^{1}b^{\bar{1}}c^{\infty m}1$ | t3 |
| 94.135.1.1 | 135.491 | $\{-1\|0\}$ | $P^{1}4_2/^{\bar{1}}m^{\bar{1}}b^{\bar{1}}c^{\infty m}1$ | t3 |
| 102.136.1.1 | 136.497 | $\{-1\|0\}$ | $P^{1}4_2/^{\bar{1}}m^{1}n^{1}m^{\infty m}1$ | t3 |
| 113.136.1.1 | 136.500 | $\{-1\|0\}$ | $P^{\bar{1}}4_2/^{\bar{1}}m^{\bar{1}}n^{1}m^{\infty m}1$ | t3 |
| 118.136.1.1 | 136.502 | $\{-1\|0\}$ | $P^{\bar{1}}4_2/^{\bar{1}}m^{1}n^{\bar{1}}m^{\infty m}1$ | t3 |
| 94.136.1.1 | 136.503 | $\{-1\|0\}$ | $P^{1}4_2/^{\bar{1}}m^{\bar{1}}n^{\bar{1}}m^{\infty m}1$ | t3 |
| 105.137.1.1 | 137.509 | $\{-1\|0\}$ | $P^{1}4_2/^{\bar{1}}n^{1}m^{1}c^{\infty m}1$ | t3 |
| 114.137.1.1 | 137.512 | $\{-1\|0\}$ | $P^{\bar{1}}4_2/^{\bar{1}}n^{\bar{1}}m^{1}c^{\infty m}1$ | t3 |
| 115.137.1.1 | 137.514 | $\{-1\|0\}$ | $P^{\bar{1}}4_2/^{\bar{1}}n^{1}m^{\bar{1}}c^{\infty m}1$ | t3 |
| 94.137.1.1 | 137.515 | $\{-1\|0\}$ | $P^{1}4_2/^{\bar{1}}n^{\bar{1}}m^{\bar{1}}c^{\infty m}1$ | t3 |
| 101.138.1.1 | 138.521 | $\{-1\|0\}$ | $P^{1}4_2/^{\bar{1}}n^{1}c^{1}m^{\infty m}1$ | t3 |
| 113.138.1.1 | 138.524 | $\{-1\|0\}$ | $P^{\bar{1}}4_2/^{\bar{1}}n^{\bar{1}}c^{1}m^{\infty m}1$ | t3 |
| 116.138.1.1 | 138.526 | $\{-1\|0\}$ | $P^{\bar{1}}4_2/^{\bar{1}}n^{1}c^{\bar{1}}m^{\infty m}1$ | t3 |
| 94.138.1.1 | 138.527 | $\{-1\|0\}$ | $P^{1}4_2/^{\bar{1}}n^{\bar{1}}c^{\bar{1}}m^{\infty m}1$ | t3 |
| 107.139.1.1 | 139.533 | $\{-1\|0\}$ | $I^{1}4/^{\bar{1}}m^{1}m^{1}m^{\infty m}1$ | t4 |
| 121.139.1.1 | 139.536 | $\{-1\|0\}$ | $I^{\bar{1}}4/^{\bar{1}}m^{\bar{1}}m^{1}m^{\infty m}1$ | t4 |
| 119.139.1.1 | 139.538 | $\{-1\|0\}$ | $I^{\bar{1}}4/^{\bar{1}}m^{1}m^{\bar{1}}m^{\infty m}1$ | t4 |
| 97.139.1.1 | 139.539 | $\{-1\|0\}$ | $I^{1}4/^{\bar{1}}m^{\bar{1}}m^{\bar{1}}m^{\infty m}1$ | t4 |
| 108.140.1.1 | 140.543 | $\{-1\|0\}$ | $I^{1}4/^{\bar{1}}m^{1}c^{1}m^{\infty m}1$ | t4 |
| 121.140.1.1 | 140.546 | $\{-1\|0\}$ | $I^{\bar{1}}4/^{\bar{1}}m^{\bar{1}}c^{1}m^{\infty m}1$ | t4 |
| 120.140.1.1 | 140.548 | $\{-1\|0\}$ | $I^{\bar{1}}4/^{\bar{1}}m^{1}c^{\bar{1}}m^{\infty m}1$ | t4 |
| 97.140.1.1 | 140.549 | $\{-1\|0\}$ | $I^{1}4/^{\bar{1}}m^{\bar{1}}c^{\bar{1}}m^{\infty m}1$ | t4 |
| 109.141.1.1 | 141.553 | $\{-1\|0\}$ | $I^{1}4_1/^{\bar{1}}a^{1}m^{1}d^{\infty m}1$ | t4 |
| 122.141.1.1 | 141.556 | $\{-1\|0\}$ | $I^{\bar{1}}4_1/^{\bar{1}}a^{\bar{1}}m^{1}d^{\infty m}1$ | t4 |
| 119.141.1.1 | 141.558 | $\{-1\|0\}$ | $I^{\bar{1}}4_1/^{\bar{1}}a^{1}m^{\bar{1}}d^{\infty m}1$ | t4 |
| 98.141.1.1 | 141.559 | $\{-1\|0\}$ | $I^{1}4_1/^{\bar{1}}a^{\bar{1}}m^{\bar{1}}d^{\infty m}1$ | t4 |
| 110.142.1.1 | 142.563 | $\{-1\|0\}$ | $I^{1}4_1/^{\bar{1}}a^{1}c^{1}d^{\infty m}1$ | t4 |
| 122.142.1.1 | 142.566 | $\{-1\|0\}$ | $I^{\bar{1}}4_1/^{\bar{1}}a^{\bar{1}}c^{1}d^{\infty m}1$ | t4 |
| 120.142.1.1 | 142.568 | $\{-1\|0\}$ | $I^{\bar{1}}4_1/^{\bar{1}}a^{1}c^{\bar{1}}d^{\infty m}1$ | t4 |
| 98.142.1.1 | 142.569 | $\{-1\|0\}$ | $I^{1}4_1/^{\bar{1}}a^{\bar{1}}c^{\bar{1}}d^{\infty m}1$ | t4 |
| 143.147.1.1 | 147.15 | $\{-1\|0\}$ | $P^{\bar{1}}\bar{3}^{\infty m}1$ | hP1 |
| 146.148.1.1 | 148.19 | $\{-1\|0\}$ | $R^{\bar{1}}\bar{3}^{\infty m}1$ | hR1 |
| 157.162.1.1 | 162.75 | $\{-1\|0\}$ | $P^{\bar{1}}\bar{3}^{1}1^{1}m^{\infty m}1$ | hP2 |
| 149.162.1.1 | 162.76 | $\{-1\|0\}$ | $P^{\bar{1}}\bar{3}^{1}1^{\bar{1}}m^{\infty m}1$ | hP2 |
| 159.163.1.1 | 163.81 | $\{-1\|0\}$ | $P^{\bar{1}}\bar{3}^{1}1^{1}c^{\infty m}1$ | hP2 |
| 149.163.1.1 | 163.82 | $\{-1\|0\}$ | $P^{\bar{1}}\bar{3}^{1}1^{\bar{1}}c^{\infty m}1$ | hP2 |
| 156.164.1.1 | 164.87 | $\{-1\|0\}$ | $P^{\bar{1}}\bar{3}^{1}m^{1}1^{\infty m}1$ | hP3 |
| 150.164.1.1 | 164.88 | $\{-1\|0\}$ | $P^{\bar{1}}\bar{3}^{\bar{1}}m^{1}1^{\infty m}1$ | hP3 |

| 158.165.1.1 | 165.93 | $\{-1\mid 0\}$ | $P^{\bar{1}}\bar{3}^{1}c^{1}1^{\infty m}1$ | hP3 |
| --- | --- | --- | --- | --- |
| 150.165.1.1 | 165.94 | $\{-1\mid 0\}$ | $P^{\bar{1}}\bar{3}^{\bar{1}}c^{1}1^{\infty m}1$ | hP3 |
| 160.166.1.1 | 166.99 | $\{-1\mid 0\}$ | $R^{\bar{1}}\bar{3}^{1}m^{\infty m}1$ | hR2 |
| 155.166.1.1 | 166.100 | $\{-1\mid 0\}$ | $R^{\bar{1}}\bar{3}^{\bar{1}}m^{\infty m}1$ | hR2 |
| 161.167.1.1 | 167.105 | $\{-1\mid 0\}$ | $R^{\bar{1}}\bar{3}^{1}c^{\infty m}1$ | hR2 |
| 155.167.1.1 | 167.106 | $\{-1\mid 0\}$ | $R^{\bar{1}}\bar{3}^{\bar{1}}c^{\infty m}1$ | hR2 |
| 174.175.1.1 | 175.139 | $\{-1\mid 0\}$ | $P^{\bar{1}}6/^{1}m^{\infty m}1$ | hP4 |
| 168.175.1.1 | 175.140 | $\{-1\mid 0\}$ | $P^{1}6/^{\bar{1}}m^{\infty m}1$ | hP4 |
| 174.176.1.1 | 176.145 | $\{-1\mid 0\}$ | $P^{\bar{1}}6_{3}/^{1}m^{\infty m}1$ | hP4 |
| 173.176.1.1 | 176.146 | $\{-1\mid 0\}$ | $P^{1}6_{3}/^{\bar{1}}m^{\infty m}1$ | hP4 |
| 183.191.1.1 | 191.235 | $\{-1\mid 0\}$ | $P^{1}6/^{\bar{1}}m^{1}m^{1}m^{\infty m}1$ | hP5 |
| 189.191.1.1 | 191.236 | $\{-1\mid 0\}$ | $P^{\bar{1}}6/^{1}m^{\bar{1}}m^{1}m^{\infty m}1$ | hP5 |
| 187.191.1.1 | 191.237 | $\{-1\mid 0\}$ | $P^{\bar{1}}6/^{1}m^{1}m^{\bar{1}}m^{\infty m}1$ | hP5 |
| 177.191.1.1 | 191.241 | $\{-1\mid 0\}$ | $P^{1}6/^{\bar{1}}m^{\bar{1}}m^{\bar{1}}m^{\infty m}1$ | hP5 |
| 184.192.1.1 | 192.245 | $\{-1\mid 0\}$ | $P^{1}6/^{\bar{1}}m^{1}c^{1}c^{\infty m}1$ | hP5 |
| 190.192.1.1 | 192.246 | $\{-1\mid 0\}$ | $P^{\bar{1}}6/^{1}m^{\bar{1}}c^{1}c^{\infty m}1$ | hP5 |
| 188.192.1.1 | 192.247 | $\{-1\mid 0\}$ | $P^{\bar{1}}6/^{1}m^{1}c^{\bar{1}}c^{\infty m}1$ | hP5 |
| 177.192.1.1 | 192.251 | $\{-1\mid 0\}$ | $P^{1}6/^{\bar{1}}m^{\bar{1}}c^{\bar{1}}c^{\infty m}1$ | hP5 |
| 185.193.1.1 | 193.255 | $\{-1\mid 0\}$ | $P^{1}6_{3}/^{\bar{1}}m^{1}c^{1}m^{\infty m}1$ | hP5 |
| 189.193.1.1 | 193.256 | $\{-1\mid 0\}$ | $P^{\bar{1}}6_{3}/^{1}m^{\bar{1}}c^{1}m^{\infty m}1$ | hP5 |
| 188.193.1.1 | 193.257 | $\{-1\mid 0\}$ | $P^{\bar{1}}6_{3}/^{1}m^{1}c^{\bar{1}}m^{\infty m}1$ | hP5 |
| 182.193.1.1 | 193.261 | $\{-1\mid 0\}$ | $P^{1}6_{3}/^{\bar{1}}m^{\bar{1}}c^{\bar{1}}m^{\infty m}1$ | hP5 |
| 186.194.1.1 | 194.265 | $\{-1\mid 0\}$ | $P^{1}6_{3}/^{\bar{1}}m^{1}m^{1}c^{\infty m}1$ | hP5 |
| 190.194.1.1 | 194.266 | $\{-1\mid 0\}$ | $P^{\bar{1}}6_{3}/^{1}m^{\bar{1}}m^{1}c^{\infty m}1$ | hP5 |
| 187.194.1.1 | 194.267 | $\{-1\mid 0\}$ | $P^{\bar{1}}6_{3}/^{1}m^{1}m^{\bar{1}}c^{\infty m}1$ | hP5 |
| 182.194.1.1 | 194.271 | $\{-1\mid 0\}$ | $P^{1}6_{3}/^{\bar{1}}m^{\bar{1}}m^{\bar{1}}c^{\infty m}1$ | hP5 |
| 195.200.1.1 | 200.16 | $\{-1\mid 0\}$ | $P^{\bar{1}}m^{\bar{1}}\bar{3}^{\infty m}1$ | c1 |
| 195.201.1.1 | 201.20 | $\{-1\mid 0\}$ | $P^{\bar{1}}n^{\bar{1}}\bar{3}^{\infty m}1$ | c1 |
| 196.202.1.1 | 202.24 | $\{-1\mid 0\}$ | $F^{\bar{1}}m^{\bar{1}}\bar{3}^{\infty m}1$ | c2 |
| 196.203.1.1 | 203.28 | $\{-1\mid 0\}$ | $F^{\bar{1}}d^{\bar{1}}\bar{3}^{\infty m}1$ | c2 |
| 197.204.1.1 | 204.32 | $\{-1\mid 0\}$ | $I^{\bar{1}}m^{\bar{1}}\bar{3}^{\infty m}1$ | c3 |
| 198.205.1.1 | 205.35 | $\{-1\mid 0\}$ | $P^{\bar{1}}a^{\bar{1}}\bar{3}^{\infty m}1$ | c1 |
| 199.206.1.1 | 206.39 | $\{-1\mid 0\}$ | $I^{\bar{1}}a^{\bar{1}}\bar{3}^{\infty m}1$ | c3 |
| 215.221.1.1 | 221.94 | $\{-1\mid 0\}$ | $P^{\bar{1}}m^{\bar{1}}\bar{3}^{1}m^{\infty m}1$ | c4 |
| 207.221.1.1 | 221.96 | $\{-1\mid 0\}$ | $P^{\bar{1}}m^{\bar{1}}\bar{3}^{\bar{1}}m^{\infty m}1$ | c4 |
| 218.222.1.1 | 222.100 | $\{-1\mid 0\}$ | $P^{\bar{1}}n^{\bar{1}}\bar{3}^{1}n^{\infty m}1$ | c4 |
| 207.222.1.1 | 222.102 | $\{-1\mid 0\}$ | $P^{\bar{1}}n^{\bar{1}}\bar{3}^{\bar{1}}n^{\infty m}1$ | c4 |
| 218.223.1.1 | 223.106 | $\{-1\mid 0\}$ | $P^{\bar{1}}m^{\bar{1}}\bar{3}^{1}n^{\infty m}1$ | c4 |
| 208.223.1.1 | 223.108 | $\{-1\mid 0\}$ | $P^{\bar{1}}m^{\bar{1}}\bar{3}^{\bar{1}}n^{\infty m}1$ | c4 |
| 215.224.1.1 | 224.112 | $\{-1\mid 0\}$ | $P^{\bar{1}}n^{\bar{1}}\bar{3}^{1}m^{\infty m}1$ | c4 |
| 208.224.1.1 | 224.114 | $\{-1\mid 0\}$ | $P^{\bar{1}}n^{\bar{1}}\bar{3}^{\bar{1}}m^{\infty m}1$ | c4 |
| 216.225.1.1 | 225.118 | $\{-1\mid 0\}$ | $F^{\bar{1}}m^{\bar{1}}\bar{3}^{1}m^{\infty m}1$ | c5 |
| 209.225.1.1 | 225.120 | $\{-1\mid 0\}$ | $F^{\bar{1}}m^{\bar{1}}\bar{3}^{\bar{1}}m^{\infty m}1$ | c5 |

| 219.226.1.1 | 226.124 | $\{-1\|0\}$ | $F^{\bar{1}}m^{\bar{1}}\bar{3}^{1}c^{\infty m}1$ | c5 |
|---|---|---|---|---|
| 209.226.1.1 | 226.126 | $\{-1\|0\}$ | $F^{\bar{1}}m^{\bar{1}}\bar{3}^{\bar{1}}c^{\infty m}1$ | c5 |
| 216.227.1.1 | 227.130 | $\{-1\|0\}$ | $F^{\bar{1}}d^{\bar{1}}\bar{3}^{1}m^{\infty m}1$ | c5 |
| 210.227.1.1 | 227.132 | $\{-1\|0\}$ | $F^{\bar{1}}d^{\bar{1}}\bar{3}^{\bar{1}}m^{\infty m}1$ | c5 |
| 219.228.1.1 | 228.136 | $\{-1\|0\}$ | $F^{\bar{1}}d^{\bar{1}}\bar{3}^{1}c^{\infty m}1$ | c5 |
| 210.228.1.1 | 228.138 | $\{-1\|0\}$ | $F^{\bar{1}}d^{\bar{1}}\bar{3}^{\bar{1}}c^{\infty m}1$ | c5 |
| 217.229.1.1 | 229.142 | $\{-1\|0\}$ | $I^{\bar{1}}m^{\bar{1}}\bar{3}^{1}m^{\infty m}1$ | c6 |
| 211.229.1.1 | 229.144 | $\{-1\|0\}$ | $I^{\bar{1}}m^{\bar{1}}\bar{3}^{\bar{1}}m^{\infty m}1$ | c6 |
| 220.230.1.1 | 230.147 | $\{-1\|0\}$ | $I^{\bar{1}}a^{\bar{1}}\bar{3}^{1}d^{\infty m}1$ | c6 |
| 214.230.1.1 | 230.149 | $\{-1\|0\}$ | $I^{\bar{1}}a^{\bar{1}}\bar{3}^{\bar{1}}d^{\infty m}1$ | c6 |

## S2.4. Type-III collinear SSGs

Supplementary Table XVI. Full tabulation of 422 type-III collinear SSGs describing altermagnets. The five columns represent the group number using a four-index nomenclature[8], the one-to-one correspondence with MSGs, the representative lattice-space element $A$ that connects the two opposite-spin sublattices, the international notation of collinear SSGs and the spin Brillouin zone, respectively. For detailed information about the notation used in the spin Brillouin zone, please refer to Supplementary Table VIII in Supplementary Information S1.

| $G_{NSS}$ | MSG | $A$ | $G_S$ | SBZ |
|---|---|---|---|---|
| 1.3.1.1 | 3.3 | $\{2_{010}\|0\}$ | $P^{\bar{1}}2^{\infty m}1$ | m1 |
| 1.4.1.1 | 4.9 | $\{2_{010}\|0\ 1/2\ 0\}$ | $P^{\bar{1}}2_1{}^{\infty m}1$ | m1 |
| 1.5.1.1 | 5.15 | $\{2_{010}\|0\}$ | $C^{\bar{1}}2^{\infty m}1$ | m2 |
| 1.6.1.1 | 6.20 | $\{m_{010}\|0\}$ | $P^{\bar{1}}m^{\infty m}1$ | m1 |
| 1.7.1.1 | 7.26 | $\{\mathrm{m}_{010}\|0\ 0\ 1/2\}$ | $P^{\bar{1}}c^{\infty m}1$ | m1 |
| 1.8.1.1 | 8.34 | $\{m_{010}\|0\}$ | $C^{\bar{1}}m^{\infty m}1$ | m2 |
| 1.9.1.1 | 9.39 | $\{\mathrm{m}_{010}\|0\ 0\ 1/2\}$ | $C^{\bar{1}}c^{\infty m}1$ | m2 |
| 2.10.1.1 | 10.46 | $\{2_{010}\|0\}$ | $P^{\bar{1}}2/^{\bar{1}}m^{\infty m}1$ | m1 |
| 2.11.1.1 | 11.54 | $\{2_{010}\|0\ 1/2\ 0\}$ | $P^{\bar{1}}2_1/^{\bar{1}}m^{\infty m}1$ | m1 |
| 2.12.1.1 | 12.62 | $\{2_{010}\|0\}$ | $C^{\bar{1}}2/^{\bar{1}}m^{\infty m}1$ | m2 |
| 2.13.1.1 | 13.69 | $\{2_{010}\|0\ 0\ 1/2\}$ | $P^{\bar{1}}2/^{\bar{1}}c^{\infty m}1$ | m1 |
| 2.14.1.1 | 14.79 | $\{2_{010}\|0\ 1/2\ 1/2\}$ | $P^{\bar{1}}2_1/^{\bar{1}}c^{\infty m}1$ | m1 |
| 2.15.1.1 | 15.89 | $\{2_{010}\|0\ 0\ 1/2\}$ | $C^{\bar{1}}2/^{\bar{1}}c^{\infty m}1$ | m2 |
| 3.16.1.1 | 16.3 | $\{2_{100}\|0\}$ | $P^{\bar{1}}2^{\bar{1}}2^{1}2^{\infty m}1$ | o1 |
| 4.17.1.1 | 17.9 | $\{2_{100}\|0\}$ | $P^{\bar{1}}2^{\bar{1}}2^{1}2_1{}^{\infty m}1$ | o1 |
| 3.17.1.1 | 17.10 | $\{2_{010}\|0\ 0\ 1/2\}$ | $P^{1}2^{\bar{1}}2^{\bar{1}}2_1{}^{\infty m}1$ | o1 |
| 3.18.1.1 | 18.18 | $\{2_{100}\|1/2\ 1/2\ 0\}$ | $P^{\bar{1}}2_1{}^{\bar{1}}2_1{}^{1}2^{\infty m}1$ | o1 |
| 4.18.1.1 | 18.19 | $\{2_{001}\|0\}$ | $P^{1}2_1{}^{\bar{1}}2_1{}^{\bar{1}}2^{\infty m}1$ | o1 |
| 4.19.1.1 | 19.27 | $\{2_{100}\|1/2\ 1/2\ 0\}$ | $P^{\bar{1}}2_1{}^{\bar{1}}2_1{}^{1}2_1{}^{\infty m}1$ | o1 |

| 4.20.1.1 | 20.33 | $\{2_{100}\vert 0\}$ | $C^{\bar{1}}2^{\bar{1}}2^{1}2_{1}{}^{\infty m}1$ | o2C |
|---|---|---|---|---|
| 5.20.1.1 | 20.34 | $\{2_{010}\vert 0\ 0\ 1/2\}$ | $C^{1}2^{\bar{1}}2^{\bar{1}}2_{1}{}^{\infty m}1$ | o2C |
| 3.21.1.1 | 21.40 | $\{2_{100}\vert 0\}$ | $C^{\bar{1}}2^{\bar{1}}2^{1}2^{\infty m}1$ | o2C |
| 5.21.1.1 | 21.41 | $\{2_{010}\vert 0\}$ | $C^{1}2^{\bar{1}}2^{\bar{1}}2^{\infty m}1$ | o2C |
| 5.22.1.1 | 22.47 | $\{2_{100}\vert 0\}$ | $F^{\bar{1}}2^{\bar{1}}2^{1}2^{\infty m}1$ | o3 |
| 5.23.1.1 | 23.51 | $\{2_{100}\vert 0\}$ | $I^{\bar{1}}2^{\bar{1}}2^{1}2^{\infty m}1$ | o4 |
| 5.24.1.1 | 24.55 | $\{2_{100}\vert 0\ 0\ 1/2\}$ | $I^{\bar{1}}2_{1}{}^{\bar{1}}2_{1}{}^{1}2_{1}{}^{\infty m}1$ | o4 |
| 6.25.1.1 | 25.59 | $\{m_{100}\vert 0\}$ | $P^{\bar{1}}m^{1}m^{\bar{1}}2^{\infty m}1$ | o1 |
| 3.25.1.1 | 25.60 | $\{m_{100}\vert 0\}$ | $P^{\bar{1}}m^{\bar{1}}m^{1}2^{\infty m}1$ | o1 |
| 7.26.1.1 | 26.68 | $\{m_{100}\vert 0\}$ | $P^{\bar{1}}m^{1}c^{\bar{1}}2_{1}{}^{\infty m}1$ | o1 |
| 6.26.1.1 | 26.69 | $\{m_{010}\vert 0\ 0\ 1/2\}$ | $P^{1}m^{\bar{1}}c^{\bar{1}}2_{1}{}^{\infty m}1$ | o1 |
| 4.26.1.1 | 26.70 | $\{m_{100}\vert 0\}$ | $P^{\bar{1}}m^{\bar{1}}c^{1}2_{1}{}^{\infty m}1$ | o1 |
| 7.27.1.1 | 27.80 | $\{m_{100}\vert 0\ 0\ 1/2\}$ | $P^{\bar{1}}c^{1}c^{\bar{1}}2^{\infty m}1$ | o1 |
| 3.27.1.1 | 27.81 | $\{m_{010}\vert 0\ 0\ 1/2\}$ | $P^{\bar{1}}c^{\bar{1}}c^{1}2^{\infty m}1$ | o1 |
| 7.28.1.1 | 28.89 | $\{m_{100}\vert 1/2\ 0\ 0\}$ | $P^{\bar{1}}m^{1}a^{\bar{1}}2^{\infty m}1$ | o1 |
| 6.28.1.1 | 28.90 | $\{m_{010}\vert 1/2\ 0\ 0\}$ | $P^{1}m^{\bar{1}}a^{\bar{1}}2^{\infty m}1$ | o1 |
| 3.28.1.1 | 28.91 | $\{m_{100}\vert 1/2\ 0\ 0\}$ | $P^{\bar{1}}m^{\bar{1}}a^{1}2^{\infty m}1$ | o1 |
| 7.29.1.1 | 29.101 | $\{m_{100}\vert 1/2\ 0\ 1/2\}$ | $P^{\bar{1}}c^{1}a^{\bar{1}}2_{1}{}^{\infty m}1$ | o1 |
| 7.29.1.4 | 29.102 | $\{m_{010}\vert 1/2\ 0\ 0\}$ | $P^{1}c^{\bar{1}}a^{\bar{1}}2_{1}{}^{\infty m}1$ | o1 |
| 4.29.1.1 | 29.103 | $\{m_{100}\vert 1/2\ 0\ 1/2\}$ | $P^{\bar{1}}c^{\bar{1}}a^{1}2_{1}{}^{\infty m}1$ | o1 |
| 7.30.1.1 | 30.113 | $\{m_{100}\vert 0\ 1/2\ 1/2\}$ | $P^{\bar{1}}n^{1}c^{\bar{1}}2^{\infty m}1$ | o1 |
| 7.30.1.4 | 30.114 | $\{m_{010}\vert 0\ 1/2\ 1/2\}$ | $P^{1}n^{\bar{1}}c^{\bar{1}}2^{\infty m}1$ | o1 |
| 3.30.1.1 | 30.115 | $\{m_{100}\vert 0\ 1/2\ 1/2\}$ | $P^{\bar{1}}n^{\bar{1}}c^{1}2^{\infty m}1$ | o1 |
| 7.31.1.1 | 31.125 | $\{m_{100}\vert 0\}$ | $P^{\bar{1}}m^{1}n^{\bar{1}}2_{1}{}^{\infty m}1$ | o1 |
| 6.31.1.1 | 31.126 | $\{m_{010}\vert 1/2\ 0\ 1/2\}$ | $P^{1}m^{\bar{1}}n^{\bar{1}}2_{1}{}^{\infty m}1$ | o1 |
| 4.31.1.1 | 31.127 | $\{m_{100}\vert 0\}$ | $P^{\bar{1}}m^{\bar{1}}n^{1}2_{1}{}^{\infty m}1$ | o1 |
| 7.32.1.1 | 32.137 | $\{m_{100}\vert 1/2\ 1/2\ 0\}$ | $P^{\bar{1}}b^{1}a^{\bar{1}}2^{\infty m}1$ | o1 |
| 3.32.1.1 | 32.138 | $\{m_{100}\vert 1/2\ 1/2\ 0\}$ | $P^{\bar{1}}b^{\bar{1}}a^{1}2^{\infty m}1$ | o1 |
| 7.33.1.1 | 33.146 | $\{m_{100}\vert 1/2\ 1/2\ 1/2\}$ | $P^{\bar{1}}n^{1}a^{\bar{1}}2_{1}{}^{\infty m}1$ | o1 |
| 7.33.1.4 | 33.147 | $\{m_{010}\vert 1/2\ 1/2\ 0\}$ | $P^{1}n^{\bar{1}}a^{\bar{1}}2_{1}{}^{\infty m}1$ | o1 |
| 4.33.1.1 | 33.148 | $\{m_{100}\vert 1/2\ 1/2\ 1/2\}$ | $P^{\bar{1}}n^{\bar{1}}a^{1}2_{1}{}^{\infty m}1$ | o1 |
| 7.34.1.1 | 34.158 | $\{m_{100}\vert 1/2\ 1/2\ 1/2\}$ | $P^{\bar{1}}n^{1}n^{\bar{1}}2^{\infty m}1$ | o1 |
| 3.34.1.1 | 34.159 | $\{m_{100}\vert 1/2\ 1/2\ 1/2\}$ | $P^{\bar{1}}n^{\bar{1}}n^{1}2^{\infty m}1$ | o1 |
| 8.35.1.1 | 35.167 | $\{m_{100}\vert 0\}$ | $C^{\bar{1}}m^{1}m^{\bar{1}}2^{\infty m}1$ | o2C |
| 3.35.1.1 | 35.168 | $\{m_{100}\vert 0\}$ | $C^{\bar{1}}m^{\bar{1}}m^{1}2^{\infty m}1$ | o2C |
| 9.36.1.1 | 36.174 | $\{m_{100}\vert 0\}$ | $C^{\bar{1}}m^{1}c^{\bar{1}}2_{1}{}^{\infty m}1$ | o2C |
| 8.36.1.1 | 36.175 | $\{m_{010}\vert 0\ 0\ 1/2\}$ | $C^{1}m^{\bar{1}}c^{\bar{1}}2_{1}{}^{\infty m}1$ | o2C |
| 4.36.1.1 | 36.176 | $\{m_{100}\vert 0\}$ | $C^{\bar{1}}m^{\bar{1}}c^{1}2_{1}{}^{\infty m}1$ | o2C |
| 9.37.1.1 | 37.182 | $\{m_{100}\vert 0\ 0\ 1/2\}$ | $C^{\bar{1}}c^{1}c^{\bar{1}}2^{\infty m}1$ | o2C |
| 3.37.1.1 | 37.183 | $\{m_{100}\vert 0\ 0\ 1/2\}$ | $C^{\bar{1}}c^{\bar{1}}c^{1}2^{\infty m}1$ | o2C |
| 8.38.1.1 | 38.189 | $\{m_{100}\vert 0\}$ | $A^{\bar{1}}m^{1}m^{\bar{1}}2^{\infty m}1$ | o2A |
| 6.38.1.1 | 38.190 | $\{m_{010}\vert 0\}$ | $A^{1}m^{\bar{1}}m^{\bar{1}}2^{\infty m}1$ | o2A |
| 5.38.1.1 | 38.191 | $\{m_{100}\vert 0\}$ | $A^{\bar{1}}m^{\bar{1}}m^{1}2^{\infty m}1$ | o2A |

| 8.39.1.1 | 39.197 | $\{m_{100}\|0\ 0\ 1/2\}$ | $A^{\bar{1}}e^{1}m^{\bar{1}}2^{\infty m}1$ | o2A |
|---|---|---|---|---|
| 7.39.1.1 | 39.198 | $\{m_{010}\|0\ 0\ 1/2\}$ | $A^{1}e^{\bar{1}}m^{\bar{1}}2^{\infty m}1$ | o2A |
| 5.39.1.1 | 39.199 | $\{m_{100}\|0\ 0\ 1/2\}$ | $A^{\bar{1}}e^{\bar{1}}m^{1}2^{\infty m}1$ | o2A |
| 9.40.1.1 | 40.205 | $\{m_{100}\|1/2\ 0\ 0\}$ | $A^{\bar{1}}m^{1}a^{\bar{1}}2^{\infty m}1$ | o2A |
| 6.40.1.1 | 40.206 | $\{m_{010}\|1/2\ 0\ 0\}$ | $A^{1}m^{\bar{1}}a^{\bar{1}}2^{\infty m}1$ | o2A |
| 5.40.1.1 | 40.207 | $\{m_{100}\|1/2\ 0\ 0\}$ | $A^{\bar{1}}m^{\bar{1}}a^{1}2^{\infty m}1$ | o2A |
| 9.41.1.1 | 41.213 | $\{m_{100}\|1/2\ 0\ 1/2\}$ | $A^{\bar{1}}e^{1}a^{\bar{1}}2^{\infty m}1$ | o2A |
| 7.41.1.1 | 41.214 | $\{m_{010}\|1/2\ 0\ 1/2\}$ | $A^{1}e^{\bar{1}}a^{\bar{1}}2^{\infty m}1$ | o2A |
| 5.41.1.1 | 41.215 | $\{m_{100}\|1/2\ 0\ 1/2\}$ | $A^{\bar{1}}e^{\bar{1}}a^{1}2^{\infty m}1$ | o2A |
| 8.42.1.1 | 42.221 | $\{m_{100}\|0\}$ | $F^{\bar{1}}m^{1}m^{\bar{1}}2^{\infty m}1$ | o3 |
| 5.42.1.1 | 42.222 | $\{m_{100}\|0\}$ | $F^{\bar{1}}m^{\bar{1}}m^{1}2^{\infty m}1$ | o3 |
| 9.43.1.1 | 43.226 | $\{m_{100}\|1/4\ 1/4\ 1/4\}$ | $F^{\bar{1}}d^{1}d^{\bar{1}}2^{\infty m}1$ | o3 |
| 5.43.1.1 | 43.227 | $\{m_{100}\|1/4\ 1/4\ 1/4\}$ | $F^{\bar{1}}d^{\bar{1}}d^{1}2^{\infty m}1$ | o3 |
| 8.44.1.1 | 44.231 | $\{m_{100}\|0\}$ | $I^{\bar{1}}m^{1}m^{\bar{1}}2^{\infty m}1$ | o4 |
| 5.44.1.1 | 44.232 | $\{m_{100}\|0\}$ | $I^{\bar{1}}m^{\bar{1}}m^{1}2^{\infty m}1$ | o4 |
| 9.45.1.1 | 45.237 | $\{m_{100}\|0\ 0\ 1/2\}$ | $I^{\bar{1}}b^{1}a^{\bar{1}}2^{\infty m}1$ | o4 |
| 5.45.1.1 | 45.238 | $\{m_{100}\|0\ 0\ 1/2\}$ | $I^{\bar{1}}b^{\bar{1}}a^{1}2^{\infty m}1$ | o4 |
| 9.46.1.1 | 46.243 | $\{m_{100}\|1/2\ 0\ 0\}$ | $I^{\bar{1}}m^{1}a^{\bar{1}}2^{\infty m}1$ | o4 |
| 8.46.1.1 | 46.244 | $\{m_{010}\|1/2\ 0\ 0\}$ | $I^{1}m^{\bar{1}}a^{\bar{1}}2^{\infty m}1$ | o4 |
| 5.46.1.1 | 46.245 | $\{m_{100}\|1/2\ 0\ 0\}$ | $I^{\bar{1}}m^{\bar{1}}a^{1}2^{\infty m}1$ | o4 |
| 10.47.1.1 | 47.252 | $\{2_{100}\|0\}$ | $P^{\bar{1}}m^{\bar{1}}m^{1}m^{\infty m}1$ | o1 |
| 13.48.1.1 | 48.260 | $\{2_{100}\|0\ 1/2\ 1/2\}$ | $P^{\bar{1}}n^{\bar{1}}n^{1}n^{\infty m}1$ | o1 |
| 10.49.1.1 | 49.269 | $\{2_{100}\|0\ 0\ 1/2\}$ | $P^{\bar{1}}c^{\bar{1}}c^{1}m^{\infty m}1$ | o1 |
| 13.49.1.1 | 49.270 | $\{2_{100}\|0\ 0\ 1/2\}$ | $P^{\bar{1}}c^{1}c^{\bar{1}}m^{\infty m}1$ | o1 |
| 13.50.1.1 | 50.281 | $\{2_{100}\|0\ 1/2\ 0\}$ | $P^{\bar{1}}b^{\bar{1}}a^{1}n^{\infty m}1$ | o1 |
| 13.50.1.4 | 50.282 | $\{2_{100}\|0\ 1/2\ 0\}$ | $P^{\bar{1}}b^{1}a^{\bar{1}}n^{\infty m}1$ | o1 |
| 13.51.1.1 | 51.294 | $\{2_{100}\|1/2\ 0\ 0\}$ | $P^{\bar{1}}m^{\bar{1}}m^{1}a^{\infty m}1$ | o1 |
| 11.51.1.1 | 51.295 | $\{2_{010}\|0\}$ | $P^{1}m^{\bar{1}}m^{\bar{1}}a^{\infty m}1$ | o1 |
| 10.51.1.1 | 51.296 | $\{2_{100}\|1/2\ 0\ 0\}$ | $P^{\bar{1}}m^{1}m^{\bar{1}}a^{\infty m}1$ | o1 |
| 13.52.1.1 | 52.310 | $\{2_{100}\|0\ 1/2\ 1/2\}$ | $P^{\bar{1}}n^{\bar{1}}n^{1}a^{\infty m}1$ | o1 |
| 13.52.1.4 | 52.311 | $\{2_{010}\|1/2\ 1/2\ 1/2\}$ | $P^{1}n^{\bar{1}}n^{\bar{1}}a^{\infty m}1$ | o1 |
| 14.52.1.1 | 52.312 | $\{2_{100}\|0\ 1/2\ 1/2\}$ | $P^{\bar{1}}n^{1}n^{\bar{1}}a^{\infty m}1$ | o1 |
| 14.53.1.1 | 53.326 | $\{2_{100}\|0\}$ | $P^{\bar{1}}m^{\bar{1}}n^{1}a^{\infty m}1$ | o1 |
| 10.53.1.1 | 53.327 | $\{2_{010}\|1/2\ 0\ 1/2\}$ | $P^{1}m^{\bar{1}}n^{\bar{1}}a^{\infty m}1$ | o1 |
| 13.53.1.1 | 53.328 | $\{2_{100}\|0\}$ | $P^{\bar{1}}m^{1}n^{\bar{1}}a^{\infty m}1$ | o1 |
| 13.54.1.1 | 54.342 | $\{2_{100}\|1/2\ 0\ 1/2\}$ | $P^{\bar{1}}c^{\bar{1}}c^{1}a^{\infty m}1$ | o1 |
| 14.54.1.1 | 54.343 | $\{2_{010}\|0\ 0\ 1/2\}$ | $P^{1}c^{\bar{1}}c^{\bar{1}}a^{\infty m}1$ | o1 |
| 13.54.1.4 | 54.344 | $\{2_{100}\|1/2\ 0\ 1/2\}$ | $P^{\bar{1}}c^{1}c^{\bar{1}}a^{\infty m}1$ | o1 |
| 10.55.1.1 | 55.357 | $\{2_{100}\|1/2\ 1/2\ 0\}$ | $P^{\bar{1}}b^{\bar{1}}a^{1}m^{\infty m}1$ | o1 |
| 14.55.1.1 | 55.358 | $\{2_{100}\|1/2\ 1/2\ 0\}$ | $P^{\bar{1}}b^{1}a^{\bar{1}}m^{\infty m}1$ | o1 |
| 13.56.1.1 | 56.369 | $\{2_{100}\|1/2\ 0\ 1/2\}$ | $P^{\bar{1}}c^{\bar{1}}c^{1}n^{\infty m}1$ | o1 |
| 14.56.1.1 | 56.370 | $\{2_{100}\|1/2\ 0\ 1/2\}$ | $P^{\bar{1}}c^{1}c^{\bar{1}}n^{\infty m}1$ | o1 |
| 11.57.1.1 | 57.382 | $\{2_{100}\|0\ 1/2\ 0\}$ | $P^{\bar{1}}b^{\bar{1}}c^{1}m^{\infty m}1$ | o1 |

| 13.57.1.1 | 57.383 | $\{2_{010}\vert 0\ 1/2\ 1/2\}$ | $P^{1}b^{\bar{1}}c^{\bar{1}}m^{\infty m}1$ | o1 |
|---|---|---|---|---|
| 14.57.1.1 | 57.384 | $\{2_{100}\vert 0\ 1/2\ 0\}$ | $P^{\bar{1}}b^{1}c^{\bar{1}}m^{\infty m}1$ | o1 |
| 10.58.1.1 | 58.397 | $\{2_{100}\vert 1/2\ 1/2\ 1/2\}$ | $P^{\bar{1}}n^{\bar{1}}n^{1}m^{\infty m}1$ | o1 |
| 14.58.1.1 | 58.398 | $\{2_{010}\vert 1/2\ 1/2\ 1/2\}$ | $P^{1}n^{\bar{1}}n^{\bar{1}}m^{\infty m}1$ | o1 |
| 13.59.1.1 | 59.409 | $\{2_{100}\vert 1/2\ 0\ 0\}$ | $P^{\bar{1}}m^{\bar{1}}m^{1}n^{\infty m}1$ | o1 |
| 11.59.1.1 | 59.410 | $\{2_{010}\vert 0\ 1/2\ 0\}$ | $P^{1}m^{\bar{1}}m^{\bar{1}}n^{\infty m}1$ | o1 |
| 14.60.1.1 | 60.422 | $\{2_{100}\vert 1/2\ 1/2\ 0\}$ | $P^{\bar{1}}b^{\bar{1}}c^{1}n^{\infty m}1$ | o1 |
| 14.60.1.4 | 60.423 | $\{2_{010}\vert 0\ 0\ 1/2\}$ | $P^{1}b^{\bar{1}}c^{\bar{1}}n^{\infty m}1$ | o1 |
| 13.60.1.1 | 60.424 | $\{2_{100}\vert 1/2\ 1/2\ 0\}$ | $P^{\bar{1}}b^{1}c^{\bar{1}}n^{\infty m}1$ | o1 |
| 14.61.1.1 | 61.436 | $\{2_{100}\vert 1/2\ 1/2\ 0\}$ | $P^{\bar{1}}b^{\bar{1}}c^{1}a^{\infty m}1$ | o1 |
| 14.62.1.1 | 62.446 | $\{2_{100}\vert 1/2\ 1/2\ 1/2\}$ | $P^{\bar{1}}n^{\bar{1}}m^{1}a^{\infty m}1$ | o1 |
| 14.62.1.4 | 62.447 | $\{2_{010}\vert 0\ 1/2\ 0\}$ | $P^{1}n^{\bar{1}}m^{\bar{1}}a^{\infty m}1$ | o1 |
| 11.62.1.1 | 62.448 | $\{2_{100}\vert 1/2\ 1/2\ 1/2\}$ | $P^{\bar{1}}n^{1}m^{\bar{1}}a^{\infty m}1$ | o1 |
| 11.63.1.1 | 63.462 | $\{2_{100}\vert 0\}$ | $C^{\bar{1}}m^{\bar{1}}c^{1}m^{\infty m}1$ | o2C |
| 12.63.1.1 | 63.463 | $\{2_{010}\vert 0\ 0\ 1/2\}$ | $C^{1}m^{\bar{1}}c^{\bar{1}}m^{\infty m}1$ | o2C |
| 15.63.1.1 | 63.464 | $\{2_{100}\vert 0\}$ | $C^{\bar{1}}m^{1}c^{\bar{1}}m^{\infty m}1$ | o2C |
| 14.64.1.1 | 64.474 | $\{2_{100}\vert 0\}$ | $C^{\bar{1}}m^{\bar{1}}c^{1}e^{\infty m}1$ | o2C |
| 12.64.1.1 | 64.475 | $\{2_{010}\vert 1/2\ 0\ 1/2\}$ | $C^{1}m^{\bar{1}}c^{\bar{1}}e^{\infty m}1$ | o2C |
| 15.64.1.1 | 64.476 | $\{2_{100}\vert 0\}$ | $C^{\bar{1}}m^{1}c^{\bar{1}}e^{\infty m}1$ | o2C |
| 10.65.1.1 | 65.485 | $\{2_{100}\vert 0\}$ | $C^{\bar{1}}m^{\bar{1}}m^{1}m^{\infty m}1$ | o2C |
| 12.65.1.1 | 65.486 | $\{2_{010}\vert 0\}$ | $C^{1}m^{\bar{1}}m^{\bar{1}}m^{\infty m}1$ | o2C |
| 10.66.1.1 | 66.495 | $\{2_{100}\vert 0\ 0\ 1/2\}$ | $C^{\bar{1}}c^{\bar{1}}c^{1}m^{\infty m}1$ | o2C |
| 15.66.1.1 | 66.496 | $\{2_{010}\vert 0\ 0\ 1/2\}$ | $C^{1}c^{\bar{1}}c^{\bar{1}}m^{\infty m}1$ | o2C |
| 13.67.1.1 | 67.505 | $\{2_{100}\vert 0\}$ | $C^{\bar{1}}m^{\bar{1}}m^{1}e^{\infty m}1$ | o2C |
| 12.67.1.1 | 67.506 | $\{2_{010}\vert 1/2\ 0\ 0\}$ | $C^{1}m^{\bar{1}}m^{\bar{1}}e^{\infty m}1$ | o2C |
| 13.68.1.1 | 68.515 | $\{2_{100}\vert 0\ 1/2\ 1/2\}$ | $C^{\bar{1}}c^{\bar{1}}c^{1}e^{\infty m}1$ | o2C |
| 15.68.1.1 | 68.516 | $\{2_{010}\vert 0\ 0\ 1/2\}$ | $C^{1}c^{\bar{1}}c^{\bar{1}}e^{\infty m}1$ | o2C |
| 12.69.1.1 | 69.524 | $\{2_{100}\vert 0\}$ | $F^{\bar{1}}m^{\bar{1}}m^{1}m^{\infty m}1$ | o3 |
| 15.70.1.1 | 70.530 | $\{2_{100}\vert 0\ 3/4\ 3/4\}$ | $F^{\bar{1}}d^{\bar{1}}d^{1}d^{\infty m}1$ | o3 |
| 12.71.1.1 | 71.536 | $\{2_{100}\vert 0\}$ | $I^{\bar{1}}m^{\bar{1}}m^{1}m^{\infty m}1$ | o4 |
| 12.72.1.1 | 72.543 | $\{2_{100}\vert 0\ 0\ 1/2\}$ | $I^{\bar{1}}b^{\bar{1}}a^{1}m^{\infty m}1$ | o4 |
| 15.72.1.1 | 72.544 | $\{2_{010}\vert 0\ 0\ 1/2\}$ | $I^{1}b^{\bar{1}}a^{\bar{1}}m^{\infty m}1$ | o4 |
| 15.73.1.1 | 73.551 | $\{2_{100}\vert 0\ 0\ 1/2\}$ | $I^{\bar{1}}b^{\bar{1}}c^{1}a^{\infty m}1$ | o4 |
| 15.74.1.1 | 74.558 | $\{2_{100}\vert 0\}$ | $I^{\bar{1}}m^{\bar{1}}m^{1}a^{\infty m}1$ | o4 |
| 12.74.1.1 | 74.559 | $\{2_{010}\vert 0\ 1/2\ 0\}$ | $I^{1}m^{\bar{1}}m^{\bar{1}}a^{\infty m}1$ | o4 |
| 3.75.1.1 | 75.3 | $\{4^{+}_{001}\vert 0\}$ | $P^{\bar{1}}4^{\infty m}1$ | t1 |
| 4.76.1.1 | 76.9 | $\{4^{+}_{001}\vert 0\ 0\ 1/4\}$ | $P^{\bar{1}}{4_1}^{\infty m}1$ | t1 |
| 3.77.1.1 | 77.15 | $\{4^{+}_{001}\vert 0\ 0\ 1/2\}$ | $P^{\bar{1}}{4_2}^{\infty m}1$ | t1 |
| 4.78.1.1 | 78.21 | $\{4^{+}_{001}\vert 0\ 0\ 3/4\}$ | $P^{\bar{1}}{4_3}^{\infty m}1$ | t1 |
| 5.79.1.1 | 79.27 | $\{4^{+}_{001}\vert 0\}$ | $I^{\bar{1}}4^{\infty m}1$ | t2 |
| 5.80.1.1 | 80.31 | $\{4^{+}_{001}\vert 0\ 1/2\ 1/4\}$ | $I^{\bar{1}}{4_1}^{\infty m}1$ | t2 |
| 3.81.1.1 | 81.35 | $\{-4^{+}_{001}\vert 0\}$ | $P^{\bar{1}}\bar{4}^{\infty m}1$ | t1 |
| 5.82.1.1 | 82.41 | $\{-4^{+}_{001}\vert 0\}$ | $I^{\bar{1}}\bar{4}^{\infty m}1$ | t2 |

| 10.83.1.1 | 83.45 | $\{4^{+}_{001}\vert 0\}$ | $P^{\bar{1}}4/^{1}m^{\infty m}1$ | t1 |
|---|---|---|---|---|
| 10.84.1.1 | 84.53 | $\{4^{+}_{001}\vert 0\ 0\ 1/2\}$ | $P^{\bar{1}}4_2/^{1}m^{\infty m}1$ | t1 |
| 13.85.1.1 | 85.61 | $\{4^{+}_{001}\vert 1/2\ 0\ 0\}$ | $P^{\bar{1}}4/^{1}n^{\infty m}1$ | t1 |
| 13.86.1.1 | 86.69 | $\{4^{+}_{001}\vert 0\ 1/2\ 1/2\}$ | $P^{\bar{1}}4_2/^{1}n^{\infty m}1$ | t1 |
| 12.87.1.1 | 87.77 | $\{4^{+}_{001}\vert 0\}$ | $I^{\bar{1}}4/^{1}m^{\infty m}1$ | t2 |
| 15.88.1.1 | 88.83 | $\{4^{+}_{001}\vert 3/4\ 1/4\ 1/4\}$ | $I^{\bar{1}}4_1/^{1}a^{\infty m}1$ | t2 |
| 16.89.1.1 | 89.89 | $\{2_{110}\vert 0\}$ | $P^{\bar{1}}4^{1}2^{\bar{1}}2^{\infty m}1$ | t3 |
| 75.89.1.1 | 89.90 | $\{2_{100}\vert 0\}$ | $P^{1}4^{\bar{1}}2^{\bar{1}}2^{\infty m}1$ | t3 |
| 21.89.1.1 | 89.91 | $\{2_{100}\vert 0\}$ | $P^{\bar{1}}4^{\bar{1}}2^{1}2^{\infty m}1$ | t3 |
| 18.90.1.1 | 90.97 | $\{2_{110}\vert 0\}$ | $P^{\bar{1}}4^{1}2_1{}^{\bar{1}}2^{\infty m}1$ | t3 |
| 75.90.1.1 | 90.98 | $\{2_{100}\vert 1/2\ 1/2\ 0\}$ | $P^{1}4^{\bar{1}}2_1{}^{\bar{1}}2^{\infty m}1$ | t3 |
| 21.90.1.1 | 90.99 | $\{2_{100}\vert 1/2\ 1/2\ 0\}$ | $P^{\bar{1}}4^{\bar{1}}2_1{}^{1}2^{\infty m}1$ | t3 |
| 17.91.1.1 | 91.105 | $\{2_{110}\vert 0\ 0\ 3/4\}$ | $P^{\bar{1}}4_1{}^{1}2^{\bar{1}}2^{\infty m}1$ | t3 |
| 76.91.1.1 | 91.106 | $\{2_{100}\vert 0\ 0\ 1/2\}$ | $P^{1}4_1{}^{\bar{1}}2^{\bar{1}}2^{\infty m}1$ | t3 |
| 20.91.1.1 | 91.107 | $\{2_{100}\vert 0\ 0\ 1/2\}$ | $P^{\bar{1}}4_1{}^{\bar{1}}2^{1}2^{\infty m}1$ | t3 |
| 19.92.1.1 | 92.113 | $\{2_{110}\vert 0\}$ | $P^{\bar{1}}4_1{}^{1}2_1{}^{\bar{1}}2^{\infty m}1$ | t3 |
| 76.92.1.1 | 92.114 | $\{2_{100}\vert 1/2\ 1/2\ 3/4\}$ | $P^{1}4_1{}^{\bar{1}}2_1{}^{\bar{1}}2^{\infty m}1$ | t3 |
| 20.92.1.1 | 92.115 | $\{2_{100}\vert 1/2\ 1/2\ 3/4\}$ | $P^{\bar{1}}4_1{}^{\bar{1}}2_1{}^{1}2^{\infty m}1$ | t3 |
| 16.93.1.1 | 93.121 | $\{2_{110}\vert 0\ 0\ 1/2\}$ | $P^{\bar{1}}4_2{}^{1}2^{\bar{1}}2^{\infty m}1$ | t3 |
| 77.93.1.1 | 93.122 | $\{2_{100}\vert 0\}$ | $P^{1}4_2{}^{\bar{1}}2^{\bar{1}}2^{\infty m}1$ | t3 |
| 21.93.1.1 | 93.123 | $\{2_{100}\vert 0\}$ | $P^{\bar{1}}4_2{}^{\bar{1}}2^{1}2^{\infty m}1$ | t3 |
| 18.94.1.1 | 94.129 | $\{2_{110}\vert 0\}$ | $P^{\bar{1}}4_2{}^{1}2_1{}^{\bar{1}}2^{\infty m}1$ | t3 |
| 77.94.1.1 | 94.130 | $\{2_{100}\vert 1/2\ 1/2\ 1/2\}$ | $P^{1}4_2{}^{\bar{1}}2_1{}^{\bar{1}}2^{\infty m}1$ | t3 |
| 21.94.1.1 | 94.131 | $\{2_{100}\vert 1/2\ 1/2\ 1/2\}$ | $P^{\bar{1}}4_2{}^{\bar{1}}2_1{}^{1}2^{\infty m}1$ | t3 |
| 17.95.1.1 | 95.137 | $\{2_{110}\vert 0\ 0\ 1/4\}$ | $P^{\bar{1}}4_3{}^{1}2^{\bar{1}}2^{\infty m}1$ | t3 |
| 78.95.1.1 | 95.138 | $\{2_{100}\vert 0\ 0\ 1/2\}$ | $P^{1}4_3{}^{\bar{1}}2^{\bar{1}}2^{\infty m}1$ | t3 |
| 20.95.1.1 | 95.139 | $\{2_{100}\vert 0\ 0\ 1/2\}$ | $P^{\bar{1}}4_3{}^{\bar{1}}2^{1}2^{\infty m}1$ | t3 |
| 19.96.1.1 | 96.145 | $\{2_{110}\vert 0\}$ | $P^{\bar{1}}4_3{}^{1}2_1{}^{\bar{1}}2^{\infty m}1$ | t3 |
| 78.96.1.1 | 96.146 | $\{2_{100}\vert 1/2\ 1/2\ 1/4\}$ | $P^{1}4_3{}^{\bar{1}}2_1{}^{\bar{1}}2^{\infty m}1$ | t3 |
| 20.96.1.1 | 96.147 | $\{2_{100}\vert 1/2\ 1/2\ 1/4\}$ | $P^{\bar{1}}4_3{}^{\bar{1}}2_1{}^{1}2^{\infty m}1$ | t3 |
| 23.97.1.1 | 97.153 | $\{2_{110}\vert 0\}$ | $I^{\bar{1}}4^{1}2^{\bar{1}}2^{\infty m}1$ | t4 |
| 79.97.1.1 | 97.154 | $\{2_{100}\vert 0\}$ | $I^{1}4^{\bar{1}}2^{\bar{1}}2^{\infty m}1$ | t4 |
| 22.97.1.1 | 97.155 | $\{2_{100}\vert 0\}$ | $I^{\bar{1}}4^{\bar{1}}2^{1}2^{\infty m}1$ | t4 |
| 24.98.1.1 | 98.159 | $\{2_{110}\vert 0\}$ | $I^{\bar{1}}4_1{}^{1}2^{\bar{1}}2^{\infty m}1$ | t4 |
| 80.98.1.1 | 98.160 | $\{2_{100}\vert 0\ 1/2\ 1/4\}$ | $I^{1}4_1{}^{\bar{1}}2^{\bar{1}}2^{\infty m}1$ | t4 |
| 22.98.1.1 | 98.161 | $\{2_{100}\vert 0\ 1/2\ 1/4\}$ | $I^{\bar{1}}4_1{}^{\bar{1}}2^{1}2^{\infty m}1$ | t4 |
| 35.99.1.1 | 99.165 | $\{m_{100}\vert 0\}$ | $P^{\bar{1}}4^{\bar{1}}m^{1}m^{\infty m}1$ | t3 |
| 25.99.1.1 | 99.166 | $\{m_{110}\vert 0\}$ | $P^{\bar{1}}4^{1}m^{\bar{1}}m^{\infty m}1$ | t3 |
| 75.99.1.1 | 99.167 | $\{m_{100}\vert 0\}$ | $P^{1}4^{\bar{1}}m^{\bar{1}}m^{\infty m}1$ | t3 |
| 35.100.1.1 | 100.173 | $\{m_{100}\vert 1/2\ 1/2\ 0\}$ | $P^{\bar{1}}4^{\bar{1}}b^{1}m^{\infty m}1$ | t3 |
| 32.100.1.1 | 100.174 | $\{m_{110}\vert 1/2\ 1/2\ 0\}$ | $P^{\bar{1}}4^{1}b^{\bar{1}}m^{\infty m}1$ | t3 |
| 75.100.1.1 | 100.175 | $\{m_{100}\vert 1/2\ 1/2\ 0\}$ | $P^{1}4^{\bar{1}}b^{\bar{1}}m^{\infty m}1$ | t3 |
| 35.101.1.1 | 101.181 | $\{m_{100}\vert 0\ 0\ 1/2\}$ | $P^{\bar{1}}4_2{}^{\bar{1}}c^{1}m^{\infty m}1$ | t3 |

| 27.101.1.1 | 101.182 | $\{m_{110}\vert 0\}$ | $P^{\bar{1}}4_2{}^{1}c^{\bar{1}}m^{\infty m}1$ | t3 |
|---|---|---|---|---|
| 77.101.1.1 | 101.183 | $\{m_{100}\vert 0\ 0\ 1/2\}$ | $P^{1}4_2{}^{\bar{1}}c^{\bar{1}}m^{\infty m}1$ | t3 |
| 35.102.1.1 | 102.189 | $\{m_{100}\vert 1/2\ 1/2\ 1/2\}$ | $P^{\bar{1}}4_2{}^{\bar{1}}n^{1}m^{\infty m}1$ | t3 |
| 34.102.1.1 | 102.190 | $\{m_{110}\vert 0\}$ | $P^{\bar{1}}4_2{}^{1}n^{\bar{1}}m^{\infty m}1$ | t3 |
| 77.102.1.1 | 102.191 | $\{m_{100}\vert 1/2\ 1/2\ 1/2\}$ | $P^{1}4_2{}^{\bar{1}}n^{\bar{1}}m^{\infty m}1$ | t3 |
| 37.103.1.1 | 103.197 | $\{m_{100}\vert 0\ 0\ 1/2\}$ | $P^{\bar{1}}4^{\bar{1}}c^{1}c^{\infty m}1$ | t3 |
| 27.103.1.1 | 103.198 | $\{m_{110}\vert 0\ 0\ 1/2\}$ | $P^{\bar{1}}4^{1}c^{\bar{1}}c^{\infty m}1$ | t3 |
| 75.103.1.1 | 103.199 | $\{m_{100}\vert 0\ 0\ 1/2\}$ | $P^{1}4^{\bar{1}}c^{\bar{1}}c^{\infty m}1$ | t3 |
| 37.104.1.1 | 104.205 | $\{m_{100}\vert 1/2\ 1/2\ 1/2\}$ | $P^{\bar{1}}4^{\bar{1}}n^{1}c^{\infty m}1$ | t3 |
| 34.104.1.1 | 104.206 | $\{m_{110}\vert 1/2\ 1/2\ 1/2\}$ | $P^{\bar{1}}4^{1}n^{\bar{1}}c^{\infty m}1$ | t3 |
| 75.104.1.1 | 104.207 | $\{m_{100}\vert 1/2\ 1/2\ 1/2\}$ | $P^{1}4^{\bar{1}}n^{\bar{1}}c^{\infty m}1$ | t3 |
| 37.105.1.1 | 105.213 | $\{m_{100}\vert 0\}$ | $P^{\bar{1}}4_2{}^{\bar{1}}m^{1}c^{\infty m}1$ | t3 |
| 25.105.1.1 | 105.214 | $\{m_{110}\vert 0\ 0\ 1/2\}$ | $P^{\bar{1}}4_2{}^{1}m^{\bar{1}}c^{\infty m}1$ | t3 |
| 77.105.1.1 | 105.215 | $\{m_{100}\vert 0\}$ | $P^{1}4_2{}^{\bar{1}}m^{\bar{1}}c^{\infty m}1$ | t3 |
| 37.106.1.1 | 106.221 | $\{m_{100}\vert 1/2\ 1/2\ 0\}$ | $P^{\bar{1}}4_2{}^{\bar{1}}b^{1}c^{\infty m}1$ | t3 |
| 32.106.1.1 | 106.222 | $\{m_{110}\vert 1/2\ 1/2\ 1/2\}$ | $P^{\bar{1}}4_2{}^{1}b^{\bar{1}}c^{\infty m}1$ | t3 |
| 77.106.1.1 | 106.223 | $\{m_{100}\vert 1/2\ 1/2\ 0\}$ | $P^{1}4_2{}^{\bar{1}}b^{\bar{1}}c^{\infty m}1$ | t3 |
| 42.107.1.1 | 107.229 | $\{m_{100}\vert 0\}$ | $I^{\bar{1}}4^{\bar{1}}m^{1}m^{\infty m}1$ | t4 |
| 44.107.1.1 | 107.230 | $\{m_{110}\vert 0\}$ | $I^{\bar{1}}4^{1}m^{\bar{1}}m^{\infty m}1$ | t4 |
| 79.107.1.1 | 107.231 | $\{m_{100}\vert 0\}$ | $I^{1}4^{\bar{1}}m^{\bar{1}}m^{\infty m}1$ | t4 |
| 42.108.1.1 | 108.235 | $\{m_{100}\vert 0\ 0\ 1/2\}$ | $I^{\bar{1}}4^{\bar{1}}c^{1}m^{\infty m}1$ | t4 |
| 45.108.1.1 | 108.236 | $\{m_{110}\vert 0\ 0\ 1/2\}$ | $I^{\bar{1}}4^{1}c^{\bar{1}}m^{\infty m}1$ | t4 |
| 79.108.1.1 | 108.237 | $\{m_{100}\vert 0\ 0\ 1/2\}$ | $I^{1}4^{\bar{1}}c^{\bar{1}}m^{\infty m}1$ | t4 |
| 43.109.1.1 | 109.241 | $\{m_{100}\vert 0\}$ | $I^{\bar{1}}4_1{}^{\bar{1}}m^{1}d^{\infty m}1$ | t4 |
| 44.109.1.1 | 109.242 | $\{m_{110}\vert 0\ 1/2\ 1/4\}$ | $I^{\bar{1}}4_1{}^{1}m^{\bar{1}}d^{\infty m}1$ | t4 |
| 80.109.1.1 | 109.243 | $\{m_{100}\vert 0\}$ | $I^{1}4_1{}^{\bar{1}}m^{\bar{1}}d^{\infty m}1$ | t4 |
| 43.110.1.1 | 110.247 | $\{m_{100}\vert 0\ 0\ 1/2\}$ | $I^{\bar{1}}4_1{}^{\bar{1}}c^{1}d^{\infty m}1$ | t4 |
| 45.110.1.1 | 110.248 | $\{m_{110}\vert 1/2\ 0\ 1/4\}$ | $I^{\bar{1}}4_1{}^{1}c^{\bar{1}}d^{\infty m}1$ | t4 |
| 80.110.1.1 | 110.249 | $\{m_{100}\vert 0\ 0\ 1/2\}$ | $I^{1}4_1{}^{\bar{1}}c^{\bar{1}}d^{\infty m}1$ | t4 |
| 35.111.1.1 | 111.253 | $\{2_{100}\vert 0\}$ | $P^{\bar{1}}\bar{4}^{\bar{1}}2^{1}m^{\infty m}1$ | t3 |
| 16.111.1.1 | 111.254 | $\{m_{110}\vert 0\}$ | $P^{\bar{1}}\bar{4}^{1}2^{\bar{1}}m^{\infty m}1$ | t3 |
| 81.111.1.1 | 111.255 | $\{2_{100}\vert 0\}$ | $P^{1}\bar{4}^{\bar{1}}2^{\bar{1}}m^{\infty m}1$ | t3 |
| 37.112.1.1 | 112.261 | $\{2_{100}\vert 0\ 0\ 1/2\}$ | $P^{\bar{1}}\bar{4}^{\bar{1}}2^{1}c^{\infty m}1$ | t3 |
| 16.112.1.1 | 112.262 | $\{m_{110}\vert 0\ 0\ 1/2\}$ | $P^{\bar{1}}\bar{4}^{1}2^{\bar{1}}c^{\infty m}1$ | t3 |
| 81.112.1.1 | 112.263 | $\{2_{100}\vert 0\ 0\ 1/2\}$ | $P^{1}\bar{4}^{\bar{1}}2^{\bar{1}}c^{\infty m}1$ | t3 |
| 35.113.1.1 | 113.269 | $\{2_{100}\vert 1/2\ 1/2\ 0\}$ | $P^{\bar{1}}\bar{4}^{\bar{1}}2_1{}^{1}m^{\infty m}1$ | t3 |
| 18.113.1.1 | 113.270 | $\{m_{110}\vert 1/2\ 1/2\ 0\}$ | $P^{\bar{1}}\bar{4}^{1}2_1{}^{\bar{1}}m^{\infty m}1$ | t3 |
| 81.113.1.1 | 113.271 | $\{2_{100}\vert 1/2\ 1/2\ 0\}$ | $P^{1}\bar{4}^{\bar{1}}2_1{}^{\bar{1}}m^{\infty m}1$ | t3 |
| 37.114.1.1 | 114.277 | $\{2_{100}\vert 1/2\ 1/2\ 1/2\}$ | $P^{\bar{1}}\bar{4}^{\bar{1}}2_1{}^{1}c^{\infty m}1$ | t3 |
| 18.114.1.1 | 114.278 | $\{m_{110}\vert 1/2\ 1/2\ 1/2\}$ | $P^{\bar{1}}\bar{4}^{1}2_1{}^{\bar{1}}c^{\infty m}1$ | t3 |
| 81.114.1.1 | 114.279 | $\{2_{100}\vert 1/2\ 1/2\ 1/2\}$ | $P^{1}\bar{4}^{\bar{1}}2_1{}^{\bar{1}}c^{\infty m}1$ | t3 |
| 21.115.1.1 | 115.285 | $\{m_{100}\vert 0\}$ | $P^{\bar{1}}\bar{4}^{\bar{1}}m^{1}2^{\infty m}1$ | t3 |
| 25.115.1.1 | 115.286 | $\{2_{110}\vert 0\}$ | $P^{\bar{1}}\bar{4}^{1}m^{\bar{1}}2^{\infty m}1$ | t3 |
| 81.115.1.1 | 115.287 | $\{2_{110}\|0\}$ | $P^{1}\bar{4}^{\bar{1}}m^{\bar{1}}2^{\infty m}1$ | t3 |
| 21.116.1.1 | 116.293 | $\{m_{100}\|0\ 0\ 1/2\}$ | $P^{\bar{1}}\bar{4}^{\bar{1}}c^{1}2^{\infty m}1$ | t3 |
| 27.116.1.1 | 116.294 | $\{2_{110}\|0\ 0\ 1/2\}$ | $P^{\bar{1}}\bar{4}^{1}c^{\bar{1}}2^{\infty m}1$ | t3 |
| 81.116.1.1 | 116.295 | $\{2_{110}\|0\ 0\ 1/2\}$ | $P^{1}\bar{4}^{\bar{1}}c^{\bar{1}}2^{\infty m}1$ | t3 |
| 21.117.1.1 | 117.301 | $\{m_{100}\|1/2\ 1/2\ 0\}$ | $P^{\bar{1}}\bar{4}^{\bar{1}}b^{1}2^{\infty m}1$ | t3 |
| 32.117.1.1 | 117.302 | $\{2_{110}\|1/2\ 1/2\ 0\}$ | $P^{\bar{1}}\bar{4}^{1}b^{\bar{1}}2^{\infty m}1$ | t3 |
| 81.117.1.1 | 117.303 | $\{2_{110}\|1/2\ 1/2\ 0\}$ | $P^{1}\bar{4}^{\bar{1}}b^{\bar{1}}2^{\infty m}1$ | t3 |
| 21.118.1.1 | 118.309 | $\{m_{100}\|1/2\ 1/2\ 1/2\}$ | $P^{\bar{1}}\bar{4}^{\bar{1}}n^{1}2^{\infty m}1$ | t3 |
| 34.118.1.1 | 118.310 | $\{2_{110}\|1/2\ 1/2\ 1/2\}$ | $P^{\bar{1}}\bar{4}^{1}n^{\bar{1}}2^{\infty m}1$ | t3 |
| 81.118.1.1 | 118.311 | $\{2_{110}\|1/2\ 1/2\ 1/2\}$ | $P^{1}\bar{4}^{\bar{1}}n^{\bar{1}}2^{\infty m}1$ | t3 |
| 22.119.1.1 | 119.317 | $\{m_{100}\|0\}$ | $I^{\bar{1}}\bar{4}^{\bar{1}}m^{1}2^{\infty m}1$ | t4 |
| 44.119.1.1 | 119.318 | $\{2_{110}\|0\}$ | $I^{\bar{1}}\bar{4}^{1}m^{\bar{1}}2^{\infty m}1$ | t4 |
| 82.119.1.1 | 119.319 | $\{2_{110}\|0\}$ | $I^{1}\bar{4}^{\bar{1}}m^{\bar{1}}2^{\infty m}1$ | t4 |
| 22.120.1.1 | 120.323 | $\{m_{100}\|0\ 0\ 1/2\}$ | $I^{\bar{1}}\bar{4}^{\bar{1}}c^{1}2^{\infty m}1$ | t4 |
| 45.120.1.1 | 120.324 | $\{2_{110}\|0\ 0\ 1/2\}$ | $I^{\bar{1}}\bar{4}^{1}c^{\bar{1}}2^{\infty m}1$ | t4 |
| 82.120.1.1 | 120.325 | $\{2_{110}\|0\ 0\ 1/2\}$ | $I^{1}\bar{4}^{\bar{1}}c^{\bar{1}}2^{\infty m}1$ | t4 |
| 42.121.1.1 | 121.329 | $\{2_{100}\|0\}$ | $I^{\bar{1}}\bar{4}^{\bar{1}}2^{1}m^{\infty m}1$ | t4 |
| 23.121.1.1 | 121.330 | $\{m_{110}\|0\}$ | $I^{\bar{1}}\bar{4}^{1}2^{\bar{1}}m^{\infty m}1$ | t4 |
| 82.121.1.1 | 121.331 | $\{2_{100}\|0\}$ | $I^{1}\bar{4}^{\bar{1}}2^{\bar{1}}m^{\infty m}1$ | t4 |
| 43.122.1.1 | 122.335 | $\{2_{100}\|0\ 1/2\ 1/4\}$ | $I^{\bar{1}}\bar{4}^{\bar{1}}2^{1}d^{\infty m}1$ | t4 |
| 24.122.1.1 | 122.336 | $\{m_{110}\|0\ 1/2\ 1/4\}$ | $I^{\bar{1}}\bar{4}^{1}2^{\bar{1}}d^{\infty m}1$ | t4 |
| 82.122.1.1 | 122.337 | $\{2_{100}\|0\ 1/2\ 1/4\}$ | $I^{1}\bar{4}^{\bar{1}}2^{\bar{1}}d^{\infty m}1$ | t4 |
| 65.123.1.1 | 123.342 | $\{2_{100}\|0\}$ | $P^{\bar{1}}4/^{1}m^{\bar{1}}m^{1}m^{\infty m}1$ | t3 |
| 47.123.1.1 | 123.343 | $\{2_{110}\|0\}$ | $P^{\bar{1}}4/^{1}m^{1}m^{\bar{1}}m^{\infty m}1$ | t3 |
| 83.123.1.1 | 123.345 | $\{2_{100}\|0\}$ | $P^{1}4/^{1}m^{\bar{1}}m^{\bar{1}}m^{\infty m}1$ | t3 |
| 66.124.1.1 | 124.354 | $\{2_{100}\|0\ 0\ 1/2\}$ | $P^{\bar{1}}4/^{1}m^{\bar{1}}c^{1}c^{\infty m}1$ | t3 |
| 49.124.1.1 | 124.355 | $\{2_{110}\|0\ 0\ 1/2\}$ | $P^{\bar{1}}4/^{1}m^{1}c^{\bar{1}}c^{\infty m}1$ | t3 |
| 83.124.1.1 | 124.357 | $\{2_{100}\|0\ 0\ 1/2\}$ | $P^{1}4/^{1}m^{\bar{1}}c^{\bar{1}}c^{\infty m}1$ | t3 |
| 67.125.1.1 | 125.366 | $\{2_{100}\|0\ 1/2\ 0\}$ | $P^{\bar{1}}4/^{1}n^{\bar{1}}b^{1}m^{\infty m}1$ | t3 |
| 50.125.1.1 | 125.367 | $\{2_{110}\|0\}$ | $P^{\bar{1}}4/^{1}n^{1}b^{\bar{1}}m^{\infty m}1$ | t3 |
| 85.125.1.1 | 125.369 | $\{2_{100}\|0\ 1/2\ 0\}$ | $P^{1}4/^{1}n^{\bar{1}}b^{\bar{1}}m^{\infty m}1$ | t3 |
| 68.126.1.1 | 126.378 | $\{2_{100}\|0\ 1/2\ 1/2\}$ | $P^{\bar{1}}4/^{1}n^{\bar{1}}n^{1}c^{\infty m}1$ | t3 |
| 48.126.1.1 | 126.379 | $\{2_{110}\|0\ 0\ 1/2\}$ | $P^{\bar{1}}4/^{1}n^{1}n^{\bar{1}}c^{\infty m}1$ | t3 |
| 85.126.1.1 | 126.381 | $\{2_{100}\|0\ 1/2\ 1/2\}$ | $P^{1}4/^{1}n^{\bar{1}}n^{\bar{1}}c^{\infty m}1$ | t3 |
| 65.127.1.1 | 127.390 | $\{2_{100}\|1/2\ 1/2\ 0\}$ | $P^{\bar{1}}4/^{1}m^{\bar{1}}b^{1}m^{\infty m}1$ | t3 |
| 55.127.1.1 | 127.391 | $\{2_{110}\|1/2\ 1/2\ 0\}$ | $P^{\bar{1}}4/^{1}m^{1}b^{\bar{1}}m^{\infty m}1$ | t3 |
| 83.127.1.1 | 127.393 | $\{2_{100}\|1/2\ 1/2\ 0\}$ | $P^{1}4/^{1}m^{\bar{1}}b^{\bar{1}}m^{\infty m}1$ | t3 |
| 66.128.1.1 | 128.402 | $\{2_{100}\|1/2\ 1/2\ 1/2\}$ | $P^{\bar{1}}4/^{1}m^{\bar{1}}n^{1}c^{\infty m}1$ | t3 |
| 58.128.1.1 | 128.403 | $\{2_{110}\|1/2\ 1/2\ 1/2\}$ | $P^{\bar{1}}4/^{1}m^{1}n^{\bar{1}}c^{\infty m}1$ | t3 |
| 83.128.1.1 | 128.405 | $\{2_{100}\|1/2\ 1/2\ 1/2\}$ | $P^{1}4/^{1}m^{\bar{1}}n^{\bar{1}}c^{\infty m}1$ | t3 |
| 67.129.1.1 | 129.414 | $\{2_{100}\|1/2\ 0\ 0\}$ | $P^{\bar{1}}4/^{1}n^{\bar{1}}m^{1}m^{\infty m}1$ | t3 |
| 59.129.1.1 | 129.415 | $\{2_{110}\|1/2\ 1/2\ 0\}$ | $P^{\bar{1}}4/^{1}n^{1}m^{\bar{1}}m^{\infty m}1$ | t3 |
| 85.129.1.1 | 129.417 | $\{2_{100}\|1/2\ 0\ 0\}$ | $P^{1}4/^{1}n^{\bar{1}}m^{\bar{1}}m^{\infty m}1$ | t3 |

| 68.130.1.1 | 130.426 | $\{2_{100}\|1/2\ 0\ 1/2\}$ | $P^{\bar{1}}4/{}^{1}n^{\bar{1}}c^{1}c^{\infty m}1$ | t3 |
|---|---|---|---|---|
| 56.130.1.1 | 130.427 | $\{2_{110}\|1/2\ 1/2\ 1/2\}$ | $P^{\bar{1}}4/{}^{1}n^{1}c^{\bar{1}}c^{\infty m}1$ | t3 |
| 85.130.1.1 | 130.429 | $\{2_{100}\|1/2\ 0\ 1/2\}$ | $P^{1}4/{}^{1}n^{\bar{1}}c^{\bar{1}}c^{\infty m}1$ | t3 |
| 66.131.1.1 | 131.438 | $\{2_{100}\|0\}$ | $P^{\bar{1}}4_2/{}^{1}m^{\bar{1}}m^{1}c^{\infty m}1$ | t3 |
| 47.131.1.1 | 131.439 | $\{2_{110}\|0\ 0\ 1/2\}$ | $P^{\bar{1}}4_2/{}^{1}m^{1}m^{\bar{1}}c^{\infty m}1$ | t3 |
| 84.131.1.1 | 131.441 | $\{2_{100}\|0\}$ | $P^{1}4_2/{}^{1}m^{\bar{1}}m^{\bar{1}}c^{\infty m}1$ | t3 |
| 65.132.1.1 | 132.450 | $\{2_{100}\|0\ 0\ 1/2\}$ | $P^{\bar{1}}4_2/{}^{1}m^{\bar{1}}c^{1}m^{\infty m}1$ | t3 |
| 49.132.1.1 | 132.451 | $\{2_{110}\|0\}$ | $P^{\bar{1}}4_2/{}^{1}m^{1}c^{\bar{1}}m^{\infty m}1$ | t3 |
| 84.132.1.1 | 132.453 | $\{2_{100}\|0\ 0\ 1/2\}$ | $P^{1}4_2/{}^{1}m^{\bar{1}}c^{\bar{1}}m^{\infty m}1$ | t3 |
| 68.133.1.1 | 133.462 | $\{2_{100}\|0\ 1/2\ 0\}$ | $P^{\bar{1}}4_2/{}^{1}n^{\bar{1}}b^{1}c^{\infty m}1$ | t3 |
| 50.133.1.1 | 133.463 | $\{2_{110}\|0\ 0\ 1/2\}$ | $P^{\bar{1}}4_2/{}^{1}n^{1}b^{\bar{1}}c^{\infty m}1$ | t3 |
| 86.133.1.1 | 133.465 | $\{2_{100}\|0\ 1/2\ 0\}$ | $P^{1}4_2/{}^{1}n^{\bar{1}}b^{\bar{1}}c^{\infty m}1$ | t3 |
| 67.134.1.1 | 134.474 | $\{2_{100}\|0\ 1/2\ 1/2\}$ | $P^{\bar{1}}4_2/{}^{1}n^{\bar{1}}n^{1}m^{\infty m}1$ | t3 |
| 48.134.1.1 | 134.475 | $\{2_{110}\|0\}$ | $P^{\bar{1}}4_2/{}^{1}n^{1}n^{\bar{1}}m^{\infty m}1$ | t3 |
| 86.134.1.1 | 134.477 | $\{2_{100}\|0\ 1/2\ 1/2\}$ | $P^{1}4_2/{}^{1}n^{\bar{1}}n^{\bar{1}}m^{\infty m}1$ | t3 |
| 66.135.1.1 | 135.486 | $\{2_{100}\|1/2\ 1/2\ 0\}$ | $P^{\bar{1}}4_2/{}^{1}m^{\bar{1}}b^{1}c^{\infty m}1$ | t3 |
| 55.135.1.1 | 135.487 | $\{2_{110}\|1/2\ 1/2\ 1/2\}$ | $P^{\bar{1}}4_2/{}^{1}m^{1}b^{\bar{1}}c^{\infty m}1$ | t3 |
| 84.135.1.1 | 135.489 | $\{2_{100}\|1/2\ 1/2\ 0\}$ | $P^{1}4_2/{}^{1}m^{\bar{1}}b^{\bar{1}}c^{\infty m}1$ | t3 |
| 65.136.1.1 | 136.498 | $\{2_{100}\|1/2\ 1/2\ 1/2\}$ | $P^{\bar{1}}4_2/{}^{1}m^{\bar{1}}n^{1}m^{\infty m}1$ | t3 |
| 58.136.1.1 | 136.499 | $\{2_{110}\|0\}$ | $P^{\bar{1}}4_2/{}^{1}m^{1}n^{\bar{1}}m^{\infty m}1$ | t3 |
| 84.136.1.1 | 136.501 | $\{2_{100}\|1/2\ 1/2\ 1/2\}$ | $P^{1}4_2/{}^{1}m^{\bar{1}}n^{\bar{1}}m^{\infty m}1$ | t3 |
| 68.137.1.1 | 137.510 | $\{2_{100}\|1/2\ 0\ 0\}$ | $P^{\bar{1}}4_2/{}^{1}n^{\bar{1}}m^{1}c^{\infty m}1$ | t3 |
| 59.137.1.1 | 137.511 | $\{2_{110}\|1/2\ 1/2\ 1/2\}$ | $P^{\bar{1}}4_2/{}^{1}n^{1}m^{\bar{1}}c^{\infty m}1$ | t3 |
| 86.137.1.1 | 137.513 | $\{2_{100}\|1/2\ 0\ 0\}$ | $P^{1}4_2/{}^{1}n^{\bar{1}}m^{\bar{1}}c^{\infty m}1$ | t3 |
| 67.138.1.1 | 138.522 | $\{2_{100}\|1/2\ 0\ 1/2\}$ | $P^{\bar{1}}4_2/{}^{1}n^{\bar{1}}c^{1}m^{\infty m}1$ | t3 |
| 56.138.1.1 | 138.523 | $\{2_{110}\|1/2\ 1/2\ 0\}$ | $P^{\bar{1}}4_2/{}^{1}n^{1}c^{\bar{1}}m^{\infty m}1$ | t3 |
| 86.138.1.1 | 138.525 | $\{2_{100}\|1/2\ 0\ 1/2\}$ | $P^{1}4_2/{}^{1}n^{\bar{1}}c^{\bar{1}}m^{\infty m}1$ | t3 |
| 69.139.1.1 | 139.534 | $\{2_{100}\|0\}$ | $I^{\bar{1}}4/{}^{1}m^{\bar{1}}m^{1}m^{\infty m}1$ | t4 |
| 71.139.1.1 | 139.535 | $\{2_{110}\|0\}$ | $I^{\bar{1}}4/{}^{1}m^{1}m^{\bar{1}}m^{\infty m}1$ | t4 |
| 87.139.1.1 | 139.537 | $\{2_{100}\|0\}$ | $I^{1}4/{}^{1}m^{\bar{1}}m^{\bar{1}}m^{\infty m}1$ | t4 |
| 69.140.1.1 | 140.544 | $\{2_{100}\|0\ 0\ 1/2\}$ | $I^{\bar{1}}4/{}^{1}m^{\bar{1}}c^{1}m^{\infty m}1$ | t4 |
| 72.140.1.1 | 140.545 | $\{2_{110}\|0\ 0\ 1/2\}$ | $I^{\bar{1}}4/{}^{1}m^{1}c^{\bar{1}}m^{\infty m}1$ | t4 |
| 87.140.1.1 | 140.547 | $\{2_{100}\|0\ 0\ 1/2\}$ | $I^{1}4/{}^{1}m^{\bar{1}}c^{\bar{1}}m^{\infty m}1$ | t4 |
| 70.141.1.1 | 141.554 | $\{2_{100}\|0\}$ | $I^{\bar{1}}4_1/{}^{1}a^{\bar{1}}m^{1}d^{\infty m}1$ | t4 |
| 74.141.1.1 | 141.555 | $\{2_{110}\|3/4\ 1/4\ 3/4\}$ | $I^{\bar{1}}4_1/{}^{1}a^{1}m^{\bar{1}}d^{\infty m}1$ | t4 |
| 88.141.1.1 | 141.557 | $\{2_{100}\|0\}$ | $I^{1}4_1/{}^{1}a^{\bar{1}}m^{\bar{1}}d^{\infty m}1$ | t4 |
| 70.142.1.1 | 142.564 | $\{2_{100}\|1/2\ 1/2\ 0\}$ | $I^{\bar{1}}4_1/{}^{1}a^{\bar{1}}c^{1}d^{\infty m}1$ | t4 |
| 73.142.1.1 | 142.565 | $\{2_{110}\|3/4\ 1/4\ 1/4\}$ | $I^{\bar{1}}4_1/{}^{1}a^{1}c^{\bar{1}}d^{\infty m}1$ | t4 |
| 88.142.1.1 | 142.567 | $\{2_{100}\|1/2\ 1/2\ 0\}$ | $I^{1}4_1/{}^{1}a^{\bar{1}}c^{\bar{1}}d^{\infty m}1$ | t4 |
| 143.149.1.1 | 149.23 | $\{2_{210}\|0\}$ | $P^{1}3^{1}1^{\bar{1}}2^{\infty m}1$ | hP2 |
| 143.150.1.1 | 150.27 | $\{2_{100}\|0\}$ | $P^{1}3^{\bar{1}}2^{1}1^{\infty m}1$ | hP3 |
| 144.151.1.1 | 151.31 | $\{2_{210}\|0\}$ | $P^{1}3_1{}^{1}1^{\bar{1}}2^{\infty m}1$ | hP2 |
| 144.152.1.1 | 152.35 | $\{2_{110}\|0\}$ | $P^{1}3_1{}^{\bar{1}}2^{1}1^{\infty m}1$ | hP3 |

| 145.153.1.1 | 153.39 | $\{2_{210}\|0\}$ | $P^{1}3_{2}{}^{1}1^{\bar{1}}2^{\infty m}1$ | hP2 |
|---|---|---|---|---|
| 145.154.1.1 | 154.43 | $\{2_{110}\|0\}$ | $P^{1}3_{2}{}^{\bar{1}}2^{1}1^{\infty m}1$ | hP3 |
| 146.155.1.1 | 155.47 | $\{2_{100}\|0\}$ | $R^{1}3^{\bar{1}}2^{\infty m}1$ | hR2 |
| 143.156.1.1 | 156.51 | $\{\mathrm{m}_{100}\|0\}$ | $P^{1}3^{\bar{1}}m^{1}1^{\infty}1$ | hP3 |
| 143.157.1.1 | 157.55 | $\{m_{210}\|0\}$ | $P^{1}3^{1}1^{\bar{1}}m^{\infty m}1$ | hP2 |
| 143.158.1.1 | 158.59 | $\{m_{100}\|0\ 0\ 1/2\}$ | $P^{1}3^{\bar{1}}c^{1}1^{\infty m}1$ | hP3 |
| 143.159.1.1 | 159.63 | $\{m_{210}\|0\ 0\ 1/2\}$ | $P^{1}3^{1}1^{\bar{1}}c^{\infty m}1$ | hP2 |
| 146.160.1.1 | 160.67 | $\{m_{100}\|0\}$ | $R^{1}3^{\bar{1}}m^{\infty m}1$ | hR2 |
| 146.161.1.1 | 161.71 | $\{m_{100}\|0\ 0\ 1/2\}$ | $R^{1}3^{\bar{1}}c^{\infty m}1$ | hR2 |
| 147.162.1.1 | 162.77 | $\{2_{210}\|0\}$ | $P^{1}\bar{3}^{1}1^{\bar{1}}m^{\infty m}1$ | hP2 |
| 147.163.1.1 | 163.83 | $\{2_{210}\|0\ 0\ 1/2\}$ | $P^{1}\bar{3}^{1}1^{\bar{1}}c^{\infty m}1$ | hP2 |
| 147.164.1.1 | 164.89 | $\{2_{100}\|0\}$ | $P^{1}\bar{3}^{\bar{1}}m^{1}1^{\infty m}1$ | hP3 |
| 147.165.1.1 | 165.95 | $\{2_{100}\|0\ 0\ 1/2\}$ | $P^{1}\bar{3}^{\bar{1}}c^{1}1^{\infty m}1$ | hP3 |
| 148.166.1.1 | 166.101 | $\{2_{100}\|0\}$ | $R^{1}\bar{3}^{\bar{1}}m^{\infty m}1$ | hR2 |
| 148.167.1.1 | 167.107 | $\{2_{100}\|0\ 0\ 1/2\}$ | $R^{1}\bar{3}^{\bar{1}}c^{\infty m}1$ | hR2 |
| 143.168.1.1 | 168.111 | $\{2_{001}\|0\}$ | $P^{\bar{1}}6^{\infty m}1$ | hP4 |
| 144.169.1.1 | 169.115 | $\{2_{001}\|0\ 0\ 1/2\}$ | $P^{\bar{1}}6_{1}{}^{\infty m}1$ | hP4 |
| 145.170.1.1 | 170.119 | $\{2_{001}\|0\ 0\ 1/2\}$ | $P^{\bar{1}}6_{5}{}^{\infty m}1$ | hP4 |
| 145.171.1.1 | 171.123 | $\{2_{001}\|0\}$ | $P^{\bar{1}}6_{2}{}^{\infty m}1$ | hP4 |
| 144.172.1.1 | 172.127 | $\{2_{001}\|0\}$ | $P^{\bar{1}}6_{4}{}^{\infty m}1$ | hP4 |
| 143.173.1.1 | 173.131 | $\{2_{001}\|0\ 0\ 1/2\}$ | $P^{\bar{1}}6_{3}{}^{\infty m}1$ | hP4 |
| 143.174.1.1 | 174.135 | $\{m_{001}\|0\}$ | $P^{\bar{1}}\bar{6}^{\infty m}1$ | hP4 |
| 147.175.1.1 | 175.141 | $\{2_{001}\|0\}$ | $P^{\bar{1}}6/^{\bar{1}}m^{\infty m}1$ | hP4 |
| 147.176.1.1 | 176.147 | $\{2_{001}\|0\ 0\ 1/2\}$ | $P^{\bar{1}}6_{3}/^{\bar{1}}m^{\infty m}1$ | hP4 |
| 149.177.1.1 | 177.151 | $\{2_{100}\|0\}$ | $P^{\bar{1}}6^{\bar{1}}2^{1}2^{\infty m}1$ | hP5 |
| 150.177.1.1 | 177.152 | $\{2_{001}\|0\}$ | $P^{\bar{1}}6^{1}2^{\bar{1}}2^{\infty m}1$ | hP5 |
| 168.177.1.1 | 177.153 | $\{2_{100}\|0\}$ | $P^{1}6^{\bar{1}}2^{\bar{1}}2^{\infty m}1$ | hP5 |
| 151.178.1.1 | 178.157 | $\{2_{100}\|0\}$ | $P^{\bar{1}}6_{1}{}^{\bar{1}}2^{1}2^{\infty m}1$ | hP5 |
| 152.178.1.1 | 178.158 | $\{2_{120}\|0\ 0\ 1/2\}$ | $P^{\bar{1}}6_{1}{}^{1}2^{\bar{1}}2^{\infty m}1$ | hP5 |
| 169.178.1.1 | 178.159 | $\{2_{100}\|0\}$ | $P^{1}6_{1}{}^{\bar{1}}2^{\bar{1}}2^{\infty m}1$ | hP5 |
| 153.179.1.1 | 179.163 | $\{2_{100}\|0\}$ | $P^{\bar{1}}6_{5}{}^{\bar{1}}2^{1}2^{\infty m}1$ | hP5 |
| 154.179.1.1 | 179.164 | $\{2_{001}\|0\ 0\ 1/2\}$ | $P^{\bar{1}}6_{5}{}^{1}2^{\bar{1}}2^{\infty m}1$ | hP5 |
| 170.179.1.1 | 179.165 | $\{2_{100}\|0\}$ | $P^{1}6_{5}{}^{\bar{1}}2^{\bar{1}}2^{\infty m}1$ | hP5 |
| 153.180.1.1 | 180.169 | $\{2_{100}\|0\}$ | $P^{\bar{1}}6_{2}{}^{\bar{1}}2^{1}2^{\infty m}1$ | hP5 |
| 154.180.1.1 | 180.170 | $\{2_{001}\|0\}$ | $P^{\bar{1}}6_{2}{}^{1}2^{\bar{1}}2^{\infty m}1$ | hP5 |
| 171.180.1.1 | 180.171 | $\{2_{100}\|0\}$ | $P^{1}6_{2}{}^{\bar{1}}2^{\bar{1}}2^{\infty m}1$ | hP5 |
| 151.181.1.1 | 181.175 | $\{2_{100}\|0\}$ | $P^{\bar{1}}6_{4}{}^{\bar{1}}2^{1}2^{\infty m}1$ | hP5 |
| 152.181.1.1 | 181.176 | $\{2_{001}\|0\}$ | $P^{\bar{1}}6_{4}{}^{1}2^{\bar{1}}2^{\infty m}1$ | hP5 |
| 172.181.1.1 | 181.177 | $\{2_{100}\|0\}$ | $P^{1}6_{4}{}^{\bar{1}}2^{\bar{1}}2^{\infty m}1$ | hP5 |
| 149.182.1.1 | 182.181 | $\{2_{100}\|0\}$ | $P^{\bar{1}}6_{3}{}^{\bar{1}}2^{1}2^{\infty m}1$ | hP5 |
| 150.182.1.1 | 182.182 | $\{2_{001}\|0\ 0\ 1/2\}$ | $P^{\bar{1}}6_{3}{}^{1}2^{\bar{1}}2^{\infty m}1$ | hP5 |
| 173.182.1.1 | 182.183 | $\{2_{100}\|0\}$ | $P^{1}6_{3}{}^{\bar{1}}2^{\bar{1}}2^{\infty m}1$ | hP5 |
| 157.183.1.1 | 183.187 | $\{2_{001}\|0\}$ | $P^{\bar{1}}6^{\bar{1}}m^{1}m^{\infty m}1$ | hP5 |

| 156.183.1.1 | 183.188 | $\{2_{001}\|0\}$ | $P^{\bar{1}}6^{1}m^{\bar{1}}m^{\infty m}1$ | hP5 |
|---|---|---|---|---|
| 168.183.1.1 | 183.189 | $\{m_{100}\|0\}$ | $P^{1}6^{\bar{1}}m^{\bar{1}}m^{\infty m}1$ | hP5 |
| 159.184.1.1 | 184.193 | $\{2_{001}\|0\}$ | $P^{\bar{1}}6^{\bar{1}}c^{1}c^{\infty m}1$ | hP5 |
| 158.184.1.1 | 184.194 | $\{2_{001}\|0\}$ | $P^{\bar{1}}6^{1}c^{\bar{1}}c^{\infty m}1$ | hP5 |
| 168.184.1.1 | 184.195 | $\{m_{100}\|0\ 0\ 1/2\}$ | $P^{1}6^{\bar{1}}c^{\bar{1}}c^{\infty m}1$ | hP5 |
| 157.185.1.1 | 185.199 | $\{2_{001}\|0\ 0\ 1/2\}$ | $P^{\bar{1}}6_{3}{}^{\bar{1}}c^{1}m^{\infty m}1$ | hP5 |
| 158.185.1.1 | 185.200 | $\{2_{001}\|0\ 0\ 1/2\}$ | $P^{\bar{1}}6_{3}{}^{1}c^{\bar{1}}m^{\infty m}1$ | hP5 |
| 173.185.1.1 | 185.201 | $\{m_{100}\|0\ 0\ 1/2\}$ | $P^{1}6_{3}{}^{\bar{1}}c^{\bar{1}}m^{\infty m}1$ | hP5 |
| 159.186.1.1 | 186.205 | $\{2_{001}\|0\ 0\ 1/2\}$ | $P^{\bar{1}}6_{3}{}^{\bar{1}}m^{1}c^{\infty m}1$ | hP5 |
| 156.186.1.1 | 186.206 | $\{2_{001}\|0\ 0\ 1/2\}$ | $P^{\bar{1}}6_{3}{}^{1}m^{\bar{1}}c^{\infty m}1$ | hP5 |
| 173.186.1.1 | 186.207 | $\{m_{100}\|0\}$ | $P^{1}6_{3}{}^{\bar{1}}m^{\bar{1}}c^{\infty m}1$ | hP5 |
| 149.187.1.1 | 187.211 | $\{m_{100}\|0\}$ | $P^{\bar{1}}\bar{6}^{\bar{1}}m^{1}2^{\infty m}1$ | hP5 |
| 156.187.1.1 | 187.212 | $\{m_{001}\|0\}$ | $P^{\bar{1}}\bar{6}^{1}m^{\bar{1}}2^{\infty m}1$ | hP5 |
| 174.187.1.1 | 187.213 | $\{m_{100}\|0\}$ | $P^{1}\bar{6}^{\bar{1}}m^{\bar{1}}2^{\infty m}1$ | hP5 |
| 149.188.1.1 | 188.217 | $\{m_{100}\|0\ 0\ 1/2\}$ | $P^{\bar{1}}\bar{6}^{\bar{1}}c^{1}2^{\infty m}1$ | hP5 |
| 158.188.1.1 | 188.218 | $\{m_{001}\|0\ 0\ 1/2\}$ | $P^{\bar{1}}\bar{6}^{1}c^{\bar{1}}2^{\infty m}1$ | hP5 |
| 174.188.1.1 | 188.219 | $\{m_{100}\|0\ 0\ 1/2\}$ | $P^{1}\bar{6}^{\bar{1}}c^{\bar{1}}2^{\infty m}1$ | hP5 |
| 157.189.1.1 | 189.223 | $\{2_{100}\|0\}$ | $P^{\bar{1}}\bar{6}^{\bar{1}}2^{1}m^{\infty m}1$ | hP5 |
| 150.189.1.1 | 189.224 | $\{m_{001}\|0\}$ | $P^{\bar{1}}\bar{6}^{1}2^{\bar{1}}m^{\infty m}1$ | hP5 |
| 174.189.1.1 | 189.225 | $\{2_{100}\|0\}$ | $P^{1}\bar{6}^{\bar{1}}2^{\bar{1}}m^{\infty m}1$ | hP5 |
| 159.190.1.1 | 190.229 | $\{2_{100}\|0\}$ | $P^{\bar{1}}\bar{6}^{\bar{1}}2^{1}c^{\infty m}1$ | hP5 |
| 150.190.1.1 | 190.230 | $\{m_{001}\|0\ 0\ 1/2\}$ | $P^{\bar{1}}\bar{6}^{1}2^{\bar{1}}c^{\infty m}1$ | hP5 |
| 174.190.1.1 | 190.231 | $\{2_{100}\|0\}$ | $P^{1}\bar{6}^{\bar{1}}2^{\bar{1}}c^{\infty m}1$ | hP5 |
| 162.191.1.1 | 191.238 | $\{2_{001}\|0\}$ | $P^{\bar{1}}6/{}^{\bar{1}}m^{\bar{1}}m^{1}m^{\infty m}1$ | hP5 |
| 164.191.1.1 | 191.239 | $\{2_{001}\|0\}$ | $P^{\bar{1}}6/{}^{\bar{1}}m^{1}m^{\bar{1}}m^{\infty m}1$ | hP5 |
| 175.191.1.1 | 191.240 | $\{2_{100}\|0\}$ | $P^{1}6/{}^{1}m^{\bar{1}}m^{\bar{1}}m^{\infty m}1$ | hP5 |
| 163.192.1.1 | 192.248 | $\{2_{001}\|0\}$ | $P^{\bar{1}}6/{}^{\bar{1}}m^{\bar{1}}c^{1}c^{\infty m}1$ | hP5 |
| 165.192.1.1 | 192.249 | $\{2_{001}\|0\}$ | $P^{\bar{1}}6/{}^{\bar{1}}m^{1}c^{\bar{1}}c^{\infty m}1$ | hP5 |
| 175.192.1.1 | 192.250 | $\{2_{100}\|0\ 0\ 1/2\}$ | $P^{1}6/{}^{1}m^{\bar{1}}c^{\bar{1}}c^{\infty m}1$ | hP5 |
| 162.193.1.1 | 193.258 | $\{2_{100}\|0\ 0\ 1/2\}$ | $P^{\bar{1}}6_{3}/{}^{\bar{1}}m^{\bar{1}}c^{1}m^{\infty m}1$ | hP5 |
| 165.193.1.1 | 193.259 | $\{2_{001}\|0\ 0\ 1/2\}$ | $P^{\bar{1}}6_{3}/{}^{\bar{1}}m^{1}c^{\bar{1}}m^{\infty m}1$ | hP5 |
| 176.193.1.1 | 193.260 | $\{2_{100}\|0\ 0\ 1/2\}$ | $P^{1}6_{3}/{}^{1}m^{\bar{1}}c^{\bar{1}}m^{\infty m}1$ | hP5 |
| 163.194.1.1 | 194.268 | $\{2_{100}\|0\}$ | $P^{\bar{1}}6_{3}/{}^{\bar{1}}m^{\bar{1}}m^{1}c^{\infty m}1$ | hP5 |
| 164.194.1.1 | 194.269 | $\{2_{001}\|0\ 0\ 1/2\}$ | $P^{\bar{1}}6_{3}/{}^{\bar{1}}m^{1}m^{\bar{1}}c^{\infty m}1$ | hP5 |
| 176.194.1.1 | 194.270 | $\{2_{100}\|0\}$ | $P^{1}6_{3}/{}^{1}m^{\bar{1}}m^{\bar{1}}c^{\infty m}1$ | hP5 |
| 195.207.1.1 | 207.42 | $\{2_{110}\|0\}$ | $P^{\bar{1}}4^{1}3^{\bar{1}}2^{\infty m}1$ | c4 |
| 195.208.1.1 | 208.46 | $\{2_{110}\|1/2\ 1/2\ 1/2\}$ | $P^{\bar{1}}4_{2}{}^{1}3^{\bar{1}}2^{\infty m}1$ | c4 |
| 196.209.1.1 | 209.50 | $\{2_{110}\|0\}$ | $F^{\bar{1}}4^{1}3^{\bar{1}}2^{\infty m}1$ | c5 |
| 196.210.1.1 | 210.54 | $\{2_{110}\|1/4\ 1/4\ 1/4\}$ | $F^{\bar{1}}4_{1}{}^{1}3^{\bar{1}}2^{\infty m}1$ | c5 |
| 197.211.1.1 | 211.58 | $\{2_{110}\|0\}$ | $I^{\bar{1}}4^{1}3^{\bar{1}}2^{\infty m}1$ | c6 |
| 198.212.1.1 | 212.61 | $\{2_{110}\|1/4\ 3/4\ 3/4\}$ | $P^{\bar{1}}4_{3}{}^{1}3^{\bar{1}}2^{\infty m}1$ | c4 |
| 198.213.1.1 | 213.65 | $\{2_{110}\|3/4\ 1/4\ 1/4\}$ | $P^{\bar{1}}4_{1}{}^{1}3^{\bar{1}}2^{\infty m}1$ | c4 |
| 199.214.1.1 | 214.69 | $\{2_{110}\|3/4\ 1/4\ 1/4\}$ | $I^{\bar{1}}4_{1}{}^{1}3^{\bar{1}}2^{\infty m}1$ | c6 |

| 195.215.1.1 | 215.72 | $\{m_{110}\|0\}$ | $P^{\bar{1}}\bar{4}^{1}3^{\bar{1}}m^{\infty m}1$ | c4 |
|---|---|---|---|---|
| 196.216.1.1 | 216.76 | $\{m_{110}\|0\}$ | $F^{\bar{1}}\bar{4}^{1}3^{\bar{1}}m^{\infty m}1$ | c5 |
| 197.217.1.1 | 217.80 | $\{m_{110}\|0\}$ | $I^{\bar{1}}\bar{4}^{1}3^{\bar{1}}m^{\infty m}1$ | c6 |
| 195.218.1.1 | 218.83 | $\{m_{110}\|1/2\ 1/2\ 1/2\}$ | $P^{\bar{1}}\bar{4}^{1}3^{\bar{1}}n^{\infty m}1$ | c4 |
| 196.219.1.1 | 219.87 | $\{m_{110}\|1/2\ 1/2\ 1/2\}$ | $F^{\bar{1}}\bar{4}^{1}3^{\bar{1}}c^{\infty m}1$ | c5 |
| 199.220.1.1 | 220.91 | $\{m_{110}\|3/4\ 1/4\ 1/4\}$ | $I^{\bar{1}}\bar{4}^{1}3^{\bar{1}}d^{\infty m}1$ | c6 |
| 200.221.1.1 | 221.95 | $\{2_{110}\|0\}$ | $P^{1}m^{1}\bar{3}^{\bar{1}}m^{\infty m}1$ | c4 |
| 201.222.1.1 | 222.101 | $\{2_{110}\|0\ 0\ 1/2\}$ | $P^{1}n^{1}\bar{3}^{\bar{1}}n^{\infty m}1$ | c4 |
| 200.223.1.1 | 223.107 | $\{2_{110}\|1/2\ 1/2\ 1/2\}$ | $P^{1}m^{1}\bar{3}^{\bar{1}}n^{\infty m}1$ | c4 |
| 201.224.1.1 | 224.113 | $\{2_{110}\|1/2\ 1/2\ 0\}$ | $P^{1}n^{1}\bar{3}^{\bar{1}}m^{\infty m}1$ | c4 |
| 202.225.1.1 | 225.119 | $\{2_{110}\|0\}$ | $F^{1}m^{1}\bar{3}^{\bar{1}}m^{\infty m}1$ | c5 |
| 202.226.1.1 | 226.125 | $\{2_{110}\|1/2\ 1/2\ 1/2\}$ | $F^{1}m^{1}\bar{3}^{\bar{1}}c^{\infty m}1$ | c5 |
| 203.227.1.1 | 227.131 | $\{2_{110}\|1/4\ 1/4\ 0\}$ | $F^{1}d^{1}\bar{3}^{\bar{1}}m^{\infty m}1$ | c5 |
| 203.228.1.1 | 228.137 | $\{2_{110}\|1/4\ 1/4\ 1/2\}$ | $F^{1}d^{1}\bar{3}^{\bar{1}}c^{\infty m}1$ | c5 |
| 204.229.1.1 | 229.143 | $\{2_{110}\|0\}$ | $I^{1}m^{1}\bar{3}^{\bar{1}}m^{\infty m}1$ | c6 |
| 206.230.1.1 | 230.148 | $\{2_{110}\|3/4\ 1/4\ 1/4\}$ | $I^{1}a^{1}\bar{3}^{\bar{1}}d^{\infty m}1$ | c6 |

## S2.5. Type-IV collinear SSGs

Supplementary Table XVII. Full tabulation of 517 type-IV collinear SSGs describing collinear Uτ AFMs. The five columns represent the group number using a four-index nomenclature[8], the one-to-one correspondence with MSGs, the representative lattice-space element $A$ that connects the two opposite-spin sublattices, the international notation of collinear SSGs and the spin Brillouin zone, respectively. For detailed information about the notation used in the spin Brillouin zone, please refer to Supplementary Table VIII in Supplementary Information S1. Note that the collinear type-IV SSGs, due to the existence of spin-only groups, Uτ, Tτ, and TUτ are equivalent. Therefore, we inherit the notation from MSGs to denote the fractional translations connecting two sub-lattices in lattice space.

| $G_{NSS}$ | MSG | $A$ | $G_S$ | SBZ |
|---|---|---|---|---|
| 1.1.2.1 | 1.3 | $\{1\|0\ 0\ 1/2\}$ | $P_S{}^{1}1^{\infty m}1$ | a1 |
| 2.2.2.1 | 2.7 | $\{1\|0\ 0\ 1/2\}$ | $P_S{}^{1}\bar{1}^{\infty m}1$ | a1 |
| 3.3.2.4 | 3.4 | $\{1\|1/2\ 0\ 0\}$ | $P_a{}^{1}2^{\infty m}1$ | m1 |
| 3.3.2.1 | 3.5 | $\{1\|0\ 1/2\ 0\}$ | $P_b{}^{1}2^{\infty m}1$ | m1 |
| 3.5.2.1 | 3.6 | $\{1\|1/2\ 1/2\ 0\}$ | $P_C{}^{1}2^{\infty m}1$ | m1 |
| 4.4.2.1 | 4.10 | $\{1\|1/2\ 0\ 0\}$ | $P_a{}^{1}2_1{}^{\infty m}1$ | m1 |
| 4.3.2.1 | 4.11 | $\{1\|0\ 1/2\ 0\}$ | $P_b{}^{1}2_1{}^{\infty m}1$ | m1 |
| 4.5.2.1 | 4.12 | $\{1\|1/2\ 1/2\ 0\}$ | $P_C{}^{1}2_1{}^{\infty m}1$ | m1 |
| 5.5.2.1 | 5.16 | $\{1\|0\ 0\ 1/2\}$ | $C_c{}^{1}2^{\infty m}1$ | m2 |
| 5.3.2.1 | 5.17 | $\{1\|1/2\ 0\ 0\}$ | $C_a{}^{1}2^{\infty m}1$ | m2 |

<table>
<tr><td>6.6.2.4</td><td>6.21</td><td>{1|1/2 0 0}</td><td>$P_a{}^{1}m^{\infty m}1$</td><td>m1</td></tr>
<tr><td>6.6.2.1</td><td>6.22</td><td>{1|0 1/2 0}</td><td>$P_b{}^{1}m^{\infty m}1$</td><td>m1</td></tr>
<tr><td>6.8.2.1</td><td>6.23</td><td>{1|1/2 1/2 0}</td><td>$P_C{}^{1}m^{\infty m}1$</td><td>m1</td></tr>
<tr><td>7.7.2.4</td><td>7.27</td><td>{1|1/2 0 0}</td><td>$P_a{}^{1}c^{\infty m}1$</td><td>m1</td></tr>
<tr><td>7.6.2.1</td><td>7.28</td><td>{1|0 0 1/2}</td><td>$P_c{}^{1}c^{\infty m}1$</td><td>m1</td></tr>
<tr><td>7.7.2.1</td><td>7.29</td><td>{1|0 1/2 0}</td><td>$P_b{}^{1}c^{\infty m}1$</td><td>m1</td></tr>
<tr><td>7.9.2.1</td><td>7.30</td><td>{1|1/2 1/2 0}</td><td>$P_C{}^{1}c^{\infty m}1$</td><td>m1</td></tr>
<tr><td>7.8.2.1</td><td>7.31</td><td>{1|0 1/2 1/2}</td><td>$P_A{}^{1}c^{\infty m}1$</td><td>m1</td></tr>
<tr><td>8.8.2.1</td><td>8.35</td><td>{1|0 0 1/2}</td><td>$C_c{}^{1}m^{\infty m}1$</td><td>m2</td></tr>
<tr><td>8.6.2.1</td><td>8.36</td><td>{1|1/2 0 0}</td><td>$C_a{}^{1}m^{\infty m}1$</td><td>m2</td></tr>
<tr><td>9.8.2.1</td><td>9.40</td><td>{1|0 0 1/2}</td><td>$C_c{}^{1}c^{\infty m}1$</td><td>m2</td></tr>
<tr><td>9.7.2.1</td><td>9.41</td><td>{1|1/2 0 0}</td><td>$C_a{}^{1}c^{\infty m}1$</td><td>m2</td></tr>
<tr><td>10.10.2.4</td><td>10.47</td><td>{1|1/2 0 0}</td><td>$P_a{}^{1}2/{}^{1}m^{\infty m}1$</td><td>m1</td></tr>
<tr><td>10.10.2.1</td><td>10.48</td><td>{1|0 1/2 0}</td><td>$P_b{}^{1}2/{}^{1}m^{\infty m}1$</td><td>m1</td></tr>
<tr><td>10.12.2.1</td><td>10.49</td><td>{1|1/2 1/2 0}</td><td>$P_C{}^{1}2/{}^{1}m^{\infty m}1$</td><td>m1</td></tr>
<tr><td>11.11.2.1</td><td>11.55</td><td>{1|1/2 0 0}</td><td>$P_a{}^{1}2_1/{}^{1}m^{\infty m}1$</td><td>m1</td></tr>
<tr><td>11.10.2.1</td><td>11.56</td><td>{1|0 1/2 0}</td><td>$P_b{}^{1}2_1/{}^{1}m^{\infty m}1$</td><td>m1</td></tr>
<tr><td>11.12.2.1</td><td>11.57</td><td>{1|1/2 1/2 0}</td><td>$P_C{}^{1}2_1/{}^{1}m^{\infty m}1$</td><td>m1</td></tr>
<tr><td>12.12.2.1</td><td>12.63</td><td>{1|0 0 1/2}</td><td>$C_c{}^{1}2_1/{}^{1}m^{\infty m}1$</td><td>m2</td></tr>
<tr><td>12.10.2.1</td><td>12.64</td><td>{1|1/2 0 0}</td><td>$C_a{}^{1}2_1/{}^{1}m^{\infty m}1$</td><td>m2</td></tr>
<tr><td>13.13.2.4</td><td>13.70</td><td>{1|1/2 0 0}</td><td>$P_a{}^{1}2/{}^{1}c^{\infty m}1$</td><td>m1</td></tr>
<tr><td>13.13.2.1</td><td>13.71</td><td>{1|0 1/2 0}</td><td>$P_b{}^{1}2/{}^{1}c^{\infty m}1$</td><td>m1</td></tr>
<tr><td>13.10.2.1</td><td>13.72</td><td>{1|0 0 1/2}</td><td>$P_c{}^{1}2/{}^{1}c^{\infty m}1$</td><td>m1</td></tr>
<tr><td>13.12.2.1</td><td>13.73</td><td>{1|0 1/2 1/2}</td><td>$P_A{}^{1}2/{}^{1}c^{\infty m}1$</td><td>m1</td></tr>
<tr><td>13.15.2.1</td><td>13.74</td><td>{1|1/2 1/2 0}</td><td>$P_C{}^{1}2/{}^{1}c^{\infty m}1$</td><td>m1</td></tr>
<tr><td>14.14.2.1</td><td>14.80</td><td>{1|1/2 0 0}</td><td>$P_a{}^{1}2_1/{}^{1}c^{\infty m}1$</td><td>m1</td></tr>
<tr><td>14.13.2.1</td><td>14.81</td><td>{1|0 1/2 0}</td><td>$P_b{}^{1}2_1/{}^{1}c^{\infty m}1$</td><td>m1</td></tr>
<tr><td>14.11.2.1</td><td>14.82</td><td>{1|0 0 1/2}</td><td>$P_c{}^{1}2_1/{}^{1}c^{\infty m}1$</td><td>m1</td></tr>
<tr><td>14.12.2.1</td><td>14.83</td><td>{1|0 1/2 1/2}</td><td>$P_A{}^{1}2_1/{}^{1}c^{\infty m}1$</td><td>m1</td></tr>
<tr><td>14.15.2.1</td><td>14.84</td><td>{1|1/2 1/2 0}</td><td>$P_C{}^{1}2_1/{}^{1}c^{\infty m}1$</td><td>m1</td></tr>
<tr><td>15.12.2.1</td><td>15.90</td><td>{1|0 0 1/2}</td><td>$C_c{}^{1}2_1/{}^{1}c^{\infty m}1$</td><td>m2</td></tr>
<tr><td>15.13.2.1</td><td>15.91</td><td>{1|1/2 0 0}</td><td>$C_a{}^{1}2_1/{}^{1}c^{\infty m}1$</td><td>m2</td></tr>
<tr><td>16.16.2.1</td><td>16.4</td><td>{1|1/2 0 0}</td><td>$P_a{}^{1}2{}^{1}2{}^{1}2^{\infty m}1$</td><td>o1</td></tr>
<tr><td>16.21.2.1</td><td>16.5</td><td>{1|1/2 1/2 0}</td><td>$P_C{}^{1}2{}^{1}2{}^{1}2^{\infty m}1$</td><td>o1</td></tr>
<tr><td>16.23.2.1</td><td>16.6</td><td>{1|1/2 1/2 1/2}</td><td>$P_I{}^{1}2{}^{1}2{}^{1}2^{\infty m}1$</td><td>o1</td></tr>
<tr><td>17.17.2.1</td><td>17.11</td><td>{1|1/2 0 0}</td><td>$P_a{}^{1}2{}^{1}2{}^{1}2_1{}^{\infty m}1$</td><td>o1</td></tr>
<tr><td>17.16.2.1</td><td>17.12</td><td>{1|0 0 1/2}</td><td>$P_c{}^{1}2{}^{1}2{}^{1}2_1{}^{\infty m}1$</td><td>o1</td></tr>
<tr><td>17.21.2.1</td><td>17.13</td><td>{1|1/2 0 1/2}</td><td>$P_B{}^{1}2{}^{1}2{}^{1}2_1{}^{\infty m}1$</td><td>o1</td></tr>
<tr><td>17.20.2.1</td><td>17.14</td><td>{1|1/2 1/2 0}</td><td>$P_C{}^{1}2{}^{1}2{}^{1}2_1{}^{\infty m}1$</td><td>o1</td></tr>
<tr><td>17.24.2.1</td><td>17.15</td><td>{1|1/2 1/2 1/2}</td><td>$P_I{}^{1}2{}^{1}2{}^{1}2_1{}^{\infty m}1$</td><td>o1</td></tr>
<tr><td>18.17.2.1</td><td>18.20</td><td>{1|0 1/2 0}</td><td>$P_b{}^{1}2_1{}^{1}2_1{}^{1}2^{\infty m}1$</td><td>o1</td></tr>
<tr><td>18.18.2.1</td><td>18.21</td><td>{1|0 0 1/2}</td><td>$P_c{}^{1}2_1{}^{1}2_1{}^{1}2^{\infty m}1$</td><td>o1</td></tr>
<tr><td>18.20.2.1</td><td>18.22</td><td>{1|1/2 0 1/2}</td><td>$P_B{}^{1}2_1{}^{1}2_1{}^{1}2^{\infty m}1$</td><td>o1</td></tr>
</table>

| 18.21.2.1 | 18.23 | {1\|1/2 1/2 0} | $P_C{}^{1}2_1{}^{1}2_1{}^{1}2^{\infty m}1$ | o1 |
|---|---|---|---|---|
| 18.23.2.1 | 18.24 | {1\|1/2 1/2 1/2} | $P_I{}^{1}2_1{}^{1}2_1{}^{1}2^{\infty m}1$ | o1 |
| 19.18.2.1 | 19.28 | {1\|0 0 1/2} | $P_c{}^{1}2_1{}^{1}2_1{}^{1}2_1{}^{\infty m}1$ | o1 |
| 19.20.2.1 | 19.29 | {1\|1/2 1/2 0} | $P_C{}^{1}2_1{}^{1}2_1{}^{1}2_1{}^{\infty m}1$ | o1 |
| 19.24.2.1 | 19.30 | {1\|1/2 1/2 1/2} | $P_I{}^{1}2_1{}^{1}2_1{}^{1}2_1{}^{\infty m}1$ | o1 |
| 20.21.2.1 | 20.35 | {1\|0 0 1/2} | $C_c{}^{1}2{}^{1}2{}^{1}2_1{}^{\infty m}1$ | o2C |
| 20.17.2.1 | 20.36 | {1\|1/2 0 0} | $C_a{}^{1}2{}^{1}2{}^{1}2_1{}^{\infty m}1$ | o2C |
| 20.22.2.1 | 20.37 | {1\|0 1/2 1/2} | $C_A{}^{1}2{}^{1}2{}^{1}2_1{}^{\infty m}1$ | o2C |
| 21.21.2.1 | 21.42 | {1\|0 0 1/2} | $C_c{}^{1}2{}^{1}2{}^{1}2^{\infty m}1$ | o2C |
| 21.16.2.1 | 21.43 | {1\|1/2 0 0} | $C_a{}^{1}2{}^{1}2{}^{1}2^{\infty m}1$ | o2C |
| 21.22.2.1 | 21.44 | {1\|0 1/2 1/2} | $C_A{}^{1}2{}^{1}2{}^{1}2^{\infty m}1$ | o2C |
| 22.16.2.1 | 22.48 | {1\|1/2 1/2 1/2} | $F_s{}^{1}2{}^{1}2{}^{1}2^{\infty m}1$ | o3 |
| 23.21.2.1 | 23.52 | {1\|1/2 1/2 1/2} | $I_c{}^{1}2{}^{1}2{}^{1}2^{\infty m}1$ | o4 |
| 24.21.2.1 | 24.56 | {1\|1/2 1/2 1/2} | $I_c{}^{1}2_1{}^{1}2_1{}^{1}2_1{}^{\infty m}1$ | o4 |
| 25.25.2.4 | 25.61 | {1\|0 0 1/2} | $P_c{}^{1}m{}^{1}m{}^{1}2^{\infty m}1$ | o1 |
| 25.25.2.1 | 25.62 | {1\|1/2 0 0} | $P_a{}^{1}m{}^{1}m{}^{1}2^{\infty m}1$ | o1 |
| 25.35.2.1 | 25.63 | {1\|1/2 1/2 0} | $P_C{}^{1}m{}^{1}m{}^{1}2^{\infty m}1$ | o1 |
| 25.38.2.1 | 25.64 | {1\|0 1/2 1/2} | $P_A{}^{1}m{}^{1}m{}^{1}2^{\infty m}1$ | o1 |
| 25.44.2.1 | 25.65 | {1\|1/2 1/2 1/2} | $P_I{}^{1}m{}^{1}m{}^{1}2^{\infty m}1$ | o1 |
| 26.26.2.1 | 26.71 | {1\|1/2 0 0} | $P_a{}^{1}m{}^{1}c{}^{1}2_1{}^{\infty m}1$ | o1 |
| 26.26.2.4 | 26.72 | {1\|0 1/2 0} | $P_b{}^{1}m{}^{1}c{}^{1}2_1{}^{\infty m}1$ | o1 |
| 26.25.2.1 | 26.73 | {1\|0 0 1/2} | $P_c{}^{1}m{}^{1}c{}^{1}2_1{}^{\infty m}1$ | o1 |
| 26.38.2.1 | 26.74 | {1\|0 1/2 1/2} | $P_A{}^{1}m{}^{1}c{}^{1}2_1{}^{\infty m}1$ | o1 |
| 26.39.2.1 | 26.75 | {1\|1/2 0 1/2} | $P_B{}^{1}m{}^{1}c{}^{1}2_1{}^{\infty m}1$ | o1 |
| 26.36.2.1 | 26.76 | {1\|1/2 1/2 0} | $P_C{}^{1}m{}^{1}c{}^{1}2_1{}^{\infty m}1$ | o1 |
| 26.46.2.1 | 26.77 | {1\|1/2 1/2 1/2} | $P_I{}^{1}m{}^{1}c{}^{1}2_1{}^{\infty m}1$ | o1 |
| 27.25.2.1 | 27.82 | {1\|0 0 1/2} | $P_c{}^{1}c{}^{1}c{}^{1}2^{\infty m}1$ | o1 |
| 27.27.2.1 | 27.83 | {1\|1/2 0 0} | $P_a{}^{1}c{}^{1}c{}^{1}2^{\infty m}1$ | o1 |
| 27.37.2.1 | 27.84 | {1\|1/2 1/2 0} | $P_C{}^{1}c{}^{1}c{}^{1}2^{\infty m}1$ | o1 |
| 27.39.2.1 | 27.85 | {1\|0 1/2 1/2} | $P_A{}^{1}c{}^{1}c{}^{1}2^{\infty m}1$ | o1 |
| 27.45.2.1 | 27.86 | {1\|1/2 1/2 1/2} | $P_I{}^{1}c{}^{1}c{}^{1}2^{\infty m}1$ | o1 |
| 28.25.2.1 | 28.92 | {1\|1/2 0 0} | $P_a{}^{1}m{}^{1}a{}^{1}2^{\infty m}1$ | o1 |
| 28.28.2.1 | 28.93 | {1\|0 1/2 0} | $P_b{}^{1}m{}^{1}a{}^{1}2^{\infty m}1$ | o1 |
| 28.28.2.4 | 28.94 | {1\|0 0 1/2} | $P_c{}^{1}m{}^{1}a{}^{1}2^{\infty m}1$ | o1 |
| 28.40.2.1 | 28.95 | {1\|0 1/2 1/2} | $P_A{}^{1}m{}^{1}a{}^{1}2^{\infty m}1$ | o1 |
| 28.39.2.1 | 28.96 | {1\|1/2 0 1/2} | $P_B{}^{1}m{}^{1}a{}^{1}2^{\infty m}1$ | o1 |
| 28.35.2.1 | 28.97 | {1\|1/2 1/2 0} | $P_C{}^{1}m{}^{1}a{}^{1}2^{\infty m}1$ | o1 |
| 28.46.2.1 | 28.98 | {1\|1/2 1/2 1/2} | $P_I{}^{1}m{}^{1}a{}^{1}2^{\infty m}1$ | o1 |
| 29.26.2.1 | 29.104 | {1\|1/2 0 0} | $P_a{}^{1}c{}^{1}a{}^{1}2_1{}^{\infty m}1$ | o1 |
| 29.29.2.1 | 29.105 | {1\|0 1/2 0} | $P_b{}^{1}c{}^{1}a{}^{1}2_1{}^{\infty m}1$ | o1 |
| 29.28.2.1 | 29.106 | {1\|0 0 1/2} | $P_c{}^{1}c{}^{1}a{}^{1}2_1{}^{\infty m}1$ | o1 |
| 29.41.2.1 | 29.107 | {1\|0 1/2 1/2} | $P_A{}^{1}c{}^{1}a{}^{1}2_1{}^{\infty m}1$ | o1 |
| 29.39.2.1 | 29.108 | {1\|1/2 0 1/2} | $P_B{}^{1}c{}^{1}a{}^{1}2_1{}^{\infty m}1$ | o1 |

| 29.36.2.1 | 29.109 | {1\|1/2 1/2 0} | $P_{C}{}^{1}c^{1}a^{1}2_{1}{}^{\infty m}1$ | o1 |
|---|---|---|---|---|
| 29.45.2.1 | 29.110 | {1\|1/2 1/2 1/2} | $P_{I}{}^{1}c^{1}a^{1}2_{1}{}^{\infty m}1$ | o1 |
| 30.30.2.1 | 30.116 | {1\|1/2 0 0} | $P_{a}{}^{1}n^{1}c^{1}2^{\infty m}1$ | o1 |
| 30.27.2.1 | 30.117 | {1\|0 1/2 0} | $P_{b}{}^{1}n^{1}c^{1}2^{\infty m}1$ | o1 |
| 30.28.2.1 | 30.118 | {1\|0 0 1/2} | $P_{c}{}^{1}n^{1}c^{1}2^{\infty m}1$ | o1 |
| 30.38.2.1 | 30.119 | {1\|0 1/2 1/2} | $P_{A}{}^{1}n^{1}c^{1}2^{\infty m}1$ | o1 |
| 30.41.2.1 | 30.120 | {1\|1/2 0 1/2} | $P_{B}{}^{1}n^{1}c^{1}2^{\infty m}1$ | o1 |
| 30.37.2.1 | 30.121 | {1\|1/2 1/2 0} | $P_{C}{}^{1}n^{1}c^{1}2^{\infty m}1$ | o1 |
| 30.46.2.1 | 30.122 | {1\|1/2 1/2 1/2} | $P_{I}{}^{1}n^{1}c^{1}2^{\infty m}1$ | o1 |
| 31.26.2.1 | 31.128 | {1\|1/2 0 0} | $P_{a}{}^{1}m^{1}n^{1}2_{1}{}^{\infty m}1$ | o1 |
| 31.31.2.1 | 31.129 | {1\|0 1/2 0} | $P_{b}{}^{1}m^{1}n^{1}2_{1}{}^{\infty m}1$ | o1 |
| 31.28.2.1 | 31.130 | {1\|0 0 1/2} | $P_{c}{}^{1}m^{1}n^{1}2_{1}{}^{\infty m}1$ | o1 |
| 31.40.2.1 | 31.131 | {1\|0 1/2 1/2} | $P_{A}{}^{1}m^{1}n^{1}2_{1}{}^{\infty m}1$ | o1 |
| 31.38.2.1 | 31.132 | {1\|1/2 0 1/2} | $P_{B}{}^{1}m^{1}n^{1}2_{1}{}^{\infty m}1$ | o1 |
| 31.36.2.1 | 31.133 | {1\|1/2 1/2 0} | $P_{C}{}^{1}m^{1}n^{1}2_{1}{}^{\infty m}1$ | o1 |
| 31.44.2.1 | 31.134 | {1\|1/2 1/2 1/2} | $P_{I}{}^{1}m^{1}n^{1}2_{1}{}^{\infty m}1$ | o1 |
| 32.32.2.1 | 32.139 | {1\|0 0 1/2} | $P_{c}{}^{1}b^{1}a^{1}2^{\infty m}1$ | o1 |
| 32.28.2.1 | 32.140 | {1\|0 1/2 0} | $P_{b}{}^{1}b^{1}a^{1}2^{\infty m}1$ | o1 |
| 32.35.2.1 | 32.141 | {1\|1/2 1/2 0} | $P_{C}{}^{1}b^{1}a^{1}2^{\infty m}1$ | o1 |
| 32.41.2.1 | 32.142 | {1\|0 1/2 1/2} | $P_{A}{}^{1}b^{1}a^{1}2^{\infty m}1$ | o1 |
| 32.45.2.1 | 32.143 | {1\|1/2 1/2 1/2} | $P_{I}{}^{1}b^{1}a^{1}2^{\infty m}1$ | o1 |
| 33.31.2.1 | 33.149 | {1\|1/2 0 0} | $P_{a}{}^{1}n^{1}a^{1}2_{1}{}^{\infty m}1$ | o1 |
| 33.29.2.1 | 33.150 | {1\|0 1/2 0} | $P_{b}{}^{1}n^{1}a^{1}2_{1}{}^{\infty m}1$ | o1 |
| 33.32.2.1 | 33.151 | {1\|0 0 1/2} | $P_{c}{}^{1}n^{1}a^{1}2_{1}{}^{\infty m}1$ | o1 |
| 33.40.2.1 | 33.152 | {1\|0 1/2 1/2} | $P_{A}{}^{1}n^{1}a^{1}2_{1}{}^{\infty m}1$ | o1 |
| 33.41.2.1 | 33.153 | {1\|1/2 0 1/2} | $P_{B}{}^{1}n^{1}a^{1}2_{1}{}^{\infty m}1$ | o1 |
| 33.36.2.1 | 33.154 | {1\|1/2 1/2 0} | $P_{C}{}^{1}n^{1}a^{1}2_{1}{}^{\infty m}1$ | o1 |
| 33.46.2.1 | 33.155 | {1\|1/2 1/2 1/2} | $P_{I}{}^{1}n^{1}a^{1}2_{1}{}^{\infty m}1$ | o1 |
| 34.30.2.1 | 34.160 | {1\|1/2 0 0} | $P_{a}{}^{1}n^{1}n^{1}2^{\infty m}1$ | o1 |
| 34.32.2.1 | 34.161 | {1\|0 0 1/2} | $P_{c}{}^{1}n^{1}n^{1}2^{\infty m}1$ | o1 |
| 34.40.2.1 | 34.162 | {1\|0 1/2 1/2} | $P_{A}{}^{1}n^{1}n^{1}2^{\infty m}1$ | o1 |
| 34.37.2.1 | 34.163 | {1\|1/2 1/2 0} | $P_{C}{}^{1}n^{1}n^{1}2^{\infty m}1$ | o1 |
| 34.44.2.1 | 34.164 | {1\|1/2 1/2 1/2} | $P_{I}{}^{1}n^{1}n^{1}2^{\infty m}1$ | o1 |
| 35.35.2.1 | 35.169 | {1\|0 0 1/2} | $C_{c}{}^{1}m^{1}m^{1}2^{\infty m}1$ | o2C |
| 35.25.2.1 | 35.170 | {1\|1/2 0 0} | $C_{a}{}^{1}m^{1}m^{1}2^{\infty m}1$ | o2C |
| 35.42.2.1 | 35.171 | {1\|0 1/2 1/2} | $C_{A}{}^{1}m^{1}m^{1}2^{\infty m}1$ | o2C |
| 36.35.2.1 | 36.177 | {1\|0 0 1/2} | $C_{c}{}^{1}m^{1}c^{1}2_{1}{}^{\infty m}1$ | o2C |
| 36.26.2.1 | 36.178 | {1\|1/2 0 0} | $C_{a}{}^{1}m^{1}c^{1}2_{1}{}^{\infty m}1$ | o2C |
| 36.42.2.1 | 36.179 | {1\|0 1/2 1/2} | $C_{A}{}^{1}m^{1}c^{1}2_{1}{}^{\infty m}1$ | o2C |
| 37.35.2.1 | 37.184 | {1\|0 0 1/2} | $C_{c}{}^{1}c^{1}c^{1}2^{\infty m}1$ | o2C |
| 37.27.2.1 | 37.185 | {1\|1/2 0 0} | $C_{a}{}^{1}c^{1}c^{1}2^{\infty m}1$ | o2C |
| 37.42.2.1 | 37.186 | {1\|0 1/2 1/2} | $C_{A}{}^{1}c^{1}c^{1}2^{\infty m}1$ | o2C |
| 38.38.2.1 | 38.192 | {1\|1/2 0 0} | $A_{a}{}^{1}m^{1}m^{1}2^{\infty m}1$ | o2A |
| 38.25.2.1 | 38.193 | {1\|0 1/2 0} | $A_{b}{}^{1}m{}^{1}m{}^{1}2^{\infty m}1$ | o2A |
| 38.42.2.1 | 38.194 | {1\|1/2 0 1/2} | $A_{B}{}^{1}m{}^{1}m{}^{1}2^{\infty m}1$ | o2A |
| 39.39.2.1 | 39.200 | {1\|1/2 0 0} | $A_{a}{}^{1}e{}^{1}m{}^{1}2^{\infty m}1$ | o2A |
| 39.25.2.1 | 39.201 | {1\|0 1/2 0} | $A_{b}{}^{1}e{}^{1}m{}^{1}2^{\infty m}1$ | o2A |
| 39.42.2.1 | 39.202 | {1\|1/2 0 1/2} | $A_{B}{}^{1}e{}^{1}m{}^{1}2^{\infty m}1$ | o2A |
| 40.38.2.1 | 40.208 | {1\|1/2 0 0} | $A_{a}{}^{1}m{}^{1}a{}^{1}2^{\infty m}1$ | o2A |
| 40.28.2.1 | 40.209 | {1\|0 1/2 0} | $A_{b}{}^{1}m{}^{1}a{}^{1}2^{\infty m}1$ | o2A |
| 40.42.2.1 | 40.210 | {1\|1/2 0 1/2} | $A_{B}{}^{1}m{}^{1}a{}^{1}2^{\infty m}1$ | o2A |
| 41.39.2.1 | 41.216 | {1\|1/2 0 0} | $A_{a}{}^{1}e{}^{1}a{}^{1}2^{\infty m}1$ | o2A |
| 41.28.2.1 | 41.217 | {1\|0 1/2 0} | $A_{b}{}^{1}e{}^{1}a{}^{1}2^{\infty m}1$ | o2A |
| 41.42.2.1 | 41.218 | {1\|1/2 0 1/2} | $A_{B}{}^{1}e{}^{1}a{}^{1}2^{\infty m}1$ | o2A |
| 42.25.2.1 | 42.223 | {1\|1/2 1/2 1/2} | $F_{s}{}^{1}m{}^{1}m{}^{1}2^{\infty m}1$ | o3 |
| 43.34.2.1 | 43.228 | {1\|1/2 1/2 1/2} | $F_{s}{}^{1}d{}^{1}d{}^{1}2^{\infty m}1$ | o3 |
| 44.35.2.1 | 44.233 | {1\|0 0 1/2} | $I_{c}{}^{1}m{}^{1}m{}^{1}2^{\infty m}1$ | o4 |
| 44.38.2.1 | 44.234 | {1\|1/2 0 0} | $I_{a}{}^{1}m{}^{1}m{}^{1}2^{\infty m}1$ | o4 |
| 45.35.2.1 | 45.239 | {1\|0 0 1/2} | $I_{c}{}^{1}b{}^{1}a{}^{1}2^{\infty m}1$ | o4 |
| 45.39.2.1 | 45.240 | {1\|1/2 0 0} | $I_{a}{}^{1}b{}^{1}a{}^{1}2^{\infty m}1$ | o4 |
| 46.35.2.1 | 46.246 | {1\|0 0 1/2} | $I_{c}{}^{1}m{}^{1}a{}^{1}2^{\infty m}1$ | o4 |
| 46.38.2.1 | 46.247 | {1\|1/2 0 0} | $I_{a}{}^{1}m{}^{1}a{}^{1}2^{\infty m}1$ | o4 |
| 46.39.2.1 | 46.248 | {1\|0 1/2 0} | $I_{b}{}^{1}m{}^{1}a{}^{1}2^{\infty m}1$ | o4 |
| 47.47.2.1 | 47.254 | {1\|1/2 0 0} | $P_{a}{}^{1}m{}^{1}m{}^{1}m^{\infty m}1$ | o1 |
| 47.65.2.1 | 47.255 | {1\|1/2 1/2 0} | $P_{C}{}^{1}m{}^{1}m{}^{1}m^{\infty m}1$ | o1 |
| 47.71.2.1 | 47.256 | {1\|1/2 1/2 1/2} | $P_{I}{}^{1}m{}^{1}m{}^{1}m^{\infty m}1$ | o1 |
| 48.50.2.1 | 48.262 | {1\|0 0 1/2} | $P_{c}{}^{1}n{}^{1}n{}^{1}n^{\infty m}1$ | o1 |
| 48.66.2.1 | 48.263 | {1\|1/2 1/2 0} | $P_{C}{}^{1}n{}^{1}n{}^{1}n^{\infty m}1$ | o1 |
| 48.71.2.1 | 48.264 | {1\|1/2 1/2 1/2} | $P_{I}{}^{1}n{}^{1}n{}^{1}n^{\infty m}1$ | o1 |
| 49.49.2.1 | 49.272 | {1\|1/2 0 0} | $P_{a}{}^{1}c{}^{1}c{}^{1}m^{\infty m}1$ | o1 |
| 49.47.2.1 | 49.273 | {1\|0 0 1/2} | $P_{c}{}^{1}c{}^{1}c{}^{1}m^{\infty m}1$ | o1 |
| 49.67.2.1 | 49.274 | {1\|1/2 0 1/2} | $P_{B}{}^{1}c{}^{1}c{}^{1}m^{\infty m}1$ | o1 |
| 49.66.2.1 | 49.275 | {1\|1/2 1/2 0} | $P_{C}{}^{1}c{}^{1}c{}^{1}m^{\infty m}1$ | o1 |
| 49.72.2.1 | 49.276 | {1\|1/2 1/2 1/2} | $P_{I}{}^{1}c{}^{1}c{}^{1}m^{\infty m}1$ | o1 |
| 50.49.2.1 | 50.284 | {1\|1/2 0 0} | $P_{a}{}^{1}b{}^{1}a{}^{1}n^{\infty m}1$ | o1 |
| 50.50.2.1 | 50.285 | {1\|0 0 1/2} | $P_{c}{}^{1}b{}^{1}a{}^{1}n^{\infty m}1$ | o1 |
| 50.68.2.1 | 50.286 | {1\|0 1/2 1/2} | $P_{A}{}^{1}b{}^{1}a{}^{1}n^{\infty m}1$ | o1 |
| 50.65.2.1 | 50.287 | {1\|1/2 1/2 0} | $P_{C}{}^{1}b{}^{1}a{}^{1}n^{\infty m}1$ | o1 |
| 50.72.2.1 | 50.288 | {1\|1/2 1/2 1/2} | $P_{I}{}^{1}b{}^{1}a{}^{1}n^{\infty m}1$ | o1 |
| 51.47.2.1 | 51.298 | {1\|1/2 0 0} | $P_{a}{}^{1}m{}^{1}m{}^{1}a^{\infty m}1$ | o1 |
| 51.51.2.1 | 51.299 | {1\|0 1/2 0} | $P_{b}{}^{1}m{}^{1}m{}^{1}a^{\infty m}1$ | o1 |
| 51.51.2.4 | 51.300 | {1\|0 0 1/2} | $P_{c}{}^{1}m{}^{1}m{}^{1}a^{\infty m}1$ | o1 |
| 51.63.2.1 | 51.301 | {1\|0 1/2 1/2} | $P_{A}{}^{1}m{}^{1}m{}^{1}a^{\infty m}1$ | o1 |
| 51.65.2.1 | 51.302 | {1\|1/2 0 1/2} | $P_{B}{}^{1}m{}^{1}m{}^{1}a^{\infty m}1$ | o1 |
| 51.67.2.1 | 51.303 | {1\|1/2 1/2 0} | $P_{C}{}^{1}m{}^{1}m{}^{1}a^{\infty m}1$ | o1 |
| 51.74.2.1 | 51.304 | {1\|1/2 1/2 1/2} | $P_{I}{}^{1}m{}^{1}m{}^{1}a^{\infty m}1$ | o1 |

| 52.53.2.1 | 52.314 | {1\|1/2 0 0} | $P_{a}{}^{1}n^{1}n^{1}a^{\infty m}1$ | o1 |
|---|---|---|---|---|
| 52.50.2.1 | 52.315 | {1\|0 1/2 0} | $P_{b}{}^{1}n^{1}n^{1}a^{\infty m}1$ | o1 |
| 52.54.2.1 | 52.316 | {1\|0 0 1/2} | $P_{c}{}^{1}n^{1}n^{1}a^{\infty m}1$ | o1 |
| 52.66.2.1 | 52.317 | {1\|0 1/2 1/2} | $P_{A}{}^{1}n^{1}n^{1}a^{\infty m}1$ | o1 |
| 52.63.2.1 | 52.318 | {1\|1/2 0 1/2} | $P_{B}{}^{1}n^{1}n^{1}a^{\infty m}1$ | o1 |
| 52.68.2.1 | 52.319 | {1\|1/2 1/2 0} | $P_{C}{}^{1}n^{1}n^{1}a^{\infty m}1$ | o1 |
| 52.74.2.1 | 52.320 | {1\|1/2 1/2 1/2} | $P_{I}{}^{1}n^{1}n^{1}a^{\infty m}1$ | o1 |
| 53.51.2.1 | 53.330 | {1\|1/2 0 0} | $P_{a}{}^{1}m^{1}n^{1}a^{\infty m}1$ | o1 |
| 53.53.2.1 | 53.331 | {1\|0 1/2 0} | $P_{b}{}^{1}m^{1}n^{1}a^{\infty m}1$ | o1 |
| 53.49.2.1 | 53.332 | {1\|0 0 1/2} | $P_{c}{}^{1}m^{1}n^{1}a^{\infty m}1$ | o1 |
| 53.66.2.1 | 53.333 | {1\|0 1/2 1/2} | $P_{A}{}^{1}m^{1}n^{1}a^{\infty m}1$ | o1 |
| 53.65.2.1 | 53.334 | {1\|1/2 0 1/2} | $P_{B}{}^{1}m^{1}n^{1}a^{\infty m}1$ | o1 |
| 53.64.2.1 | 53.335 | {1\|1/2 1/2 0} | $P_{C}{}^{1}m^{1}n^{1}a^{\infty m}1$ | o1 |
| 53.74.2.1 | 53.336 | {1\|1/2 1/2 1/2} | $P_{I}{}^{1}m^{1}n^{1}a^{\infty m}1$ | o1 |
| 54.49.2.1 | 54.346 | {1\|1/2 0 0} | $P_{a}{}^{1}c^{1}c^{1}a^{\infty m}1$ | o1 |
| 54.54.2.1 | 54.347 | {1\|0 1/2 0} | $P_{b}{}^{1}c^{1}c^{1}a^{\infty m}1$ | o1 |
| 54.51.2.1 | 54.348 | {1\|0 0 1/2} | $P_{c}{}^{1}c^{1}c^{1}a^{\infty m}1$ | o1 |
| 54.64.2.1 | 54.349 | {1\|0 1/2 1/2} | $P_{A}{}^{1}c^{1}c^{1}a^{\infty m}1$ | o1 |
| 54.67.2.1 | 54.350 | {1\|1/2 0 1/2} | $P_{B}{}^{1}c^{1}c^{1}a^{\infty m}1$ | o1 |
| 54.68.2.1 | 54.351 | {1\|1/2 1/2 0} | $P_{C}{}^{1}c^{1}c^{1}a^{\infty m}1$ | o1 |
| 54.73.2.1 | 54.352 | {1\|1/2 1/2 1/2} | $P_{I}{}^{1}c^{1}c^{1}a^{\infty m}1$ | o1 |
| 55.51.2.1 | 55.360 | {1\|1/2 0 0} | $P_{a}{}^{1}b^{1}a^{1}m^{\infty m}1$ | o1 |
| 55.55.2.1 | 55.361 | {1\|0 0 1/2} | $P_{c}{}^{1}b^{1}a^{1}m^{\infty m}1$ | o1 |
| 55.64.2.1 | 55.362 | {1\|0 1/2 1/2} | $P_{A}{}^{1}b^{1}a^{1}m^{\infty m}1$ | o1 |
| 55.65.2.1 | 55.363 | {1\|1/2 1/2 0} | $P_{C}{}^{1}b^{1}a^{1}m^{\infty m}1$ | o1 |
| 55.72.2.1 | 55.364 | {1\|1/2 1/2 1/2} | $P_{I}{}^{1}b^{1}a^{1}m^{\infty m}1$ | o1 |
| 56.54.2.1 | 56.372 | {1\|0 1/2 0} | $P_{b}{}^{1}c^{1}c^{1}n^{\infty m}1$ | o1 |
| 56.59.2.1 | 56.373 | {1\|0 0 1/2} | $P_{c}{}^{1}c^{1}c^{1}n^{\infty m}1$ | o1 |
| 56.64.2.1 | 56.374 | {1\|0 1/2 1/2} | $P_{A}{}^{1}c^{1}c^{1}n^{\infty m}1$ | o1 |
| 56.66.2.1 | 56.375 | {1\|1/2 1/2 0} | $P_{C}{}^{1}c^{1}c^{1}n^{\infty m}1$ | o1 |
| 56.72.2.1 | 56.376 | {1\|1/2 1/2 1/2} | $P_{I}{}^{1}c^{1}c^{1}n^{\infty m}1$ | o1 |
| 57.57.2.1 | 57.386 | {1\|1/2 0 0} | $P_{a}{}^{1}b^{1}c^{1}m^{\infty m}1$ | o1 |
| 57.51.2.4 | 57.387 | {1\|0 1/2 0} | $P_{b}{}^{1}b^{1}c^{1}m^{\infty m}1$ | o1 |
| 57.51.2.1 | 57.388 | {1\|0 0 1/2} | $P_{c}{}^{1}b^{1}c^{1}m^{\infty m}1$ | o1 |
| 57.67.2.1 | 57.389 | {1\|0 1/2 1/2} | $P_{A}{}^{1}b^{1}c^{1}m^{\infty m}1$ | o1 |
| 57.64.2.1 | 57.390 | {1\|1/2 0 1/2} | $P_{B}{}^{1}b^{1}c^{1}m^{\infty m}1$ | o1 |
| 57.63.2.1 | 57.391 | {1\|1/2 1/2 0} | $P_{C}{}^{1}b^{1}c^{1}m^{\infty m}1$ | o1 |
| 57.72.2.1 | 57.392 | {1\|1/2 1/2 1/2} | $P_{I}{}^{1}b^{1}c^{1}m^{\infty m}1$ | o1 |
| 58.53.2.1 | 58.400 | {1\|1/2 0 0} | $P_{a}{}^{1}n^{1}n^{1}m^{\infty m}1$ | o1 |
| 58.55.2.1 | 58.401 | {1\|0 0 1/2} | $P_{c}{}^{1}n^{1}n^{1}m^{\infty m}1$ | o1 |
| 58.63.2.1 | 58.402 | {1\|1/2 0 1/2} | $P_{B}{}^{1}n^{1}n^{1}m^{\infty m}1$ | o1 |
| 58.66.2.1 | 58.403 | {1\|1/2 1/2 0} | $P_{C}{}^{1}n^{1}n^{1}m^{\infty m}1$ | o1 |
| 58.71.2.1 | 58.404 | {1\|1/2 1/2 1/2} | $P_{I}{}^{1}n^{1}n^{1}m^{\infty m}1$ | o1 |

| 59.51.2.1 | 59.412 | $\{1\vert 0\ 1/2\ 0\}$ | $P_b{}^1m^1m^1n^{\infty m}1$ | o1 |
|---|---|---|---|---|
| 59.59.2.1 | 59.413 | $\{1\vert 0\ 0\ 1/2\}$ | $P_c{}^1m^1m^1n^{\infty m}1$ | o1 |
| 59.63.2.1 | 59.414 | $\{1\vert 1/2\ 0\ 1/2\}$ | $P_B{}^1m^1m^1n^{\infty m}1$ | o1 |
| 59.65.2.1 | 59.415 | $\{1\vert 1/2\ 1/2\ 0\}$ | $P_C{}^1m^1m^1n^{\infty m}1$ | o1 |
| 59.71.2.1 | 59.416 | $\{1\vert 1/2\ 1/2\ 1/2\}$ | $P_I{}^1m^1m^1n^{\infty m}1$ | o1 |
| 60.54.2.1 | 60.426 | $\{1\vert 1/2\ 0\ 0\}$ | $P_a{}^1b^1c^1n^{\infty m}1$ | o1 |
| 60.57.2.1 | 60.427 | $\{1\vert 0\ 1/2\ 0\}$ | $P_b{}^1b^1c^1n^{\infty m}1$ | o1 |
| 60.53.2.1 | 60.428 | $\{1\vert 0\ 0\ 1/2\}$ | $P_c{}^1b^1c^1n^{\infty m}1$ | o1 |
| 60.64.2.1 | 60.429 | $\{1\vert 0\ 1/2\ 1/2\}$ | $P_A{}^1b^1c^1n^{\infty m}1$ | o1 |
| 60.68.2.1 | 60.430 | $\{1\vert 1/2\ 0\ 1/2\}$ | $P_B{}^1b^1c^1n^{\infty m}1$ | o1 |
| 60.63.2.1 | 60.431 | $\{1\vert 1/2\ 1/2\ 0\}$ | $P_C{}^1b^1c^1n^{\infty m}1$ | o1 |
| 60.72.2.1 | 60.432 | $\{1\vert 1/2\ 1/2\ 1/2\}$ | $P_I{}^1b^1c^1n^{\infty m}1$ | o1 |
| 61.57.2.1 | 61.438 | $\{1\vert 1/2\ 0\ 0\}$ | $P_a{}^1b^1c^1a^{\infty m}1$ | o1 |
| 61.64.2.1 | 61.439 | $\{1\vert 1/2\ 1/2\ 0\}$ | $P_C{}^1b^1c^1a^{\infty m}1$ | o1 |
| 61.73.2.1 | 61.440 | $\{1\vert 1/2\ 1/2\ 1/2\}$ | $P_I{}^1b^1c^1a^{\infty m}1$ | o1 |
| 62.59.2.1 | 62.450 | $\{1\vert 1/2\ 0\ 0\}$ | $P_a{}^1n^1m^1a^{\infty m}1$ | o1 |
| 62.55.2.1 | 62.451 | $\{1\vert 0\ 1/2\ 0\}$ | $P_b{}^1n^1m^1a^{\infty m}1$ | o1 |
| 62.57.2.1 | 62.452 | $\{1\vert 0\ 0\ 1/2\}$ | $P_c{}^1n^1m^1a^{\infty m}1$ | o1 |
| 62.63.2.4 | 62.453 | $\{1\vert 0\ 1/2\ 1/2\}$ | $P_A{}^1n^1m^1a^{\infty m}1$ | o1 |
| 62.63.2.1 | 62.454 | $\{1\vert 1/2\ 0\ 1/2\}$ | $P_B{}^1n^1m^1a^{\infty m}1$ | o1 |
| 62.64.2.1 | 62.455 | $\{1\vert 1/2\ 1/2\ 0\}$ | $P_C{}^1n^1m^1a^{\infty m}1$ | o1 |
| 62.74.2.1 | 62.456 | $\{1\vert 1/2\ 1/2\ 1/2\}$ | $P_I{}^1n^1m^1a^{\infty m}1$ | o1 |
| 63.65.2.1 | 63.466 | $\{1\vert 0\ 0\ 1/2\}$ | $C_c{}^1m^1c^1m^{\infty m}1$ | o2C |
| 63.51.2.1 | 63.467 | $\{1\vert 1/2\ 0\ 0\}$ | $C_a{}^1m^1c^1m^{\infty m}1$ | o2C |
| 63.69.2.1 | 63.468 | $\{1\vert 0\ 1/2\ 1/2\}$ | $C_A{}^1m^1c^1m^{\infty m}1$ | o2C |
| 64.67.2.1 | 64.478 | $\{1\vert 0\ 0\ 1/2\}$ | $C_c{}^1m^1c^1e^{\infty m}1$ | o2C |
| 64.51.2.1 | 64.479 | $\{1\vert 1/2\ 0\ 0\}$ | $C_a{}^1m^1c^1e^{\infty m}1$ | o2C |
| 64.69.2.1 | 64.480 | $\{1\vert 0\ 1/2\ 1/2\}$ | $C_A{}^1m^1c^1e^{\infty m}1$ | o2C |
| 65.65.2.1 | 65.488 | $\{1\vert 0\ 0\ 1/2\}$ | $C_c{}^1m^1m^1m^{\infty m}1$ | o2C |
| 65.47.2.1 | 65.489 | $\{1\vert 1/2\ 0\ 0\}$ | $C_a{}^1m^1m^1m^{\infty m}1$ | o2C |
| 65.69.2.1 | 65.490 | $\{1\vert 0\ 1/2\ 1/2\}$ | $C_A{}^1m^1m^1m^{\infty m}1$ | o2C |
| 66.65.2.1 | 66.498 | $\{1\vert 0\ 0\ 1/2\}$ | $C_c{}^1c^1c^1m^{\infty m}1$ | o2C |
| 66.49.2.1 | 66.499 | $\{1\vert 1/2\ 0\ 0\}$ | $C_a{}^1c^1c^1m^{\infty m}1$ | o2C |
| 66.69.2.1 | 66.500 | $\{1\vert 0\ 1/2\ 1/2\}$ | $C_A{}^1c^1c^1m^{\infty m}1$ | o2C |
| 67.67.2.1 | 67.508 | $\{1\vert 0\ 0\ 1/2\}$ | $C_c{}^1m^1m^1e^{\infty m}1$ | o2C |
| 67.47.2.1 | 67.509 | $\{1\vert 1/2\ 0\ 0\}$ | $C_a{}^1m^1m^1e^{\infty m}1$ | o2C |
| 67.69.2.1 | 67.510 | $\{1\vert 0\ 1/2\ 1/2\}$ | $C_A{}^1m^1m^1e^{\infty m}1$ | o2C |
| 68.67.2.1 | 68.518 | $\{1\vert 0\ 0\ 1/2\}$ | $C_c{}^1c^1c^1e^{\infty m}1$ | o2C |
| 68.49.2.1 | 68.519 | $\{1\vert 1/2\ 0\ 0\}$ | $C_a{}^1c^1c^1e^{\infty m}1$ | o2C |
| 68.69.2.1 | 68.520 | $\{1\vert 0\ 1/2\ 1/2\}$ | $C_A{}^1c^1c^1e^{\infty m}1$ | o2C |
| 69.47.2.1 | 69.526 | $\{1\vert 1/2\ 1/2\ 1/2\}$ | $F_S{}^1m^1m^1m^{\infty m}1$ | o3 |
| 70.48.2.1 | 70.532 | $\{1\vert 1/2\ 1/2\ 1/2\}$ | $F_S{}^1d^1d^1d^{\infty m}1$ | o3 |
| 71.65.2.1 | 71.538 | $\{1\vert 0\ 0\ 1/2\}$ | $I_c{}^1m^1m^1m^{\infty m}1$ | o4 |

| 72.65.2.1 | 72.546 | {1\|0 0 1/2} | $I_c{}^{1}b^{1}a^{1}m^{\infty m}1$ | o4 |
|---|---|---|---|---|
| 72.67.2.1 | 72.547 | {1\|0 1/2 0} | $I_b{}^{1}b^{1}a^{1}m^{\infty m}1$ | o4 |
| 73.67.2.1 | 73.553 | {1\|0 0 1/2} | $I_c{}^{1}b^{1}c^{1}a^{\infty m}1$ | o4 |
| 74.67.2.1 | 74.561 | {1\|0 0 1/2} | $I_c{}^{1}m^{1}m^{1}a^{\infty m}1$ | o4 |
| 74.65.2.1 | 74.562 | {1\|0 1/2 0} | $I_b{}^{1}m^{1}m^{1}a^{\infty m}1$ | o4 |
| 75.75.2.1 | 75.4 | {1\|0 0 1/2} | $P_c{}^{1}4^{\infty m}1$ | t1 |
| 75.75.2.4 | 75.5 | {1\|1/2 1/2 0} | $P_C{}^{1}4^{\infty m}1$ | t1 |
| 75.79.2.1 | 75.6 | {1\|1/2 1/2 1/2} | $P_I{}^{1}4^{\infty m}1$ | t1 |
| 76.77.2.1 | 76.10 | {1\|0 0 1/2} | $P_c{}^{1}4_1{}^{\infty m}1$ | t1 |
| 76.76.2.1 | 76.11 | {1\|1/2 1/2 0} | $P_C{}^{1}4_1{}^{\infty m}1$ | t1 |
| 76.80.2.1 | 76.12 | {1\|1/2 1/2 1/2} | $P_I{}^{1}4_1{}^{\infty m}1$ | t1 |
| 77.75.2.1 | 77.16 | {1\|0 0 1/2} | $P_c{}^{1}4_2{}^{\infty m}1$ | t1 |
| 77.77.2.1 | 77.17 | {1\|1/2 1/2 0} | $P_C{}^{1}4_2{}^{\infty m}1$ | t1 |
| 77.79.2.1 | 77.18 | {1\|1/2 1/2 1/2} | $P_I{}^{1}4_2{}^{\infty m}1$ | t1 |
| 78.77.2.1 | 78.22 | {1\|0 0 1/2} | $P_c{}^{1}4_3{}^{\infty m}1$ | t1 |
| 78.78.2.1 | 78.23 | {1\|1/2 1/2 0} | $P_C{}^{1}4_3{}^{\infty m}1$ | t1 |
| 78.80.2.1 | 78.24 | {1\|1/2 1/2 1/2} | $P_I{}^{1}4_3{}^{\infty m}1$ | t1 |
| 79.75.2.1 | 79.28 | {1\|0 0 1/2} | $I_c{}^{1}4^{\infty m}1$ | t2 |
| 80.77.2.1 | 80.32 | {1\|0 0 1/2} | $I_c{}^{1}4_1{}^{\infty m}1$ | t2 |
| 81.81.2.1 | 81.36 | {1\|0 0 1/2} | $P_c{}^{1}\bar{4}^{\infty m}1$ | t1 |
| 81.81.2.4 | 81.37 | {1\|1/2 1/2 0} | $P_C{}^{1}\bar{4}^{\infty m}1$ | t1 |
| 81.82.2.1 | 81.38 | {1\|1/2 1/2 1/2} | $P_I{}^{1}\bar{4}^{\infty m}1$ | t1 |
| 82.81.2.1 | 82.42 | {1\|1/2 1/2 1/2} | $I_c{}^{1}\bar{4}^{\infty m}1$ | t2 |
| 83.83.2.1 | 83.48 | {1\|0 0 1/2} | $P_c{}^{1}4/{}^{1}m^{\infty m}1$ | t1 |
| 83.83.2.4 | 83.49 | {1\|1/2 1/2 0} | $P_C{}^{1}4/{}^{1}m^{\infty m}1$ | t1 |
| 83.87.2.1 | 83.50 | {1\|1/2 1/2 1/2} | $P_I{}^{1}4/{}^{1}m^{\infty m}1$ | t1 |
| 84.83.2.1 | 84.56 | {1\|0 0 1/2} | $P_c{}^{1}4_2/{}^{1}m^{\infty m}1$ | t1 |
| 84.84.2.1 | 84.57 | {1\|1/2 1/2 0} | $P_C{}^{1}4_2/{}^{1}m^{\infty m}1$ | t1 |
| 84.87.2.1 | 84.58 | {1\|1/2 1/2 1/2} | $P_I{}^{1}4_2/{}^{1}m^{\infty m}1$ | t1 |
| 85.85.2.1 | 85.64 | {1\|0 0 1/2} | $P_c{}^{1}4/{}^{1}n^{\infty m}1$ | t1 |
| 85.83.2.1 | 85.65 | {1\|1/2 1/2 0} | $P_C{}^{1}4/{}^{1}n^{\infty m}1$ | t1 |
| 85.87.2.1 | 85.66 | {1\|1/2 1/2 1/2} | $P_I{}^{1}4/{}^{1}n^{\infty m}1$ | t1 |
| 86.85.2.1 | 86.72 | {1\|0 0 1/2} | $P_c{}^{1}4_2/{}^{1}n^{\infty m}1$ | t1 |
| 86.84.2.1 | 86.73 | {1\|1/2 1/2 0} | $P_C{}^{1}4_2/{}^{1}n^{\infty m}1$ | t1 |
| 86.87.2.1 | 86.74 | {1\|1/2 1/2 1/2} | $P_I{}^{1}4_2/{}^{1}n^{\infty m}1$ | t1 |
| 87.83.2.1 | 87.80 | {1\|0 0 1/2} | $I_c{}^{1}4/{}^{1}m^{\infty m}1$ | t2 |
| 88.86.2.1 | 88.86 | {1\|0 0 1/2} | $I_c{}^{1}4_1/{}^{1}a^{\infty m}1$ | t2 |
| 89.89.2.1 | 89.92 | {1\|0 0 1/2} | $P_c{}^{1}4^{1}2^{1}2^{\infty m}1$ | t3 |
| 89.89.2.4 | 89.93 | {1\|1/2 1/2 0} | $P_C{}^{1}4^{1}2^{1}2^{\infty m}1$ | t3 |
| 89.97.2.1 | 89.94 | {1\|1/2 1/2 1/2} | $P_I{}^{1}4^{1}2^{1}2^{\infty m}1$ | t3 |
| 90.90.2.1 | 90.100 | {1\|0 0 1/2} | $P_c{}^{1}4^{1}2_1{}^{1}2^{\infty m}1$ | t3 |
| 90.89.2.1 | 90.101 | {1\|1/2 1/2 0} | $P_C{}^{1}4^{1}2_1{}^{1}2^{\infty m}1$ | t3 |
| 90.97.2.1 | 90.102 | {1\|1/2 1/2 1/2} | $P_I{}^{1}4^{1}2_1{}^{1}2^{\infty m}1$ | t3 |

| 91.93.2.1 | 91.108 | {1\|0 0 1/2} | $P_c{}^{1}4_1{}^{1}2^{1}2^{\infty m}1$ | t3 |
|---|---|---|---|---|
| 91.91.2.1 | 91.109 | {1\|1/2 1/2 0} | $P_C{}^{1}4_1{}^{1}2^{1}2^{\infty m}1$ | t3 |
| 91.98.2.1 | 91.110 | {1\|1/2 1/2 1/2} | $P_I{}^{1}4_1{}^{1}2^{1}2^{\infty m}1$ | t3 |
| 92.94.2.1 | 92.116 | {1\|0 0 1/2} | $P_c{}^{1}4_1{}^{1}2_1{}^{1}2^{\infty m}1$ | t3 |
| 92.91.2.1 | 92.117 | {1\|1/2 1/2 0} | $P_C{}^{1}4_1{}^{1}2_1{}^{1}2^{\infty m}1$ | t3 |
| 92.98.2.1 | 92.118 | {1\|1/2 1/2 1/2} | $P_I{}^{1}4_1{}^{1}2_1{}^{1}2^{\infty m}1$ | t3 |
| 93.89.2.1 | 93.124 | {1\|0 0 1/2} | $P_c{}^{1}4_2{}^{1}2^{1}2^{\infty m}1$ | t3 |
| 93.93.2.1 | 93.125 | {1\|1/2 1/2 0} | $P_C{}^{1}4_2{}^{1}2^{1}2^{\infty m}1$ | t3 |
| 93.97.2.1 | 93.126 | {1\|1/2 1/2 1/2} | $P_I{}^{1}4_2{}^{1}2^{1}2^{\infty m}1$ | t3 |
| 94.90.2.1 | 94.132 | {1\|0 0 1/2} | $P_c{}^{1}4_2{}^{1}2_1{}^{1}2^{\infty m}1$ | t3 |
| 94.93.2.1 | 94.133 | {1\|1/2 1/2 0} | $P_C{}^{1}4_2{}^{1}2_1{}^{1}2^{\infty m}1$ | t3 |
| 94.97.2.1 | 94.134 | {1\|1/2 1/2 1/2} | $P_I{}^{1}4_2{}^{1}2_1{}^{1}2^{\infty m}1$ | t3 |
| 95.93.2.1 | 95.140 | {1\|0 0 1/2} | $P_c{}^{1}4_3{}^{1}2^{1}2^{\infty m}1$ | t3 |
| 95.95.2.1 | 95.141 | {1\|1/2 1/2 0} | $P_C{}^{1}4_3{}^{1}2^{1}2^{\infty m}1$ | t3 |
| 95.98.2.1 | 95.142 | {1\|1/2 1/2 1/2} | $P_I{}^{1}4_3{}^{1}2^{1}2^{\infty m}1$ | t3 |
| 96.94.2.1 | 96.148 | {1\|0 0 1/2} | $P_c{}^{1}4_3{}^{1}2_1{}^{1}2^{\infty m}1$ | t3 |
| 96.95.2.1 | 96.149 | {1\|1/2 1/2 0} | $P_C{}^{1}4_3{}^{1}2_1{}^{1}2^{\infty m}1$ | t3 |
| 96.98.2.1 | 96.150 | {1\|1/2 1/2 1/2} | $P_I{}^{1}4_3{}^{1}2_1{}^{1}2^{\infty m}1$ | t3 |
| 97.89.2.1 | 97.156 | {1\|0 0 1/2} | $I_c{}^{1}4^{1}2^{1}2^{\infty m}1$ | t4 |
| 98.93.2.1 | 98.162 | {1\|0 0 1/2} | $I_c{}^{1}4_1{}^{1}2^{1}2^{\infty m}1$ | t4 |
| 99.99.2.1 | 99.168 | {1\|0 0 1/2} | $P_c{}^{1}4^{1}m^{1}m^{\infty m}1$ | t3 |
| 99.99.2.4 | 99.169 | {1\|1/2 1/2 0} | $P_C{}^{1}4^{1}m^{1}m^{\infty m}1$ | t3 |
| 99.107.2.1 | 99.170 | {1\|1/2 1/2 1/2} | $P_I{}^{1}4^{1}m^{1}m^{\infty m}1$ | t3 |
| 100.100.2.1 | 100.176 | {1\|0 0 1/2} | $P_c{}^{1}4^{1}b^{1}m^{\infty m}1$ | t3 |
| 100.99.2.1 | 100.177 | {1\|1/2 1/2 0} | $P_C{}^{1}4^{1}b^{1}m^{\infty m}1$ | t3 |
| 100.108.2.1 | 100.178 | {1\|1/2 1/2 1/2} | $P_I{}^{1}4^{1}b^{1}m^{\infty m}1$ | t3 |
| 101.99.2.1 | 101.184 | {1\|0 0 1/2} | $P_c{}^{1}4_2{}^{1}c^{1}m^{\infty m}1$ | t3 |
| 101.105.2.1 | 101.185 | {1\|1/2 1/2 0} | $P_C{}^{1}4_2{}^{1}c^{1}m^{\infty m}1$ | t3 |
| 101.108.2.1 | 101.186 | {1\|1/2 1/2 1/2} | $P_I{}^{1}4_2{}^{1}c^{1}m^{\infty m}1$ | t3 |
| 102.100.2.1 | 102.192 | {1\|0 0 1/2} | $P_c{}^{1}4_2{}^{1}n^{1}m^{\infty m}1$ | t3 |
| 102.105.2.1 | 102.193 | {1\|1/2 1/2 0} | $P_C{}^{1}4_2{}^{1}n^{1}m^{\infty m}1$ | t3 |
| 102.107.2.1 | 102.194 | {1\|1/2 1/2 1/2} | $P_I{}^{1}4_2{}^{1}n^{1}m^{\infty m}1$ | t3 |
| 103.99.2.1 | 103.200 | {1\|0 0 1/2} | $P_c{}^{1}4^{1}c^{1}c^{\infty m}1$ | t3 |
| 103.103.2.1 | 103.201 | {1\|1/2 1/2 0} | $P_C{}^{1}4^{1}c^{1}c^{\infty m}1$ | t3 |
| 103.108.2.1 | 103.202 | {1\|1/2 1/2 1/2} | $P_I{}^{1}4^{1}c^{1}c^{\infty m}1$ | t3 |
| 104.100.2.1 | 104.208 | {1\|0 0 1/2} | $P_c{}^{1}4^{1}n^{1}c^{\infty m}1$ | t3 |
| 104.103.2.1 | 104.209 | {1\|1/2 1/2 0} | $P_C{}^{1}4^{1}n^{1}c^{\infty m}1$ | t3 |
| 104.107.2.1 | 104.210 | {1\|1/2 1/2 1/2} | $P_I{}^{1}4^{1}\mathrm{n}^{1}c^{\infty m}1$ | t3 |
| 105.99.2.1 | 105.216 | {1\|0 0 1/2} | $P_c{}^{1}4_2{}^{1}m^{1}c^{\infty m}1$ | t3 |
| 105.101.2.1 | 105.217 | {1\|1/2 1/2 0} | $P_C{}^{1}4_2{}^{1}m^{1}c^{\infty m}1$ | t3 |
| 105.107.2.1 | 105.218 | {1\|1/2 1/2 1/2} | $P_I{}^{1}4_2{}^{1}m^{1}c^{\infty m}1$ | t3 |
| 106.100.2.1 | 106.224 | {1\|0 0 1/2} | $P_c{}^{1}4_2{}^{1}b^{1}c^{\infty m}1$ | t3 |
| 106.101.2.1 | 106.225 | {1\|1/2 1/2 0} | $P_C{}^{1}4_2{}^{1}b^{1}c^{\infty m}1$ | t3 |

| 106.108.2.1 | 106.226 | {1\|1/2 1/2 1/2} | $P_I{}^14_2{}^1b^1c^{\infty m}1$ | t3 |
|---|---|---|---|---|
| 107.99.2.1 | 107.232 | {1\|0 0 1/2} | $I_c{}^14^1m^1m^{\infty m}1$ | t4 |
| 108.99.2.1 | 108.238 | {1\|0 0 1/2} | $I_c{}^14^1c^1m^{\infty m}1$ | t4 |
| 109.102.2.1 | 109.244 | {1\|0 0 1/2} | $I_c{}^14_1{}^1m^1d^{\infty m}1$ | t4 |
| 110.102.2.1 | 110.250 | {1\|0 0 1/2} | $I_c{}^14_1{}^1c^1d^{\infty m}1$ | t4 |
| 111.111.2.1 | 111.256 | {1\|0 0 1/2} | $P_c{}^1\bar{4}^12^1m^{\infty m}1$ | t3 |
| 111.115.2.1 | 111.257 | {1\|1/2 1/2 0} | $P_C{}^1\bar{4}^12^1m^{\infty m}1$ | t3 |
| 111.121.2.1 | 111.258 | {1\|1/2 1/2 1/2} | $P_I{}^1\bar{4}^12^1m^{\infty m}1$ | t3 |
| 112.111.2.1 | 112.264 | {1\|0 0 1/2} | $P_c{}^1\bar{4}^12^1c^{\infty m}1$ | t3 |
| 112.116.2.1 | 112.265 | {1\|1/2 1/2 0} | $P_C{}^1\bar{4}^12^1c^{\infty m}1$ | t3 |
| 112.121.2.1 | 112.266 | {1\|1/2 1/2 1/2} | $P_I{}^1\bar{4}^12^1c^{\infty m}1$ | t3 |
| 113.113.2.1 | 113.272 | {1\|0 0 1/2} | $P_c{}^1\bar{4}^12_1{}^1m^{\infty m}1$ | t3 |
| 113.115.2.1 | 113.273 | {1\|1/2 1/2 0} | $P_C{}^1\bar{4}^12_1{}^1m^{\infty m}1$ | t3 |
| 113.121.2.1 | 113.274 | {1\|1/2 1/2 1/2} | $P_I{}^1\bar{4}^12_1{}^1m^{\infty m}1$ | t3 |
| 114.113.2.1 | 114.280 | {1\|0 0 1/2} | $P_c{}^1\bar{4}^12_1{}^1c^{\infty m}1$ | t3 |
| 114.116.2.1 | 114.281 | {1\|1/2 1/2 0} | $P_C{}^1\bar{4}^12_1{}^1c^{\infty m}1$ | t3 |
| 114.121.2.1 | 114.282 | {1\|1/2 1/2 1/2} | $P_I{}^1\bar{4}^12_1{}^1c^{\infty m}1$ | t3 |
| 115.115.2.1 | 115.288 | {1\|0 0 1/2} | $P_c{}^1\bar{4}^1m^12^{\infty m}1$ | t3 |
| 115.111.2.1 | 115.289 | {1\|1/2 1/2 0} | $P_C{}^1\bar{4}^1m^12^{\infty m}1$ | t3 |
| 115.119.2.1 | 115.290 | {1\|1/2 1/2 1/2} | $P_I{}^1\bar{4}^1m^12^{\infty m}1$ | t3 |
| 116.115.2.1 | 116.296 | {1\|0 0 1/2} | $P_c{}^1\bar{4}^1c^12^{\infty m}1$ | t3 |
| 116.112.2.1 | 116.297 | {1\|1/2 1/2 0} | $P_C{}^1\bar{4}^1c^12^{\infty m}1$ | t3 |
| 116.120.2.1 | 116.298 | {1\|1/2 1/2 1/2} | $P_I{}^1\bar{4}^1c^12^{\infty m}1$ | t3 |
| 117.117.2.1 | 117.304 | {1\|0 0 1/2} | $P_c{}^1\bar{4}^1b^12^{\infty m}1$ | t3 |
| 117.111.2.1 | 117.305 | {1\|1/2 1/2 0} | $P_C{}^1\bar{4}^1b^12^{\infty m}1$ | t3 |
| 117.120.2.1 | 117.306 | {1\|1/2 1/2 1/2} | $P_I{}^1\bar{4}^1b^12^{\infty m}1$ | t3 |
| 118.117.2.1 | 118.312 | {1\|0 0 1/2} | $P_c{}^1\bar{4}^1n^12^{\infty m}1$ | t3 |
| 118.112.2.1 | 118.313 | {1\|1/2 1/2 0} | $P_C{}^1\bar{4}^1n^12^{\infty m}1$ | t3 |
| 118.119.2.1 | 118.314 | {1\|1/2 1/2 1/2} | $P_I{}^1\bar{4}^1n^12^{\infty m}1$ | t3 |
| 119.111.2.1 | 119.320 | {1\|0 0 1/2} | $I_c{}^1\bar{4}^1m^12^{\infty m}1$ | t4 |
| 120.111.2.1 | 120.326 | {1\|0 0 1/2} | $I_c{}^1\bar{4}^1c^12^{\infty m}1$ | t4 |
| 121.115.2.1 | 121.332 | {1\|0 0 1/2} | $I_c{}^1\bar{4}^12^1m^{\infty m}1$ | t4 |
| 122.118.2.1 | 122.338 | {1\|0 0 1/2} | $I_c{}^1\bar{4}^12^1d^{\infty m}1$ | t4 |
| 123.123.2.1 | 123.348 | {1\|0 0 1/2} | $P_c{}^14/{}^1m^1m^1m^{\infty m}1$ | t3 |
| 123.123.2.4 | 123.349 | {1\|1/2 1/2 0} | $P_C{}^14/{}^1m^1m^1m^{\infty m}1$ | t3 |
| 123.139.2.1 | 123.350 | {1\|1/2 1/2 1/2} | $P_I{}^14/{}^1m^1m^1m^{\infty m}1$ | t3 |
| 124.123.2.1 | 124.360 | {1\|0 0 1/2} | $P_c{}^14/{}^1m^1c^1c^{\infty m}1$ | t3 |
| 124.124.2.1 | 124.361 | {1\|1/2 1/2 0} | $P_C{}^14/{}^1m^1c^1c^{\infty m}1$ | t3 |
| 124.140.2.1 | 124.362 | {1\|1/2 1/2 1/2} | $P_I{}^14/{}^1m^1c^1c^{\infty m}1$ | t3 |
| 125.125.2.1 | 125.372 | {1\|0 0 1/2} | $P_c{}^14/{}^1n^1b^1m^{\infty m}1$ | t3 |
| 125.123.2.1 | 125.373 | {1\|1/2 1/2 0} | $P_C{}^14/{}^1n^1b^1m^{\infty m}1$ | t3 |
| 125.140.2.1 | 125.374 | {1\|1/2 1/2 1/2} | $P_I{}^14/{}^1n^1b^1m^{\infty m}1$ | t3 |
| 126.125.2.1 | 126.384 | {1\|0 0 1/2} | $P_c{}^14/{}^1n^1n^1c^{\infty m}1$ | t3 |

| 126.124.2.1 | 126.385 | {1\|1/2 1/2 0} | $P_C{}^14/{}^1n{}^1n{}^1c^{\infty m}1$ | t3 |
|---|---|---|---|---|
| 126.139.2.1 | 126.386 | {1\|1/2 1/2 1/2} | $P_I{}^14/{}^1n{}^1n{}^1c^{\infty m}1$ | t3 |
| 127.127.2.1 | 127.396 | {1\|0 0 1/2} | $P_c{}^14/{}^1m{}^1b{}^1m^{\infty m}1$ | t3 |
| 127.123.2.1 | 127.397 | {1\|1/2 1/2 0} | $P_C{}^14/{}^1m{}^1b{}^1m^{\infty m}1$ | t3 |
| 127.140.2.1 | 127.398 | {1\|1/2 1/2 1/2} | $P_I{}^14/{}^1m{}^1b{}^1m^{\infty m}1$ | t3 |
| 128.127.2.1 | 128.408 | {1\|0 0 1/2} | $P_c{}^14/{}^1m{}^1n{}^1c^{\infty m}1$ | t3 |
| 128.124.2.1 | 128.409 | {1\|1/2 1/2 0} | $P_C{}^14/{}^1m{}^1n{}^1c^{\infty m}1$ | t3 |
| 128.139.2.1 | 128.410 | {1\|1/2 1/2 1/2} | $P_I{}^14/{}^1m{}^1n{}^1c^{\infty m}1$ | t3 |
| 129.129.2.1 | 129.420 | {1\|0 0 1/2} | $P_c{}^14/{}^1n{}^1m{}^1m^{\infty m}1$ | t3 |
| 129.123.2.1 | 129.421 | {1\|1/2 1/2 0} | $P_C{}^14/{}^1n{}^1m{}^1m^{\infty m}1$ | t3 |
| 129.139.2.1 | 129.422 | {1\|1/2 1/2 1/2} | $P_I{}^14/{}^1n{}^1m{}^1m^{\infty m}1$ | t3 |
| 130.129.2.1 | 130.432 | {1\|0 0 1/2} | $P_c{}^14/{}^1n{}^1c{}^1c^{\infty m}1$ | t3 |
| 130.124.2.1 | 130.433 | {1\|1/2 1/2 0} | $P_C{}^14/{}^1n{}^1c{}^1c^{\infty m}1$ | t3 |
| 130.140.2.1 | 130.434 | {1\|1/2 1/2 1/2} | $P_I{}^14/{}^1n{}^1c{}^1c^{\infty m}1$ | t3 |
| 131.123.2.1 | 131.444 | {1\|0 0 1/2} | $P_c{}^14_2/{}^1m{}^1m{}^1c^{\infty m}1$ | t3 |
| 131.132.2.1 | 131.445 | {1\|1/2 1/2 0} | $P_C{}^14_2/{}^1m{}^1m{}^1c^{\infty m}1$ | t3 |
| 131.139.2.1 | 131.446 | {1\|1/2 1/2 1/2} | $P_I{}^14_2/{}^1m{}^1m{}^1c^{\infty m}1$ | t3 |
| 132.123.2.1 | 132.456 | {1\|0 0 1/2} | $P_c{}^14_2/{}^1m{}^1c{}^1m^{\infty m}1$ | t3 |
| 132.131.2.1 | 132.457 | {1\|1/2 1/2 0} | $P_C{}^14_2/{}^1m{}^1c{}^1m^{\infty m}1$ | t3 |
| 132.140.2.1 | 132.458 | {1\|1/2 1/2 1/2} | $P_I{}^14_2/{}^1m{}^1c{}^1m^{\infty m}1$ | t3 |
| 133.125.2.1 | 133.468 | {1\|0 0 1/2} | $P_c{}^14_2/{}^1n{}^1b{}^1c^{\infty m}1$ | t3 |
| 133.132.2.1 | 133.469 | {1\|1/2 1/2 0} | $P_C{}^14_2/{}^1n{}^1b{}^1c^{\infty m}1$ | t3 |
| 133.140.2.1 | 133.470 | {1\|1/2 1/2 1/2} | $P_I{}^14_2/{}^1n{}^1b{}^1c^{\infty m}1$ | t3 |
| 134.125.2.1 | 134.480 | {1\|0 0 1/2} | $P_c{}^14_2/{}^1n{}^1n{}^1m^{\infty m}1$ | t3 |
| 134.131.2.1 | 134.481 | {1\|1/2 1/2 0} | $P_C{}^14_2/{}^1n{}^1n{}^1m^{\infty m}1$ | t3 |
| 134.139.2.1 | 134.482 | {1\|1/2 1/2 1/2} | $P_I{}^14_2/{}^1n{}^1n{}^1m^{\infty m}1$ | t3 |
| 135.127.2.1 | 135.492 | {1\|0 0 1/2} | $P_c{}^14_2/{}^1m{}^1b{}^1c^{\infty m}1$ | t3 |
| 135.132.2.1 | 135.493 | {1\|1/2 1/2 0} | $P_C{}^14_2/{}^1m{}^1b{}^1c^{\infty m}1$ | t3 |
| 135.140.2.1 | 135.494 | {1\|1/2 1/2 1/2} | $P_I{}^14_2/{}^1m{}^1b{}^1c^{\infty m}1$ | t3 |
| 136.127.2.1 | 136.504 | {1\|0 0 1/2} | $P_c{}^14_2/{}^1m{}^1n{}^1m^{\infty m}1$ | t3 |
| 136.131.2.1 | 136.505 | {1\|1/2 1/2 0} | $P_C{}^14_2/{}^1m{}^1n{}^1m^{\infty m}1$ | t3 |
| 136.139.2.1 | 136.506 | {1\|1/2 1/2 1/2} | $P_I{}^14_2/{}^1m{}^1n{}^1m^{\infty m}1$ | t3 |
| 137.129.2.1 | 137.516 | {1\|0 0 1/2} | $P_c{}^14_2/{}^1n{}^1m{}^1c^{\infty m}1$ | t3 |
| 137.132.2.1 | 137.517 | {1\|1/2 1/2 0} | $P_C{}^14_2/{}^1n{}^1m{}^1c^{\infty m}1$ | t3 |
| 137.139.2.1 | 137.518 | {1\|1/2 1/2 1/2} | $P_I{}^14_2/{}^1n{}^1m{}^1c^{\infty m}1$ | t3 |
| 138.129.2.1 | 138.528 | {1\|0 0 1/2} | $P_c{}^14_2/{}^1n{}^1c{}^1m^{\infty m}1$ | t3 |
| 138.131.2.1 | 138.529 | {1\|1/2 1/2 0} | $P_C{}^14_2/{}^1n{}^1c{}^1m^{\infty m}1$ | t3 |
| 138.140.2.1 | 138.530 | {1\|1/2 1/2 1/2} | $P_I{}^14_2/{}^1n{}^1c{}^1m^{\infty m}1$ | t3 |
| 139.123.2.1 | 139.540 | {1\|0 0 1/2} | $I_c{}^14/{}^1m{}^1m{}^1m^{\infty m}1$ | t4 |
| 140.123.2.1 | 140.550 | {1\|0 0 1/2} | $I_c{}^14/{}^1m{}^1c{}^1m^{\infty m}1$ | t4 |
| 141.134.2.1 | 141.560 | {1\|0 0 1/2} | $I_c{}^14_1/{}^1a{}^1m{}^1d^{\infty m}1$ | t4 |
| 142.134.2.1 | 142.570 | {1\|0 0 1/2} | $I_c{}^14_1/{}^1a{}^1c{}^1d^{\infty m}1$ | t4 |
| 143.143.2.1 | 143.3 | {1\|0 0 1/2} | $P_c{}^13^{\infty m}1$ | hP1 |

| 144.145.2.1 | 144.6 | $\{1\vert 0\ 0\ 1/2\}$ | $P_c{}^{1}3_1{}^{\infty m}1$ | hP1 |
|---|---|---|---|---|
| 145.144.2.1 | 145.9 | $\{1\vert 0\ 0\ 1/2\}$ | $P_c{}^{1}3_2{}^{\infty m}1$ | hP1 |
| 146.146.2.1 | 146.12 | $\{1\vert 0\ 0\ 1/2\}$ | $R_I{}^{1}3{}^{\infty m}1$ | hR1 |
| 147.147.2.1 | 147.16 | $\{1\vert 0\ 0\ 1/2\}$ | $P_c{}^{1}\bar{3}{}^{\infty m}1$ | hP1 |
| 148.148.2.1 | 148.20 | $\{1\vert 0\ 0\ 1/2\}$ | $R_I{}^{1}\bar{3}{}^{\infty m}1$ | hR1 |
| 149.149.2.1 | 149.24 | $\{1\vert 0\ 0\ 1/2\}$ | $P_c{}^{1}3{}^{1}1{}^{1}2{}^{\infty m}1$ | hP2 |
| 150.150.2.1 | 150.28 | $\{1\vert 0\ 0\ 1/2\}$ | $P_c{}^{1}3{}^{1}2{}^{1}1{}^{\infty m}1$ | hP3 |
| 151.153.2.1 | 151.32 | $\{1\vert 0\ 0\ 1/2\}$ | $P_c{}^{1}3_1{}^{1}1{}^{1}2{}^{\infty m}1$ | hP2 |
| 152.154.2.1 | 152.36 | $\{1\vert 0\ 0\ 1/2\}$ | $P_c{}^{1}3_1{}^{1}2{}^{1}1{}^{\infty m}1$ | hP3 |
| 153.151.2.1 | 153.40 | $\{1\vert 0\ 0\ 1/2\}$ | $P_c{}^{1}3_2{}^{1}1{}^{1}2{}^{\infty m}1$ | hP2 |
| 154.152.2.1 | 154.44 | $\{1\vert 0\ 0\ 1/2\}$ | $P_c{}^{1}3_2{}^{1}2{}^{1}1{}^{\infty m}1$ | hP3 |
| 155.155.2.1 | 155.48 | $\{1\vert 0\ 0\ 1/2\}$ | $R_I{}^{1}3{}^{1}2{}^{\infty m}1$ | hR2 |
| 156.156.2.1 | 156.52 | $\{1\vert 0\ 0\ 1/2\}$ | $P_c{}^{1}3{}^{1}m{}^{1}1{}^{\infty m}1$ | hP3 |
| 157.157.2.1 | 157.56 | $\{1\vert 0\ 0\ 1/2\}$ | $P_c{}^{1}3{}^{1}1{}^{1}m{}^{\infty m}1$ | hP2 |
| 158.156.2.1 | 158.60 | $\{1\vert 0\ 0\ 1/2\}$ | $P_c{}^{1}3{}^{1}c{}^{1}1{}^{\infty m}1$ | hP3 |
| 159.157.2.1 | 159.64 | $\{1\vert 0\ 0\ 1/2\}$ | $P_c{}^{1}3{}^{1}1{}^{1}c{}^{\infty m}1$ | hP2 |
| 160.160.2.1 | 160.68 | $\{1\vert 0\ 0\ 1/2\}$ | $R_I{}^{1}3{}^{1}m{}^{\infty m}1$ | hR2 |
| 161.160.2.1 | 161.72 | $\{1\vert 0\ 0\ 1/2\}$ | $R_I{}^{1}3{}^{1}c{}^{\infty m}1$ | hR2 |
| 162.162.2.1 | 162.78 | $\{1\vert 0\ 0\ 1/2\}$ | $P_c{}^{1}\bar{3}{}^{1}1{}^{1}m{}^{\infty m}1$ | hP2 |
| 163.162.2.1 | 163.84 | $\{1\vert 0\ 0\ 1/2\}$ | $P_c{}^{1}\bar{3}{}^{1}1{}^{1}c{}^{\infty m}1$ | hP2 |
| 164.164.2.1 | 164.90 | $\{1\vert 0\ 0\ 1/2\}$ | $P_c{}^{1}\bar{3}{}^{1}m{}^{1}1{}^{\infty m}1$ | hP3 |
| 165.164.2.1 | 165.96 | $\{1\vert 0\ 0\ 1/2\}$ | $P_c{}^{1}\bar{3}{}^{1}c{}^{1}1{}^{\infty m}1$ | hP3 |
| 166.166.2.1 | 166.102 | $\{1\vert 0\ 0\ 1/2\}$ | $R_I{}^{1}\bar{3}{}^{1}m{}^{\infty m}1$ | hR2 |
| 167.166.2.1 | 167.108 | $\{1\vert 0\ 0\ 1/2\}$ | $R_I{}^{1}\bar{3}{}^{1}c{}^{\infty m}1$ | hR2 |
| 168.168.2.1 | 168.112 | $\{1\vert 0\ 0\ 1/2\}$ | $P_c{}^{1}6{}^{\infty m}1$ | hP4 |
| 169.171.2.1 | 169.116 | $\{1\vert 0\ 0\ 1/2\}$ | $P_c{}^{1}6_1{}^{\infty m}1$ | hP4 |
| 170.172.2.1 | 170.120 | $\{1\vert 0\ 0\ 1/2\}$ | $P_c{}^{1}6_5{}^{\infty m}1$ | hP4 |
| 171.172.2.1 | 171.124 | $\{1\vert 0\ 0\ 1/2\}$ | $P_c{}^{1}6_2{}^{\infty m}1$ | hP4 |
| 172.171.2.1 | 172.128 | $\{1\vert 0\ 0\ 1/2\}$ | $P_c{}^{1}6_4{}^{\infty m}1$ | hP4 |
| 173.168.2.1 | 173.132 | $\{1\vert 0\ 0\ 1/2\}$ | $P_c{}^{1}6_3{}^{\infty m}1$ | hP4 |
| 174.174.2.1 | 174.136 | $\{1\vert 0\ 0\ 1/2\}$ | $P_c{}^{1}\bar{6}{}^{\infty m}1$ | hP4 |
| 175.175.2.1 | 175.142 | $\{1\vert 0\ 0\ 1/2\}$ | $P_c{}^{1}6/{}^{1}m{}^{\infty m}1$ | hP4 |
| 176.175.2.1 | 176.148 | $\{1\vert 0\ 0\ 1/2\}$ | $P_c{}^{1}6_3/{}^{1}m{}^{\infty m}1$ | hP4 |
| 177.177.2.1 | 177.154 | $\{1\vert 0\ 0\ 1/2\}$ | $P_c{}^{1}6{}^{1}2{}^{1}2{}^{\infty m}1$ | hP5 |
| 178.180.2.1 | 178.160 | $\{1\vert 0\ 0\ 1/2\}$ | $P_c{}^{1}6_1{}^{1}2{}^{1}2{}^{\infty m}1$ | hP5 |
| 179.181.2.1 | 179.166 | $\{1\vert 0\ 0\ 1/2\}$ | $P_c{}^{1}6_5{}^{1}2{}^{1}2{}^{\infty m}1$ | hP5 |
| 180.181.2.1 | 180.172 | $\{1\vert 0\ 0\ 1/2\}$ | $P_c{}^{1}6_2{}^{1}2{}^{1}2{}^{\infty m}1$ | hP5 |
| 181.180.2.1 | 181.178 | $\{1\vert 0\ 0\ 1/2\}$ | $P_c{}^{1}6_4{}^{1}2{}^{1}2{}^{\infty m}1$ | hP5 |
| 182.177.2.1 | 182.184 | $\{1\vert 0\ 0\ 1/2\}$ | $P_c{}^{1}6_3{}^{1}2{}^{1}2{}^{\infty m}1$ | hP5 |
| 183.183.2.1 | 183.190 | $\{1\vert 0\ 0\ 1/2\}$ | $P_c{}^{1}6{}^{1}m{}^{1}m{}^{\infty m}1$ | hP5 |
| 184.183.2.1 | 184.196 | $\{1\vert 0\ 0\ 1/2\}$ | $P_c{}^{1}6{}^{1}c{}^{1}c{}^{\infty m}1$ | hP5 |
| 185.183.2.1 | 185.202 | $\{1\vert 0\ 0\ 1/2\}$ | $P_c{}^{1}6_3{}^{1}c{}^{1}m{}^{\infty m}1$ | hP5 |
| 186.183.2.1 | 186.208 | $\{1\vert 0\ 0\ 1/2\}$ | $P_c{}^{1}6_3{}^{1}m{}^{1}c{}^{\infty m}1$ | hP5 |

| 187.187.2.1 | 187.214 | {1\|0 0 1/2} | $P_c{}^1\bar{6}{}^1m{}^12^{\infty m}1$ | hP5 |
|---|---|---|---|---|
| 188.187.2.1 | 188.220 | {1\|0 0 1/2} | $P_c{}^1\bar{6}{}^1c{}^12^{\infty m}1$ | hP5 |
| 189.189.2.1 | 189.226 | {1\|0 0 1/2} | $P_c{}^1\bar{6}{}^12{}^1m^{\infty m}1$ | hP5 |
| 190.189.2.1 | 190.232 | {1\|0 0 1/2} | $P_c{}^1\bar{6}{}^12{}^1c^{\infty m}1$ | hP5 |
| 191.191.2.1 | 191.242 | {1\|0 0 1/2} | $P_c{}^16/{}^1m{}^1m{}^1m^{\infty m}1$ | hP5 |
| 192.191.2.1 | 192.252 | {1\|0 0 1/2} | $P_c{}^16/{}^1m{}^1c{}^1c^{\infty m}1$ | hP5 |
| 193.191.2.1 | 193.262 | {1\|0 0 1/2} | $P_c{}^16_3/{}^1m{}^1c{}^1m^{\infty m}1$ | hP5 |
| 194.191.2.1 | 194.272 | {1\|0 0 1/2} | $P_c{}^16_3/{}^1m{}^1m{}^1c^{\infty m}1$ | hP5 |
| 195.197.2.1 | 195.3 | {1\|1/2 1/2 1/2} | $P_I{}^12{}^13^{\infty m}1$ | c1 |
| 196.195.2.1 | 196.6 | {1\|1/2 1/2 1/2} | $F_s{}^12{}^13^{\infty m}1$ | c2 |
| 198.199.2.1 | 198.11 | {1\|1/2 1/2 1/2} | $P_I{}^12_1{}^13^{\infty m}1$ | c1 |
| 200.204.2.1 | 200.17 | {1\|1/2 1/2 1/2} | $P_I{}^1m{}^1\bar{3}^{\infty m}1$ | c1 |
| 201.204.2.1 | 201.21 | {1\|1/2 1/2 1/2} | $P_I{}^1n{}^1\bar{3}^{\infty m}1$ | c1 |
| 202.200.2.1 | 202.25 | {1\|1/2 1/2 1/2} | $F_s{}^1m{}^1\bar{3}^{\infty m}1$ | c2 |
| 203.201.2.1 | 203.29 | {1\|1/2 1/2 1/2} | $F_s{}^1d{}^1\bar{3}^{\infty m}1$ | c2 |
| 205.206.2.1 | 205.36 | {1\|1/2 1/2 1/2} | $P_I{}^1a{}^1\bar{3}^{\infty m}1$ | c1 |
| 207.211.2.1 | 207.43 | {1\|1/2 1/2 1/2} | $P_I{}^14{}^13{}^12^{\infty m}1$ | c4 |
| 208.211.2.1 | 208.47 | {1\|1/2 1/2 1/2} | $P_I{}^14_2{}^13{}^12^{\infty m}1$ | c4 |
| 209.207.2.1 | 209.51 | {1\|1/2 1/2 1/2} | $F_s{}^14{}^13{}^12^{\infty m}1$ | c5 |
| 210.208.2.1 | 210.55 | {1\|1/2 1/2 1/2} | $F_s{}^14_1{}^13{}^12^{\infty m}1$ | c5 |
| 212.214.2.1 | 212.62 | {1\|1/2 1/2 1/2} | $P_I{}^14_3{}^13{}^12^{\infty m}1$ | c4 |
| 213.214.2.1 | 213.66 | {1\|1/2 1/2 1/2} | $P_I{}^14_1{}^13{}^12^{\infty m}1$ | c4 |
| 215.217.2.1 | 215.73 | {1\|1/2 1/2 1/2} | $P_I{}^1\bar{4}{}^13{}^1m^{\infty m}1$ | c4 |
| 216.215.2.1 | 216.77 | {1\|1/2 1/2 1/2} | $F_s{}^1\bar{4}{}^13{}^1m^{\infty m}1$ | c5 |
| 218.217.2.1 | 218.84 | {1\|1/2 1/2 1/2} | $P_I{}^1\bar{4}{}^13{}^1n^{\infty m}1$ | c4 |
| 219.215.2.1 | 219.88 | {1\|1/2 1/2 1/2} | $F_s{}^1\bar{4}{}^13{}^1c^{\infty m}1$ | c5 |
| 221.229.2.1 | 221.97 | {1\|1/2 1/2 1/2} | $P_I{}^1m{}^1\bar{3}{}^1m^{\infty m}1$ | c4 |
| 222.229.2.1 | 222.103 | {1\|1/2 1/2 1/2} | $P_I{}^1n{}^1\bar{3}{}^1n^{\infty m}1$ | c4 |
| 223.229.2.1 | 223.109 | {1\|1/2 1/2 1/2} | $P_I{}^1m{}^1\bar{3}{}^1n^{\infty m}1$ | c4 |
| 224.229.2.1 | 224.115 | {1\|1/2 1/2 1/2} | $P_I{}^1n{}^1\bar{3}{}^1m^{\infty m}1$ | c4 |
| 225.221.2.1 | 225.121 | {1\|1/2 1/2 1/2} | $F_s{}^1m{}^1\bar{3}{}^1m^{\infty m}1$ | c5 |
| 226.221.2.1 | 226.127 | {1\|1/2 1/2 1/2} | $F_s{}^1m{}^1\bar{3}{}^1c^{\infty m}1$ | c5 |
| 227.224.2.1 | 227.133 | {1\|1/2 1/2 1/2} | $F_s{}^1d{}^1\bar{3}{}^1m^{\infty m}1$ | c5 |
| 228.224.2.1 | 228.139 | {1\|1/2 1/2 1/2} | $F_s{}^1d{}^1\bar{3}{}^1c^{\infty m}1$ | c5 |

## S3. Band representations hosting unconventional magnons

Here we introduce the band representations hosting unconventional magnons in four types of collinear SSGs. For the sake of convenience, we have arranged the content of high-dimensional band representations and spin non-degenerate band representations separately. They are listed in sections S3.2~S3.5 and S3.6, respectively.

### S3.1. Group statistics

For collinear FMs, there are 36 and 5 SSGs with triple and sextuple nodal point, respectively. For collinear AFM, we list all the *k* points and the band representations with high-dimensional band degeneracies, including quadruple nodal planes, sextuple nodal points, octuple nodal points and lines, duodecuple nodal points. Besides, the momentum positions of the 422 altermagnetic SSGs with AFM-induced spin splitting (in electronic band structure) and chirality splitting (in magnon band structure) are also tabulated. In Supplementary Table XVIII, statistics of the number of SSGs (band representations) with different nodal characters for collinear SSGs are given. Detailed information on band representations that accommodate unconventional magnons can be checked in Supplementary Table XIX~XXII.

Supplementary Table XVIII. Statistics of SSGs (band representations) hosting unconventional magnons in collinear SSGs.

| Nodal feature | Type-I | Type-II | Type-III | Type-IV |
|---|---|---|---|---|
| Duodecuple point (DCP) | × | 3 (3) | × | 3 (3) |
| Octuple nodal line (ONL) | × | × | × | 25 (31) |
| Octuple point (OP) | × | 37 (48) | 20 (25) | 112 (164) |
| Sextuple point (SP) | 5 (5) | 27 (73) | 24 (57) | 26 (71) |
| Quadruple nodal plane (QNPL) | × | 56 (66) | × | 205 (276) |
| Triple point (TP) | 36 (129) | × | × | × |

Except for triple, sextuple, and duodecuple nodal points that only appear in the cubic lattice systems, we also complete the statistics of other high-dimensional nodal features in different lattice systems and in different types of collinear SSGs in Supplementary Fig. 5.

For altermagnets and chiral magnon, we also provide diagrams illustrating the momentum distribution of chiral magnons/electronic spin splitting in the collinear SSGs with respect to the different lattice systems in Supplementary Fig. 6. Note that the numbers refer to the highest-symmetric momentum positions with magnon chirality splitting/electronic spin splitting. For example, if a collinear SSG $G_1$ has magnonic chirality splitting/electronic spin splitting at high-symmetry points, they are not included in the statistics of high-symmetry lines, planes and general points.

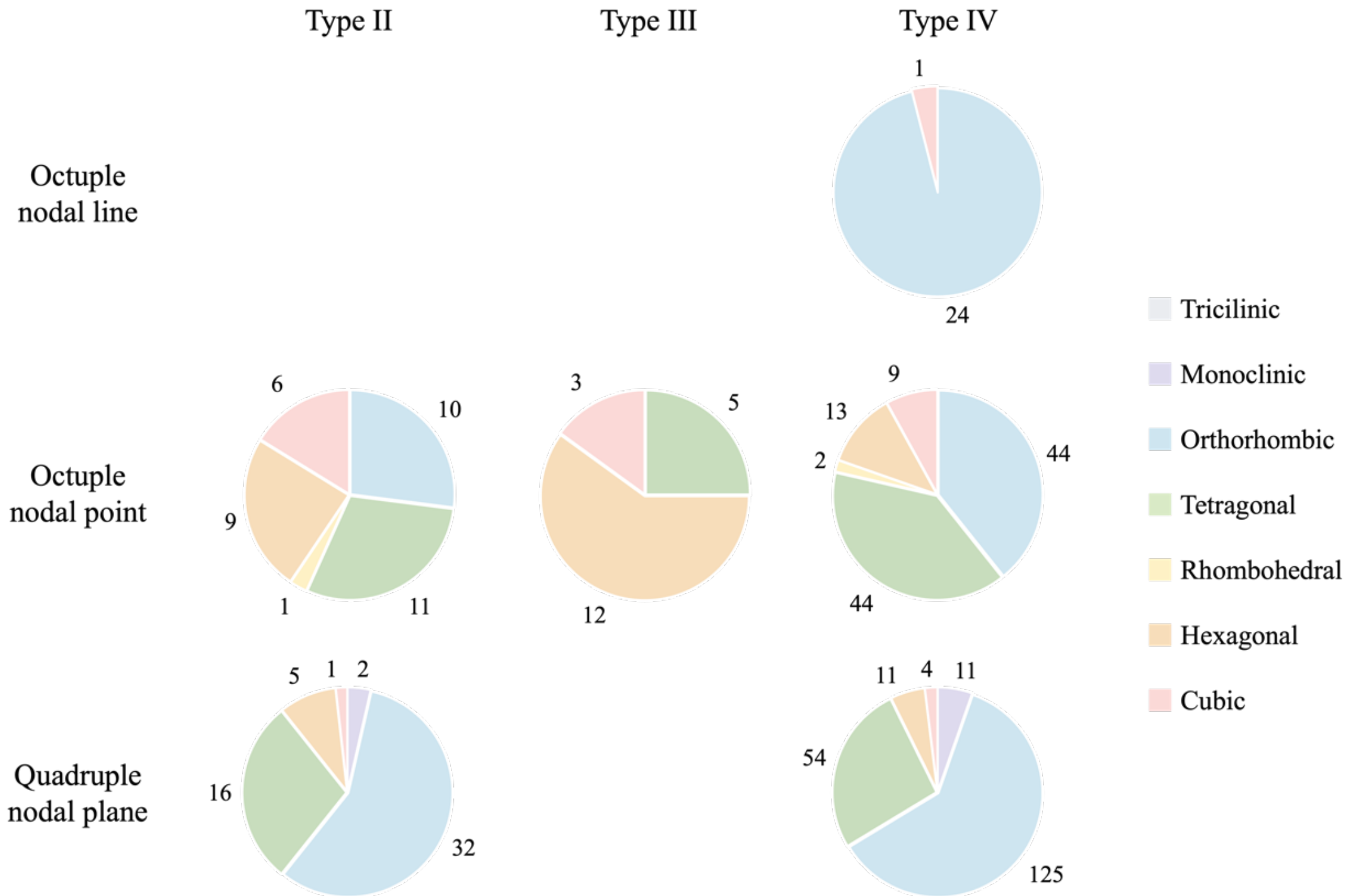

Supplementary Fig. 5. Statistics on the lattice systems of different high-dimensional nodal features in collinear SSGs.

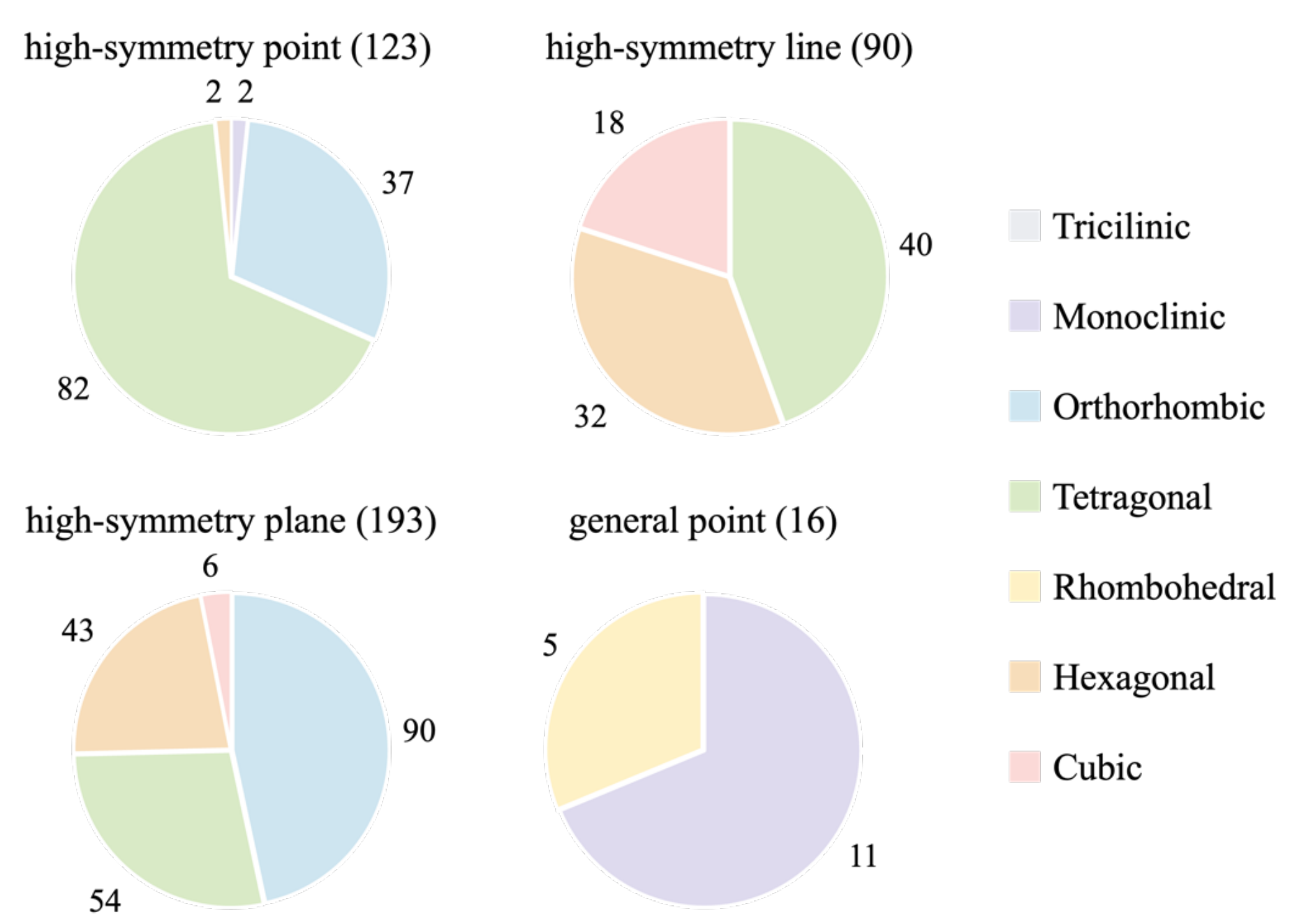

Supplementary Fig. 6. Statistics on the lattice systems of different momentum positions that host chiral magnons in collinear SSGs.

## S3.2. Type-I collinear SSGs

Supplementary Table XIX. Tabulation of type-I collinear SSGs that accommodate unconventional magnons with high-dimensional band representations. The five columns illustrate the group number, the international notation of collinear SSGs, the momentum-position, band representations, and the quasiparticle type for unconventional magnons, respectively. "$C$–$n$" means that the topological charge $C$ is $|C| = n$.

| SSG | $G_S$ | ***k*** point | Rep | Quasiparticle |
| --- | --- | --- | --- | --- |
| 195.195.1.1 | $P^{1}2^{1}3^{\infty m}1$ | Γ:(0,0,0) | $\Gamma_4^S(3)$ | $C$–2 TP |
| | | R:(1/2,1/2,1/2) | $R_4^S(3)$ | $C$–2 TP |
| 196.196.1.1 | $F^{1}2^{1}3^{\infty m}1$ | Γ:(0,0,0) | $\Gamma_4^S(3)$ | $C$–2 TP |
| 197.197.1.1 | $I^{1}2^{1}3^{\infty m}1$ | Γ:(0,0,0) | $\Gamma_4^S(3)$ | $C$–2 TP |
| | | H:(1,1,1) | $H_4^S(3)$ | $C$–2 TP |
| | | P:(1/2,1/2,1/2) | $P_4^S(3)$ | $C$–2 TP |
| 198.198.1.1 | $P^{1}2_1{}^{1}3^{\infty m}1$ | Γ:(0,0,0) | $\Gamma_4^S(3)$ | $C$–2 TP |
| 199.199.1.1 | $I^{1}2_1{}^{1}3^{\infty m}1$ | Γ:(0,0,0) | $\Gamma_4^S(3)$ | $C$–2 TP |
| | | H:(1,1,1) | $H_4^S(3)$ | $C$–2 TP |
| 200.200.1.1 | $P^{1}m^{1}\bar{3}^{\infty m}1$ | Γ:(0,0,0) | $\Gamma_4^{S,+}(3), \Gamma_4^{S,-}(3)$ | TP |
| | | R:(1/2,1/2,1/2) | $R_4^{S,+}(3), R_4^{S,-}(3)$ | TP |
| 201.201.1.1 | $P^{1}n^{1}\bar{3}^{\infty m}1$ | Γ:(0,0,0) | $\Gamma_4^{S,+}(3), \Gamma_4^{S,-}(3)$ | TP |
| | | R:(1/2,1/2,1/2) | $R_4^{S,+}(3), R_4^{S,-}(3)$ | TP |
| 202.202.1.1 | $F^{1}m^{1}\bar{3}^{\infty m}1$ | Γ:(0,0,0) | $\Gamma_4^{S,+}(3), \Gamma_4^{S,-}(3)$ | TP |
| 203.203.1.1 | $F^{1}d^{1}\bar{3}^{\infty m}1$ | Γ:(0,0,0) | $\Gamma_4^{S,+}(3), \Gamma_4^{S,-}(3)$ | TP |
| 204.204.1.1 | $I^{1}m^{1}\bar{3}^{\infty m}1$ | Γ:(0,0,0) | $\Gamma_4^{S,+}(3), \Gamma_4^{S,-}(3)$ | TP |
| | | H:(1,1,1) | $H_4^{S,+}(3), H_4^{S,-}(3)$ | TP |
| | | P:(1/2,1/2,1/2) | $P_4^S(3)$ | TP |
| 205.205.1.1 | $P^{1}a^{1}\bar{3}^{\infty m}1$ | Γ:(0,0,0) | $\Gamma_4^{S,+}(3), \Gamma_4^{S,-}(3)$ | TP |
| 206.206.1.1 | $I^1a^1\bar{3}^{\infty m}1$ | Γ:(0,0,0) | $\Gamma_4^{S,+}(3), \Gamma_4^{S,-}(3)$ | TP |
| | | H:(1,1,1) | $H_4^{S,+}(3), H_4^{S,-}(3)$ | TP |
| 207.207.1.1 | $P^14^13^12^{\infty m}1$ | Γ:(0,0,0) | $\Gamma_4^S(3), \Gamma_5^S(3)$ | $C$–2 TP |
| | | R:(1/2,1/2,1/2) | $R_4^S(3), R_5^S(3)$ | $C$–2 TP |
| 208.208.1.1 | $P^14_2{}^13^12^{\infty m}1$ | Γ:(0,0,0) | $\Gamma_4^S(3), \Gamma_5^S(3)$ | $C$–2 TP |
| | | R:(1/2,1/2,1/2) | $R_4^S(3), R_5^S(3)$ | $C$–2 TP |
| 209.209.1.1 | $F^14^13^12^{\infty m}1$ | Γ:(0,0,0) | $\Gamma_4^S(3), \Gamma_5^S(3)$ | $C$–2 TP |
| 210.210.1.1 | $F^14_1{}^13^12^{\infty m}1$ | Γ:(0,0,0) | $\Gamma_4^S(3), \Gamma_5^S(3)$ | $C$–2 TP |
| 211.211.1.1 | $I^14^13^12^{\infty m}1$ | Γ:(0,0,0) | $\Gamma_4^S(3), \Gamma_5^S(3)$ | $C$–2 TP |
| | | H:(1,1,1) | $H_4^S(3), H_5^S(3)$ | $C$–2 TP |
| | | P:(1/2,1/2,1/2) | $P_4^S(3)$ | $C$–2 TP |
| 212.212.1.1 | $P^14_3{}^13^12^{\infty m}1$ | Γ:(0,0,0) | $\Gamma_4^S(3), \Gamma_5^S(3)$ | $C$–2 TP |
| 213.213.1.1 | $P^14_1{}^13^12^{\infty m}1$ | Γ:(0,0,0) | $\Gamma_4^S(3), \Gamma_5^S(3)$ | $C$–2 TP |
| 214.214.1.1 | $I^14_1{}^13^12^{\infty m}1$ | Γ:(0,0,0) | $\Gamma_4^S(3), \Gamma_5^S(3)$ | $C$–2 TP |
| | | H:(1,1,1) | $H_4^S(3), H_5^S(3)$ | $C$–2 TP |
| 215.215.1.1 | $P^1\bar{4}^13^1m^{\infty m}1$ | Γ:(0,0,0) | $\Gamma_4^S(3), \Gamma_5^S(3)$ | TP |
| | | R:(1/2,1/2,1/2) | $R_4^S(3), R_5^S(3)$ | TP |
| 216.216.1.1 | $F^1\bar{4}^13^1m^{\infty m}1$ | Γ:(0,0,0) | $\Gamma_4^S(3), \Gamma_5^S(3)$ | TP |
| 217.217.1.1 | $I^1\bar{4}^13^1m^{\infty m}1$ | Γ:(0,0,0) | $\Gamma_4^S(3), \Gamma_5^S(3)$ | TP |
| | | H:(1,1,1) | $H_4^S(3), H_5^S(3)$ | TP |
| | | P:(1/2,1/2,1/2) | $P_4^S(3), P_5^S(3)$ | TP |
| 218.218.1.1 | $P^1\bar{4}^13^1n^{\infty m}1$ | Γ:(0,0,0) | $\Gamma_4^S(3), \Gamma_5^S(3)$ | TP |
| | | R:(1/2,1/2,1/2) | $R_4^SR_5^S(6)$ | SP |
| 219.219.1.1 | $F^1\bar{4}^13^1c^{\infty m}1$ | Γ:(0,0,0) | $\Gamma_4^S(3), \Gamma_5^S(3)$ | TP |
| 220.220.1.1 | $I^1\bar{4}^13^1d^{\infty m}1$ | Γ:(0,0,0) | $\Gamma_4^S(3), \Gamma_5^S(3)$ | TP |
| | | H:(1,1,1) | $H_4^SH_5^S(6)$ | SP |
| 221.221.1.1 | $P^1m^1\bar{3}^1m^{\infty m}1$ | Γ:(0,0,0) | $\Gamma_4^{S,+}(3), \Gamma_4^{S,-}(3), \Gamma_5^{S,+}(3), \Gamma_5^{S,-}(3)$ | TP |
| | | R:(1/2,1/2,1/2) | $R_4^{S,+}(3), R_4^{S,-}(3), R_5^{S,+}(3), R_5^{S,-}(3)$ | TP |
| 222.222.1.1 | $P^1n^1\bar{3}^1n^{\infty m}1$ | Γ:(0,0,0) | $\Gamma_4^{S,+}(3), \Gamma_4^{S,-}(3), \Gamma_5^{S,+}(3), \Gamma_5^{S,-}(3)$ | TP |
| | | R:(1/2,1/2,1/2) | $R_4^S(6)$ | SP |
| 223.223.1.1 | $P^1m^1\bar{3}^1n^{\infty m}1$ | Γ:(0,0,0) | $\Gamma_4^{S,+}(3), \Gamma_4^{S,-}(3), \Gamma_5^{S,+}(3), \Gamma_5^{S,-}(3)$ | TP |
| | | R:(1/2,1/2,1/2) | $R_4^S(6)$ | SP |
| 224.224.1.1 | $P^1n^1\bar{3}^1m^{\infty m}1$ | Γ:(0,0,0) | $\Gamma_4^{S,+}(3), \Gamma_4^{S,-}(3), \Gamma_5^{S,+}(3), \Gamma_5^{S,-}(3)$ | TP |
| | | R:(1/2,1/2,1/2) | $R_4^{S,+}(3), R_4^{S,-}(3), R_5^{S,+}(3), R_5^{S,-}(3)$ | TP |

| 225.225.1.1 | $F^{1}m^{1}\bar{3}^{1}m^{\infty m}1$ | Γ:(0,0,0) | $\Gamma_4^{S,+}(3), \Gamma_4^{S,-}(3), \Gamma_5^{S,+}(3), \Gamma_5^{S,-}(3)$ | TP |
|---|---|---|---|---|
| 226.226.1.1 | $F^{1}m^{1}\bar{3}^{1}c^{\infty m}1$ | Γ:(0,0,0) | $\Gamma_4^{S,+}(3), \Gamma_4^{S,-}(3), \Gamma_5^{S,+}(3), \Gamma_5^{S,-}(3)$ | TP |
| 227.227.1.1 | $F^{1}d^{1}\bar{3}^{1}m^{\infty m}1$ | Γ:(0,0,0) | $\Gamma_4^{S,+}(3), \Gamma_4^{S,-}(3), \Gamma_5^{S,+}(3), \Gamma_5^{S,-}(3)$ | TP |
| 228.228.1.1 | $F^{1}d^{1}\bar{3}^{1}c^{\infty m}1$ | Γ:(0,0,0) | $\Gamma_4^{S,+}(3), \Gamma_4^{S,-}(3), \Gamma_5^{S,+}(3), \Gamma_5^{S,-}(3)$ | TP |
| 229.229.1.1 | $I^{1}m^{1}\bar{3}^{1}m^{\infty m}1$ | Γ:(0,0,0) | $\Gamma_4^{S,+}(3), \Gamma_4^{S,-}(3), \Gamma_5^{S,+}(3), \Gamma_5^{S,-}(3)$ | TP |
| | | H:(1,1,1) | $H_4^{S,+}(3), H_4^{S,-}(3), H_5^{S,+}(3), H_5^{S,-}(3)$ | TP |
| | | P:(1/2,1/2,1/2) | $P_4^S(3), P_5^S(3)$ | TP |
| 230.230.1.1 | $I^{1}a^{1}\bar{3}^{1}d^{\infty m}1$ | Γ:(0,0,0) | $\Gamma_4^{S,+}(3), \Gamma_4^{S,-}(3), \Gamma_5^{S,+}(3), \Gamma_5^{S,-}(3)$ | TP |
| | | H:(1,1,1) | $H_4^S(6)$ | SP |

## S3.3. Type-II collinear SSGs

Supplementary Table XX. Tabulation of type-II collinear SSGs that accommodate unconventional magnons with high-dimensional band representations. The five columns illustrate the group number, the international notation of collinear SSGs, the momentum-position, band representations, and the quasiparticle type for unconventional magnons, respectively. "$C_2$–*m*" means that the monopole charge $C_2 = (C_\uparrow - C_\downarrow)/2$ is $|C_2| = m$.

| SSG | $G_S$ | ***k*** point | Rep | Quasiparticle |
|---|---|---|---|---|
| 4.11.1.1 | $P^{1}2_1/^{\bar{1}}m^{\infty m}1$ | G:(*u*, 1/2, *w*) | $G_1^S G_1^S(4)$ | QNPL |
| 4.14.1.1 | $P^{1}2_1/^{\bar{1}}c^{\infty m}1$ | G:(*u*, 1/2, *w*) | $G_1^S G_1^S(4)$ | QNPL |
| 17.51.1.1 | $P^{\bar{1}}m^{\bar{1}}m^{\bar{1}}a^{\infty m}1$ | L:(1/2,*v*,*w*) | $(L)W_1^S W_1^S(4)$ | QNPL |
| 17.52.1.1 | $P^{\bar{1}}n^{\bar{1}}n^{\bar{1}}a^{\infty m}1$ | N:(*u*,1/2,*w*) | $(N)W_1^S W_1^S(4)$ | QNPL |
| 17.53.1.1 | $P^{\bar{1}}m^{\bar{1}}n^{\bar{1}}a^{\infty m}1$ | W:(*u*,*v*,1/2) | $W_1^S W_1^S(4)$ | QNPL |
| 17.54.1.1 | $P^{\bar{1}}c^{\bar{1}}c^{\bar{1}}a^{\infty m}1$ | L:(1/2,*v*,*w*) | $(L)W_1^S W_1^S(4)$ | QNPL |
| 18.55.1.1 | $P^{\bar{1}}b^{\bar{1}}a^{\bar{1}}m^{\infty m}1$ | L:(1/2,*v*,*w*) | $L_1^S L_1^S(4)$ | QNPL |
| | | N:(*u*,1/2,*w*) | $N_1^S N_1^S(4)$ | QNPL |
| 18.56.1.1 | $P^{\bar{1}}c^{\bar{1}}c^{\bar{1}}n^{\infty m}1$ | L:(1/2,*v*,*w*) | $L_1^S L_1^S(4)$ | QNPL |
| | | N:(*u*,1/2,*w*) | $N_1^S N_1^S(4)$ | QNPL |
| 18.57.1.1 | $P^{\bar{1}}b^{\bar{1}}c^{\bar{1}}m^{\infty m}1$ | W:(*u*,*v*,1/2) | $(W)L_1^S L_1^S(4)$ | QNPL |
| | | N:(*u*,1/2,*w*) | $N_1^S N_1^S(4)$ | QNPL |
| 18.58.1.1 | $P^{\bar{1}}n^{\bar{1}}n^{\bar{1}}m^{\infty m}1$ | L:(1/2,*v*,*w*) | $L_1^S L_1^S(4)$ | QNPL |
| | | N:(*u*,1/2,*w*) | $N_1^S N_1^S(4)$ | QNPL |
| 18.59.1.1 | $P^{\bar{1}}m^{\bar{1}}m^{\bar{1}}n^{\infty m}1$ | L:(1/2,*v*,*w*) | $L_1^S L_1^S(4)$ | QNPL |
| | | N:($u$,1/2,$w$) | $N_1^S N_1^S(4)$ | QNPL |
| 18.60.1.1 | $P^{\bar{1}}b^{\bar{1}}c^{\bar{1}}n^{\infty m}1$ | L:(1/2,$v$,$w$) | $L_1^S L_1^S(4)$ | QNPL |
| | | W:($u$,$v$,1/2) | $(W)N_1^S N_1^S(4)$ | QNPL |
| 19.61.1.1 | $P^{\bar{1}}b^{\bar{1}}c^{\bar{1}}a^{\infty m}1$ | R:(1/2,1/2,1/2) | $R_1^S R_1^S(8)$ | $C_2$−2 OP |
| | | L:(1/2,$v$,$w$) | $L_1^S L_1^S(4)$ | QNPL |
| | | N:($u$,1/2,$w$) | $N_1^S N_1^S(4)$ | QNPL |
| | | W:($u$,$v$,1/2) | $W_1^S W_1^S(4)$ | QNPL |
| 19.62.1.1 | $P^{\bar{1}}n^{\bar{1}}m^{\bar{1}}a^{\infty m}1$ | R:(1/2,1/2,1/2) | $R_1^S R_1^S(8)$ | $C_2$−2 OP |
| | | L:(1/2,$v$,$w$) | $L_1^S L_1^S(4)$ | QNPL |
| | | N:($u$,1/2,$w$) | $N_1^S N_1^S(4)$ | QNPL |
| | | W:($u$,$v$,1/2) | $W_1^S W_1^S(4)$ | QNPL |
| 20.63.1.1 | $C^{\bar{1}}m^{\bar{1}}c^{\bar{1}}m^{\infty m}1$ | Q:($u$,$v$,1/2) | $Q_1^S Q_1^S(4)$ | QNPL |
| 20.64.1.1 | $C^{\bar{1}}m^{\bar{1}}c^{\bar{1}}e^{\infty m}1$ | Q:($u$,$v$,1/2) | $Q_1^S Q_1^S(4)$ | QNPL |
| 26.51.1.1 | $P^{\bar{1}}m^{1}m^{1}a^{\infty m}1$ | L:(1/2,$v$,$w$) | $(L)W_1^S W_1^S(4)$ | QNPL |
| 26.55.1.1 | $P^{\bar{1}}b^{1}a^{1}m^{\infty m}1$ | L:(1/2,$v$,$w$) | $(L)W_1^S W_1^S(4)$ | QNPL |
| 26.57.1.1 | $P^{1}b^{\bar{1}}c^{1}m^{\infty m}1$ | N:($u$,1/2,$w$) | $(N)W_1^S W_1^S(4)$ | QNPL |
| 26.62.1.1 | $P^{\bar{1}}n^{1}m^{1}a^{\infty m}1$ | L:(1/2,$v$,$w$) | $(L)W_1^S W_1^S(4)$ | QNPL |
| 29.54.1.1 | $P^{\bar{1}}c^{1}c^{1}a^{\infty m}1$ | R:(1/2,1/2,1/2) | $R_1^S R_1^S(8)$ | OP |
| | | U:(1/2,0,1/2) | $U_1^S U_1^S(8)$ | OP |
| | | L:(1/2,$v$,$w$) | $(L)W_1^S W_1^S(4)$ | QNPL |
| 29.57.1.1 | $P^{1}b^{1}c^{\bar{1}}m^{\infty m}1$ | R:(1/2,1/2,1/2) | $R_1^S R_1^S(8)$ | OP |
| | | T:(0,1/2,1/2) | $(T)U_1^S U_1^S(8)$ | OP |
| | | W:($u$,$v$,1/2) | $W_1^S W_1^S(4)$ | QNPL |
| 29.60.1.1 | $P^{1}b^{1}c^{\bar{1}}n^{\infty m}1$ | R:(1/2,1/2,1/2) | $R_1^S R_1^S(8)$ | OP |
| | | T:(0,1/2,1/2) | $(T)U_1^S U_1^S(8)$ | OP |
| | | W:($u$,$v$,1/2) | $W_1^S W_1^S(4)$ | QNPL |
| 29.61.1.1 | $P^{\bar{1}}b^{1}c^{1}a^{\infty m}1$ | R:(1/2,1/2,1/2) | $R_1^S R_1^S(8)$ | OP |
| | | U:(1/2,0,1/2) | $U_1^S U_1^S(8)$ | OP |
| | | L:(1/2,$v$,$w$) | $(L)W_1^S W_1^S(4)$ | QNPL |
| 31.53.1.1 | $P^{1}m^{1}n^{\bar{1}}a^{\infty m}1$ | W:($u$,$v$,1/2) | $W_1^S W_1^S(4)$ | QNPL |
| 31.58.1.1 | $P^{\bar{1}}n^{1}n^{1}m^{\infty m}1$ | L:(1/2,$v$,$w$) | $(L)W_1^S W_1^S(4)$ | QNPL |
| 31.59.1.1 | $P^{\bar{1}}m^{1}m^{1}n^{\infty m}1$ | L:(1/2,$v$,$w$) | $(L)W_1^S W_1^S(4)$ | QNPL |
| 31.62.1.1 | $P^{1}n^{1}m^{\bar{1}}a^{\infty m}1$ | W:($u$,$v$,1/2) | $W_1^S W_1^S(4)$ | QNPL |
| 33.52.1.1 | $P^{1}n^{\bar{1}}n^{1}a^{\infty m}1$ | S:(1/2,1/2,0) | $(S)U_1^S U_1^S(8)$ | OP |
| | | N:($u$,1/2,$w$) | $(N)W_1^S W_1^S(4)$ | QNPL |
| 33.56.1.1 | $P^{\bar{1}}c^{1}c^{1}n^{\infty m}1$ | U:(1/2,0,1/2) | $U_1^S U_1^S(8)$ | OP |
| | | L:(1/2,$v$,$w$) | $(L)W_1^S W_1^S(4)$ | QNPL |
| 33.60.1.1 | $P^{\bar{1}}b^{1}c^{1}n^{\infty m}1$ | U:(1/2,0,1/2) | $U_1^S U_1^S(8)$ | OP |
| | | L:(1/2,$v$,$w$) | $(L)W_1^S W_1^S(4)$ | QNPL |
| 33.62.1.1 | $P^{1}n^{\bar{1}}m^{1}a^{\infty m}1$ | S:(1/2,1/2,0) | $(S)U_1^S U_1^S(8)$ | OP |
| | | N:($u$,1/2,$w$) | $(N)W_1^S W_1^S(4)$ | QNPL |
| 36.63.1.1 | $C^{1}m^{1}c^{\bar{1}}m^{\infty m}1$ | Q:($u$,$v$,1/2) | $Q_1^S Q_1^S(4)$ | QNPL |
| 36.64.1.1 | $C^{1}m^{1}c^{\bar{1}}e^{\infty m}1$ | Q:($u$,$v$,1/2) | $Q_1^S Q_1^S(4)$ | QNPL |

| 90.127.1.1 | $P^{1}4/^{\bar{1}}m^{\bar{1}}b^{\bar{1}}m^{\infty m}1$ | F:($u$,1/2,$w$) | $F_1^S F_1^S(4)$ | QNPL |
|---|---|---|---|---|
| 90.128.1.1 | $P^{1}4/^{\bar{1}}m^{\bar{1}}n^{\bar{1}}c^{\infty m}1$ | F:($u$,1/2,$w$) | $F_1^S F_1^S(4)$ | QNPL |
| 90.129.1.1 | $P^{1}4/^{\bar{1}}n^{\bar{1}}m^{\bar{1}}m^{\infty m}1$ | F:($u$,1/2,$w$) | $F_1^S F_1^S(4)$ | QNPL |
| 90.130.1.1 | $P^{1}4/^{\bar{1}}n^{\bar{1}}c^{\bar{1}}c^{\infty m}1$ | F:($u$,1/2,$w$) | $F_1^S F_1^S(4)$ | QNPL |
| 94.135.1.1 | $P^{1}4_2/^{\bar{1}}m^{\bar{1}}b^{\bar{1}}c^{\infty m}1$ | F:($u$,1/2,$w$) | $F_1^S F_1^S(4)$ | QNPL |
| 94.136.1.1 | $P^{1}4_2/^{\bar{1}}m^{\bar{1}}n^{\bar{1}}m^{\infty m}1$ | F:($u$,1/2,$w$) | $F_1^S F_1^S(4)$ | QNPL |
| 94.137.1.1 | $P^{1}4_2/^{\bar{1}}n^{\bar{1}}m^{\bar{1}}c^{\infty m}1$ | F:($u$,1/2,$w$) | $F_1^S F_1^S(4)$ | QNPL |
| 94.138.1.1 | $P^{1}4_2/^{\bar{1}}n^{\bar{1}}c^{\bar{1}}m^{\infty m}1$ | F:($u$,1/2,$w$) | $F_1^S F_1^S(4)$ | QNPL |
| 103.124.1.1 | $P^{1}4/^{\bar{1}}m^{1}c^{1}c^{\infty m}1$ | A:(1/2,1/2,1/2) | $A_5^S A_5^S(8)$ | OP |
| | | Z:(0,0,1/2) | $Z_5^S Z_5^S(8)$ | OP |
| 103.130.1.1 | $P^{1}4/^{\bar{1}}n^{1}c^{1}c^{\infty m}1$ | A:(1/2,1/2,1/2) | $A_5^S A_5^S(8)$ | OP |
| | | Z:(0,0,1/2) | $Z_5^S Z_5^S(8)$ | OP |
| 104.126.1.1 | $P^{1}4/^{\bar{1}}n^{1}n^{1}c^{\infty m}1$ | Z:(0,0,1/2) | $Z_5^S Z_5^S(8)$ | OP |
| 104.128.1.1 | $P^{1}4/^{\bar{1}}m^{1}n^{1}c^{\infty m}1$ | Z:(0,0,1/2) | $Z_5^S Z_5^S(8)$ | OP |
| 106.133.1.1 | $P^{1}4_2/^{\bar{1}}n^{1}b^{1}c^{\infty m}1$ | A:(1/2,1/2,1/2) | $A_5^S A_5^S(8)$ | OP |
| 106.135.1.1 | $P^{1}4_2/^{\bar{1}}n^{1}b^{1}c^{\infty m}1$ | A:(1/2,1/2,1/2) | $A_5^S A_5^S(8)$ | OP |
| 110.142.1.1 | $I^{1}4_1/^{\bar{1}}a^{1}c^{1}d^{\infty m}1$ | P:(1/2,1/2,1/2) | $P_1^S P_1^S(8)$ | OP |
| 113.127.1.1 | $P^{\bar{1}}4/^{\bar{1}}m^{\bar{1}}b^{1}m^{\infty m}1$ | F:($u$,1/2,$w$) | $F_1^S F_1^S(4)$ | QNPL |
| 113.129.1.1 | $P^{\bar{1}}4/^{\bar{1}}n^{\bar{1}}m^{1}m^{\infty m}1$ | F:($u$,1/2,$w$) | $F_1^S F_1^S(4)$ | QNPL |
| 113.136.1.1 | $P^{\bar{1}}4_2/^{\bar{1}}m^{\bar{1}}n^{1}m^{\infty m}1$ | F:($u$,1/2,$w$) | $F_1^S F_1^S(4)$ | QNPL |
| 113.138.1.1 | $P^{\bar{1}}4_2/^{\bar{1}}n^{\bar{1}}c^{1}m^{\infty m}1$ | F:($u$,1/2,$w$) | $F_1^S F_1^S(4)$ | QNPL |
| 114.128.1.1 | $P^{\bar{1}}4/^{\bar{1}}m^{\bar{1}}n^{1}c^{\infty m}1$ | A:(1/2,1/2,1/2) | $A_5^S A_5^S(8)$ | OP |
| | | F:($u$,1/2,$w$) | $F_1^S F_1^S(4)$ | QNPL |
| 114.130.1.1 | $P^{\bar{1}}4/^{\bar{1}}n^{\bar{1}}c^{1}c^{\infty m}1$ | A:(1/2,1/2,1/2) | $A_5^S A_5^S(8)$ | OP |
| | | F:($u$,1/2,$w$) | $F_1^S F_1^S(4)$ | QNPL |
| 114.135.1.1 | $P^{\bar{1}}4_2/^{\bar{1}}m^{\bar{1}}b^{1}c^{\infty m}1$ | A:(1/2,1/2,1/2) | $A_5^S A_5^S(8)$ | OP |
| | | F:($u$,1/2,$w$) | $F_1^S F_1^S(4)$ | QNPL |
| 114.137.1.1 | $P^{\bar{1}}4_2/^{\bar{1}}n^{\bar{1}}m^{1}c^{\infty m}1$ | A:(1/2,1/2,1/2) | $A_5^S A_5^S(8)$ | OP |
| | | F:($u$,1/2,$w$) | $F_1^S F_1^S(4)$ | QNPL |
| 158.165.1.1 | $P^{\bar{1}}\bar{3}^{1}c^{1}1^{\infty m}1$ | A:(0,0,1/2) | $A_3^S A_3^S(8)$ | OP |
| 159.163.1.1 | $P^{\bar{1}}\bar{3}^{1}1^{1}c^{\infty m}1$ | A:(0,0,1/2) | $A_3^S A_3^S(8)$ | OP |
| 161.167.1.1 | $R^{\bar{1}}\bar{3}^{1}c^{\infty m}1$ | T:(0,0,3/2) | $T_3^S T_3^S(8)$ | OP |
| 173.176.1.1 | $P^{1}6_3/^{\bar{1}}m^{\infty m}1$ | E:($u$,$v$,1/2) | $E_1^S E_1^S(4)$ | QNPL |
| 182.193.1.1 | $P^{1}6_3/^{\bar{1}}m^{\bar{1}}c^{\bar{1}}m^{\infty m}1$ | E:($u$,$v$,1/2) | $E_1^S E_1^S(4)$ | QNPL |
| 182.194.1.1 | $P^{1}6_3/^{\bar{1}}m^{\bar{1}}m^{\bar{1}}c^{\infty m}1$ | E:($u$,$v$,1/2) | $E_1^S E_1^S(4)$ | QNPL |
| 184.192.1.1 | $P^{1}6/^{\bar{1}}m^{1}c^{1}c^{\infty m}1$ | A:(0,0,1/2) | $A_5^S A_5^S(8)$<br>$A_6^S A_6^S(8)$ | OP |
| | | H:(1/3,1/3,1/2) | $H_3^S H_3^S(8)$ | OP |
| 185.193.1.1 | $P^{1}6_3/^{\bar{1}}m^{1}c^{1}m^{\infty m}1$ | A:(0,0,1/2) | $A_5^S A_6^S(8)$ | OP |
| | | H:(1/3,1/3,1/2) | $H_3^S H_3^S(8)$ | OP |
| | | E:($u$,$v$,1/2) | $E_1^S E_1^S(4)$ | QNPL |
| 186.194.1.1 | $P^{1}6_3/^{\bar{1}}m^{1}m^{1}c^{\infty m}1$ | A:(0,0,1/2) | $A_5^S A_6^S(8)$ | OP |
| | | E:($u$,$v$,1/2) | $E_1^S E_1^S(4)$ | QNPL |

| 188.192.1.1 | $P^{\bar{1}}6/^{1}m^{1}c^{\bar{1}}c^{\infty m}1$ | A:(0,0,1/2) | $A_2^S A_3^S(8)$ | OP |
|---|---|---|---|---|
| 188.193.1.1 | $P^{\bar{1}}6_3/^{1}m^{1}c^{\bar{1}}m^{\infty m}1$ | A:(0,0,1/2) | $A_2^S A_3^S(8)$ | OP |
| 190.192.1.1 | $P^{\bar{1}}6/^{1}m^{\bar{1}}c^{1}c^{\infty m}1$ | A:(0,0,1/2) | $A_1^S A_2^S(8)$ | OP |
| 190.194.1.1 | $P^{\bar{1}}6_3/^{1}m^{\bar{1}}m^{1}c^{\infty m}1$ | A:(0,0,1/2) | $A_1^S A_2^S(8)$ | OP |
| 195.200.1.1 | $P^{\bar{1}}m^{\bar{1}}\bar{3}^{\infty m}1$ | Γ:(0,0,0) | $\Gamma_4^S(6)$ | $C_2$−2 SP |
| | | R:(1/2,1/2,1/2) | $R_4^S(6)$ | $C_2$−2 SP |
| 195.201.1.1 | $P^{\bar{1}}n^{\bar{1}}\bar{3}^{\infty m}1$ | Γ:(0,0,0) | $\Gamma_4^S(6)$ | $C_2$−2 SP |
| | | R:(1/2,1/2,1/2) | $R_4^S(6)$ | $C_2$−2 SP |
| 196.202.1.1 | $F^{\bar{1}}m^{\bar{1}}\bar{3}^{\infty m}1$ | Γ:(0,0,0) | $\Gamma_4^S(6)$ | $C_2$−2 SP |
| 196.203.1.1 | $F^{\bar{1}}d^{\bar{1}}\bar{3}^{\infty m}1$ | Γ:(0,0,0) | $\Gamma_4^S(6)$ | $C_2$−2 SP |
| 197.204.1.1 | $I^{\bar{1}}m^{\bar{1}}\bar{3}^{\infty m}1$ | Γ:(0,0,0) | $\Gamma_4^S(6)$ | $C_2$−2 SP |
| | | H:(1,1,1) | $H_4^S(6)$ | $C_2$−2 SP |
| | | P:(1/2,1/2,1/2) | $P_4^S(6)$ | $C_2$−2 SP |
| 198.205.1.1 | $P^{\bar{1}}a^{\bar{1}}\bar{3}^{\infty m}1$ | Γ:(0,0,0) | $\Gamma_4^S(6)$ | $C_2$−2 SP |
| | | R:(1/2,1/2,1/2) | $R_1^S R_3^S(8)$<br>$R_2^S R_2^S(8)$ | $C_2$−2 OP |
| | | B:($u$,1/2,$w$) | $B_1^S B_1^S(4)$ | QNPL |
| 199.206.1.1 | $I^{\bar{1}}a^{\bar{1}}\bar{3}^{\infty m}1$ | Γ:(0,0,0) | $\Gamma_4^S(6)$ | $C_2$−2 SP |
| | | H:(1,1,1) | $H_4^S(6)$ | $C_2$−2 SP |
| 207.221.1.1 | $P^{\bar{1}}m^{\bar{1}}\bar{3}^{\bar{1}}m^{\infty m}1$ | Γ:(0,0,0) | $\Gamma_4^S(6)$, $\Gamma_5^S(6)$ | $C_2$−2 SP |
| | | R:(1/2,1/2,1/2) | $R_4^S(6)$, $R_5^S(6)$ | $C_2$−2 SP |
| 207.222.1.1 | $P^{\bar{1}}n^{\bar{1}}\bar{3}^{\bar{1}}n^{\infty m}1$ | Γ:(0,0,0) | $\Gamma_4^S(6)$, $\Gamma_5^S(6)$ | $C_2$−2 SP |
| | | R:(1/2,1/2,1/2) | $R_4^S(6)$, $R_5^S(6)$ | $C_2$−2 SP |
| 208.223.1.1 | $P^{\bar{1}}m^{\bar{1}}\bar{3}^{\bar{1}}n^{\infty m}1$ | Γ:(0,0,0) | $\Gamma_4^S(6)$, $\Gamma_5^S(6)$ | $C_2$−2 SP |
| | | R:(1/2,1/2,1/2) | $R_4^S(6)$, $R_5^S(6)$ | $C_2$−2 SP |
| 208.224.1.1 | $P^{\bar{1}}n^{\bar{1}}\bar{3}^{\bar{1}}m^{\infty m}1$ | Γ:(0,0,0) | $\Gamma_4^S(6)$, $\Gamma_5^S(6)$ | $C_2$−2 SP |
| | | R:(1/2,1/2,1/2) | $R_4^S(6)$, $R_5^S(6)$ | $C_2$−2 SP |
| 209.225.1.1 | $F^{\bar{1}}m^{\bar{1}}\bar{3}^{\bar{1}}m^{\infty m}1$ | Γ:(0,0,0) | $\Gamma_4^S(6)$, $\Gamma_5^S(6)$ | $C_2$−2 SP |
| 209.226.1.1 | $F^{\bar{1}}m^{\bar{1}}\bar{3}^{\bar{1}}c^{\infty m}1$ | Γ:(0,0,0) | $\Gamma_4^S(6)$, $\Gamma_5^S(6)$ | $C_2$−2 SP |
| 210.227.1.1 | $F^{\bar{1}}d^{\bar{1}}\bar{3}^{\bar{1}}m^{\infty m}1$ | Γ:(0,0,0) | $\Gamma_4^S(6)$, $\Gamma_5^S(6)$ | $C_2$−2 SP |
| 210.228.1.1 | $F^{\bar{1}}d^{\bar{1}}\bar{3}^{\bar{1}}c^{\infty m}1$ | Γ:(0,0,0) | $\Gamma_4^S(6)$, $\Gamma_5^S(6)$ | $C_2$−2 SP |
| 211.229.1.1 | $I^{\bar{1}}m^{\bar{1}}\bar{3}^{\bar{1}}m^{\infty m}1$ | Γ:(0,0,0) | $\Gamma_4^S(6)$, $\Gamma_5^S(6)$ | $C_2$−2 SP |
| | | H:(1,1,1) | $H_4^S(6)$, $H_5^S(6)$ | $C_2$−2 SP |
| | | P:(1/2,1/2,1/2) | $P_4^S(6)$ | $C_2$−2 SP |
| 214.230.1.1 | $I^{\bar{1}}a^{\bar{1}}\bar{3}^{\bar{1}}d^{\infty m}1$ | Γ:(0,0,0) | $\Gamma_4^S(6)$, $\Gamma_5^S(6)$ | $C_2$−2 SP |
| | | H:(1,1,1) | $H_4^S(6)$, $H_5^S(6)$ | $C_2$−2 SP |
| 215.221.1.1 | $P^{\bar{1}}m^{\bar{1}}\bar{3}^{1}m^{\infty m}1$ | Γ:(0,0,0) | $\Gamma_4^S(6)$, $\Gamma_5^S(6)$ | SP |
| | | R:(1/2,1/2,1/2) | $R_4^S(6)$, $R_5^S(6)$ | SP |
| 215.224.1.1 | $P^{\bar{1}}n^{\bar{1}}\bar{3}^{1}m^{\infty m}1$ | Γ:(0,0,0) | $\Gamma_4^S(6)$, $\Gamma_5^S(6)$ | SP |
| | | R:(1/2,1/2,1/2) | $R_4^S(6)$, $R_5^S(6)$ | SP |
| 216.225.1.1 | $F^{\bar{1}}m^{\bar{1}}\bar{3}^{1}m^{\infty m}1$ | Γ:(0,0,0) | $\Gamma_4^S(6)$, $\Gamma_5^S(6)$ | SP |
| 216.227.1.1 | $F^{\bar{1}}d^{\bar{1}}\bar{3}^{1}m^{\infty m}1$ | Γ:(0,0,0) | $\Gamma_4^S(6)$, $\Gamma_5^S(6)$ | SP |
| 217.229.1.1 | $I^{\bar{1}}m^{\bar{1}}\bar{3}^{1}m^{\infty m}1$ | Γ:(0,0,0) | $\Gamma_4^S(6)$, $\Gamma_5^S(6)$ | SP |

|  |  | H:(1,1,1) | $H_4^S(6)$, $H_5^S(6)$ | SP |
|---|---|---|---|---|
|  |  | P:(1/2,1/2,1/2) | $P_4^S(6)$, $P_5^S(6)$ | SP |
| 218.222.1.1 | $P^{\bar{1}}n^{\bar{1}}\bar{3}^{1}n^{\infty m}1$ | Γ:(0,0,0) | $\Gamma_4^S(6)$, $\Gamma_5^S(6)$ | SP |
|  |  | R:(1/2,1/2,1/2) | $R_3^S R_3^S(8)$ | OP |
|  |  |  | $R_4^S R_5^S(12)$ | DCP |
| 218.223.1.1 | $P^{\bar{1}}m^{\bar{1}}\bar{3}^{1}n^{\infty m}1$ | Γ:(0,0,0) | $\Gamma_4^S(6)$, $\Gamma_5^S(6)$ | SP |
|  |  | R:(1/2,1/2,1/2) | $R_3^S R_3^S(8)$ | OP |
|  |  |  | $R_4^S R_5^S(12)$ | DCP |
| 219.226.1.1 | $F^{\bar{1}}m^{\bar{1}}\bar{3}^{1}c^{\infty m}1$ | Γ:(0,0,0) | $\Gamma_4^S(6)$, $\Gamma_5^S(6)$ | SP |
|  |  | L:(1/2,1/2,1/2) | $L_3^S L_3^S(8)$ | OP |
| 219.228.1.1 | $F^{\bar{1}}d^{\bar{1}}\bar{3}^{1}c^{\infty m}1$ | Γ:(0,0,0) | $\Gamma_4^S(6)$, $\Gamma_5^S(6)$ | SP |
|  |  | L:(1/2,1/2,1/2) | $L_3^S L_3^S(8)$ | OP |
| 220.230.1.1 | $I^{\bar{1}}a^{\bar{1}}\bar{3}^{1}d^{\infty m}1$ | Γ:(0,0,0) | $\Gamma_4^S(6)$, $\Gamma_5^S(6)$ | SP |
|  |  | H:(1,1,1) | $H_3^S H_3^S(8)$ | OP |
|  |  |  | $H_4^S H_5^S(12)$ | DCP |
|  |  | P:(1/2,1/2,1/2) | $P_3^S(8)$ | OP |

## S3.4. Type-III collinear SSGs

Supplementary Table XXI. Tabulation of type-III collinear SSGs that accommodate unconventional magnons with high-dimensional band representations. The five columns illustrate the group number, the international notation of collinear SSGs, the momentum-position, band representations, and the quasiparticle type for unconventional magnons, respectively. "$C$–$n$" means that the topological charge $C$ is $|C| = n$. "$C_2$–$m$" means that the monopole charge $C_2 = (C_\uparrow - C_\downarrow)/2$ is $|C_2| = m$. We also include new quasiparticles with higher topological charge that go beyond MSGs. These new quasiparticles, namely $C$–4 OP, $C$–4 SP, and $C$–8 DP, are also tabulated.

| SSG | $G_S$ | ***k*** point | Rep | Quasiparticle |
|---|---|---|---|---|
| 19.92.1.1 | $P^{\bar{1}}4_1{}^{1}2_1{}^{\bar{1}}2^{\infty m}1$ | A:(1/2,1/2,1/2) | $(A)R_1^S R_1^S(8)$ | $C$–4 OP |
| 19.96.1.1 | $P^{\bar{1}}4_3{}^{1}2_1{}^{\bar{1}}2^{\infty m}1$ | A:(1/2,1/2,1/2) | $(A)R_1^S R_1^S(8)$ | $C$–4 OP |
| 56.130.1.1 | $P^{\bar{1}}4/^{1}n^{1}c^{\bar{1}}c^{\infty m}1$ | R:(0,1/2,1/2) | $(R)T_1^S T_2^S(8)$ | OP |
| 56.138.1.1 | $P^{\bar{1}}4_2/^{1}n^{1}c^{\bar{1}}m^{\infty m}1$ | R:(0,1/2,1/2) | $(R)T_1^S T_2^S(8)$ | OP |
| 73.142.1.1 | $I^{\bar{1}}4_1/^{1}a^{1}c^{\bar{1}}d^{\infty m}1$ | P:(1/2,1/2,1/2) | $(P)W_1^S W_1^S(8)$ | OP |
| 158.184.1.1 | $P^{\bar{1}}6^{1}c^{\bar{1}}c^{\infty m}1$ | A:(0,0,1/2) | $A_3^S A_3^S(8)$ | OP |
| 158.185.1.1 | $P^{\bar{1}}6_3{}^{1}c^{\bar{1}}m^{\infty m}1$ | A:(0,0,1/2) | $A_3^S A_3^S(8)$ | OP |
| 158.188.1.1 | $P^{\bar{1}}\bar{6}^{1}c^{\bar{1}}2^{\infty m}1$ | A:(0,0,1/2) | $A_3^S A_3^S(8)$ | OP |
| 159.184.1.1 | $P^{\bar{1}}6^{\bar{1}}c^{1}c^{\infty m}1$ | A:(0,0,1/2) | $A_3^S A_3^S(8)$ | OP |
| 159.186.1.1 | $P^{\bar{1}}6_3{}^{\bar{1}}m^{1}c^{\infty m}1$ | A:(0,0,1/2) | $A_3^S A_3^S(8)$ | OP |
| 159.190.1.1 | $P^{\bar{1}}\bar{6}^{\bar{1}}2^{1}c^{\infty m}1$ | A:(0,0,1/2) | $A_3^S A_3^S(8)$ | OP |

| 163.192.1.1 | $P^{\bar{1}}6/{}^{\bar{1}}m^{\bar{1}}c^{1}c^{\infty m}1$ | A:(0,0,1/2) | $A_3^S A_3^S(8)$ | OP |
|---|---|---|---|---|
| 163.194.1.1 | $P^{\bar{1}}6_3/{}^{\bar{1}}m^{\bar{1}}m^{1}c^{\infty m}1$ | A:(0,0,1/2) | $A_3^S A_3^S(8)$ | OP |
| 165.192.1.1 | $P^{\bar{1}}6/{}^{\bar{1}}m^{1}c^{\bar{1}}c^{\infty m}1$ | A:(0,0,1/2) | $A_1^S A_2^S(8)$ | OP |
| | | H:(1/3,1/3,1/2) | $H_3^S H_3^S(8)$ | OP |
| 165.193.1.1 | $P^{\bar{1}}6_3/{}^{\bar{1}}m^{1}c^{\bar{1}}m^{\infty m}1$ | A:(0,0,1/2) | $A_1^S A_2^S(8)$ | OP |
| | | H:(1/3,1/3,1/2) | $H_3^S H_3^S(8)$ | OP |
| 176.193.1.1 | $P^{1}6_3/{}^{1}m^{\bar{1}}c^{\bar{1}}m^{\infty m}1$ | A:(0,0,1/2) | $A_2^S A_3^S(8)$ | OP |
| 176.194.1.1 | $P^{1}6_3/{}^{1}m^{\bar{1}}m^{\bar{1}}c^{\infty m}1$ | A:(0,0,1/2) | $A_2^S A_3^S(8)$ | OP |
| 195.207.1.1 | $P^{\bar{1}}4^{1}3^{\bar{1}}2^{\infty m}1$ | Γ:(0,0,0) | $\Gamma_4^S(6)$ | $C$−4 SP |
| | | R:(1/2,1/2,1/2) | $R_4^S(6)$ | $C$−4 SP |
| | | Γ:(0,0,0) | $\Gamma_2^S \Gamma_3^S(4)$ | $C$−8 DP |
| | | R:(1/2,1/2,1/2) | $R_2^S R_3^S(4)$ | $C$−8 DP |
| 195.208.1.1 | $P^{\bar{1}}4_2{}^{1}3^{\bar{1}}2^{\infty m}1$ | Γ:(0,0,0) | $\Gamma_4^S(6)$ | $C$−4 SP |
| | | R:(1/2,1/2,1/2) | $R_4^S(6)$ | $C$−4 SP |
| | | Γ:(0,0,0) | $\Gamma_2^S \Gamma_3^S(4)$ | $C$−8 DP |
| | | R:(1/2,1/2,1/2) | $R_2^S R_3^S(4)$ | $C$−8 DP |
| 195.215.1.1 | $P^{\bar{1}}\bar{4}^{1}3^{\bar{1}}m^{\infty m}1$ | Γ:(0,0,0) | $\Gamma_4^S(6)$ | $C_2$−2 SP |
| | | R:(1/2,1/2,1/2) | $R_4^S(6)$ | $C_2$−2 SP |
| 195.218.1.1 | $P^{\bar{1}}\bar{4}^{1}3^{\bar{1}}n^{\infty m}1$ | Γ:(0,0,0) | $\Gamma_4^S(6)$ | $C_2$−2 SP |
| | | R:(1/2,1/2,1/2) | $R_4^S(6)$ | $C_2$−2 SP |
| 196.209.1.1 | $F^{\bar{1}}4^{1}3^{\bar{1}}2^{\infty m}1$ | Γ:(0,0,0) | $\Gamma_4^S(6)$ | $C$−4 SP |
| | | Γ:(0,0,0) | $\Gamma_2^S \Gamma_3^S(4)$ | $C$−8 DP |
| 196.210.1.1 | $F^{\bar{1}}4_1{}^{1}3^{\bar{1}}2^{\infty m}1$ | Γ:(0,0,0) | $\Gamma_4^S(6)$ | $C$−4 SP |
| | | Γ:(0,0,0) | $\Gamma_2^S \Gamma_3^S(4)$ | $C$−8 DP |
| 196.216.1.1 | $F^{\bar{1}}\bar{4}^{1}3^{\bar{1}}m^{\infty m}1$ | Γ:(0,0,0) | $\Gamma_4^S(6)$ | $C_2$−2 SP |
| 196.219.1.1 | $F^{\bar{1}}\bar{4}^{1}3^{\bar{1}}c^{\infty m}1$ | Γ:(0,0,0) | $\Gamma_4^S(6)$ | $C_2$−2 SP |
| 197.211.1.1 | $I^{\bar{1}}4^{1}3^{\bar{1}}2^{\infty m}1$ | Γ:(0,0,0) | $\Gamma_4^S(6)$ | $C$−4 SP |
| | | H:(1,1,1) | $H_4^S(6)$ | $C$−4 SP |
| | | P:(1/2,1/2,1/2) | $P_4^S(6)$ | $C$−4 SP |
| | | Γ:(0,0,0) | $\Gamma_2^S \Gamma_3^S(4)$ | $C$−8 DP |
| | | H:(1,1,1) | $H_2^S H_3^S(4)$ | $C$−8 DP |
| 197.217.1.1 | $I^{\bar{1}}\bar{4}^{1}3^{\bar{1}}m^{\infty m}1$ | Γ:(0,0,0) | $\Gamma_4^S(6)$ | $C_2$−2 SP |
| | | H:(1,1,1) | $H_4^S(6)$ | $C_2$−2 SP |
| | | P:(1/2,1/2,1/2) | $P_4^S(6)$ | $C_2$−2 SP |
| 198.212.1.1 | $P^{\bar{1}}4_3{}^{1}3^{\bar{1}}2^{\infty m}1$ | Γ:(0,0,0) | $\Gamma_4^S(6)$ | $C$−4 SP |
| | | Γ:(0,0,0) | $\Gamma_2^S \Gamma_3^S(4)$ | $C$−8 DP |
| | | R:(1/2,1/2,1/2) | $R_1^S R_3^S(8)$<br>$R_2^S R_2^S(8)$ | $C$−4 OP |
| 198.213.1.1 | $P^{\bar{1}}4_1{}^{1}3^{\bar{1}}2^{\infty m}1$ | Γ:(0,0,0) | $\Gamma_4^S(6)$ | $C$−4 SP |
| | | Γ:(0,0,0) | $\Gamma_2^S \Gamma_3^S(4)$ | $C$−8 DP |
| | | R:(1/2,1/2,1/2) | $R_1^S R_3^S(8)$<br>$R_2^S R_2^S(8)$ | $C$−4 OP |
| 199.214.1.1 | $I^{\bar{1}}4_1{}^{1}3^{\bar{1}}2^{\infty m}1$ | Γ:(0,0,0) | $\Gamma_4^S(6)$ | $C$−4 SP |
| | | H:(1,1,1) | $H_4^S(6)$ | $C$–4 SP |
| | | Γ:(0,0,0) | $\Gamma_2^S\Gamma_3^S(4)$ | $C$–8 DP |
| | | H:(1,1,1) | $H_2^S H_3^S(4)$ | $C$–8 DP |
| 199.220.1.1 | $I^{\bar{1}}\bar{4}^{1}3^{\bar{1}}d^{\infty m}1$ | Γ:(0,0,0) | $\Gamma_4^S(6)$ | $C_2$–2 SP |
| | | H:(1,1,1) | $H_4^S(6)$ | $C_2$–2 SP |
| 200.221.1.1 | $P^{1}m^{1}\bar{3}^{\bar{1}}m^{\infty m}1$ | Γ:(0,0,0) | $\Gamma_4^{S,+}(6)$, $\Gamma_4^{S,-}(6)$ | SP |
| | | R:(1/2,1/2,1/2) | $R_4^{S,+}(6)$, $R_4^{S,-}(6)$ | SP |
| 200.223.1.1 | $P^{1}m^{1}\bar{3}^{\bar{1}}n^{\infty m}1$ | Γ:(0,0,0) | $\Gamma_4^{S,+}(6)$, $\Gamma_4^{S,-}(6)$ | SP |
| | | R:(1/2,1/2,1/2) | $R_4^{S,+}(6)$, $R_4^{S,-}(6)$ | SP |
| 201.222.1.1 | $P^{1}n^{1}\bar{3}^{\bar{1}}n^{\infty m}1$ | Γ:(0,0,0) | $\Gamma_4^{S,+}(6)$, $\Gamma_4^{S,-}(6)$ | SP |
| | | R:(1/2,1/2,1/2) | $R_4^{S,+}(6)$, $R_4^{S,-}(6)$ | SP |
| 201.224.1.1 | $P^{1}n^{1}\bar{3}^{\bar{1}}m^{\infty m}1$ | Γ:(0,0,0) | $\Gamma_4^{S,+}(6)$, $\Gamma_4^{S,-}(6)$ | SP |
| | | R:(1/2,1/2,1/2) | $R_4^{S,+}(6)$, $R_4^{S,-}(6)$ | SP |
| 202.225.1.1 | $F^{1}m^{1}\bar{3}^{\bar{1}}m^{\infty m}1$ | Γ:(0,0,0) | $\Gamma_4^{S,+}(6)$, $\Gamma_4^{S,-}(6)$ | SP |
| 202.226.1.1 | $F^{1}m^{1}\bar{3}^{\bar{1}}c^{\infty m}1$ | Γ:(0,0,0) | $\Gamma_4^{S,+}(6)$, $\Gamma_4^{S,-}(6)$ | SP |
| 203.227.1.1 | $F^{1}d^{1}\bar{3}^{\bar{1}}m^{\infty m}1$ | Γ:(0,0,0) | $\Gamma_4^{S,+}(6)$, $\Gamma_4^{S,-}(6)$ | SP |
| 203.228.1.1 | $F^{1}d^{1}\bar{3}^{\bar{1}}c^{\infty m}1$ | Γ:(0,0,0) | $\Gamma_4^{S,+}(6)$, $\Gamma_4^{S,-}(6)$ | SP |
| 204.229.1.1 | $I^{1}m^{1}\bar{3}^{\bar{1}}m^{\infty m}1$ | Γ:(0,0,0) | $\Gamma_4^{S,+}(6)$, $\Gamma_4^{S,-}(6)$ | SP |
| | | H:(1,1,1) | $H_4^{S,+}(6)$, $H_4^{S,-}(6)$ | SP |
| | | P:(1/2,1/2,1/2) | $P_4^S(6)$ | SP |
| 206.230.1.1 | $I^{1}a^{1}\bar{3}^{\bar{1}}d^{\infty m}1$ | Γ:(0,0,0) | $\Gamma_4^{S,+}(6)$, $\Gamma_4^{S,-}(6)$ | SP |
| | | H:(1,1,1) | $H_4^{S,+}(6)$, $H_4^{S,-}(6)$ | SP |
| | | P:(1/2,1/2,1/2) | $P_1^S P_3^S(8)$, $P_2^S P_2^S(8)$ | OP |

## S3.5. Type-IV collinear SSGs

Supplementary Table XXII. Tabulation of type-IV collinear SSGs that accommodate unconventional

magnons with high-dimensional band representations. The five columns illustrate the group number, the international notation of collinear SSGs, the momentum-position, band representations, and the quasiparticle type for unconventional magnons, respectively. "$C$–$n$" means that the topological charge $C$ is $|C| = n$. We also include new quasiparticles with higher topological charge that go beyond MSGs. These new quasiparticles, namely $C$–4 OP, $C$–4 SP, and $C$–8 DP, are also tabulated.

| SSG | $G_S$ | ***k*** point | Rep | Quasiparticle |
|---|---|---|---|---|
| 4.4.2.1 | $P_a{}^{1}2_1{}^{\infty m}1$ | G:($u$, 1/2, $w$) | $G_1^S G_1^S(4)$ | QNPL |
| 4.3.2.1 | $P_b{}^{1}2_1{}^{\infty m}1$ | G:($u$, 1/2, $w$) | $G_1^S G_1^S(4)$ | QNPL |
| 4.5.2.1 | $P_C{}^{1}2_1{}^{\infty m}1$ | G:($u$, 1/2, $w$) | $G_1^S G_1^S(4)$ | QNPL |
| 11.11.2.1 | $P_a{}^{1}2_1/{}^{1}m{}^{\infty m}1$ | G:($u$, 1/2, $w$) | $G_1^S G_2^S(4)$ | QNPL |
| 11.10.2.1 | $P_b{}^{1}2_1/{}^{1}m{}^{\infty m}1$ | G:($u$, 1/2, $w$) | $G_1^S G_2^S(4)$ | QNPL |
| 11.12.2.1 | $P_C{}^{1}2_1/{}^{1}m{}^{\infty m}1$ | G:($u$, 1/2, $w$) | $G_1^S G_2^S(4)$ | QNPL |
| 14.14.2.1 | $P_a{}^{1}2_1/{}^{1}c{}^{\infty m}1$ | G:($u$, 1/2, $w$) | $G_1^S G_1^S(4)$ | QNPL |
| 14.13.2.1 | $P_b{}^{1}2_1/{}^{1}c{}^{\infty m}1$ | G:($u$, 1/2, $w$) | $G_1^S G_1^S(4)$ | QNPL |
| 14.11.2.1 | $P_c{}^{1}2_1/{}^{1}c{}^{\infty m}1$ | G:($u$, 1/2, $w$) | $G_1^S G_1^S(4)$ | QNPL |
| 14.12.2.1 | $P_A{}^{1}2_1/{}^{1}c{}^{\infty m}1$ | G:($u$, 1/2, $w$) | $G_1^S G_1^S(4)$ | QNPL |
| 14.15.2.1 | $P_C{}^{1}2_1/{}^{1}c{}^{\infty m}1$ | G:($u$, 1/2, $w$) | $G_1^S G_1^S(4)$ | QNPL |
| 17.17.2.1 | $P_a{}^{1}2{}^{1}2{}^{1}2_1{}^{\infty m}1$ | W:($u$,$v$,1/2) | $W_1^S W_1^S(4)$ | QNPL |
| 17.16.2.1 | $P_c{}^{1}2{}^{1}2{}^{1}2_1{}^{\infty m}1$ | W:($u$,$v$,1/2) | $W_1^S W_1^S(4)$ | QNPL |
| 17.21.2.1 | $P_A{}^{1}2{}^{1}2{}^{1}2_1{}^{\infty m}1$ | W:($u$,$v$,1/2) | $W_1^S W_1^S(4)$ | QNPL |
| 17.20.2.1 | $P_C{}^{1}2{}^{1}2{}^{1}2_1{}^{\infty m}1$ | W:($u$,$v$,1/2) | $W_1^S W_1^S(4)$ | QNPL |
| 17.24.2.1 | $P_I{}^{1}2{}^{1}2{}^{1}2_1{}^{\infty m}1$ | W:($u$,$v$,1/2) | $W_1^S W_1^S(4)$ | QNPL |
| 18.17.2.1 | $P_b{}^{1}2_1{}^{1}2_1{}^{1}2{}^{\infty m}1$ | L:(1/2,$v$,$w$) | $L_1^S L_1^S(4)$ | QNPL |
| | | N:($u$,1/2,$w$) | $N_1^S N_1^S(4)$ | QNPL |
| 18.18.2.1 | $P_c{}^{1}2_1{}^{1}2_1{}^{1}2{}^{\infty m}1$ | L:(1/2,$v$,$w$) | $L_1^S L_1^S(4)$ | QNPL |
| | | N:($u$,1/2,$w$) | $N_1^S N_1^S(4)$ | QNPL |
| 18.20.2.1 | $P_B{}^{1}2_1{}^{1}2_1{}^{1}2{}^{\infty m}1$ | L:(1/2,$v$,$w$) | $L_1^S L_1^S(4)$ | QNPL |
| | | N:($u$,1/2,$w$) | $N_1^S N_1^S(4)$ | QNPL |
| 18.21.2.1 | $P_C{}^{1}2_1{}^{1}2_1{}^{1}2{}^{\infty m}1$ | L:(1/2,$v$,$w$) | $L_1^S L_1^S(4)$ | QNPL |
| | | N:($u$,1/2,$w$) | $N_1^S N_1^S(4)$ | QNPL |
| 18.23.2.1 | $P_I{}^{1}2_1{}^{1}2_1{}^{1}2{}^{\infty m}1$ | L:(1/2,$v$,$w$) | $L_1^S L_1^S(4)$ | QNPL |
| | | N:($u$,1/2,$w$) | $N_1^S N_1^S(4)$ | QNPL |
| 19.18.2.1 | $P_c{}^{1}2_1{}^{1}2_1{}^{1}2_1{}^{\infty m}1$ | R:(1/2,1/2,1/2) | $R_1^S R_1^S(8)$ | $C$–4 OP |
| | | L:(1/2,$v$,$w$) | $L_1^S L_1^S(4)$ | QNPL |
| | | N:($u$,1/2,$w$) | $N_1^S N_1^S(4)$ | QNPL |
| | | W:($u$,$v$,1/2) | $W_1^S W_1^S(4)$ | QNPL |
| 19.20.2.1 | $P_C{}^{1}2_1{}^{1}2_1{}^{1}2_1{}^{\infty m}1$ | R:(1/2,1/2,1/2) | $R_1^S R_1^S(8)$ | $C$–4 OP |
| | | L:(1/2,$v$,$w$) | $L_1^S L_1^S(4)$ | QNPL |
| | | N:($u$,1/2,$w$) | $N_1^S N_1^S(4)$ | QNPL |
| | | W:($u$,$v$,1/2) | $W_1^S W_1^S(4)$ | QNPL |
| 19.24.2.1 | $P_I{}^{1}2_1{}^{1}2_1{}^{1}2_1{}^{\infty m}1$ | R:(1/2,1/2,1/2) | $R_1^S R_1^S(8)$ | $C$−4 OP |
| | | L:(1/2,$v$,$w$) | $L_1^S L_1^S(4)$ | QNPL |
| | | N:($u$,1/2,$w$) | $N_1^S N_1^S(4)$ | QNPL |
| | | W:($u$,$v$,1/2) | $W_1^S W_1^S(4)$ | QNPL |
| 20.21.2.1 | $C_c{}^{1}2{}^{1}2{}^{1}2_1{}^{\infty m}1$ | Q:($u$,$v$,1/2) | $Q_1^S Q_1^S(4)$ | QNPL |
| 20.17.2.1 | $C_a{}^{1}2{}^{1}2{}^{1}2_1{}^{\infty m}1$ | Q:($u$,$v$,1/2) | $Q_1^S Q_1^S(4)$ | QNPL |
| 20.22.2.1 | $C_A{}^{1}2{}^{1}2{}^{1}2_1{}^{\infty m}1$ | Q:($u$,$v$,1/2) | $Q_1^S Q_1^S(4)$ | QNPL |
| 26.26.2.1 | $P_a{}^{1}m{}^{1}c{}^{1}2_1{}^{\infty m}1$ | W:($u$,$v$,1/2) | $W_1^S W_1^S(4)$ | QNPL |
| 26.26.2.4 | $P_b{}^{1}m{}^{1}c{}^{1}2_1{}^{\infty m}1$ | W:($u$,$v$,1/2) | $W_1^S W_1^S(4)$ | QNPL |
| 26.25.2.1 | $P_c{}^{1}m{}^{1}c{}^{1}2_1{}^{\infty m}1$ | W:($u$,$v$,1/2) | $W_1^S W_1^S(4)$ | QNPL |
| 26.38.2.1 | $P_A{}^{1}m{}^{1}c{}^{1}2_1{}^{\infty m}1$ | W:($u$,$v$,1/2) | $W_1^S W_1^S(4)$ | QNPL |
| 26.39.2.1 | $P_B{}^{1}m{}^{1}c{}^{1}2_1{}^{\infty m}1$ | W:($u$,$v$,1/2) | $W_1^S W_1^S(4)$ | QNPL |
| 26.36.2.1 | $P_C{}^{1}m{}^{1}c{}^{1}2_1{}^{\infty m}1$ | W:($u$,$v$,1/2) | $W_1^S W_1^S(4)$ | QNPL |
| 26.46.2.1 | $P_I{}^{1}m{}^{1}c{}^{1}2_1{}^{\infty m}1$ | W:($u$,$v$,1/2) | $W_1^S W_1^S(4)$ | QNPL |
| 29.26.2.1 | $P_a{}^{1}c{}^{1}a{}^{1}2_1{}^{\infty m}1$ | R:(1/2,1/2,1/2) | $R_1^S R_1^S(8)$ | OP |
| | | U:(1/2,0,1/2) | $U_1^S U_1^S(8)$ | OP |
| | | W:($u$,$v$,1/2) | $W_1^S W_1^S(4)$ | QNPL |
| 29.29.2.1 | $P_b{}^{1}c{}^{1}a{}^{1}2_1{}^{\infty m}1$ | R:(1/2,1/2,1/2) | $R_1^S R_1^S(8)$ | OP |
| | | U:(1/2,0,1/2) | $U_1^S U_1^S(8)$ | OP |
| | | W:($u$,$v$,1/2) | $W_1^S W_1^S(4)$ | QNPL |
| 29.28.2.1 | $P_c{}^{1}c{}^{1}a{}^{1}2_1{}^{\infty m}1$ | R:(1/2,1/2,1/2) | $R_1^S R_1^S(8)$ | OP |
| | | U:(1/2,0,1/2) | $U_1^S U_1^S(8)$ | OP |
| | | W:($u$,$v$,1/2) | $W_1^S W_1^S(4)$ | QNPL |
| 29.41.2.1 | $P_A{}^{1}c{}^{1}a{}^{1}2_1{}^{\infty m}1$ | R:(1/2,1/2,1/2) | $R_1^S R_1^S(8)$ | OP |
| | | U:(1/2,0,1/2) | $U_1^S U_1^S(8)$ | OP |
| | | W:($u$,$v$,1/2) | $W_1^S W_1^S(4)$ | QNPL |
| 29.39.2.1 | $P_B{}^{1}c{}^{1}a{}^{1}2_1{}^{\infty m}1$ | R:(1/2,1/2,1/2) | $R_1^S R_1^S(8)$ | OP |
| | | U:(1/2,0,1/2) | $U_1^S U_1^S(8)$ | OP |
| | | W:($u$,$v$,1/2) | $W_1^S W_1^S(4)$ | QNPL |
| 29.36.2.1 | $P_C{}^{1}c{}^{1}a{}^{1}2_1{}^{\infty m}1$ | R:(1/2,1/2,1/2) | $R_1^S R_1^S(8)$ | OP |
| | | U:(1/2,0,1/2) | $U_1^S U_1^S(8)$ | OP |
| | | W:($u$,$v$,1/2) | $W_1^S W_1^S(4)$ | QNPL |
| 29.45.2.1 | $P_I{}^{1}c{}^{1}a{}^{1}2_1{}^{\infty m}1$ | R:(1/2,1/2,1/2) | $R_1^S R_1^S(8)$ | OP |
| | | U:(1/2,0,1/2) | $U_1^S U_1^S(8)$ | OP |
| | | W:($u$,$v$,1/2) | $W_1^S W_1^S(4)$ | QNPL |
| 31.26.2.1 | $P_a{}^{1}m{}^{1}n{}^{1}2_1{}^{\infty m}1$ | W:($u$,$v$,1/2) | $W_1^S W_1^S(4)$ | QNPL |
| 31.31.2.1 | $P_b{}^{1}m{}^{1}n{}^{1}2_1{}^{\infty m}1$ | W:($u$,$v$,1/2) | $W_1^S W_1^S(4)$ | QNPL |
| 31.28.2.1 | $P_c{}^{1}m{}^{1}n{}^{1}2_1{}^{\infty m}1$ | W:($u$,$v$,1/2) | $W_1^S W_1^S(4)$ | QNPL |
| 31.40.2.1 | $P_A{}^{1}m{}^{1}n{}^{1}2_1{}^{\infty m}1$ | W:($u$,$v$,1/2) | $W_1^S W_1^S(4)$ | QNPL |
| 31.38.2.1 | $P_B{}^{1}m{}^{1}n{}^{1}2_1{}^{\infty m}1$ | W:($u$,$v$,1/2) | $W_1^S W_1^S(4)$ | QNPL |
| 31.36.2.1 | $P_C{}^{1}m{}^{1}n{}^{1}2_1{}^{\infty m}1$ | W:($u$,$v$,1/2) | $W_1^S W_1^S(4)$ | QNPL |
| 31.44.2.1 | $P_I{}^{1}m{}^{1}n{}^{1}2_1{}^{\infty m}1$ | W:($u$,$v$,1/2) | $W_1^S W_1^S(4)$ | QNPL |
| 33.31.2.1 | $P_a{}^{1}n{}^{1}a{}^{1}2_1{}^{\infty m}1$ | U:(1/2,0,1/2) | $U_1^S U_1^S(8)$ | OP |

| | | W:$(u,v,1/2)$ | $W_1^S W_1^S(4)$ | QNPL |
|---|---|---|---|---|
| 33.29.2.1 | $P_{b}{}^{1}n^{1}a^{1}2_{1}{}^{\infty m}1$ | U:$(1/2,0,1/2)$ | $U_1^S U_1^S(8)$ | OP |
| | | W:$(u,v,1/2)$ | $W_1^S W_1^S(4)$ | QNPL |
| 33.32.2.1 | $P_{c}{}^{1}n^{1}a^{1}2_{1}{}^{\infty m}1$ | U:$(1/2,0,1/2)$ | $U_1^S U_1^S(8)$ | OP |
| | | W:$(u,v,1/2)$ | $W_1^S W_1^S(4)$ | QNPL |
| 33.40.2.1 | $P_{A}{}^{1}n^{1}a^{1}2_{1}{}^{\infty m}1$ | U:$(1/2,0,1/2)$ | $U_1^S U_1^S(8)$ | OP |
| | | W:$(u,v,1/2)$ | $W_1^S W_1^S(4)$ | QNPL |
| 33.41.2.1 | $P_{B}{}^{1}n^{1}a^{1}2_{1}{}^{\infty m}1$ | U:$(1/2,0,1/2)$ | $U_1^S U_1^S(8)$ | OP |
| | | W:$(u,v,1/2)$ | $W_1^S W_1^S(4)$ | QNPL |
| 33.36.2.1 | $P_{C}{}^{1}n^{1}a^{1}2_{1}{}^{\infty m}1$ | U:$(1/2,0,1/2)$ | $U_1^S U_1^S(8)$ | OP |
| | | W:$(u,v,1/2)$ | $W_1^S W_1^S(4)$ | QNPL |
| 33.46.2.1 | $P_{I}{}^{1}n^{1}a^{1}2_{1}{}^{\infty m}1$ | U:$(1/2,0,1/2)$ | $U_1^S U_1^S(8)$ | OP |
| | | W:$(u,v,1/2)$ | $W_1^S W_1^S(4)$ | QNPL |
| 36.35.2.1 | $C_{c}{}^{1}m^{1}c^{1}2_{1}{}^{\infty m}1$ | Q:$(u,v,1/2)$ | $Q_1^S Q_1^S(4)$ | QNPL |
| 36.26.2.1 | $C_{a}{}^{1}m^{1}c^{1}2_{1}{}^{\infty m}1$ | Q:$(u,v,1/2)$ | $Q_1^S Q_1^S(4)$ | QNPL |
| 36.42.2.1 | $C_{A}{}^{1}m^{1}c^{1}2_{1}{}^{\infty m}1$ | Q:$(u,v,1/2)$ | $Q_1^S Q_1^S(4)$ | QNPL |
| 51.47.2.1 | $P_{a}{}^{1}m^{1}m^{1}a^{\infty m}1$ | L:$(1/2,v,w)$ | $L_1^S L_2^S(4)$ | QNPL |
| 51.51.2.1 | $P_{b}{}^{1}m^{1}m^{1}a^{\infty m}1$ | L:$(1/2,v,w)$ | $L_1^S L_2^S(4)$ | QNPL |
| 51.51.2.4 | $P_{c}{}^{1}m^{1}m^{1}a^{\infty m}1$ | L:$(1/2,v,w)$ | $L_1^S L_2^S(4)$ | QNPL |
| 51.63.2.1 | $P_{A}{}^{1}m^{1}m^{1}a^{\infty m}1$ | L:$(1/2,v,w)$ | $L_1^S L_2^S(4)$ | QNPL |
| 51.65.2.1 | $P_{B}{}^{1}m^{1}m^{1}a^{\infty m}1$ | L:$(1/2,v,w)$ | $L_1^S L_2^S(4)$ | QNPL |
| 51.67.2.1 | $P_{C}{}^{1}m^{1}m^{1}a^{\infty m}1$ | L:$(1/2,v,w)$ | $L_1^S L_2^S(4)$ | QNPL |
| 51.74.2.1 | $P_{I}{}^{1}m^{1}m^{1}a^{\infty m}1$ | L:$(1/2,v,w)$ | $L_1^S L_2^S(4)$ | QNPL |
| 52.53.2.1 | $P_{a}{}^{1}n^{1}n^{1}a^{\infty m}1$ | S:$(1/2,1/2,0)$ | $S_1^S S_2^S(8)$ | OP |
| | | N:$(u,1/2,w)$ | $N_1^S N_2^S(4)$ | QNPL |
| 52.50.2.1 | $P_{b}{}^{1}n^{1}n^{1}a^{\infty m}1$ | S:$(1/2,1/2,0)$ | $S_1^S S_2^S(8)$ | OP |
| | | N:$(u,1/2,w)$ | $N_1^S N_2^S(4)$ | QNPL |
| 52.54.2.1 | $P_{c}{}^{1}n^{1}n^{1}a^{\infty m}1$ | S:$(1/2,1/2,0)$ | $S_1^S S_2^S(8)$ | OP |
| | | N:$(u,1/2,w)$ | $N_1^S N_2^S(4)$ | QNPL |
| 52.66.2.1 | $P_{A}{}^{1}n^{1}n^{1}a^{\infty m}1$ | S:$(1/2,1/2,0)$ | $S_1^S S_2^S(8)$ | OP |
| | | N:$(u,1/2,w)$ | $N_1^S N_2^S(4)$ | QNPL |
| 52.63.2.1 | $P_{B}{}^{1}n^{1}n^{1}a^{\infty m}1$ | S:$(1/2,1/2,0)$ | $S_1^S S_2^S(8)$ | OP |
| | | N:$(u,1/2,w)$ | $N_1^S N_2^S(4)$ | QNPL |
| 52.68.2.1 | $P_{C}{}^{1}n^{1}n^{1}a^{\infty m}1$ | S:$(1/2,1/2,0)$ | $S_1^S S_2^S(8)$ | OP |
| | | N:$(u,1/2,w)$ | $N_1^S N_2^S(4)$ | QNPL |
| 52.74.2.1 | $P_{I}{}^{1}n^{1}n^{1}a^{\infty m}1$ | S:$(1/2,1/2,0)$ | $S_1^S S_2^S(8)$ | OP |
| | | N:$(u,1/2,w)$ | $N_1^S N_2^S(4)$ | QNPL |
| 53.51.2.1 | $P_{a}{}^{1}m^{1}n^{1}a^{\infty m}1$ | W:$(u,v,1/2)$ | $W_1^S W_2^S(4)$ | QNPL |
| 53.53.2.1 | $P_{b}{}^{1}m^{1}n^{1}a^{\infty m}1$ | W:$(u,v,1/2)$ | $W_1^S W_2^S(4)$ | QNPL |
| 53.49.2.1 | $P_{c}{}^{1}m^{1}n^{1}a^{\infty m}1$ | W:$(u,v,1/2)$ | $W_1^S W_2^S(4)$ | QNPL |
| 53.66.2.1 | $P_{A}{}^{1}m^{1}n^{1}a^{\infty m}1$ | W:$(u,v,1/2)$ | $W_1^S W_2^S(4)$ | QNPL |
| 53.65.2.1 | $P_{B}{}^{1}m^{1}n^{1}a^{\infty m}1$ | W:$(u,v,1/2)$ | $W_1^S W_2^S(4)$ | QNPL |
| 53.64.2.1 | $P_{C}{}^{1}m^{1}n^{1}a^{\infty m}1$ | W:$(u,v,1/2)$ | $W_1^S W_2^S(4)$ | QNPL |

| 53.74.2.1 | $P_I{}^1m^1n^1a^{\infty m}1$ | W:($u$,$v$,1/2) | $W_1^SW_2^S$(4) | QNPL |
|---|---|---|---|---|
| 54.49.2.1 | $P_a{}^1c^1c^1a^{\infty m}1$ | R:(1/2,1/2,1/2) | $R_1^SR_2^S$(8) | OP |
| | | U:(1/2,0,1/2) | $U_1^SU_2^S$(8) | OP |
| | | L:(1/2,$v$,$w$) | $L_1^SL_2^S$(4) | QNPL |
| 54.54.2.1 | $P_b{}^1c^1c^1a^{\infty m}1$ | R:(1/2,1/2,1/2) | $R_1^SR_2^S$(8) | OP |
| | | U:(1/2,0,1/2) | $U_1^SU_2^S$(8) | OP |
| | | L:(1/2,$v$,$w$) | $L_1^SL_2^S$(4) | QNPL |
| 54.51.2.1 | $P_c{}^1c^1c^1a^{\infty m}1$ | R:(1/2,1/2,1/2) | $R_1^SR_2^S$(8) | OP |
| | | U:(1/2,0,1/2) | $U_1^SU_2^S$(8) | OP |
| | | L:(1/2,$v$,$w$) | $L_1^SL_2^S$(4) | QNPL |
| 54.64.2.1 | $P_A{}^1c^1c^1a^{\infty m}1$ | R:(1/2,1/2,1/2) | $R_1^SR_2^S$(8) | OP |
| | | U:(1/2,0,1/2) | $U_1^SU_2^S$(8) | OP |
| | | L:(1/2,$v$,$w$) | $L_1^SL_2^S$(4) | QNPL |
| 54.67.2.1 | $P_B{}^1c^1c^1a^{\infty m}1$ | R:(1/2,1/2,1/2) | $R_1^SR_2^S$(8) | OP |
| | | U:(1/2,0,1/2) | $U_1^SU_2^S$(8) | OP |
| | | L:(1/2,$v$,$w$) | $L_1^SL_2^S$(4) | QNPL |
| 54.68.2.1 | $P_C{}^1c^1c^1a^{\infty m}1$ | R:(1/2,1/2,1/2) | $R_1^SR_2^S$(8) | OP |
| | | U:(1/2,0,1/2) | $U_1^SU_2^S$(8) | OP |
| | | L:(1/2,$v$,$w$) | $L_1^SL_2^S$(4) | QNPL |
| 54.73.2.1 | $P_I{}^1c^1c^1a^{\infty m}1$ | R:(1/2,1/2,1/2) | $R_1^SR_2^S$(8) | OP |
| | | U:(1/2,0,1/2) | $U_1^SU_2^S$(8) | OP |
| | | L:(1/2,$v$,$w$) | $L_1^SL_2^S$(4) | QNPL |
| 55.51.2.1 | $P_a{}^1b^1a^1m^{\infty m}1$ | L:(1/2,$v$,$w$) | $L_1^SL_2^S$(4) | QNPL |
| | | N:($u$,1/2,$w$) | $N_1^SN_2^S$(4) | QNPL |
| 55.55.2.1 | $P_c{}^1b^1a^1m^{\infty m}1$ | L:(1/2,$v$,$w$) | $L_1^SL_2^S$(4) | QNPL |
| | | N:($u$,1/2,$w$) | $N_1^SN_2^S$(4) | QNPL |
| 55.64.2.1 | $P_A{}^1b^1a^1m^{\infty m}1$ | L:(1/2,$v$,$w$) | $L_1^SL_2^S$(4) | QNPL |
| | | N:($u$,1/2,$w$) | $N_1^SN_2^S$(4) | QNPL |
| 55.65.2.1 | $P_C{}^1b^1a^1m^{\infty m}1$ | L:(1/2,$v$,$w$) | $L_1^SL_2^S$(4) | QNPL |
| | | N:($u$,1/2,$w$) | $N_1^SN_2^S$(4) | QNPL |
| 55.72.2.1 | $P_I{}^1b^1a^1m^{\infty m}1$ | L:(1/2,$v$,$w$) | $L_1^SL_2^S$(4) | QNPL |
| | | N:($u$,1/2,$w$) | $N_1^SN_2^S$(4) | QNPL |
| 56.54.2.1 | $P_b{}^1c^1c^1n^{\infty m}1$ | T:(0,1/2,1/2) | $T_1^ST_2^S$(8) | OP |
| | | U:(1/2,0,1/2) | $U_1^SU_2^S$(8) | OP |
| | | L:(1/2,$v$,$w$) | $L_1^SL_2^S$(4) | QNPL |
| | | N:($u$,1/2,$w$) | $N_1^SN_2^S$(4) | QNPL |
| 56.59.2.1 | $P_c{}^1c^1c^1n^{\infty m}1$ | T:(0,1/2,1/2) | $T_1^ST_2^S$(8) | OP |
| | | U:(1/2,0,1/2) | $U_1^SU_2^S$(8) | OP |
| | | L:(1/2,$v$,$w$) | $L_1^SL_2^S$(4) | QNPL |
| | | N:($u$,1/2,$w$) | $N_1^SN_2^S$(4) | QNPL |
| 56.64.2.1 | $P_A{}^1c^1c^1n^{\infty m}1$ | T:(0,1/2,1/2) | $T_1^ST_2^S$(8) | OP |
| | | U:(1/2,0,1/2) | $U_1^SU_2^S$(8) | OP |
| | | L:(1/2,$v$,$w$) | $L_1^SL_2^S$(4) | QNPL |

| | | N:($u$,1/2,$w$) | $N_1^S N_2^S(4)$ | QNPL |
|---|---|---|---|---|
| 56.66.2.1 | $P_C{}^1c^1c^1n^{\infty m}1$ | T:(0,1/2,1/2) | $T_1^S T_2^S(8)$ | OP |
| | | U:(1/2,0,1/2) | $U_1^S U_2^S(8)$ | OP |
| | | L:(1/2,$v$,$w$) | $L_1^S L_2^S(4)$ | QNPL |
| | | N:($u$,1/2,$w$) | $N_1^S N_2^S(4)$ | QNPL |
| 56.72.2.1 | $P_I{}^1c^1c^1n^{\infty m}1$ | T:(0,1/2,1/2) | $T_1^S T_2^S(8)$ | OP |
| | | U:(1/2,0,1/2) | $U_1^S U_2^S(8)$ | OP |
| | | L:(1/2,$v$,$w$) | $L_1^S L_2^S(4)$ | QNPL |
| | | N:($u$,1/2,$w$) | $N_1^S N_2^S(4)$ | QNPL |
| 57.57.2.1 | $P_a{}^1b^1c^1m^{\infty m}1$ | E:($u$,1/2,1/2) | $E_1^S E_1^S(8)$ | ONL |
| | | N:($u$,1/2,$w$) | $N_1^S N_2^S(4)$ | QNPL |
| | | W:($u$,$v$,1/2) | $W_1^S W_2^S(4)$ | QNPL |
| 57.51.2.4 | $P_b{}^1b^1c^1m^{\infty m}1$ | E:($u$,1/2,1/2) | $E_1^S E_1^S(8)$ | ONL |
| | | N:($u$,1/2,$w$) | $N_1^S N_2^S(4)$ | QNPL |
| | | W:($u$,$v$,1/2) | $W_1^S W_2^S(4)$ | QNPL |
| 57.51.2.1 | $P_c{}^1b^1c^1m^{\infty m}1$ | E:($u$,1/2,1/2) | $E_1^S E_1^S(8)$ | ONL |
| | | N:($u$,1/2,$w$) | $N_1^S N_2^S(4)$ | QNPL |
| | | W:($u$,$v$,1/2) | $W_1^S W_2^S(4)$ | QNPL |
| 57.67.2.1 | $P_A{}^1b^1c^1m^{\infty m}1$ | E:($u$,1/2,1/2) | $E_1^S E_1^S(8)$ | ONL |
| | | N:($u$,1/2,$w$) | $N_1^S N_2^S(4)$ | QNPL |
| | | W:($u$,$v$,1/2) | $W_1^S W_2^S(4)$ | QNPL |
| 57.64.2.1 | $P_B{}^1b^1c^1m^{\infty m}1$ | E:($u$,1/2,1/2) | $E_1^S E_1^S(8)$ | ONL |
| | | N:($u$,1/2,$w$) | $N_1^S N_2^S(4)$ | QNPL |
| | | W:($u$,$v$,1/2) | $W_1^S W_2^S(4)$ | QNPL |
| 57.63.2.1 | $P_C{}^1b^1c^1m^{\infty m}1$ | E:($u$,1/2,1/2) | $E_1^S E_1^S(8)$ | ONL |
| | | N:($u$,1/2,$w$) | $N_1^S N_2^S(4)$ | QNPL |
| | | W:($u$,$v$,1/2) | $W_1^S W_2^S(4)$ | QNPL |
| 57.72.2.1 | $P_I{}^1b^1c^1m^{\infty m}1$ | E:($u$,1/2,1/2) | $E_1^S E_1^S(8)$ | ONL |
| | | N:($u$,1/2,$w$) | $N_1^S N_2^S(4)$ | QNPL |
| | | W:($u$,$v$,1/2) | $W_1^S W_2^S(4)$ | QNPL |
| 58.53.2.1 | $P_a{}^1n^1n^1m^{\infty m}1$ | L:(1/2,$v$,$w$) | $L_1^S L_2^S(4)$ | QNPL |
| | | N:($u$,1/2,$w$) | $N_1^S N_2^S(4)$ | QNPL |
| 58.55.2.1 | $P_c{}^1n^1n^1m^{\infty m}1$ | L:(1/2,$v$,$w$) | $L_1^S L_2^S(4)$ | QNPL |
| | | N:($u$,1/2,$w$) | $N_1^S N_2^S(4)$ | QNPL |
| 58.63.2.1 | $P_B{}^1n^1n^1m^{\infty m}1$ | L:(1/2,$v$,$w$) | $L_1^S L_2^S(4)$ | QNPL |
| | | N:($u$,1/2,$w$) | $N_1^S N_2^S(4)$ | QNPL |
| 58.66.2.1 | $P_C{}^1n^1n^1m^{\infty m}1$ | L:(1/2,$v$,$w$) | $L_1^S L_2^S(4)$ | QNPL |
| | | N:($u$,1/2,$w$) | $N_1^S N_2^S(4)$ | QNPL |
| 58.71.2.1 | $P_I{}^1n^1n^1m^{\infty m}1$ | L:(1/2,$v$,$w$) | $L_1^S L_2^S(4)$ | QNPL |
| | | N:($u$,1/2,$w$) | $N_1^S N_2^S(4)$ | QNPL |
| 59.51.2.1 | $P_b{}^1m^1m^1n^{\infty m}1$ | L:(1/2,$v$,$w$) | $L_1^S L_2^S(4)$ | QNPL |
| | | N:($u$,1/2,$w$) | $N_1^S N_2^S(4)$ | QNPL |
| 59.59.2.1 | $P_c{}^1m^1m^1n^{\infty m}1$ | L:(1/2,$v$,$w$) | $L_1^S L_2^S(4)$ | QNPL |

| | | N:$(u,1/2,w)$ | $N_1^S N_2^S(4)$ | QNPL |
|---|---|---|---|---|
| 59.63.2.1 | $P_B{}^1m{}^1m{}^1n^{\infty m}1$ | L:$(1/2,v,w)$ | $L_1^S L_2^S(4)$ | QNPL |
| | | N:$(u,1/2,w)$ | $N_1^S N_2^S(4)$ | QNPL |
| 59.65.2.1 | $P_C{}^1m{}^1m{}^1n^{\infty m}1$ | L:$(1/2,v,w)$ | $L_1^S L_2^S(4)$ | QNPL |
| | | N:$(u,1/2,w)$ | $N_1^S N_2^S(4)$ | QNPL |
| 59.71.2.1 | $P_I{}^1m{}^1m{}^1n^{\infty m}1$ | L:$(1/2,v,w)$ | $L_1^S L_2^S(4)$ | QNPL |
| | | N:$(u,1/2,w)$ | $N_1^S N_2^S(4)$ | QNPL |
| 60.54.2.1 | $P_a{}^1b{}^1c{}^1n^{\infty m}1$ | P:$(1/2,v,1/2)$ | $P_1^S P_1^S(8)$ | ONL |
| | | T:$(0,1/2,1/2)$ | $T_1^S T_2^S(8)$ | OP |
| | | L:$(1/2,v,w)$ | $L_1^S L_2^S(4)$ | QNPL |
| | | W:$(u,v,1/2)$ | $W_1^S W_2^S(4)$ | QNPL |
| 60.57.2.1 | $P_b{}^1b{}^1c{}^1n^{\infty m}1$ | P:$(1/2,v,1/2)$ | $P_1^S P_1^S(8)$ | ONL |
| | | T:$(0,1/2,1/2)$ | $T_1^S T_2^S(8)$ | OP |
| | | L:$(1/2,v,w)$ | $L_1^S L_2^S(4)$ | QNPL |
| | | W:$(u,v,1/2)$ | $W_1^S W_2^S(4)$ | QNPL |
| 60.53.2.1 | $P_c{}^1b{}^1c{}^1n^{\infty m}1$ | P:$(1/2,v,1/2)$ | $P_1^S P_1^S(8)$ | ONL |
| | | T:$(0,1/2,1/2)$ | $T_1^S T_2^S(8)$ | OP |
| | | L:$(1/2,v,w)$ | $L_1^S L_2^S(4)$ | QNPL |
| | | W:$(u,v,1/2)$ | $W_1^S W_2^S(4)$ | QNPL |
| 60.64.2.1 | $P_A{}^1b{}^1c{}^1n^{\infty m}1$ | P:$(1/2,v,1/2)$ | $P_1^S P_1^S(8)$ | ONL |
| | | T:$(0,1/2,1/2)$ | $T_1^S T_2^S(8)$ | OP |
| | | L:$(1/2,v,w)$ | $L_1^S L_2^S(4)$ | QNPL |
| | | W:$(u,v,1/2)$ | $W_1^S W_2^S(4)$ | QNPL |
| 60.68.2.1 | $P_B{}^1b{}^1c{}^1n^{\infty m}1$ | P:$(1/2,v,1/2)$ | $P_1^S P_1^S(8)$ | ONL |
| | | T:$(0,1/2,1/2)$ | $T_1^S T_2^S(8)$ | OP |
| | | L:$(1/2,v,w)$ | $L_1^S L_2^S(4)$ | QNPL |
| | | W:$(u,v,1/2)$ | $W_1^S W_2^S(4)$ | QNPL |
| 60.63.2.1 | $P_C{}^1b{}^1c{}^1n^{\infty m}1$ | P:$(1/2,v,1/2)$ | $P_1^S P_1^S(8)$ | ONL |
| | | T:$(0,1/2,1/2)$ | $T_1^S T_2^S(8)$ | OP |
| | | L:$(1/2,v,w)$ | $L_1^S L_2^S(4)$ | QNPL |
| | | W:$(u,v,1/2)$ | $W_1^S W_2^S(4)$ | QNPL |
| 60.72.2.1 | $P_I{}^1b{}^1c{}^1n^{\infty m}1$ | P:$(1/2,v,1/2)$ | $P_1^S P_1^S(8)$ | ONL |
| | | T:$(0,1/2,1/2)$ | $T_1^S T_2^S(8)$ | OP |
| | | L:$(1/2,v,w)$ | $L_1^S L_2^S(4)$ | QNPL |
| | | W:$(u,v,1/2)$ | $W_1^S W_2^S(4)$ | QNPL |
| 61.57.2.1 | $P_a{}^1b{}^1c{}^1a^{\infty m}1$ | E:$(u,1/2,1/2)$ | $E_1^S E_1^S(8)$ | ONL |
| | | P:$(1/2,v,1/2)$ | $P_1^S P_1^S(8)$ | ONL |
| | | Q:$(1/2,1/2,w)$ | $Q_1^S Q_1^S(8)$ | ONL |
| | | L:$(1/2,v,w)$ | $L_1^S L_2^S(4)$ | QNPL |
| | | N:$(u,1/2,w)$ | $N_1^S N_2^S(4)$ | QNPL |
| | | W:$(u,v,1/2)$ | $W_1^S W_2^S(4)$ | QNPL |
| 61.64.2.1 | $P_C{}^1b{}^1c{}^1a^{\infty m}1$ | E:$(u,1/2,1/2)$ | $E_1^S E_1^S(8)$ | ONL |
| | | P:$(1/2,v,1/2)$ | $P_1^S P_1^S(8)$ | ONL |

| | | Q:(1/2,1/2,$w$) | $Q_1^SQ_1^S(8)$ | ONL |
|---|---|---|---|---|
| | | L:(1/2,$v$,$w$) | $L_1^SL_2^S(4)$ | QNPL |
| | | N:($u$,1/2,$w$) | $N_1^SN_2^S(4)$ | QNPL |
| | | W:($u$,$v$,1/2) | $W_1^SW_2^S(4)$ | QNPL |
| 61.73.2.1 | $P_I{}^1b^1c^1a^{\infty m}1$ | E:($u$,1/2,1/2) | $E_1^SE_1^S(8)$ | ONL |
| | | P:(1/2,$v$,1/2) | $P_1^SP_1^S(8)$ | ONL |
| | | Q:(1/2,1/2,$w$) | $Q_1^SQ_1^S(8)$ | ONL |
| | | L:(1/2,$v$,$w$) | $L_1^SL_2^S(4)$ | QNPL |
| | | N:($u$,1/2,$w$) | $N_1^SN_2^S(4)$ | QNPL |
| | | W:($u$,$v$,1/2) | $W_1^SW_2^S(4)$ | QNPL |
| 62.59.2.1 | $P_a{}^1n^1m^1a^{\infty m}1$ | Q:(1/2,1/2,$w$) | $Q_1^SQ_1^S(8)$ | ONL |
| | | L:(1/2,$v$,$w$) | $L_1^SL_2^S(4)$ | QNPL |
| | | N:($u$,1/2,$w$) | $N_1^SN_2^S(4)$ | QNPL |
| | | W:($u$,$v$,1/2) | $W_1^SW_2^S(4)$ | QNPL |
| 62.55.2.1 | $P_b{}^1n^1m^1a^{\infty m}1$ | Q:(1/2,1/2,$w$) | $Q_1^SQ_1^S(8)$ | ONL |
| | | L:(1/2,$v$,$w$) | $L_1^SL_2^S(4)$ | QNPL |
| | | N:($u$,1/2,$w$) | $N_1^SN_2^S(4)$ | QNPL |
| | | W:($u$,$v$,1/2) | $W_1^SW_2^S(4)$ | QNPL |
| 62.57.2.1 | $P_c{}^1n^1m^1a^{\infty m}1$ | Q:(1/2,1/2,$w$) | $Q_1^SQ_1^S(8)$ | ONL |
| | | L:(1/2,$v$,$w$) | $L_1^SL_2^S(4)$ | QNPL |
| | | N:($u$,1/2,$w$) | $N_1^SN_2^S(4)$ | QNPL |
| | | W:($u$,$v$,1/2) | $W_1^SW_2^S(4)$ | QNPL |
| 62.63.2.4 | $P_A{}^1n^1m^1a^{\infty m}1$ | Q:(1/2,1/2,$w$) | $Q_1^SQ_1^S(8)$ | ONL |
| | | L:(1/2,$v$,$w$) | $L_1^SL_2^S(4)$ | QNPL |
| | | N:($u$,1/2,$w$) | $N_1^SN_2^S(4)$ | QNPL |
| | | W:($u$,$v$,1/2) | $W_1^SW_2^S(4)$ | QNPL |
| 62.63.2.1 | $P_B{}^1n^1m^1a^{\infty m}1$ | Q:(1/2,1/2,$w$) | $Q_1^SQ_1^S(8)$ | ONL |
| | | L:(1/2,$v$,$w$) | $L_1^SL_2^S(4)$ | QNPL |
| | | N:($u$,1/2,$w$) | $N_1^SN_2^S(4)$ | QNPL |
| | | W:($u$,$v$,1/2) | $W_1^SW_2^S(4)$ | QNPL |
| 62.64.2.1 | $P_C{}^1n^1m^1a^{\infty m}1$ | Q:(1/2,1/2,$w$) | $Q_1^SQ_1^S(8)$ | ONL |
| | | L:(1/2,$v$,$w$) | $L_1^SL_2^S(4)$ | QNPL |
| | | N:($u$,1/2,$w$) | $N_1^SN_2^S(4)$ | QNPL |
| | | W:($u$,$v$,1/2) | $W_1^SW_2^S(4)$ | QNPL |
| 62.74.2.1 | $P_I{}^1n^1m^1a^{\infty m}1$ | Q:(1/2,1/2,$w$) | $Q_1^SQ_1^S(8)$ | ONL |
| | | L:(1/2,$v$,$w$) | $L_1^SL_2^S(4)$ | QNPL |
| | | N:($u$,1/2,$w$) | $N_1^SN_2^S(4)$ | QNPL |
| | | W:($u$,$v$,1/2) | $W_1^SW_2^S(4)$ | QNPL |
| 63.65.2.1 | $C_c{}^1m^1c^1m^{\infty m}1$ | Q:($u$,$v$,1/2) | $Q_1^SQ_2^S(4)$ | QNPL |
| 63.51.2.1 | $C_a{}^1m^1c^1m^{\infty m}1$ | Q:($u$,$v$,1/2) | $Q_1^SQ_2^S(4)$ | QNPL |
| 63.69.2.1 | $C_A{}^1m^1c^1m^{\infty m}1$ | Q:($u$,$v$,1/2) | $Q_1^SQ_2^S(4)$ | QNPL |
| 64.67.2.1 | $C_c{}^1m^1c^1e^{\infty m}1$ | Q:($u$,$v$,1/2) | $Q_1^SQ_2^S(4)$ | QNPL |
| 64.51.2.1 | $C_a{}^1m^1c^1e^{\infty m}1$ | Q:($u$,$v$,1/2) | $Q_1^SQ_2^S(4)$ | QNPL |

| 64.69.2.1 | $C_A{}^1m{}^1c{}^1e^{\infty m}1$ | Q:($u$,$v$,1/2) | $Q_1^SQ_2^S(4)$ | QNPL |
|---|---|---|---|---|
| 73.67.2.1 | $I_c{}^1b{}^1c{}^1a^{\infty m}1$ | W:(1/2,1/2,1/2) | $W_1^SW_1^S(8)$ | OP |
| 76.77.2.1 | $P_c{}^14_1{}^{\infty m}1$ | E:($u$,$v$,1/2) | $E_1^SE_1^S(4)$ | QNPL |
| 76.76.2.1 | $P_C{}^14_1{}^{\infty m}1$ | E:($u$,$v$,1/2) | $E_1^SE_1^S(4)$ | QNPL |
| 76.80.2.1 | $P_I{}^14_1{}^{\infty m}1$ | E:($u$,$v$,1/2) | $E_1^SE_1^S(4)$ | QNPL |
| 78.77.2.1 | $P_c{}^14_3{}^{\infty m}1$ | E:($u$,$v$,1/2) | $E_1^SE_1^S(4)$ | QNPL |
| 78.78.2.1 | $P_C{}^14_3{}^{\infty m}1$ | E:($u$,$v$,1/2) | $E_1^SE_1^S(4)$ | QNPL |
| 78.80.2.1 | $P_I{}^14_3{}^{\infty m}1$ | E:($u$,$v$,1/2) | $E_1^SE_1^S(4)$ | QNPL |
| 90.90.2.1 | $P_c{}^14{}^12_1{}^12^{\infty m}1$ | F:($u$,1/2,$w$) | $F_1^SF_1^S(4)$ | QNPL |
| 90.89.2.1 | $P_C{}^14{}^12_1{}^12^{\infty m}1$ | F:($u$,1/2,$w$) | $F_1^SF_1^S(4)$ | QNPL |
| 90.97.2.1 | $P_I{}^14{}^12_1{}^12^{\infty m}1$ | F:($u$,1/2,$w$) | $F_1^SF_1^S(4)$ | QNPL |
| 91.93.2.1 | $P_c{}^14_1{}^12{}^12^{\infty m}1$ | E:($u$,$v$,1/2) | $E_1^SE_1^S(4)$ | QNPL |
| 91.91.2.1 | $P_C{}^14_1{}^12{}^12^{\infty m}1$ | E:($u$,$v$,1/2) | $E_1^SE_1^S(4)$ | QNPL |
| 91.98.2.1 | $P_I{}^14_1{}^12{}^12^{\infty m}1$ | E:($u$,$v$,1/2) | $E_1^SE_1^S(4)$ | QNPL |
| 92.94.2.1 | $P_c{}^14_1{}^12_1{}^12^{\infty m}1$ | A:(1/2,1/2,1/2) | $A_1^SA_2^S(8)$ | $C-4$ OP |
| | | E:($u$,$v$,1/2) | $E_1^SE_1^S(4)$ | QNPL |
| | | F:($u$,1/2,$w$) | $F_1^SF_1^S(4)$ | QNPL |
| 92.91.2.1 | $P_C{}^14_1{}^12_1{}^12^{\infty m}1$ | A:(1/2,1/2,1/2) | $A_1^SA_2^S(8)$ | $C-4$ OP |
| | | E:($u$,$v$,1/2) | $E_1^SE_1^S(4)$ | QNPL |
| | | F:($u$,1/2,$w$) | $F_1^SF_1^S(4)$ | QNPL |
| 92.98.2.1 | $P_I{}^14_1{}^12_1{}^12^{\infty m}1$ | A:(1/2,1/2,1/2) | $A_1^SA_2^S(8)$ | $C-4$ OP |
| | | E:($u$,$v$,1/2) | $E_1^SE_1^S(4)$ | QNPL |
| | | F:($u$,1/2,$w$) | $F_1^SF_1^S(4)$ | QNPL |
| 94.90.2.1 | $P_c{}^14_2{}^12_1{}^12^{\infty m}1$ | F:($u$,1/2,$w$) | $F_1^SF_1^S(4)$ | QNPL |
| 94.93.2.1 | $P_C{}^14_2{}^12_1{}^12^{\infty m}1$ | F:($u$,1/2,$w$) | $F_1^SF_1^S(4)$ | QNPL |
| 94.97.2.1 | $P_I{}^14_2{}^12_1{}^12^{\infty m}1$ | F:($u$,1/2,$w$) | $F_1^SF_1^S(4)$ | QNPL |
| 95.93.2.1 | $P_c{}^14_3{}^12{}^12^{\infty m}1$ | E:($u$,$v$,1/2) | $E_1^SE_1^S(4)$ | QNPL |
| 95.95.2.1 | $P_C{}^14_3{}^12{}^12^{\infty m}1$ | E:($u$,$v$,1/2) | $E_1^SE_1^S(4)$ | QNPL |
| 95.98.2.1 | $P_I{}^14_3{}^12{}^12^{\infty m}1$ | E:($u$,$v$,1/2) | $E_1^SE_1^S(4)$ | QNPL |
| 96.94.2.1 | $P_c{}^14_3{}^12_1{}^12^{\infty m}1$ | A:(1/2,1/2,1/2) | $A_1^SA_2^S(8)$ | $C-4$ OP |
| | | E:($u$,$v$,1/2) | $E_1^SE_1^S(4)$ | QNPL |
| | | F:($u$,1/2,$w$) | $F_1^SF_1^S(4)$ | QNPL |
| 96.95.2.1 | $P_C{}^14_3{}^12_1{}^12^{\infty m}1$ | A:(1/2,1/2,1/2) | $A_1^SA_2^S(8)$ | $C-4$ OP |
| | | E:($u$,$v$,1/2) | $E_1^SE_1^S(4)$ | QNPL |
| | | F:($u$,1/2,$w$) | $F_1^SF_1^S(4)$ | QNPL |
| 96.98.2.1 | $P_I{}^14_3{}^12_1{}^12^{\infty m}1$ | A:(1/2,1/2,1/2) | $A_1^SA_2^S(8)$ | $C-4$ OP |
| | | E:($u$,$v$,1/2) | $E_1^SE_1^S(4)$ | QNPL |
| | | F:($u$,1/2,$w$) | $F_1^SF_1^S(4)$ | QNPL |
| 103.99.2.1 | $P_c{}^14{}^1c{}^1c^{\infty m}1$ | A:(1/2,1/2,1/2) | $A_5^SA_5^S(8)$ | OP |
| | | Z:(0,0,1/2) | $Z_5^SZ_5^S(8)$ | OP |
| 103.103.2.1 | $P_C{}^14{}^1c{}^1c^{\infty m}1$ | A:(1/2,1/2,1/2) | $A_5^SA_5^S(8)$ | OP |
| | | Z:(0,0,1/2) | $Z_5^SZ_5^S(8)$ | OP |
| 103.108.2.1 | $P_I{}^14{}^1c{}^1c^{\infty m}1$ | A:(1/2,1/2,1/2) | $A_5^SA_5^S(8)$ | OP |

| | | Z:(0,0,1/2) | $Z_5^SZ_5^S(8)$ | OP |
|---|---|---|---|---|
| 104.100.2.1 | $P_c{}^14{}^1n{}^1c^{\infty m}1$ | Z:(0,0,1/2) | $Z_5^SZ_5^S(8)$ | OP |
| 104.103.2.1 | $P_C{}^14{}^1n{}^1c^{\infty m}1$ | Z:(0,0,1/2) | $Z_5^SZ_5^S(8)$ | OP |
| 104.107.2.1 | $P_I{}^14{}^1\mathrm{n}{}^1c^{\infty m}1$ | Z:(0,0,1/2) | $Z_5^SZ_5^S(8)$ | OP |
| 106.100.2.1 | $P_c{}^14_2{}^1b{}^1c^{\infty m}1$ | A:(1/2,1/2,1/2) | $A_5^SA_5^S(8)$ | OP |
| 106.101.2.1 | $P_C{}^14_2{}^1b{}^1c^{\infty m}1$ | A:(1/2,1/2,1/2) | $A_5^SA_5^S(8)$ | OP |
| 106.108.2.1 | $P_I{}^14_2{}^1b{}^1c^{\infty m}1$ | A:(1/2,1/2,1/2) | $A_5^SA_5^S(8)$ | OP |
| 110.102.2.1 | $I_c{}^14_1{}^1c{}^1d^{\infty m}1$ | P:(1/2,1/2,1/2) | $P_1^SP_1^S(8)$ | OP |
| 113.113.2.1 | $P_c{}^1\bar{4}{}^12_1{}^1m^{\infty m}1$ | F:($u$,1/2,$w$) | $F_1^SF_1^S(4)$ | QNPL |
| 113.115.2.1 | $P_C{}^1\bar{4}{}^12_1{}^1m^{\infty m}1$ | F:($u$,1/2,$w$) | $F_1^SF_1^S(4)$ | QNPL |
| 113.121.2.1 | $P_I{}^1\bar{4}{}^12_1{}^1m^{\infty m}1$ | F:($u$,1/2,$w$) | $F_1^SF_1^S(4)$ | QNPL |
| 114.113.2.1 | $P_c{}^1\bar{4}{}^12_1{}^1c^{\infty m}1$ | A:(1/2,1/2,1/2) | $A_5^SA_5^S(8)$ | OP |
| | | F:($u$,1/2,$w$) | $F_1^SF_1^S(4)$ | QNPL |
| 114.116.2.1 | $P_C{}^1\bar{4}{}^12_1{}^1c^{\infty m}1$ | A:(1/2,1/2,1/2) | $A_5^SA_5^S(8)$ | OP |
| | | F:($u$,1/2,$w$) | $F_1^SF_1^S(4)$ | QNPL |
| 114.121.2.1 | $P_I{}^1\bar{4}{}^12_1{}^1c^{\infty m}1$ | A:(1/2,1/2,1/2) | $A_5^SA_5^S(8)$ | OP |
| | | F:($u$,1/2,$w$) | $F_1^SF_1^S(4)$ | QNPL |
| 124.123.2.1 | $P_c{}^14/{}^1m{}^1c{}^1c^{\infty m}1$ | A:(1/2,1/2,1/2) | $A_3^SA_4^S(8)$ | OP |
| | | Z:(0,0,1/2) | $Z_3^SZ_4^S(8)$ | OP |
| 124.124.2.1 | $P_C{}^14/{}^1m{}^1c{}^1c^{\infty m}1$ | A:(1/2,1/2,1/2) | $A_3^SA_4^S(8)$ | OP |
| | | Z:(0,0,1/2) | $Z_3^SZ_4^S(8)$ | OP |
| 124.140.2.1 | $P_I{}^14/{}^1m{}^1c{}^1c^{\infty m}1$ | A:(1/2,1/2,1/2) | $A_3^SA_4^S(8)$ | OP |
| | | Z:(0,0,1/2) | $Z_3^SZ_4^S(8)$ | OP |
| 126.125.2.1 | $P_c{}^14/{}^1n{}^1n{}^1c^{\infty m}1$ | Z:(0,0,1/2) | $Z_3^SZ_4^S(8)$ | OP |
| 126.124.2.1 | $P_C{}^14/{}^1n{}^1n{}^1c^{\infty m}1$ | Z:(0,0,1/2) | $Z_3^SZ_4^S(8)$ | OP |
| 126.139.2.1 | $P_I{}^14/{}^1n{}^1n{}^1c^{\infty m}1$ | Z:(0,0,1/2) | $Z_3^SZ_4^S(8)$ | OP |
| 127.127.2.1 | $P_c{}^14/{}^1m{}^1b{}^1m^{\infty m}1$ | F:($u$,1/2,$w$) | $F_1^SF_2^S(4)$ | QNPL |
| 127.123.2.1 | $P_C{}^14/{}^1m{}^1b{}^1m^{\infty m}1$ | F:($u$,1/2,$w$) | $F_1^SF_2^S(4)$ | QNPL |
| 127.140.2.1 | $P_I{}^14/{}^1m{}^1b{}^1m^{\infty m}1$ | F:($u$,1/2,$w$) | $F_1^SF_2^S(4)$ | QNPL |
| 128.127.2.1 | $P_c{}^14/{}^1m{}^1n{}^1c^{\infty m}1$ | Z:(0,0,1/2) | $Z_3^SZ_4^S(8)$ | OP |
| | | A:(1/2,1/2,1/2) | $A_3^SA_4^S(8)$ | OP |
| | | F:($u$,1/2,$w$) | $F_1^SF_2^S(4)$ | QNPL |
| 128.124.2.1 | $P_C{}^14/{}^1m{}^1n{}^1c^{\infty m}1$ | Z:(0,0,1/2) | $Z_3^SZ_4^S(8)$ | OP |
| | | A:(1/2,1/2,1/2) | $A_3^SA_4^S(8)$ | OP |
| | | F:($u$,1/2,$w$) | $F_1^SF_2^S(4)$ | QNPL |
| 128.139.2.1 | $P_I{}^14/{}^1m{}^1n{}^1c^{\infty m}1$ | Z:(0,0,1/2) | $Z_3^SZ_4^S(8)$ | OP |
| | | A:(1/2,1/2,1/2) | $A_3^SA_4^S(8)$ | OP |
| | | F:($u$,1/2,$w$) | $F_1^SF_2^S(4)$ | QNPL |
| 129.129.2.1 | $P_c{}^14/{}^1n{}^1m{}^1m^{\infty m}1$ | F:($u$,1/2,$w$) | $F_1^SF_2^S(4)$ | QNPL |
| 129.123.2.1 | $P_C{}^14/{}^1n{}^1m{}^1m^{\infty m}1$ | F:($u$,1/2,$w$) | $F_1^SF_2^S(4)$ | QNPL |
| 129.139.2.1 | $P_I{}^14/{}^1n{}^1m{}^1m^{\infty m}1$ | F:($u$,1/2,$w$) | $F_1^SF_2^S(4)$ | QNPL |
| 130.129.2.1 | $P_c{}^14/{}^1n{}^1c{}^1c^{\infty m}1$ | Z:(0,0,1/2) | $Z_3^SZ_4^S(8)$ | OP |
| | | A:(1/2,1/2,1/2) | $A_1^SA_2^S(8)$, $A_3^SA_4^S(8)$ | OP |
| | | R:(0,1/2,1/2) | $R_1^S R_2^S(8)$ | OP |
| | | F:($u$,1/2,$w$) | $F_1^S F_2^S(4)$ | QNPL |
| 130.124.2.1 | $P_C{}^14/{}^1n^1c^1c^{\infty m}1$ | Z:(0,0,1/2) | $Z_3^S Z_4^S(8)$ | OP |
| | | A:(1/2,1/2,1/2) | $A_1^S A_2^S(8)$, $A_3^S A_4^S(8)$ | OP |
| | | R:(0,1/2,1/2) | $R_1^S R_2^S(8)$ | OP |
| | | F:($u$,1/2,$w$) | $F_1^S F_2^S(4)$ | QNPL |
| 130.140.2.1 | $P_I{}^14/{}^1n^1c^1c^{\infty m}1$ | Z:(0,0,1/2) | $Z_3^S Z_4^S(8)$ | OP |
| | | A:(1/2,1/2,1/2) | $A_1^S A_2^S(8)$, $A_3^S A_4^S(8)$ | OP |
| | | R:(0,1/2,1/2) | $R_1^S R_2^S(8)$ | OP |
| | | F:($u$,1/2,$w$) | $F_1^S F_2^S(4)$ | QNPL |
| 133.125.2.1 | $P_c{}^14_2/{}^1n^1b^1c^{\infty m}1$ | A:(1/2,1/2,1/2) | $A_3^S A_4^S(8)$ | OP |
| 133.132.2.1 | $P_C{}^14_2/{}^1n^1b^1c^{\infty m}1$ | A:(1/2,1/2,1/2) | $A_3^S A_4^S(8)$ | OP |
| 133.140.2.1 | $P_I{}^14_2/{}^1n^1b^1c^{\infty m}1$ | A:(1/2,1/2,1/2) | $A_3^S A_4^S(8)$ | OP |
| 135.127.2.1 | $P_c{}^14_2/{}^1m^1b^1c^{\infty m}1$ | A:(1/2,1/2,1/2) | $A_1^S A_2^S(8)$, $A_3^S A_4^S(8)$ | OP |
| | | F:($u$,1/2,$w$) | $F_1^S F_2^S(4)$ | QNPL |
| 135.132.2.1 | $P_C{}^14_2/{}^1m^1b^1c^{\infty m}1$ | A:(1/2,1/2,1/2) | $A_1^S A_2^S(8)$, $A_3^S A_4^S(8)$ | OP |
| | | F:($u$,1/2,$w$) | $F_1^S F_2^S(4)$ | QNPL |
| 135.140.2.1 | $P_I{}^14_2/{}^1m^1b^1c^{\infty m}1$ | A:(1/2,1/2,1/2) | $A_1^S A_2^S(8)$, $A_3^S A_4^S(8)$ | OP |
| | | F:($u$,1/2,$w$) | $F_1^S F_2^S(4)$ | QNPL |
| 136.127.2.1 | $P_c{}^14_2/{}^1m^1n^1m^{\infty m}1$ | F:($u$,1/2,$w$) | $F_1^S F_2^S(4)$ | QNPL |
| 136.131.2.1 | $P_C{}^14_2/{}^1m^1n^1m^{\infty m}1$ | F:($u$,1/2,$w$) | $F_1^S F_2^S(4)$ | QNPL |
| 136.139.2.1 | $P_I{}^14_2/{}^1m^1n^1m^{\infty m}1$ | F:($u$,1/2,$w$) | $F_1^S F_2^S(4)$ | QNPL |
| 137.129.2.1 | $P_c{}^14_2/{}^1n^1m^1c^{\infty m}1$ | A:(1/2,1/2,1/2) | $A_3^S A_4^S(8)$ | OP |
| | | F:($u$,1/2,$w$) | $F_1^S F_2^S(4)$ | QNPL |
| 137.132.2.1 | $P_C{}^14_2/{}^1n^1m^1c^{\infty m}1$ | A:(1/2,1/2,1/2) | $A_3^S A_4^S(8)$ | OP |
| | | F:($u$,1/2,$w$) | $F_1^S F_2^S(4)$ | QNPL |
| 137.139.2.1 | $P_I{}^14_2/{}^1n^1m^1c^{\infty m}1$ | A:(1/2,1/2,1/2) | $A_3^S A_4^S(8)$ | OP |
| | | F:($u$,1/2,$w$) | $F_1^S F_2^S(4)$ | QNPL |
| 138.129.2.1 | $P_c{}^14_2/{}^1n^1c^1m^{\infty m}1$ | R:(0,1/2,1/2) | $R_1^S R_2^S(8)$ | OP |
| | | F:($u$,1/2,$w$) | $F_1^S F_2^S(4)$ | QNPL |
| 138.131.2.1 | $P_C{}^14_2/{}^1n^1c^1m^{\infty m}1$ | R:(0,1/2,1/2) | $R_1^S R_2^S(8)$ | OP |
| | | F:($u$,1/2,$w$) | $F_1^S F_2^S(4)$ | QNPL |
| 138.140.2.1 | $P_I{}^14_2/{}^1n^1c^1m^{\infty m}1$ | R:(0,1/2,1/2) | $R_1^S R_2^S(8)$ | OP |
| | | F:($u$,1/2,$w$) | $F_1^S F_2^S(4)$ | QNPL |
| 142.134.2.1 | $I_c{}^14_1/{}^1a^1c^1d^{\infty m}1$ | P:(1/2,1/2,1/2) | $P_1^S P_2^S(8)$ | OP |
| 158.156.2.1 | $P_c{}^13^1c^11^{\infty m}1$ | A:(0,0,1/2) | $A_3^S A_3^S(8)$ | OP |
| 159.157.2.1 | $P_c{}^13^11^1c^{\infty m}1$ | A:(0,0,1/2) | $A_3^S A_3^S(8)$ | OP |
| 161.160.2.1 | $R_I{}^13^1c^{\infty m}1$ | T:(0,0,3/2) | $T_3^S T_3^S(8)$ | OP |
| 163.162.2.1 | $P_c{}^1\bar{3}^11^1c^{\infty m}1$ | A:(0,0,1/2) | $A_1^S A_2^S(8)$ | OP |
| 165.164.2.1 | $P_c{}^1\bar{3}^1c^11^{\infty m}1$ | A:(0,0,1/2) | $A_1^S A_2^S(8)$ | OP |
| | | H:(1/3,1/3,1/2) | $H_3^S H_3^S(8)$ | OP |
| 167.166.2.1 | $R_I{}^1\bar{3}^1c^{\infty m}1$ | T:(0,0,3/2) | $T_1^S T_2^S(8)$ | OP |
| 169.171.2.1 | $P_c{}^16_1{}^{\infty m}1$ | E:($u$,$v$,1/2) | $E_1^S E_1^S(4)$ | QNPL |

| 170.172.2.1 | $P_c{}^{1}6_5{}^{\infty m}1$ | E:($u$,$v$,1/2) | $E_1^S E_1^S(4)$ | QNPL |
|---|---|---|---|---|
| 173.168.2.1 | $P_c{}^{1}6_3{}^{\infty m}1$ | E:($u$,$v$,1/2) | $E_1^S E_1^S(4)$ | QNPL |
| 176.175.2.1 | $P_c{}^{1}6_3/{}^{1}m{}^{\infty m}1$ | A:(0,0,1/2) | $A_2^S A_3^S(8)$ | OP |
| | | E:($u$,$v$,1/2) | $E_1^S E_1^S(4)$ | QNPL |
| 178.180.2.1 | $P_c{}^{1}6_1{}^{1}2{}^{1}2{}^{\infty m}1$ | E:($u$,$v$,1/2) | $E_1^S E_1^S(4)$ | QNPL |
| 179.181.2.1 | $P_c{}^{1}6_5{}^{1}2{}^{1}2{}^{\infty m}1$ | E:($u$,$v$,1/2) | $E_1^S E_1^S(4)$ | QNPL |
| 182.177.2.1 | $P_c{}^{1}6_3{}^{1}2{}^{1}2{}^{\infty m}1$ | E:($u$,$v$,1/2) | $E_1^S E_1^S(4)$ | QNPL |
| 184.183.2.1 | $P_c{}^{1}6{}^{1}c{}^{1}c{}^{\infty m}1$ | A:(0,0,1/2) | $A_5^S A_5^S(8)$, $A_6^S A_6^S(8)$ | OP |
| | | H:(1/3,1/3,1/2) | $H_3^S H_3^S(8)$ | OP |
| 185.183.2.1 | $P_c{}^{1}6_3{}^{1}c{}^{1}m{}^{\infty m}1$ | A:(0,0,1/2) | $A_5^S A_6^S(8)$ | OP |
| | | H:(1/3,1/3,1/2) | $H_3^S H_3^S(8)$ | OP |
| | | E:($u$,$v$,1/2) | $E_1^S E_1^S(4)$ | QNPL |
| 186.183.2.1 | $P_c{}^{1}6_3{}^{1}m{}^{1}c{}^{\infty m}1$ | A:(0,0,1/2) | $A_5^S A_6^S(8)$ | OP |
| | | E:($u$,$v$,1/2) | $E_1^S E_1^S(4)$ | QNPL |
| 188.187.2.1 | $P_c{}^{1}\bar{6}{}^{1}c{}^{1}2{}^{\infty m}1$ | A:(0,0,1/2) | $A_2^S A_3^S(8)$ | OP |
| 190.189.2.1 | $P_c{}^{1}\bar{6}{}^{1}2{}^{1}c{}^{\infty m}1$ | A:(0,0,1/2) | $A_1^S A_2^S(8)$ | OP |
| 192.191.2.1 | $P_c{}^{1}6/{}^{1}m{}^{1}c{}^{1}c{}^{\infty m}1$ | A:(0,0,1/2) | $A_1^S A_2^S(8)$, $A_3^S A_4^S(8)$ | OP |
| | | H:(1/3,1/3,1/2) | $H_1^S H_2^S(8)$ | OP |
| 193.191.2.1 | $P_c{}^{1}6_3/{}^{1}m{}^{1}c{}^{1}m{}^{\infty m}1$ | A:(0,0,1/2) | $A_3^S(8)$ | OP |
| | | H:(1/3,1/3,1/2) | $H_5^S H_6^S(8)$ | OP |
| | | E:($u$,$v$,1/2) | $E_1^S E_2^S(4)$ | QNPL |
| 194.191.2.1 | $P_c{}^{1}6_3/{}^{1}m{}^{1}m{}^{1}c{}^{\infty m}1$ | A:(0,0,1/2) | $A_3^S(8)$ | OP |
| | | E:($u$,$v$,1/2) | $E_1^S E_2^S(4)$ | QNPL |
| 195.197.2.1 | $P_I{}^{1}2{}^{1}3{}^{\infty m}1$ | Γ:(0,0,0) | $\Gamma_4^S(6)$ | $C$−4 SP |
| | | Γ:(0,0,0) | $\Gamma_2^S \Gamma_3^S(4)$ | $C$−8 DP |
| | | R:(1/2,1/2,1/2) | $R_4^S(6)$ | $C$−4 SP |
| | | R:(1/2,1/2,1/2) | $R_2^S R_3^S(4)$ | $C$−8 DP |
| 196.195.2.1 | $F_s{}^{1}2{}^{1}3{}^{\infty m}1$ | Γ:(0,0,0) | $\Gamma_4^S(6)$ | $C$−4 SP |
| | | Γ:(0,0,0) | $\Gamma_2^S \Gamma_3^S(4)$ | $C$−8 DP |
| 198.199.2.1 | $P_I{}^{1}2_1{}^{1}3{}^{\infty m}1$ | R:(1/2,1/2,1/2) | $R_1^S R_3^S(8)$, $R_2^S R_2^S(8)$ | $C$−4 OP |
| | | Γ:(0,0,0) | $\Gamma_4^S(6)$ | $C$−4 SP |
| | | Γ:(0,0,0) | $\Gamma_2^S \Gamma_3^S(4)$ | $C$−8 DP |
| | | B:($u$,1/2,$w$) | $B_1^S B_1^S(4)$ | QNPL |
| 200.204.2.1 | $P_I{}^{1}m{}^{1}\bar{3}{}^{\infty m}1$ | Γ:(0,0,0) | $\Gamma_4^{S,+}(6)$, $\Gamma_4^{S,-}(6)$ | SP |
| | | R:(1/2,1/2,1/2) | $R_4^{S,+}(6)$, $R_4^{S,-}(6)$ | SP |
| 201.204.2.1 | $P_I{}^{1}n{}^{1}\bar{3}{}^{\infty m}1$ | Γ:(0,0,0) | $\Gamma_4^{S,+}(6)$, $\Gamma_4^{S,-}(6)$ | SP |
| | | R:(1/2,1/2,1/2) | $R_4^{S,+}(6)$, $R_4^{S,-}(6)$ | SP |
| 202.200.2.1 | $F_s{}^{1}m{}^{1}\bar{3}{}^{\infty m}1$ | Γ:(0,0,0) | $\Gamma_4^{S,+}(6)$, $\Gamma_4^{S,-}(6)$ | SP |

| 203.201.2.1 | $F_s{}^1d^1\bar{3}^{\infty m}1$ | Γ:(0,0,0) | $\Gamma_4^{S,+}(6),\ \Gamma_4^{S,-}(6)$ | SP |
|---|---|---|---|---|
| 205.206.2.1 | $P_I{}^1a^1\bar{3}^{\infty m}1$ | Γ:(0,0,0) | $\Gamma_4^{S,+}(6),\ \Gamma_4^{S,-}(6)$ | SP |
| | | T:(1/2,1/2,$w$) | $T_1^S T_1^S(8)$ | ONL |
| | | B:($u$,1/2,$w$) | $B_1^S B_2^S(4)$ | QNPL |
| 207.211.2.1 | $P_I{}^14^13^12^{\infty m}1$ | Γ:(0,0,0) | $\Gamma_4^S(6),\ \Gamma_5^S(6)$ | $C$−4 SP |
| | | Γ:(0,0,0) | $\Gamma_3^S(4)$ | $C$−8 DP |
| | | R:(1/2,1/2,1/2) | $R_4^S(6),\ R_5^S(6)$ | $C$−4 SP |
| | | R:(1/2,1/2,1/2) | $R_3^S(4)$ | $C$−8 DP |
| 208.211.2.1 | $P_I{}^14_2{}^13^12^{\infty m}1$ | Γ:(0,0,0) | $\Gamma_4^S(6),\ \Gamma_5^S(6)$ | $C$−4 SP |
| | | Γ:(0,0,0) | $\Gamma_3^S(4)$ | $C$−8 DP |
| | | R:(1/2,1/2,1/2) | $R_4^S(6),\ R_5^S(6)$ | $C$−4 SP |
| | | R:(1/2,1/2,1/2) | $R_3^S(4)$ | $C$−8 DP |
| 209.207.2.1 | $F_s{}^14^13^12^{\infty m}1$ | Γ:(0,0,0) | $\Gamma_4^S(6),\ \Gamma_5^S(6)$ | $C$−4 SP |
| | | Γ:(0,0,0) | $\Gamma_3^S(4)$ | $C$−8 DP |
| 210.208.2.1 | $F_s{}^14_1{}^13^12^{\infty m}1$ | Γ:(0,0,0) | $\Gamma_4^S(6),\ \Gamma_5^S(6)$ | $C$−4 SP |
| | | Γ:(0,0,0) | $\Gamma_3^S(4)$ | $C$−8 DP |
| 212.214.2.1 | $P_I{}^14_3{}^13^12^{\infty m}1$ | R:(1/2,1/2,1/2) | $R_1^S R_2^S(8),\ R_3^S(8)$ | $C$−4 OP |
| | | Γ:(0,0,0) | $\Gamma_4^S(6),\ \Gamma_5^S(6)$ | $C$−4 SP |
| | | Γ:(0,0,0) | $\Gamma_3^S(4)$ | $C$−8 DP |
| | | B:($u$,1/2,$w$) | $B_1^S B_1^S(4)$ | QNPL |
| 213.214.2.1 | $P_I{}^14_1{}^13^12^{\infty m}1$ | R:(1/2,1/2,1/2) | $R_1^S R_2^S(8),\ R_3^S(8)$ | $C$−4 OP |
| | | Γ:(0,0,0) | $\Gamma_4^S(6),\ \Gamma_5^S(6)$ | $C$−4 SP |
| | | Γ:(0,0,0) | $\Gamma_3^S(4)$ | $C$−8 DP |
| | | B:($u$,1/2,$w$) | $B_1^S B_1^S(4)$ | QNPL |
| 215.217.2.1 | $P_I{}^1\bar{4}^13^1m^{\infty m}1$ | Γ:(0,0,0) | $\Gamma_4^S(6),\ \Gamma_5^S(6)$ | SP |
| | | R:(1/2,1/2,1/2) | $R_4^S(6),\ R_5^S(6)$ | SP |
| 216.215.2.1 | $F_s{}^1\bar{4}^13^1m^{\infty m}1$ | Γ:(0,0,0) | $\Gamma_4^S(6),\ \Gamma_5^S(6)$ | SP |
| 218.217.2.1 | $P_I{}^1\bar{4}^13^1n^{\infty m}1$ | R:(1/2,1/2,1/2) | $R_4^S R_5^S(12)$ | DCP |
| | | R:(1/2,1/2,1/2) | $R_3^S R_3^S(8)$ | OP |
| | | Γ:(0,0,0) | $\Gamma_4^S(6),\ \Gamma_5^S(6)$ | SP |
| 219.215.2.1 | $F_s{}^1\bar{4}^13^1c^{\infty m}1$ | L:(1/2,1/2,1/2) | $L_3^S L_3^S(8)$ | OP |
| | | Γ:(0,0,0) | $\Gamma_4^S(6),\ \Gamma_5^S(6)$ | SP |
| 221.229.2.1 | $P_I{}^1m^1\bar{3}^1m^{\infty m}1$ | Γ:(0,0,0) | $\Gamma_4^{S,+}(6),\ \Gamma_4^{S,-}(6),$<br>$\Gamma_5^{S,+}(6),\ \Gamma_5^{S,-}(6)$ | SP |
| | | R:(1/2,1/2,1/2) | $R_4^{S,+}(6),\ R_4^{S,-}(6),$<br>$R_5^{S,+}(6),\ R_5^{S,-}(6)$ | SP |
| 222.229.2.1 | $P_I{}^1n^1\bar{3}^1n^{\infty m}1$ | R:(1/2,1/2,1/2) | $R_4^S(12)$ | DCP |
| | | X:(0,1/2,0) | $X_3^S X_4^S(8)$ | OP |
| | | R:(1/2,1/2,1/2) | $R_2^S R_3^S(8)$ | OP |
| | | Γ:(0,0,0) | $\Gamma_4^{S,+}(6)$, $\Gamma_4^{S,-}(6)$, $\Gamma_5^{S,+}(6)$, $\Gamma_5^{S,-}(6)$ | SP |
| 223.229.2.1 | $P_I{}^1m{}^1\bar{3}{}^1n{}^{\infty m}1$ | R:(1/2,1/2,1/2) | $R_4^S(12)$ | DCP |
| | | R:(1/2,1/2,1/2) | $R_2^S R_3^S(8)$ | OP |
| | | Γ:(0,0,0) | $\Gamma_4^{S,+}(6)$, $\Gamma_4^{S,-}(6)$, $\Gamma_5^{S,+}(6)$, $\Gamma_5^{S,-}(6)$ | SP |
| 224.229.2.1 | $P_I{}^1n{}^1\bar{3}{}^1m{}^{\infty m}1$ | Γ:(0,0,0) | $\Gamma_4^{S,+}(6)$, $\Gamma_4^{S,-}(6)$, $\Gamma_5^{S,+}(6)$, $\Gamma_5^{S,-}(6)$ | SP |
| | | R:(1/2,1/2,1/2) | $R_4^{S,+}(6)$, $R_4^{S,-}(6)$, $R_5^{S,+}(6)$, $R_5^{S,-}(6)$ | SP |
| 225.221.2.1 | $F_S{}^1m{}^1\bar{3}{}^1m{}^{\infty m}1$ | Γ:(0,0,0) | $\Gamma_4^{S,+}(6)$, $\Gamma_4^{S,-}(6)$, $\Gamma_5^{S,+}(6)$, $\Gamma_5^{S,-}(6)$ | SP |
| 226.221.2.1 | $F_S{}^1m{}^1\bar{3}{}^1c{}^{\infty m}1$ | L:(1/2,1/2,1/2) | $L_1^S L_2^S(8)$ | OP |
| | | Γ:(0,0,0) | $\Gamma_4^{S,+}(6)$, $\Gamma_4^{S,-}(6)$, $\Gamma_5^{S,+}(6)$, $\Gamma_5^{S,-}(6)$ | SP |
| 227.224.2.1 | $F_S{}^1d{}^1\bar{3}{}^1m{}^{\infty m}1$ | Γ:(0,0,0) | $\Gamma_4^{S,+}(6)$, $\Gamma_4^{S,-}(6)$, $\Gamma_5^{S,+}(6)$, $\Gamma_5^{S,-}(6)$ | SP |
| 228.224.2.1 | $F_S{}^1d{}^1\bar{3}{}^1c{}^{\infty m}1$ | L:(1/2,1/2,1/2) | $L_1^S L_2^S(8)$ | OP |
| | | W:(1/2,1,0) | $W_1^S W_2^S(8)$ | OP |
| | | Γ:(0,0,0) | $\Gamma_4^{S,+}(6)$, $\Gamma_4^{S,-}(6)$, $\Gamma_5^{S,+}(6)$, $\Gamma_5^{S,-}(6)$ | SP |

## S3.6. Chiral magnon (momentum-position)

Supplementary Table XXIII. The momentum-position of type-III collinear SSGs that accommodate

unconventional magnons with spin non-degeneracy (chiral magnons). The three columns illustrate the group number, the international notation of collinear SSGs, the momentum-position that allows magnonic chirality splitting (electronic spin splitting also). For detailed information about the *k*-point and the spin Brillouin zones for collinear SSGs, please refer to Supplementary Table VIII in Supplementary Information S1.

| $G_{NS}$ | $G_S$ | ***k*-point** |
|---|---|---|
| 1.3.1.1 | $P^{\bar{1}}2^{\infty m}1$ | GP:$(u,v,w)$ |
| 1.4.1.1 | $P^{\bar{1}}2_1{}^{\infty m}1$ | GP:$(u,v,w)$ |
| 1.5.1.1 | $C^{\bar{1}}2^{\infty m}1$ | GP:$(u,v,w)$ |
| 1.6.1.1 | $P^{\bar{1}}m^{\infty m}1$ | GP:$(u,v,w)$ |
| 1.7.1.1 | $P^{\bar{1}}c^{\infty m}1$ | GP:$(u,v,w)$ |
| 1.8.1.1 | $C^{\bar{1}}m^{\infty m}1$ | GP:$(u,v,w)$ |
| 1.9.1.1 | $C^{\bar{1}}c^{\infty m}1$ | GP:$(u,v,w)$ |
| 2.10.1.1 | $P^{\bar{1}}2/^{\bar{1}}m^{\infty m}1$ | GP:$(u,v,w)$ |
| 2.11.1.1 | $P^{\bar{1}}2_1/^{\bar{1}}m^{\infty m}1$ | GP:$(u,v,w)$ |
| 2.12.1.1 | $C^{\bar{1}}2/^{\bar{1}}m^{\infty m}1$ | L:(1/2,1/2,1/2) V:(1/2,1/2,0) GP:$(u,v,w)$ |
| 2.13.1.1 | $P^{\bar{1}}2/^{\bar{1}}c^{\infty m}1$ | GP:$(u,v,w)$ |
| 2.14.1.1 | $P^{\bar{1}}2_1/^{\bar{1}}c^{\infty m}1$ | GP:$(u,v,w)$ |
| 2.15.1.1 | $C^{\bar{1}}2/^{\bar{1}}c^{\infty m}1$ | L:(1/2,1/2,1/2) V:(1/2,1/2,0) GP:$(u,v,w)$ |
| 3.16.1.1 | $P^{\bar{1}}2^{\bar{1}}2^{1}2^{\infty m}1$ | V:$(u,v,0)$ W:$(u,v,1/2)$ GP:$(u,v,w)$ |
| 4.17.1.1 | $P^{\bar{1}}2^{\bar{1}}2^{1}2_1{}^{\infty m}1$ | V:$(u,v,0)$ W:$(u,v,1/2)$ GP:$(u,v,w)$ |
| 3.17.1.1 | $P^{1}2^{\bar{1}}2^{\bar{1}}2_1{}^{\infty m}1$ | K:$(0,v,w)$ L:$(1/2,v,w)$ GP:$(u,v,w)$ |
| 3.18.1.1 | $P^{\bar{1}}2_1{}^{\bar{1}}2_1{}^{1}2^{\infty m}1$ | V:$(u,v,0)$ W:$(u,v,1/2)$ GP:$(u,v,w)$ |
| 4.18.1.1 | $P^{1}2_1{}^{\bar{1}}2_1{}^{\bar{1}}2^{\infty m}1$ | K:$(0,v,w)$ L:$(1/2,v,w)$ GP:$(u,v,w)$ |
| 4.19.1.1 | $P^{\bar{1}}2_1{}^{\bar{1}}2_1{}^{1}2_1{}^{\infty m}1$ | V:$(u,v,0)$ W:$(u,v,1/2)$ GP:$(u,v,w)$ |
| 4.20.1.1 | $C^{\bar{1}}2^{\bar{1}}2^{1}2_1{}^{\infty m}1$ | R:(1/2,1/2,1/2) S:(1/2,1/2,0) D:$(1/2,1/2,w)$ P:$(u,v,0)$ Q:$(u,v,1/2)$ GP:$(u,v,w)$ |
| 5.20.1.1 | $C^{1}2^{\bar{1}}2^{\bar{1}}2_1{}^{\infty m}1$ | K:$(0,v,w)$ GP:$(u,v,w)$ |
| 3.21.1.1 | $C^{\bar{1}}2^{\bar{1}}2^{1}2^{\infty m}1$ | R:(1/2,1/2,1/2) S:(1/2,1/2,0) D:$(1/2,1/2,w)$ P:$(u,v,0)$ Q:$(u,v,1/2)$ GP:$(u,v,w)$ |
| 5.21.1.1 | $C^{1}2^{\bar{1}}2^{\bar{1}}2^{\infty m}1$ | K:$(0,v,w)$ GP:$(u,v,w)$ |
| 5.22.1.1 | $F^{\bar{1}}2^{\bar{1}}2^{1}2^{\infty m}1$ | L:(1/2,1/2,1/2) M:$(u,v,0)$ GP:$(u,v,w)$ |
| 5.23.1.1 | $I^{\bar{1}}2^{\bar{1}}2^{1}2^{\infty m}1$ | T:(1/2,1/2,0) C:$(u,v,0)$ P:$(1/2,1/2,w)$ GP:$(u,v,w)$ |
| 5.24.1.1 | $I^{\bar{1}}2_1{}^{\bar{1}}2_1{}^{1}2_1{}^{\infty m}1$ | T:(1/2,1/2,0) C:$(u,v,0)$ P:$(1/2,1/2,w)$ GP:$(u,v,w)$ |
| 6.25.1.1 | $P^{\bar{1}}m^{1}m^{\bar{1}}2^{\infty m}1$ | M:$(u,0,w)$ N:$(u,1/2,w)$ GP:$(u,v,w)$ |
| 3.25.1.1 | $P^{\bar{1}}m^{\bar{1}}m^{1}2^{\infty m}1$ | V:$(u,v,0)$ W:$(u,v,1/2)$ GP:$(u,v,w)$ |
| 7.26.1.1 | $P^{\bar{1}}m^{1}c^{\bar{1}}2_1{}^{\infty m}1$ | M:$(u,0,w)$ N:$(u,1/2,w)$ GP:$(u,v,w)$ |
| 6.26.1.1 | $P^{1}m^{\bar{1}}c^{\bar{1}}2_1{}^{\infty m}1$ | K:$(0,v,w)$ L:$(1/2,v,w)$ GP:$(u,v,w)$ |
| 4.26.1.1 | $P^{\bar{1}}m^{\bar{1}}c^{1}2_1{}^{\infty m}1$ | V:$(u,v,0)$ W:$(u,v,1/2)$ GP:$(u,v,w)$ |
| 7.27.1.1 | $P^{\bar{1}}c^{1}c^{\bar{1}}2^{\infty m}1$ | M:$(u,0,w)$ N:$(u,1/2,w)$ GP:$(u,v,w)$ |
| 3.27.1.1 | $P^{\bar{1}}c^{\bar{1}}c^{1}2^{\infty m}1$ | V:$(u,v,0)$ W:$(u,v,1/2)$ GP:$(u,v,w)$ |
| 7.28.1.1 | $P^{\bar{1}}m^{1}a^{\bar{1}}2^{\infty m}1$ | M:$(u,0,w)$ N:$(u,1/2,w)$ GP:$(u,v,w)$ |

| 6.28.1.1 | $P^{1}m^{\bar{1}}a^{\bar{1}}2^{\infty m}1$ | K:(0,*v*,*w*) L:(1/2,*v*,*w*) GP:(*u*,*v*,*w*) |
|---|---|---|
| 3.28.1.1 | $P^{\bar{1}}m^{\bar{1}}a^{1}2^{\infty m}1$ | V:(*u*,*v*,0) W:(*u*,*v*,1/2) GP:(*u*,*v*,*w*) |
| 7.29.1.1 | $P^{\bar{1}}c^{1}a^{\bar{1}}2_{1}{}^{\infty m}1$ | M:(*u*,0,*w*) N:(*u*,1/2,*w*) GP:(*u*,*v*,*w*) |
| 7.29.1.4 | $P^{1}c^{\bar{1}}a^{\bar{1}}2_{1}{}^{\infty m}1$ | K:(0,*v*,*w*) L:(1/2,*v*,*w*) GP:(*u*,*v*,*w*) |
| 4.29.1.1 | $P^{\bar{1}}c^{\bar{1}}a^{1}2_{1}{}^{\infty m}1$ | V:(*u*,*v*,0) W:(*u*,*v*,1/2) GP:(*u*,*v*,*w*) |
| 7.30.1.1 | $P^{\bar{1}}n^{1}c^{\bar{1}}2^{\infty m}1$ | M:(*u*,0,*w*) N:(*u*,1/2,*w*) GP:(*u*,*v*,*w*) |
| 7.30.1.4 | $P^{1}n^{\bar{1}}c^{\bar{1}}2^{\infty m}1$ | K:(0,*v*,*w*) L:(1/2,*v*,*w*) GP:(*u*,*v*,*w*) |
| 3.30.1.1 | $P^{\bar{1}}n^{\bar{1}}c^{1}2^{\infty m}1$ | V:(*u*,*v*,0) W:(*u*,*v*,1/2) GP:(*u*,*v*,*w*) |
| 7.31.1.1 | $P^{\bar{1}}m^{1}n^{\bar{1}}2_{1}{}^{\infty m}1$ | M:(*u*,0,*w*) N:(*u*,1/2,*w*) GP:(*u*,*v*,*w*) |
| 6.31.1.1 | $P^{1}m^{\bar{1}}n^{\bar{1}}2_{1}{}^{\infty m}1$ | K:(0,*v*,*w*) L:(1/2,*v*,*w*) GP:(*u*,*v*,*w*) |
| 4.31.1.1 | $P^{\bar{1}}m^{\bar{1}}n^{1}2_{1}{}^{\infty m}1$ | V:(*u*,*v*,0) W:(*u*,*v*,1/2) GP:(*u*,*v*,*w*) |
| 7.32.1.1 | $P^{\bar{1}}b^{1}a^{\bar{1}}2^{\infty m}1$ | M:(*u*,0,*w*) N:(*u*,1/2,*w*) GP:(*u*,*v*,*w*) |
| 3.32.1.1 | $P^{\bar{1}}b^{\bar{1}}a^{1}2^{\infty m}1$ | V:(*u*,*v*,0) W:(*u*,*v*,1/2) GP:(*u*,*v*,*w*) |
| 7.33.1.1 | $P^{\bar{1}}n^{1}a^{\bar{1}}2_{1}{}^{\infty m}1$ | M:(*u*,0,*w*) N:(*u*,1/2,*w*) GP:(*u*,*v*,*w*) |
| 7.33.1.4 | $P^{1}n^{\bar{1}}a^{\bar{1}}2_{1}{}^{\infty m}1$ | K:(0,*v*,*w*) L:(1/2,*v*,*w*) GP:(*u*,*v*,*w*) |
| 4.33.1.1 | $P^{\bar{1}}n^{\bar{1}}a^{1}2_{1}{}^{\infty m}1$ | V:(*u*,*v*,0) W:(*u*,*v*,1/2) GP:(*u*,*v*,*w*) |
| 7.34.1.1 | $P^{\bar{1}}n^{1}n^{\bar{1}}2^{\infty m}1$ | M:(*u*,0,*w*) N:(*u*,1/2,*w*) GP:(*u*,*v*,*w*) |
| 3.34.1.1 | $P^{\bar{1}}n^{\bar{1}}n^{1}2^{\infty m}1$ | V:(*u*,*v*,0) W:(*u*,*v*,1/2) GP:(*u*,*v*,*w*) |
| 8.35.1.1 | $C^{\bar{1}}m^{1}m^{\bar{1}}2^{\infty m}1$ | M:(*u*,0,*w*) GP:(*u*,*v*,*w*) |
| 3.35.1.1 | $C^{\bar{1}}m^{\bar{1}}m^{1}2^{\infty m}1$ | R:(1/2,1/2,1/2) S:(1/2,1/2,0) D:(1/2,1/2,*w*) P:(*u*,*v*,0) Q:(*u*,*v*,1/2) GP:(*u*,*v*,*w*) |
| 9.36.1.1 | $C^{\bar{1}}m^{1}c^{\bar{1}}2_{1}{}^{\infty m}1$ | M:(*u*,0,*w*) GP:(*u*,*v*,*w*) |
| 8.36.1.1 | $C^{1}m^{\bar{1}}c^{\bar{1}}2_{1}{}^{\infty m}1$ | K:(0,*v*,*w*) GP:(*u*,*v*,*w*) |
| 4.36.1.1 | $C^{\bar{1}}m^{\bar{1}}c^{1}2_{1}{}^{\infty m}1$ | R:(1/2,1/2,1/2) S:(1/2,1/2,0) D:(1/2,1/2,*w*) P:(*u*,*v*,0) Q:(*u*,*v*,1/2) GP:(*u*,*v*,*w*) |
| 9.37.1.1 | $C^{\bar{1}}c^{1}c^{\bar{1}}2^{\infty m}1$ | M:(*u*,0,*w*) GP:(*u*,*v*,*w*) |
| 3.37.1.1 | $C^{\bar{1}}c^{\bar{1}}c^{1}2^{\infty m}1$ | R:(1/2,1/2,1/2) S:(1/2,1/2,0) D:(1/2,1/2,*w*) P:(*u*,*v*,0) Q:(*u*,*v*,1/2) GP:(*u*,*v*,*w*) |
| 8.38.1.1 | $A^{\bar{1}}m^{1}m^{\bar{1}}2^{\infty m}1$ | M:(*u*,0,*w*) GP:(*u*,*v*,*w*) |
| 6.38.1.1 | $A^{1}m^{\bar{1}}m^{\bar{1}}2^{\infty m}1$ | K:(0,*v*,*w*) GP:(*u*,*v*,*w*) |
| 5.38.1.1 | $A^{\bar{1}}m^{\bar{1}}m^{1}2^{\infty m}1$ | R:(1/2,1/2,1/2) S:(1/2,1/2,0) D:(1/2,1/2,*w*) P:(*u*,*v*,0) Q:(*u*,*v*,1/2) GP:(*u*,*v*,*w*) |
| 8.39.1.1 | $A^{\bar{1}}e^{1}m^{\bar{1}}2^{\infty m}1$ | M:(*u*,0,*w*) GP:(*u*,*v*,*w*) |
| 7.39.1.1 | $A^{1}e^{\bar{1}}m^{\bar{1}}2^{\infty m}1$ | K:(0,*v*,*w*) GP:(*u*,*v*,*w*) |
| 5.39.1.1 | $A^{\bar{1}}e^{\bar{1}}m^{1}2^{\infty m}1$ | R:(1/2,1/2,1/2) S:(1/2,1/2,0) D:(1/2,1/2,*w*) P:(*u*,*v*,0) Q:(*u*,*v*,1/2) GP:(*u*,*v*,*w*) |
| 9.40.1.1 | $A^{\bar{1}}m^{1}a^{\bar{1}}2^{\infty m}1$ | M:(*u*,0,*w*) GP:(*u*,*v*,*w*) |
| 6.40.1.1 | $A^{1}m^{\bar{1}}a^{\bar{1}}2^{\infty m}1$ | K:(0,*v*,*w*) GP:(*u*,*v*,*w*) |
| 5.40.1.1 | $A^{\bar{1}}m^{\bar{1}}a^{1}2^{\infty m}1$ | R:(1/2,1/2,1/2) S:(1/2,1/2,0) D:(1/2,1/2,*w*) P:(*u*,*v*,0) Q:(*u*,*v*,1/2) GP:(*u*,*v*,*w*) |
| 9.41.1.1 | $A^{\bar{1}}e^{1}a^{\bar{1}}2^{\infty m}1$ | M:(*u*,0,*w*) GP:(*u*,*v*,*w*) |
| 7.41.1.1 | $A^{1}e^{\bar{1}}a^{\bar{1}}2^{\infty m}1$ | K:(0,*v*,*w*) GP:(*u*,*v*,*w*) |
| 5.41.1.1 | $A^{\bar{1}}e^{\bar{1}}a^{1}2^{\infty m}1$ | R:(1/2,1/2,1/2) S:(1/2,1/2,0) D:(1/2,1/2,*w*) P:(*u*,*v*,0) Q:(*u*,*v*,1/2) |

| | | GP:(*u*,*v*,*w*) |
|---|---|---|
| 8.42.1.1 | $F^{\bar{1}}m^{1}m^{\bar{1}}2^{\infty m}1$ | L:(1/2,1/2,1/2) J:(*u*,0,*w*) GP:(*u*,*v*,*w*) |
| 5.42.1.1 | $F^{\bar{1}}m^{\bar{1}}m^{1}2^{\infty m}1$ | L:(1/2,1/2,1/2) M:(*u*,*v*,0) GP:(*u*,*v*,*w*) |
| 9.43.1.1 | $F^{\bar{1}}d^{1}d^{\bar{1}}2^{\infty m}1$ | L:(1/2,1/2,1/2) J:(*u*,0,*w*) GP:(*u*,*v*,*w*) |
| 5.43.1.1 | $F^{\bar{1}}d^{\bar{1}}d^{1}2^{\infty m}1$ | L:(1/2,1/2,1/2) M:(*u*,*v*,0) GP:(*u*,*v*,*w*) |
| 8.44.1.1 | $I^{\bar{1}}m^{1}m^{\bar{1}}2^{\infty m}1$ | R:(1/2,0,1/2) B:(*u*,0,*w*) Q:(1/2,*v*,1/2) GP:(*u*,*v*,*w*) |
| 5.44.1.1 | $I^{\bar{1}}m^{\bar{1}}m^{1}2^{\infty m}1$ | T:(1/2,1/2,0) P:(1/2,1/2,*w*) GP:(*u*,*v*,*w*) |
| 9.45.1.1 | $I^{\bar{1}}b^{1}a^{\bar{1}}2^{\infty m}1$ | R:(1/2,0,1/2) B:(*u*,0,*w*) Q:(1/2,*v*,1/2) GP:(*u*,*v*,*w*) |
| 5.45.1.1 | $I^{\bar{1}}b^{\bar{1}}a^{1}2^{\infty m}1$ | T:(1/2,1/2,0) P:(1/2,1/2,*w*) GP:(*u*,*v*,*w*) |
| 9.46.1.1 | $I^{\bar{1}}m^{1}a^{\bar{1}}2^{\infty m}1$ | R:(1/2,0,1/2) B:(*u*,0,*w*) Q:(1/2,*v*,1/2) GP:(*u*,*v*,*w*) |
| 8.46.1.1 | $I^{1}m^{\bar{1}}a^{\bar{1}}2^{\infty m}1$ | S:(0,1/2,1/2) A:(0,*v*,*w*) D:(*u*,1/2,1/2) GP:(*u*,*v*,*w*) |
| 5.46.1.1 | $I^{\bar{1}}m^{\bar{1}}a^{1}2^{\infty m}1$ | T:(1/2,1/2,0) P:(1/2,1/2,*w*) GP:(*u*,*v*,*w*) |
| 10.47.1.1 | $P^{\bar{1}}m^{\bar{1}}m^{1}m^{\infty m}1$ | V:(*u*,*v*,0) W:(*u*,*v*,1/2) GP:(*u*,*v*,*w*) |
| 13.48.1.1 | $P^{\bar{1}}n^{\bar{1}}n^{1}n^{\infty m}1$ | V:(*u*,*v*,0) W:(*u*,*v*,1/2) GP:(*u*,*v*,*w*) |
| 10.49.1.1 | $P^{\bar{1}}c^{\bar{1}}c^{1}m^{\infty m}1$ | V:(*u*,*v*,0) W:(*u*,*v*,1/2) GP:(*u*,*v*,*w*) |
| 13.49.1.1 | $P^{\bar{1}}c^{1}c^{\bar{1}}m^{\infty m}1$ | M:(*u*,0,*w*) N:(*u*,1/2,*w*) GP:(*u*,*v*,*w*) |
| 13.50.1.1 | $P^{\bar{1}}b^{\bar{1}}a^{1}n^{\infty m}1$ | V:(*u*,*v*,0) W:(*u*,*v*,1/2) GP:(*u*,*v*,*w*) |
| 13.50.1.4 | $P^{\bar{1}}b^{1}a^{\bar{1}}n^{\infty m}1$ | M:(*u*,0,*w*) N:(*u*,1/2,*w*) GP:(*u*,*v*,*w*) |
| 13.51.1.1 | $P^{\bar{1}}m^{\bar{1}}m^{1}a^{\infty m}1$ | V:(*u*,*v*,0) W:(*u*,*v*,1/2) GP:(*u*,*v*,*w*) |
| 11.51.1.1 | $P^{1}m^{\bar{1}}m^{\bar{1}}a^{\infty m}1$ | K:(0,*v*,*w*) L:(1/2,*v*,*w*) GP:(*u*,*v*,*w*) |
| 10.51.1.1 | $P^{\bar{1}}m^{1}m^{\bar{1}}a^{\infty m}1$ | M:(*u*,0,*w*) N:(*u*,1/2,*w*) GP:(*u*,*v*,*w*) |
| 13.52.1.1 | $P^{\bar{1}}n^{\bar{1}}n^{1}a^{\infty m}1$ | V:(*u*,*v*,0) W:(*u*,*v*,1/2) GP:(*u*,*v*,*w*) |
| 13.52.1.4 | $P^{1}n^{\bar{1}}n^{\bar{1}}a^{\infty m}1$ | K:(0,*v*,*w*) L:(1/2,*v*,*w*) GP:(*u*,*v*,*w*) |
| 14.52.1.1 | $P^{\bar{1}}n^{1}n^{\bar{1}}a^{\infty m}1$ | M:(*u*,0,*w*) N:(*u*,1/2,*w*) GP:(*u*,*v*,*w*) |
| 14.53.1.1 | $P^{\bar{1}}m^{\bar{1}}n^{1}a^{\infty m}1$ | V:(*u*,*v*,0) W:(*u*,*v*,1/2) GP:(*u*,*v*,*w*) |
| 10.53.1.1 | $P^{1}m^{\bar{1}}n^{\bar{1}}a^{\infty m}1$ | K:(0,*v*,*w*) L:(1/2,*v*,*w*) GP:(*u*,*v*,*w*) |
| 13.53.1.1 | $P^{\bar{1}}m^{1}n^{\bar{1}}a^{\infty m}1$ | M:(*u*,0,*w*) N:(*u*,1/2,*w*) GP:(*u*,*v*,*w*) |
| 13.54.1.1 | $P^{\bar{1}}c^{\bar{1}}c^{1}a^{\infty m}1$ | V:(*u*,*v*,0) W:(*u*,*v*,1/2) GP:(*u*,*v*,*w*) |
| 14.54.1.1 | $P^{1}c^{\bar{1}}c^{\bar{1}}a^{\infty m}1$ | K:(0,*v*,*w*) L:(1/2,*v*,*w*) GP:(*u*,*v*,*w*) |
| 13.54.1.4 | $P^{\bar{1}}c^{1}c^{\bar{1}}a^{\infty m}1$ | M:(*u*,0,*w*) N:(*u*,1/2,*w*) GP:(*u*,*v*,*w*) |
| 10.55.1.1 | $P^{\bar{1}}b^{\bar{1}}a^{1}m^{\infty m}1$ | V:(*u*,*v*,0) W:(*u*,*v*,1/2) GP:(*u*,*v*,*w*) |
| 14.55.1.1 | $P^{\bar{1}}b^{1}a^{\bar{1}}m^{\infty m}1$ | M:(*u*,0,*w*) N:(*u*,1/2,*w*) GP:(*u*,*v*,*w*) |
| 13.56.1.1 | $P^{\bar{1}}c^{\bar{1}}c^{1}n^{\infty m}1$ | V:(*u*,*v*,0) W:(*u*,*v*,1/2) GP:(*u*,*v*,*w*) |
| 14.56.1.1 | $P^{\bar{1}}c^{1}c^{\bar{1}}n^{\infty m}1$ | M:(*u*,0,*w*) N:(*u*,1/2,*w*) GP:(*u*,*v*,*w*) |
| 11.57.1.1 | $P^{\bar{1}}b^{\bar{1}}c^{1}m^{\infty m}1$ | V:(*u*,*v*,0) W:(*u*,*v*,1/2) GP:(*u*,*v*,*w*) |
| 13.57.1.1 | $P^{1}b^{\bar{1}}c^{\bar{1}}m^{\infty m}1$ | K:(0,*v*,*w*) L:(1/2,*v*,*w*) GP:(*u*,*v*,*w*) |
| 14.57.1.1 | $P^{\bar{1}}b^{1}c^{\bar{1}}m^{\infty m}1$ | M:(*u*,0,*w*) N:(*u*,1/2,*w*) GP:(*u*,*v*,*w*) |
| 10.58.1.1 | $P^{\bar{1}}n^{\bar{1}}n^{1}m^{\infty m}1$ | V:(*u*,*v*,0) W:(*u*,*v*,1/2) GP:(*u*,*v*,*w*) |
| 14.58.1.1 | $P^{1}n^{\bar{1}}n^{\bar{1}}m^{\infty m}1$ | K:(0,*v*,*w*) L:(1/2,*v*,*w*) GP:(*u*,*v*,*w*) |
| 13.59.1.1 | $P^{\bar{1}}m^{\bar{1}}m^{1}n^{\infty m}1$ | V:(*u*,*v*,0) W:(*u*,*v*,1/2) GP:(*u*,*v*,*w*) |
| 11.59.1.1 | $P^{1}m^{\bar{1}}m^{\bar{1}}n^{\infty m}1$ | K:(0,*v*,*w*) L:(1/2,*v*,*w*) GP:(*u*,*v*,*w*) |
| 14.60.1.1 | $P^{\bar{1}}b^{\bar{1}}c^{1}n^{\infty m}1$ | V:(*u*,*v*,0) W:(*u*,*v*,1/2) GP:(*u*,*v*,*w*) |
| 14.60.1.4 | $P^{1}b^{\bar{1}}c^{\bar{1}}n^{\infty m}1$ | K:(0,*v*,*w*) L:(1/2,*v*,*w*) GP:(*u*,*v*,*w*) |

| 13.60.1.1 | $P^{\bar{1}}b^{1}c^{\bar{1}}n^{\infty m}1$ | M:($u$,0,$w$) N:($u$,1/2,$w$) GP:($u$,$v$,$w$) |
|---|---|---|
| 14.61.1.1 | $P^{\bar{1}}b^{\bar{1}}c^{1}a^{\infty m}1$ | V:($u$,$v$,0) W:($u$,$v$,1/2) GP:($u$,$v$,$w$) |
| 14.62.1.1 | $P^{\bar{1}}n^{\bar{1}}m^{1}a^{\infty m}1$ | V:($u$,$v$,0) W:($u$,$v$,1/2) GP:($u$,$v$,$w$) |
| 14.62.1.4 | $P^{1}n^{\bar{1}}m^{\bar{1}}a^{\infty m}1$ | K:(0,$v$,$w$) L:(1/2,$v$,$w$) GP:($u$,$v$,$w$) |
| 11.62.1.1 | $P^{\bar{1}}n^{1}m^{\bar{1}}a^{\infty m}1$ | M:($u$,0,$w$) N:($u$,1/2,$w$) GP:($u$,$v$,$w$) |
| 11.63.1.1 | $C^{\bar{1}}m^{\bar{1}}c^{1}m^{\infty m}1$ | R:(1/2,1/2,1/2) S:(1/2,1/2,0) D:(1/2,1/2,$w$) P:($u$,$v$,0) Q:($u$,$v$,1/2)<br>GP:($u$,$v$,$w$) |
| 12.63.1.1 | $C^{1}m^{\bar{1}}c^{\bar{1}}m^{\infty m}1$ | K:(0,$v$,$w$) GP:($u$,$v$,$w$) |
| 15.63.1.1 | $C^{\bar{1}}m^{1}c^{\bar{1}}m^{\infty m}1$ | M:($u$,0,$w$) P:($u$,$v$,0) GP:($u$,$v$,$w$) |
| 14.64.1.1 | $C^{\bar{1}}m^{\bar{1}}c^{1}e^{\infty m}1$ | R:(1/2,1/2,1/2) S:(1/2,1/2,0) D:(1/2,1/2,$w$) P:($u$,$v$,0) Q:($u$,$v$,1/2)<br>GP:($u$,$v$,$w$) |
| 12.64.1.1 | $C^{1}m^{\bar{1}}c^{\bar{1}}e^{\infty m}1$ | K:(0,$v$,$w$) GP:($u$,$v$,$w$) |
| 15.64.1.1 | $C^{\bar{1}}m^{1}c^{\bar{1}}e^{\infty m}1$ | M:($u$,0,$w$) P:($u$,$v$,0) GP:($u$,$v$,$w$) |
| 10.65.1.1 | $C^{\bar{1}}m^{\bar{1}}m^{1}m^{\infty m}1$ | R:(1/2,1/2,1/2) S:(1/2,1/2,0) D:(1/2,1/2,$w$) P:($u$,$v$,0) Q:($u$,$v$,1/2)<br>GP:($u$,$v$,$w$) |
| 12.65.1.1 | $C^{1}m^{\bar{1}}m^{\bar{1}}m^{\infty m}1$ | K:(0,$v$,$w$) GP:($u$,$v$,$w$) |
| 10.66.1.1 | $C^{\bar{1}}c^{\bar{1}}c^{1}m^{\infty m}1$ | R:(1/2,1/2,1/2) S:(1/2,1/2,0) D:(1/2,1/2,$w$) P:($u$,$v$,0) Q:($u$,$v$,1/2)<br>GP:($u$,$v$,$w$) |
| 15.66.1.1 | $C^{1}c^{\bar{1}}c^{\bar{1}}m^{\infty m}1$ | K:(0,$v$,$w$) GP:($u$,$v$,$w$) |
| 13.67.1.1 | $C^{\bar{1}}m^{\bar{1}}m^{1}e^{\infty m}1$ | R:(1/2,1/2,1/2) S:(1/2,1/2,0) D:(1/2,1/2,$w$) P:($u$,$v$,0) Q:($u$,$v$,1/2)<br>GP:($u$,$v$,$w$) |
| 12.67.1.1 | $C^{1}m^{\bar{1}}m^{\bar{1}}e^{\infty m}1$ | K:(0,$v$,$w$) GP:($u$,$v$,$w$) |
| 13.68.1.1 | $C^{\bar{1}}c^{\bar{1}}c^{1}e^{\infty m}1$ | R:(1/2,1/2,1/2) S:(1/2,1/2,0) D:(1/2,1/2,$w$) P:($u$,$v$,0) Q:($u$,$v$,1/2)<br>GP:($u$,$v$,$w$) |
| 15.68.1.1 | $C^{1}c^{\bar{1}}c^{\bar{1}}e^{\infty m}1$ | K:(0,$v$,$w$) GP:($u$,$v$,$w$) |
| 12.69.1.1 | $F^{\bar{1}}m^{\bar{1}}m^{1}m^{\infty m}1$ | L:(1/2,1/2,1/2) M:($u$,$v$,0) GP:($u$,$v$,$w$) |
| 15.70.1.1 | $F^{\bar{1}}d^{\bar{1}}d^{1}d^{\infty m}1$ | L:(1/2,1/2,1/2) M:($u$,$v$,0) GP:($u$,$v$,$w$) |
| 12.71.1.1 | $I^{\bar{1}}m^{\bar{1}}m^{1}m^{\infty m}1$ | T:(1/2,1/2,0) C:($u$,$v$,0) P:(1/2,1/2,$w$) GP:($u$,$v$,$w$) |
| 12.72.1.1 | $I^{\bar{1}}b^{\bar{1}}a^{1}m^{\infty m}1$ | T:(1/2,1/2,0) C:($u$,$v$,0) P:(1/2,1/2,$w$) GP:($u$,$v$,$w$) |
| 15.72.1.1 | $I^{1}b^{\bar{1}}a^{\bar{1}}m^{\infty m}1$ | S:(0,1/2,1/2) A:(0,$v$,$w$) D:($u$,1/2,1/2) GP:($u$,$v$,$w$) |
| 15.73.1.1 | $I^{\bar{1}}b^{\bar{1}}c^{1}a^{\infty m}1$ | T:(1/2,1/2,0) C:($u$,$v$,0) P:(1/2,1/2,$w$) GP:($u$,$v$,$w$) |
| 15.74.1.1 | $I^{\bar{1}}m^{\bar{1}}m^{1}a^{\infty m}1$ | T:(1/2,1/2,0) C:($u$,$v$,0) P:(1/2,1/2,$w$) GP:($u$,$v$,$w$) |
| 12.74.1.1 | $I^{1}m^{\bar{1}}m^{\bar{1}}a^{\infty m}1$ | S:(0,1/2,1/2) A:(0,$v$,$w$) D:($u$,1/2,1/2) GP:($u$,$v$,$w$) |
| 3.75.1.1 | $P^{\bar{1}}4^{\infty m}1$ | R:(0,1/2,1/2) X:(0,1/2,0) D:($u$,$v$,0) Δ:(0,$v$,0) E:($u$,$v$,1/2)<br>S:($u$,$u$,1/2) Σ:($u$,$u$,0) T:($u$,1/2,1/2) U:(0,$v$,1/2) W:(0,1/2,$w$)<br>Y:($u$,1/2,0) B:(0,$v$,$w$) C:($u$,$u$,$w$) F:($u$,1/2,$w$) GP:($u$,$v$,$w$) |
| 4.76.1.1 | $P^{\bar{1}}4_{1}{}^{\infty m}1$ | R:(0,1/2,1/2) X:(0,1/2,0) D:($u$,$v$,0) Δ:(0,$v$,0) E:($u$,$v$,1/2)<br>S:($u$,$u$,1/2) Σ:($u$,$u$,0) T:($u$,1/2,1/2) U:(0,$v$,1/2) W:(0,1/2,$w$)<br>Y:($u$,1/2,0) B:(0,$v$,$w$) C:($u$,$u$,$w$) F:($u$,1/2,$w$) GP:($u$,$v$,$w$) |
| 3.77.1.1 | $P^{\bar{1}}4_{2}{}^{\infty m}1$ | R:(0,1/2,1/2) X:(0,1/2,0) D:($u$,$v$,0) Δ:(0,$v$,0) E:($u$,$v$,1/2)<br>S:($u$,$u$,1/2) Σ:($u$,$u$,0) T:($u$,1/2,1/2) U:(0,$v$,1/2) W:(0,1/2,$w$)<br>Y:($u$,1/2,0) B:(0,$v$,$w$) C:($u$,$u$,$w$) F:($u$,1/2,$w$) GP:($u$,$v$,$w$) |
| 4.78.1.1 | $P^{\bar{1}}4_{3}{}^{\infty m}1$ | R:(0,1/2,1/2) X:(0,1/2,0) D:($u$,$v$,0) Δ:(0,$v$,0) E:($u$,$v$,1/2) |
| | | S:(*u*,*u*,1/2) Σ:(*u*,*u*,0) T:(*u*,1/2,1/2) U:(0,*v*,1/2) W:(0,1/2,*w*)<br>Y:(*u*,1/2,0) B:(0,*v*,*w*) C:(*u*,*u*,*w*) F:(*u*,1/2,*w*) GP:(*u*,*v*,*w*) |
| 5.79.1.1 | $I^{\bar{1}}4^{\infty m}1$ | X:(1/2,1/2,0) N:(1/2,0,1/2) W:(1/2,1/2,*w*) Q:(1/2,*v*,1/2) Σ:(*u*,0,0)<br>Y:(*u*,1-*u*,0) Δ:(*u*,*u*,0) A:(*u*,*u*,*w*) B:(*u*,0,*w*) C:(*u*,*v*,0) GP:(*u*,*v*,*w*) |
| 5.80.1.1 | $I^{\bar{1}}4_{1}{}^{\infty m}1$ | X:(1/2,1/2,0) N:(1/2,0,1/2) W:(1/2,1/2,*w*) Q:(1/2,*v*,1/2) Σ:(*u*,0,0)<br>Y:(*u*,1-*u*,0) Δ:(*u*,*u*,0) A:(*u*,*u*,*w*) B:(*u*,0,*w*) C:(*u*,*v*,0) GP:(*u*,*v*,*w*) |
| 3.81.1.1 | $P^{\bar{1}}\bar{4}^{\infty m}1$ | R:(0,1/2,1/2) X:(0,1/2,0) D:(*u*,*v*,0) Δ:(0,*v*,0) E:(*u*,*v*,1/2)<br>S:(*u*,*u*,1/2) Σ:(*u*,*u*,0) T:(*u*,1/2,1/2) U:(0,*v*,1/2) W:(0,1/2,*w*)<br>Y:(*u*,1/2,0) B:(0,*v*,*w*) C:(*u*,*u*,*w*) F:(*u*,1/2,*w*) GP:(*u*,*v*,*w*) |
| 5.82.1.1 | $I^{\bar{1}}\bar{4}^{\infty m}1$ | X:(1/2,1/2,0) N:(1/2,0,1/2) W:(1/2,1/2,*w*) Q:(1/2,*v*,1/2) Σ:(*u*,0,0)<br>Y:(*u*,1-*u*,0) Δ:(*u*,*u*,0) A:(*u*,*u*,*w*) B:(*u*,0,*w*) C:(*u*,*v*,0) GP:(*u*,*v*,*w*) |
| 10.83.1.1 | $P^{\bar{1}}4/{}^{1}m^{\infty m}1$ | R:(0,1/2,1/2) X:(0,1/2,0) D:(*u*,*v*,0) Δ:(0,*v*,0) E:(*u*,*v*,1/2)<br>S:(*u*,*u*,1/2) Σ:(*u*,*u*,0) T:(*u*,1/2,1/2) U:(0,*v*,1/2) W:(0,1/2,*w*)<br>Y:(*u*,1/2,0) B:(0,*v*,*w*) C:(*u*,*u*,*w*) F:(*u*,1/2,*w*) GP:(*u*,*v*,*w*) |
| 10.84.1.1 | $P^{\bar{1}}4_{2}/{}^{1}m^{\infty m}1$ | R:(0,1/2,1/2) X:(0,1/2,0) D:(*u*,*v*,0) Δ:(0,*v*,0) E:(*u*,*v*,1/2)<br>S:(*u*,*u*,1/2) Σ:(*u*,*u*,0) T:(*u*,1/2,1/2) U:(0,*v*,1/2) W:(0,1/2,*w*)<br>Y:(*u*,1/2,0) B:(0,*v*,*w*) C:(*u*,*u*,*w*) F:(*u*,1/2,*w*) GP:(*u*,*v*,*w*) |
| 13.85.1.1 | $P^{\bar{1}}4/{}^{1}n^{\infty m}1$ | R:(0,1/2,1/2) X:(0,1/2,0) D:(*u*,*v*,0) Δ:(0,*v*,0) E:(*u*,*v*,1/2)<br>S:(*u*,*u*,1/2) Σ:(*u*,*u*,0) T:(*u*,1/2,1/2) U:(0,*v*,1/2) W:(0,1/2,*w*)<br>Y:(*u*,1/2,0) B:(0,*v*,*w*) C:(*u*,*u*,*w*) F:(*u*,1/2,*w*) GP:(*u*,*v*,*w*) |
| 13.86.1.1 | $P^{\bar{1}}4_{2}/{}^{1}n^{\infty m}1$ | R:(0,1/2,1/2) X:(0,1/2,0) D:(*u*,*v*,0) Δ:(0,*v*,0) E:(*u*,*v*,1/2)<br>S:(*u*,*u*,1/2) Σ:(*u*,*u*,0) T:(*u*,1/2,1/2) U:(0,*v*,1/2) W:(0,1/2,*w*)<br>Y:(*u*,1/2,0) B:(0,*v*,*w*) C:(*u*,*u*,*w*) F:(*u*,1/2,*w*) GP:(*u*,*v*,*w*) |
| 12.87.1.1 | $I^{\bar{1}}4/{}^{1}m^{\infty m}1$ | X:(1/2,1/2,0) N:(1/2,0,1/2) W:(1/2,1/2,*w*) Q:(1/2,*v*,1/2) Σ:(*u*,0,0)<br>Y:(*u*,1-*u*,0) Δ:(*u*,*u*,0) A:(*u*,*u*,*w*) B:(*u*,0,*w*) C:(*u*,*v*,0) GP:(*u*,*v*,*w*) |
| 15.88.1.1 | $I^{\bar{1}}4_{1}/{}^{1}a^{\infty m}1$ | X:(1/2,1/2,0) N:(1/2,0,1/2) W:(1/2,1/2,*w*) Q:(1/2,*v*,1/2) Σ:(*u*,0,0)<br>Y:(*u*,1-*u*,0) Δ:(*u*,*u*,0) A:(*u*,*u*,*w*) B:(*u*,0,*w*) C:(*u*,*v*,0) GP:(*u*,*v*,*w*) |
| 16.89.1.1 | $P^{\bar{1}}4^{1}2^{\bar{1}}2^{\infty m}1$ | R:(0,1/2,1/2) X:(0,1/2,0) Δ:(0,*v*,0) T:(*u*,1/2,1/2) U:(0,*v*,1/2)<br>W:(0,1/2,*w*) Y:(*u*,1/2,0) B:(0,*v*,*w*) D:(*u*,*v*,0) E:(*u*,*v*,1/2)<br>F:(*u*,1/2,*w*) GP:(*u*,*v*,*w*) |
| 75.89.1.1 | $P^{1}4^{\bar{1}}2^{\bar{1}}2^{\infty m}1$ | D:(*u*,*v*,0) E:(*u*,*v*,1/2) GP:(*u*,*v*,*w*) |
| 21.89.1.1 | $P^{\bar{1}}4^{\bar{1}}2^{1}2^{\infty m}1$ | S:(*u*,*u*,1/2) Σ:(*u*,*u*,0) C:(*u*,*u*,*w*) D:(*u*,*v*,0) E:(*u*,*v*,1/2) GP:(*u*,*v*,*w*) |
| 18.90.1.1 | $P^{\bar{1}}4^{1}2_{1}{}^{\bar{1}}2^{\infty m}1$ | R:(0,1/2,1/2) X:(0,1/2,0) Δ:(0,*v*,0) T:(*u*,1/2,1/2) U:(0,*v*,1/2)<br>W:(0,1/2,*w*) Y:(*u*,1/2,0) B:(0,*v*,*w*) D:(*u*,*v*,0) E:(*u*,*v*,1/2)<br>F:(*u*,1/2,*w*) GP:(*u*,*v*,*w*) |
| 75.90.1.1 | $P^{1}4^{\bar{1}}2_{1}{}^{\bar{1}}2^{\infty m}1$ | D:(*u*,*v*,0) E:(*u*,*v*,1/2) GP:(*u*,*v*,*w*) |
| 21.90.1.1 | $P^{\bar{1}}4^{\bar{1}}2_{1}{}^{1}2^{\infty m}1$ | S:(*u*,*u*,1/2) Σ:(*u*,*u*,0) C:(*u*,*u*,*w*) D:(*u*,*v*,0) E:(*u*,*v*,1/2) GP:(*u*,*v*,*w*) |
| 17.91.1.1 | $P^{\bar{1}}4_{1}{}^{1}2^{\bar{1}}2^{\infty m}1$ | R:(0,1/2,1/2) X:(0,1/2,0) Δ:(0,*v*,0) T:(*u*,1/2,1/2) U:(0,*v*,1/2)<br>W:(0,1/2,*w*) Y:(*u*,1/2,0) B:(0,*v*,*w*) D:(*u*,*v*,0) E:(*u*,*v*,1/2)<br>F:(*u*,1/2,*w*) GP:(*u*,*v*,*w*) |
| 76.91.1.1 | $P^{1}4_{1}{}^{\bar{1}}2^{\bar{1}}2^{\infty m}1$ | D:(*u*,*v*,0) E:(*u*,*v*,1/2) GP:(*u*,*v*,*w*) |
| 20.91.1.1 | $P^{\bar{1}}4_{1}{}^{\bar{1}}2^{1}2^{\infty m}1$ | S:(*u*,*u*,1/2) Σ:(*u*,*u*,0) C:(*u*,*u*,*w*) D:(*u*,*v*,0) E:(*u*,*v*,1/2) GP:(*u*,*v*,*w*) |
| 19.92.1.1 | $P^{\bar{1}}4_{1}{}^{1}2_{1}{}^{\bar{1}}2^{\infty m}1$ | R:(0,1/2,1/2) X:(0,1/2,0) Δ:(0,*v*,0) T:(*u*,1/2,1/2) U:(0,*v*,1/2)<br>W:(0,1/2,*w*) Y:(*u*,1/2,0) B:(0,*v*,*w*) D:(*u*,*v*,0) E:(*u*,*v*,1/2) |

| | | F:(*u*,1/2,*w*) GP:(*u*,*v*,*w*) |
|---|---|---|
| 76.92.1.1 | $P^{1}4_{1}{}^{\bar{1}}2_{1}{}^{\bar{1}}2^{\infty m}1$ | D:(*u*,*v*,0) E:(*u*,*v*,1/2) GP:(*u*,*v*,*w*) |
| 20.92.1.1 | $P^{\bar{1}}4_{1}{}^{\bar{1}}2_{1}{}^{1}2^{\infty m}1$ | S:(*u*,*u*,1/2) Σ:(*u*,*u*,0) C:(*u*,*u*,*w*) D:(*u*,*v*,0) E:(*u*,*v*,1/2) GP:(*u*,*v*,*w*) |
| 16.93.1.1 | $P^{\bar{1}}4_{2}{}^{1}2^{\bar{1}}2^{\infty m}1$ | R:(0,1/2,1/2) X:(0,1/2,0) Δ:(0,*v*,0) T:(*u*,1/2,1/2) U:(0,*v*,1/2)<br>W:(0,1/2,*w*) Y:(*u*,1/2,0) B:(0,*v*,*w*) D:(*u*,*v*,0) E:(*u*,*v*,1/2)<br>F:(*u*,1/2,*w*) GP:(*u*,*v*,*w*) |
| 77.93.1.1 | $P^{1}4_{2}{}^{\bar{1}}2^{\bar{1}}2^{\infty m}1$ | D:(*u*,*v*,0) E:(*u*,*v*,1/2) GP:(*u*,*v*,*w*) |
| 21.93.1.1 | $P^{\bar{1}}4_{2}{}^{\bar{1}}2^{1}2^{\infty m}1$ | S:(*u*,*u*,1/2) Σ:(*u*,*u*,0) C:(*u*,*u*,*w*) D:(*u*,*v*,0) E:(*u*,*v*,1/2) GP:(*u*,*v*,*w*) |
| 18.94.1.1 | $P^{\bar{1}}4_{2}{}^{1}2_{1}{}^{\bar{1}}2^{\infty m}1$ | R:(0,1/2,1/2) X:(0,1/2,0) Δ:(0,*v*,0) T:(*u*,1/2,1/2) U:(0,*v*,1/2)<br>W:(0,1/2,*w*) Y:(*u*,1/2,0) B:(0,*v*,*w*) D:(*u*,*v*,0) E:(*u*,*v*,1/2)<br>F:(*u*,1/2,*w*) GP:(*u*,*v*,*w*) |
| 77.94.1.1 | $P^{1}4_{2}{}^{\bar{1}}2_{1}{}^{\bar{1}}2^{\infty m}1$ | D:(*u*,*v*,0) E:(*u*,*v*,1/2) GP:(*u*,*v*,*w*) |
| 21.94.1.1 | $P^{\bar{1}}4_{2}{}^{\bar{1}}2_{1}{}^{1}2^{\infty m}1$ | S:(*u*,*u*,1/2) Σ:(*u*,*u*,0) C:(*u*,*u*,*w*) D:(*u*,*v*,0) E:(*u*,*v*,1/2) GP:(*u*,*v*,*w*) |
| 17.95.1.1 | $P^{\bar{1}}4_{3}{}^{1}2^{\bar{1}}2^{\infty m}1$ | R:(0,1/2,1/2) X:(0,1/2,0) Δ:(0,*v*,0) T:(*u*,1/2,1/2) U:(0,*v*,1/2)<br>W:(0,1/2,*w*) Y:(*u*,1/2,0) B:(0,*v*,*w*) D:(*u*,*v*,0) E:(*u*,*v*,1/2)<br>F:(*u*,1/2,*w*) GP:(*u*,*v*,*w*) |
| 78.95.1.1 | $P^{1}4_{3}{}^{\bar{1}}2^{\bar{1}}2^{\infty m}1$ | D:(*u*,*v*,0) E:(*u*,*v*,1/2) GP:(*u*,*v*,*w*) |
| 20.95.1.1 | $P^{\bar{1}}4_{3}{}^{\bar{1}}2^{1}2^{\infty m}1$ | S:(*u*,*u*,1/2) Σ:(*u*,*u*,0) C:(*u*,*u*,*w*) D:(*u*,*v*,0) E:(*u*,*v*,1/2) GP:(*u*,*v*,*w*) |
| 19.96.1.1 | $P^{\bar{1}}4_{3}{}^{1}2_{1}{}^{\bar{1}}2^{\infty m}1$ | R:(0,1/2,1/2) X:(0,1/2,0) Δ:(0,*v*,0) T:(*u*,1/2,1/2) U:(0,*v*,1/2)<br>W:(0,1/2,*w*) Y:(*u*,1/2,0) B:(0,*v*,*w*) D:(*u*,*v*,0) E:(*u*,*v*,1/2)<br>F:(*u*,1/2,*w*) GP:(*u*,*v*,*w*) |
| 78.96.1.1 | $P^{1}4_{3}{}^{\bar{1}}2_{1}{}^{\bar{1}}2^{\infty m}1$ | D:(*u*,*v*,0) E:(*u*,*v*,1/2) GP:(*u*,*v*,*w*) |
| 20.96.1.1 | $P^{\bar{1}}4_{3}{}^{\bar{1}}2_{1}{}^{1}2^{\infty m}1$ | S:(*u*,*u*,1/2) Σ:(*u*,*u*,0) C:(*u*,*u*,*w*) D:(*u*,*v*,0) E:(*u*,*v*,1/2) GP:(*u*,*v*,*w*) |
| 23.97.1.1 | $I^{\bar{1}}4^{1}2^{\bar{1}}2^{\infty m}1$ | N:(1/2,0,1/2) Q:(1/2,*v*,1/2) Σ:(*u*,0,0) B:(*u*,0,*w*) C:(*u*,*v*,0)<br>GP:(*u*,*v*,*w*) |
| 79.97.1.1 | $I^{1}4^{\bar{1}}2^{\bar{1}}2^{\infty m}1$ | C:(*u*,*v*,0) GP:(*u*,*v*,*w*) |
| 22.97.1.1 | $I^{\bar{1}}4^{\bar{1}}2^{1}2^{\infty m}1$ | X:(1/2,1/2,0) Δ:(*u*,*u*,0) W:(1/2,1/2,*w*) Y:(*u*,1-*u*,0) A:(*u*,*u*,*w*)<br>C:(*u*,*v*,0) GP:(*u*,*v*,*w*) |
| 24.98.1.1 | $I^{\bar{1}}4_{1}{}^{1}2^{\bar{1}}2^{\infty m}1$ | N:(1/2,0,1/2) Q:(1/2,*v*,1/2) Σ:(*u*,0,0) B:(*u*,0,*w*) C:(*u*,*v*,0)<br>GP:(*u*,*v*,*w*) |
| 80.98.1.1 | $I^{1}4_{1}{}^{\bar{1}}2^{\bar{1}}2^{\infty m}1$ | C:(*u*,*v*,0) GP:(*u*,*v*,*w*) |
| 22.98.1.1 | $I^{\bar{1}}4_{1}{}^{\bar{1}}2^{1}2^{\infty m}1$ | X:(1/2,1/2,0) Δ:(*u*,*u*,0) W:(1/2,1/2,*w*) Y:(*u*,1-*u*,0) A:(*u*,*u*,*w*)<br>C:(*u*,*v*,0) GP:(*u*,*v*,*w*) |
| 35.99.1.1 | $P^{\bar{1}}4^{\bar{1}}m^{1}m^{\infty m}1$ | S:(*u*,*u*,1/2) Σ:(*u*,*u*,0) C:(*u*,*u*,*w*) D:(*u*,*v*,0) E:(*u*,*v*,1/2) GP:(*u*,*v*,*w*) |
| 25.99.1.1 | $P^{\bar{1}}4^{1}m^{\bar{1}}m^{\infty m}1$ | R:(0,1/2,1/2) X:(0,1/2,0) Δ:(0,*v*,0) T:(*u*,1/2,1/2) U:(0,*v*,1/2)<br>W:(0,1/2,*w*) Y:(*u*,1/2,0) B:(0,*v*,*w*) D:(*u*,*v*,0) E:(*u*,*v*,1/2)<br>F:(*u*,1/2,*w*) GP:(*u*,*v*,*w*) |
| 75.99.1.1 | $P^{1}4^{\bar{1}}m^{\bar{1}}m^{\infty m}1$ | D:(*u*,*v*,0) E:(*u*,*v*,1/2) GP:(*u*,*v*,*w*) |
| 35.100.1.1 | $P^{\bar{1}}4^{\bar{1}}b^{1}m^{\infty m}1$ | S:(*u*,*u*,1/2) Σ:(*u*,*u*,0) C:(*u*,*u*,*w*) D:(*u*,*v*,0) E:(*u*,*v*,1/2) GP:(*u*,*v*,*w*) |
| 32.100.1.1 | $P^{\bar{1}}4^{1}b^{\bar{1}}m^{\infty m}1$ | R:(0,1/2,1/2) X:(0,1/2,0) Δ:(0,*v*,0) T:(*u*,1/2,1/2) U:(0,*v*,1/2)<br>W:(0,1/2,*w*) Y:(*u*,1/2,0) B:(0,*v*,*w*) D:(*u*,*v*,0) E:(*u*,*v*,1/2)<br>F:(*u*,1/2,*w*) GP:(*u*,*v*,*w*) |
| 75.100.1.1 | $P^{1}4^{\bar{1}}b^{\bar{1}}m^{\infty m}1$ | D:(*u*,*v*,0) E:(*u*,*v*,1/2) GP:(*u*,*v*,*w*) |

| 35.101.1.1 | $P^{\bar{1}}4_2{}^{\bar{1}}c^{1}m^{\infty m}1$ | S:$(u,u,1/2)$ Σ:$(u,u,0)$ C:$(u,u,w)$ D:$(u,v,0)$ E:$(u,v,1/2)$ GP:$(u,v,w)$ |
|---|---|---|
| 27.101.1.1 | $P^{\bar{1}}4_2{}^{1}c^{\bar{1}}m^{\infty m}1$ | R:$(0,1/2,1/2)$ X:$(0,1/2,0)$ Δ:$(0,v,0)$ T:$(u,1/2,1/2)$ U:$(0,v,1/2)$ W:$(0,1/2,w)$ Y:$(u,1/2,0)$ B:$(0,v,w)$ D:$(u,v,0)$ E:$(u,v,1/2)$ F:$(u,1/2,w)$ GP:$(u,v,w)$ |
| 77.101.1.1 | $P^{1}4_2{}^{\bar{1}}c^{\bar{1}}m^{\infty m}1$ | D:$(u,v,0)$ E:$(u,v,1/2)$ GP:$(u,v,w)$ |
| 35.102.1.1 | $P^{\bar{1}}4_2{}^{\bar{1}}n^{1}m^{\infty m}1$ | S:$(u,u,1/2)$ Σ:$(u,u,0)$ C:$(u,u,w)$ D:$(u,v,0)$ E:$(u,v,1/2)$ GP:$(u,v,w)$ |
| 34.102.1.1 | $P^{\bar{1}}4_2{}^{1}n^{\bar{1}}m^{\infty m}1$ | R:$(0,1/2,1/2)$ X:$(0,1/2,0)$ Δ:$(0,v,0)$ T:$(u,1/2,1/2)$ U:$(0,v,1/2)$ W:$(0,1/2,w)$ Y:$(u,1/2,0)$ B:$(0,v,w)$ D:$(u,v,0)$ E:$(u,v,1/2)$ F:$(u,1/2,w)$ GP:$(u,v,w)$ |
| 77.102.1.1 | $P^{1}4_2{}^{\bar{1}}n^{\bar{1}}m^{\infty m}1$ | D:$(u,v,0)$ E:$(u,v,1/2)$ GP:$(u,v,w)$ |
| 37.103.1.1 | $P^{\bar{1}}4^{\bar{1}}c^{1}c^{\infty m}1$ | S:$(u,u,1/2)$ Σ:$(u,u,0)$ C:$(u,u,w)$ D:$(u,v,0)$ E:$(u,v,1/2)$ GP:$(u,v,w)$ |
| 27.103.1.1 | $P^{\bar{1}}4^{1}c^{\bar{1}}c^{\infty m}1$ | R:$(0,1/2,1/2)$ X:$(0,1/2,0)$ Δ:$(0,v,0)$ T:$(u,1/2,1/2)$ U:$(0,v,1/2)$ W:$(0,1/2,w)$ Y:$(u,1/2,0)$ B:$(0,v,w)$ D:$(u,v,0)$ E:$(u,v,1/2)$ F:$(u,1/2,w)$ GP:$(u,v,w)$ |
| 75.103.1.1 | $P^{1}4^{\bar{1}}c^{\bar{1}}c^{\infty m}1$ | D:$(u,v,0)$ E:$(u,v,1/2)$ GP:$(u,v,w)$ |
| 37.104.1.1 | $P^{\bar{1}}4^{\bar{1}}n^{1}c^{\infty m}1$ | S:$(u,u,1/2)$ Σ:$(u,u,0)$ C:$(u,u,w)$ D:$(u,v,0)$ E:$(u,v,1/2)$ GP:$(u,v,w)$ |
| 34.104.1.1 | $P^{\bar{1}}4^{1}n^{\bar{1}}c^{\infty m}1$ | R:$(0,1/2,1/2)$ X:$(0,1/2,0)$ Δ:$(0,v,0)$ T:$(u,1/2,1/2)$ U:$(0,v,1/2)$ W:$(0,1/2,w)$ Y:$(u,1/2,0)$ B:$(0,v,w)$ D:$(u,v,0)$ E:$(u,v,1/2)$ F:$(u,1/2,w)$ GP:$(u,v,w)$ |
| 75.104.1.1 | $P^{1}4^{\bar{1}}n^{\bar{1}}c^{\infty m}1$ | D:$(u,v,0)$ E:$(u,v,1/2)$ GP:$(u,v,w)$ |
| 37.105.1.1 | $P^{\bar{1}}4_2{}^{\bar{1}}m^{1}c^{\infty m}1$ | S:$(u,u,1/2)$ Σ:$(u,u,0)$ C:$(u,u,w)$ D:$(u,v,0)$ E:$(u,v,1/2)$ GP:$(u,v,w)$ |
| 25.105.1.1 | $P^{\bar{1}}4_2{}^{1}m^{\bar{1}}c^{\infty m}1$ | R:$(0,1/2,1/2)$ X:$(0,1/2,0)$ Δ:$(0,v,0)$ T:$(u,1/2,1/2)$ U:$(0,v,1/2)$ W:$(0,1/2,w)$ Y:$(u,1/2,0)$ B:$(0,v,w)$ D:$(u,v,0)$ E:$(u,v,1/2)$ F:$(u,1/2,w)$ GP:$(u,v,w)$ |
| 77.105.1.1 | $P^{1}4_2{}^{\bar{1}}m^{\bar{1}}c^{\infty m}1$ | D:$(u,v,0)$ E:$(u,v,1/2)$ GP:$(u,v,w)$ |
| 37.106.1.1 | $P^{\bar{1}}4_2{}^{\bar{1}}b^{1}c^{\infty m}1$ | S:$(u,u,1/2)$ Σ:$(u,u,0)$ C:$(u,u,w)$ D:$(u,v,0)$ E:$(u,v,1/2)$ GP:$(u,v,w)$ |
| 32.106.1.1 | $P^{\bar{1}}4_2{}^{1}b^{\bar{1}}c^{\infty m}1$ | R:$(0,1/2,1/2)$ X:$(0,1/2,0)$ Δ:$(0,v,0)$ T:$(u,1/2,1/2)$ U:$(0,v,1/2)$ W:$(0,1/2,w)$ Y:$(u,1/2,0)$ B:$(0,v,w)$ D:$(u,v,0)$ E:$(u,v,1/2)$ F:$(u,1/2,w)$ GP:$(u,v,w)$ |
| 77.106.1.1 | $P^{1}4_2{}^{\bar{1}}b^{\bar{1}}c^{\infty m}1$ | D:$(u,v,0)$ E:$(u,v,1/2)$ GP:$(u,v,w)$ |
| 42.107.1.1 | $I^{\bar{1}}4^{\bar{1}}m^{1}m^{\infty m}1$ | X:$(1/2,1/2,0)$ Δ:$(u,u,0)$ W:$(1/2,1/2,w)$ Y:$(u,1\text{-}u,0)$ A:$(u,u,w)$ C:$(u,v,0)$ GP:$(u,v,w)$ |
| 44.107.1.1 | $I^{\bar{1}}4^{1}m^{\bar{1}}m^{\infty m}1$ | N:$(1/2,0,1/2)$ Q:$(1/2,v,1/2)$ Σ:$(u,0,0)$ B:$(u,0,w)$ C:$(u,v,0)$ GP:$(u,v,w)$ |
| 79.107.1.1 | $I^{1}4^{\bar{1}}m^{\bar{1}}m^{\infty m}1$ | C:$(u,v,0)$ GP:$(u,v,w)$ |
| 42.108.1.1 | $I^{\bar{1}}4^{\bar{1}}c^{1}m^{\infty m}1$ | X:$(1/2,1/2,0)$ Δ:$(u,u,0)$ W:$(1/2,1/2,w)$ Y:$(u,1\text{-}u,0)$ A:$(u,u,w)$ C:$(u,v,0)$ GP:$(u,v,w)$ |
| 45.108.1.1 | $I^{\bar{1}}4^{1}c^{\bar{1}}m^{\infty m}1$ | N:$(1/2,0,1/2)$ Q:$(1/2,v,1/2)$ Σ:$(u,0,0)$ B:$(u,0,w)$ C:$(u,v,0)$ GP:$(u,v,w)$ |
| 79.108.1.1 | $I^{1}4^{\bar{1}}c^{\bar{1}}m^{\infty m}1$ | C:$(u,v,0)$ GP:$(u,v,w)$ |
| 43.109.1.1 | $I^{\bar{1}}4_1{}^{\bar{1}}m^{1}d^{\infty m}1$ | X:$(1/2,1/2,0)$ Δ:$(u,u,0)$ W:$(1/2,1/2,w)$ Y:$(u,1\text{-}u,0)$ A:$(u,u,w)$ C:$(u,v,0)$ GP:$(u,v,w)$ |
| 44.109.1.1 | $I^{\bar{1}}4_1{}^{1}m^{\bar{1}}d^{\infty m}1$ | N:$(1/2,0,1/2)$ Q:$(1/2,v,1/2)$ Σ:$(u,0,0)$ B:$(u,0,w)$ C:$(u,v,0)$ |

| | | GP:(u,v,w) |
|---|---|---|
| 80.109.1.1 | $I^{1}4_{1}{}^{\bar{1}}m^{\bar{1}}d^{\infty m}1$ | C:(u,v,0) GP:(u,v,w) |
| 43.110.1.1 | $I^{\bar{1}}4_{1}{}^{\bar{1}}c^{1}d^{\infty m}1$ | X:(1/2,1/2,0) Δ:(u,u,0) W:(1/2,1/2,w) Y:(u,1-u,0) A:(u,u,w) C:(u,v,0) GP:(u,v,w) |
| 45.110.1.1 | $I^{\bar{1}}4_{1}{}^{1}c^{\bar{1}}d^{\infty m}1$ | N:(1/2,0,1/2) Q:(1/2,v,1/2) Σ:(u,0,0) B:(u,0,w) C:(u,v,0) GP:(u,v,w) |
| 80.110.1.1 | $I^{1}4_{1}{}^{\bar{1}}c^{\bar{1}}d^{\infty m}1$ | C:(u,v,0) GP:(u,v,w) |
| 35.111.1.1 | $P^{\bar{1}}\bar{4}^{\bar{1}}2^{1}m^{\infty m}1$ | S:(u,u,1/2) Σ:(u,u,0) C:(u,u,w) D:(u,v,0) E:(u,v,1/2) GP:(u,v,w) |
| 16.111.1.1 | $P^{\bar{1}}\bar{4}^{1}2^{\bar{1}}m^{\infty m}1$ | R:(0,1/2,1/2) X:(0,1/2,0) Δ:(0,v,0) T:(u,1/2,1/2) U:(0,v,1/2) W:(0,1/2,w) Y:(u,1/2,0) B:(0,v,w) D:(u,v,0) E:(u,v,1/2) F:(u,1/2,w) GP:(u,v,w) |
| 81.111.1.1 | $P^{1}\bar{4}^{\bar{1}}2^{\bar{1}}m^{\infty m}1$ | D:(u,v,0) E:(u,v,1/2) GP:(u,v,w) |
| 37.112.1.1 | $P^{\bar{1}}\bar{4}^{\bar{1}}2^{1}c^{\infty m}1$ | S:(u,u,1/2) Σ:(u,u,0) C:(u,u,w) D:(u,v,0) E:(u,v,1/2) GP:(u,v,w) |
| 16.112.1.1 | $P^{\bar{1}}\bar{4}^{1}2^{\bar{1}}c^{\infty m}1$ | R:(0,1/2,1/2) X:(0,1/2,0) Δ:(0,v,0) T:(u,1/2,1/2) U:(0,v,1/2) W:(0,1/2,w) Y:(u,1/2,0) B:(0,v,w) D:(u,v,0) E:(u,v,1/2) F:(u,1/2,w) GP:(u,v,w) |
| 81.112.1.1 | $P^{1}\bar{4}^{\bar{1}}2^{\bar{1}}c^{\infty m}1$ | D:(u,v,0) E:(u,v,1/2) GP:(u,v,w) |
| 35.113.1.1 | $P^{\bar{1}}\bar{4}^{\bar{1}}2_{1}{}^{1}m^{\infty m}1$ | S:(u,u,1/2) Σ:(u,u,0) C:(u,u,w) D:(u,v,0) E:(u,v,1/2) GP:(u,v,w) |
| 18.113.1.1 | $P^{\bar{1}}\bar{4}^{1}2_{1}{}^{\bar{1}}m^{\infty m}1$ | R:(0,1/2,1/2) X:(0,1/2,0) Δ:(0,v,0) T:(u,1/2,1/2) U:(0,v,1/2) W:(0,1/2,w) Y:(u,1/2,0) B:(0,v,w) D:(u,v,0) E:(u,v,1/2) F:(u,1/2,w) GP:(u,v,w) |
| 81.113.1.1 | $P^{1}\bar{4}^{\bar{1}}2_{1}{}^{\bar{1}}m^{\infty m}1$ | D:(u,v,0) E:(u,v,1/2) GP:(u,v,w) |
| 37.114.1.1 | $P^{\bar{1}}\bar{4}^{\bar{1}}2_{1}{}^{1}c^{\infty m}1$ | S:(u,u,1/2) Σ:(u,u,0) C:(u,u,w) D:(u,v,0) E:(u,v,1/2) GP:(u,v,w) |
| 18.114.1.1 | $P^{\bar{1}}\bar{4}^{1}2_{1}{}^{\bar{1}}c^{\infty m}1$ | R:(0,1/2,1/2) X:(0,1/2,0) Δ:(0,v,0) T:(u,1/2,1/2) U:(0,v,1/2) W:(0,1/2,w) Y:(u,1/2,0) B:(0,v,w) D:(u,v,0) E:(u,v,1/2) F:(u,1/2,w) GP:(u,v,w) |
| 81.114.1.1 | $P^{1}\bar{4}^{\bar{1}}2_{1}{}^{\bar{1}}c^{\infty m}1$ | D:(u,v,0) E:(u,v,1/2) GP:(u,v,w) |
| 21.115.1.1 | $P^{\bar{1}}\bar{4}^{\bar{1}}m^{1}2^{\infty m}1$ | S:(u,u,1/2) Σ:(u,u,0) C:(u,u,w) D:(u,v,0) E:(u,v,1/2) GP:(u,v,w) |
| 25.115.1.1 | $P^{\bar{1}}\bar{4}^{1}m^{\bar{1}}2^{\infty m}1$ | R:(0,1/2,1/2) X:(0,1/2,0) Δ:(0,v,0) T:(u,1/2,1/2) U:(0,v,1/2) W:(0,1/2,w) Y:(u,1/2,0) B:(0,v,w) D:(u,v,0) E:(u,v,1/2) F:(u,1/2,w) GP:(u,v,w) |
| 81.115.1.1 | $P^{1}\bar{4}^{\bar{1}}m^{\bar{1}}2^{\infty m}1$ | D:(u,v,0) E:(u,v,1/2) GP:(u,v,w) |
| 21.116.1.1 | $P^{\bar{1}}\bar{4}^{\bar{1}}c^{1}2^{\infty m}1$ | S:(u,u,1/2) Σ:(u,u,0) C:(u,u,w) D:(u,v,0) E:(u,v,1/2) GP:(u,v,w) |
| 27.116.1.1 | $P^{\bar{1}}\bar{4}^{1}c^{\bar{1}}2^{\infty m}1$ | R:(0,1/2,1/2) X:(0,1/2,0) Δ:(0,v,0) T:(u,1/2,1/2) U:(0,v,1/2) W:(0,1/2,w) Y:(u,1/2,0) B:(0,v,w) D:(u,v,0) E:(u,v,1/2) F:(u,1/2,w) GP:(u,v,w) |
| 81.116.1.1 | $P^{1}\bar{4}^{\bar{1}}c^{\bar{1}}2^{\infty m}1$ | D:(u,v,0) E:(u,v,1/2) GP:(u,v,w) |
| 21.117.1.1 | $P^{\bar{1}}\bar{4}^{\bar{1}}b^{1}2^{\infty m}1$ | S:(u,u,1/2) Σ:(u,u,0) C:(u,u,w) D:(u,v,0) E:(u,v,1/2) GP:(u,v,w) |
| 32.117.1.1 | $P^{\bar{1}}\bar{4}^{1}b^{\bar{1}}2^{\infty m}1$ | R:(0,1/2,1/2) X:(0,1/2,0) Δ:(0,v,0) T:(u,1/2,1/2) U:(0,v,1/2) W:(0,1/2,w) Y:(u,1/2,0) B:(0,v,w) D:(u,v,0) E:(u,v,1/2) F:(u,1/2,w) GP:(u,v,w) |
| 81.117.1.1 | $P^{1}\bar{4}^{\bar{1}}b^{\bar{1}}2^{\infty m}1$ | D:(u,v,0) E:(u,v,1/2) GP:(u,v,w) |
| 21.118.1.1 | $P^{\bar{1}}\bar{4}^{\bar{1}}n^{1}2^{\infty m}1$ | S:(u,u,1/2) Σ:(u,u,0) C:(u,u,w) D:(u,v,0) E:(u,v,1/2) GP:(u,v,w) |

| 34.118.1.1 | $P^{\bar{1}}\bar{4}^{1}n^{\bar{1}}2^{\infty m}1$ | R:(0,1/2,1/2) X:(0,1/2,0) Δ:(0,$v$,0) T:($u$,1/2,1/2) U:(0,$v$,1/2) W:(0,1/2,$w$) Y:($u$,1/2,0) B:(0,$v$,$w$) D:($u$,$v$,0) E:($u$,$v$,1/2) F:($u$,1/2,$w$) GP:($u$,$v$,$w$) |
|---|---|---|
| 81.118.1.1 | $P^{1}\bar{4}^{\bar{1}}n^{\bar{1}}2^{\infty m}1$ | D:($u$,$v$,0) E:($u$,$v$,1/2) GP:($u$,$v$,$w$) |
| 22.119.1.1 | $I^{\bar{1}}\bar{4}^{\bar{1}}m^{1}2^{\infty m}1$ | X:(1/2,1/2,0) Δ:($u$,$u$,0) W:(1/2,1/2,$w$) Y:($u$,1-$u$,0) A:($u$,$u$,$w$) C:($u$,$v$,0) GP:($u$,$v$,$w$) |
| 44.119.1.1 | $I^{\bar{1}}\bar{4}^{1}m^{\bar{1}}2^{\infty m}1$ | N:(1/2,0,1/2) Q:(1/2,$v$,1/2) Σ:($u$,0,0) B:($u$,0,$w$) C:($u$,$v$,0) GP:($u$,$v$,$w$) |
| 82.119.1.1 | $I^{1}\bar{4}^{\bar{1}}m^{\bar{1}}2^{\infty m}1$ | C:($u$,$v$,0) GP:($u$,$v$,$w$) |
| 22.120.1.1 | $I^{\bar{1}}\bar{4}^{\bar{1}}c^{1}2^{\infty m}1$ | X:(1/2,1/2,0) Δ:($u$,$u$,0) W:(1/2,1/2,$w$) Y:($u$,1-$u$,0) A:($u$,$u$,$w$) C:($u$,$v$,0) GP:($u$,$v$,$w$) |
| 45.120.1.1 | $I^{\bar{1}}\bar{4}^{1}c^{\bar{1}}2^{\infty m}1$ | N:(1/2,0,1/2) Q:(1/2,$v$,1/2) Σ:($u$,0,0) B:($u$,0,$w$) C:($u$,$v$,0) GP:($u$,$v$,$w$) |
| 82.120.1.1 | $I^{1}\bar{4}^{\bar{1}}c^{\bar{1}}2^{\infty m}1$ | C:($u$,$v$,0) GP:($u$,$v$,$w$) |
| 42.121.1.1 | $I^{\bar{1}}\bar{4}^{\bar{1}}2^{1}m^{\infty m}1$ | X:(1/2,1/2,0) Δ:($u$,$u$,0) W:(1/2,1/2,$w$) Y:($u$,1-$u$,0) A:($u$,$u$,$w$) C:($u$,$v$,0) GP:($u$,$v$,$w$) |
| 23.121.1.1 | $I^{\bar{1}}\bar{4}^{1}2^{\bar{1}}m^{\infty m}1$ | N:(1/2,0,1/2) Q:(1/2,$v$,1/2) Σ:($u$,0,0) B:($u$,0,$w$) C:($u$,$v$,0) GP:($u$,$v$,$w$) |
| 82.121.1.1 | $I^{1}\bar{4}^{\bar{1}}2^{\bar{1}}m^{\infty m}1$ | C:($u$,$v$,0) GP:($u$,$v$,$w$) |
| 43.122.1.1 | $I^{\bar{1}}\bar{4}^{\bar{1}}2^{1}d^{\infty m}1$ | X:(1/2,1/2,0) Δ:($u$,$u$,0) W:(1/2,1/2,$w$) Y:($u$,1-$u$,0) A:($u$,$u$,$w$) C:($u$,$v$,0) GP:($u$,$v$,$w$) |
| 24.122.1.1 | $I^{\bar{1}}\bar{4}^{1}2^{\bar{1}}d^{\infty m}1$ | N:(1/2,0,1/2) Q:(1/2,$v$,1/2) Σ:($u$,0,0) B:($u$,0,$w$) C:($u$,$v$,0) GP:($u$,$v$,$w$) |
| 82.122.1.1 | $I^{1}\bar{4}^{\bar{1}}2^{\bar{1}}d^{\infty m}1$ | C:($u$,$v$,0) GP:($u$,$v$,$w$) |
| 65.123.1.1 | $P^{\bar{1}}4/^{1}m^{\bar{1}}m^{1}m^{\infty m}1$ | S:($u$,$u$,1/2) Σ:($u$,$u$,0) C:($u$,$u$,$w$) D:($u$,$v$,0) E:($u$,$v$,1/2) GP:($u$,$v$,$w$) |
| 47.123.1.1 | $P^{\bar{1}}4/^{1}m^{1}m^{\bar{1}}m^{\infty m}1$ | R:(0,1/2,1/2) X:(0,1/2,0) Δ:(0,$v$,0) T:($u$,1/2,1/2) U:(0,$v$,1/2) W:(0,1/2,$w$) Y:($u$,1/2,0) B:(0,$v$,$w$) D:($u$,$v$,0) E:($u$,$v$,1/2) F:($u$,1/2,$w$) GP:($u$,$v$,$w$) |
| 83.123.1.1 | $P^{1}4/^{1}m^{\bar{1}}m^{\bar{1}}m^{\infty m}1$ | D:($u$,$v$,0) E:($u$,$v$,1/2) GP:($u$,$v$,$w$) |
| 66.124.1.1 | $P^{\bar{1}}4/^{1}m^{\bar{1}}c^{1}c^{\infty m}1$ | S:($u$,$u$,1/2) Σ:($u$,$u$,0) C:($u$,$u$,$w$) D:($u$,$v$,0) E:($u$,$v$,1/2) GP:($u$,$v$,$w$) |
| 49.124.1.1 | $P^{\bar{1}}4/^{1}m^{1}c^{\bar{1}}c^{\infty m}1$ | R:(0,1/2,1/2) X:(0,1/2,0) Δ:(0,$v$,0) T:($u$,1/2,1/2) U:(0,$v$,1/2) W:(0,1/2,$w$) Y:($u$,1/2,0) B:(0,$v$,$w$) D:($u$,$v$,0) E:($u$,$v$,1/2) F:($u$,1/2,$w$) GP:($u$,$v$,$w$) |
| 83.124.1.1 | $P^{1}4/^{1}m^{\bar{1}}c^{\bar{1}}c^{\infty m}1$ | D:($u$,$v$,0) E:($u$,$v$,1/2) GP:($u$,$v$,$w$) |
| 67.125.1.1 | $P^{\bar{1}}4/^{1}n^{\bar{1}}b^{1}m^{\infty m}1$ | S:($u$,$u$,1/2) Σ:($u$,$u$,0) C:($u$,$u$,$w$) D:($u$,$v$,0) E:($u$,$v$,1/2) GP:($u$,$v$,$w$) |
| 50.125.1.1 | $P^{\bar{1}}4/^{1}n^{1}b^{\bar{1}}m^{\infty m}1$ | R:(0,1/2,1/2) X:(0,1/2,0) Δ:(0,$v$,0) T:($u$,1/2,1/2) U:(0,$v$,1/2) W:(0,1/2,$w$) Y:($u$,1/2,0) B:(0,$v$,$w$) D:($u$,$v$,0) E:($u$,$v$,1/2) F:($u$,1/2,$w$) GP:($u$,$v$,$w$) |
| 85.125.1.1 | $P^{1}4/^{1}n^{\bar{1}}b^{\bar{1}}m^{\infty m}1$ | D:($u$,$v$,0) E:($u$,$v$,1/2) GP:($u$,$v$,$w$) |
| 68.126.1.1 | $P^{\bar{1}}4/^{1}n^{\bar{1}}n^{1}c^{\infty m}1$ | S:($u$,$u$,1/2) Σ:($u$,$u$,0) C:($u$,$u$,$w$) D:($u$,$v$,0) E:($u$,$v$,1/2) GP:($u$,$v$,$w$) |
| 48.126.1.1 | $P^{\bar{1}}4/^{1}n^{1}n^{\bar{1}}c^{\infty m}1$ | R:(0,1/2,1/2) X:(0,1/2,0) Δ:(0,$v$,0) T:($u$,1/2,1/2) U:(0,$v$,1/2) W:(0,1/2,$w$) Y:($u$,1/2,0) B:(0,$v$,$w$) D:($u$,$v$,0) E:($u$,$v$,1/2) F:($u$,1/2,$w$) GP:($u$,$v$,$w$) |
| 85.126.1.1 | $P^{1}4/^{1}n^{\bar{1}}n^{\bar{1}}c^{\infty m}1$ | D:(*u*,*v*,0) E:(*u*,*v*,1/2) GP:(*u*,*v*,*w*) |
| 65.127.1.1 | $P^{\bar{1}}4/^{1}m^{\bar{1}}b^{1}m^{\infty m}1$ | S:(*u*,*u*,1/2) Σ:(*u*,*u*,0) C:(*u*,*u*,*w*) D:(*u*,*v*,0) E:(*u*,*v*,1/2) GP:(*u*,*v*,*w*) |
| 55.127.1.1 | $P^{\bar{1}}4/^{1}m^{1}b^{\bar{1}}m^{\infty m}1$ | R:(0,1/2,1/2) X:(0,1/2,0) Δ:(0,*v*,0) T:(*u*,1/2,1/2) U:(0,*v*,1/2) W:(0,1/2,*w*) Y:(*u*,1/2,0) B:(0,*v*,*w*) D:(*u*,*v*,0) E:(*u*,*v*,1/2) F:(*u*,1/2,*w*) GP:(*u*,*v*,*w*) |
| 83.127.1.1 | $P^{1}4/^{1}m^{\bar{1}}b^{\bar{1}}m^{\infty m}1$ | D:(*u*,*v*,0) E:(*u*,*v*,1/2) GP:(*u*,*v*,*w*) |
| 66.128.1.1 | $P^{\bar{1}}4/^{1}m^{\bar{1}}n^{1}c^{\infty m}1$ | S:(*u*,*u*,1/2) Σ:(*u*,*u*,0) C:(*u*,*u*,*w*) D:(*u*,*v*,0) E:(*u*,*v*,1/2) GP:(*u*,*v*,*w*) |
| 58.128.1.1 | $P^{\bar{1}}4/^{1}m^{1}n^{\bar{1}}c^{\infty m}1$ | R:(0,1/2,1/2) X:(0,1/2,0) Δ:(0,*v*,0) T:(*u*,1/2,1/2) U:(0,*v*,1/2) W:(0,1/2,*w*) Y:(*u*,1/2,0) B:(0,*v*,*w*) D:(*u*,*v*,0) E:(*u*,*v*,1/2) F:(*u*,1/2,*w*) GP:(*u*,*v*,*w*) |
| 83.128.1.1 | $P^{1}4/^{1}m^{\bar{1}}n^{\bar{1}}c^{\infty m}1$ | D:(*u*,*v*,0) E:(*u*,*v*,1/2) GP:(*u*,*v*,*w*) |
| 67.129.1.1 | $P^{\bar{1}}4/^{1}n^{\bar{1}}m^{1}m^{\infty m}1$ | S:(*u*,*u*,1/2) Σ:(*u*,*u*,0) C:(*u*,*u*,*w*) D:(*u*,*v*,0) E:(*u*,*v*,1/2) GP:(*u*,*v*,*w*) |
| 59.129.1.1 | $P^{\bar{1}}4/^{1}n^{1}m^{\bar{1}}m^{\infty m}1$ | R:(0,1/2,1/2) X:(0,1/2,0) Δ:(0,*v*,0) T:(*u*,1/2,1/2) U:(0,*v*,1/2) W:(0,1/2,*w*) Y:(*u*,1/2,0) B:(0,*v*,*w*) D:(*u*,*v*,0) E:(*u*,*v*,1/2) F:(*u*,1/2,*w*) GP:(*u*,*v*,*w*) |
| 85.129.1.1 | $P^{1}4/^{1}n^{\bar{1}}m^{\bar{1}}m^{\infty m}1$ | D:(*u*,*v*,0) E:(*u*,*v*,1/2) GP:(*u*,*v*,*w*) |
| 68.130.1.1 | $P^{\bar{1}}4/^{1}n^{\bar{1}}c^{1}c^{\infty m}1$ | S:(*u*,*u*,1/2) Σ:(*u*,*u*,0) C:(*u*,*u*,*w*) D:(*u*,*v*,0) E:(*u*,*v*,1/2) GP:(*u*,*v*,*w*) |
| 56.130.1.1 | $P^{\bar{1}}4/^{1}n^{1}c^{\bar{1}}c^{\infty m}1$ | R:(0,1/2,1/2) X:(0,1/2,0) Δ:(0,*v*,0) T:(*u*,1/2,1/2) U:(0,*v*,1/2) W:(0,1/2,*w*) Y:(*u*,1/2,0) B:(0,*v*,*w*) D:(*u*,*v*,0) E:(*u*,*v*,1/2) F:(*u*,1/2,*w*) GP:(*u*,*v*,*w*) |
| 85.130.1.1 | $P^{1}4/^{1}n^{\bar{1}}c^{\bar{1}}c^{\infty m}1$ | D:(*u*,*v*,0) E:(*u*,*v*,1/2) GP:(*u*,*v*,*w*) |
| 66.131.1.1 | $P^{\bar{1}}4_2/^{1}m^{\bar{1}}m^{1}c^{\infty m}1$ | S:(*u*,*u*,1/2) Σ:(*u*,*u*,0) C:(*u*,*u*,*w*) D:(*u*,*v*,0) E:(*u*,*v*,1/2) GP:(*u*,*v*,*w*) |
| 47.131.1.1 | $P^{\bar{1}}4_2/^{1}m^{1}m^{\bar{1}}c^{\infty m}1$ | R:(0,1/2,1/2) X:(0,1/2,0) Δ:(0,*v*,0) T:(*u*,1/2,1/2) U:(0,*v*,1/2) W:(0,1/2,*w*) Y:(*u*,1/2,0) B:(0,*v*,*w*) D:(*u*,*v*,0) E:(*u*,*v*,1/2) F:(*u*,1/2,*w*) GP:(*u*,*v*,*w*) |
| 84.131.1.1 | $P^{1}4_2/^{1}m^{\bar{1}}m^{\bar{1}}c^{\infty m}1$ | D:(*u*,*v*,0) E:(*u*,*v*,1/2) GP:(*u*,*v*,*w*) |
| 65.132.1.1 | $P^{\bar{1}}4_2/^{1}m^{\bar{1}}c^{1}m^{\infty m}1$ | S:(*u*,*u*,1/2) Σ:(*u*,*u*,0) C:(*u*,*u*,*w*) D:(*u*,*v*,0) E:(*u*,*v*,1/2) GP:(*u*,*v*,*w*) |
| 49.132.1.1 | $P^{\bar{1}}4_2/^{1}m^{1}c^{\bar{1}}m^{\infty m}1$ | R:(0,1/2,1/2) X:(0,1/2,0) Δ:(0,*v*,0) T:(*u*,1/2,1/2) U:(0,*v*,1/2) W:(0,1/2,*w*) Y:(*u*,1/2,0) B:(0,*v*,*w*) D:(*u*,*v*,0) E:(*u*,*v*,1/2) F:(*u*,1/2,*w*) GP:(*u*,*v*,*w*) |
| 84.132.1.1 | $P^{1}4_2/^{1}m^{\bar{1}}c^{\bar{1}}m^{\infty m}1$ | D:(*u*,*v*,0) E:(*u*,*v*,1/2) GP:(*u*,*v*,*w*) |
| 68.133.1.1 | $P^{\bar{1}}4_2/^{1}n^{\bar{1}}b^{1}c^{\infty m}1$ | S:(*u*,*u*,1/2) Σ:(*u*,*u*,0) C:(*u*,*u*,*w*) D:(*u*,*v*,0) E:(*u*,*v*,1/2) GP:(*u*,*v*,*w*) |
| 50.133.1.1 | $P^{\bar{1}}4_2/^{1}n^{1}b^{\bar{1}}c^{\infty m}1$ | R:(0,1/2,1/2) X:(0,1/2,0) Δ:(0,*v*,0) T:(*u*,1/2,1/2) U:(0,*v*,1/2) W:(0,1/2,*w*) Y:(*u*,1/2,0) B:(0,*v*,*w*) D:(*u*,*v*,0) E:(*u*,*v*,1/2) F:(*u*,1/2,*w*) GP:(*u*,*v*,*w*) |
| 86.133.1.1 | $P^{1}4_2/^{1}n^{\bar{1}}b^{\bar{1}}c^{\infty m}1$ | D:(*u*,*v*,0) E:(*u*,*v*,1/2) GP:(*u*,*v*,*w*) |
| 67.134.1.1 | $P^{\bar{1}}4_2/^{1}n^{\bar{1}}n^{1}m^{\infty m}1$ | S:(*u*,*u*,1/2) Σ:(*u*,*u*,0) C:(*u*,*u*,*w*) D:(*u*,*v*,0) E:(*u*,*v*,1/2) GP:(*u*,*v*,*w*) |
| 48.134.1.1 | $P^{\bar{1}}4_2/^{1}n^{1}n^{\bar{1}}m^{\infty m}1$ | R:(0,1/2,1/2) X:(0,1/2,0) Δ:(0,*v*,0) T:(*u*,1/2,1/2) U:(0,*v*,1/2) W:(0,1/2,*w*) Y:(*u*,1/2,0) B:(0,*v*,*w*) D:(*u*,*v*,0) E:(*u*,*v*,1/2) F:(*u*,1/2,*w*) GP:(*u*,*v*,*w*) |
| 86.134.1.1 | $P^{1}4_2/^{1}n^{\bar{1}}n^{\bar{1}}m^{\infty m}1$ | D:(*u*,*v*,0) E:(*u*,*v*,1/2) GP:(*u*,*v*,*w*) |
| 66.135.1.1 | $P^{\bar{1}}4_2/^{1}m^{\bar{1}}b^{1}c^{\infty m}1$ | S:(*u*,*u*,1/2) Σ:(*u*,*u*,0) C:(*u*,*u*,*w*) D:(*u*,*v*,0) E:(*u*,*v*,1/2) GP:(*u*,*v*,*w*) |
| 55.135.1.1 | $P^{\bar{1}}4_2/^{1}m^{1}b^{\bar{1}}c^{\infty m}1$ | R:(0,1/2,1/2) X:(0,1/2,0) Δ:(0,*v*,0) T:(*u*,1/2,1/2) U:(0,*v*,1/2) |
| | | W:(0,1/2,*w*) Y:(*u*,1/2,0) B:(0,*v*,*w*) D:(*u*,*v*,0) E:(*u*,*v*,1/2) F:(*u*,1/2,*w*) GP:(*u*,*v*,*w*) |
| 84.135.1.1 | $P^{1}4_{2}/{}^{1}m^{\bar{1}}b^{\bar{1}}c^{\infty m}1$ | D:(*u*,*v*,0) E:(*u*,*v*,1/2) GP:(*u*,*v*,*w*) |
| 65.136.1.1 | $P^{\bar{1}}4_{2}/{}^{1}m^{\bar{1}}n^{1}m^{\infty m}1$ | S:(*u*,*u*,1/2) Σ:(*u*,*u*,0) C:(*u*,*u*,*w*) D:(*u*,*v*,0) E:(*u*,*v*,1/2) GP:(*u*,*v*,*w*) |
| 58.136.1.1 | $P^{\bar{1}}4_{2}/{}^{1}m^{1}n^{\bar{1}}m^{\infty m}1$ | R:(0,1/2,1/2) X:(0,1/2,0) Δ:(0,*v*,0) T:(*u*,1/2,1/2) U:(0,*v*,1/2) W:(0,1/2,*w*) Y:(*u*,1/2,0) B:(0,*v*,*w*) D:(*u*,*v*,0) E:(*u*,*v*,1/2) F:(*u*,1/2,*w*) GP:(*u*,*v*,*w*) |
| 84.136.1.1 | $P^{1}4_{2}/{}^{1}m^{\bar{1}}n^{\bar{1}}m^{\infty m}1$ | D:(*u*,*v*,0) E:(*u*,*v*,1/2) GP:(*u*,*v*,*w*) |
| 68.137.1.1 | $P^{\bar{1}}4_{2}/{}^{1}n^{\bar{1}}m^{1}c^{\infty m}1$ | S:(*u*,*u*,1/2) Σ:(*u*,*u*,0) C:(*u*,*u*,*w*) D:(*u*,*v*,0) E:(*u*,*v*,1/2) GP:(*u*,*v*,*w*) |
| 59.137.1.1 | $P^{\bar{1}}4_{2}/{}^{1}n^{1}m^{\bar{1}}c^{\infty m}1$ | R:(0,1/2,1/2) X:(0,1/2,0) Δ:(0,*v*,0) T:(*u*,1/2,1/2) U:(0,*v*,1/2) W:(0,1/2,*w*) Y:(*u*,1/2,0) B:(0,*v*,*w*) D:(*u*,*v*,0) E:(*u*,*v*,1/2) F:(*u*,1/2,*w*) GP:(*u*,*v*,*w*) |
| 86.137.1.1 | $P^{1}4_{2}/{}^{1}n^{\bar{1}}m^{\bar{1}}c^{\infty m}1$ | D:(*u*,*v*,0) E:(*u*,*v*,1/2) GP:(*u*,*v*,*w*) |
| 67.138.1.1 | $P^{\bar{1}}4_{2}/{}^{1}n^{\bar{1}}c^{1}m^{\infty m}1$ | S:(*u*,*u*,1/2) Σ:(*u*,*u*,0) C:(*u*,*u*,*w*) D:(*u*,*v*,0) E:(*u*,*v*,1/2) GP:(*u*,*v*,*w*) |
| 56.138.1.1 | $P^{\bar{1}}4_{2}/{}^{1}n^{1}c^{\bar{1}}m^{\infty m}1$ | R:(0,1/2,1/2) X:(0,1/2,0) Δ:(0,*v*,0) T:(*u*,1/2,1/2) U:(0,*v*,1/2) W:(0,1/2,*w*) Y:(*u*,1/2,0) B:(0,*v*,*w*) D:(*u*,*v*,0) E:(*u*,*v*,1/2) F:(*u*,1/2,*w*) GP:(*u*,*v*,*w*) |
| 86.138.1.1 | $P^{1}4_{2}/{}^{1}n^{\bar{1}}c^{\bar{1}}m^{\infty m}1$ | D:(*u*,*v*,0) E:(*u*,*v*,1/2) GP:(*u*,*v*,*w*) |
| 69.139.1.1 | $I^{\bar{1}}4/{}^{1}m^{\bar{1}}m^{1}m^{\infty m}1$ | X:(1/2,1/2,0) Δ:(*u*,*u*,0) W:(1/2,1/2,*w*) Y:(*u*,1-*u*,0) A:(*u*,*u*,*w*) C:(*u*,*v*,0) GP:(*u*,*v*,*w*) |
| 71.139.1.1 | $I^{\bar{1}}4/{}^{1}m^{1}m^{\bar{1}}m^{\infty m}1$ | N:(1/2,0,1/2) Q:(1/2,*v*,1/2) Σ:(*u*,0,0) B:(*u*,0,*w*) C:(*u*,*v*,0) GP:(*u*,*v*,*w*) |
| 87.139.1.1 | $I^{1}4/{}^{1}m^{\bar{1}}m^{\bar{1}}m^{\infty m}1$ | C:(*u*,*v*,0) GP:(*u*,*v*,*w*) |
| 69.140.1.1 | $I^{\bar{1}}4/{}^{1}m^{\bar{1}}c^{1}m^{\infty m}1$ | X:(1/2,1/2,0) Δ:(*u*,*u*,0) W:(1/2,1/2,*w*) Y:(*u*,1-*u*,0) A:(*u*,*u*,*w*) C:(*u*,*v*,0) GP:(*u*,*v*,*w*) |
| 72.140.1.1 | $I^{\bar{1}}4/{}^{1}m^{1}c^{\bar{1}}m^{\infty m}1$ | N:(1/2,0,1/2) Q:(1/2,*v*,1/2) Σ:(*u*,0,0) B:(*u*,0,*w*) C:(*u*,*v*,0) GP:(*u*,*v*,*w*) |
| 87.140.1.1 | $I^{1}4/{}^{1}m^{\bar{1}}c^{\bar{1}}m^{\infty m}1$ | C:(*u*,*v*,0) GP:(*u*,*v*,*w*) |
| 70.141.1.1 | $I^{\bar{1}}4_{1}/{}^{1}a^{\bar{1}}m^{1}d^{\infty m}1$ | X:(1/2,1/2,0) Δ:(*u*,*u*,0) W:(1/2,1/2,*w*) Y:(*u*,1-*u*,0) A:(*u*,*u*,*w*) C:(*u*,*v*,0) GP:(*u*,*v*,*w*) |
| 74.141.1.1 | $I^{\bar{1}}4_{1}/{}^{1}a^{1}m^{\bar{1}}d^{\infty m}1$ | N:(1/2,0,1/2) Q:(1/2,*v*,1/2) Σ:(*u*,0,0) B:(*u*,0,*w*) C:(*u*,*v*,0) GP:(*u*,*v*,*w*) |
| 88.141.1.1 | $I^{1}4_{1}/{}^{1}a^{\bar{1}}m^{\bar{1}}d^{\infty m}1$ | C:(*u*,*v*,0) GP:(*u*,*v*,*w*) |
| 70.142.1.1 | $I^{\bar{1}}4_{1}/{}^{1}a^{\bar{1}}c^{1}d^{\infty m}1$ | X:(1/2,1/2,0) Δ:(*u*,*u*,0) W:(1/2,1/2,*w*) Y:(*u*,1-*u*,0) A:(*u*,*u*,*w*) C:(*u*,*v*,0) GP:(*u*,*v*,*w*) |
| 73.142.1.1 | $I^{\bar{1}}4_{1}/{}^{1}a^{1}c^{\bar{1}}d^{\infty m}1$ | N:(1/2,0,1/2) Q:(1/2,*v*,1/2) Σ:(*u*,0,0) B:(*u*,0,*w*) C:(*u*,*v*,0) GP:(*u*,*v*,*w*) |
| 88.142.1.1 | $I^{1}4_{1}/{}^{1}a^{\bar{1}}c^{\bar{1}}d^{\infty m}1$ | C:(*u*,*v*,0) GP:(*u*,*v*,*w*) |
| 143.149.1.1 | $P^{1}3^{1}1^{\bar{1}}2^{\infty m}1$ | B:(*u*,*v*,0) D:(*u*,0,*w*) E:(*u*,*v*,1/2) GP:(*u*,*v*,*w*) |
| 143.150.1.1 | $P^{1}3^{\bar{1}}2^{1}1^{\infty m}1$ | P:(1/3,1/3,*w*) B:(*u*,*v*,0) C:(*u*,*u*,*w*) E:(*u*,*v*,1/2) GP:(*u*,*v*,*w*) |
| 144.151.1.1 | $P^{1}3_{1}{}^{1}1^{\bar{1}}2^{\infty m}1$ | B:(*u*,*v*,0) D:(*u*,0,*w*) E:(*u*,*v*,1/2) GP:(*u*,*v*,*w*) |
| 144.152.1.1 | $P^{1}3_{1}{}^{\bar{1}}2^{1}1^{\infty m}1$ | P:(1/3,1/3,*w*) B:(*u*,*v*,0) C:(*u*,*u*,*w*) E:(*u*,*v*,1/2) GP:(*u*,*v*,*w*) |
| 145.153.1.1 | $P^{1}3_{2}{}^{1}1^{\bar{1}}2^{\infty m}1$ | B:(*u*,*v*,0) D:(*u*,0,*w*) E:(*u*,*v*,1/2) GP:(*u*,*v*,*w*) |
| 145.154.1.1 | $P^{1}3_{2}{}^{\bar{1}}2^{1}1^{\infty m}1$ | P:(1/3,1/3,*w*) B:(*u*,*v*,0) C:(*u*,*u*,*w*) E:(*u*,*v*,1/2) GP:(*u*,*v*,*w*) |
| 146.155.1.1 | $R^{1}3^{\bar{1}}2^{\infty m}1$ | GP:(*u*,*v*,*w*) |
| 143.156.1.1 | $P^{1}3^{\bar{1}}m^{1}1^{\infty}1$ | P:(1/3,1/3,*w*) B:(*u*,*v*,0) C:(*u*,*u*,*w*) E:(*u*,*v*,1/2) GP:(*u*,*v*,*w*) |
| 143.157.1.1 | $P^{1}3^{1}1^{\bar{1}}m^{\infty m}1$ | B:(*u*,*v*,0) D:(*u*,0,*w*) E:(*u*,*v*,1/2) GP:(*u*,*v*,*w*) |
| 143.158.1.1 | $P^{1}3^{\bar{1}}c^{1}1^{\infty m}1$ | P:(1/3,1/3,*w*) B:(*u*,*v*,0) C:(*u*,*u*,*w*) E:(*u*,*v*,1/2) GP:(*u*,*v*,*w*) |
| 143.159.1.1 | $P^{1}3^{1}1^{\bar{1}}c^{\infty m}1$ | B:(*u*,*v*,0) D:(*u*,0,*w*) E:(*u*,*v*,1/2) GP:(*u*,*v*,*w*) |
| 146.160.1.1 | $R^{1}3^{\bar{1}}m^{\infty m}1$ | GP:(*u*,*v*,*w*) |
| 146.161.1.1 | $R^{1}3^{\bar{1}}c^{\infty m}1$ | GP:(*u*,*v*,*w*) |
| 147.162.1.1 | $P^{1}\bar{3}^{1}1^{\bar{1}}m^{\infty m}1$ | B:(*u*,*v*,0) D:(*u*,0,*w*) E:(*u*,*v*,1/2) GP:(*u*,*v*,*w*) |
| 147.163.1.1 | $P^{1}\bar{3}^{1}1^{\bar{1}}c^{\infty m}1$ | B:(*u*,*v*,0) D:(*u*,0,*w*) E:(*u*,*v*,1/2) GP:(*u*,*v*,*w*) |
| 147.164.1.1 | $P^{1}\bar{3}^{\bar{1}}m^{1}1^{\infty m}1$ | L:(1/2,0,1/2) M:(1/2,0,0) P:(1/3,1/3,*w*) B:(*u*,*v*,0) C:(*u*,*u*,*w*)<br>E:(*u*,*v*,1/2) GP:(*u*,*v*,*w*) |
| 147.165.1.1 | $P^{1}\bar{3}^{\bar{1}}c^{1}1^{\infty m}1$ | L:(1/2,0,1/2) M:(1/2,0,0) P:(1/3,1/3,*w*) B:(*u*,*v*,0) C:(*u*,*u*,*w*)<br>E:(*u*,*v*,1/2) GP:(*u*,*v*,*w*) |
| 148.166.1.1 | $R^{1}\bar{3}^{\bar{1}}m^{\infty m}1$ | GP:(*u*,*v*,*w*) |
| 148.167.1.1 | $R^{1}\bar{3}^{\bar{1}}c^{\infty m}1$ | GP:(*u*,*v*,*w*) |
| 143.168.1.1 | $P^{\bar{1}}6^{\infty m}1$ | P:(1/3,1/3,*w*) C:(*u*,*u*,*w*) D:(*u*,0,*w*) GP:(*u*,*v*,*w*) |
| 144.169.1.1 | $P^{\bar{1}}6_{1}{}^{\infty m}1$ | P:(1/3,1/3,*w*) C:(*u*,*u*,*w*) D:(*u*,0,*w*) GP:(*u*,*v*,*w*) |
| 145.170.1.1 | $P^{\bar{1}}6_{5}{}^{\infty m}1$ | P:(1/3,1/3,*w*) C:(*u*,*u*,*w*) D:(*u*,0,*w*) GP:(*u*,*v*,*w*) |
| 145.171.1.1 | $P^{\bar{1}}6_{2}{}^{\infty m}1$ | P:(1/3,1/3,*w*) C:(*u*,*u*,*w*) D:(*u*,0,*w*) GP:(*u*,*v*,*w*) |
| 144.172.1.1 | $P^{\bar{1}}6_{4}{}^{\infty m}1$ | P:(1/3,1/3,*w*) C:(*u*,*u*,*w*) D:(*u*,0,*w*) GP:(*u*,*v*,*w*) |
| 143.173.1.1 | $P^{\bar{1}}6_{3}{}^{\infty m}1$ | P:(1/3,1/3,*w*) C:(*u*,*u*,*w*) D:(*u*,0,*w*) GP:(*u*,*v*,*w*) |
| 143.174.1.1 | $P^{\bar{1}}\bar{6}^{\infty m}1$ | P:(1/3,1/3,*w*) C:(*u*,*u*,*w*) D:(*u*,0,*w*) GP:(*u*,*v*,*w*) |
| 147.175.1.1 | $P^{\bar{1}}6/{}^{\bar{1}}m^{\infty m}1$ | P:(1/3,1/3,*w*) C:(*u*,*u*,*w*) D:(*u*,0,*w*) GP:(*u*,*v*,*w*) |
| 147.176.1.1 | $P^{\bar{1}}6_{3}/{}^{\bar{1}}m^{\infty m}1$ | P:(1/3,1/3,*w*) C:(*u*,*u*,*w*) D:(*u*,0,*w*) GP:(*u*,*v*,*w*) |
| 149.177.1.1 | $P^{\bar{1}}6^{\bar{1}}2^{1}2^{\infty m}1$ | P:(1/3,1/3,*w*) C:(*u*,*u*,*w*) GP:(*u*,*v*,*w*) |
| 150.177.1.1 | $P^{\bar{1}}6^{1}2^{\bar{1}}2^{\infty m}1$ | D:(*u*,0,*w*) GP:(*u*,*v*,*w*) |
| 168.177.1.1 | $P^{1}6^{\bar{1}}2^{\bar{1}}2^{\infty m}1$ | B:(*u*,*v*,0) E:(*u*,*v*,1/2) GP:(*u*,*v*,*w*) |
| 151.178.1.1 | $P^{\bar{1}}6_{1}{}^{\bar{1}}2^{1}2^{\infty m}1$ | P:(1/3,1/3,*w*) C:(*u*,*u*,*w*) GP:(*u*,*v*,*w*) |
| 152.178.1.1 | $P^{\bar{1}}6_{1}{}^{1}2^{\bar{1}}2^{\infty m}1$ | D:(*u*,0,*w*) GP:(*u*,*v*,*w*) |
| 169.178.1.1 | $P^{1}6_{1}{}^{\bar{1}}2^{\bar{1}}2^{\infty m}1$ | B:(*u*,*v*,0) E:(*u*,*v*,1/2) GP:(*u*,*v*,*w*) |
| 153.179.1.1 | $P^{\bar{1}}6_{5}{}^{\bar{1}}2^{1}2^{\infty m}1$ | P:(1/3,1/3,*w*) C:(*u*,*u*,*w*) GP:(*u*,*v*,*w*) |
| 154.179.1.1 | $P^{\bar{1}}6_{5}{}^{1}2^{\bar{1}}2^{\infty m}1$ | D:(*u*,0,*w*) GP:(*u*,*v*,*w*) |
| 170.179.1.1 | $P^{1}6_{5}{}^{\bar{1}}2^{\bar{1}}2^{\infty m}1$ | B:(*u*,*v*,0) E:(*u*,*v*,1/2) GP:(*u*,*v*,*w*) |
| 153.180.1.1 | $P^{\bar{1}}6_{2}{}^{\bar{1}}2^{1}2^{\infty m}1$ | P:(1/3,1/3,*w*) C:(*u*,*u*,*w*) GP:(*u*,*v*,*w*) |
| 154.180.1.1 | $P^{\bar{1}}6_{2}{}^{1}2^{\bar{1}}2^{\infty m}1$ | D:(*u*,0,*w*) GP:(*u*,*v*,*w*) |
| 171.180.1.1 | $P^{1}6_{2}{}^{\bar{1}}2^{\bar{1}}2^{\infty m}1$ | B:(*u*,*v*,0) E:(*u*,*v*,1/2) GP:(*u*,*v*,*w*) |
| 151.181.1.1 | $P^{\bar{1}}6_{4}{}^{\bar{1}}2^{1}2^{\infty m}1$ | P:(1/3,1/3,*w*) C:(*u*,*u*,*w*) GP:(*u*,*v*,*w*) |
| 152.181.1.1 | $P^{\bar{1}}6_{4}{}^{1}2^{\bar{1}}2^{\infty m}1$ | D:(*u*,0,*w*) GP:(*u*,*v*,*w*) |
| 172.181.1.1 | $P^{1}6_{4}{}^{\bar{1}}2^{\bar{1}}2^{\infty m}1$ | B:(*u*,*v*,0) E:(*u*,*v*,1/2) GP:(*u*,*v*,*w*) |
| 149.182.1.1 | $P^{\bar{1}}6_{3}{}^{\bar{1}}2^{1}2^{\infty m}1$ | P:(1/3,1/3,*w*) C:(*u*,*u*,*w*) GP:(*u*,*v*,*w*) |
| 150.182.1.1 | $P^{\bar{1}}6_{3}{}^{1}2^{\bar{1}}2^{\infty m}1$ | D:(*u*,0,*w*) GP:(*u*,*v*,*w*) |
| 173.182.1.1 | $P^{1}6_{3}{}^{\bar{1}}2^{\bar{1}}2^{\infty m}1$ | B:(*u*,*v*,0) E:(*u*,*v*,1/2) GP:(*u*,*v*,*w*) |
| 157.183.1.1 | $P^{\bar{1}}6^{\bar{1}}m^{1}m^{\infty m}1$ | P:(1/3,1/3,*w*) C:(*u*,*u*,*w*) GP:(*u*,*v*,*w*) |
| 156.183.1.1 | $P^{\bar{1}}6^{1}m^{\bar{1}}m^{\infty m}1$ | D:(*u*,0,*w*) GP:(*u*,*v*,*w*) |
| 168.183.1.1 | $P^{1}6^{\bar{1}}m^{\bar{1}}m^{\infty m}1$ | B:(*u*,*v*,0) E:(*u*,*v*,1/2) GP:(*u*,*v*,*w*) |
| 159.184.1.1 | $P^{\bar{1}}6^{\bar{1}}c^{1}c^{\infty m}1$ | P:(1/3,1/3,*w*) C:(*u*,*u*,*w*) GP:(*u*,*v*,*w*) |
| 158.184.1.1 | $P^{\bar{1}}6^{1}c^{\bar{1}}c^{\infty m}1$ | D:(*u*,0,*w*) GP:(*u*,*v*,*w*) |
| 168.184.1.1 | $P^{1}6^{\bar{1}}c^{\bar{1}}c^{\infty m}1$ | B:(*u*,*v*,0) E:(*u*,*v*,1/2) GP:(*u*,*v*,*w*) |
| 157.185.1.1 | $P^{\bar{1}}6_3{}^{\bar{1}}c^{1}m^{\infty m}1$ | P:(1/3,1/3,*w*) C:(*u*,*u*,*w*) GP:(*u*,*v*,*w*) |
| 158.185.1.1 | $P^{\bar{1}}6_3{}^{1}c^{\bar{1}}m^{\infty m}1$ | D:(*u*,0,*w*) GP:(*u*,*v*,*w*) |
| 173.185.1.1 | $P^{1}6_3{}^{\bar{1}}c^{\bar{1}}m^{\infty m}1$ | B:(*u*,*v*,0) E:(*u*,*v*,1/2) GP:(*u*,*v*,*w*) |
| 159.186.1.1 | $P^{\bar{1}}6_3{}^{\bar{1}}m^{1}c^{\infty m}1$ | P:(1/3,1/3,*w*) C:(*u*,*u*,*w*) GP:(*u*,*v*,*w*) |
| 156.186.1.1 | $P^{\bar{1}}6_3{}^{1}m^{\bar{1}}c^{\infty m}1$ | D:(*u*,0,*w*) GP:(*u*,*v*,*w*) |
| 173.186.1.1 | $P^{1}6_3{}^{\bar{1}}m^{\bar{1}}c^{\infty m}1$ | B:(*u*,*v*,0) E:(*u*,*v*,1/2) GP:(*u*,*v*,*w*) |
| 149.187.1.1 | $P^{\bar{1}}\bar{6}^{\bar{1}}m^{1}2^{\infty m}1$ | P:(1/3,1/3,*w*) C:(*u*,*u*,*w*) D:(*u*,0,*w*) GP:(*u*,*v*,*w*) |
| 156.187.1.1 | $P^{\bar{1}}\bar{6}^{1}m^{\bar{1}}2^{\infty m}1$ | D:(*u*,0,*w*) GP:(*u*,*v*,*w*) |
| 174.187.1.1 | $P^{1}\bar{6}^{\bar{1}}m^{\bar{1}}2^{\infty m}1$ | B:(*u*,*v*,0) E:(*u*,*v*,1/2) GP:(*u*,*v*,*w*) |
| 149.188.1.1 | $P^{\bar{1}}\bar{6}^{\bar{1}}c^{1}2^{\infty m}1$ | P:(1/3,1/3,*w*) C:(*u*,*u*,*w*) D:(*u*,0,*w*) GP:(*u*,*v*,*w*) |
| 158.188.1.1 | $P^{\bar{1}}\bar{6}^{1}c^{\bar{1}}2^{\infty m}1$ | D:(*u*,0,*w*) GP:(*u*,*v*,*w*) |
| 174.188.1.1 | $P^{1}\bar{6}^{\bar{1}}c^{\bar{1}}2^{\infty m}1$ | B:(*u*,*v*,0) E:(*u*,*v*,1/2) GP:(*u*,*v*,*w*) |
| 157.189.1.1 | $P^{\bar{1}}\bar{6}^{\bar{1}}2^{1}m^{\infty m}1$ | P:(1/3,1/3,*w*) C:(*u*,*u*,*w*) D:(*u*,0,*w*) GP:(*u*,*v*,*w*) |
| 150.189.1.1 | $P^{\bar{1}}\bar{6}^{1}2^{\bar{1}}m^{\infty m}1$ | D:(*u*,0,*w*) GP:(*u*,*v*,*w*) |
| 174.189.1.1 | $P^{1}\bar{6}^{\bar{1}}2^{\bar{1}}m^{\infty m}1$ | B:(*u*,*v*,0) E:(*u*,*v*,1/2) GP:(*u*,*v*,*w*) |
| 159.190.1.1 | $P^{\bar{1}}\bar{6}^{\bar{1}}2^{1}c^{\infty m}1$ | P:(1/3,1/3,*w*) C:(*u*,*u*,*w*) D:(*u*,0,*w*) GP:(*u*,*v*,*w*) |
| 150.190.1.1 | $P^{\bar{1}}\bar{6}^{1}2^{\bar{1}}c^{\infty m}1$ | D:(*u*,0,*w*) GP:(*u*,*v*,*w*) |
| 174.190.1.1 | $P^{1}\bar{6}^{\bar{1}}2^{\bar{1}}c^{\infty m}1$ | B:(*u*,*v*,0) E:(*u*,*v*,1/2) GP:(*u*,*v*,*w*) |
| 162.191.1.1 | $P^{\bar{1}}6/{}^{\bar{1}}m^{\bar{1}}m^{1}m^{\infty m}1$ | P:(1/3,1/3,*w*) C:(*u*,*u*,*w*) GP:(*u*,*v*,*w*) |
| 164.191.1.1 | $P^{\bar{1}}6/{}^{\bar{1}}m^{1}m^{\bar{1}}m^{\infty m}1$ | D:(*u*,0,*w*) GP:(*u*,*v*,*w*) |
| 175.191.1.1 | $P^{1}6/{}^{1}m^{\bar{1}}m^{\bar{1}}m^{\infty m}1$ | B:(*u*,*v*,0) E:(*u*,*v*,1/2) GP:(*u*,*v*,*w*) |
| 163.192.1.1 | $P^{\bar{1}}6/{}^{\bar{1}}m^{\bar{1}}c^{1}c^{\infty m}1$ | P:(1/3,1/3,*w*) C:(*u*,*u*,*w*) GP:(*u*,*v*,*w*) |
| 165.192.1.1 | $P^{\bar{1}}6/{}^{\bar{1}}m^{1}c^{\bar{1}}c^{\infty m}1$ | D:(*u*,0,*w*) GP:(*u*,*v*,*w*) |
| 175.192.1.1 | $P^{1}6/{}^{1}m^{\bar{1}}c^{\bar{1}}c^{\infty m}1$ | B:(*u*,*v*,0) E:(*u*,*v*,1/2) GP:(*u*,*v*,*w*) |
| 162.193.1.1 | $P^{\bar{1}}6_3/{}^{\bar{1}}m^{\bar{1}}c^{1}m^{\infty m}1$ | P:(1/3,1/3,*w*) C:(*u*,*u*,*w*) GP:(*u*,*v*,*w*) |
| 165.193.1.1 | $P^{\bar{1}}6_3/{}^{\bar{1}}m^{1}c^{\bar{1}}m^{\infty m}1$ | D:(*u*,0,*w*) GP:(*u*,*v*,*w*) |
| 176.193.1.1 | $P^{1}6_3/{}^{1}m^{\bar{1}}c^{\bar{1}}m^{\infty m}1$ | B:(*u*,*v*,0) E:(*u*,*v*,1/2) GP:(*u*,*v*,*w*) |
| 163.194.1.1 | $P^{\bar{1}}6_3/{}^{\bar{1}}m^{\bar{1}}m^{1}c^{\infty m}1$ | P:(1/3,1/3,*w*) C:(*u*,*u*,*w*) GP:(*u*,*v*,*w*) |
| 164.194.1.1 | $P^{\bar{1}}6_3/{}^{\bar{1}}m^{1}m^{\bar{1}}c^{\infty m}1$ | D:(*u*,0,*w*) GP:(*u*,*v*,*w*) |
| 176.194.1.1 | $P^{1}6_3/{}^{1}m^{\bar{1}}m^{\bar{1}}c^{\infty m}1$ | B:(*u*,*v*,0) E:(*u*,*v*,1/2) GP:(*u*,*v*,*w*) |
| 195.207.1.1 | $P^{\bar{1}}4^{1}3^{\bar{1}}2^{\infty m}1$ | Z:(*u*,1/2,0) A:(*u*,*v*,0) B:(*u*,1/2,*w*) GP:(*u*,*v*,*w*) |
| 195.208.1.1 | $P^{\bar{1}}4_2{}^{1}3^{\bar{1}}2^{\infty m}1$ | Z:(*u*,1/2,0) A:(*u*,*v*,0) B:(*u*,1/2,*w*) GP:(*u*,*v*,*w*) |
| 196.209.1.1 | $F^{\bar{1}}4^{1}3^{\bar{1}}2^{\infty m}1$ | V:(*u*,1,0) A:(*u*,*v*,0) GP:(*u*,*v*,*w*) |
| 196.210.1.1 | $F^{\bar{1}}4_1{}^{1}3^{\bar{1}}2^{\infty m}1$ | V:(*u*,1,0) A:(*u*,*v*,0) GP:(*u*,*v*,*w*) |
| 197.211.1.1 | $I^{\bar{1}}4^{1}3^{\bar{1}}2^{\infty m}1$ | A:(*u*,*v*,0) GP:(*u*,*v*,*w*) |
| 198.212.1.1 | $P^{\bar{1}}4_3{}^{1}3^{\bar{1}}2^{\infty m}1$ | Z:(*u*,1/2,0) A:(*u*,*v*,0) B:(*u*,1/2,*w*) GP:(*u*,*v*,*w*) |
| 198.213.1.1 | $P^{\bar{1}}4_1{}^{1}3^{\bar{1}}2^{\infty m}1$ | Z:(*u*,1/2,0) A:(*u*,*v*,0) B:(*u*,1/2,*w*) GP:(*u*,*v*,*w*) |

| 199.214.1.1 | $I^{\bar{1}}4_1{}^{1}3^{\bar{1}}2^{\infty m}1$ | A:$(u,v,0)$ GP:$(u,v,w)$ |
| --- | --- | --- |
| 195.215.1.1 | $P^{\bar{1}}\bar{4}^{1}3^{\bar{1}}m^{\infty m}1$ | Z:$(u,1/2,0)$ A:$(u,v,0)$ B:$(u,1/2,w)$ GP:$(u,v,w)$ |
| 196.216.1.1 | $F^{\bar{1}}\bar{4}^{1}3^{\bar{1}}m^{\infty m}1$ | V:$(u,1,0)$ A:$(u,v,0)$ GP:$(u,v,w)$ |
| 197.217.1.1 | $I^{\bar{1}}\bar{4}^{1}3^{\bar{1}}m^{\infty m}1$ | A:$(u,v,0)$ GP:$(u,v,w)$ |
| 195.218.1.1 | $P^{\bar{1}}\bar{4}^{1}3^{\bar{1}}n^{\infty m}1$ | Z:$(u,1/2,0)$ A:$(u,v,0)$ B:$(u,1/2,w)$ GP:$(u,v,w)$ |
| 196.219.1.1 | $F^{\bar{1}}\bar{4}^{1}3^{\bar{1}}c^{\infty m}1$ | V:$(u,1,0)$ A:$(u,v,0)$ GP:$(u,v,w)$ |
| 199.220.1.1 | $I^{\bar{1}}\bar{4}^{1}3^{\bar{1}}d^{\infty m}1$ | A:$(u,v,0)$ GP:$(u,v,w)$ |
| 200.221.1.1 | $P^{1}m^{1}\bar{3}^{\bar{1}}m^{\infty m}1$ | Z:$(u,1/2,0)$ A:$(u,v,0)$ B:$(u,1/2,w)$ GP:$(u,v,w)$ |
| 201.222.1.1 | $P^{1}n^{1}\bar{3}^{\bar{1}}n^{\infty m}1$ | Z:$(u,1/2,0)$ A:$(u,v,0)$ B:$(u,1/2,w)$ GP:$(u,v,w)$ |
| 200.223.1.1 | $P^{1}m^{1}\bar{3}^{\bar{1}}n^{\infty m}1$ | Z:$(u,1/2,0)$ A:$(u,v,0)$ B:$(u,1/2,w)$ GP:$(u,v,w)$ |
| 201.224.1.1 | $P^{1}n^{1}\bar{3}^{\bar{1}}m^{\infty m}1$ | Z:$(u,1/2,0)$ A:$(u,v,0)$ B:$(u,1/2,w)$ GP:$(u,v,w)$ |
| 202.225.1.1 | $F^{1}m^{1}\bar{3}^{\bar{1}}m^{\infty m}1$ | V:$(u,1,0)$ A:$(u,v,0)$ GP:$(u,v,w)$ |
| 202.226.1.1 | $F^{1}m^{1}\bar{3}^{\bar{1}}c^{\infty m}1$ | V:$(u,1,0)$ A:$(u,v,0)$ GP:$(u,v,w)$ |
| 203.227.1.1 | $F^{1}d^{1}\bar{3}^{\bar{1}}m^{\infty m}1$ | V:$(u,1,0)$ A:$(u,v,0)$ GP:$(u,v,w)$ |
| 203.228.1.1 | $F^{1}d^{1}\bar{3}^{\bar{1}}c^{\infty m}1$ | V:$(u,1,0)$ A:$(u,v,0)$ GP:$(u,v,w)$ |
| 204.229.1.1 | $I^{1}m^{1}\bar{3}^{\bar{1}}m^{\infty m}1$ | A:$(u,v,0)$ GP:$(u,v,w)$ |
| 206.230.1.1 | $I^{1}a^{1}\bar{3}^{\bar{1}}d^{\infty m}1$ | A:$(u,v,0)$ GP:$(u,v,w)$ |

## S4. Material candidates hosting unconventional magnons

### S4.1. Materials statistics

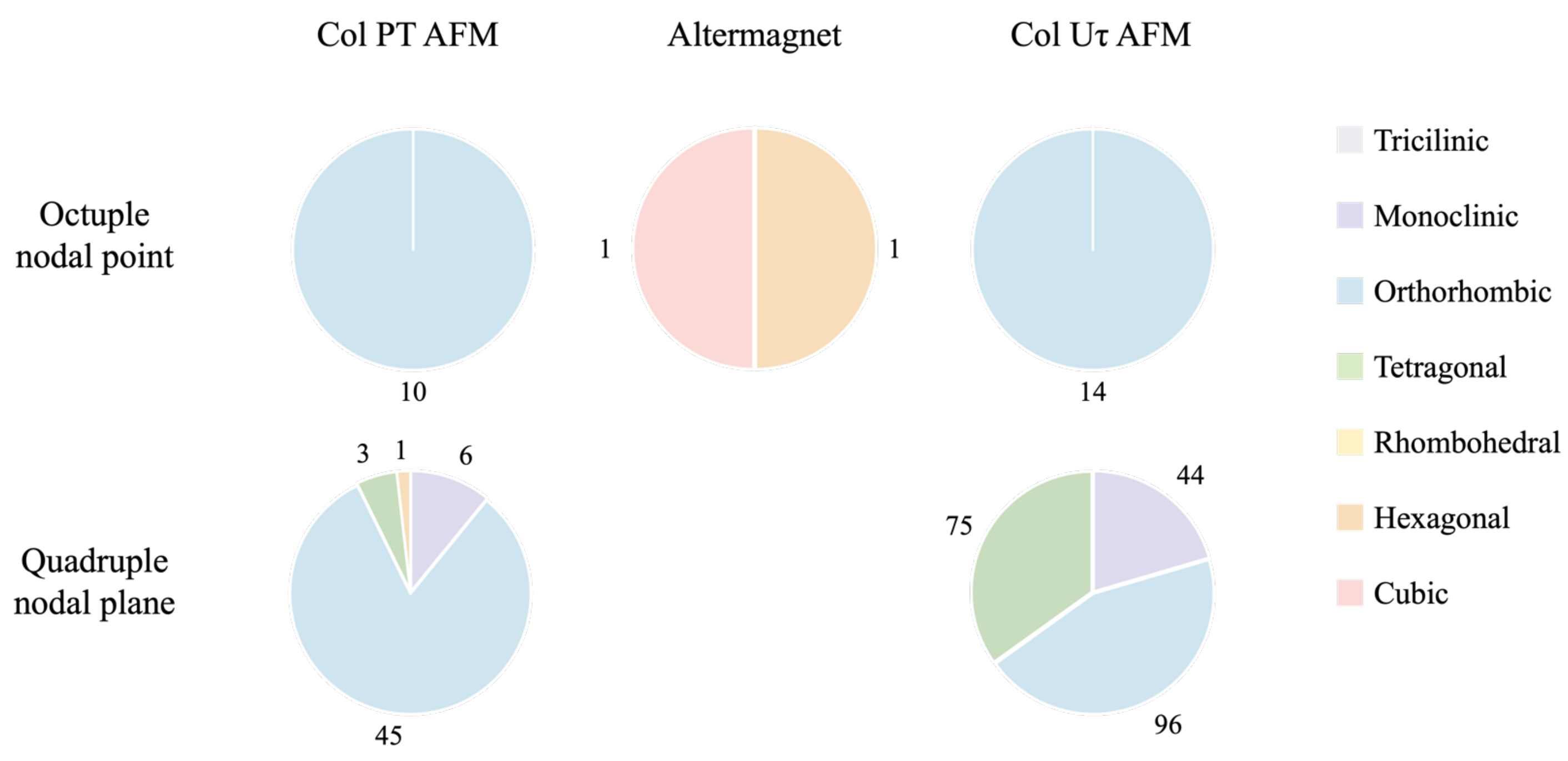

Supplementary Fig. 7. Statistics on the lattice systems of different high-dimensional nodal features in candidate collinear magnets.

Through a high-throughput search based on the band representations of collinear SSGs, we found 498 candidate materials with unconventional magnons, including 22 collinear FMs/FiMs, 56 collinear PT AFMs, 203 altermagnets and 217 collinear Uτ AFMs, which are given in Supplementary Tables XXIV~XXIX. For collinear FMs/FiMs, all candidate materials have triple nodal points, with three of them having sextuple nodal points. For collinear PT AFMs, there are 10, 1 and 55 candidate materials with octuple nodal points, sextuple nodal point and quadruple nodal plane, respectively. For altermagnets, all of them have chiral magnons, with 2 and 1 of them having octuple nodal points and sextuple nodal points, respectively. For collinear Uτ AFMs, there are 14, 14, 2, 215 candidate materials with octuple nodal lines, octuple nodal points, sextuple nodal points and quadruple nodal planes, respectively.

Through statistical analysis of the candidate materials, we found that triple and sextuple nodal points only appear in cubic lattices, while octuple nodal lines only appear in simple orthogonal lattices. In addition, the statistics of other high-dimensional nodal features in different lattice systems and in different types of collinear magnets are shown in Supplementary Fig. 7.

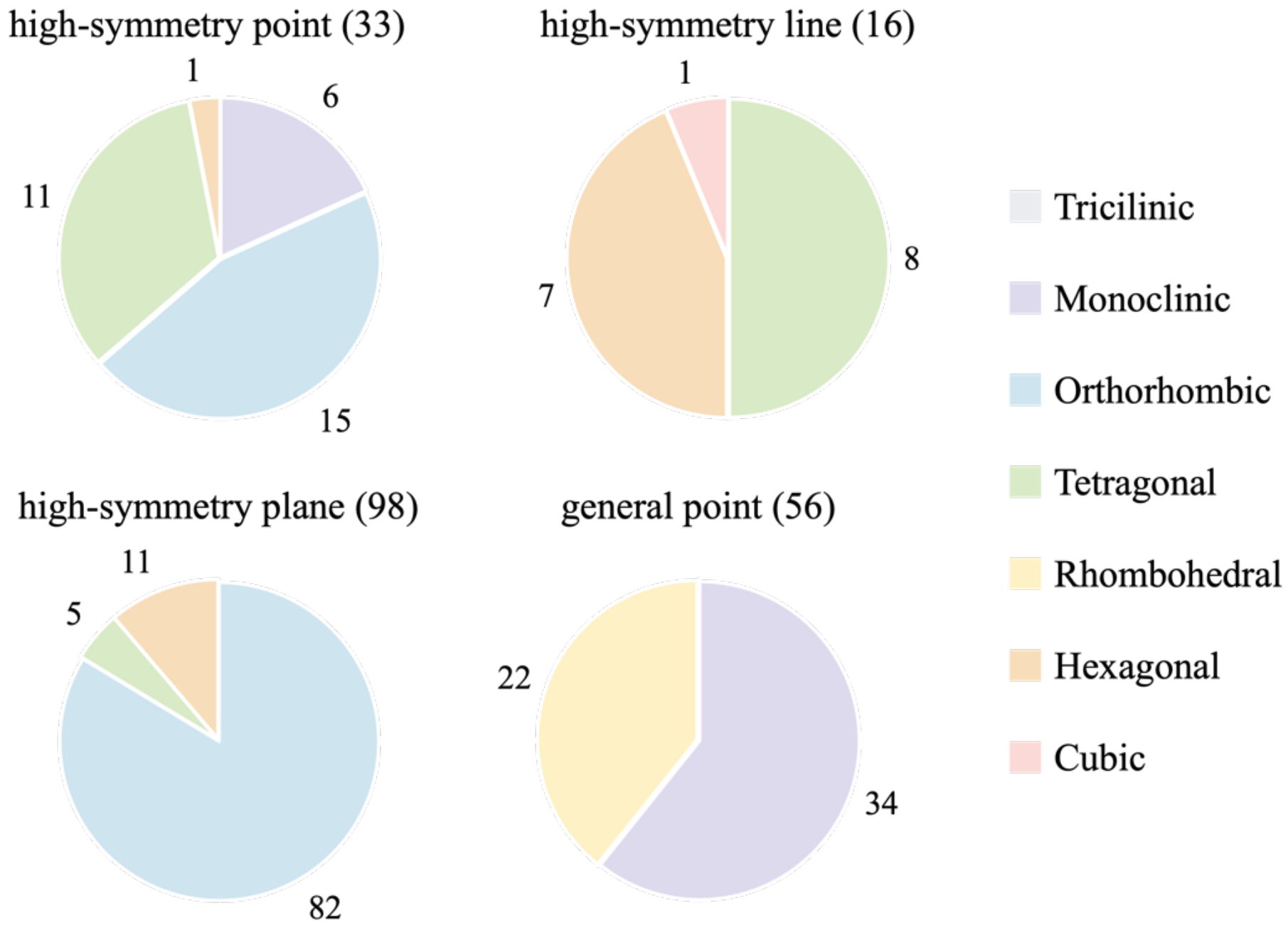

Supplementary Fig. 8. Statistics on the lattice systems of different momentum positions that host chiral magnons in candidate collinear magnets.

For altermagnets with chiral magnon, we also provide diagrams illustrating the momentum distribution of chiral magnons/electronic spin splitting in the collinear magnets with respect to the different lattice systems in Supplementary Fig. 8. Note that the numbers refer to the highest-symmetric momentum positions with magnon chirality splitting/electronic spin splitting. For example, if a collinear SSG $G_1$ has magnonic chirality splitting/electronic spin splitting at high-symmetry points, they are not included in the statistics of high-symmetry lines, planes and general points.

## S4.2. Collinear FMs and FiMs

Supplementary Table XXIV. Tabulation of collinear FMs that accommodate unconventional magnons. The ten columns illustrate the material ID in MAGNDATA, chemical formula, space group, spin space groups, spin Wyckoff Positions, the momentum-position, band representations, and the quasiparticle type for unconventional magnons, the orientation of magnetic materials and the corresponding magnetic space groups. The spin axis is written under the lattice basis, and the "$[a_1b_1c_1]$ +θ $[a_2b_2c_2]$" labels the axis by rotating from the direction of "$[a_1b_1c_1]$" towards the direction of "$[a_2b_2c_2]$" by θ degrees.

| ID | Formula | SG | SSG | WP | ***k*** point | Rep | Quasiparticle | Spin Axis | MSG |
|---|---|---|---|---|---|---|---|---|---|
| 0.275 | $Mn_3AlN$ | 221 ($Pm\bar{3}m$) | 221.221.1.1 ($P^1m^1\bar{3}^1m^{\infty m}1$) | Mn 3c | R:(1/2,1/2,1/2) | $R_5^{S,+}(3)$ | TP | [111] | 166.101 ($R\bar{3}m'$) |
| MP-20831 | $Gd_4Sb_3$[9] | 220 ($I\bar{4}3d$) | 220.220.1.1 ($I^1\bar{4}^13^1d^{\infty m}1$) | Gd 16c | Γ:(0,0,0) | $\Gamma_4^S(3), \Gamma_5^S(3)$ | TP | [001] | 122.337 ($I\bar{4}2'd'$) |
| | | | | | H:(1,1,1) | $H_4^S H_5^S(6)$ | SP | | |
| MP-530 | $Nd_4Sb_3$[9] | 220 ($I\bar{4}3d$) | 220.220.1.1 ($I^1\bar{4}^13^1d^{\infty m}1$) | Nd 16c | Γ:(0,0,0) | $\Gamma_4^S(3), \Gamma_5^S(3)$ | TP | [001] | 122.337 ($I\bar{4}2'd'$) |
| | | | | | H:(1,1,1) | $H_4^S H_5^S(6)$ | SP | | |
| MP-367 | $Sm_4Sb_3$[9] | 220 ($I\bar{4}3d$) | 220.220.1.1 ($I^1\bar{4}^13^1d^{\infty m}1$) | Sm 16c | Γ:(0,0,0) | $\Gamma_4^S(3), \Gamma_5^S(3)$ | TP | [001] | 122.337 ($I\bar{4}2'd'$) |
| | | | | | H:(1,1,1) | $H_4^S H_5^S(6)$ | SP | | |
| | $Lu_2V_2O_7$[10] | 227 ($Fd\bar{3}m$) | 227.227.1.1 ($F^1d^1\bar{3}^1m^{\infty m}1$) | V 16c | Γ:(0,0,0) | $\Gamma_5^{S,+}(3)$ | TP | [001] | 141.557 ($I4_1/am'd'$) |

Supplementary Table XXV. Tabulation of collinear FiMs that accommodate unconventional magnons. The ten columns illustrate the material ID in MAGNDATA, chemical formula, space group, spin space groups, spin Wyckoff Positions, the momentum-position, band representations, and the quasiparticle type for unconventional magnons, the orientation of magnetic materials and the corresponding magnetic space groups. The spin axis is written under the lattice basis, and the "$[a_1b_1c_1]$ +θ $[a_2b_2c_2]$" labels the axis by rotating from the direction of "$[a_1b_1c_1]$" towards the direction of "$[a_2b_2c_2]$" by θ degrees.

| ID | Formula | SG | SSG | WP | ***k*** point | Rep | Quasi-particle | Spin Axis | MSG |
|---|---|---|---|---|---|---|---|---|---|
| 0.35 | $Cu_2OSeO_3$ | 198 ($P2_13$) | 198.198.1.1 ($P^12_1{}^13^{\infty m}1$) | Cu 4a, 12b | Γ:(0,0,0) | $\Gamma_4^S(3)$ | $C$–2 TP | [111] | 146.10 ($R3$) |
| 0.274 | $Mn_4N$ | 221 ($Pm\bar{3}m$) | 221.221.1.1 ($P^1m^1\bar{3}^1m^{\infty m}1$) | Mn 1a, 3c | R:(1/2,1/2,1/2) | $R_5^{S,+}(3)$ | TP | [111] | 166.101 ($R\bar{3}m'$) |
| 0.508 | $FeMnO_3$ | 206 ($Ia\bar{3}$) | 206.206.1.1 ($I^1a^1\bar{3}^{\infty m}1$) | Fe 8b, Mn 24d | Γ:(0,0,0) | $\Gamma_4^{S,+}(3)$, $\Gamma_4^{S,-}(3)$ | TP | [001] | 73.551 ($Ib'c'a$) |
| | | | | | H:(1,1,1) | $H_4^{S,+}(3)$, $H_4^{S,-}(3)$ | TP | | |
| 0.570 | $Li_{0.5}FeCr_{1.5}O_4$ | 227 ($Fd\bar{3}m$) | 227.227.1.1 ($F^1d^1\bar{3}^1m^{\infty m}1$) | Fe 8b, 16c | Γ:(0,0,0) | $\Gamma_5^{S,+}(3)$ | TP | [001] | 141.557 ($I4_1/am'd'$) |
| 0.613 | $FeCr_2S_4$ | 227 ($Fd\bar{3}m$) | 227.227.1.1 ($F^1d^1\bar{3}^1m^{\infty m}1$) | Fe 8b, Cr 16c | Γ:(0,0,0) | $\Gamma_5^{S,+}(3)$ | TP | [001] | 141.557 ($I4_1/am'd'$) |
| 0.614 | $FeCr_2S_4$ | 227 ($Fd\bar{3}m$) | 227.227.1.1 ($F^1d^1\bar{3}^1m^{\infty m}1$) | Fe 8b, Cr 16c | Γ:(0,0,0) | $\Gamma_5^{S,+}(3)$ | TP | [001] | 141.557 ($I4_1/am'd'$) |
| 0.615 | $FeCr_2S_4$ | 227 ($Fd\bar{3}m$) | 227.227.1.1 ($F^1d^1\bar{3}^1m^{\infty m}1$) | Fe 8b, Cr 16c | Γ:(0,0,0) | $\Gamma_5^{S,+}(3)$ | TP | [001] | 141.557 ($I4_1/am'd'$) |
| 0.672 | $CaCu_3Fe_2Sb_2O_{12}$ | 201 ($Pn\bar{3}$) | 201.201.1.1 ($P^1n^1\bar{3}^{\infty m}1$) | Fe 4c, Cu 6d | Γ:(0,0,0) | $\Gamma_4^{S,+}(3)$ | TP | [001] | 48.260 ($Pn'n'n$) |
| | | | | | R:(1/2,1/2,1/2) | $R_4^{S,+}(3)$, $R_4^{S,-}(3)$ | TP | | |
| 0.713 | $NiFe_2O_4$ | 227 ($Fd\bar{3}m$) | 227.227.1.1 ($F^1d^1\bar{3}^1m^{\infty m}1$) | Ni 8b, Fe 16c | Γ:(0,0,0) | $\Gamma_5^{S,+}(3)$ | TP | [001] | 141.557 ($I4_1/am'd'$) |
| 0.725 | $Co_5TeO_8$ | 227 $(Fd\bar{3}m)$ | 227.227.1.1 $(F^1d^1\bar{3}^1m^{\infty m}1)$ | Co 8b, 16c | Γ:(0,0,0) | $\Gamma_5^{S,+}(3)$ | TP | [001] | 141.557 $(I4_1/am'd')$ |
| 0.886 | $SnCo_2O_4$ | 227 $(Fd\bar{3}m)$ | 227.227.1.1 $(F^1d^1\bar{3}^1m^{\infty m}1)$ | Co 8b, 16c | Γ:(0,0,0) | $\Gamma_5^{S,+}(3)$ | TP | [100] | 141.557 $(I4_1/am'd')$ |
| 0.887 | $MnCo_2O_4$ | 227 $(Fd\bar{3}m)$ | 227.227.1.1 $(F^1d^1\bar{3}^1m^{\infty m}1)$ | Co 8b, 16c | Γ:(0,0,0) | $\Gamma_5^{S,+}(3)$ | TP | [100] | 141.557 $(I4_1/am'd')$ |
| 0.888 | $Mn_{0.6}Co_{2.4}O_4$ | 227 $(Fd\bar{3}m)$ | 227.227.1.1 $(F^1d^1\bar{3}^1m^{\infty m}1)$ | Co 8b, Co/Mn 16c | Γ:(0,0,0) | $\Gamma_5^{S,+}(3)$ | TP | [100] | 141.557 $(I4_1/am'd')$ |
| 0.889 | $Mn_{0.8}Co_{2.2}O_4$ | 227 $(Fd\bar{3}m)$ | 227.227.1.1 $(F^1d^1\bar{3}^1m^{\infty m}1)$ | Co 8b, Co/Mn 16c | Γ:(0,0,0) | $\Gamma_5^{S,+}(3)$ | TP | [100] | 141.557 $(I4_1/am'd')$ |
| 0.890 | $Mn_{1.2}Co_{1.8}O_4$ | 227 $(Fd\bar{3}m)$ | 227.227.1.1 $(F^1d^1\bar{3}^1m^{\infty m}1)$ | Co 8b, Co/Mn 16c | Γ:(0,0,0) | $\Gamma_5^{S,+}(3)$ | TP | [100] | 141.557 $(I4_1/am'd')$ |
| 1.277 | $LiFeCr_4O_8$ | 216 $(F\bar{4}3m)$ | 216.216.1.1 $(F^1\bar{4}^13^1m^{\infty m}1)$ | Fe 4d, Cr 16e | Γ:(0,0,0) | $\Gamma_4^S(3)$ | TP | [001] | 119.319 $(I\bar{4}m'2')$ |
| | $LiFe_5O_8$[11] | 212 $(P4_332)$ | 212.212.1.1 $(P^14_3{}^13^12^{\infty m}1)$ | Fe 8c, 12d | Γ:(0,0,0) | $\Gamma_4^S(3)$, $\Gamma_5^S(3)$ | $C$–2 TP | [001] | 92.114 $(P4_12_1'2')$ |

## S4.3. Collinear PT AFMs

Supplementary Table XXVI. Tabulation of collinear PT AFMs that accommodate unconventional magnons. The ten columns illustrate the material ID in MAGNDATA, chemical formula, space group, spin space groups, spin Wyckoff Positions, the momentum-position, band representations, and the quasiparticle type for unconventional magnons, the orientation of magnetic materials and the corresponding magnetic space groups. The spin axis is written under the lattice basis, and the "[$a_1b_1c_1$] +θ [$a_2b_2c_2$]" labels the axis by rotating from the direction of "[$a_1b_1c_1$]" towards the direction of "[$a_2b_2c_2$]" by θ degrees.

| ID | Formula | SG | SSG | WP | ***k*** point | Rep | Quasi-particle | Spin Axis | MSG |
|---|---|---|---|---|---|---|---|---|---|
| 0.14 | $Gd_5Ge_4$ | 62 $(Pnma)$ | 26.62.1.1 $(P^{\bar{1}}n^1m^1a^{\infty m}1)$ | Gd 4c, 8d, 8d | L:(1/2,$v$,$w$) | $(L)W_1^SW_1^S(4)$ | QNPL | [001] | 62.444 $(Pnm'a)$ |
| 0.22 | $DyB_4$ | 127 $(P4/mbm)$ | 26.55.1.1 $(P^{\bar{1}}b^1a^1m^{\infty m}1)$ | Dy 4g | L:(1/2,$v$,$w$) | $(L)W_1^SW_1^S(4)$ | QNPL | [001] | 55.355 $(Pb'am)$ |

| 0.24 | $LiMnPO_4$ | 62<br>$(Pnma)$ | 26.62.1.1<br>$(P^{\bar{1}}n^{1}m^{1}a^{\infty m}1)$ | Mn 4c | L:(1/2,$v$,$w$) | $(L)W_1^SW_1^S(4)$ | QNPL | [100] | 62.449<br>$(Pn'm'a')$ |
|---|---|---|---|---|---|---|---|---|---|
| 0.71 | $Li_2Ni(SO_4)_2$ | 61<br>$(Pbca)$ | 29.61.1.1<br>$(P^{\bar{1}}b^{1}c^{1}a^{\infty m}1)$ | Ni 8c | R:(1/2,1/2,1/2) | $R_1^SR_1^S(8)$ | OP | [001] | 61.437<br>$(Pb'c'a')$ |
| | | | | | U:(1/2,0,1/2) | $U_1^SU_1^S(8)$ | OP | | |
| | | | | | L:(1/2,$v$,$w$) | $(L)W_1^SW_1^S(4)$ | QNPL | | |
| 0.75 | $Cr_2WO_6$ | 136<br>$(P4_2/mnm)$ | 113.136.1.1<br>$(P^{\bar{1}}4_2/^{\bar{1}}m^{\bar{1}}n^{1}m^{\infty m}1)$ | Cr 4e | F:($u$,1/2,$w$) | $F_1^SF_1^S(4)$ | QNPL | [010] | 58.395<br>$(Pn'nm)$ |
| 0.86 | $KMn_4(PO_4)_3$ | 62<br>$(Pnma)$ | 26.62.1.1<br>$(P^{\bar{1}}n^{1}m^{1}a^{\infty m}1)$ | Mn 4c,<br>4c, 8d | L:(1/2,$v$,$w$) | $(L)W_1^SW_1^S(4)$ | QNPL | [001] | 62.445<br>$(Pnma')$ |
| 0.87 | $NaFePO_4$ | 62<br>$(Pnma)$ | 26.62.1.1<br>$(P^{\bar{1}}n^{1}m^{1}a^{\infty m}1)$ | Fe 4c | L:(1/2,$v$,$w$) | $(L)W_1^SW_1^S(4)$ | QNPL | [010] | 62.445<br>$(Pnma')$ |
| 0.95 | $LiFePO_4$ | 62<br>$(Pnma)$ | 26.62.1.1<br>$(P^{\bar{1}}n^{1}m^{1}a^{\infty m}1)$ | Fe 4c | L:(1/2,$v$,$w$) | $(L)W_1^SW_1^S(4)$ | QNPL | [010] | 62.445<br>$(Pnma')$ |
| 0.144 | $Cr_2WO_6$ | 136<br>$(P4_2/mnm)$ | 113.136.1.1<br>$(P^{\bar{1}}4_2/^{\bar{1}}m^{\bar{1}}n^{1}m^{\infty m}1)$ | Cr 4e | F:($u$,1/2,$w$) | $F_1^SF_1^S(4)$ | QNPL | [010] | 58.395<br>$(Pn'nm)$ |
| 0.146 | $EuZrO_3$ | 62<br>$(Pnma)$ | 31.62.1.1<br>$(P^{1}n^{1}m^{\bar{1}}a^{\infty m}1)$ | Eu 4c | W:($u$,$v$,1/2) | $W_1^SW_1^S(4)$ | QNPL | [100] | 62.444<br>$(Pnm'a)$ |
| 0.147 | $EuZrO_3$ | 62<br>$(Pnma)$ | 31.62.1.1<br>$(P^{1}n^{1}m^{\bar{1}}a^{\infty m}1)$ | Eu 4c | W:($u$,$v$,1/2) | $W_1^SW_1^S(4)$ | QNPL | [001] | 62.449<br>$(Pn'm'a')$ |
| 0.161 | $CoSe_2O_5$ | 60<br>$(Pbcn)$ | 18.60.1.1<br>$(P^{\bar{1}}b^{\bar{1}}c^{\bar{1}}n^{\infty m}1)$ | Co 4c | L:(1/2,$v$,$w$) | $L_1^SL_1^S(4)$ | QNPL | [100]<br>+14°<br>([001]) | 60.419<br>$(Pb'cn)$ |
| | | | | | W:($u$,$v$,1/2) | $(W)N_1^SN_1^S(4)$ | QNPL | | |
| 0.193 | $LiCoPO_4$ | 62<br>$(Pnma)$ | 26.62.1.1<br>$(P^{\bar{1}}n^{1}m^{1}a^{\infty m}1)$ | Co 4c | L:(1/2,$v$,$w$) | $(L)W_1^SW_1^S(4)$ | QNPL | [010] | 62.445<br>$(Pnma')$ |
| 0.216 | $SrEr_2O_4$ | 62<br>$(Pnma)$ | 26.62.1.1<br>$(P^{\bar{1}}n^{1}m^{1}a^{\infty m}1)$ | Er 4c | L:(1/2,$v$,$w$) | $(L)W_1^SW_1^S(4)$ | QNPL | [001] | 62.445<br>$(Pnma')$ |
| 0.217 | $LiCrGe_2O_6$ | 14<br>$(P2_1/c)$ | 4.14.1.1<br>$(P^{1}2_1/^{\bar{1}}c^{\infty m}1)$ | Cr 4e | G:($u$, 1/2, $w$) | $G_1^SG_1^S(4)$ | QNPL | [100]<br>+30°<br>([001]) | 14.77<br>$(P2_1'/c)$ |
| 0.223 | $Cu_{0.95}MnAs$ | 62<br>$(Pnma)$ | 31.62.1.1<br>$(P^{1}n^{1}m^{\bar{1}}a^{\infty m}1)$ | Mn 4c | W:($u$,$v$,1/2) | $W_1^SW_1^S(4)$ | QNPL | [010] | 62.443<br>$(Pn'ma)$ |
| 0.243 | $Li_2Fe(SO_4)_2$ | 61<br>$(Pbca)$ | 29.61.1.1<br>$(P^{\bar{1}}b^{1}c^{1}a^{\infty m}1)$ | Fe 8c | R:(1/2,1/2,1/2)<br>U:(1/2,0,1/2)<br>L:(1/2,$v$,$w$) | $R_1^S R_1^S(8)$<br>$U_1^S U_1^S(8)$<br>$(L)W_1^S W_1^S(4)$ | OP<br>OP<br>QNPL | [100]<br>+25°<br>([010]) | 14.77<br>$(P2_1'/c)$ |
| 0.244 | $Li_2Co(SO_4)_2$ | 61<br>$(Pbca)$ | 29.61.1.1<br>$(P^{\bar{1}}b^{1}c^{1}a^{\infty m}1)$ | Co 8c | R:(1/2,1/2,1/2)<br>U:(1/2,0,1/2)<br>L:(1/2,$v$,$w$) | $R_1^S R_1^S(8)$<br>$U_1^S U_1^S(8)$<br>$(L)W_1^S W_1^S(4)$ | OP<br>OP<br>QNPL | [001] | 61.437<br>$(Pb'c'a')$ |
| 0.245 | $Li_{1.5}Fe(SO_4)_2$ | 61<br>$(Pbca)$ | 29.61.1.1<br>$(P^{\bar{1}}b^{1}c^{1}a^{\infty m}1)$ | Fe 8c | R:(1/2,1/2,1/2)<br>U:(1/2,0,1/2)<br>L:(1/2,$v$,$w$) | $R_1^S R_1^S(8)$<br>$U_1^S U_1^S(8)$<br>$(L)W_1^S W_1^S(4)$ | OP<br>OP<br>QNPL | [100]<br>+43°<br>([010]) | 14.77<br>$(P2_1'/c)$ |
| 0.246 | $LiFe(SO_4)_2$ | 61<br>$(Pbca)$ | 29.61.1.1<br>$(P^{\bar{1}}b^{1}c^{1}a^{\infty m}1)$ | Fe 8c | R:(1/2,1/2,1/2)<br>U:(1/2,0,1/2)<br>L:(1/2,$v$,$w$) | $R_1^S R_1^S(8)$<br>$U_1^S U_1^S(8)$<br>$(L)W_1^S W_1^S(4)$ | OP<br>OP<br>QNPL | [001] | 61.437<br>$(Pb'c'a')$ |
| 0.346 | $Tb_2ReC_2$ | 62<br>$(Pnma)$ | 26.62.1.1<br>$(P^{\bar{1}}n^{1}m^{1}a^{\infty m}1)$ | Tb 4c,<br>4c | L:(1/2,$v$,$w$) | $(L)W_1^S W_1^S(4)$ | QNPL | [010] | 62.445<br>$(Pnma')$ |
| 0.362 | $RbFeCl_5(D_2O)$ | 62<br>$(Pnma)$ | 26.62.1.1<br>$(P^{\bar{1}}n^{1}m^{1}a^{\infty m}1)$ | Fe 4c | L:(1/2,$v$,$w$) | $(L)W_1^S W_1^S(4)$ | QNPL | [100] | 62.449<br>$(Pn'm'a')$ |
| 0.363 | $KFeCl_5(D_2O)$ | 62<br>$(Pnma)$ | 26.62.1.1<br>$(P^{\bar{1}}n^{1}m^{1}a^{\infty m}1)$ | Fe 4c | L:(1/2,$v$,$w$) | $(L)W_1^S W_1^S(4)$ | QNPL | [100] | 62.449<br>$(Pn'm'a')$ |
| 0.382 | $LiMnPO_4$ | 62<br>$(Pnma)$ | 26.62.1.1<br>$(P^{\bar{1}}n^{1}m^{1}a^{\infty m}1)$ | Mn 4c | L:(1/2,$v$,$w$) | $(L)W_1^S W_1^S(4)$ | QNPL | [100] | 62.449<br>$(Pn'm'a')$ |
| 0.383 | $LiCoPO_4$ | 62<br>$(Pnma)$ | 26.62.1.1<br>$(P^{\bar{1}}n^{1}m^{1}a^{\infty m}1)$ | Co 4c | L:(1/2,$v$,$w$) | $(L)W_1^S W_1^S(4)$ | QNPL | [010] | 62.445<br>$(Pnma')$ |
| 0.384 | $LiCoPO_4$ | 62<br>$(Pnma)$ | 26.62.1.1<br>$(P^{\bar{1}}n^{1}m^{1}a^{\infty m}1)$ | Co 4c | L:(1/2,$v$,$w$) | $(L)W_1^S W_1^S(4)$ | QNPL | [010]<br>+4°<br>([001]) | 14.77<br>$(P2_1'/c)$ |
| 0.386 | $Fe_3BO_5$ | 62<br>$(Pnma)$ | 31.62.1.1<br>$(P^{1}n^{1}m^{\bar{1}}a^{\infty m}1)$ | Fe 4c,<br>4c, 4c | W:($u$,$v$,1/2) | $W_1^S W_1^S(4)$ | QNPL | [010] | 62.444<br>$(Pnm'a)$ |
| 0.399 | FeOOH | 62<br>$(Pnma)$ | 26.62.1.1<br>$(P^{\bar{1}}n^{1}m^{1}a^{\infty m}1)$ | Fe 4c | L:(1/2,$v$,$w$) | $(L)W_1^S W_1^S(4)$ | QNPL | [001] | 62.445<br>$(Pnma')$ |
| 0.410 | $GdAlO_3$ | 62<br>$(Pnma)$ | 31.62.1.1<br>$(P^{1}n^{1}m^{\bar{1}}a^{\infty m}1)$ | Gd 4c | W:($u$,$v$,1/2) | $W_1^S W_1^S(4)$ | QNPL | [100] | 62.449<br>$(Pn'm'a')$ |

| 0.421 | $EuMnSb_2$ | 62 ($Pnma$) | 26.62.1.1 ($P^{\bar{1}}n^{1}m^{1}a^{\infty m}1$) | Mn 4c | L:(1/2,$v$,$w$) | $(L)W_1^SW_1^S(4)$ | QNPL | [100] | 62.449 ($Pn'm'a'$) |
|---|---|---|---|---|---|---|---|---|---|
| 0.423 | $EuMnSb_2$ | 62 ($Pnma$) | 26.62.1.1 ($P^{\bar{1}}n^{1}m^{1}a^{\infty m}1$) | Mn 4c | L:(1/2,$v$,$w$) | $(L)W_1^SW_1^S(4)$ | QNPL | [100] | 62.449 ($Pn'm'a'$) |
| 0.429 | $CaCr_{0.86}Fe_{3.14}As_3$ | 62 ($Pnma$) | 31.62.1.1 ($P^{1}n^{1}m^{\bar{1}}a^{\infty m}1$) | Fe 4c, 4c, 4c, 4c | W:($u$,$v$,1/2) | $W_1^SW_1^S(4)$ | QNPL | [010] | 62.443 ($Pn'ma$) |
| 0.455 | $RbFeO_2$ | 61 ($Pbca$) | 29.61.1.1 ($P^{\bar{1}}b^{1}c^{1}a^{\infty m}1$) | Fe 8c, 8c | R:(1/2,1/2,1/2) | $R_1^SR_1^S(8)$ | OP | [001] | 61.437 ($Pb'c'a'$) |
| | | | | | U:(1/2,0,1/2) | $U_1^SU_1^S(8)$ | OP | | |
| | | | | | L:(1/2,$v$,$w$) | $(L)W_1^SW_1^S(4)$ | QNPL | | |
| 0.457 | $CsFeO_2$ | 61 ($Pbca$) | 29.61.1.1 ($P^{\bar{1}}b^{1}c^{1}a^{\infty m}1$) | Fe 8c, 8c | R:(1/2,1/2,1/2) | $R_1^SR_1^S(8)$ | OP | [001] | 61.437 ($Pb'c'a'$) |
| | | | | | U:(1/2,0,1/2) | $U_1^SU_1^S(8)$ | OP | | |
| | | | | | L:(1/2,$v$,$w$) | $(L)W_1^SW_1^S(4)$ | QNPL | | |
| 0.459 | $KFeO_2$ | 61 ($Pbca$) | 29.61.1.1 ($P^{\bar{1}}b^{1}c^{1}a^{\infty m}1$) | Fe 8c, 8c | R:(1/2,1/2,1/2) | $R_1^SR_1^S(8)$ | OP | [010] | 61.435 ($Pb'ca$) |
| | | | | | U:(1/2,0,1/2) | $U_1^SU_1^S(8)$ | OP | | |
| | | | | | L:(1/2,$v$,$w$) | $(L)W_1^SW_1^S(4)$ | QNPL | | |
| 0.460 | $KFeO_2$ | 61 ($Pbca$) | 29.61.1.1 ($P^{\bar{1}}b^{1}c^{1}a^{\infty m}1$) | Fe 8c, 8c | R:(1/2,1/2,1/2) | $R_1^SR_1^S(8)$ | OP | [010] | 61.435 ($Pb'ca$) |
| | | | | | U:(1/2,0,1/2) | $U_1^SU_1^S(8)$ | OP | | |
| | | | | | L:(1/2,$v$,$w$) | $(L)W_1^SW_1^S(4)$ | QNPL | | |
| 0.468 | $ErB_4$ | 127 ($P4/mbm$) | 26.55.1.1 ($P^{\bar{1}}b^{1}a^{1}m^{\infty m}1$) | Er 4g | L:(1/2,$v$,$w$) | $(L)W_1^SW_1^S(4)$ | QNPL | [001] | 55.355 ($Pb'am$) |
| 0.469 | $TbB_4$ | 127 ($P4/mbm$) | 26.55.1.1 ($P^{\bar{1}}b^{1}a^{1}m^{\infty m}1$) | Tb 4g | L:(1/2,$v$,$w$) | $(L)W_1^SW_1^S(4)$ | QNPL | [010] | 55.359 ($Pb'a'm'$) |
| 0.505 | $Pb_2VO(PO_4)_2$ | 14 ($P2_1/c$) | 4.14.1.1 ($P^{1}2_1/^{\bar{1}}c^{\infty m}1$) | V 4e | G:($u$, 1/2, $w$) | $G_1^SG_1^S(4)$ | QNPL | [010] | 14.78 ($P2_1/c'$) |
| 0.692 | $Ba_4Ru_3O_{10}$ | 64 ($Cmce$) | 36.64.1.1 ($C^{1}m^{1}c^{\bar{1}}e^{\infty m}1$) | Ru 8f | Q:($u$,$v$,1/2) | $Q_1^SQ_1^S(4)$ | QNPL | [010] | 64.471 ($Cm'ca$) |
| 0.693 | $Ba_4Ru_3O_{10}$ | 64 ($Cmce$) | 36.64.1.1 ($C^{1}m^{1}c^{\bar{1}}e^{\infty m}1$) | Ru 8f | Q:($u$,$v$,1/2) | $Q_1^SQ_1^S(4)$ | QNPL | [100] | 64.472 ($Cmc'a$) |
| 0.798 | $MnPd_2$ | 62 ($Pnma$) | 26.62.1.1 ($P^{\bar{1}}n^{1}m^{1}a^{\infty m}1$) | Mn 4c | L:(1/2,$v$,$w$) | $(L)W_1^SW_1^S(4)$ | QNPL | [010] | 62.445 ($Pnma'$) |

| 0.801 | $Tl_3Fe_2S_4$ | 62<br>$(Pnma)$ | 33.62.1.1<br>$(P^{1}n^{\bar{1}}m^{1}a^{\infty m}1)$ | Fe 8d | S:(1/2,1/2,0)<br>N:($u$,1/2,$w$) | $(S)U_1^SU_1^S(8)$<br>$(N)W_1^SW_1^S(4)$ | OP<br>QNPL | [100] | 62.445<br>$(Pnma')$ |
|---|---|---|---|---|---|---|---|---|---|
| 0.815 | $MnNb_2O_6$ | 60<br>$(Pbcn)$ | 18.60.1.1<br>$(P^{\bar{1}}b^{\bar{1}}c^{\bar{1}}n^{\infty m}1)$ | Mn 4c | L:(1/2,$v$,$w$)<br>W:($u$,$v$,1/2) | $L_1^SL_1^S(4)$<br>$(W)N_1^SN_1^S(4)$ | QNPL<br>QNPL | [100] | 60.419<br>$(Pb'cn)$ |
| 0.816 | $MnTa_2O_6$ | 60<br>$(Pbcn)$ | 18.60.1.1<br>$(P^{\bar{1}}b^{\bar{1}}c^{\bar{1}}n^{\infty m}1)$ | Mn 4c | L:(1/2,$v$,$w$)<br>W:($u$,$v$,1/2) | $L_1^SL_1^S(4)$<br>$(W)N_1^SN_1^S(4)$ | QNPL<br>QNPL | [100] | 60.419<br>$(Pb'cn)$ |
| 0.821 | $SrGd_2O_4$ | 62<br>$(Pnma)$ | 26.62.1.1<br>$(P^{\bar{1}}n^{1}m^{1}a^{\infty m}1)$ | Gd 4c,<br>4c | L:(1/2,$v$,$w$) | $(L)W_1^SW_1^S(4)$ | QNPL | [001] | 62.445<br>$(Pnma')$ |
| 0.909 | $Er_2PtGe_6$ | 64<br>$(Cmce)$ | 36.64.1.1<br>$(C^{1}m^{1}c^{\bar{1}}e^{\infty m}1)$ | Er 16g | Q:($u$,$v$,1/2) | $Q_1^SQ_1^S(4)$ | QNPL | [100] | 64.472<br>$(Cmc'a)$ |
| 0.932 | $Er_2PtGe_6$ | 64<br>$(Cmce)$ | 36.64.1.1<br>$(C^{1}m^{1}c^{\bar{1}}e^{\infty m}1)$ | Er 16g | Q:($u$,$v$,1/2) | $Q_1^SQ_1^S(4)$ | QNPL | [100] | 64.472<br>$(Cmc'a)$ |
| 0.961 | $LiCrGe_2O_6$ | 14<br>$(P2_1/c)$ | 4.14.1.1<br>$(P^{1}2_1/^{\bar{1}}c^{\infty m}1)$ | Cr 4e | G:($u$, 1/2, $w$) | $G_1^SG_1^S(4)$ | QNPL | [100]<br>+31°<br>([001]) | 14.77<br>$(P2_1'/c)$ |
| 0.962 | $LiCrGe_2O_6$ | 14<br>$(P2_1/c)$ | 4.14.1.1<br>$(P^{1}2_1/^{\bar{1}}c^{\infty m}1)$ | Cr 4e | G:($u$, 1/2, $w$) | $G_1^SG_1^S(4)$ | QNPL | [100]<br>+30°<br>([001]) | 14.77<br>$(P2_1'/c)$ |
| 0.963 | $LiCrGe_2O_6$ | 14<br>$(P2_1/c)$ | 4.14.1.1<br>$(P^{1}2_1/^{\bar{1}}c^{\infty m}1)$ | Cr 4e | G:($u$, 1/2, $w$) | $G_1^SG_1^S(4)$ | QNPL | [100]<br>+30°<br>([001]) | 14.77<br>$(P2_1'/c)$ |
| 0.964 | $LiCrGe_2O_6$ | 14<br>$(P2_1/c)$ | 4.14.1.1<br>$(P^{1}2_1/^{\bar{1}}c^{\infty m}1)$ | Cr 4e | G:($u$, 1/2, $w$) | $G_1^SG_1^S(4)$ | QNPL | [100]<br>+31°<br>([001]) | 14.77<br>$(P2_1'/c)$ |
| 0.966 | $V_2WO_6$ | 136<br>$(P4_2/mnm)$ | 113.136.1.1<br>$(P^{\bar{1}}4_2/^{\bar{1}}m^{\bar{1}}n^{1}m^{\infty m}1)$ | V 4e | F:($u$,1/2,$w$) | $F_1^SF_1^S(4)$ | QNPL | [100] | 58.395<br>$(Pn'nm)$ |
| 0.969 | $CaFe_2O_4$ | 62<br>$(Pnma)$ | 26.62.1.1<br>$(P^{\bar{1}}n^{1}m^{1}a^{\infty m}1)$ | Fe 4c,<br>4c | L:(1/2,$v$,$w$) | $(L)W_1^SW_1^S(4)$ | QNPL | [001] | 62.445<br>$(Pnma')$ |
| 0.1001 | $PbMn_2Ni_6Te_3O_{18}$ | 176<br>$(P6_3/m)$ | 173.176.1.1<br>$(P^{1}6_3/^{\bar{1}}m^{\infty m}1)$ | Mn 4f,<br>Ni 12i | E:($u$,$v$,1/2) | $E_1^SE_1^S(4)$ | QNPL | [001] | 176.146<br>$(P6_3/m')$ |
|  | $Cu_3TeO_6$[12] | 206 | 199.206.1.1 | Cu 24d | Γ:(0,0,0) | $\Gamma_4^S(6)$ | $C_2$−2 SP | [111] | 148.19 |

| | | $(Ia\bar{3})$ | $(I^{\bar{1}}a^{\bar{1}}\bar{3}^{\infty m}1)$ | | H:(1,1,1) | $H_4^S(6)$ | $C_2$−2 SP | | $(R\bar{3}')$ |
|---|---|---|---|---|---|---|---|---|---|

## S4.4. Altermagnets

Supplementary Table XXVII. Tabulation of altermagnets that accommodate unconventional magnons. The ten columns illustrate the material ID in MAGNDATA, chemical formula, space group, spin space groups, spin Wyckoff Positions, the momentum-position, band representations, and the quasiparticle type for unconventional magnons, the orientation of magnetic materials and the corresponding magnetic space groups. The spin axis is written under the lattice basis, and the "$[a_1b_1c_1]$ +θ $[a_2b_2c_2]$" labels the axis by rotating from the direction of "$[a_1b_1c_1]$" towards the direction of "$[a_2b_2c_2]$" by θ degrees.

| ID | Formula | SG | SSG | WP | ***k*** point | Rep | Quasi-particle | Spin Axis | MSG |
|---|---|---|---|---|---|---|---|---|---|
| | FeS[13] | 190 $(P\bar{6}2c)$ | 159.190.1.1 $(P^{\bar{1}}\bar{6}^{\bar{1}}2^{1}c^{\infty m}1)$ | Fe 12i | A:(0,0,1/2) | $A_3^SA_3^S(8)$ | OP | [001] | 190.230 $(P\bar{6}'2c')$ |
| | $Co_2Mo_3N$[14] | 213 $(P4_132)$ | 198.213.1.1 $(P^{\bar{1}}4_1{}^{1}3^{\bar{1}}2^{\infty m}1)$ | Co 8c | Γ:(0,0,0) | $\Gamma_4^S(6)$ | $C$−4 SP | $[10\bar{1}]$ | 20.34 $(C22'2_1')$ |
| | | | | | R:(1/2,1/2,1/2) | $R_1^SR_3^S(8)$ | $C$−4 OP | | |

## S4.5. Collinear Uτ AFMs

Supplementary Table XXVIII. Tabulation of collinear Uτ AFMs that accommodate unconventional magnons. The ten columns illustrate the material ID in MAGNDATA, chemical formula, space group, spin space groups, spin Wyckoff Positions, the momentum-position, band representations, and the quasiparticle type for unconventional magnons, the orientation of magnetic materials and the corresponding magnetic space groups. The spin axis is written under the lattice basis, and the "$[a_1b_1c_1]$ +θ $[a_2b_2c_2]$" labels the axis by rotating from the direction of "$[a_1b_1c_1]$" towards the direction of "$[a_2b_2c_2]$" by θ degrees.

| ID | Formula | SG | SSG | WP | ***k*** point | Rep | Quasi-particle | Spin Axis | MSG |
|---|---|---|---|---|---|---|---|---|---|
| 0.435 | $Pb_5Fe_3TiO_{11}Cl$ | 123 $(P4/mmm)$ | 129.123.2.1 $(P_C{}^{1}4/{}^{1}n^{1}m^{1}m^{\infty m}1)$ | Fe 4g | F:($u$,1/2,$w$) | $F_1^SF_2^S(4)$ | QNPL | $[1\bar{1}0]$ | 51.302 $(P_Bmma)$ |
| 0.563 | $Ce_2Ni_3Ge_5$ | 72 $(Ibam)$ | 57.72.2.1 $(P_I{}^{1}b^{1}c^{1}m^{\infty m}1)$ | Ce 8j | E:($u$,1/2,1/2) | $E_1^SE_1^S(8)$ | ONL | [100] | 56.376 $(P_Iccn)$ |
| | | | | | N:($u$,1/2,$w$) | $N_1^SN_2^S(4)$ | QNPL | | |

| | | | | | W:$(u,v,1/2)$ | $W_1^S W_2^S(4)$ | QNPL | | |
|---|---|---|---|---|---|---|---|---|---|
| 0.691 | $CaCo_{1.86}As_2$ | 139 $(I4/mmm)$ | 129.139.2.1 $(P_I{}^{1}4/{}^{1}n{}^{1}m{}^{\infty m}1)$ | Co 4d | F:$(u,1/2,w)$ | $F_1^S F_2^S(4)$ | QNPL | [001] | 126.386 $(P_I4/nnc)$ |
| 0.749 | $Ba_3CoRu_2O_9$ | 63 $(Cmcm)$ | 58.63.2.1 $(P_B{}^{1}n{}^{1}n{}^{1}m{}^{\infty m}1)$ | Co 4a, Ru 8f | L:$(1/2,v,w)$ | $L_1^S L_2^S(4)$ | QNPL | [010] | 62.454 $(P_Bnma)$ |
| | | | | | N:$(u,1/2,w)$ | $N_1^S N_2^S(4)$ | QNPL | | |
| 0.750 | $Ba_3CoRu_2O_9$ | 194 $(P6_3/mmc)$ | 58.63.2.1 $(P_B{}^{1}n{}^{1}n{}^{1}m{}^{\infty m}1)$ | Co 4b, Ru 8f | L:$(1/2,v,w)$ | $L_1^S L_2^S(4)$ | QNPL | [010] | 62.454 $(P_Bnma)$ |
| | | | | | N:$(u,1/2,w)$ | $N_1^S N_2^S(4)$ | QNPL | | |
| 1.1 | $Mn_3O_4$ | 57 $(Pbcm)$ | 61.57.2.1 $(P_a{}^{1}b{}^{1}c{}^{1}a{}^{\infty m}1)$ | Mn 16e | E:$(u,1/2,1/2)$ | $E_1^S E_1^S(8)$ | ONL | [100] | 62.452 $(P_cnma)$ |
| | | | | | P:$(1/2,v,1/2)$ | $P_1^S P_1^S(8)$ | ONL | | |
| | | | | | Q:$(1/2,1/2,w)$ | $Q_1^S Q_1^S(8)$ | ONL | | |
| | | | | | L:$(1/2,v,w)$ | $L_1^S L_2^S(4)$ | QNPL | | |
| | | | | | N:$(u,1/2,w)$ | $N_1^S N_2^S(4)$ | QNPL | | |
| | | | | | W:$(u,v,1/2)$ | $W_1^S W_2^S(4)$ | QNPL | | |
| 1.3 | $Sr_2IrO_4$ | 142 $(I4_1/acd)$ | 54.73.2.1 $(P_I{}^{1}c{}^{1}c{}^{1}a{}^{\infty m}1)$ | Ir 8c | R:$(1/2,1/2,1/2)$ | $R_1^S R_2^S(8)$ | OP | [100] | 54.352 $(P_Icca)$ |
| | | | | | U:$(1/2,0,1/2)$ | $U_1^S U_2^S(8)$ | OP | | |
| | | | | | L:$(1/2,v,w)$ | $L_1^S L_2^S(4)$ | QNPL | | |
| 1.4 | $YBa_2Cu_3O_{6+d}$ | 123 $(P4/mmm)$ | 129.123.2.1 $(P_C{}^{1}4/{}^{1}n{}^{1}m{}^{1}m{}^{\infty m}1)$ | Cu 4g | F:$(u,1/2,w)$ | $F_1^S F_2^S(4)$ | QNPL | [010] | 65.489 $(C_ammm)$ |
| 1.8 | $CeRu_2Al_{10}$ | 63 $(Cmcm)$ | 59.63.2.1 $(P_B{}^{1}m{}^{1}m{}^{1}n{}^{\infty m}1)$ | Ce 4c | L:$(1/2,v,w)$ | $L_1^S L_2^S(4)$ | QNPL | [001] | 57.391 $(P_Cbcm)$ |
| | | | | | N:$(u,1/2,w)$ | $N_1^S N_2^S(4)$ | QNPL | | |
| 1.9 | $Li_2VOSiO_4$ | 129 $(P4/nmm)$ | 51.67.2.1 $(P_C{}^{1}m{}^{1}m{}^{1}a{}^{\infty m}1)$ | V 4e | L:$(1/2,v,w)$ | $L_1^S L_2^S(4)$ | QNPL | [110] | 57.389 $(P_Abcm)$ |
| 1.18 | $MnS_2$ | 205 $(Pa\bar{3})$ | 33.29.2.1 $(P_b{}^{1}n{}^{1}a{}^{1}2_1{}^{\infty m}1)$ | Mn 8a | U:$(1/2,0,1/2)$ | $U_1^S U_1^S(8)$ | OP | [010] | 29.105 $(P_bca2_1)$ |
| | | | | | W:$(u,v,1/2)$ | $W_1^S W_1^S(4)$ | QNPL | | |
| 1.20 | $HoMnO_3$ | 62 $(Pnma)$ | 33.31.2.1 $(P_a{}^{1}n{}^{1}a{}^{1}2_1{}^{\infty m}1)$ | Mn 8b | U:$(1/2,0,1/2)$ | $U_1^S U_1^S(8)$ | OP | [100] | 31.129 $(P_bmn2_1)$ |
| | | | | | W:$(u,v,1/2)$ | $W_1^S W_1^S(4)$ | QNPL | | |
| 1.23 | $La_2CuO_4$ | 64 $(Cmce)$ | 55.64.2.1 $(P_A{}^{1}b{}^{1}a{}^{1}m{}^{\infty m}1)$ | Cu 4a | L:$(1/2,v,w)$ | $L_1^S L_2^S(4)$ | QNPL | [001] | 56.374 $(P_Accn)$ |
| | | | | | N:$(u,1/2,w)$ | $N_1^S N_2^S(4)$ | QNPL | | |
| 1.24 | $ZnV_2O_4$ | 141 $(I4_1/amd)$ | 95.98.2.1 $(P_I{}^{1}4_3{}^{1}2{}^{1}2{}^{\infty m}1)$ | V 8f | E:$(u,v,1/2)$ | $E_1^S E_1^S(4)$ | QNPL | [001] | 96.150 $(P_I4_32_12)$ |
| 1.26 | $CsFe_2Se_3$ | 63 | 14.11.2.1 | Fe 8f | G:$(u, 1/2, w)$ | $G_1^S G_1^S(4)$ | QNPL | [001] | 14.82 |

| | | $(Cmcm)$ | $(P_c{}^12_1/{}^1c^{\infty m}1)$ | | | | | | $(P_c2_1/c)$ |
|---|---|---|---|---|---|---|---|---|---|
| 1.28 | CrN | 225 $(Fm\bar{3}m)$ | 59.59.2.1 $(P_c{}^1m{}^1m{}^1n^{\infty m}1)$ | Cr 4a | L:(1/2,$v$,$w$) | $L_1^SL_2^S(4)$ | QNPL | [1$\bar{1}$0] | 62.450 $(P_anma)$ |
| | | | | | N:($u$,1/2,$w$) | $N_1^SN_2^S(4)$ | QNPL | | |
| 1.30 | $BaCo_2V_2O_8$ | 142 $(I4_1/acd)$ | 61.73.2.1 $(P_I{}^1b{}^1c{}^1a^{\infty m}1)$ | Co 16f | E:($u$,1/2,1/2) | $E_1^SE_1^S(8)$ | ONL | [001] | 54.352 $(P_Icca)$ |
| | | | | | P:(1/2,$v$,1/2) | $P_1^SP_1^S(8)$ | ONL | | |
| | | | | | Q:(1/2,1/2,$w$) | $Q_1^SQ_1^S(8)$ | ONL | | |
| | | | | | L:(1/2,$v$,$w$) | $L_1^SL_2^S(4)$ | QNPL | | |
| | | | | | N:($u$,1/2,$w$) | $N_1^SN_2^S(4)$ | QNPL | | |
| | | | | | W:($u$,$v$,1/2) | $W_1^SW_2^S(4)$ | QNPL | | |
| 1.32 | $Lu_2MnCoO_6$ | 14 $(P2_1/c)$ | 4.4.2.1 $(P_a{}^12_1{}^{\infty m}1)$ | Mn 4a, Co 4a | G:($u$, 1/2, $w$) | $G_1^SG_1^S(4)$ | QNPL | [001] | 4.10 $(P_a2_1)$ |
| 1.33 | ErAuGe | 186 $(P6_3mc)$ | 26.36.2.1 $(P_C{}^1m{}^1c{}^12_1{}^{\infty m}1)$ | Er 4a | W:($u$,$v$,1/2) | $W_1^SW_1^S(4)$ | QNPL | [001] | 33.154 $(P_Cna2_1)$ |
| 1.42 | $La_2NiO_4$ | 64 $(Cmce)$ | 55.64.2.1 $(P_A{}^1b{}^1a{}^1m^{\infty m}1)$ | Ni 4a | L:(1/2,$v$,$w$) | $L_1^SL_2^S(4)$ | QNPL | [100] | 53.335 $(P_Cmna)$ |
| | | | | | N:($u$,1/2,$w$) | $N_1^SN_2^S(4)$ | QNPL | | |
| 1.43 | $PrNiO_3$ | 62 $(Pnma)$ | 36.26.2.1 $(C_a{}^1m{}^1c{}^12_1{}^{\infty m}1)$ | Ni 16c | Q:($u$,$v$,1/2) | $Q_1^SQ_1^S(4)$ | QNPL | [100] | 36.178 $(C_amc2_1)$ |
| 1.45 | $NdNiO_3$ | 62 $(Pnma)$ | 36.26.2.1 $(C_a{}^1m{}^1c{}^12_1{}^{\infty m}1)$ | Ni 16c | Q:($u$,$v$,1/2) | $Q_1^SQ_1^S(4)$ | QNPL | [100] | 36.178 $(C_amc2_1)$ |
| 1.50 | $AgNiO_2$ | 182 $(P6_322)$ | 18.20.2.1 $(P_B{}^12_1{}^12_1{}^12^{\infty m}1)$ | Ni 4a | L:(1/2,$v$,$w$) | $L_1^SL_1^S(4)$ | QNPL | [001] | 18.22 $(P_B2_12_12)$ |
| | | | | | N:($u$,1/2,$w$) | $N_1^SN_1^S(4)$ | QNPL | | |
| 1.62 | CuO | 15 $(C2/c)$ | 14.14.2.1 $(P_a{}^12_1/{}^1c^{\infty m}1)$ | Cu 8e | G:($u$, 1/2, $w$) | $G_1^SG_1^S(4)$ | QNPL | [010] | 14.80 $(P_a2_1/c)$ |
| 1.63 | $MnPb_4Sb_6S_{14}$ | 14 $(P2_1/c)$ | 14.14.2.1 $(P_a{}^12_1/{}^1c^{\infty m}1)$ | Mn 4a | G:($u$, 1/2, $w$) | $G_1^SG_1^S(4)$ | QNPL | [010] | 14.80 $(P_a2_1/c)$ |
| 1.64 | $BaNiF_4$ | 36 $(Cmc2_1)$ | 4.4.2.1 $(P_a{}^12_1{}^{\infty m}1)$ | Ni 4a | G:($u$, 1/2, $w$) | $G_1^SG_1^S(4)$ | QNPL | [010] | 4.10 $(P_a2_1)$ |
| 1.71 | $SrCo_2V_2O_8$ | 110 $(I4_1cd)$ | 29.45.2.1 $(P_I{}^1c{}^1a{}^12_1{}^{\infty m}1)$ | Co 8c, 8c | R:(1/2,1/2,1/2) | $R_1^SR_1^S(8)$ | OP | [001] | 29.110 $(P_Ica2_1)$ |
| | | | | | U:(1/2,0,1/2) | $U_1^SU_1^S(8)$ | OP | | |
| | | | | | W:($u$,$v$,1/2) | $W_1^SW_1^S(4)$ | QNPL | | |
| 1.78 | $Li_2MnSiO_4$ | 14 | 14.14.2.1 | Mn 8e | G:($u$, 1/2, $w$) | $G_1^SG_1^S(4)$ | QNPL | [001] | 14.80 |

| | | $(P2_1/c)$ | $(P_a{}^12_1/{}^1c{}^{\infty m}1)$ | | | | | −32° ([100]) | $(P_a2_1/c)$ |
|---|---|---|---|---|---|---|---|---|---|
| 1.86 | $GeV_4S_8$ | 216 $(F\bar{4}3m)$ | 31.31.2.1 $(P_b{}^1m{}^1n{}^12_1{}^{\infty m}1)$ | V 4a, 8b | W:$(u,v,1/2)$ | $W_1^SW_1^S(4)$ | QNPL | [001] | 33.149 $(P_ana2_1)$ |
| 1.94 | $Ba_3LaRu_2O_9$ | 194 $(P6_3/mmc)$ | 51.63.2.1 $(P_A{}^1m{}^1m{}^1a{}^{\infty m}1)$ | Ru 8f | L:$(1/2,v,w)$ | $L_1^SL_2^S(4)$ | QNPL | [100] +25° ([001]) | 14.84 $(P_C2_1/c)$ |
| 1.101 | $LuMnO_3$ | 62 $(Pnma)$ | 33.31.2.1 $(P_a{}^1n{}^1a{}^12_1{}^{\infty m}1)$ | Mn 8b | U:$(1/2,0,1/2)$<br>W:$(u,v,1/2)$ | $U_1^SU_1^S(8)$<br>$W_1^SW_1^S(4)$ | OP<br>QNPL | [010] | 31.129 $(P_bmn2_1)$ |
| 1.103 | $U_2Rh_2Sn$ | 127 $(P4/mbm)$ | 136.127.2.1 $(P_c{}^14_2/{}^1m{}^1n{}^1m{}^{\infty m}1)$ | U 8g | F:$(u,1/2,w)$ | $F_1^SF_2^S(4)$ | QNPL | [001] | 135.492 $(P_c4_2/mbc)$ |
| 1.112 | $NiTa_2O_6$ | 136 $(P4_2/mnm)$ | 11.11.2.1 $(P_a{}^12_1/{}^1m{}^{\infty m}1)$ | Ni 4e, 4e | G:$(u,\ 1/2,\ w)$ | $G_1^SG_2^S(4)$ | QNPL | $[1\bar{1}0]$ | 14.82 $(P_c2_1/c)$ |
| 1.113 | $NiSb_2O_6$ | 136 $(P4_2/mnm)$ | 14.14.2.1 $(P_a{}^12_1/{}^1c{}^{\infty m}1)$ | Ni 8e | G:$(u,\ 1/2,\ w)$ | $G_1^SG_1^S(4)$ | QNPL | [110] | 2.7 $(P_S\bar{1})$ |
| 1.121 | $NaFeSO_4F$ | 15 $(C2/c)$ | 14.15.2.1 $(P_C{}^12_1/{}^1c{}^{\infty m}1)$ | Fe 4c | G:$(u,\ 1/2,\ w)$ | $G_1^SG_1^S(4)$ | QNPL | [001] +20° ([100]) | 13.74 $(P_C2/c)$ |
| 1.130 | $Cr_2As$ | 129 $(P4/nmm)$ | 59.59.2.1 $(P_c{}^1m{}^1m{}^1n{}^{\infty m}1)$ | Cr 4a, 4c | L:$(1/2,v,w)$<br>N:$(u,1/2,w)$ | $L_1^SL_2^S(4)$<br>$N_1^SN_2^S(4)$ | QNPL<br>QNPL | [100] | 62.450 $(P_anma)$ |
| 1.131 | $Fe_2As$ | 129 $(P4/nmm)$ | 129.129.2.1 $(P_c{}^14/{}^1n{}^1m{}^1m{}^{\infty m}1)$ | Fe 4a, 4c | F:$(u,1/2,w)$ | $F_1^SF_2^S(4)$ | QNPL | [010] | 62.450 $(P_anma)$ |
| 1.132 | $Mn_2As$ | 129 $(P4/nmm)$ | 129.129.2.1 $(P_c{}^14/{}^1n{}^1m{}^1m{}^{\infty m}1)$ | Mn 4a, 4c | F:$(u,1/2,w)$ | $F_1^SF_2^S(4)$ | QNPL | [010] | 62.450 $(P_anma)$ |
| 1.139 | $Ho_2RhIn_8$ | 123 $(P4/mmm)$ | 51.47.2.1 $(P_a{}^1m{}^1m{}^1a{}^{\infty m}1)$ | Ho 4q | L:$(1/2,v,w)$ | $L_1^SL_2^S(4)$ | QNPL | [001] | 49.273 $(P_cccm)$ |
| 1.140 | PrMgPb | 139 $(I4/mmm)$ | 129.139.2.1 $(P_I{}^14/{}^1n{}^1m{}^1m{}^{\infty m}1)$ | Pr 4e | F:$(u,1/2,w)$ | $F_1^SF_2^S(4)$ | QNPL | [100] +32° ([001]) | 13.73 $(P_A2/c)$ |
| 1.141 | NdMgPb | 139 $(I4/mmm)$ | 129.139.2.1 $(P_I{}^14/{}^1n{}^1m{}^1m{}^{\infty m}1)$ | Nd 4e | F:$(u,1/2,w)$ | $F_1^SF_2^S(4)$ | QNPL | [100] +33° | 13.73 $(P_A2/c)$ |

| | | | | | | | | ([001]) | |
|---|---|---|---|---|---|---|---|---|---|
| 1.142 | CeMgPb | 139<br>($I4/mmm$) | 63.69.2.1<br>($C_A{}^1m{}^1c{}^1m^{\infty m}1$) | Ce 8g | Q:($u$,$v$,1/2) | $Q_1^SQ_2^S(4)$ | QNPL | [1$\bar{1}$0] | 67.510<br>($C_Amme$) |
| 1.146 | LaCrAsO | 129<br>($P4/nmm$) | 137.129.2.1<br>($P_c{}^14_2/{}^1n{}^1m{}^1c^{\infty m}1$) | Cr 4b | F:($u$,1/2,$w$) | $F_1^SF_2^S(4)$ | QNPL | [001] | 138.528<br>($P_c4_2/ncm$) |
| 1.148 | $CeOs_{1.84}Ir_{0.16}Al_{10}$ | 63<br>($Cmcm$) | 59.63.2.1<br>($P_B{}^1m{}^1m{}^1n^{\infty m}1$) | Ce 4c | L:(1/2,$v$,$w$)<br>N:($u$,1/2,$w$) | $L_1^SL_2^S(4)$<br>$N_1^SN_2^S(4)$ | QNPL<br>QNPL | [100] | 62.453<br>($P_Anma$) |
| 1.156 | $LaMn_3Cr_4O_{12}$ | 204<br>($Im\bar{3}$) | 195.197.2.1<br>($P_I{}^12{}^13^{\infty m}1$) | Mn 6b,<br>Cr 8c | Γ:(0,0,0)<br>R:(1/2,1/2,1/2)<br>Γ:(0,0,0) | $\Gamma_4^S(6)$<br>$R_4^S(6)$<br>$\Gamma_2^S\Gamma_3^S(4)$ | $C$−4 SP<br>$C$−4 SP<br>$C$−8 DP | [111] | 146.12<br>($R_I3$) |
| 1.158 | $YMn_3Al_4O_{12}$ | 204<br>($Im\bar{3}$) | 200.204.2.1<br>($P_I{}^1m{}^1\bar{3}^{\infty m}1$) | Mn 6b | R:(1/2,1/2,1/2) | $R_4^{S,+}(6)$ | SP | [001] | 58.404<br>($P_Innm$) |
| 1.179 | NdCoAsO | 129<br>($P4/nmm$) | 129.129.2.1<br>($P_c{}^14/{}^1n{}^1m{}^1m^{\infty m}1$) | Nd 4a,<br>Co 4c | F:($u$,1/2,$w$) | $F_1^SF_2^S(4)$ | QNPL | [100] | 62.450<br>($P_anma$) |
| 1.181 | $Ba_3Fe_3O_7F$ | 11<br>($P2_1/m$) | 11.11.2.1<br>($P_a{}^12_1/{}^1m^{\infty m}1$) | Fe 4e,<br>4e, 4e | G:($u$, 1/2, $w$) | $G_1^SG_2^S(4)$ | QNPL | [010] | 11.55<br>($P_a2_1/m$) |
| 1.184 | $Na_2Co_2TeO_6$ | 182<br>($P6_322$) | 18.20.2.1<br>($P_B{}^12_1{}^12_1{}^12^{\infty m}1$) | Co 4a,<br>4a | L:(1/2,$v$,$w$)<br>N:($u$,1/2,$w$) | $L_1^SL_1^S(4)$<br>$N_1^SN_1^S(4)$ | QNPL<br>QNPL | [010] | 19.29<br>($P_C2_12_12_1$) |
| 1.192 | $SmMn_2O_5$ | 55<br>($Pbam$) | 26.26.2.1<br>($P_b{}^1m{}^1c{}^12_1{}^{\infty m}1$) | Sm 4a,<br>4a; Mn<br>8c, 4b,<br>4b | W:($u$,$v$,1/2) | $W_1^SW_1^S(4)$ | QNPL | [001] | 26.72<br>($P_bmc2_1$) |
| 1.195 | $Er_2Ni_2In$ | 65<br>($Cmmm$) | 63.51.2.1<br>($C_a{}^1m{}^1c{}^1m^{\infty m}1$) | Er 16i | Q:($u$,$v$,1/2) | $Q_1^SQ_2^S(4)$ | QNPL | [010] | 63.467<br>($C_amcm$) |
| 1.199 | $Sc_2NiMnO_6$ | 14<br>($P2_1/c$) | 14.14.2.1<br>($P_a{}^12_1/{}^1c^{\infty m}1$) | Mn 4a | G:($u$, 1/2, $w$) | $G_1^SG_2^S(4)$ | QNPL | [001]<br>+45°<br>([100]) | 14.80<br>($P_a2_1/c$) |
| 1.200 | $U_2Ni_2Sn$ | 127<br>($P4/mbm$) | 136.127.2.1<br>($P_c{}^14_2/{}^1m{}^1n{}^1m^{\infty m}1$) | U 8g | F:($u$,1/2,$w$) | $F_1^SF_2^S(4)$ | QNPL | [110] | 63.466<br>($C_cmcm$) |
| 1.213 | $Ho_2O_2Se$ | 164<br>($P\bar{3}m1$) | 11.12.2.1<br>($P_C{}^12_1/{}^1m^{\infty m}1$) | Ho 4e | G:($u$, 1/2, $w$) | $G_1^SG_2^S(4)$ | QNPL | [001] | 13.73<br>($P_A2/c$) |

| 1.215 | $UP_2$ | 129 ($P4/nmm$) | 129.129.2.1 ($P_c{}^{1}4/{}^{1}n{}^{1}m{}^{1}m{}^{\infty m}1$) | U 4c | F:($u$,1/2,$w$) | $F_1^S F_2^S(4)$ | QNPL | [001] | 130.432 ($P_c4/ncc$) |
|---|---|---|---|---|---|---|---|---|---|
| 1.219 | $CuF_2$ | 14 ($P2_1/c$) | 14.14.2.1 ($P_a{}^{1}2_1/{}^{1}c{}^{\infty m}1$) | Cu 4a | G:($u$, 1/2, $w$) | $G_1^S G_2^S(4)$ | QNPL | [010] +21° ([100]) | 2.7 ($P_S\bar{1}$) |
| 1.222 | $Er_2CoGa_8$ | 123 ($P4/mmm$) | 51.47.2.1 ($P_a{}^{1}m{}^{1}m{}^{1}a{}^{\infty m}1$) | Er 4q | L:(1/2,$v$,$w$) | $L_1^S L_2^S(4)$ | QNPL | [010] | 51.298 ($P_amma$) |
| 1.228 | $RuCl_3$ | 12 ($C2/m$) | 14.12.2.1 ($P_A{}^{1}2_1/{}^{1}c{}^{\infty m}1$) | Ru 8j | G:($u$, 1/2, $w$) | $G_1^S G_2^S(4)$ | QNPL | [100] +40° ([001]) | 10.49 ($P_C2/m$) |
| 1.252 | $CaCo_2P_2$ | 139 ($I4/mmm$) | 129.139.2.1 ($P_I{}^{1}4/{}^{1}n{}^{1}m{}^{1}m{}^{\infty m}1$) | Co 4d | F:($u$,1/2,$w$) | $F_1^S F_2^S(4)$ | QNPL | [010] | 59.416 ($P_Immn$) |
| 1.253 | $CeCo_2P_2$ | 139 ($I4/mmm$) | 129.139.2.1 ($P_I{}^{1}4/{}^{1}n{}^{1}m{}^{1}m{}^{\infty m}1$) | Co 4d | F:($u$,1/2,$w$) | $F_1^S F_2^S(4)$ | QNPL | [001] | 126.386 ($P_I4/nnc$) |
| 1.258 | $Cu_3Co_2SbO_6$ | 15 ($C2/c$) | 14.15.2.1 ($P_C{}^{1}2_1/{}^{1}c{}^{\infty m}1$) | Co 8f | G:($u$, 1/2, $w$) | $G_1^S G_2^S(4)$ | QNPL | [010] | 14.84 ($P_C2_1/c$) |
| 1.259 | $Cu_3Ni_2SbO_6$ | 15 ($C2/c$) | 14.15.2.1 ($P_C{}^{1}2_1/{}^{1}c{}^{\infty m}1$) | Ni 8f | G:($u$, 1/2, $w$) | $G_1^S G_2^S(4)$ | QNPL | [001] +17° ([100]) | 13.74 ($P_C2/c$) |
| 1.263 | $Ca_3Ru_2O_7$ | 36 ($Cmc2_1$) | 26.36.2.1 ($P_C{}^{1}m{}^{1}c{}^{1}2_1{}^{\infty m}1$) | Ru 8b | W:($u$,$v$,1/2) | $W_1^S W_1^S(4)$ | QNPL | [010] | 33.154 ($P_Cna2_1$) |
| 1.271 | CeSbTe | 129 ($P4/nmm$) | 129.129.2.1 ($P_c{}^{1}4/{}^{1}n{}^{1}m{}^{1}m{}^{\infty m}1$) | Ce 4c | F:($u$,1/2,$w$) | $F_1^S F_2^S(4)$ | QNPL | [001] | 130.432 ($P_c4/ncc$) |
| 1.276 | $Na_{0.5}Li_{0.5}FeGe_2O_6$ | 14 ($P2_1/c$) | 14.14.2.1 ($P_a{}^{1}2_1/{}^{1}c{}^{\infty m}1$) | Fe 8e | G:($u$, 1/2, $w$) | $G_1^S G_2^S(4)$ | QNPL | [100] | 14.80 ($P_a2_1/c$) |
| 1.297 | $CuFe_2(P_2O_7)_2$ | 14 ($P2_1/c$) | 14.14.2.1 ($P_a{}^{1}2_1/{}^{1}c{}^{\infty m}1$) | Cu 4a, Fe 8e | G:($u$, 1/2, $w$) | $G_1^S G_2^S(4)$ | QNPL | [010] | 14.80 ($P_a2_1/c$) |
| 1.298 | $BaCdVO(PO_4)_2$ | 61 ($Pbca$) | 29.29.2.1 ($P_b{}^{1}c{}^{1}a{}^{1}2_1{}^{\infty m}1$) | V 8a, 8a | R:(1/2,1/2,1/2) | $R_1^S R_1^S(8)$ | OP | [100] | 33.150 ($P_bna2_1$) |
| | | | | | U:(1/2,0,1/2) | $U_1^S U_1^S(8)$ | OP | | |
| | | | | | W:($u$,$v$,1/2) | $W_1^S W_1^S(4)$ | QNPL | | |
| 1.305 | $Mn_5Si_3$ | 193 | 62.63.2.1 | Mn 8g | Q:(1/2,1/2,$w$) | $Q_1^S Q_1^S(8)$ | ONL | [010] | 60.431 |

| | | $(P6_3/mcm)$ | $(P_B{}^1n^1m^1a^{\infty m}1)$ | | L:(1/2,$v$,$w$)<br>N:($u$,1/2,$w$)<br>W:($u$,$v$,1/2) | $L_1^S L_2^S(4)$<br>$N_1^S N_2^S(4)$<br>$W_1^S W_2^S(4)$ | QNPL<br>QNPL<br>QNPL | | $(P_cbcn)$ |
|---|---|---|---|---|---|---|---|---|---|
| 1.331 | $Li_{0.31}Na_{0.69}FeGe_2O_6$ | 14<br>$(P2_1/c)$ | 14.14.2.1<br>$(P_a{}^12_1/{}^1c^{\infty m}1)$ | Fe 8e | G:($u$, 1/2, $w$) | $G_1^S G_2^S(4)$ | QNPL | [100]<br>−4°<br>([001]) | 14.80<br>$(P_a2_1/c)$ |
| 1.334 | $Pr_2Pd_2In$ | 127<br>$(P4/mbm)$ | 62.55.2.1<br>$(P_b{}^1n^1m^1a^{\infty m}1)$ | Pr 8g | Q:(1/2,1/2,$w$)<br>L:(1/2,$v$,$w$)<br>N:($u$,1/2,$w$)<br>W:($u$,$v$,1/2) | $Q_1^S Q_1^S(8)$<br>$L_1^S L_2^S(4)$<br>$N_1^S N_2^S(4)$<br>$W_1^S W_2^S(4)$ | ONL<br>QNPL<br>QNPL<br>QNPL | [001] | 62.451<br>$(P_bnma)$ |
| 1.335 | $Nd_2Pd_2In$ | 127<br>$(P4/mbm)$ | 51.65.2.1<br>$(P_B{}^1m^1m^1a^{\infty m}1)$ | Nd 4f,<br>4f, 8q | L:(1/2,$v$,$w$) | $L_1^S L_2^S(4)$ | QNPL | [001] | 26.73<br>$(P_cmc2_1)$ |
| 1.337 | $U_2Pd_{2.35}Sn_{0.65}$ | 127<br>$(P4/mbm)$ | 127.127.2.1<br>$(P_c{}^14/{}^1m^1b^1m^{\infty m}1)$ | U 8g | F:($u$,1/2,$w$) | $F_1^S F_2^S(4)$ | QNPL | [001] | 128.408<br>$(P_c4/mnc)$ |
| 1.341 | $TmMnO_3$ | 62<br>$(Pnma)$ | 33.31.2.1<br>$(P_a{}^1n^1a^12_1{}^{\infty m}1)$ | Mn 8b | U:(1/2,0,1/2)<br>W:($u$,$v$,1/2) | $U_1^S U_1^S(8)$<br>$W_1^S W_1^S(4)$ | OP<br>QNPL | [100] | 31.129<br>$(P_bmn2_1)$ |
| 1.349 | $CoNb_3S_6$ | 182<br>$(P6_322)$ | 17.20.2.1<br>$(P_C{}^12^12^12_1{}^{\infty m}1)$ | Co 4a | W:($u$,$v$,1/2) | $W_1^S W_1^S(4)$ | QNPL | [100] | 18.22<br>$(P_B2_12_12)$ |
| 1.351 | $Ba_2Co_2F_7Cl$ | 11<br>$(P2_1/m)$ | 11.11.2.1<br>$(P_a{}^12_1/{}^1m^{\infty m}1)$ | Co 4e,<br>4e | G:($u$, 1/2, $w$) | $G_1^S G_2^S(4)$ | QNPL | [010] | 11.55<br>$(P_a2_1/m)$ |
| 1.353 | $SmNiO_3$ | 62<br>$(Pnma)$ | 36.26.2.1<br>$(C_a{}^1m^1c^12_1{}^{\infty m}1)$ | Sm 8a,<br>Ni 16c | Q:($u$,$v$,1/2) | $Q_1^S Q_1^S(4)$ | QNPL | [100] | 36.178<br>$(C_amc2_1)$ |
| 1.354 | $EuNiO_3$ | 62<br>$(Pnma)$ | 36.26.2.1<br>$(C_a{}^1m^1c^12_1{}^{\infty m}1)$ | Ni 16c | Q:($u$,$v$,1/2) | $Q_1^S Q_1^S(4)$ | QNPL | [100] | 36.178<br>$(C_amc2_1)$ |
| 1.355 | $DyGe_3$ | 63<br>$(Cmcm)$ | 11.11.2.1<br>$(P_a{}^12_1/{}^1m^{\infty m}1)$ | Dy 4e,<br>8f | G:($u$, 1/2, $w$) | $G_1^S G_2^S(4)$ | QNPL | [001] | 11.55<br>$(P_a2_1/m)$ |
| 1.356 | $Ho_3Ge_4$ | 63<br>$(Cmcm)$ | 62.63.2.4<br>$(P_A{}^1n^1m^1a^{\infty m}1)$ | Ho 8f | Q:(1/2,1/2,$w$)<br>L:(1/2,$v$,$w$)<br>N:($u$,1/2,$w$)<br>W:($u$,$v$,1/2) | $Q_1^S Q_1^S(8)$<br>$L_1^S L_2^S(4)$<br>$N_1^S N_2^S(4)$<br>$W_1^S W_2^S(4)$ | ONL<br>QNPL<br>QNPL<br>QNPL | [001] | 52.318<br>$(P_Bnna)$ |
| 1.371 | $Nd_2NiO_4$ | 64 | 55.64.2.1 | Ni 4a | L:(1/2,$v$,$w$) | $L_1^S L_2^S(4)$ | QNPL | [100] | 53.335 |

| | | $(Cmce)$ | $(P_A{}^1b^1a^1m^{\infty m}1)$ | | N:$(u,1/2,w)$ | $N_1^S N_2^S(4)$ | QNPL | | $(P_C mna)$ |
|---|---|---|---|---|---|---|---|---|---|
| 1.376 | CeScGe | 139<br>$(I4/mmm)$ | 129.139.2.1<br>$(P_I{}^14/{}^1n^1m^1m^{\infty m}1)$ | Ce 4e | F:$(u,1/2,w)$ | $F_1^S F_2^S(4)$ | QNPL | [110] | 63.468<br>$(C_A mcm)$ |
| 1.378 | CeScSi | 139<br>$(I4/mmm)$ | 129.139.2.1<br>$(P_I{}^14/{}^1n^1m^1m^{\infty m}1)$ | Ce 4e | F:$(u,1/2,w)$ | $F_1^S F_2^S(4)$ | QNPL | [110] | 63.468<br>$(C_A mcm)$ |
| 1.379 | ErNiGe | 62<br>$(Pnma)$ | 14.14.2.1<br>$(P_a{}^12_1/{}^1c^{\infty m}1)$ | Er 8e | G:$(u,\ 1/2,\ w)$ | $G_1^S G_2^S(4)$ | QNPL | [010]<br>+25°<br>([100]) | 14.80<br>$(P_a 2_1/c)$ |
| 1.384 | $USb_2$ | 129<br>$(P4/nmm)$ | 129.129.2.1<br>$(P_c{}^14/{}^1n^1m^1m^{\infty m}1)$ | U 4c | F:$(u,1/2,w)$ | $F_1^S F_2^S(4)$ | QNPL | [001] | 130.432<br>$(P_c 4/ncc)$ |
| 1.388 | $La_2NiO_3F_2$ | 66<br>$(Cccm)$ | 53.66.2.1<br>$(P_A{}^1m^1n^1a^{\infty m}1)$ | Ni 4e | W:$(u,v,1/2)$ | $W_1^S W_2^S(4)$ | QNPL | [001] | 53.333<br>$(P_A mna)$ |
| 1.389 | $Sr_2CoO_3Cl$ | 129<br>$(P4/nmm)$ | 51.67.2.1<br>$(P_C{}^1m^1m^1a^{\infty m}1)$ | Co 4e | L:$(1/2,v,w)$ | $L_1^S L_2^S(4)$ | QNPL | [100] | 49.274<br>$(P_B ccm)$ |
| 1.391 | $Fe_2MnBO_5$ | 55<br>$(Pbam)$ | 58.55.2.1<br>$(P_c{}^1n^1n^1m^{\infty m}1)$ | Fe 4c,<br>8g | L:$(1/2,v,w)$<br>N:$(u,1/2,w)$ | $L_1^S L_2^S(4)$<br>$N_1^S N_2^S(4)$ | QNPL<br>QNPL | [010] | 62.451<br>$(P_b nma)$ |
| 1.401 | $Nd_5Pb_3$ | 193<br>$(P6_3/mcm)$ | 51.63.2.1<br>$(P_A{}^1m^1m^1a^{\infty m}1)$ | Nd 4c,<br>8e, 8g | L:$(1/2,v,w)$ | $L_1^S L_2^S(4)$ | QNPL | [001] | 62.454<br>$(P_B nma)$ |
| 1.402 | $Nd_5Pb_3$ | 193<br>$(P6_3/mcm)$ | 51.63.2.1<br>$(P_A{}^1m^1m^1a^{\infty m}1)$ | Nd 4c,<br>8e | L:$(1/2,v,w)$ | $L_1^S L_2^S(4)$ | QNPL | [001] | 62.454<br>$(P_B nma)$ |
| 1.403 | $La_2CoO_4$ | 64<br>$(Cmce)$ | 55.64.2.1<br>$(P_A{}^1b^1a^1m^{\infty m}1)$ | Co 4a | L:$(1/2,v,w)$<br>N:$(u,1/2,w)$ | $L_1^S L_2^S(4)$<br>$N_1^S N_2^S(4)$ | QNPL<br>QNPL | [100] | 53.335<br>$(P_C mna)$ |
| 1.405 | $La_2CuO_4$ | 64<br>$(Cmce)$ | 55.64.2.1<br>$(P_A{}^1b^1a^1m^{\infty m}1)$ | Cu 4a | L:$(1/2,v,w)$<br>N:$(u,1/2,w)$ | $L_1^S L_2^S(4)$<br>$N_1^S N_2^S(4)$ | QNPL<br>QNPL | [010] | 56.374<br>$(P_A ccn)$ |
| 1.410 | $Sr_2Fe_{1.9}Cr_{0.1}O_5$ | 74<br>$(Imma)$ | 51.74.2.1<br>$(P_I{}^1m^1m^1a^{\infty m}1)$ | Fe 8i,<br>Fe/Co<br>4a | L:$(1/2,v,w)$ | $L_1^S L_2^S(4)$ | QNPL | [001] | 53.336<br>$(P_I mna)$ |
| 1.413 | $Ce_3Ni_2Ge_7$ | 65<br>$(Cmmm)$ | 51.65.2.1<br>$(P_B{}^1m^1m^1a^{\infty m}1)$ | Ce 4e | L:$(1/2,v,w)$ | $L_1^S L_2^S(4)$ | QNPL | [100] | 59.415<br>$(P_C mmn)$ |
| 1.414 | $CeNiGe_3$ | 65<br>$(Cmmm)$ | 51.65.2.1<br>$(P_B{}^1m^1m^1a^{\infty m}1)$ | Ce 4f | L:$(1/2,v,w)$ | $L_1^S L_2^S(4)$ | QNPL | [010] | 59.415<br>$(P_C mmn)$ |

| 1.415 | $Tb_2Pd_2In$ | 127 $(P4/mbm)$ | 63.51.2.1 $(C_a{}^1m^1c^1m^{\infty m}1)$ | Td 8e, 8e, 16i | Q:$(u,v,1/2)$ | $Q_1^SQ_2^S(4)$ | QNPL | [001] | 64.479 $(C_amce)$ |
|---|---|---|---|---|---|---|---|---|---|
| 1.420 | $YBa_2Cu_3O_6$ | 123 $(P4/mmm)$ | 129.123.2.1 $(P_C{}^14/{}^1n^1m^1m^{\infty m}1)$ | Cu 4g | F:$(u,1/2,w)$ | $F_1^SF_2^S(4)$ | QNPL | [010] | 65.489 $(C_ammm)$ |
| 1.425 | UGeTe | 139 $(I4/mmm)$ | 129.139.2.1 $(P_I{}^14/{}^1n^1m^1m^{\infty m}1)$ | U 4e | F:$(u,1/2,w)$ | $F_1^SF_2^S(4)$ | QNPL | [001] | 126.386 $(P_I4/nnc)$ |
| 1.426 | UGeS | 129 $(P4/nmm)$ | 129.129.2.1 $(P_c{}^14/{}^1n^1m^1m^{\infty m}1)$ | U 4c | F:$(u,1/2,w)$ | $F_1^SF_2^S(4)$ | QNPL | [001] | 130.432 $(P_c4/ncc)$ |
| 1.434 | $Fe_{1.05}Te$ | 129 $(P4/nmm)$ | 11.11.2.1 $(P_a{}^12_1/{}^1m^{\infty m}1)$ | Fe 4e | G:$(u,\ 1/2,\ w)$ | $G_1^SG_2^S(4)$ | QNPL | [010] | 11.55 $(P_a2_1/m)$ |
| 1.435 | $Fe_{1.05}Te$ | 129 $(P4/nmm)$ | 11.11.2.1 $(P_a{}^12_1/{}^1m^{\infty m}1)$ | Fe 4e, 4e | G:$(u,\ 1/2,\ w)$ | $G_1^SG_2^S(4)$ | QNPL | [010] | 11.55 $(P_a2_1/m)$ |
| 1.436 | $Fe_{1.125}Te$ | 129 $(P4/nmm)$ | 11.11.2.1 $(P_a{}^12_1/{}^1m^{\infty m}1)$ | Fe 4e | G:$(u,\ 1/2,\ w)$ | $G_1^SG_2^S(4)$ | QNPL | $[\bar{1}40]$ + 41° ([001]) | 2.7 $(P_S\bar{1})$ |
| 1.437 | $Fe_{1.068}Te$ | 129 $(P4/nmm)$ | 11.11.2.1 $(P_a{}^12_1/{}^1m^{\infty m}1)$ | Fe 4e | G:$(u,\ 1/2,\ w)$ | $G_1^SG_2^S(4)$ | QNPL | $[\bar{1}01]$ +74° ([010]) | 2.7 $(P_S\bar{1})$ |
| 1.438 | $BaCoF_4$ | 36 $(Cmc2_1)$ | 4.4.2.1 $(P_a{}^12_1{}^{\infty m}1)$ | Co 4a | G:$(u,\ 1/2,\ w)$ | $G_1^SG_2^S(4)$ | QNPL | [100] | 4.10 $(P_a2_1)$ |
| 1.439 | $BaCoF_4$ | 36 $(Cmc2_1)$ | 33.29.2.1 $(P_b{}^1n^1a^12_1{}^{\infty m}1)$ | Co 8a | U:$(1/2,0,1/2)$<br>W:$(u,v,1/2)$ | $U_1^SU_1^S(8)$<br>$W_1^SW_1^S(4)$ | OP<br>QNPL | [100] | 29.105 $(P_bca2_1)$ |
| 1.445 | $Y_2BaCuO_5$ | 62 $(Pnma)$ | 14.14.2.1 $(P_a{}^12_1/{}^1c^{\infty m}1)$ | Cu 8e | G:$(u,\ 1/2,\ w)$ | $G_1^SG_2^S(4)$ | QNPL | [001] | 14.80 $(P_a2_1/c)$ |
| 1.446 | $CeCoAl_4$ | 51 $(Pmma)$ | 63.51.2.1 $(C_a{}^1m^1c^1m^{\infty m}1)$ | Ce 8e | Q:$(u,v,1/2)$ | $Q_1^SQ_2^S(4)$ | QNPL | [010] | 64.479 $(C_amce)$ |
| 1.447 | $Ce_3Ni_2Sn_7$ | 65 $(Cmmm)$ | 51.65.2.1 $(P_B{}^1m^1m^1a^{\infty m}1)$ | Ce 4e | L:$(1/2,v,w)$ | $L_1^SL_2^S(4)$ | QNPL | [100] | 59.415 $(P_Cmmn)$ |
| 1.450 | $Pr_6Fe_{13}Sn$ | 140 $(I4/mcm)$ | 127.140.2.1 $(P_I{}^14/{}^1m^1b^1m^{\infty m}1)$ | Pr 8f, 16l; Fe 4d, 16k, | F:$(u,1/2,w)$ | $F_1^SF_2^S(4)$ | QNPL | [010] | 60.432 $(P_Ibcn)$ |

| | | | | 16l, 16l | | | | | |
|---|---|---|---|---|---|---|---|---|---|
| 1.451 | $Nd_6Fe_{13}Sn$ | 140 $(I4/mcm)$ | 127.140.2.1 $(P_I{}^14/{}^1m{}^1b{}^1m{}^{\infty m}1)$ | Nd 8f, 16l; Fe 4d, 16k, 16l, 16l | F:$(u$,1/2,$w)$ | $F_1^SF_2^S(4)$ | QNPL | [001] | 124.362 $(P_I4/mcc)$ |
| 1.453 | $EuMn_2Si_2$ | 139 $(I4/mmm)$ | 129.139.2.1 $(P_I{}^14/{}^1n{}^1m{}^1m{}^{\infty m}1)$ | Mn 4d | F:$(u$,1/2,$w)$ | $F_1^SF_2^S(4)$ | QNPL | [001] | 126.386 $(P_I4/nnc)$ |
| 1.454 | $Mn_6Ni_{16}Si_7$ | 225 $(Fm\bar{3}m)$ | 136.139.2.1 $(P_I{}^14_2/{}^1m{}^1n{}^1m{}^{\infty m}1)$ | Mn 8h | F:$(u$,1/2,$w)$ | $F_1^SF_2^S(4)$ | QNPL | [010] | 64.480 $(C_Amce)$ |
| 1.457 | $NdNiMg_{15}$ | 129 $(P4/nmm)$ | 51.67.2.1 $(P_C{}^1m{}^1m{}^1a{}^{\infty m}1)$ | Nd 4e | L:(1/2,$v$,$w)$ | $L_1^SL_2^S(4)$ | QNPL | [001] | 54.350 $(P_Bcca)$ |
| 1.458 | $CsCo_2Se_2$ | 139 $(I4/mmm)$ | 129.139.2.1 $(P_I{}^14/{}^1n{}^1m{}^1m{}^{\infty m}1)$ | Co 4d | F:$(u$,1/2,$w)$ | $F_1^SF_2^S(4)$ | QNPL | [110] | 63.468 $(C_Amcm)$ |
| 1.460 | PrCuSi | 194 $(P6_3/mmc)$ | 51.63.2.1 $(P_A{}^1m{}^1m{}^1a{}^{\infty m}1)$ | Pr 4a | L:(1/2,$v$,$w)$ | $L_1^SL_2^S(4)$ | QNPL | [1$\bar{1}$0] | 60.431 $(P_Cbcn)$ |
| 1.462 | $La_2CoPtO_6$ | 14 $(P2_1/c)$ | 14.14.2.1 $(P_a{}^12_1/{}^1c{}^{\infty m}1)$ | Co 4a | G:$(u$, 1/2, $w)$ | $G_1^SG_2^S(4)$ | QNPL | [001] +16° ([100]) | 14.80 $(P_a2_1/c)$ |
| 1.468 | $TbMn_2Si_2$ | 139 $(I4/mmm)$ | 129.139.2.1 $(P_I{}^14/{}^1n{}^1m{}^1m{}^{\infty m}1)$ | Mn 4d | F:$(u$,1/2,$w)$ | $F_1^SF_2^S(4)$ | QNPL | [001] | 126.386 $(P_I4/nnc)$ |
| 1.469 | $YMn_2Si_2$ | 139 $(I4/mmm)$ | 129.139.2.1 $(P_I{}^14/{}^1n{}^1m{}^1m{}^{\infty m}1)$ | Mn 4d | F:$(u$,1/2,$w)$ | $F_1^SF_2^S(4)$ | QNPL | [001] | 126.386 $(P_I4/nnc)$ |
| 1.472 | CaOFeS | 186 $(P6_3mc)$ | 31.36.2.1 $(P_C{}^1m{}^1n{}^12_1{}^{\infty m}1)$ | Fe 4a | W:$(u$,$v$,1/2) | $W_1^SW_1^S(4)$ | QNPL | [001] | 29.109 $(P_Cca2_1)$ |
| 1.475 | $DyNiAl_4$ | 63 $(Cmcm)$ | 59.63.2.1 $(P_B{}^1m{}^1m{}^1n{}^{\infty m}1)$ | Dy 4c | L:(1/2,$v$,$w)$<br>N:$(u$,1/2,$w)$ | $L_1^SL_2^S(4)$<br>$N_1^SN_2^S(4)$ | QNPL<br>QNPL | [100] | 62.453 $(P_Anma)$ |
| 1.478 | $CoTi_2O_5$ | 63 $(Cmcm)$ | 11.11.2.1 $(P_a{}^12_1/{}^1m{}^{\infty m}1)$ | Co 4e | G:$(u$, 1/2, $w)$ | $G_1^SG_2^S(4)$ | QNPL | [001] | 11.55 $(P_a2_1/m)$ |
| 1.479 | $U_2Ni_2Sn$ | 127 $(P4/mbm)$ | 136.127.2.1 $(P_c{}^14_2/{}^1m{}^1n{}^1m{}^{\infty m}1)$ | U 8g | F:$(u$,1/2,$w)$ | $F_1^SF_2^S(4)$ | QNPL | [001] | 135.492 $(P_c4_2/mbc)$ |
| 1.481 | $LaSr_3Fe_3O_9$ | 63 | 59.63.2.1 | Fe 4c, | L:(1/2,$v$,$w)$ | $L_1^SL_2^S(4)$ | QNPL | [001] | 57.391 |

| | | $(Cmcm)$ | $(P_B{}^1m{}^1m{}^1n{}^{\infty m}1)$ | 8f | N:$(u,1/2,w)$ | $N_1^S N_2^S(4)$ | QNPL | | $(P_Cbcm)$ |
|---|---|---|---|---|---|---|---|---|---|
| 1.485 | $Mn_3TeO_6$ | 14 $(P2_1/c)$ | 14.14.2.1 $(P_a{}^12_1/{}^1c{}^{\infty m}1)$ | Mn 4a, 8e | G:$(u,$ $1/2,$ $w)$ | $G_1^S G_2^S(4)$ | QNPL | [100] +26° ([001]) | 14.80 $(P_a2_1/c)$ |
| 1.488 | $CeMn_2Si_2$ | 139 $(I4/mmm)$ | 129.139.2.1 $(P_I{}^14/{}^1n{}^1m{}^1m{}^{\infty m}1)$ | Mn 4d | F:$(u,1/2,w)$ | $F_1^S F_2^S(4)$ | QNPL | [001] | 126.386 $(P_I4/nnc)$ |
| 1.489 | $CeMn_2Si_2$ | 139 $(I4/mmm)$ | 129.139.2.1 $(P_I{}^14/{}^1n{}^1m{}^1m{}^{\infty m}1)$ | Mn 4d | F:$(u,1/2,w)$ | $F_1^S F_2^S(4)$ | QNPL | [001] | 126.386 $(P_I4/nnc)$ |
| 1.490 | $CeMn_2Si_2$ | 139 $(I4/mmm)$ | 129.139.2.1 $(P_I{}^14/{}^1n{}^1m{}^1m{}^{\infty m}1)$ | Mn 4d | F:$(u,1/2,w)$ | $F_1^S F_2^S(4)$ | QNPL | [001] | 126.386 $(P_I4/nnc)$ |
| 1.491 | $PrMn_2Si_2$ | 139 $(I4/mmm)$ | 129.139.2.1 $(P_I{}^14/{}^1n{}^1m{}^1m{}^{\infty m}1)$ | Mn 4d | F:$(u,1/2,w)$ | $F_1^S F_2^S(4)$ | QNPL | [001] | 126.386 $(P_I4/nnc)$ |
| 1.492 | $PrMn_2Si_2$ | 139 $(I4/mmm)$ | 129.139.2.1 $(P_I{}^14/{}^1n{}^1m{}^1m{}^{\infty m}1)$ | Mn 4d | F:$(u,1/2,w)$ | $F_1^S F_2^S(4)$ | QNPL | [001] | 126.386 $(P_I4/nnc)$ |
| 1.493 | $NdMn_2Si_2$ | 139 $(I4/mmm)$ | 129.139.2.1 $(P_I{}^14/{}^1n{}^1m{}^1m{}^{\infty m}1)$ | Mn 4d | F:$(u,1/2,w)$ | $F_1^S F_2^S(4)$ | QNPL | [001] | 126.386 $(P_I4/nnc)$ |
| 1.494 | $NdMn_2Si_2$ | 139 $(I4/mmm)$ | 129.139.2.1 $(P_I{}^14/{}^1n{}^1m{}^1m{}^{\infty m}1)$ | Mn 4d | F:$(u,1/2,w)$ | $F_1^S F_2^S(4)$ | QNPL | [001] | 126.386 $(P_I4/nnc)$ |
| 1.495 | $YMn_2Si_2$ | 139 $(I4/mmm)$ | 129.139.2.1 $(P_I{}^14/{}^1n{}^1m{}^1m{}^{\infty m}1)$ | Mn 4d | F:$(u,1/2,w)$ | $F_1^S F_2^S(4)$ | QNPL | [001] | 126.386 $(P_I4/nnc)$ |
| 1.496 | $YMn_2Ge_2$ | 139 $(I4/mmm)$ | 129.139.2.1 $(P_I{}^14/{}^1n{}^1m{}^1m{}^{\infty m}1)$ | Mn 4d | F:$(u,1/2,w)$ | $F_1^S F_2^S(4)$ | QNPL | [001] | 126.386 $(P_I4/nnc)$ |
| 1.503 | $NdScSiC_{0.5}H_{0.2}$ | 139 $(I4/mmm)$ | 129.139.2.1 $(P_I{}^14/{}^1n{}^1m{}^1m{}^{\infty m}1)$ | Nd 4e | F:$(u,1/2,w)$ | $F_1^S F_2^S(4)$ | QNPL | [001] | 126.386 $(P_I4/nnc)$ |
| 1.504 | GdCuSn | 186 $(P6_3mc)$ | 26.36.2.1 $(P_C{}^1m{}^1c{}^12_1{}^{\infty m}1)$ | Gd 4a | W:$(u,v,1/2)$ | $W_1^S W_1^S(4)$ | QNPL | [001] | 33.154 $(P_Cna2_1)$ |
| 1.505 | GdAgSn | 186 $(P6_3mc)$ | 26.36.2.1 $(P_C{}^1m{}^1c{}^12_1{}^{\infty m}1)$ | Gd 4a | W:$(u,v,1/2)$ | $W_1^S W_1^S(4)$ | QNPL | [001] | 33.154 $(P_Cna2_1)$ |
| 1.506 | GdAuSn | 186 $(P6_3mc)$ | 26.36.2.1 $(P_C{}^1m{}^1c{}^12_1{}^{\infty m}1)$ | Gd 4a | W:$(u,v,1/2)$ | $W_1^S W_1^S(4)$ | QNPL | [001] | 33.154 $(P_Cna2_1)$ |
| 1.507 | $NdPd_5Al_2$ | 139 | 59.59.2.1 | Nd 4a | L:$(1/2,v,w)$ | $L_1^S L_2^S(4)$ | QNPL | [001] | 62.450 |

| | | $(I4/mmm)$ | $(P_c{}^1m^1m^1n^{\infty m}1)$ | | N:$(u,1/2,w)$ | $N_1^S N_2^S(4)$ | QNPL | | $(P_a nma)$ |
|---|---|---|---|---|---|---|---|---|---|
| 1.520 | $NiSO_4$ | 63 $(Cmcm)$ | 51.63.2.1 $(P_A{}^1m^1m^1a^{\infty m}1)$ | Ni 4a | L:$(1/2,v,w)$ | $L_1^S L_2^S(4)$ | QNPL | [010] | 60.431 $(P_c bcn)$ |
| 1.521 | $FeSO_4$ | 63 $(Cmcm)$ | 51.63.2.1 $(P_A{}^1m^1m^1a^{\infty m}1)$ | Fe 4a | L:$(1/2,v,w)$ | $L_1^S L_2^S(4)$ | QNPL | [010] | 60.431 $(P_c bcn)$ |
| 1.523 | $VPO_4$ | 63 $(Cmcm)$ | 60.57.2.1 $(P_b{}^1b^1c^1n^{\infty m}1)$ | V 8c | P:$(1/2,v,1/2)$ | $P_1^S P_1^S(8)$ | ONL | [010] | 62.452 $(P_c nma)$ |
| | | | | | T:$(0,1/2,1/2)$ | $T_1^S T_2^S(8)$ | OP | | |
| | | | | | L:$(1/2,v,w)$ | $L_1^S L_2^S(4)$ | QNPL | | |
| | | | | | W:$(u,v,1/2)$ | $W_1^S W_2^S(4)$ | QNPL | | |
| 1.526 | $LiCoF_4$ | 14 $(P2_1/c)$ | 14.14.2.1 $(P_a{}^12_1/{}^1c^{\infty m}1)$ | Co 4a | G:$(u,\ 1/2,\ w)$ | $G_1^S G_2^S(4)$ | QNPL | [001] +90° ([100]) | 14.80 $(P_a 2_1/c)$ |
| 1.527 | $CsNiF_3$ | 194 $(P6_3/mmc)$ | 51.63.2.1 $(P_A{}^1m^1m^1a^{\infty m}1)$ | Ni 4a | L:$(1/2,v,w)$ | $L_1^S L_2^S(4)$ | QNPL | [010] | 58.402 $(P_B nnm)$ |
| 1.535 | $UPd_2Ge_2$ | 139 $(I4/mmm)$ | 129.129.2.1 $(P_c{}^14/{}^1n^1m^1m^{\infty m}1)$ | U 4c, 4c | F:$(u,1/2,w)$ | $F_1^S F_2^S(4)$ | QNPL | [001] | 130.432 $(P_c 4/ncc)$ |
| 1.539 | KMnP | 129 $(P4/nmm)$ | 137.129.2.1 $(P_c{}^14_2/{}^1n^1m^1c^{\infty m}1)$ | Mn 4b | F:$(u,1/2,w)$ | $F_1^S F_2^S(4)$ | QNPL | [001] | 138.528 $(P_c 4_2/ncm)$ |
| 1.540 | KMnP | 129 $(P4/nmm)$ | 137.129.2.1 $(P_c{}^14_2/{}^1n^1m^1c^{\infty m}1)$ | Mn 4b | F:$(u,1/2,w)$ | $F_1^S F_2^S(4)$ | QNPL | [001] | 138.528 $(P_c 4_2/ncm)$ |
| 1.541 | RbMnP | 129 $(P4/nmm)$ | 137.129.2.1 $(P_c{}^14_2/{}^1n^1m^1c^{\infty m}1)$ | Mn 4b | F:$(u,1/2,w)$ | $F_1^S F_2^S(4)$ | QNPL | [001] | 138.528 $(P_c 4_2/ncm)$ |
| 1.542 | RbMnP | 129 $(P4/nmm)$ | 137.129.2.1 $(P_c{}^14_2/{}^1n^1m^1c^{\infty m}1)$ | Mn 4b | F:$(u,1/2,w)$ | $F_1^S F_2^S(4)$ | QNPL | [001] | 138.528 $(P_c 4_2/ncm)$ |
| 1.543 | RbMnAs | 129 $(P4/nmm)$ | 137.129.2.1 $(P_c{}^14_2/{}^1n^1m^1c^{\infty m}1)$ | Mn 4b | F:$(u,1/2,w)$ | $F_1^S F_2^S(4)$ | QNPL | [001] | 138.528 $(P_c 4_2/ncm)$ |
| 1.544 | RbMnAs | 129 $(P4/nmm)$ | 137.129.2.1 $(P_c{}^14_2/{}^1n^1m^1c^{\infty m}1)$ | Mn 4b | F:$(u,1/2,w)$ | $F_1^S F_2^S(4)$ | QNPL | [001] | 138.528 $(P_c 4_2/ncm)$ |
| 1.545 | RbMnBi | 129 $(P4/nmm)$ | 137.129.2.1 $(P_c{}^14_2/{}^1n^1m^1c^{\infty m}1)$ | Mn 4b | F:$(u,1/2,w)$ | $F_1^S F_2^S(4)$ | QNPL | [001] | 138.528 $(P_c 4_2/ncm)$ |
| 1.546 | CsMnBi | 129 | 137.129.2.1 | Mn 4b | F:$(u,1/2,w)$ | $F_1^S F_2^S(4)$ | QNPL | [001] | 138.528 |

|  |  | $(P4/nmm)$ | $(P_c{}^{1}4_2/{}^{1}n{}^{1}m{}^{1}c^{\infty m}1)$ |  |  |  |  |  | $(P_c4_2/ncm)$ |
|---|---|---|---|---|---|---|---|---|---|
| 1.547 | CsMnP | 129<br>$(P4/nmm)$ | 137.129.2.1<br>$(P_c{}^{1}4_2/{}^{1}n{}^{1}m{}^{1}c^{\infty m}1)$ | Mn 4b | F:($u$,1/2,$w$) | $F_1^SF_2^S(4)$ | QNPL | [001] | 138.528<br>$(P_c4_2/ncm)$ |
| 1.548 | CsMnP | 129<br>$(P4/nmm)$ | 137.129.2.1<br>$(P_c{}^{1}4_2/{}^{1}n{}^{1}m{}^{1}c^{\infty m}1)$ | Mn 4b | F:($u$,1/2,$w$) | $F_1^SF_2^S(4)$ | QNPL | [001] | 138.528<br>$(P_c4_2/ncm)$ |
| 1.550 | LiMnAs | 129<br>$(P4/nmm)$ | 137.129.2.1<br>$(P_c{}^{1}4_2/{}^{1}n{}^{1}m{}^{1}c^{\infty m}1)$ | Mn 4b | F:($u$,1/2,$w$) | $F_1^SF_2^S(4)$ | QNPL | [001] | 138.528<br>$(P_c4_2/ncm)$ |
| 1.551 | LiMnAs | 129<br>$(P4/nmm)$ | 137.129.2.1<br>$(P_c{}^{1}4_2/{}^{1}n{}^{1}m{}^{1}c^{\infty m}1)$ | Mn 4b | F:($u$,1/2,$w$) | $F_1^SF_2^S(4)$ | QNPL | [001] | 138.528<br>$(P_c4_2/ncm)$ |
| 1.552 | LiMnAs | 129<br>$(P4/nmm)$ | 137.129.2.1<br>$(P_c{}^{1}4_2/{}^{1}n{}^{1}m{}^{1}c^{\infty m}1)$ | Mn 4b | F:($u$,1/2,$w$) | $F_1^SF_2^S(4)$ | QNPL | [001] | 138.528<br>$(P_c4_2/ncm)$ |
| 1.553 | KMnAs | 129<br>$(P4/nmm)$ | 137.129.2.1<br>$(P_c{}^{1}4_2/{}^{1}n{}^{1}m{}^{1}c^{\infty m}1)$ | Mn 4b | F:($u$,1/2,$w$) | $F_1^SF_2^S(4)$ | QNPL | [001] | 138.528<br>$(P_c4_2/ncm)$ |
| 1.554 | KMnAs | 129<br>$(P4/nmm)$ | 137.129.2.1<br>$(P_c{}^{1}4_2/{}^{1}n{}^{1}m{}^{1}c^{\infty m}1)$ | Mn 4b | F:($u$,1/2,$w$) | $F_1^SF_2^S(4)$ | QNPL | [001] | 138.528<br>$(P_c4_2/ncm)$ |
| 1.570 | $La_3OsO_7$ | 63<br>$(Cmcm)$ | 14.11.2.1<br>$(P_c{}^{1}2_1/{}^{1}c^{\infty m}1)$ | Os 4c | G:($u$, 1/2, $w$) | $G_1^SG_2^S(4)$ | QNPL | [100]<br>+22°<br>([010]) | 11.55<br>$(P_a2_1/m)$ |
| 1.571 | $La_3OsO_7$ | 63<br>$(Cmcm)$ | 14.11.2.1<br>$(P_c{}^{1}2_1/{}^{1}c^{\infty m}1)$ | Os 4c | G:($u$, 1/2, $w$) | $G_1^SG_2^S(4)$ | QNPL | [100]<br>+27°<br>([010]) | 11.55<br>$(P_a2_1/m)$ |
| 1.572 | $La_{2.8}Ca_{0.2}OsO_7$ | 63<br>$(Cmcm)$ | 14.11.2.1<br>$(P_c{}^{1}2_1/{}^{1}c^{\infty m}1)$ | Os 4c | G:($u$, 1/2, $w$) | $G_1^SG_2^S(4)$ | QNPL | [100]<br>+30°<br>([010]) | 11.55<br>$(P_a2_1/m)$ |
| 1.573 | $FeSO_4$ | 62<br>$(Pnma)$ | 11.11.2.1<br>$(P_a{}^{1}2_1/{}^{1}m^{\infty m}1)$ | Fe 4a,<br>4c | G:($u$, 1/2, $w$) | $G_1^SG_2^S(4)$ | QNPL | [100]<br>+49°<br>([010]) | 14.82<br>$(P_c2_1/c)$ |
| 1.589 | $Fe_{0.967}Nb_3S_6$ | 182<br>$(P6_322)$ | 19.18.2.1<br>$(P_c{}^{1}2_1{}^{1}2_1{}^{1}2_1{}^{\infty m}1)$ | Fe 8c | R:(1/2,1/2,1/2) | $R_1^SR_1^S(8)$ | $C$−4 OP | [001] | 18.21<br>$(P_c2_12_12)$ |
|  |  |  |  |  | L:(1/2,$v$,$w$) | $L_1^SL_1^S(4)$ | QNPL |  |  |
|  |  |  |  |  | N:($u$,1/2,$w$) | $N_1^SN_1^S(4)$ | QNPL |  |  |
|  |  |  |  |  | W:($u$,$v$,1/2) | $W_1^SW_1^S(4)$ | QNPL |  |  |
| 1.593 | BaCoSO | 63 $(Cmcm)$ | 62.57.2.1 $(P_c{}^1n^1m^1a^{\infty m}1)$ | Co 8d | Q:(1/2,1/2,$w$) | $Q_1^SQ_1^S(8)$ | ONL | [001] | 57.386 $(P_abcm)$ |
| | | | | | L:(1/2,$v$,$w$) | $L_1^SL_2^S(4)$ | QNPL | | |
| | | | | | N:($u$,1/2,$w$) | $N_1^SN_2^S(4)$ | QNPL | | |
| | | | | | W:($u$,$v$,1/2) | $W_1^SW_2^S(4)$ | QNPL | | |
| 1.594 | BaCoSO | 63 $(Cmcm)$ | 62.57.2.1 $(P_c{}^1n^1m^1a^{\infty m}1)$ | Co 8d | Q:(1/2,1/2,$w$) | $Q_1^SQ_1^S(8)$ | ONL | [001] | 57.386 $(P_abcm)$ |
| | | | | | L:(1/2,$v$,$w$) | $L_1^SL_2^S(4)$ | QNPL | | |
| | | | | | N:($u$,1/2,$w$) | $N_1^SN_2^S(4)$ | QNPL | | |
| | | | | | W:($u$,$v$,1/2) | $W_1^SW_2^S(4)$ | QNPL | | |
| 1.596 | $TbCuSb_2$ | 129 $(P4/nmm)$ | 11.11.2.1 $(P_a{}^12_1/{}^1m^{\infty m}1)$ | Tb 4e | G:($u$, 1/2, $w$) | $G_1^SG_2^S(4)$ | QNPL | [110] | 2.7 $(P_S\bar{1})$ |
| 1.597 | $TbCuSb_2$ | 129 $(P4/nmm)$ | 11.11.2.1 $(P_a{}^12_1/{}^1m^{\infty m}1)$ | Tb 4e, 4e | G:($u$, 1/2, $w$) | $G_1^SG_2^S(4)$ | QNPL | [100] | 11.55 $(P_a2_1/m)$ |
| 1.620 | $NdCu_2$ | 74 $(Imma)$ | 51.74.2.1 $(P_I{}^1m^1m^1a^{\infty m}1)$ | Nd 4e, 8i, 8i | L:(1/2,$v$,$w$) | $L_1^SL_2^S(4)$ | QNPL | [010] | 62.456 $(P_Inma)$ |
| 1.621 | $La(Fe_{0.91}Al_{0.09})_{13}$ | 226 $(Fm\bar{3}c)$ | 127.140.2.1 $(P_I{}^14/{}^1m^1b^1m^{\infty m}1)$ | Fe 8f, 32m, 32m, 32m | F:($u$,1/2,$w$) | $F_1^SF_2^S(4)$ | QNPL | [001] | 124.362 $(P_I4/mcc)$ |
| 1.625 | $Sr_2Fe_3S_2O_3$ | 55 $(Pbam)$ | 62.55.2.1 $(P_b{}^1n^1m^1a^{\infty m}1)$ | Fe 8g | Q:(1/2,1/2,$w$) | $Q_1^SQ_1^S(8)$ | ONL | [001] | 62.451 $(P_bnma)$ |
| | | | | | L:(1/2,$v$,$w$) | $L_1^SL_2^S(4)$ | QNPL | | |
| | | | | | N:($u$,1/2,$w$) | $N_1^SN_2^S(4)$ | QNPL | | |
| | | | | | W:($u$,$v$,1/2) | $W_1^SW_2^S(4)$ | QNPL | | |
| 1.628 | $PrMnSi_2$ | 63 $(Cmcm)$ | 59.63.2.1 $(P_B{}^1m^1m^1n^{\infty m}1)$ | Pr 4a, Mn 4a | L:(1/2,$v$,$w$) | $L_1^SL_2^S(4)$ | QNPL | [010] | 52.318 $(P_Bnna)$ |
| | | | | | N:($u$,1/2,$w$) | $N_1^SN_2^S(4)$ | QNPL | | |
| 1.632 | $ErFe_6Ge_6$ | 71 $(Immm)$ | 59.71.2.1 $(P_I{}^1m^1m^1n^{\infty m}1)$ | Fe 4f, 8k, 8n, 16o | L:(1/2,$v$,$w$) | $L_1^SL_2^S(4)$ | QNPL | [100] | 48.264 $(P_Innn)$ |
| | | | | | N:($u$,1/2,$w$) | $N_1^SN_2^S(4)$ | QNPL | | |
| 1.633 | $YFe_6Sn_6$ | 71 $(Immm)$ | 59.71.2.1 $(P_I{}^1m^1m^1n^{\infty m}1)$ | Fe 4f, 8k, 8n, 16o | L:(1/2,$v$,$w$) | $L_1^SL_2^S(4)$ | QNPL | [100] | 48.264 $(P_Innn)$ |
| | | | | | N:($u$,1/2,$w$) | $N_1^SN_2^S(4)$ | QNPL | | |
| 1.634 | $YFe_6Ge_6$ | 63 | 57.63.2.1 | Fe 8d, | E:($u$,1/2,1/2) | $E_1^SE_1^S(8)$ | ONL | [100] | 52.318 |
| | | $(Cmcm)$ | $(P_C{}^1b{}^1c{}^1m^{\infty m}1)$ | 8e, 8g | N:$(u,1/2,w)$<br>W:$(u,v,1/2)$ | $N_1^SN_2^S(4)$<br>$W_1^SW_2^S(4)$ | QNPL<br>QNPL | | $(P_Bnna)$ |
| 1.635 | $ErFe_2Si_2$ | 139 $(I4/mmm)$ | 129.129.2.1 $(P_c{}^14/{}^1n{}^1m{}^1m^{\infty m}1)$ | Er 4c | F:$(u,1/2,w)$ | $F_1^SF_2^S(4)$ | QNPL | [100] | 62.450 $(P_anma)$ |
| 1.636 | $ErMn_2Si_2$ | 139 $(I4/mmm)$ | 129.139.2.1 $(P_I{}^14/{}^1n{}^1m{}^1m^{\infty m}1)$ | Mn 4d | F:$(u,1/2,w)$ | $F_1^SF_2^S(4)$ | QNPL | [001] | 126.386 $(P_I4/nnc)$ |
| 1.637 | $ErMn_2Si_2$ | 139 $(I4/mmm)$ | 129.139.2.1 $(P_I{}^14/{}^1n{}^1m{}^1m^{\infty m}1)$ | Mn 4d | F:$(u,1/2,w)$ | $F_1^SF_2^S(4)$ | QNPL | [001] | 126.386 $(P_I4/nnc)$ |
| 1.638 | $ErMn_2Ge_2$ | 139 $(I4/mmm)$ | 129.139.2.1 $(P_I{}^14/{}^1n{}^1m{}^1m^{\infty m}1)$ | Mn 4d | F:$(u,1/2,w)$ | $F_1^SF_2^S(4)$ | QNPL | [001] | 126.386 $(P_I4/nnc)$ |
| 1.639 | $ErMn_2Ge_2$ | 139 $(I4/mmm)$ | 129.139.2.1 $(P_I{}^14/{}^1n{}^1m{}^1m^{\infty m}1)$ | Mn 4d | F:$(u,1/2,w)$ | $F_1^SF_2^S(4)$ | QNPL | [001] | 126.386 $(P_I4/nnc)$ |
| 1.640 | $ErMn_2Ge_2$ | 139 $(I4/mmm)$ | 129.139.2.1 $(P_I{}^14/{}^1n{}^1m{}^1m^{\infty m}1)$ | Mn 4d | F:$(u,1/2,w)$ | $F_1^SF_2^S(4)$ | QNPL | [001] | 126.386 $(P_I4/nnc)$ |
| 1.641 | $Ba_2FeSi_2O_7$ | 113 $(P\bar{4}2_1m)$ | 114.113.2.1 $(P_c{}^1\bar{4}{}^12_1{}^1c^{\infty m}1)$ | Fe 4b | F:$(u,1/2,w)$ | $F_1^SF_2^S(4)$ | QNPL | [110] | 36.177 $(C_cmc2_1)$ |
| 1.643 | DyOCl | 129 $(P4/nmm)$ | 129.129.2.1 $(P_c{}^14/{}^1n{}^1m{}^1m^{\infty m}1)$ | Dy 4c | F:$(u,1/2,w)$ | $F_1^SF_2^S(4)$ | QNPL | [100] | 62.450 $(P_anma)$ |
| 1.652 | $Tb_2Ni_{1.78}In$ | 127 $(P4/mbm)$ | 63.51.2.1 $(C_a{}^1m{}^1c{}^1m^{\infty m}1)$ | Td 8e, 8e, 16i | Q:$(u,v,1/2)$ | $Q_1^SQ_2^S(4)$ | QNPL | [001] | 64.479 $(C_amce)$ |
| 1.655 | $FeNb_2O_6$ | 60 $(Pbcn)$ | 18.18.2.1 $(P_c{}^12_1{}^12_1{}^12^{\infty m}1)$ | Fe 4a, 4b | L:$(1/2,v,w)$<br>N:$(u,1/2,w)$ | $L_1^SL_1^S(4)$<br>$N_1^SN_1^S(4)$ | QNPL<br>QNPL | [100] | 19.28 $(P_c2_12_12_1)$ |
| 1.656 | $CoNb_2O_6$ | 60 $(Pbcn)$ | 14.14.2.1 $(P_a{}^12_1/{}^1c^{\infty m}1)$ | Co 8e | G:$(u,\ 1/2,\ w)$ | $G_1^SG_2^S(4)$ | QNPL | [100] | 14.80 $(P_a2_1/c)$ |
| 1.664 | $DyVO_4$ | 141 $(I4_1/amd)$ | 51.74.2.1 $(P_I{}^1m{}^1m{}^1a^{\infty m}1)$ | Dy 4b | L:$(1/2,v,w)$ | $L_1^SL_2^S(4)$ | QNPL | [010] | 62.456 $(P_Inma)$ |
| 1.676 | $Fe_{0.32}NbS_2$ | 182 $(P6_322)$ | 17.20.2.1 $(P_C{}^12{}^12{}^12_1{}^{\infty m}1)$ | Fe 4a | W:$(u,v,1/2)$ | $W_1^SW_1^S(4)$ | QNPL | [001] | 19.29 $(P_c2_12_12_1)$ |
| 1.677 | $Fe_{0.35}NbS_2$ | 182 $(P6_322)$ | 19.18.2.1 $(P_c{}^12_1{}^12_1{}^12_1{}^{\infty m}1)$ | Fe 8c | R:$(1/2,1/2,1/2)$<br>L:$(1/2,v,w)$<br>N:$(u,1/2,w)$ | $R_1^SR_1^S(8)$<br>$L_1^SL_1^S(4)$<br>$N_1^SN_1^S(4)$ | $C$−4 OP<br>QNPL<br>QNPL | [001] | 18.21 $(P_c2_12_12)$ |
| | | | | | W:($u$,$v$,1/2) | $W_1^SW_1^S(4)$ | QNPL | | |
| 1.678 | CrN | 225<br>$(Fm\bar{3}m)$ | 59.59.2.1<br>$(P_c{}^1m{}^1m{}^1n{}^{\infty m}1)$ | Cr 4a | L:(1/2,$v$,$w$)<br>N:($u$,1/2,$w$) | $L_1^SL_2^S(4)$<br>$N_1^SN_2^S(4)$ | QNPL<br>QNPL | [010] | 62.450<br>$(P_anma)$ |
| 1.682 | $Na_2CuSO_4Cl_2$ | 62<br>$(Pnma)$ | 31.31.2.1<br>$(P_b{}^1m{}^1n{}^12_1{}^{\infty m}1)$ | Cu 4a, 4a | W:($u$,$v$,1/2) | $W_1^SW_1^S(4)$ | QNPL | [010] | 31.129<br>$(P_bmn2_1)$ |
| 1.685 | $NiCr_2O_4$ | 141<br>$(I4_1/amd)$ | 20.22.2.1<br>$(C_A{}^12{}^12{}^12_1{}^{\infty m}1)$ | Cr 16k | Q:($u$,$v$,1/2) | $Q_1^SQ_1^S(4)$ | QNPL | [001] | 20.37<br>$(C_A222_1)$ |
| 1.689 | $LuMn_2Ge_2$ | 139<br>$(I4/mmm)$ | 129.139.2.1<br>$(P_I{}^14/{}^1n{}^1m{}^1m{}^{\infty m}1)$ | Mn 4d | F:($u$,1/2,$w$) | $F_1^SF_2^S(4)$ | QNPL | [001] | 126.386<br>$(P_I4/nnc)$ |
| 1.690 | $TmMn_2Ge_2$ | 139<br>$(I4/mmm)$ | 129.139.2.1<br>$(P_I{}^14/{}^1n{}^1m{}^1m{}^{\infty m}1)$ | Mn 4d | F:($u$,1/2,$w$) | $F_1^SF_2^S(4)$ | QNPL | [001] | 126.386<br>$(P_I4/nnc)$ |
| 1.691 | $YMn_2Ge_2$ | 139<br>$(I4/mmm)$ | 129.139.2.1<br>$(P_I{}^14/{}^1n{}^1m{}^1m{}^{\infty m}1)$ | Mn 4d | F:($u$,1/2,$w$) | $F_1^SF_2^S(4)$ | QNPL | [001] | 126.386<br>$(P_I4/nnc)$ |
| 1.692 | $YMn_2Ge_2$ | 139<br>$(I4/mmm)$ | 129.139.2.1<br>$(P_I{}^14/{}^1n{}^1m{}^1m{}^{\infty m}1)$ | Mn 4d | F:($u$,1/2,$w$) | $F_1^SF_2^S(4)$ | QNPL | [001] | 126.386<br>$(P_I4/nnc)$ |
| 1.693 | $DyMn_2Ge_2$ | 139<br>$(I4/mmm)$ | 129.139.2.1<br>$(P_I{}^14/{}^1n{}^1m{}^1m{}^{\infty m}1)$ | Mn 4d | F:($u$,1/2,$w$) | $F_1^SF_2^S(4)$ | QNPL | [001] | 126.386<br>$(P_I4/nnc)$ |
| 1.694 | $TbMn_2Ge_2$ | 139<br>$(I4/mmm)$ | 129.139.2.1<br>$(P_I{}^14/{}^1n{}^1m{}^1m{}^{\infty m}1)$ | Mn 4d | F:($u$,1/2,$w$) | $F_1^SF_2^S(4)$ | QNPL | [001] | 126.386<br>$(P_I4/nnc)$ |
| 1.696 | $HoNiSi_2$ | 63<br>$(Cmcm)$ | 59.63.2.1<br>$(P_B{}^1m{}^1m{}^1n{}^{\infty m}1)$ | Ho 4c | L:(1/2,$v$,$w$)<br>N:($u$,1/2,$w$) | $L_1^SL_2^S(4)$<br>$N_1^SN_2^S(4)$ | QNPL<br>QNPL | [001] | 57.391<br>$(P_cbcm)$ |
| 1.697 | $Tb_5Pd_2In_4$ | 55<br>$(Pbam)$ | 26.26.2.1<br>$(P_b{}^1m{}^1c{}^12_1{}^{\infty m}1)$ | Tb 4a, 4a, 4a | W:($u$,$v$,1/2) | $W_1^SW_1^S(4)$ | QNPL | [001] | 26.72<br>$(P_bmc2_1)$ |
| 1.703 | $YBaCo_2O_5$ | 123<br>$(P4/mmm)$ | 51.51.2.1<br>$(P_b{}^1m{}^1m{}^1a{}^{\infty m}1)$ | Co 4e, 4e | L:(1/2,$v$,$w$) | $L_1^SL_2^S(4)$ | QNPL | [010] | 53.330<br>$(P_amna)$ |
| 1.704 | $TaBaFe_2O_5$ | 123<br>$(P4/mmm)$ | 51.51.2.1<br>$(P_b{}^1m{}^1m{}^1a{}^{\infty m}1)$ | Fe 4e, 4e | L:(1/2,$v$,$w$) | $L_1^SL_2^S(4)$ | QNPL | [010] | 53.330<br>$(P_amna)$ |
| 1.714 | $CeAuBi_2$ | 129<br>$(P4/nmm)$ | 129.129.2.1<br>$(P_c{}^14/{}^1n{}^1m{}^1m{}^{\infty m}1)$ | Ce 4c | F:($u$,1/2,$w$) | $F_1^SF_2^S(4)$ | QNPL | [001] | 130.432<br>$(P_c4/ncc)$ |
| 1.715 | $Sr_2CoWO_6$ | 14<br>$(P2_1/c)$ | 14.14.2.1<br>$(P_a{}^12_1/{}^1c{}^{\infty m}1)$ | Co 4c | G:($u$, 1/2, $w$) | $G_1^SG_2^S(4)$ | QNPL | [110] | 2.7<br>$(P_S\bar{1})$ |

| 1.722 | $Ba_3LaRu_2O_9$ | 194<br>$(P6_3/mmc)$ | 51.63.2.1<br>$(P_A{}^1m^1m^1a^{\infty m}1)$ | Ru 8f | L:(1/2,$v$,$w$) | $L_1^SL_2^S(4)$ | QNPL | [010]<br>+22°<br>([001]) | 14.84<br>$(P_C2_1/c)$ |
|---|---|---|---|---|---|---|---|---|---|
| 1.727 | $Tm_3Cu_4Ge_4$ | 71<br>$(Immm)$ | 59.59.2.1<br>$(P_C{}^1m^1m^1n^{\infty m}1)$ | Tm 4a | L:(1/2,$v$,$w$)<br>N:($u$,1/2,$w$) | $L_1^SL_2^S(4)$<br>$N_1^SN_2^S(4)$ | QNPL<br>QNPL | [001] | 62.450<br>$(P_anma)$ |
| 1.744 | PrPdSn | 62<br>$(Pnma)$ | 14.14.2.1<br>$(P_a{}^12_1/{}^1c^{\infty m}1)$ | Pr 8e | G:($u$, 1/2, $w$) | $G_1^SG_2^S(4)$ | QNPL | [010] | 14.80<br>$(P_a2_1/c)$ |
| 1.746 | $YMn_2$ | 227<br>$(Fd\bar{3}m)$ | 95.98.2.1<br>$(P_I{}^14_3{}^12^12^{\infty m}1)$ | Mn 8f | E:($u$,$v$,1/2) | $E_1^SE_1^S(4)$ | QNPL | [111] | 4.12<br>$(P_C2_1)$ |
| 1.749 | HoSbTe | 129<br>$(P4/nmm)$ | 11.11.2.1<br>$(P_a{}^12_1/{}^1m^{\infty m}1)$ | Ho 4e | G:($u$, 1/2, $w$) | $G_1^SG_2^S(4)$ | QNPL | [010] | 11.55<br>$(P_a2_1/m)$ |
| 1.750 | HoSbTe | 129<br>$(P4/nmm)$ | 11.11.2.1<br>$(P_a{}^12_1/{}^1m^{\infty m}1)$ | Ho 4e | G:($u$, 1/2, $w$) | $G_1^SG_2^S(4)$ | QNPL | [010] | 11.55<br>$(P_a2_1/m)$ |
| 1.754 | $BaFe_2O_4$ | 36<br>$(Cmc2_1)$ | 33.36.2.1<br>$(P_C{}^1n^1a^12_1{}^{\infty m}1)$ | Fe 8b,<br>8b | U:(1/2,0,1/2)<br>W:($u$,$v$,1/2) | $U_1^SU_1^S(8)$<br>$W_1^SW_1^S(4)$ | OP<br>QNPL | [010] | 29.109<br>$(P_Cca2_1)$ |
| 1.756 | $CaFe_3O_5$ | 63<br>$(Cmcm)$ | 60.57.2.1<br>$(P_b{}^1b^1c^1n^{\infty m}1)$ | Fe 8c,<br>16e | P:(1/2,$v$,1/2)<br>T:(0,1/2,1/2)<br>L:(1/2,$v$,$w$)<br>W:($u$,$v$,1/2) | $P_1^SP_1^S(8)$<br>$T_1^ST_2^S(8)$<br>$L_1^SL_2^S(4)$<br>$W_1^SW_2^S(4)$ | ONL<br>OP<br>QNPL<br>QNPL | [001] | 61.438<br>$(P_abca)$ |
| 1.762 | $CaFe_3O_5$ | 63<br>$(Cmcm)$ | 60.57.2.1<br>$(P_b{}^1b^1c^1n^{\infty m}1)$ | Fe 8c,<br>16e | P:(1/2,$v$,1/2)<br>T:(0,1/2,1/2)<br>L:(1/2,$v$,$w$)<br>W:($u$,$v$,1/2) | $P_1^SP_1^S(8)$<br>$T_1^ST_2^S(8)$<br>$L_1^SL_2^S(4)$<br>$W_1^SW_2^S(4)$ | ONL<br>OP<br>QNPL<br>QNPL | [001] | 61.438<br>$(P_abca)$ |
| 1.765 | DySbTe | 129<br>$(P4/nmm)$ | 129.129.2.1<br>$(P_C{}^14/{}^1n^1m^1m^{\infty m}1)$ | Dy 4c | F:($u$,1/2,$w$) | $F_1^SF_2^S(4)$ | QNPL | [010] | 62.450<br>$(P_anma)$ |
| 1.766 | DySbTe | 129<br>$(P4/nmm)$ | 11.11.2.1<br>$(P_a{}^12_1/{}^1m^{\infty m}1)$ | Dy 4e | G:($u$, 1/2, $w$) | $G_1^SG_2^S(4)$ | QNPL | [010] | 11.55<br>$(P_a2_1/m)$ |
| 2.1 | $EuFe_2As_2$ | 139<br>$(I4/mmm)$ | 53.64.2.1<br>$(P_C{}^1m^1n^1a^{\infty m}1)$ | Eu 4b,<br>Fe 8e | W:($u$,$v$,1/2) | $W_1^SW_2^S(4)$ | QNPL | [100] | 61.439<br>$(P_Cbca)$ |
| 2.30 | $CeRh_2Si_2$ | 139<br>$(I4/mmm)$ | 51.67.2.1<br>$(P_C{}^1m^1m^1a^{\infty m}1)$ | Ce 4e,<br>4f, 4e, | L:(1/2,$v$,$w$) | $L_1^SL_2^S(4)$ | QNPL | [001] | 54.350<br>$(P_Bcca)$ |

<table>
<tr><td></td><td></td><td></td><td></td><td>4f</td><td></td><td></td><td></td><td></td><td></td></tr>
<tr><td rowspan="3">2.36</td><td rowspan="3">$TbGe_3$</td><td rowspan="3">63<br>($Cmcm$)</td><td rowspan="3">57.57.2.1<br>($P_a{}^1b{}^1c{}^1m{}^{\infty m}1$)</td><td rowspan="3">Tb 8d,<br>16e</td><td>E:($u$,1/2,1/2)</td><td>$E_1^S E_1^S(8)$</td><td>ONL</td><td rowspan="3">[001]</td><td rowspan="3">62.452<br>($P_cnma$)</td></tr>
<tr><td>N:($u$,1/2,$w$)</td><td>$N_1^S N_2^S(4)$</td><td>QNPL</td></tr>
<tr><td>W:($u$,$v$,1/2)</td><td>$W_1^S W_2^S(4)$</td><td>QNPL</td></tr>
</table>

## S4.6. Chiral magnons in altermagnets

Supplementary Table XXIX. Tabulation of altermagnet that accommodate chiral magnons. The seven columns illustrate the material ID in MAGNDATA, chemical formula, space group, spin space groups, the momentum-position that accommodate chiral magnon, the orientation of magnetic materials and the corresponding magnetic space groups. The spin axis is written under the lattice basis, and the "[$a_1b_1c_1$] +θ [$a_2b_2c_2$]" labels the axis by rotating from the direction of "[$a_1b_1c_1$]" towards the direction of "[$a_2b_2c_2$]" by θ degrees.

| ID | Formula | SG | SSG | ***k***-point | Spin Axis | MSG |
|---|---|---|---|---|---|---|
| 0.1 | $LaMnO_3$ | 62 ($Pnma$) | 14.62.1.1 ($P^{\bar{1}}n^{\bar{1}}m^{1}a^{\infty m}1$) | V:($u$,$v$,0) W:($u$,$v$,1/2) GP:($u$,$v$,$w$) | [100] | 62.448 ($Pn'ma'$) |
| 0.3 | $Ca_3LiOsO_6$ | 167 ($R\bar{3}c$) | 148.167.1.1 ($R^{1}\bar{3}^{\bar{1}}c^{\infty m}1$) | GP:($u$,$v$,$w$) | [100] | 15.89 ($C2'/c'$) |
| 0.13 | $Ca_3Co_{2-x}Mn_xO_6$ | 167 ($R\bar{3}c$) | 146.161.1.1 ($R^{1}3^{\bar{1}}c^{\infty m}1$) | GP:($u$,$v$,$w$) | [001] | 161.69 ($R3c$) |
| 0.15 | $MnF_2$ | 136 ($P4_2/mnm$) | 65.136.1.1 ($P^{\bar{1}}4_2/{}^{1}m^{\bar{1}}n^{1}m^{\infty m}1$) | S:($u$,$u$,1/2) Σ:($u$,$u$,0) C:($u$,$u$,$w$) D:($u$,$v$,0) E:($u$,$v$,1/2) GP:($u$,$v$,$w$) | [001] | 136.499 ($P4_2'/mnm'$) |
| 0.21 | $PbNiO_3$ | 161 ($R3c$) | 146.161.1.1 ($R^{1}3^{\bar{1}}c^{\infty m}1$) | GP:($u$,$v$,$w$) | [001] | 161.69 ($R3c$) |
| 0.23 | $Ca_3Mn_2O_7$ | 36 ($Cmc2_1$) | 4.36.1.1 ($C^{\bar{1}}m^{\bar{1}}c^{1}2_1{}^{\infty m}1$) | R:(1/2,1/2,1/2) S:(1/2,1/2,0) D:(1/2,1/2,$w$) P:($u$,$v$,0) Q:($u$,$v$,1/2) GP:($u$,$v$,$w$) | [100] | 36.174 ($Cm'c2_1'$) |
| 0.25 | $NaOsO_3$ | 62 ($Pnma$) | 14.62.1.4 ($P^{1}n^{\bar{1}}m^{\bar{1}}a^{\infty m}1$) | K:(0,$v$,$w$) L:(1/2,$v$,$w$) GP:($u$,$v$,$w$) | [001] | 62.448 ($Pn'ma'$) |
| 0.34 | $La_{0.5}Sr_{0.5}FeO_{2.5}F_{0.5}$ | 62 ($Pnma$) | 14.62.1.4 ($P^{1}n^{\bar{1}}m^{\bar{1}}a^{\infty m}1$) | K:(0,$v$,$w$) L:(1/2,$v$,$w$) GP:($u$,$v$,$w$) | [001] | 62.448 ($Pn'ma'$) |
| 0.45 | $La_2NiO_4$ | 138 ($P4_2/ncm$) | 14.56.1.1 ($P^{\bar{1}}c^{1}c^{\bar{1}}n^{\infty m}1$) | M:($u$,0,$w$) N:($u$,1/2,$w$) GP:($u$,$v$,$w$) | [100] | 56.369 ($Pc'c'n$) |
| 0.50 | $MnTiO_3$ | 161 ($R3c$) | 146.161.1.1 ($R^{1}3^{\bar{1}}c^{\infty m}1$) | GP:($u$,$v$,$w$) | [100] | 9.39 ($Cc'$) |
| 0.52 | $K_yFe_{2-x}Se_2$ | 87 ($I4/m$) | 2.12.1.1 ($C^{\bar{1}}2/^{\bar{1}}m^{\infty m}1$) | L:(1/2,1/2,1/2) V:(1/2,1/2,0) GP:($u$,$v$,$w$) | [001] | 12.62 ($C2'/m'$) |
| 0.53 | $Rb_yFe_{2-x}Se_2$ | 87 ($I4/m$) | 2.12.1.1 ($C^{\bar{1}}2/^{\bar{1}}m^{\infty m}1$) | L:(1/2,1/2,1/2) V:(1/2,1/2,0) GP:($u$,$v$,$w$) | [001] | 12.62 ($C2'/m'$) |
| 0.56 | $Ba_2CoGe_2O_7$ | 113 ($P\bar{4}2_1m$) | 81.113.1.1 ($P^{1}\bar{4}^{\bar{1}}2_1{}^{\bar{1}}m^{\infty m}1$) | D:($u$,$v$,0) E:($u$,$v$,1/2) GP:($u$,$v$,$w$) | [110] | 35.167 ($Cm'm2'$) |
| 0.57 | $ScFeO_3$ | 161 ($R3c$) | 146.161.1.1 ($R^{1}3^{\bar{1}}c^{\infty m}1$) | GP:($u$,$v$,$w$) | [100] | 9.39 ($Cc'$) |
| 0.62 | $SrMn_2V_2O_8$ | 110 ($I4_1cd$) | 45.110.1.1 ($I^{\bar{1}}4_1{}^{1}c^{\bar{1}}d^{\infty m}1$) | N:(1/2,0,1/2) Q:(1/2,$v$,1/2) Σ:($u$,0,0) B:($u$,0,$w$) C:($u$,$v$,0) GP:($u$,$v$,$w$) | [010] | 45.237 ($Ib'a2'$) |
| 0.65 | $Fe_2O_3$−alpha | 167 ($R\bar{3}c$) | 148.167.1.1 ($R^{1}\bar{3}^{\bar{1}}c^{\infty m}1$) | GP:($u$,$v$,$w$) | [100] | 15.89 ($C2'/c'$) |
| 0.66 | $Fe_2O_3$−alpha | 167 ($R\bar{3}c$) | 148.167.1.1 ($R^{1}\bar{3}^{\bar{1}}c^{\infty m}1$) | GP:($u$,$v$,$w$) | [001] +11° ([100]) | 2.4 ($P\bar{1}$) |
| 0.67 | $BiFe_{0.5}Sc_{0.5}O_3$ | 46 ($Ima2$) | 5.46.1.1 ($I^{\bar{1}}m^{\bar{1}}a^{1}2^{\infty m}1$) | T:(1/2,1/2,0) P:(1/2,1/2,$w$) GP:($u$,$v$,$w$) | [100] | 46.243 ($Im'a2'$) |
| 0.68 | $BiFe_{0.5}Sc_{0.5}O_3$ | 62 ($Pnma$) | 14.62.1.4 ($P^{1}n^{\bar{1}}m^{\bar{1}}a^{\infty m}1$) | K:(0,$v$,$w$) L:(1/2,$v$,$w$) GP:($u$,$v$,$w$) | [010] | 62.446 ($Pn'm'a$) |
| 0.79 | $CaIrO_3$ | 63 ($Cmcm$) | 12.63.1.1 ($C^{1}m^{\bar{1}}c^{\bar{1}}m^{\infty m}1$) | K:(0,$v$,$w$) GP:($u$,$v$,$w$) | [001] | 63.464 ($Cm'cm'$) |
| 0.83 | $LiFeP_2O_7$ | 4 ($P2_1$) | 1.4.1.1 ($P^{\bar{1}}2_1{}^{\infty m}1$) | GP:($u$,$v$,$w$) | [100] +11° ([001]) | 4.7 ($P2_1$) |
| 0.98 | $YBaMn_2O_{5.5}$ | 72 ($Ibam$) | 15.72.1.1 ($I^{1}b^{\bar{1}}a^{\bar{1}}m^{\infty m}1$) | S:(0,1/2,1/2) A:(0,$v$,$w$) D:($u$,1/2,1/2) GP:($u$,$v$,$w$) | [100] | 72.543 ($Ib'a'm$) |

| 0.99 | $YBaMn_2O_{5.5}$ | 72<br>$(Ibam)$ | 15.72.1.1<br>$(I^{1}b^{\bar{1}}a^{\bar{1}}m^{\infty m}1)$ | S:(0,1/2,1/2) A:(0,$v$,$w$) D:($u$,1/2,1/2) GP:($u$,$v$,$w$) | [100]<br>+28°<br>([001]) | 12.58<br>$(C2/m)$ |
|---|---|---|---|---|---|---|
| 0.105 | $ErVO_3$ | 62<br>$(Pnma)$ | 2.14.1.1<br>$(P^{\bar{1}}2_1/^{\bar{1}}c^{\infty m}1)$ | GP:($u$,$v$,$w$) | [001] | 14.75<br>$(P2_1/c)$ |
| 0.112 | $FeBO_3$ | 167<br>$(R\bar{3}c)$ | 148.167.1.1<br>$(R^{1}\bar{3}^{\bar{1}}c^{\infty m}1)$ | GP:($u$,$v$,$w$) | [100] | 15.89<br>$(C2'/c')$ |
| 0.113 | $NiCO_3$ | 167<br>$(R\bar{3}c)$ | 148.167.1.1<br>$(R^{1}\bar{3}^{\bar{1}}c^{\infty m}1)$ | GP:($u$,$v$,$w$) | [120] | 15.85<br>$(C2/c)$ |
| 0.114 | $CoCO_3$ | 167<br>$(R\bar{3}c)$ | 148.167.1.1<br>$(R^{1}\bar{3}^{\bar{1}}c^{\infty m}1)$ | GP:($u$,$v$,$w$) | [120] | 15.85<br>$(C2/c)$ |
| 0.115 | $MnCO_3$ | 167<br>$(R\bar{3}c)$ | 148.167.1.1<br>$(R^{1}\bar{3}^{\bar{1}}c^{\infty m}1)$ | GP:($u$,$v$,$w$) | [120] | 15.85<br>$(C2/c)$ |
| 0.116 | $FeCO_3$ | 167<br>$(R\bar{3}c)$ | 148.167.1.1<br>$(R^{1}\bar{3}^{\bar{1}}c^{\infty m}1)$ | GP:($u$,$v$,$w$) | [001] | 167.103<br>$(R\bar{3}c)$ |
| 0.118 | $Ba_5Co_5ClO_{13}$ | 194<br>$(P6_3/mmc)$ | 164.194.1.1<br>$(P^{\bar{1}}6_3/^{\bar{1}}m^{1}m^{\bar{1}}c^{\infty m}1)$ | D:($u$,0,$w$) GP:($u$,$v$,$w$) | [001] | 194.268<br>$(P6_3'/m'm'c)$ |
| 0.128 | $FeSO_4F$ | 15<br>$(C2/c)$ | 2.15.1.1<br>$(C^{\bar{1}}2/^{\bar{1}}c^{\infty m}1)$ | L:(1/2,1/2,1/2) V:(1/2,1/2,0) GP:($u$,$v$,$w$) | [010] | 15.89<br>$(C2'/c')$ |
| 0.131 | $Mn(N(CN_2))_2$ | 58<br>$(Pnnm)$ | 10.58.1.1<br>$(P^{\bar{1}}n^{\bar{1}}n^{1}m^{\infty m}1)$ | V:($u$,$v$,0) W:($u$,$v$,1/2) GP:($u$,$v$,$w$) | [100] | 58.398<br>$(Pnn'm')$ |
| 0.137 | $Cu_2V_2O_7$ | 43<br>$(Fdd2)$ | 9.43.1.1<br>$(F^{\bar{1}}d^{1}d^{\bar{1}}2^{\infty m}1)$ | L:(1/2,1/2,1/2) J:($u$,0,$w$) GP:($u$,$v$,$w$) | [100] | 43.227<br>$(Fd'd'2)$ |
| 0.148 | $La_2LiRuO_6$ | 14<br>$(P2_1/c)$ | 2.14.1.1<br>$(P^{\bar{1}}2_1/^{\bar{1}}c^{\infty m}1)$ | GP:($u$,$v$,$w$) | [100]<br>+39°<br>([001]) | 14.75<br>$(P2_1/c)$ |
| 0.153 | $Bi_2RuMnO_7$ | 227<br>$(Fd\bar{3}m)$ | 74.141.1.1<br>$(I^{\bar{1}}4_1/^{1}a^{1}m^{\bar{1}}d^{\infty m}1)$ | N:(1/2,0,1/2) Q:(1/2,$v$,1/2) Σ:($u$,0,0) B:($u$,0,$w$) C:($u$,$v$,0) GP:($u$,$v$,$w$) | [100] | 70.530<br>$(Fd'd'd)$ |
| 0.154 | $Er_2Ru_2O_7$ | 227<br>$(Fd\bar{3}m)$ | 74.141.1.1<br>$(I^{\bar{1}}4_1/^{1}a^{1}m^{\bar{1}}d^{\infty m}1)$ | N:(1/2,0,1/2) Q:(1/2,$v$,1/2) Σ:($u$,0,0) B:($u$,0,$w$) C:($u$,$v$,0) GP:($u$,$v$,$w$) | [100] | 141.554<br>$(I4_1'/am'd)$ |

| 0.178 | $CoF_2$ | 136 $(P4_2/mnm)$ | 65.136.1.1 $(P^{\bar{1}}4_2/{}^{1}m^{\bar{1}}n^{1}m^{\infty m}1)$ | S:$(u,u,1/2)$ Σ:$(u,u,0)$ C:$(u,u,w)$ D:$(u,v,0)$ E:$(u,v,1/2)$ GP:$(u,v,w)$ | [001] | 136.499 $(P4'_2/mnm')$ |
|---|---|---|---|---|---|---|
| 0.189 | $CeMn_2Ge_4O_{12}$ | 125 $(P4/nbm)$ | 67.125.1.1 $(P^{\bar{1}}4/{}^{1}n^{\bar{1}}b^{1}m^{\infty m}1)$ | S:$(u,u,1/2)$ Σ:$(u,u,0)$ C:$(u,u,w)$ D:$(u,v,0)$ E:$(u,v,1/2)$ GP:$(u,v,w)$ | [001] | 125.367 $(P4'/nbm')$ |
| 0.190 | $CeMnCoGe_4O_{12}$ | 125 $(P4/nbm)$ | 67.125.1.1 $(P^{\bar{1}}4/{}^{1}n^{\bar{1}}b^{1}m^{\infty m}1)$ | S:$(u,u,1/2)$ Σ:$(u,u,0)$ C:$(u,u,w)$ D:$(u,v,0)$ E:$(u,v,1/2)$ GP:$(u,v,w)$ | [100] | 50.282 $(Pb'an')$ |
| 0.201 | $Ca_2PrCr_2NbO_9$ | 62 $(Pnma)$ | 14.62.1.4 $(P^{1}n^{\bar{1}}m^{\bar{1}}a^{\infty m}1)$ | K:$(0,v,w)$ L:$(1/2,v,w)$ GP:$(u,v,w)$ | [010] | 62.446 $(Pn'm'a)$ |
| 0.202 | $Ca_2PrCr_2TaO_9$ | 62 $(Pnma)$ | 14.62.1.4 $(P^{1}n^{\bar{1}}m^{\bar{1}}a^{\infty m}1)$ | K:$(0,v,w)$ L:$(1/2,v,w)$ GP:$(u,v,w)$ | [010] | 62.446 $(Pn'm'a)$ |
| 0.206 | $Ca_2Fe_{0.875}Cr_{0.125}GaO_5$ | 62 $(Pnma)$ | 11.62.1.1 $(P^{\bar{1}}n^{1}m^{\bar{1}}a^{\infty m}1)$ | M:$(u,0,w)$ N:$(u,1/2,w)$ GP:$(u,v,w)$ | [100] | 62.446 $(Pn'm'a)$ |
| 0.229 | $Ba_2MnSi_2O_7$ | 113 $(P\bar{4}2_1m)$ | 81.113.1.1 $(P^{1}\bar{4}^{\bar{1}}2_1{}^{\bar{1}}m^{\infty m}1)$ | D:$(u,v,0)$ E:$(u,v,1/2)$ GP:$(u,v,w)$ | [001] | 113.267 $(P\bar{4}2_1m)$ |
| 0.239 | $Ca_3LiRuO_6$ | 167 $(R\bar{3}c)$ | 148.167.1.1 $(R^{1}\bar{3}^{\bar{1}}c^{\infty m}1)$ | GP:$(u,v,w)$ | [100] | 15.89 $(C2'/c')$ |
| 0.241 | $Y_2Cu_2O_5$ | 33 $(Pna2_1)$ | 7.33.1.1 $(P^{\bar{1}}n^{1}a^{\bar{1}}2_1{}^{\infty m}1)$ | M:$(u,0,w)$ N:$(u,1/2,w)$ GP:$(u,v,w)$ | [010] | 33.144 $(Pna2_1)$ |
| 0.254 | $[C(ND_2)_3]Cu(DCOO)_3$ | 33 $(Pna2_1)$ | 7.33.1.1 $(P^{\bar{1}}n^{1}a^{\bar{1}}2_1{}^{\infty m}1)$ | M:$(u,0,w)$ N:$(u,1/2,w)$ GP:$(u,v,w)$ | [010] | 33.144 $(Pna2_1)$ |
| 0.255 | $[C(ND_2)_3]Cu(DCOO)_3$ | 33 $(Pna2_1)$ | 7.33.1.1 $(P^{\bar{1}}n^{1}a^{\bar{1}}2_1{}^{\infty m}1)$ | M:$(u,0,w)$ N:$(u,1/2,w)$ GP:$(u,v,w)$ | [100] | 33.148 $(Pn'a'2_1)$ |
| 0.256 | $[C(ND_2)_3]Mn(DCOO)_3$ | 52 $(Pnna)$ | 13.52.1.4 $(P^{1}n^{\bar{1}}n^{\bar{1}}a^{\infty m}1)$ | K:$(0,v,w)$ L:$(1/2,v,w)$ GP:$(u,v,w)$ | [010] | 52.310 $(Pn'n'a)$ |
| 0.257 | $[C(ND_2)_3]Co(DCOO)_3$ | 52 $(Pnna)$ | 13.52.1.4 $(P^{1}n^{\bar{1}}n^{\bar{1}}a^{\infty m}1)$ | K:$(0,v,w)$ L:$(1/2,v,w)$ GP:$(u,v,w)$ | [001] | 52.312 $(Pn'na')$ |
| 0.260 | $CuFePO_5$ | 62 $(Pnma)$ | 11.62.1.1 $(P^{\bar{1}}n^{1}m^{\bar{1}}a^{\infty m}1)$ | M:$(u,0,w)$ N:$(u,1/2,w)$ GP:$(u,v,w)$ | [010] | 62.441 $(Pnma)$ |
| 0.261 | $NiFePO_5$ | 62 $(Pnma)$ | 11.62.1.1 $(P^{\bar{1}}n^{1}m^{\bar{1}}a^{\infty m}1)$ | M:$(u,0,w)$ N:$(u,1/2,w)$ GP:$(u,v,w)$ | [010] | 62.441 $(Pnma)$ |

| 0.263 | $Fe_2PO_5$ | 62 ($Pnma$) | 11.62.1.1 ($P^{\bar{1}}n^{1}m^{\bar{1}}a^{\infty m}1$) | M:($u$,0,$w$) N:($u$,1/2,$w$) GP:($u$,$v$,$w$) | [010] | 62.441 ($Pnma$) |
|---|---|---|---|---|---|---|
| 0.301 | $Sr_2CoTeO_6$ | 14 ($P2_1/c$) | 2.14.1.1 ($P^{\bar{1}}2_1/^{\bar{1}}c^{\infty m}1$) | GP:($u$,$v$,$w$) | [001] +26° ([100]) | 14.75 ($P2_1/c$) |
| 0.302 | $Sr_2Co_{0.9}Mg_{0.1}TeO_6$ | 14 ($P2_1/c$) | 2.14.1.1 ($P^{\bar{1}}2_1/^{\bar{1}}c^{\infty m}1$) | GP:($u$,$v$,$w$) | [001] +25° ([100]) | 14.75 ($P2_1/c$) |
| 0.303 | $BaCrF_5$ | 19 ($P2_12_12_1$) | 4.19.1.1 ($P^{\bar{1}}2_1{}^{\bar{1}}2_1{}^{1}2_1{}^{\infty m}1$) | V:($u$,$v$,0) W:($u$,$v$,1/2) GP:($u$,$v$,$w$) | [001] | 19.27 ($P2_1'2_1'2_1$) |
| 0.306 | $GaFeO_3$ | 161 ($R3c$) | 146.161.1.1 ($R^{1}3^{\bar{1}}c^{\infty m}1$) | GP:($u$,$v$,$w$) | [100] | 9.39 ($Cc'$) |
| 0.307 | $ScCrO_3$ | 62 ($Pnma$) | 11.62.1.1 ($P^{\bar{1}}n^{1}m^{\bar{1}}a^{\infty m}1$) | M:($u$,0,$w$) N:($u$,1/2,$w$) GP:($u$,$v$,$w$) | [010] | 62.441 ($Pnma$) |
| 0.308 | $InCrO_3$ | 62 ($Pnma$) | 11.62.1.1 ($P^{\bar{1}}n^{1}m^{\bar{1}}a^{\infty m}1$) | M:($u$,0,$w$) N:($u$,1/2,$w$) GP:($u$,$v$,$w$) | [010] | 62.441 ($Pnma$) |
| 0.309 | $TlCrO_3$ | 62 ($Pnma$) | 11.62.1.1 ($P^{\bar{1}}n^{1}m^{\bar{1}}a^{\infty m}1$) | M:($u$,0,$w$) N:($u$,1/2,$w$) GP:($u$,$v$,$w$) | [010] | 62.441 ($Pnma$) |
| 0.315 | $ZrMn_2Ge_4O_{12}$ | 125 ($P4/nbm$) | 67.125.1.1 ($P^{\bar{1}}4/^{1}n^{\bar{1}}b^{1}m^{\infty m}1$) | S:($u$,$u$,1/2) Σ:($u$,$u$,0) C:($u$,$u$,$w$) D:($u$,$v$,0) E:($u$,$v$,1/2) GP:($u$,$v$,$w$) | [001] | 125.367 ($P4'/nbm'$) |
| 0.323 | $LaCrO_3$ | 62 ($Pnma$) | 14.62.1.4 ($P^{1}n^{\bar{1}}m^{\bar{1}}a^{\infty m}1$) | K:(0,$v$,$w$) L:(1/2,$v$,$w$) GP:($u$,$v$,$w$) | [100] | 62.441 ($Pnma$) |
| 0.329 | $RbMnF_4$ | 14 ($P2_1/c$) | 2.14.1.1 ($P^{\bar{1}}2_1/^{\bar{1}}c^{\infty m}1$) | GP:($u$,$v$,$w$) | [110] +20° ([001]) | 2.4 ($P\bar{1}$) |
| 0.331 | $Fe_2Mo_3O_8$ | 186 ($P6_3mc$) | 156.186.1.1 ($P^{\bar{1}}6_3{}^{1}m^{\bar{1}}c^{\infty m}1$) | D:($u$,0,$w$) GP:($u$,$v$,$w$) | [001] | 186.205 ($P6_3'm'c$) |
| 0.332 | $Co_2Mo_3O_8$ | 186 ($P6_3mc$) | 156.186.1.1 ($P^{\bar{1}}6_3{}^{1}m^{\bar{1}}c^{\infty m}1$) | D:($u$,0,$w$) GP:($u$,$v$,$w$) | [001] | 186.205 ($P6_3'm'c$) |
| 0.334 | $CoF_3$ | 167 ($R\bar{3}c$) | 148.167.1.1 ($R^{1}\bar{3}^{\bar{1}}c^{\infty m}1$) | GP:($u$,$v$,$w$) | [001] | 167.103 ($R\bar{3}c$) |
| 0.335 | $FeF_3$ | 167 ($R\bar{3}c$) | 148.167.1.1 ($R^{1}\bar{3}^{\bar{1}}c^{\infty m}1$) | GP:($u$,$v$,$w$) | [100] | 15.89 ($C2'/c'$) |
| 0.336 | $NdFeO_3$ | 62 ($Pnma$) | 14.62.1.4 ($P^{1}n^{\bar{1}}m^{\bar{1}}a^{\infty m}1$) | K:(0,$v$,$w$) L:(1/2,$v$,$w$) GP:($u$,$v$,$w$) | [100] | 62.448 ($Pn'ma'$) |
| 0.338 | $Co_2Mo_3O_8$ | 186 ($P6_3mc$) | 156.186.1.1 ($P^{\bar{1}}6_3{}^{1}m^{\bar{1}}c^{\infty m}1$) | D:($u$,0,$w$) GP:($u$,$v$,$w$) | [001] | 186.205 ($P6_3'm'c$) |
| 0.344 | $ErGe_{1.83}$ | 36 ($Cmc2_1$) | 8.36.1.1 ($C^{1}m^{\bar{1}}c^{\bar{1}}2_1{}^{\infty m}1$) | K:(0,$v$,$w$) GP:($u$,$v$,$w$) | [100] | 36.172 ($Cmc2_1$) |
| 0.345 | $Tb_2C_3$ | 220 ($I\bar{4}3d$) | 5.43.1.1 ($F^{\bar{1}}d^{\bar{1}}d^{1}2^{\infty m}1$) | L:(1/2,1/2,1/2) M:($u$,$v$,0) GP:($u$,$v$,$w$) | [011] | 43.226 ($Fd'd2'$) |
| 0.351 | $TbFeO_3$ | 62 ($Pnma$) | 14.62.1.4 ($P^{1}n^{\bar{1}}m^{\bar{1}}a^{\infty m}1$) | K:(0,$v$,$w$) L:(1/2,$v$,$w$) GP:($u$,$v$,$w$) | [100] | 62.448 ($Pn'ma'$) |
| 0.354 | $TbCrO_3$ | 62 ($Pnma$) | 14.62.1.4 ($P^{1}n^{\bar{1}}m^{\bar{1}}a^{\infty m}1$) | K:(0,$v$,$w$) L:(1/2,$v$,$w$) GP:($u$,$v$,$w$) | [001] | 62.446 ($Pn'm'a$) |
| 0.358 | $CaFe_5O_7$ | 11 ($P2_1/m$) | 2.11.1.1 ($P^{\bar{1}}2_1/^{\bar{1}}m^{\infty m}1$) | GP:($u$,$v$,$w$) | [010] | 11.54 ($P2_1'/m'$) |
| 0.360 | $Mn_2ScSbO_6$ | 14 ($P2_1/c$) | 2.14.1.1 ($P^{\bar{1}}2_1/^{\bar{1}}c^{\infty m}1$) | GP:($u$,$v$,$w$) | [001] +20° ([100]) | 14.75 ($P2_1/c$) |
| 0.361 | $Sr_3LiRuO_6$ | 167 ($R\bar{3}c$) | 148.167.1.1 ($R^{1}\bar{3}^{\bar{1}}c^{\infty m}1$) | GP:($u$,$v$,$w$) | [100] | 15.89 ($C2'/c'$) |
| 0.373 | $La_{0.75}Bi_{0.25}Fe_{0.5}Cr_{0.5}O_3$ | 62 ($Pnma$) | 14.62.1.4 ($P^{1}n^{\bar{1}}m^{\bar{1}}a^{\infty m}1$) | K:(0,$v$,$w$) L:(1/2,$v$,$w$) GP:($u$,$v$,$w$) | [100] | 62.441 ($Pnma$) |
| 0.376 | $LaCaFeO_4$ | 64 ($Cmce$) | 12.64.1.1 ($C^{1}m^{\bar{1}}c^{\bar{1}}e^{\infty m}1$) | K:(0,$v$,$w$) GP:($u$,$v$,$w$) | [100] | 64.474 ($Cm'c'a$) |
| 0.379 | $SmFeO_3$ | 62 ($Pnma$) | 14.62.1.4 ($P^{1}n^{\bar{1}}m^{\bar{1}}a^{\infty m}1$) | K:(0,$v$,$w$) L:(1/2,$v$,$w$) GP:($u$,$v$,$w$) | [001] | 62.446 ($Pn'm'a$) |
| 0.380 | $SmFeO_3$ | 62 ($Pnma$) | 14.62.1.4 ($P^{1}n^{\bar{1}}m^{\bar{1}}a^{\infty m}1$) | K:(0,$v$,$w$) L:(1/2,$v$,$w$) GP:($u$,$v$,$w$) | [100] | 62.448 ($Pn'ma'$) |
| 0.389 | $Fe_{1.5}Mn_{1.5}BO_5$ | 55 ($Pbam$) | 10.55.1.1 ($P^{\bar{1}}b^{\bar{1}}a^{1}m^{\infty m}1$) | V:($u$,$v$,0) W:($u$,$v$,1/2) GP:($u$,$v$,$w$) | [001] | 55.353 ($Pbam$) |

| 0.390 | $Y_2SrCu_{0.6}Co_{1.4}O_{6.5}$ | 72 ($Ibam$) | 15.72.1.1 ($I^1b^{\bar{1}}a^{\bar{1}}m^{\infty m}1$) | S:(0,1/2,1/2) A:(0,$v$,$w$) D:($u$,1/2,1/2) GP:($u$,$v$,$w$) | [010] | 72.543 ($Ib'a'm$) |
|---|---|---|---|---|---|---|
| 0.391 | $Y_2SrCu_{0.6}Co_{1.4}O_{6.5}$ | 72 ($Ibam$) | 15.72.1.1 ($I^1b^{\bar{1}}a^{\bar{1}}m^{\infty m}1$) | S:(0,1/2,1/2) A:(0,$v$,$w$) D:($u$,1/2,1/2) GP:($u$,$v$,$w$) | [010] | 72.543 ($Ib'a'm$) |
| 0.392 | $Fe_3(PO_4)_2(OH)_2$ | 14 ($P2_1/c$) | 2.14.1.1 ($P^{\bar{1}}2_1/^{\bar{1}}c^{\infty m}1$) | GP:($u$,$v$,$w$) | [001] +18° ([100]) | 14.75 ($P2_1/c$) |
| 0.402 | $Sr_4Fe_4O_{11}$ | 65 ($Cmmm$) | 10.65.1.1 ($C^{\bar{1}}m^{\bar{1}}m^1m^{\infty m}1$) | R:(1/2,1/2,1/2) S:(1/2,1/2,0) D:(1/2,1/2,$w$) P:($u$,$v$,0) Q:($u$,$v$,1/2) GP:($u$,$v$,$w$) | [010] | 65.486 ($Cmm'm'$) |
| 0.404 | $Sr_3NaRuO_6$ | 167 ($R\bar{3}c$) | 148.167.1.1 ($R^1\bar{3}^{\bar{1}}c^{\infty m}1$) | GP:($u$,$v$,$w$) | [100] | 15.89 ($C2'/c'$) |
| 0.405 | $CsCoF_4$ | 120 ($I\bar{4}c2$) | 5.82.1.1 ($I^{\bar{1}}\bar{4}^{\infty m}1$) | X:(1/2,1/2,0) N:(1/2,0,1/2) W:(1/2,1/2,$w$) Q:(1/2,$v$,1/2) Σ:($u$,0,0) Y:($u$,1-$u$,0) Δ:($u$,$u$,0) A:($u$,$u$,$w$) B:($u$,0,$w$) C:($u$,$v$,0) GP:($u$,$v$,$w$) | [001] | 82.41 ($I\bar{4}'$) |
| 0.416 | $LaCrO_3$ | 167 ($R\bar{3}c$) | 148.167.1.1 ($R^1\bar{3}^{\bar{1}}c^{\infty m}1$) | GP:($u$,$v$,$w$) | [001] | 167.103 ($R\bar{3}c$) |
| 0.417 | $LaCrO_3$ | 62 ($Pnma$) | 14.62.1.4 ($P^1n^{\bar{1}}m^{\bar{1}}a^{\infty m}1$) | K:(0,$v$,$w$) L:(1/2,$v$,$w$) GP:($u$,$v$,$w$) | [100] | 62.448 ($Pn'ma'$) |
| 0.420 | $Sr_2LuRuO_6$ | 14 ($P2_1/c$) | 2.14.1.1 ($P^{\bar{1}}2_1/^{\bar{1}}c^{\infty m}1$) | GP:($u$,$v$,$w$) | [100] | 14.75 ($P2_1/c$) |
| 0.432 | $KMnF_3$ | 62 ($Pnma$) | 14.62.1.4 ($P^1n^{\bar{1}}m^{\bar{1}}a^{\infty m}1$) | K:(0,$v$,$w$) L:(1/2,$v$,$w$) GP:($u$,$v$,$w$) | [001] | 62.448 ($Pn'ma'$) |
| 0.433 | $KMnF_3$ | 140 ($I4/mcm$) | 87.140.1.1 ($I^14/^1m^{\bar{1}}c^{\bar{1}}m^{\infty m}1$) | C:($u$,$v$,0) GP:($u$,$v$,$w$) | [001] | 140.541 ($I4/mcm$) |
| 0.434 | $K_2ReI_6$ | 14 ($P2_1/c$) | 2.14.1.1 ($P^{\bar{1}}2_1/^{\bar{1}}c^{\infty m}1$) | GP:($u$,$v$,$w$) | [100] | 14.75 ($P2_1/c$) |
| 0.448 | $Ce_4Ge_3$ | 220 ($I\bar{4}3d$) | 82.122.1.1 ($I^1\bar{4}^{\bar{1}}2^{\bar{1}}d^{\infty m}1$) | C:($u$,$v$,0) GP:($u$,$v$,$w$) | [001] | 122.333 ($I\bar{4}2d$) |
| 0.475 | $Sr_2TbIrO_6$ | 14 ($P2_1/c$) | 2.14.1.1 ($P^{\bar{1}}2_1/^{\bar{1}}c^{\infty m}1$) | GP:($u$,$v$,$w$) | [001] | 14.75 ($P2_1/c$) |

| 0.499 | $UCr_2Si_2C$ | 123 $(P4/mmm)$ | 47.123.1.1 $(P^{\bar{1}}4/{}^{1}m^{1}m^{\bar{1}}m^{\infty m}1)$ | R:(0,1/2,1/2) X:(0,1/2,0) Δ:(0,*v*,0) T:(*u*,1/2,1/2) U:(0,*v*,1/2) W:(0,1/2,*w*) Y:(*u*,1/2,0) B:(0,*v*,*w*) D:(*u*,*v*,0) E:(*u*,*v*,1/2) F:(*u*,1/2,*w*) GP:(*u*,*v*,*w*) | [100] | 47.252 $(Pm'm'm)$ |
|---|---|---|---|---|---|---|
| 0.501 | $LiFe_2F_6$ | 136 $(P4_2/mnm)$ | 65.136.1.1 $(P^{\bar{1}}4_2/{}^{1}m^{\bar{1}}n^{1}m^{\infty m}1)$ | S:(*u*,*u*,1/2) Σ:(*u*,*u*,0) C:(*u*,*u*,*w*) D:(*u*,*v*,0) E:(*u*,*v*,1/2) GP:(*u*,*v*,*w*) | [001] | 136.499 $(P4_2'/mnm')$ |
| 0.503 | $K_{1.62}Fe_4O_{6.62}(OH)_{0.38}$ | 163 $(P\bar{3}1c)$ | 147.163.1.1 $(P^{1}\bar{3}^{1}1^{\bar{1}}c^{\infty m}1)$ | B:(*u*,*v*,0) D:(*u*,0,*w*) E:(*u*,*v*,1/2) GP:(*u*,*v*,*w*) | [001] | 163.79 $(P\bar{3}1c)$ |
| 0.513 | $YRuO_3$ | 62 $(Pnma)$ | 14.62.1.4 $(P^{1}n^{\bar{1}}m^{\bar{1}}a^{\infty m}1)$ | K:(0,*v*,*w*) L:(1/2,*v*,*w*) GP:(*u*,*v*,*w*) | [001] | 62.448 $(Pn'ma')$ |
| 0.514 | $CoFe_3O_5$ | 63 $(Cmcm)$ | 12.63.1.1 $(C^{1}m^{\bar{1}}c^{\bar{1}}m^{\infty m}1)$ | K:(0,*v*,*w*) GP:(*u*,*v*,*w*) | [001] | 63.464 $(Cm'cm')$ |
| 0.522 | $La_2O_3FeMnSe_2$ | 139 $(I4/mmm)$ | 71.139.1.1 $(I^{\bar{1}}4/{}^{1}m^{1}m^{\bar{1}}m^{\infty m}1)$ | N:(1/2,0,1/2) Q:(1/2,*v*,1/2) Σ:(*u*,0,0) B:(*u*,0,*w*) C:(*u*,*v*,0) GP:(*u*,*v*,*w*) | [100] | 71.536 $(Im'm'm)$ |
| 0.528 | CrSb | 194 $(P6_3/mmc)$ | 164.194.1.1 $(P^{\bar{1}}6_3/{}^{\bar{1}}m^{1}m^{\bar{1}}c^{\infty m}1)$ | D:(*u*,0,*w*) GP:(*u*,*v*,*w*) | [001] | 194.268 $(P6_3'/m'm'c)$ |
| 0.531 | $Sr_{0.7}Tb_{0.3}CoO_{2.9}$ | 139 $(I4/mmm)$ | 69.139.1.1 $(I^{\bar{1}}4/{}^{1}m^{\bar{1}}m^{1}m^{\infty m}1)$ | X:(1/2,1/2,0) Δ:(*u*,*u*,0) W:(1/2,1/2,*w*) Y:(*u*,1-*u*,0) A:(*u*,*u*,*w*) C:(*u*,*v*,0) GP:(*u*,*v*,*w*) | [001] | 139.535 $(I4'/mmm')$ |
| 0.532 | $Sr_{0.7}Ho_{0.3}CoO_{2.7}$ | 139 $(I4/mmm)$ | 69.139.1.1 $(I^{\bar{1}}4/{}^{1}m^{\bar{1}}m^{1}m^{\infty m}1)$ | X:(1/2,1/2,0) Δ:(*u*,*u*,0) W:(1/2,1/2,*w*) Y:(*u*,1-*u*,0) A:(*u*,*u*,*w*) C:(*u*,*v*,0) GP:(*u*,*v*,*w*) | [001] | 139.535 $(I4'/mmm')$ |
| 0.533 | $Sr_{0.7}Er_{0.3}CoO_{2.8}$ | 139 $(I4/mmm)$ | 69.139.1.1 $(I^{\bar{1}}4/{}^{1}m^{\bar{1}}m^{1}m^{\infty m}1)$ | X:(1/2,1/2,0) Δ:(*u*,*u*,0) W:(1/2,1/2,*w*) Y:(*u*,1-*u*,0) A:(*u*,*u*,*w*) C:(*u*,*v*,0) GP:(*u*,*v*,*w*) | [001] | 139.535 $(I4'/mmm')$ |
| 0.553 | $K_2ReI_6$ | 14 $(P2_1/c)$ | 2.14.1.1 $(P^{\bar{1}}2_1/{}^{\bar{1}}c^{\infty m}1)$ | GP:(*u*,*v*,*w*) | [100] | 14.75 $(P2_1/c)$ |
| 0.555 | $Ho_{0.05}Bi_{0.95}FeO_3$ | 161 $(R3c)$ | 146.161.1.1 $(R^{1}3^{\bar{1}}c^{\infty m}1)$ | GP:(*u*,*v*,*w*) | [001] | 161.69 $(R3c)$ |
| 0.556 | $Ho_{0.1}Bi_{0.9}FeO_3$ | 161 $(R3c)$ | 146.161.1.1 $(R^{1}3^{\bar{1}}c^{\infty m}1)$ | GP:(*u*,*v*,*w*) | [001] | 161.69 $(R3c)$ |
| 0.559 | $Ho_{0.15}Bi_{0.85}FeO_3$ | 62 $(Pnma)$ | 14.62.1.4 $(P^{1}n^{\bar{1}}m^{\bar{1}}a^{\infty m}1)$ | K:(0,*v*,*w*) L:(1/2,*v*,*w*) GP:(*u*,*v*,*w*) | [001] | 62.448 $(Pn'ma')$ |
| 0.560 | $Ho_{0.2}Bi_{0.8}FeO_3$ | 62 $(Pnma)$ | 14.62.1.4 $(P^{1}n^{\bar{1}}m^{\bar{1}}a^{\infty m}1)$ | K:(0,*v*,*w*) L:(1/2,*v*,*w*) GP:(*u*,*v*,*w*) | [001] | 62.448 $(Pn'ma')$ |

| 0.575 | $ZnFeF_5(H_2O)_2$ | 44<br>$(Imm2)$ | 8.44.1.1<br>$(I^{\bar{1}}m^{1}m^{\bar{1}}2^{\infty m}1)$ | R:(1/2,0,1/2) B:(*u*,0,*w*) Q:(1/2,*v*,1/2) GP:(*u*,*v*,*w*) | [010] | 44.229<br>$(Imm2)$ |
|---|---|---|---|---|---|---|
| 0.581 | $FeF_3$ | 167<br>$(R\bar{3}c)$ | 148.167.1.1<br>$(R^{1}\bar{3}^{\bar{1}}c^{\infty m}1)$ | GP:(*u*,*v*,*w*) | [100] | 15.89<br>$(C2'/c')$ |
| 0.582 | $Fe_3F_8(H_2O)_2$ | 12<br>$(C2/m)$ | 2.12.1.1<br>$(C^{\bar{1}}2/^{\bar{1}}m^{\infty m}1)$ | L:(1/2,1/2,1/2) V:(1/2,1/2,0) GP:(*u*,*v*,*w*) | [010] | 12.62<br>$(C2'/m')$ |
| 0.586 | $YCrO_3$ | 62<br>$(Pnma)$ | 14.62.1.4<br>$(P^{1}n^{\bar{1}}m^{\bar{1}}a^{\infty m}1)$ | K:(0,*v*,*w*) L:(1/2,*v*,*w*) GP:(*u*,*v*,*w*) | [100] | 62.448<br>$(Pn'ma')$ |
| 0.587 | $TmCrO_3$ | 62<br>$(Pnma)$ | 14.62.1.4<br>$(P^{1}n^{\bar{1}}m^{\bar{1}}a^{\infty m}1)$ | K:(0,*v*,*w*) L:(1/2,*v*,*w*) GP:(*u*,*v*,*w*) | [100] | 62.448<br>$(Pn'ma')$ |
| 0.591 | $ErCrO_3$ | 62<br>$(Pnma)$ | 14.62.1.4<br>$(P^{1}n^{\bar{1}}m^{\bar{1}}a^{\infty m}1)$ | K:(0,*v*,*w*) L:(1/2,*v*,*w*) GP:(*u*,*v*,*w*) | [100] | 62.448<br>$(Pn'ma')$ |
| 0.592 | $DyCrO_3$ | 62<br>$(Pnma)$ | 14.62.1.4<br>$(P^{1}n^{\bar{1}}m^{\bar{1}}a^{\infty m}1)$ | K:(0,*v*,*w*) L:(1/2,*v*,*w*) GP:(*u*,*v*,*w*) | [001] | 62.446<br>$(Pn'm'a)$ |
| 0.607 | $RuO_2$ | 136<br>$(P4_2/mnm)$ | 65.136.1.1<br>$(P^{\bar{1}}4_2/^{1}m^{\bar{1}}n^{1}m^{\infty m}1)$ | S:(*u*,*u*,1/2) Σ:(*u*,*u*,0) C:(*u*,*u*,*w*) D:(*u*,*v*,0)<br>E:(*u*,*v*,1/2) GP:(*u*,*v*,*w*) | [001] | 136.499<br>$(P4_2'/mnm')$ |
| 0.608 | $PrMnO_3$ | 62<br>$(Pnma)$ | 14.62.1.1<br>$(P^{\bar{1}}n^{\bar{1}}m^{1}a^{\infty m}1)$ | V:(*u*,*v*,0) W:(*u*,*v*,1/2) GP:(*u*,*v*,*w*) | [010] | 62.448<br>$(Pn'ma')$ |
| 0.609 | $NdMnO_3$ | 62<br>$(Pnma)$ | 14.62.1.1<br>$(P^{\bar{1}}n^{\bar{1}}m^{1}a^{\infty m}1)$ | V:(*u*,*v*,0) W:(*u*,*v*,1/2) GP:(*u*,*v*,*w*) | [010]<br>−36°<br>([100]) | 11.50<br>$(P2_1/m)$ |
| 0.642 | $LaMnO_3$ | 62<br>$(Pnma)$ | 14.62.1.1<br>$(P^{\bar{1}}n^{\bar{1}}m^{1}a^{\infty m}1)$ | V:(*u*,*v*,0) W:(*u*,*v*,1/2) GP:(*u*,*v*,*w*) | [100] | 62.448<br>$(Pn'ma')$ |
| 0.645 | $La_{0.95}Ba_{0.05}Mn_{0.95}Ti_{0.05}O_3$ | 62<br>$(Pnma)$ | 14.62.1.1<br>$(P^{\bar{1}}n^{\bar{1}}m^{1}a^{\infty m}1)$ | V:(*u*,*v*,0) W:(*u*,*v*,1/2) GP:(*u*,*v*,*w*) | [100] | 62.448<br>$(Pn'ma')$ |
| 0.669 | $Sr_2YbRuO_6$ | 14<br>$(P2_1/c)$ | 2.14.1.1<br>$(P^{\bar{1}}2_1/^{\bar{1}}c^{\infty m}1)$ | GP:(*u*,*v*,*w*) | [010]<br>+39°<br>([100]) | 14.75<br>$(P2_1/c)$ |

| 0.670 | $Sr_2YbRuO_6$ | 14<br>($P2_1/c$) | 2.14.1.1<br>($P^{\bar{1}}2_1/^{\bar{1}}c^{\infty m}1$) | GP:(*u*,*v*,*w*) | [001]<br>+23°<br>([100]) | 14.75<br>($P2_1/c$) |
|---|---|---|---|---|---|---|
| 0.671 | $Sr_2TmRuO_6$ | 14<br>($P2_1/c$) | 2.14.1.1<br>($P^{\bar{1}}2_1/^{\bar{1}}c^{\infty m}1$) | GP:(*u*,*v*,*w*) | [001] | 14.75<br>($P2_1/c$) |
| 0.679 | $TbCr_{0.5}Mn_{0.5}O_3$ | 62<br>($Pnma$) | 14.62.1.4<br>($P^{1}n^{\bar{1}}m^{\bar{1}}a^{\infty m}1$) | K:(0,*v*,*w*) L:(1/2,*v*,*w*) GP:(*u*,*v*,*w*) | [100] | 62.448<br>($Pn'ma'$) |
| 0.680 | $Bi_{0.8}La_{0.2}Fe_{0.5}Mn_{0.5}O_3$ | 74<br>($Imma$) | 12.74.1.1<br>($I^{1}m^{\bar{1}}m^{\bar{1}}a^{\infty m}1$) | S:(0,1/2,1/2) A:(0,*v*,*w*) D:(*u*,1/2,1/2) GP:(*u*,*v*,*w*) | [001] | 74.559<br>($Imm'a'$) |
| 0.681 | $Ce_4Sb_3$ | 220<br>($I\bar{4}3d$) | 43.122.1.1<br>($I^{\bar{1}}\bar{4}^{\bar{1}}2^{1}d^{\infty m}1$) | X:(1/2,1/2,0) Δ:(*u*,*u*,0) W:(1/2,1/2,*w*) Y:(*u*,1-*u*,0)<br>A:(*u*,*u*,*w*) C:(*u*,*v*,0) GP:(*u*,*v*,*w*) | [001] | 122.336<br>($I\bar{4}'2d'$) |
| 0.706 | $Tb_2Ir_3Ga_9$ | 63<br>($Cmcm$) | 11.63.1.1<br>($C^{\bar{1}}m^{\bar{1}}c^{1}m^{\infty m}1$) | R:(1/2,1/2,1/2) S:(1/2,1/2,0) D:(1/2,1/2,*w*)<br>P:(*u*,*v*,0) Q:(*u*,*v*,1/2) GP:(*u*,*v*,*w*) | [100] | 63.464<br>($Cm'cm'$) |
| 0.708 | $CrNb_4S_8$ | 194<br>($P6_3/mmc$) | 164.194.1.1<br>($P^{\bar{1}}6_3/^{\bar{1}}m^{1}m^{\bar{1}}c^{\infty m}1$) | D:(*u*,0,*w*) GP:(*u*,*v*,*w*) | [001] | 194.268<br>($P6_3'/m'm'c$) |
| 0.712 | $VNb_3S_6$ | 182<br>($P6_322$) | 149.182.1.1<br>($P^{\bar{1}}6_3{}^{\bar{1}}2^{1}2^{\infty m}1$) | P:(1/3,1/3,*w*) C:(*u*,*u*,*w*) GP:(*u*,*v*,*w*) | [100] | 20.33<br>($C2'2'2_1$) |
| 0.714 | $Li_2Ni(SO_4)_2$ | 14<br>($P2_1/c$) | 2.14.1.1<br>($P^{\bar{1}}2_1/^{\bar{1}}c^{\infty m}1$) | GP:(*u*,*v*,*w*) | [100] | 14.75<br>($P2_1/c$) |
| 0.722 | $Mn_4Nb_2O_9$ | 9<br>($Cc$) | 1.9.1.1<br>($C^{\bar{1}}c^{\infty m}1$) | GP:(*u*,*v*,*w*) | [001] | 9.37<br>($Cc$) |
| 0.747 | $Ba_3CoIr_2O_9$ | 15<br>($C2/c$) | 2.15.1.1<br>($C^{\bar{1}}2/^{\bar{1}}c^{\infty m}1$) | L:(1/2,1/2,1/2) V:(1/2,1/2,0) GP:(*u*,*v*,*w*) | [001] | 15.85<br>($C2/c$) |
| 0.748 | $Ba_3NiRu_2O_9$ | 194<br>($P6_3/mmc$) | 164.194.1.1<br>($P^{\bar{1}}6_3/^{\bar{1}}m^{1}m^{\bar{1}}c^{\infty m}1$) | D:(*u*,0,*w*) GP:(*u*,*v*,*w*) | [001] | 194.268<br>($P6_3'/m'm'c$) |
| 0.755 | $Mn_2SeO_3F_2$ | 62<br>($Pnma$) | 14.62.1.4<br>($P^{1}n^{\bar{1}}m^{\bar{1}}a^{\infty m}1$) | K:(0,*v*,*w*) L:(1/2,*v*,*w*) GP:(*u*,*v*,*w*) | [001] | 62.448<br>($Pn'ma'$) |
| 0.757 | $CeFeO_3$ | 62<br>($Pnma$) | 14.62.1.4<br>($P^{1}n^{\bar{1}}m^{\bar{1}}a^{\infty m}1$) | K:(0,*v*,*w*) L:(1/2,*v*,*w*) GP:(*u*,*v*,*w*) | [100] | 62.448<br>($Pn'ma'$) |
| 0.758 | $CeFeO_3$ | 62<br>($Pnma$) | 14.62.1.4<br>($P^{1}n^{\bar{1}}m^{\bar{1}}a^{\infty m}1$) | K:(0,*v*,*w*) L:(1/2,*v*,*w*) GP:(*u*,*v*,*w*) | [010] | 62.441<br>($Pnma$) |
| 0.760 | FeOHSO$_4$ | 15<br>$(C2/c)$ | 2.15.1.1<br>$(C^{\bar{1}}2/^{\bar{1}}c^{\infty m}1)$ | L:(1/2,1/2,1/2) V:(1/2,1/2,0) GP:$(u,v,w)$ | [010] | 15.89<br>$(C2'/c')$ |
| 0.784 | NdCoO$_3$ | 62<br>$(Pnma)$ | 11.62.1.1<br>$(P^{\bar{1}}n^{1}m^{\bar{1}}a^{\infty m}1)$ | M:$(u,0,w)$ N:$(u,1/2,w)$ GP:$(u,v,w)$ | [001] | 62.441<br>$(Pnma)$ |
| 0.786 | NdVO$_3$ | 62<br>$(Pnma)$ | 11.62.1.1<br>$(P^{\bar{1}}n^{1}m^{\bar{1}}a^{\infty m}1)$ | M:$(u,0,w)$ N:$(u,1/2,w)$ GP:$(u,v,w)$ | [010]<br>+31°<br>([100]) | 11.54<br>$(P2_1'/m')$ |
| 0.787 | YVO$_3$ | 62<br>$(Pnma)$ | 14.62.1.4<br>$(P^{1}n^{\bar{1}}m^{\bar{1}}a^{\infty m}1)$ | K:$(0,v,w)$ L:$(1/2,v,w)$ GP:$(u,v,w)$ | [001] | 62.446<br>$(Pn'm'a)$ |
| 0.790 | Sr$_2$DyRuO$_6$ | 14<br>$(P2_1/c)$ | 2.14.1.1<br>$(P^{\bar{1}}2_1/^{\bar{1}}c^{\infty m}1)$ | GP:$(u,v,w)$ | [010] | 14.79<br>$(P2_1'/c')$ |
| 0.791 | Sr$_2$TbRuO$_6$ | 14<br>$(P2_1/c)$ | 2.14.1.1<br>$(P^{\bar{1}}2_1/^{\bar{1}}c^{\infty m}1)$ | GP:$(u,v,w)$ | [001]<br>−20°<br>([100]) | 14.75<br>$(P2_1/c)$ |
| 0.792 | Sr$_2$HoRuO$_6$ | 14<br>$(P2_1/c)$ | 2.14.1.1<br>$(P^{\bar{1}}2_1/^{\bar{1}}c^{\infty m}1)$ | GP:$(u,v,w)$ | [001] | 14.75<br>$(P2_1/c)$ |
| 0.793 | Sr$_2$HoRuO$_6$ | 14<br>$(P2_1/c)$ | 2.14.1.1<br>$(P^{\bar{1}}2_1/^{\bar{1}}c^{\infty m}1)$ | GP:$(u,v,w)$ | [001] | 14.75<br>$(P2_1/c)$ |
| 0.794 | Sr$_2$HoRuO$_6$ | 14<br>$(P2_1/c)$ | 2.14.1.1<br>$(P^{\bar{1}}2_1/^{\bar{1}}c^{\infty m}1)$ | GP:$(u,v,w)$ | [001]<br>−7°<br>([100]) | 14.75<br>$(P2_1/c)$ |
| 0.795 | Sr$_2$YRuO$_6$ | 14<br>$(P2_1/c)$ | 2.14.1.1<br>$(P^{\bar{1}}2_1/^{\bar{1}}c^{\infty m}1)$ | GP:$(u,v,w)$ | [100]<br>−10°<br>([001]) | 14.75<br>$(P2_1/c)$ |
| 0.800 | MnTe | 194<br>$(P6_3/mmc)$ | 164.194.1.1<br>$(P^{\bar{1}}6_3/^{\bar{1}}m^{1}m^{\bar{1}}c^{\infty m}1)$ | D:$(u,0,w)$ GP:$(u,v,w)$ | [110] | 63.457<br>$(Cmcm)$ |
| 0.802 | CuFeS$_2$ | 122<br>$(I\bar{4}2d)$ | 82.122.1.1<br>$(I^{1}\bar{4}^{\bar{1}}2^{\bar{1}}d^{\infty m}1)$ | C:$(u,v,0)$ GP:$(u,v,w)$ | [001] | 122.333<br>$(I\bar{4}2d)$ |
| 0.811 | Fe$_2$WO$_6$ | 60<br>$(Pbcn)$ | 13.60.1.1<br>$(P^{\bar{1}}b^{1}c^{\bar{1}}n^{\infty m}1)$ | M:$(u,0,w)$ N:$(u,1/2,w)$ GP:$(u,v,w)$ | [001] | 60.423<br>$(Pbc'n')$ |

| 0.813 | $Fe_2WO_6$ | 60<br>($Pbcn$) | 13.60.1.1<br>($P^{\bar{1}}b^{1}c^{\bar{1}}n^{\infty m}1$) | M:($u$,0,$w$) N:($u$,1/2,$w$) GP:($u$,$v$,$w$) | [001] | 60.423<br>($Pbc'n'$) |
|---|---|---|---|---|---|---|
| 0.820 | $Bi_{0.85}Ca_{0.15}Fe_{0.55}Mn_{0.45}O_3$ | 62<br>($Pnma$) | 14.62.1.4<br>($P^{1}n^{\bar{1}}m^{\bar{1}}a^{\infty m}1$) | K:(0,$v$,$w$) L:(1/2,$v$,$w$) GP:($u$,$v$,$w$) | [010] | 62.446<br>($Pn'm'a$) |
| 0.823 | $Sr_2MnGaO_5$ | 46<br>($Ima2$) | 5.46.1.1<br>($I^{\bar{1}}m^{\bar{1}}a^{1}2^{\infty m}1$) | T:(1/2,1/2,0) P:(1/2,1/2,$w$) GP:($u$,$v$,$w$) | [100] | 46.243<br>($Im'a2'$) |
| 0.825 | $Ca_2MnGaO_5$ | 62<br>($Pnma$) | 14.62.1.1<br>($P^{\bar{1}}n^{\bar{1}}m^{1}a^{\infty m}1$) | V:($u$,$v$,0) W:($u$,$v$,1/2) GP:($u$,$v$,$w$) | [010] | 62.447<br>($Pnm'a'$) |
| 0.826 | $MnTeLi_{0.003}$ | 194<br>($P6_3/mmc$) | 164.194.1.1<br>($P^{\bar{1}}6_3/^{\bar{1}}m^{1}m^{\bar{1}}c^{\infty m}1$) | D:($u$,0,$w$) GP:($u$,$v$,$w$) | [1$\bar{1}$0]<br>+82°<br>([001]) | 12.62<br>($C2'/m'$) |
| 0.836 | $DyFeO_3$ | 62<br>($Pnma$) | 14.62.1.4<br>($P^{1}n^{\bar{1}}m^{\bar{1}}a^{\infty m}1$) | K:(0,$v$,$w$) L:(1/2,$v$,$w$) GP:($u$,$v$,$w$) | [100] | 62.448<br>($Pn'ma'$) |
| 0.837 | $DyFeO_3$ | 62<br>($Pnma$) | 14.62.1.4<br>($P^{1}n^{\bar{1}}m^{\bar{1}}a^{\infty m}1$) | K:(0,$v$,$w$) L:(1/2,$v$,$w$) GP:($u$,$v$,$w$) | [100] | 62.448<br>($Pn'ma'$) |
| 0.838 | $DyFeO_3$ | 62<br>($Pnma$) | 14.62.1.4<br>($P^{1}n^{\bar{1}}m^{\bar{1}}a^{\infty m}1$) | K:(0,$v$,$w$) L:(1/2,$v$,$w$) GP:($u$,$v$,$w$) | [100] | 62.448<br>($Pn'ma'$) |
| 0.839 | $DyFeO_3$ | 62<br>($Pnma$) | 14.62.1.4<br>($P^{1}n^{\bar{1}}m^{\bar{1}}a^{\infty m}1$) | K:(0,$v$,$w$) L:(1/2,$v$,$w$) GP:($u$,$v$,$w$) | [100] | 62.448<br>($Pn'ma'$) |
| 0.840 | $DyFeO_3$ | 62<br>($Pnma$) | 14.62.1.4<br>($P^{1}n^{\bar{1}}m^{\bar{1}}a^{\infty m}1$) | K:(0,$v$,$w$) L:(1/2,$v$,$w$) GP:($u$,$v$,$w$) | [010] | 62.441<br>($Pnma$) |
| 0.841 | $DyFeO_3$ | 62<br>($Pnma$) | 14.62.1.4<br>($P^{1}n^{\bar{1}}m^{\bar{1}}a^{\infty m}1$) | K:(0,$v$,$w$) L:(1/2,$v$,$w$) GP:($u$,$v$,$w$) | [010] | 62.441<br>($Pnma$) |
| 0.882 | $Bi_{0.85}Ca_{0.15}Fe_{0.55}Mn_{0.45}O_3$ | 62<br>($Pnma$) | 14.62.1.4<br>($P^{1}n^{\bar{1}}m^{\bar{1}}a^{\infty m}1$) | K:(0,$v$,$w$) L:(1/2,$v$,$w$) GP:($u$,$v$,$w$) | [010] | 62.446<br>($Pn'm'a$) |
| 0.896 | $NiCrO_4$ | 63<br>($Cmcm$) | 12.63.1.1<br>($C^{1}m^{\bar{1}}c^{\bar{1}}m^{\infty m}1$) | K:(0,$v$,$w$) GP:($u$,$v$,$w$) | [100] | 63.457<br>($Cmcm$) |
| 0.904 | $Nd_2PdGe_6$ | 64<br>($Cmce$) | 12.64.1.1<br>($C^{1}m^{\bar{1}}c^{\bar{1}}e^{\infty m}1$) | K:(0,$v$,$w$) GP:($u$,$v$,$w$) | [010] | 64.474<br>($Cm'c'a$) |
| 0.917 | $Sr_2ScOsO_6$ | 14<br>$(P2_1/c)$ | 2.14.1.1<br>$(P^{\bar{1}}2_1/^{\bar{1}}c^{\infty m}1)$ | GP:$(u,v,w)$ | [100] | 14.75<br>$(P2_1/c)$ |
| 0.927 | $Nd_2PdGe_6$ | 64<br>$(Cmce)$ | 12.64.1.1<br>$(C^{1}m^{\bar{1}}c^{\bar{1}}e^{\infty m}1)$ | K:$(0,v,w)$ GP:$(u,v,w)$ | [010] | 64.474<br>$(Cm'c'a)$ |
| 0.934 | $Sr_2NiTeO_6$ | 14<br>$(P2_1/c)$ | 2.14.1.1<br>$(P^{\bar{1}}2_1/^{\bar{1}}c^{\infty m}1)$ | GP:$(u,v,w)$ | [001]<br>+41°<br>([100]) | 14.75<br>$(P2_1/c)$ |
| 0.935 | $Sr_2Ni_{0.9}Mg_{0.1}TeO_6$ | 14<br>$(P2_1/c)$ | 2.14.1.1<br>$(P^{\bar{1}}2_1/^{\bar{1}}c^{\infty m}1)$ | GP:$(u,v,w)$ | [001]<br>+42°<br>([100]) | 14.75<br>$(P2_1/c)$ |
| 0.936 | $Sr_2MnTeO_6$ | 14<br>$(P2_1/c)$ | 2.14.1.1<br>$(P^{\bar{1}}2_1/^{\bar{1}}c^{\infty m}1)$ | GP:$(u,v,w)$ | [100] | 14.75<br>$(P2_1/c)$ |
| 0.937 | $Sr_2CoTeO_6$ | 14<br>$(P2_1/c)$ | 2.14.1.1<br>$(P^{\bar{1}}2_1/^{\bar{1}}c^{\infty m}1)$ | GP:$(u,v,w)$ | [001]<br>+32°<br>([100]) | 14.75<br>$(P2_1/c)$ |
| 0.946 | $YCr_{0.5}Fe_{0.5}O_3$ | 62<br>$(Pnma)$ | 14.62.1.4<br>$(P^{1}n^{\bar{1}}m^{\bar{1}}a^{\infty m}1)$ | K:$(0,v,w)$ L:$(1/2,v,w)$ GP:$(u,v,w)$ | [001]<br>−22°<br>([100]) | 11.50<br>$(P2_1/m)$ |
| 0.947 | $YCrO_3$ | 62<br>$(Pnma)$ | 14.62.1.4<br>$(P^{1}n^{\bar{1}}m^{\bar{1}}a^{\infty m}1)$ | K:$(0,v,w)$ L:$(1/2,v,w)$ GP:$(u,v,w)$ | [001] | 62.448<br>$(Pn'ma')$ |
| 0.955 | $Na_2Mn(H_2C_3O_4)_2(H_2O)_2$ | 61<br>$(Pbca)$ | 14.61.1.1<br>$(P^{\bar{1}}b^{\bar{1}}c^{1}a^{\infty m}1)$ | V:$(u,v,0)$ W:$(u,v,1/2)$ GP:$(u,v,w)$ | [001] | 61.433<br>$(Pbca)$ |
| 0.967 | $BaMn_2V_2O_8$ | 110<br>$(I4_1cd)$ | 45.110.1.1<br>$(I^{\bar{1}}4_1{}^{1}c^{\bar{1}}d^{\infty m}1)$ | N:$(1/2,0,1/2)$ Q:$(1/2,v,1/2)$ Σ:$(u,0,0)$ B:$(u,0,w)$<br>C:$(u,v,0)$ GP:$(u,v,w)$ | [100] | 45.237<br>$(Ib'a2')$ |
| 0.979 | $TmVO_3$ | 62<br>$(Pnma)$ | 14.62.1.4<br>$(P^{1}n^{\bar{1}}m^{\bar{1}}a^{\infty m}1)$ | K:$(0,v,w)$ L:$(1/2,v,w)$ GP:$(u,v,w)$ | [010] | 62.446<br>$(Pn'm'a)$ |
| 0.980 | $TmVO_3$ | 62<br>$(Pnma)$ | 14.62.1.4<br>$(P^{1}n^{\bar{1}}m^{\bar{1}}a^{\infty m}1)$ | K:$(0,v,w)$ L:$(1/2,v,w)$ GP:$(u,v,w)$ | [010] | 62.446<br>$(Pn'm'a)$ |
| 0.982 | $TmVO_3$ | 14<br>$(P2_1/c)$ | 2.14.1.1<br>$(P^{\bar{1}}2_1/^{\bar{1}}c^{\infty m}1)$ | GP:$(u,v,w)$ | [100]<br>+20°<br>([010]) | 14.75<br>$(P2_1/c)$ |

| 0.983 | $TmVO_3$ | 14 $(P2_1/c)$ | 2.14.1.1 $(P^{\bar{1}}2_1/^{\bar{1}}c^{\infty m}1)$ | GP:$(u,v,w)$ | [100] | 14.75 $(P2_1/c)$ |
|---|---|---|---|---|---|---|
| 0.984 | $LuVO_3$ | 62 $(Pnma)$ | 14.62.1.4 $(P^{1}n^{\bar{1}}m^{\bar{1}}a^{\infty m}1)$ | K:$(0,v,w)$ L:$(1/2,v,w)$ GP:$(u,v,w)$ | [001] | 62.446 $(Pn'm'a)$ |
| 0.986 | $CaFe_3O_5$ | 63 $(Cmcm)$ | 12.63.1.1 $(C^{1}m^{\bar{1}}c^{\bar{1}}m^{\infty m}1)$ | K:$(0,v,w)$ GP:$(u,v,w)$ | [001] | 63.464 $(Cm'cm')$ |
| 0.991 | $HoFeO_3$ | 62 $(Pnma)$ | 14.62.1.4 $(P^{1}n^{\bar{1}}m^{\bar{1}}a^{\infty m}1)$ | K:$(0,v,w)$ L:$(1/2,v,w)$ GP:$(u,v,w)$ | [100] | 62.448 $(Pn'ma')$ |
| 0.995 | $MnFe_3O_5$ | 63 $(Cmcm)$ | 12.63.1.1 $(C^{1}m^{\bar{1}}c^{\bar{1}}m^{\infty m}1)$ | K:$(0,v,w)$ GP:$(u,v,w)$ | [001] | 63.464 $(Cm'cm')$ |
| 0.996 | $MnFe_3O_5$ | 63 $(Cmcm)$ | 12.63.1.1 $(C^{1}m^{\bar{1}}c^{\bar{1}}m^{\infty m}1)$ | K:$(0,v,w)$ GP:$(u,v,w)$ | [001] | 63.464 $(Cm'cm')$ |
| 0.999 | $Fe_4O_5$ | 63 $(Cmcm)$ | 8.36.1.1 $(C^{1}m^{\bar{1}}c^{\bar{1}}2_1{}^{\infty m}1)$ | K:$(0,v,w)$ GP:$(u,v,w)$ | [001] | 36.174 $(Cm'c2_1')$ |
| 0.1007 | $Fe_{0.25}NbS_2$ | 194 $(P6_3/mmc)$ | 164.194.1.1 $(P^{\bar{1}}6_3/^{\bar{1}}m^{1}m^{\bar{1}}c^{\infty m}1)$ | D:$(u,0,w)$ GP:$(u,v,w)$ | [001] | 194.268 $(P6_3'/m'm'c)$ |
| 0.1008 | $Sr_2ErRuO_6$ | 14 $(P2_1/c)$ | 2.14.1.1 $(P^{\bar{1}}2_1/^{\bar{1}}c^{\infty m}1)$ | GP:$(u,v,w)$ | [100] | 14.75 $(P2_1/c)$ |
| 0.1013 | $Ba_2NdRuO_6$ | 14 $(P2_1/c)$ | 2.14.1.1 $(P^{\bar{1}}2_1/^{\bar{1}}c^{\infty m}1)$ | GP:$(u,v,w)$ | [1$\bar{1}$0] | 2.4 $(P\bar{1})$ |
| 0.1018 | $SrMnO_3$ | 20 $(C222_1)$ | 4.20.1.1 $(C^{\bar{1}}2^{\bar{1}}2^{1}2_1{}^{\infty m}1)$ | R:$(1/2,1/2,1/2)$ S:$(1/2,1/2,0)$ D:$(1/2,1/2,w)$ P:$(u,v,0)$ Q:$(u,v,1/2)$ GP:$(u,v,w)$ | [100] | 20.34 $(C22'2_1')$ |
| 0.1019 | $SrMnO_3$ | 20 $(C222_1)$ | 4.20.1.1 $(C^{\bar{1}}2^{\bar{1}}2^{1}2_1{}^{\infty m}1)$ | R:$(1/2,1/2,1/2)$ S:$(1/2,1/2,0)$ D:$(1/2,1/2,w)$ P:$(u,v,0)$ Q:$(u,v,1/2)$ GP:$(u,v,w)$ | [010] | 20.34 $(C22'2_1')$ |
| 1.0.9 | $CsCoCl_3$ | 194 $(P6_3/mmc)$ | 162.193.1.1 $(P^{\bar{1}}6_3/^{\bar{1}}m^{\bar{1}}c^{1}m^{\infty m}1)$ | P:$(1/3,1/3,w)$ C:$(u,u,w)$ GP:$(u,v,w)$ | [001] | 193.259 $(P6_3'/m'cm')$ |
| 1.0.16 | $La_{0.33}Sr_{0.67}FeO_3$ | 167 $(R\bar{3}c)$ | 147.165.1.1 $(P^{1}\bar{3}^{\bar{1}}c^{1}1^{\infty m}1)$ | L:$(1/2,0,1/2)$ M:$(1/2,0,0)$ P:$(1/3,1/3,w)$ B:$(u,v,0)$ C:$(u,u,w)$ E:$(u,v,1/2)$ GP:$(u,v,w)$ | [100] | 15.85 $(C2/c)$ |
| 1.0.26 | $RbCoBr_3$ | 194 $(P6_3/mmc)$ | 162.193.1.1 $(P^{\bar{1}}6_3/^{\bar{1}}m^{\bar{1}}c^{1}m^{\infty m}1)$ | P:$(1/3,1/3,w)$ C:$(u,u,w)$ GP:$(u,v,w)$ | [001] | 193.259 $(P6_3'/m'cm')$ |

| 1.0.38 | $CsCoCl_3$ | 194 ($P6_3/mmc$) | 162.193.1.1 ($P^{\bar{1}}6_3/{}^{\bar{1}}m^{\bar{1}}c^{1}m^{\infty m}1$) | P:(1/3,1/3,$w$) C:($u$,$u$,$w$) GP:($u$,$v$,$w$) | [001] | 193.259 ($P6_3'/m'cm'$) |
|---|---|---|---|---|---|---|
| 1.0.39 | $BaMnO_3$ | 194 ($P6_3/mmc$) | 162.193.1.1 ($P^{\bar{1}}6_3/{}^{\bar{1}}m^{\bar{1}}c^{1}m^{\infty m}1$) | P:(1/3,1/3,$w$) C:($u$,$u$,$w$) GP:($u$,$v$,$w$) | [001] | 193.259 ($P6_3'/m'cm'$) |
| 1.0.47 | $MnSe_2$ | 205 ($Pa\bar{3}$) | 14.61.1.1 ($P^{\bar{1}}b^{\bar{1}}c^{1}a^{\infty m}1$) | V:($u$,$v$,0) W:($u$,$v$,1/2) GP:($u$,$v$,$w$) | [100] | 61.433 ($Pbca$) |
| 1.0.48 | $MnSe_2$ | 205 ($Pa\bar{3}$) | 4.29.1.1 ($P^{\bar{1}}c^{\bar{1}}a^{1}2_1{}^{\infty m}1$) | V:($u$,$v$,0) W:($u$,$v$,1/2) GP:($u$,$v$,$w$) | [100] | 29.102 ($Pca'2_1'$) |
| 1.522 | $CrVO_4$ | 63 ($Cmcm$) | 12.63.1.1 ($C^{1}m^{\bar{1}}c^{\bar{1}}m^{\infty m}1$) | K:(0,$v$,$w$) GP:($u$,$v$,$w$) | [210] +9° ([001]) | 2.4 ($P\bar{1}$) |
| 2.17 | $Pb_2Mn_{0.6}Co_{0.4}WO_6$ | 62 ($Pnma$) | 6.26.1.1 ($P^{1}m^{\bar{1}}c^{\bar{1}}2_1{}^{\infty m}1$) | K:(0,$v$,$w$) L:(1/2,$v$,$w$) GP:($u$,$v$,$w$) | [001] | 26.68 ($Pm'c2_1'$) |
| 2.88 | UNiGa | 189 ($P\bar{6}2m$) | 157.189.1.1 ($P^{\bar{1}}\bar{6}^{\bar{1}}2^{1}m^{\infty m}1$) | P:(1/3,1/3,$w$) C:($u$,$u$,$w$) D:($u$,0,$w$) GP:($u$,$v$,$w$) | [001] | 189.224 ($P\bar{6}'2m'$) |
| | FeS[13] | 190 ($P\bar{6}2c$) | 159.190.1.1 ($P^{\bar{1}}\bar{6}^{\bar{1}}2^{1}c^{\infty m}1$) | P:(1/3,1/3,$w$) C:($u$,$u$,$w$) D:($u$,0,$w$) GP:($u$,$v$,$w$) | [001] | 190.230 ($P\bar{6}'2c'$) |
| | $Co_2Mo_3N$[14] | 213 ($P4_132$) | 198.213.1.1 ($P^{\bar{1}}4_1{}^{1}3^{\bar{1}}2^{\infty m}1$) | Z:($u$,1/2,0) A:($u$,$v$,0) B:($u$,1/2,$w$) GP:($u$,$v$,$w$) | [10$\bar{1}$] | 20.34 ($C22'2_1'$) |
| | $CuF_2$[15] | 14 ($P2_1/c$) | 2.14.1.1 ($P^{\bar{1}}2_1/{}^{\bar{1}}c^{\infty m}1$) | GP:($u$,$v$,$w$) | [010] | 14.79 ($P2_1'/c'$) |
| | $NiF_2$[16] | 136 ($P4_2/mnm$) | 65.136.1.1 ($P^{\bar{1}}4_2/{}^{1}m^{\bar{1}}n^{1}m^{\infty m}1$) | S:($u$,$u$,1/2) Σ:($u$,$u$,0) C:($u$,$u$,$w$) D:($u$,$v$,0) E:($u$,$v$,1/2) GP:($u$,$v$,$w$) | [100] | 58.398 ($Pnn'm'$) |

# S5. Magnon band structure database

We list the magnon band structures for all of the 203 collinear magnets in Supplementary Tables XXX~CLXXX. These results have been divided into five sections to correspond one-to-one with sections S4.2 to S4.6 in the previous chapter. Apart from the magnetic structure and magnon band structure for each material, the tables also include the material ID in MAGNDATA database (ID), chemical formula (Formula), the space group (SG), the transition termperature ($T_C$), the Hubbard U parameter in DFT calculations ($U$), the magnetic moment (Moment) with the form of calculated (experimental) magnetic moment, the DFT-calculated electronic gap ($E_{gap}$), the spin space group (SSG), the spin Wyckoff position of magnetic ions (SWP), the band representations hosting unconventional magnons (Unconventional magnons), the magnetic space group (MSG) and the orientation of magnetic moment (Spin axis). For metallic electronic band structure, $E_{gap} = 0$. For altermagnets that host chiral magnons, the red and yellow lines represent the magnonic spin up (S=1) and down (S=-1) channels, respectively.

We have also evaluated the SOC-related exchange interactions, and the specific calculation results are provided in Supplementary Information S7. In the tables below, we use "†" to mark systems with a significant SOC effect, where the criterion is that the one of SOC-related exchange interactions are larger than $0.1J_{max}$, where $J_{max}$ represent the largest Heisenberg exchange interaction.

## S5.1. Collinear FMs and FiMs

Supplementary Table XXX. Unconventional magnons in collinear FM $Mn_3AlN$.

| ID | Formula | SG | $T_C$ (K) | *U* (eV) | Moment ($\mu_B$) | $E_{gap}$ (eV) |
|---|---|---|---|---|---|---|
| 0.275 | $Mn_3AlN$ | 221 ($Pm\bar{3}m$) | 272 | 0 | Mn: 1.19 (1.21) | 0 |
| SSG | SWP | Unconventional magnons | | | MSG | Spin axis |
| 221.221.1.1 ($P^1m^1\bar{3}^1m^{\infty m}1$) | Mn 3c | TP | R:(1/2,1/2,1/2) | $R_5^{S,+}(3)$ | 166.101 ($R\bar{3}m'$) | [111] |
| Magnetic structure | Magnon dispersion | | | | | |

Supplementary Table XXXI. Unconventional magnons in collinear FM $Gd_4Sb_3$.

| ID | Formula | SG | $T_C$ (K) | *U* (eV) | Moment ($\mu_B$) | $E_{gap}$ (eV) |
|---|---|---|---|---|---|---|
| MP−20831 | $Gd_4Sb_3$ | 220 ($I\bar{4}3d$) | 260 | 6.0 (Gd 4*f*) | Gd: 7.35 (7.5) | 0 |
| SSG | SWP | Unconventional magnons | | | MSG | Spin axis |
| 220.220.1.1 ($I^1\bar{4}^13^1d^{\infty m}1$) | Gd 16c | TP | Γ:(0,0,0) | $\Gamma_4^S(3), \Gamma_5^S(3)$ | 122.337 ($I\bar{4}2'd'$) | [001] |
| | | SP | H:(1,1,1) | $H_4^S H_5^S(6)$ | | |
| Magnetic structure | Magnon dispersion | | | | | |

Supplementary Table XXXII. Unconventional magnons in collinear FM $Lu_2V_2O_7$.

| ID | Formula | SG | $T_C$ (K) | $U$ (eV) | Moment ($\mu_B$) | $E_{gap}$ (eV) |
| --- | --- | --- | --- | --- | --- | --- |
| | $Lu_2V_2O_7$ | 227 ($Fd\bar{3}m$) | 73 K | 3.1 (V $3d$)[17] | V: 0.98 (0.93) | 0.47 |
| SSG | SWP | Unconventional magnons | | | MSG | Spin axis |
| 227.227.1.1 ($F^1d^1\bar{3}^1m^{\infty m}1$) | V 16c | TP | Γ:(0,0,0) | $\Gamma_5^{S,+}(3)$ | 141.557 ($I4_1/am'd'$) | [001] |
| Magnetic structure | | Magnon dispersion | | | | |
| | | | | | | |

Supplementary Table XXXIII. Unconventional magnons in collinear FiM $Cu_2OSeO_3$.

| ID | Formula | SG | $T_C$ (K) | $U$ (eV) | Moment ($\mu_B$) | $E_{gap}$ (eV) |
| --- | --- | --- | --- | --- | --- | --- |
| 0.35 | $Cu_2OSeO_3$† | 198 ($P2_13$) | 60 | 6.5 (Cu $3d$)[18] | Cu: 0.71 (0.63) | 2.13 |
| SSG | SWP | Unconventional magnons | | | MSG | Spin axis |
| 198.198.1.1 ($P^12_1{}^13^{\infty m}1$) | Cu 4a, 12b | $C$–2 TP | Γ:(0,0,0) | $\Gamma_4^S(3)$ | 146.10 ($R3$) | [111] |
| Magnetic structure | | Magnon dispersion | | | | |
| | | | | | | |

Supplementary Table XXXIV. Unconventional magnons in collinear FiM $FeCr_2S_4$.

| ID | Formula | SG | $T_C$ (K) | $U$ (eV) | Moment ($\mu_B$) | $E_{gap}$ (eV) |
|---|---|---|---|---|---|---|
| 0.613, 0.614, 0.615 | $FeCr_2S_4$ | 227 ($Fd\bar{3}m$) | 180 | 3.8, 4.0 (Fe, Cr $3d$)[19] | Fe: 3.56 (3.72)<br>Cr: 3.17 (2.63) | 0.71 |
| SSG | SWP | Unconventional magnons | | | MSG | Spin axis |
| 227.227.1.1 ($F^1d^1\bar{3}^1m^{\infty m}1$) | Fe 8b,<br>Cr 16c | TP | Γ:(0,0,0) | $\Gamma_5^{S,+}(3)$ | 141.557 ($I4_1/am'd'$) | [001] |
| Magnetic structure | Magnon dispersion | | | | | |

Supplementary Table XXXV. Unconventional magnons in collinear FiM $CaCu_3Fe_2Sb_2O_{12}$.

| ID | Formula | SG | $T_C$ (K) | $U$ (eV) | Moment ($\mu_B$) | $E_{gap}$ (eV) |
|---|---|---|---|---|---|---|
| 0.672 | $CaCu_3Fe_2Sb_2O_{12}$ | 201 ($Pn\bar{3}$) | 160 | 7.0, 4.0 (Cu, Fe $3d$)[20] | Cu: -0.72 (-0.93)<br>Fe: 4.26 (4.33) | 1.86 |
| SSG | SWP | Unconventional magnons | | | MSG | Spin axis |
| 201.201.1.1 ($P^1n^1\bar{3}^{\infty m}1$) | Fe 4c,<br>Cu 6d | TP | Γ:(0,0,0) | $\Gamma_5^{S,+}(3)$ | 48.260 ($Pn'n'n$) | [001] |
| | | | R:(1/2,1/2,1/2) | $R_4^{S,+}(3)$, $R_4^{S,-}(3)$ | | |
| Magnetic structure | Magnon dispersion | | | | | |

Supplementary Table XXXVI. Unconventional magnons in collinear FiM $LiFe_5O_8$.

| ID | Formula | SG | $T_C$ (K) | $U$ (eV) | Moment ($\mu_B$) | $E_{gap}$ (eV) |
|---|---|---|---|---|---|---|
| | $LiFe_5O_8$ | 212 ($P4_332$) | 925 | 4.0 (Fe $3d$)[11] | Fe: 4.17 (4.16) -4.05 (-4.06) | 1.70 |
| SSG | SWP | Unconventional magnons | | | MSG | Spin axis |
| 212.212.1.1 ($P^14_3{}^13^12^{\infty m}1$) | Fe 8c, 12d | $C$–2 TP | Γ:(0,0,0) | $\Gamma_4^S(3)$, $\Gamma_5^S(3)$ | 92.114 ($P4_12_1'2'$) | [001] |
| Magnetic structure | Magnon dispersion | | | | | |

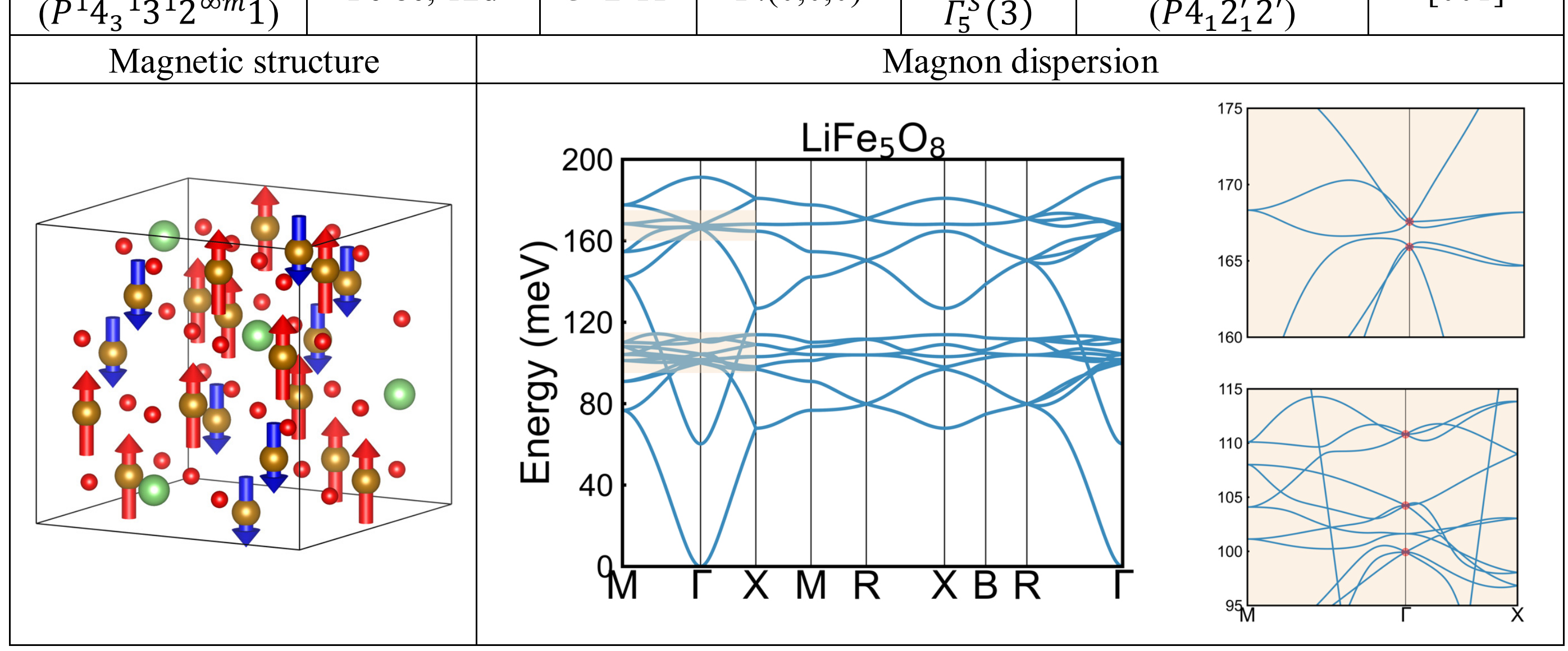

## S5.2. Collinear PT AFMs

Supplementary Table XXXVII. Unconventional magnons in collinear PT AFM $LiMnPO_4$.

<table>
<tr><td>ID</td><td>Formula</td><td colspan="2">SG</td><td>$T_C$ (K)</td><td>$U$ (eV)</td><td>Moment ($\mu_B$)</td><td>$E_{gap}$ (eV)</td></tr>
<tr><td>0.24, 0.382</td><td>$LiMnPO_4$</td><td colspan="2">62 ($Pnma$)</td><td>33.85</td><td>3.92 (Mn $3d$)[21]</td><td>Mn: 4.64 (4.28)</td><td>4.08</td></tr>
<tr><td>SSG</td><td>SWP</td><td colspan="4">Unconventional magnons</td><td>MSG</td><td>Spin axis</td></tr>
<tr><td>26.62.1.1 $(P^{\bar{1}}n^{1}m^{1}a^{\infty m}1)$</td><td>Mn 4c</td><td>L:(1/2,$v$,$w$)</td><td colspan="2">$(L)W_1^S W_1^S(4)$</td><td>QNPL</td><td>62.449 ($Pn'm'a'$)</td><td>[100]</td></tr>
<tr><td colspan="3">Magnetic structure</td><td colspan="5">Magnon dispersion</td></tr>
</table>

Supplementary Table XXXVIII. Unconventional magnons in collinear PT AFM $Cr_2WO_6$.

<table>
<tr><td>ID</td><td>Formula</td><td colspan="2">SG</td><td>$T_C$ (K)</td><td>$U$ (eV)</td><td>Moment ($\mu_B$)</td><td>$E_{gap}$ (eV)</td></tr>
<tr><td>0.75, 0.144</td><td>$Cr_2WO_6$</td><td colspan="2">136 ($P4_2/mnm$)</td><td>45</td><td>4.0 (Cr $3d$)[22]</td><td>Cr: 2.91 (2.14)</td><td>1.74</td></tr>
<tr><td>SSG</td><td>SWP</td><td colspan="4">Unconventional magnons</td><td>MSG</td><td>Spin axis</td></tr>
<tr><td>113.136.1.1 $(P^{\bar{1}}4_2/^{\bar{1}}m^{\bar{1}}n^{1}m^{\infty m}1)$</td><td>Cr 4e</td><td>F:($u$,1/2,$w$)</td><td colspan="2">$F_1^S F_1^S(4)$</td><td>QNPL</td><td>58.395 ($Pn'nm$)</td><td>[010]</td></tr>
<tr><td colspan="3">Magnetic structure</td><td colspan="5">Magnon dispersion</td></tr>
</table>

Supplementary Table XXXIX. Unconventional magnons in collinear PT AFM $NaFePO_4$.

| ID | Formula | SG | $T_C$ (K) | $U$ (eV) | Moment ($\mu_B$) | $E_{gap}$ (eV) |
|---|---|---|---|---|---|---|
| 0.87 | $NaFePO_4$ | 62 ($Pnma$) | 50 | 3.71 (Fe $3d$)[21] | Fe: 3.74 (4.55) | 3.12 |
| SSG | SWP | Unconventional magnons | | | MSG | Spin axis |
| 26.62.1.1 ($P^{\bar{1}}n^{1}m^{1}a^{\infty m}1$) | Fe 4c | L:(1/2,$v$,$w$) | $(L)W_1^S W_1^S(4)$ | QNPL | 62.445 ($Pnma'$) | [010] |
| Magnetic structure | | | Magnon dispersion | | | |

Supplementary Table XL. Unconventional magnons in collinear PT AFM $CoSe_2O_5$.

| ID | Formula | SG | $T_C$ (K) | $U$ (eV) | Moment ($\mu_B$) | $E_{gap}$ (eV) |
|---|---|---|---|---|---|---|
| 0.161 | $CoSe_2O_5$ | 60 ($Pbcn$) | 8.5 | 4.0 (Co $3d$)[23] | 2.75 (3.0) | 2.76 |
| SSG | SWP | Unconventional magnons | | | MSG | Spin axis |
| 18.60.1.1 ($P^{\bar{1}}b^{\bar{1}}c^{\bar{1}}n^{\infty m}1$) | Co 4c | L:(1/2,$v$,$w$) | $L_1^S L_1^S(4)$ | QNPL | 60.419 ($Pb'cn$) | [100] +14° ([001]) |
| | | W:($u$,$v$,1/2) | $(W)N_1^S N_1^S(4)$ | QNPL | | |
| Magnetic structure | | | Magnon dispersion | | | |

Supplementary Table XLI. Unconventional magnons in collinear PT AFM $LiCoPO_4$.

| ID | Formula | SG | $T_C$ (K) | $U$ (eV) | Moment ($\mu_B$) | $E_{gap}$ (eV) |
|---|---|---|---|---|---|---|
| 0.193, 0.383, 0.384* | $LiCoPO_4$ | 62 ($Pnma$) | 22 | 5.05 (Co $3d$)[21] | 2.81 (3.28) | 4.00 |
| SSG | SWP | Unconventional magnons | | | MSG | Spin axis |
| 26.62.1.1 ($P^{\bar{1}}n^{1}m^{1}a^{\infty m}1$) | Co 4c | L:(1/2,$v$,$w$) | $L_1^S L_1^S(4)$ | QNPL | 62.445 ($Pnma'$) | [010] |
| Magnetic structure | | | Magnon dispersion | | | |

*The spin axis of 0.384 is [010] +4° ([001]) and its corresponding MSG is 14.77 ($P2_1'/c$).

Supplementary Table XLII. Unconventional magnons in collinear PT AFM $LiCrGe_2O_6$.

| ID | Formula | SG | $T_C$ (K) | $U$ (eV) | Moment ($\mu_B$) | $E_{gap}$ (eV) |
|---|---|---|---|---|---|---|
| 0.217, 0.961, 0.962, 0.963, 0.964 | $LiCrGe_2O_6$ | 14 ($P2_1/c$) | 4.8 | 2.0 (Cr $3d$)[24] | 2.85 (2.71) | 2.78 |
| SSG | SWP | Unconventional magnons | | | MSG | Spin axis |
| 4.14.1.1 ($P^{1}2_1/^{\bar{1}}c^{\infty m}1$) | Cr 4e | G:($u$, 1/2, $w$) | $G_1^S G_1^S(4)$ | QNPL | 14.77 ($P2_1'/c$) | [100] +30° ([001]) |
| Magnetic structure | | | Magnon dispersion | | | |

Supplementary Table XLIII. Unconventional magnons in collinear PT AFM FeOOH.

| ID | Formula | SG | $T_C$ (K) | $U$ (eV) | | Moment ($\mu_B$) | $E_{gap}$ (eV) |
|---|---|---|---|---|---|---|---|
| 0.399 | FeOOH | 62 ($Pnma$) | 340 | 5.0 (Fe 3$d$)[25] | | 4.25 (4.45) | 2.30 |
| SSG | SWP | Unconventional magnons | | | | MSG | Spin axis |
| 26.62.1.1 ($P^{\bar{1}}n^{1}m^{1}a^{\infty m}1$) | Fe 4c | L:(1/2,$v$,$w$) | $(L)W_1^S W_1^S(4)$ | QNPL | | 62.445 ($Pnma'$) | [001] |
| Magnetic structure | | | Magnon dispersion | | | | |

Supplementary Table XLIV. Unconventional magnons in collinear PT AFM $GdAlO_3$.

| ID | Formula | SG | $T_C$ (K) | $U$ (eV) | | Moment ($\mu_B$) | $E_{gap}$ (eV) |
|---|---|---|---|---|---|---|---|
| 0.410 | $GdAlO_3$ | 62 ($Pnma$) | 3.87 | 6.0 (Gd 4$f$)[26] | | 7.04 (6.10) | 5.01 |
| SSG | SWP | Unconventional magnons | | | | MSG | Spin axis |
| 31.62.1.1 ($P^{1}n^{1}m^{\bar{1}}a^{\infty m}1$) | Gd 4c | W:($u$,$v$,1/2) | $W_1^S W_1^S(4)$ | QNPL | | 62.449 ($Pn'm'a'$) | [100] |
| Magnetic structure | | Magnon dispersion | | | | | |

Supplementary Table XLV. Unconventional magnons in collinear PT AFM $RbFeO_2$.

| ID | Formula | SG | $T_C$ (K) | $U$ (eV) | Moment ($\mu_B$) | $E_{gap}$ (eV) |
|---|---|---|---|---|---|---|
| 0.455 | $RbFeO_2$ | 61 ($Pbca$) | 737 | 4.0 (Fe $3d$)[27] | 3.97 (3.83) | 3.27 |
| SSG | SWP | Unconventional magnons | | | MSG | Spin axis |
| 29.61.1.1 ($P^{\bar{1}}b^{1}c^{1}a^{\infty m}1$) | Fe 8c, 8c | R:(1/2,1/2,1/2) | $R_1^S R_1^S(8)$ | OP | 61.437 ($Pb'c'a'$) | [001] |
| | | U:(1/2,0,1/2) | $U_1^S U_1^S(8)$ | OP | | |
| | | L:(1/2,$v$,$w$) | $(L)W_1^S W_1^S(4)$ | QNPL | | |
| Magnetic structure | | | Magnon dispersion | | | |

Supplementary Table XLVI. Unconventional magnons in collinear PT AFM $CsFeO_2$.

| ID | Formula | SG | $T_C$ (K) | $U$ (eV) | Moment ($\mu_B$) | $E_{gap}$ (eV) |
|---|---|---|---|---|---|---|
| 0.457 | $CsFeO_2$ | 61 ($Pbca$) | 350 | 4.0 (Fe $3d$)[27] | 3.96 (3.8) | 3.32 |
| SSG | SWP | Unconventional magnons | | | MSG | Spin axis |
| 29.61.1.1 ($P^{\bar{1}}b^{1}c^{1}a^{\infty m}1$) | Fe 8c, 8c | R:(1/2,1/2,1/2) | $R_1^S R_1^S(8)$ | OP | 61.437 ($Pb'c'a'$) | [001] |
| | | U:(1/2,0,1/2) | $U_1^S U_1^S(8)$ | OP | | |
| | | L:(1/2,$v$,$w$) | $(L)W_1^S W_1^S(4)$ | QNPL | | |
| Magnetic structure | | | Magnon dispersion | | | |

Supplementary Table XLVII. Unconventional magnons in collinear PT AFM $KFeO_2$.

| ID | Formula | SG | $T_C$ (K) | $U$ (eV) | Moment ($\mu_B$) | $E_{gap}$ (eV) |
|---|---|---|---|---|---|---|
| 0.459, 0.460 | $KFeO_2$ | 61 ($Pbca$) | 1001 (960) | 4.0 (Fe $3d$)[27] | 4.13 (3.79) | 1.76 |
| SSG | SWP | Unconventional magnons | | | MSG | Spin axis |
| 29.61.1.1 ($P^{\bar{1}}b^{1}c^{1}a^{\infty m}1$) | Fe 8c, 8c | R:(1/2,1/2,1/2) | $R_1^S R_1^S(8)$ | OP | 61.435 ($Pb'ca$) | [010], [100] |
| | | U:(1/2,0,1/2) | $U_1^S U_1^S(8)$ | OP | | |
| | | L:(1/2,$v$,$w$) | $(L)W_1^S W_1^S(4)$ | QNPL | | |
| Magnetic structure | | Magnon dispersion | | | | |

Supplementary Table XLVIII. Unconventional magnons in collinear PT AFM $Pb_2VO(PO_4)_2$.

| ID | Formula | SG | $T_C$ (K) | $U$ (eV) | Moment ($\mu_B$) | $E_{gap}$ (eV) |
|---|---|---|---|---|---|---|
| 0.505 | $Pb_2VO(PO_4)_2$ | 14 ($P2_1/c$) | 3.5 | 3.1 (V $3d$)[28] | 1.03 (0.68) | 2.62 |
| SSG | SWP | Unconventional magnons | | | MSG | Spin axis |
| 4.14.1.1 ($P^{1}2_1/^{\bar{1}}c^{\infty m}1$) | V 4e | G:($u$, 1/2, $w$) | $G_1^S G_1^S(4)$ | QNPL | 14.78 ($P2_1/c'$) | [010] |
| Magnetic structure | | Magnon dispersion | | | | |

Supplementary Table XLIX. Unconventional magnons in collinear PT AFM $MnPd_2$.

| ID | Formula | SG | $T_C$ (K) | $U$ (eV) | Moment ($\mu_B$) | $E_{gap}$ (eV) |
|---|---|---|---|---|---|---|
| 0.798 | $MnPd_2$ | 62 ($Pnma$) | 415 | 0[29] | 3.78 (4.0) | 0 |
| SSG | SWP | Unconventional magnons | | | MSG | Spin axis |
| 26.62.1.1 ($P^{\bar{1}}n^{1}m^{1}a^{\infty m}1$) | Mn 4c | L:(1/2,$v$,$w$) | $(L)W_1^S W_1^S(4)$ | QNPL | 62.445 ($Pnma'$) | [010] |
| Magnetic structure | | | Magnon dispersion | | | |
| | | | | | | |

Supplementary Table L. Unconventional magnons in collinear PT AFM $MnNb_2O_6$.

| ID | Formula | SG | $T_C$ (K) | $U$ (eV) | Moment ($\mu_B$) | $E_{gap}$ (eV) |
|---|---|---|---|---|---|---|
| 0.815 | $MnNb_2O_6$ | 60 ($Pbcn$) | 4.2 | 0[30] | 4.41 (4.0) | 2.18 |
| SSG | SWP | Unconventional magnons | | | MSG | Spin axis |
| 18.60.1.1 ($P^{\bar{1}}b^{\bar{1}}c^{\bar{1}}n^{\infty m}1$) | Mn 4c | L:(1/2,$v$,$w$) | $L_1^S L_1^S(4)$ | QNPL | 60.419 ($Pb'cn$) | [100] |
| | | W:($u$,$v$,1/2) | $(W)N_1^S N_1^S(4)$ | QNPL | | |
| Magnetic structure | | | Magnon dispersion | | | |
| | | | | | | |

Supplementary Table LI. Unconventional magnons in collinear PT AFM $MnTa_2O_6$.

| ID | Formula | SG | $T_C$ (K) | $U$ (eV) | Moment ($\mu_B$) | $E_{gap}$ (eV) |
|---|---|---|---|---|---|---|
| 0.816 | $MnTa_2O_6$ | 60 ($Pbcn$) | 4.2 | 0[30] | 4.41 (4.0) | 1.85 |
| SSG | SWP | Unconventional magnons | | | MSG | Spin axis |
| 18.60.1.1 ($P^{\bar{1}}b^{\bar{1}}c^{\bar{1}}n^{\infty m}1$) | Mn 4c | L:(1/2,$v$,$w$) | $L_1^S L_1^S(4)$ | QNPL | 60.419 ($Pb'cn$) | [100] |
| | | W:($u$,$v$,1/2) | $(W)N_1^S N_1^S(4)$ | QNPL | | |
| Magnetic structure | | | Magnon dispersion | | | |

Supplementary Table LII. Unconventional magnons in collinear PT AFM $SrGd_2O_4$.

| ID | Formula | SG | $T_C$ (K) | $U$ (eV) | Moment ($\mu_B$) | $E_{gap}$ (eV) |
|---|---|---|---|---|---|---|
| 0.821 | $SrGd_2O_4$ | 62 ($Pnma$) | 2.73 | 6.0 (Gd 4$f$)[26] | 7.03 (6.80) | 3.59 |
| SSG | SWP | Unconventional magnons | | | MSG | Spin axis |
| 26.62.1.1 ($P^{\bar{1}}n^{1}m^{1}a^{\infty m}1$) | Gd 4c, 4c | L:(1/2,$v$,$w$) | $(L)W_1^S W_1^S(4)$ | QNPL | 62.445 ($Pnma'$) | [001] |
| Magnetic structure | | | Magnon dispersion | | | |

Supplementary Table LIII. Unconventional magnons in collinear PT AFM $V_2WO_6$.

| ID | Formula | SG | $T_C$ (K) | $U$ (eV) | Moment ($\mu_B$) | $E_{gap}$ (eV) |
|---|---|---|---|---|---|---|
| 0.966 | $V_2WO_6$† | 136 ($P4_2/mnm$) | 117 | 2.85 (V $3d$)[31] | 1.83 (0.90) | 0.96 |
| SSG | SWP | Unconventional magnons | | | MSG | Spin axis |
| 113.136.1.1 ($P^{\bar{1}}4_2/^{\bar{1}}m^{\bar{1}}n^{1}m^{\infty m}1$) | V 4e | F:($u$,1/2,$w$) | $F_1^S F_1^S(4)$ | QNPL | 58.395 ($Pn'nm$) | [100] |
| Magnetic structure | | | Magnon dispersion | | | |

Supplementary Table LIV. Unconventional magnons in collinear PT AFM $CaFe_2O_4$.

| ID | Formula | SG | $T_C$ (K) | $U$ (eV) | Moment ($\mu_B$) | $E_{gap}$ (eV) |
|---|---|---|---|---|---|---|
| 0.969 | $CaFe_2O_4$ | 62 ($Pnma$) | 200 | 4.0 (Fe $3d$)[32] | 4.26 (4.26) | 1.32 |
| SSG | SWP | Unconventional magnons | | | MSG | Spin axis |
| 26.62.1.1 ($P^{\bar{1}}n^{1}m^{1}a^{\infty m}1$) | Fe 4c, 4c | L:(1/2,$v$,$w$) | $(L)W_1^S W_1^S(4)$ | QNPL | 62.445 ($Pnma'$) | [001] |
| Magnetic structure | | | Magnon dispersion | | | |

Supplementary Table LV. Unconventional magnons in collinear PT AFM $Cu_3TeO_6$.

| ID | Formula | SG | $T_C$ (K) | $U$ (eV) | Moment ($\mu_B$) | $E_{gap}$ (eV) |
| --- | --- | --- | --- | --- | --- | --- |
| | $Cu_3TeO_6$ | 206 ($Ia\bar{3}$) | 61 | | 0.67 (0.64) | 2.29 |
| SSG | SWP | Unconventional magnons | | | MSG | Spin axis |
| 199.206.1.1 ($I^{\bar{1}}a^{\bar{1}}\bar{3}^{\infty m}1$) | Cu 24d | Γ:(0,0,0) | $\Gamma_4^S(6)$ | $C_2$−2 SP | 148.19 ($R\bar{3}'$) | [111] |
| | | H:(1,1,1) | $H_4^S(6)$ | $C_2$−2 SP | | |
| Magnetic structure | Magnon dispersion | | | | | |

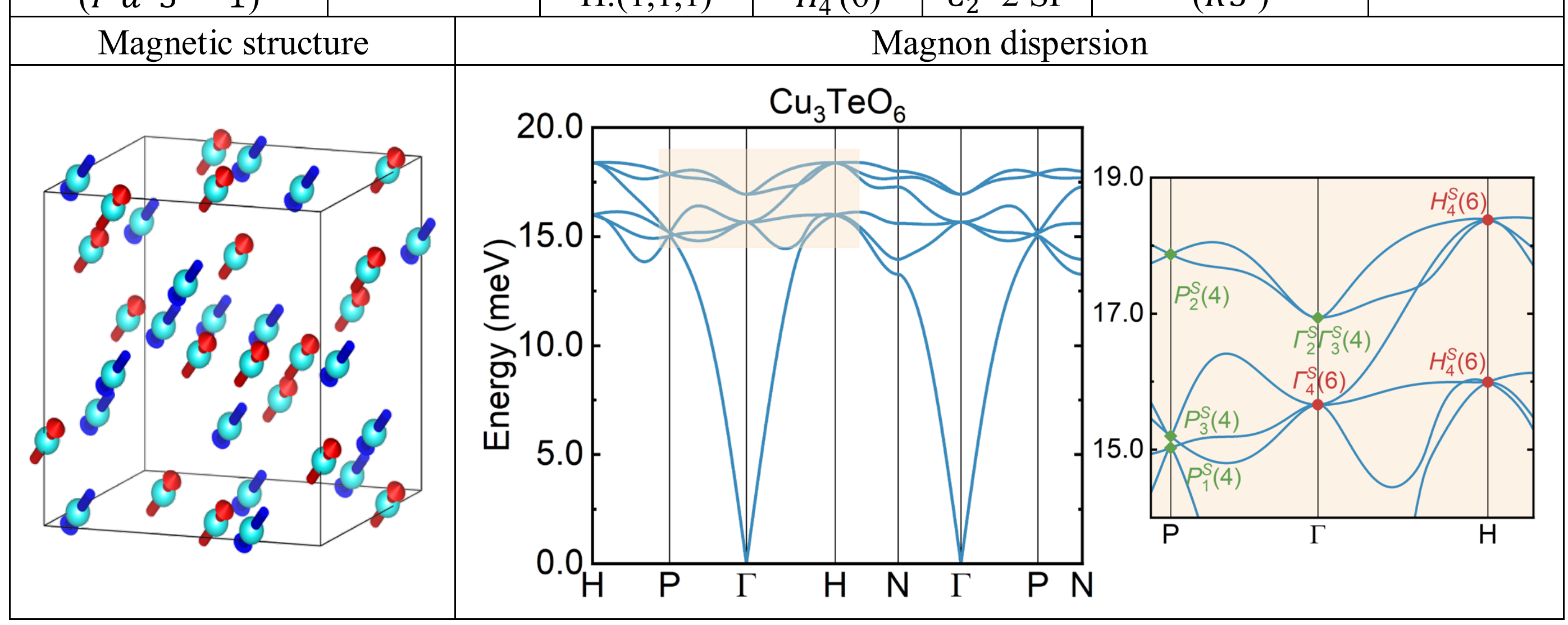

## S5.3. Altermagnets

Supplementary Table LVI. Unconventional magnons in altermagnet FeS.

| ID | Formula | SG | $T_C$ (K) | *U* (eV) | Moment ($\mu_B$) | $E_{gap}$ (eV) |
|---|---|---|---|---|---|---|
| | FeS | 190 ($P\bar{6}2c$) | 415 | 1.0 (Fe 3*d*)[13] | 2.97 (2.87) | 0.43 |
| SSG | SWP | Unconventional magnons | | | MSG | Spin axis |
| 159.190.1.1 ($P^{\bar{1}}\bar{6}^{\bar{1}}2^{1}c^{\infty m}1$) | Fe 12i | A:(0,0,1/2) | $A_3^S A_3^S(8)$ | OP | 190.230 ($P\bar{6}'2c'$) | [001] |
| Magnetic structure | Magnon dispersion | | | | | |

## S5.4. Collinear Uτ AFMs

Supplementary Table LVII. Unconventional magnons in collinear Uτ AFM $Sr_2IrO_4$.

| ID | Formula | SG | $T_C$ (K) | $U$ (eV) | Moment ($\mu_B$) | $E_{gap}$ (eV) |
|---|---|---|---|---|---|---|
| 1.3 | $Sr_2IrO_4$† | 142 ($I4_1/acd$) | 240 | 1.53 (Ir $5d$)[33] | 0.36 (0.24) | 0.14 |
| SSG | SWP | Unconventional magnons | | | MSG | Spin axis |
| 54.73.2.1 ($P_I{}^1c^1c^1a^{\infty m}1$) | Ir 8c | R:(1/2,1/2,1/2) | $R_1^S R_2^S(8)$ | OP | 54.352 ($P_Icca$) | [100] |
| | | U:(1/2,0,1/2) | $U_1^S U_2^S(8)$ | OP | | |
| | | L:(1/2,$v$,$w$) | $L_1^S L_2^S(4)$ | QNPL | | |
| Magnetic structure | Magnon dispersion | | | | | |

* $L$ is approximately octuple nodal plane, due to the weak out-of-plane exchange.

Supplementary Table LVIII. Unconventional magnons in collinear Uτ AFM $MnS_2$.

| ID | Formula | SG | $T_C$ (K) | $U$ (eV) | Moment ($\mu_B$) | $E_{gap}$ (eV) |
|---|---|---|---|---|---|---|
| 1.18 | $MnS_2$ | 205 ($Pa\bar{3}$) | 48.2 | 5.0 (Mn $3d$)[34] | 4.62 (5.0) | 1.63 |
| SSG | SWP | Unconventional magnons | | | MSG | Spin axis |
| 33.29.2.1 ($P_b{}^1n^1a^12_1{}^{\infty m}1$) | Mn 8a | U:(1/2,0,1/2) | $U_1^S U_1^S(8)$ | OP | 29.105 ($P_bca2_1$) | [010] |
| | | W:($u$,$v$,1/2) | $W_1^S W_1^S(4)$ | QNPL | | |
| Magnetic structure | Magnon dispersion | | | | | |

Supplementary Table LIX. Unconventional magnons in collinear Uτ AFM $HoMnO_3$.

| ID | Formula | SG | $T_C$ (K) | $U$ (eV) | Moment ($\mu_B$) | $E_{gap}$ (eV) |
|---|---|---|---|---|---|---|
| 1.20 | $HoMnO_3$ | 62 ($Pnma$) | 29 | 3.1 (Mn $3d$)[35] | 3.71 (3.87) | 0.21 |
| SSG | SWP | Unconventional magnons | | | MSG | Spin axis |
| 33.31.2.1 ($P_a{}^1n^1a^12_1{}^{\infty m}1$) | Mn 8b | U:(1/2,0,1/2) | $U_1^SU_1^S(8)$ | OP | 31.129 ($P_bmn2_1$) | [100] |
| | | W:($u$,$v$,1/2) | $W_1^SW_1^S(4)$ | QNPL | | |
| Magnetic structure | | Magnon dispersion | | | | |

Supplementary Table LX. Unconventional magnons in collinear Uτ AFM $La_2CuO_4$.

| ID | Formula | SG | $T_C$ (K) | $U$ (eV) | Moment ($\mu_B$) | $E_{gap}$ (eV) |
|---|---|---|---|---|---|---|
| 1.23, 1.405 | $La_2CuO_4$ | 64 ($Cmce$) | 316 | 7.2 (Cu $3d$)[36] | 0.66 (0.42) | 2.40 |
| SSG | SWP | Unconventional magnons | | | MSG | Spin axis |
| 55.64.2.1 ($P_A{}^1b^1a^1m^{\infty m}1$) | Cu 4a | L:(1/2,$v$,$w$) | $L_1^SL_2^S(4)$ | QNPL | 56.374 ($P_Accn$) | [010] |
| | | N:($u$,1/2,$w$) | $N_1^SN_2^S(4)$ | QNPL | | |
| Magnetic structure | | Magnon dispersion | | | | |
| | | * The magnon bands are approximately four-fold degenerate due to the weak out-of-plane exchange interaction. | | | | |

Supplementary Table LXI. Unconventional magnons in collinear Uτ AFM $ZnV_2O_4$.

| ID | Formula | SG | $T_C$ (K) | $U$ (eV) | Moment ($\mu_B$) | $E_{gap}$ (eV) |
| --- | --- | --- | --- | --- | --- | --- |
| 1.24 | $ZnV_2O_4$ | 141 ($I4_1/amd$) | 40 | 4.1 (V $3d$)[37] | 1.82 (0.65) | 1.57 |
| SSG | SWP | Unconventional magnons | | | MSG | Spin axis |
| 95.98.2.1 ($P_I{}^1 4_3{}^1 2{}^1 2{}^{\infty m}1$) | V 8f | E:($u$,$v$,1/2) | $E_1^S E_1^S(4)$ | QNPL | 96.150 ($P_I 4_3 2_1 2$) | [001] |
| Magnetic structure | | | Magnon dispersion | | | |

Supplementary Table LXII. Unconventional magnons in collinear Uτ AFM $CsFe_2Se_3$.

| ID | Formula | SG | $T_C$ (K) | $U$ (eV) | Moment ($\mu_B$) | $E_{gap}$ (eV) |
| --- | --- | --- | --- | --- | --- | --- |
| 1.26 | $CsFe_2Se_3$ | 63 ($Cmcm$) | 175 | 2.0 (Fe $3d$)[38] | 3.24 (2.80) | 1.15 |
| SSG | SWP | Unconventional magnons | | | MSG | Spin axis |
| 14.11.2.1 ($P_c{}^1 2_1/{}^1 c{}^{\infty m}1$) | Fe 8f | G:($u$, 1/2, $w$) | $G_1^S G_1^S(4)$ | QNPL | 14.82 ($P_c 2_1/c$) | [001] |
| Magnetic structure | | | Magnon dispersion | | | |

Supplementary Table LXIII. Unconventional magnons in collinear Uτ AFM CrN.

| ID | Formula | SG | $T_C$ (K) | $U$ (eV) | Moment ($\mu_B$) | $E_{gap}$ (eV) |
|---|---|---|---|---|---|---|
| 1.28, 1.678 | CrN | 225 ($Fm\bar{3}m$) | 273 | 0.5 (Cr 3$d$)[39] | 2.53 (2.35) | 0 |
| SSG | SWP | Unconventional magnons | | | MSG | Spin axis |
| 59.59.2.1 ($P_c{}^1m{}^1m{}^1n{}^{\infty m}1$) | Cr 4a | L:(1/2,$v$,$w$) | $L_1^S L_2^S(4)$ | QNPL | 62.450 ($P_anma$) | [1$\bar{1}$0] |
| | | N:($u$,1/2,$w$) | $N_1^S N_2^S(4)$ | QNPL | | |
| Magnetic structure | | | Magnon dispersion | | | |

Supplementary Table LXIV. Unconventional magnons in collinear Uτ AFM $La_2NiO_4$.

| ID | Formula | SG | $T_C$ (K) | $U$ (eV) | Moment ($\mu_B$) | $E_{gap}$ (eV) |
|---|---|---|---|---|---|---|
| 1.42 | $La_2NiO_4$ | 64 ($Cmce$) | 330 | 4.2 (Ni 3$d$)[40] | 1.60 (1.68) | 2.11 |
| SSG | SWP | Unconventional magnons | | | MSG | Spin axis |
| 55.64.2.1 ($P_A{}^1b{}^1a{}^1m{}^{\infty m}1$) | Ni 4a | L:(1/2,$v$,$w$) | $L_1^S L_2^S(4)$ | QNPL | 53.335 ($P_Cmna$) | [100] |
| | | N:($u$,1/2,$w$) | $N_1^S N_2^S(4)$ | QNPL | | |
| Magnetic structure | | | Magnon dispersion | | | |

Supplementary Table LXV. Unconventional magnons in collinear Uτ AFM CuO.

| ID | Formula | SG | $T_C$ (K) | $U$ (eV) | Moment ($\mu_B$) | $E_{gap}$ (eV) |
|---|---|---|---|---|---|---|
| 1.62 | CuO | 15 ($C2/c$) | 213 | 7.0 (Cu 3$d$) | 0.63 (0.65) | 1.66 |
| SSG | SWP | Unconventional magnons | | | MSG | Spin axis |
| 14.14.2.1 ($P_a{}^12_1/{}^1c^{\infty m}1$) | Cu 8e | G:($u$, 1/2, $w$) | $G_1^SG_1^S(4)$ | QNPL | 14.80 ($P_a2_1/c$) | [010] |
| Magnetic structure | | | Magnon dispersion | | | |

Supplementary Table LXVI. Unconventional magnons in collinear Uτ AFM $BaNiF_4$.

| ID | Formula | SG | $T_C$ (K) | $U$ (eV) | Moment ($\mu_B$) | $E_{gap}$ (eV) |
|---|---|---|---|---|---|---|
| 1.64 | $BaNiF_4$ | 36 ($Cmc2_1$) | 4.2 | 1.5 (Ni 3$d$)[41] | 1.66 (2.00) | 2.17 |
| SSG | SWP | Unconventional magnons | | | MSG | Spin axis |
| 4.4.2.1 ($P_a{}^12_1{}^{\infty m}1$) | Ni 4a | G:($u$, 1/2, $w$) | $G_1^SG_1^S(4)$ | QNPL | 4.10 ($P_a2_1$) | [010] |
| Magnetic structure | | | Magnon dispersion | | | |

Supplementary Table LXVII. Unconventional magnons in collinear Uτ AFM $LuMnO_3$.

| ID | Formula | SG | $T_C$ (K) | $U$ (eV) | Moment ($\mu_B$) | $E_{gap}$ (eV) |
|---|---|---|---|---|---|---|
| 1.101 | $LuMnO_3$ | 62 ($Pnma$) | 40 | 3.1 (Mn 3$d$)[35] | 3.71 (3.37) | 1.51 |
| SSG | SWP | Unconventional magnons | | | MSG | Spin axis |
| 33.31.2.1 ($P_a{}^1n^1a^12_1{}^{\infty m}1$) | Mn 8b | U:(1/2,0,1/2) | $U_1^SU_1^S(8)$ | OP | 31.129 ($P_bmn2_1$) | [010] |
| | | W:($u$,$v$,1/2) | $W_1^SW_1^S(4)$ | QNPL | | |
| Magnetic structure | | | Magnon dispersion | | | |

Supplementary Table LXVIII. Unconventional magnons in collinear Uτ AFM $NaFeSO_4F$.

| ID | Formula | SG | $T_C$ (K) | $U$ (eV) | Moment ($\mu_B$) | $E_{gap}$ (eV) |
|---|---|---|---|---|---|---|
| 1.121 | $NaFeSO_4F$ | 15 ($C2/c$) | 36 | 4.0 (Fe 3$d$)[28] | 3.78 (4.42) | 3.89 |
| SSG | SWP | Unconventional magnons | | | MSG | Spin axis |
| 14.15.2.1 ($P_c{}^12_1/{}^1c^{\infty m}1$) | Fe 4c | G:($u$, 1/2, $w$) | $G_1^SG_1^S(4)$ | QNPL | 13.74 ($P_c2/c$) | [001] +20° ([100]) |
| Magnetic structure | | | Magnon dispersion | | | |

Supplementary Table LXIX. Unconventional magnons in collinear Uτ AFM $Cr_2As$.

| ID | Formula | SG | $T_C$ (K) | $U$ (eV) | Moment ($\mu_B$) | $E_{gap}$ (eV) |
|---|---|---|---|---|---|---|
| 1.130 | $Cr_2As$ | 129 ($P4/nmm$) | 393 | 0[42] | 1.08 (0.40)<br>2.04 (1.34) | 0 |
| SSG | SWP | Unconventional magnons | | | MSG | Spin axis |
| 59.59.2.1 ($P_c{}^1m{}^1m{}^1n{}^{\infty m}1$) | Cr 4a, 4c | L:(1/2,$v$,$w$) | $L_1^SL_2^S(4)$ | QNPL | 62.450 ($P_anma$) | [100] |
| | | N:($u$,1/2,$w$) | $N_1^SN_2^S(4)$ | QNPL | | |
| Magnetic structure | | | Magnon dispersion | | | |

Supplementary Table LXX. Unconventional magnons in collinear Uτ AFM $Fe_2As$.

| ID | Formula | SG | $T_C$ (K) | $U$ (eV) | Moment ($\mu_B$) | $E_{gap}$ (eV) |
|---|---|---|---|---|---|---|
| 1.131 | $Fe_2As$ | 129 ($P4/nmm$) | 353 | 0[42] | 1.36 (0.95)<br>2.31 (1.52) | 0 |
| SSG | SWP | Unconventional magnons | | | MSG | Spin axis |
| 129.129.2.1 ($P_c{}^14/{}^1n{}^1m{}^1m{}^{\infty m}1$) | Fe 4a, 4c | F:($u$,1/2,$w$) | $F_1^SF_2^S(4)$ | QNPL | 62.450 ($P_anma$) | [010] |
| Magnetic structure | | | Magnon dispersion | | | |

Supplementary Table LXXI. Unconventional magnons in collinear Uτ AFM $Mn_2As$.

| ID | Formula | SG | $T_C$ (K) | $U$ (eV) | Moment ($\mu_B$) | $E_{gap}$ (eV) |
|---|---|---|---|---|---|---|
| 1.132 | $Mn_2As$ | 129 ($P4/nmm$) | 573 | 0[42] | 2.07 (3.70)<br>3.50 (3.50) | 0 |
| SSG | SWP | Unconventional magnons | | | MSG | Spin axis |
| 129.129.2.1 ($P_c{}^{1}4/{}^{1}n{}^{1}m{}^{1}m^{\infty m}1$) | Mn 4a, 4c | F:($u$,1/2,$w$) | $F_1^SF_2^S(4)$ | QNPL | 62.450 ($P_anma$) | [010] |
| Magnetic structure | | | Magnon dispersion | | | |

Supplementary Table LXXII. Unconventional magnons in collinear Uτ AFM $LaMn_3Cr_4O_{12}$.

| ID | Formula | SG | $T_C$ (K) | $U$ (eV) | Moment ($\mu_B$) | $E_{gap}$ (eV) |
|---|---|---|---|---|---|---|
| 1.156 | $LaMn_3Cr_4O_{12}$ | 204 ($Im\bar{3}$) | 50 | 4.0, 3.0<br>(Mn, Cr $3d$)[43] | 3.90 (3.39)<br>2.82 (2.89) | 1.74 |
| SSG | SWP | Unconventional magnons | | | MSG | Spin axis |
| 195.197.2.1 ($P_I{}^{1}2{}^{1}3^{\infty m}1$) | Mn 6b, Cr 8c | Γ:(0,0,0) | $\Gamma_4^S(6)$ | $C$−4 SP | 146.12 ($R_I3$) | [111] |
| | | Γ:(0,0,0) | $\Gamma_2^S\Gamma_3^S(4)$ | $C$−8 DP | | |
| | | R:(1/2,1/2,1/2) | $R_4^S(6)$ | $C$−4 SP | | |
| Magnetic structure | | | Magnon dispersion | | | |

Supplementary Table LXXIII. Unconventional magnons in collinear Uτ AFM $Er_2Ni_2In$.

| ID | Formula | SG | $T_C$ (K) | $U$ (eV) | Moment ($\mu_B$) | $E_{gap}$ (eV) |
|---|---|---|---|---|---|---|
| 1.195 | $Er_2Ni_2In$ | 65 ($Cmmm$) | 6 | 6.0 (Er 4$f$)[26] | 2.88 (7.71) | 0 |
| SSG | SWP | Unconventional magnons | | | MSG | Spin axis |
| 63.51.2.1 ($C_a{}^1m^1c^1m^{\infty m}1$) | Er 16i | Q:($u$,$v$,1/2) | $Q_1^S Q_2^S(4)$ | QNPL | 63.467 ($C_amcm$) | [010] |
| Magnetic structure | | | Magnon dispersion | | | |

Supplementary Table LXXIV. Unconventional magnons in collinear Uτ AFM $UP_2$.

| ID | Formula | SG | $T_C$ (K) | $U$ (eV) | Moment ($\mu_B$) | $E_{gap}$ (eV) |
|---|---|---|---|---|---|---|
| 1.215 | $UP_2$† | 129 ($P4/nmm$) | 203 | 0[44] | 0.91 (1.0) | 0 |
| SSG | SWP | Unconventional magnons | | | MSG | Spin axis |
| 129.129.2.1 ($P_c{}^1 4/{}^1n^1m^1m^{\infty m}1$) | U 4c | F:($u$,1/2,$w$) | $F_1^S F_2^S(4)$ | QNPL | 130.432 ($P_c4/ncc$) | [001] |
| Magnetic structure | | | Magnon dispersion | | | |

Supplementary Table LXXV. Unconventional magnons in collinear Uτ AFM $CuF_2$.

| ID | Formula | SG | $T_C$ (K) | $U$ (eV) | Moment ($\mu_B$) | $E_{gap}$ (eV) |
|---|---|---|---|---|---|---|
| 1.219 | $CuF_2$ | 14 ($P2_1/c$) | 69 | 4.0 (Cu 3$d$)[45] | 0.77 (0.73) | 0 |
| SSG | SWP | Unconventional magnons | | | MSG | Spin axis |
| 14.14.2.1 ($P_a{}^12_1/{}^1c^{\infty m}1$) | Cu 4a | G:($u$, 1/2, $w$) | $G_1^SG_2^S(4)$ | QNPL | 2.7 ($P_S\bar{1}$) | [010] +21° ([100]) |
| Magnetic structure | | | Magnon dispersion | | | |

Supplementary Table LXXVI. Unconventional magnons in collinear Uτ AFM $CaCo_2P_2$.

| ID | Formula | SG | $T_C$ (K) | $U$ (eV) | Moment ($\mu_B$) | $E_{gap}$ (eV) |
|---|---|---|---|---|---|---|
| 1.252 | $CaCo_2P_2$ | 139 ($I4/mmm$) | 113 | 2.0 (Co 3$d$)[44] | 0.66 (0.32) | 0 |
| SSG | SWP | Unconventional magnons | | | MSG | Spin axis |
| 129.139.2.1 ($P_I{}^14/{}^1n{}^1m{}^1m^{\infty m}1$) | Co 4d | F:($u$,1/2,$w$) | $F_1^SF_2^S(4)$ | QNPL | 59.416 ($P_Immn$) | [010] |
| Magnetic structure | | | Magnon dispersion | | | |

Supplementary Table LXXVII. Unconventional magnons in collinear Uτ AFM $CeCo_2P_2$.

| ID | Formula | SG | $T_C$ (K) | $U$ (eV) | Moment ($\mu_B$) | $E_{gap}$ (eV) |
|---|---|---|---|---|---|---|
| 1.253 | $CeCo_2P_2$ | 139 ($I4/mmm$) | 440 | 2.0 (Co $3d$)[44] | 0.88 (0.94) | 0 |
| SSG | SWP | Unconventional magnons | | | MSG | Spin axis |
| 129.139.2.1 ($P_I{}^1 4/{}^1 n^1 m^1 m^{\infty m}1$) | Co 4d | F:($u$,1/2,$w$) | $F_1^S F_2^S$(4) | QNPL | 126.386 ($P_I 4/nnc$) | [001] |
| Magnetic structure | | | Magnon dispersion | | | |

Supplementary Table LXXVIII. Unconventional magnons in collinear Uτ AFM $Ca_3Ru_2O_7$.

| ID | Formula | SG | $T_C$ (K) | $U$ (eV) | Moment ($\mu_B$) | $E_{gap}$ (eV) |
|---|---|---|---|---|---|---|
| 1.263 | $Ca_3Ru_2O_7$† | 36 ($Cmc2_1$) | 56 | 0[46] | 1.40 (1.59) | 0 |
| SSG | SWP | Unconventional magnons | | | MSG | Spin axis |
| 26.36.2.1 ($P_C{}^1 m^1 c^1 2_1{}^{\infty m}1$) | Ru 8b | W:($u$,$v$,1/2) | $W_1^S W_1^S$(4) | QNPL | 33.154 ($P_C na2_1$) | [010] |
| Magnetic structure | | | Magnon dispersion | | | |

Supplementary Table LXXIX. Unconventional magnons in collinear Uτ AFM CeSbTe.

| ID | Formula | SG | $T_C$ (K) | $U$ (eV) | Moment ($\mu_B$) | $E_{gap}$ (eV) |
|---|---|---|---|---|---|---|
| 1.271 | CeSbTe | 129 ($P4/nmm$) | 2.3 | 4.0 (Ce 4$f$)[47] | 0.91 (1.00) | 0 |
| SSG | SWP | Unconventional magnons | | | MSG | Spin axis |
| 129.129.2.1 ($P_c{}^14/{}^1n{}^1m{}^1m^{\infty m}1$) | Ce 4c | F:($u$,1/2,$w$) | $F_1^SF_2^S(4)$ | QNPL | 130.432 ($P_c4/ncc$) | [001] |
| Magnetic structure | | Magnon dispersion | | | | |

Supplementary Table LXXX. Unconventional magnons in collinear Uτ AFM $Mn_5Si_3$.

| ID | Formula | SG | $T_C$ (K) | $U$ (eV) | Moment ($\mu_B$) | $E_{gap}$ (eV) |
|---|---|---|---|---|---|---|
| 1.305 | $Mn_5Si_3$ | 193 ($P6_3/mcm$) | 99 | 0[48] | 2.67 (1.48) | 0 |
| SSG | SWP | Unconventional magnons | | | MSG | Spin axis |
| 62.63.2.1 ($P_B{}^1n{}^1m{}^1a^{\infty m}1$) | Mn 8g | Q:(1/2,1/2,$w$) | $Q_1^SQ_1^S(8)$ | ONL | 60.431 ($P_cbcn$) | [010] |
| | | L:(1/2,$v$,$w$) | $L_1^SL_2^S(4)$ | QNPL | | |
| | | N:($u$,1/2,$w$) | $N_1^SN_2^S(4)$ | QNPL | | |
| | | W:($u$,$v$,1/2) | $W_1^SW_2^S(4)$ | QNPL | | |
| Magnetic structure | | Magnon dispersion | | | | |

Supplementary Table LXXXI. Unconventional magnons in collinear Uτ AFM $TmMnO_3$.

| ID | Formula | SG | $T_C$ (K) | $U$ (eV) | Moment ($\mu_B$) | $E_{gap}$ (eV) |
|---|---|---|---|---|---|---|
| 1.341 | $TmMnO_3$ | 62 ($Pnma$) | 32 | 3.1 (Mn $3d$)[35] | 3.71 (3.75) | 1.55 |
| SSG | SWP | Unconventional magnons | | | MSG | Spin axis |
| 33.31.2.1 ($P_a{}^1n^1a^12_1{}^{\infty m}1$) | Mn 8b | U:(1/2,0,1/2) | $U_1^SU_1^S(8)$ | OP | 31.129 ($P_bmn2_1$) | [100] |
| | | W:($u$,$v$,1/2) | $W_1^SW_1^S(4)$ | QNPL | | |
| Magnetic structure | | | Magnon dispersion | | | |
| 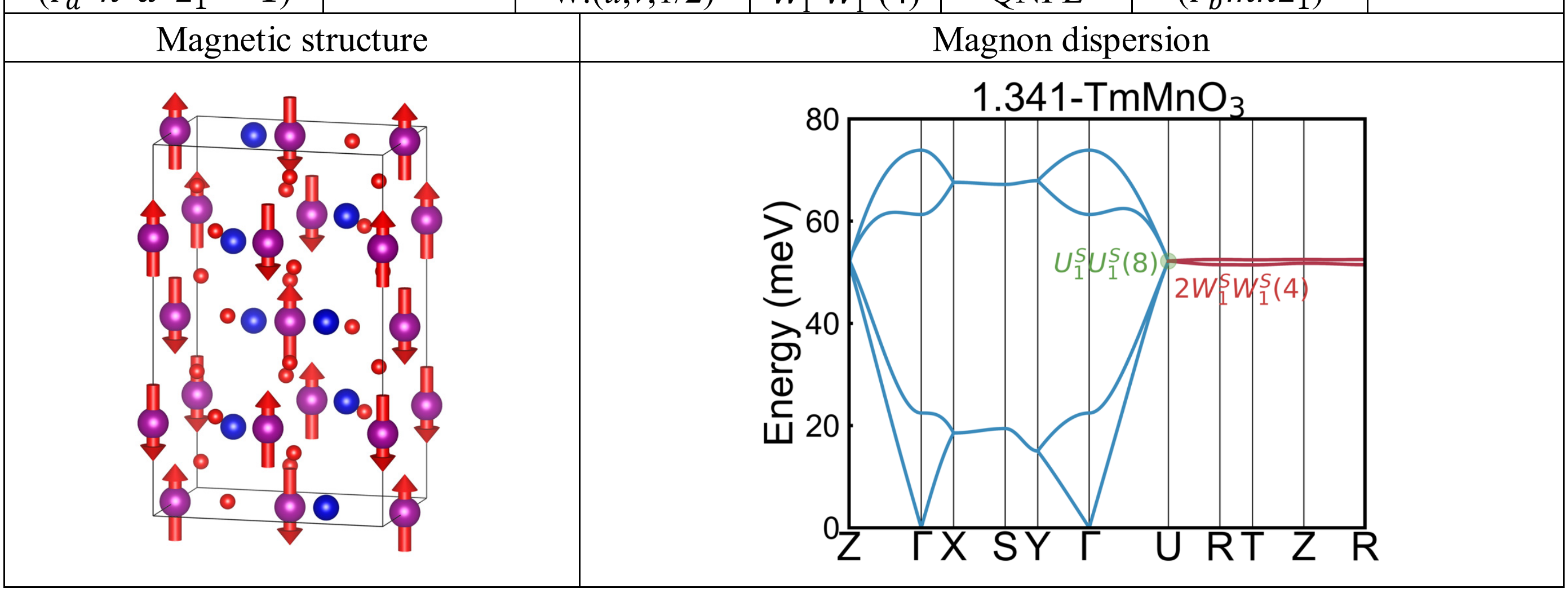 | | | | | | |

Supplementary Table LXXXII. Unconventional magnons in collinear Uτ AFM $Ba_2Co_2F_7Cl$.

| ID | Formula | SG | $T_C$ (K) | $U$ (eV) | Moment ($\mu_B$) | $E_{gap}$ (eV) |
|---|---|---|---|---|---|---|
| 1.351 | $Ba_2Co_2F_7Cl$ | 11 ($P2_1/m$) | 83 | 3.3 (Co $3d$)[49] | 2.74 (3.79) | 2.85 |
| SSG | SWP | Unconventional magnons | | | MSG | Spin axis |
| 11.11.2.1 ($P_a{}^12_1/{}^1m^{\infty m}1$) | Co 4e, 4e | G:($u$, 1/2, $w$) | $G_1^SG_2^S(4)$ | QNPL | 11.55 ($P_a2_1/m$) | [010] |
| Magnetic structure | | | Magnon dispersion | | | |
| 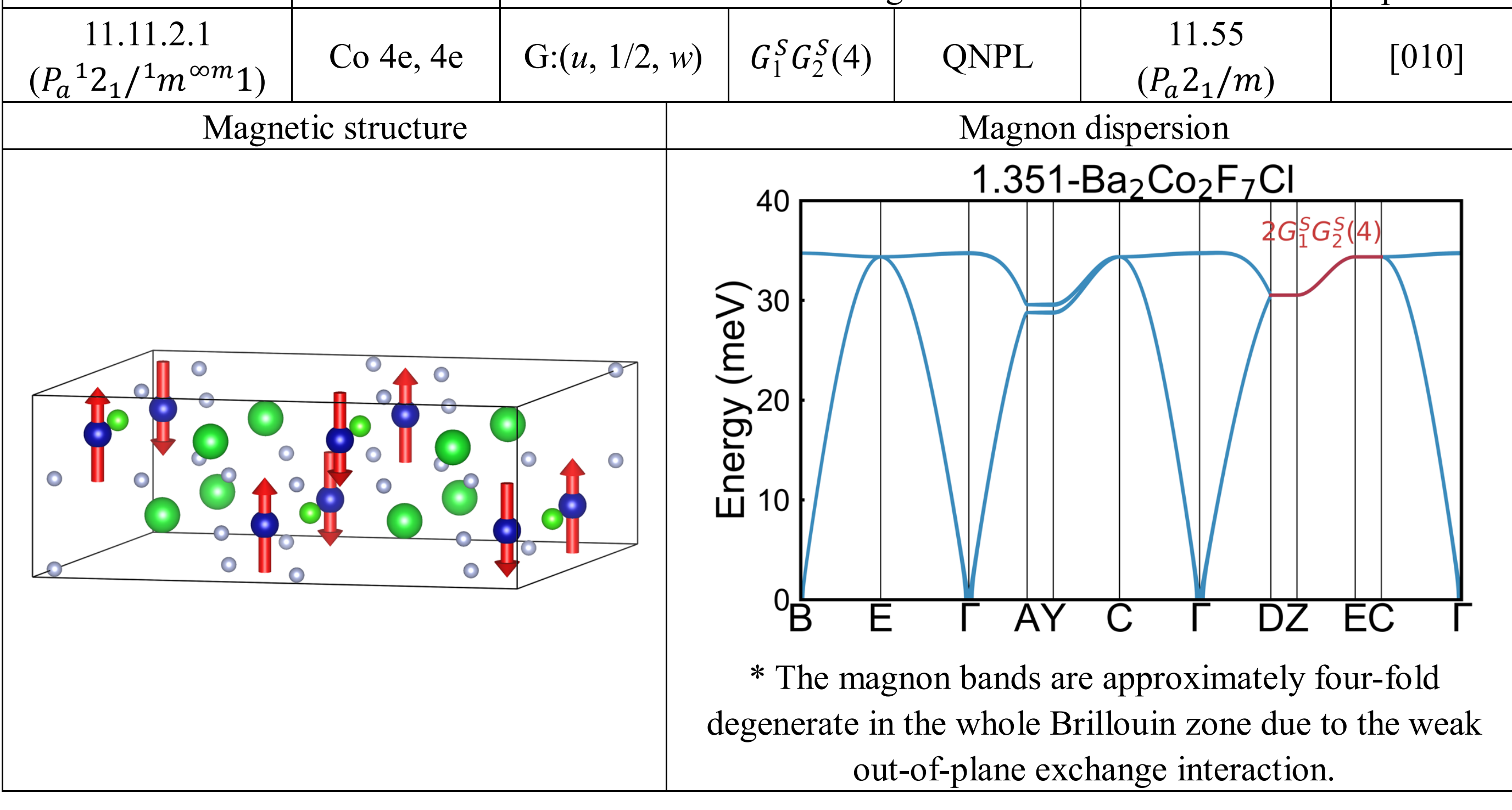 | | | | | | |

* The magnon bands are approximately four-fold degenerate in the whole Brillouin zone due to the weak out-of-plane exchange interaction.

Supplementary Table LXXXIII. Unconventional magnons in collinear Uτ AFM $Nd_2NiO_4$.

| ID | Formula | SG | $T_C$ (K) | $U$ (eV) | Moment ($\mu_B$) | $E_{gap}$ (eV) |
|---|---|---|---|---|---|---|
| 1.371 | $Nd_2NiO_4$ | 64 ($Cmce$) | 320 | 4.2 (Ni 3$d$)[40] | 1.61 (1.57) | 2.58 |
| SSG | SWP | Unconventional magnons | | | MSG | Spin axis |
| 55.64.2.1 ($P_A{}^1b^1a^1m^{\infty m}1$) | Ni 4a | L:(1/2,$v$,$w$) | $L_1^S L_2^S$(4) | QNPL | 53.335 ($P_Cmna$) | [100] |
| | | N:($u$,1/2,$w$) | $N_1^S N_2^S$(4) | QNPL | | |
| Magnetic structure | | | Magnon dispersion | | | |
| | | | | | | |

Supplementary Table LXXXIV. Unconventional magnons in collinear Uτ AFM $La_2NiO_3F_2$.

| ID | Formula | SG | $T_C$ (K) | $U$ (eV) | Moment ($\mu_B$) | $E_{gap}$ (eV) |
|---|---|---|---|---|---|---|
| 1.388 | $La_2NiO_3F_2$ | 66 ($Cccm$) | 55 | 4.2 (Ni 3$d$)[40] | 1.64 (0.70) | 1.66 |
| SSG | SWP | Unconventional magnons | | | MSG | Spin axis |
| 53.66.2.1 ($P_A{}^1m^1n^1a^{\infty m}1$) | Ni 4e | W:($u$,$v$,1/2) | $W_1^S W_2^S$(4) | QNPL | 53.333 ($P_Amna$) | [001] |
| Magnetic structure | | | Magnon dispersion | | | |
| | | | | | | |

Supplementary Table LXXXV. Unconventional magnons in collinear Uτ AFM $Sr_2CoO_3Cl$.

| ID | Formula | SG | $T_C$ (K) | $U$ (eV) | Moment ($\mu_B$) | $E_{gap}$ (eV) |
|---|---|---|---|---|---|---|
| 1.389 | $Sr_2CoO_3Cl$† | 129 ($P4/nmm$) | 330 | 3.3 (Co 3$d$)[49] | 2.86 (2.82) | 1.27 |
| SSG | SWP | Unconventional magnons | | | MSG | Spin axis |
| 51.67.2.1 ($P_C{}^1m^1m^1a^{\infty m}1$) | Co 4e | L:(1/2,$v$,$w$) | $L_1^S L_2^S(4)$ | QNPL | 49.274 ($P_Bccm$) | [100] |
| Magnetic structure | | | Magnon dispersion | | | |

Supplementary Table LXXXVI. Unconventional magnons in collinear Uτ AFM $CeNiGe_3$.

| ID | Formula | SG | $T_C$ (K) | $U$ (eV) | Moment ($\mu_B$) | $E_{gap}$ (eV) |
|---|---|---|---|---|---|---|
| 1.414 | $CeNiGe_3$† | 65 ($Cmmm$) | 5.9 | 6.0 (Ce 4$f$)[50] | 0.93 (0.22) | 0 |
| SSG | SWP | Unconventional magnons | | | MSG | Spin axis |
| 51.67.2.1 ($P_B{}^1m^1m^1a^{\infty m}1$) | Ce 4f | L:( 1/2,$v$,$w$) | $L_1^S L_2^S(4)$ | QNPL | 59.415 ($P_Cmmm$) | [010] |
| Magnetic structure | | | Magnon dispersion | | | |

Supplementary Table LXXXVII. Unconventional magnons in collinear Uτ AFM $YBa_2Cu_3O_6$.

| ID | Formula | SG | $T_C$ (K) | $U$ (eV) | Moment ($\mu_B$) | $E_{gap}$ (eV) |
|---|---|---|---|---|---|---|
| 1.420 | $YBa_2Cu_3O_6$ | 123 ($P4/mmm$) | 420 | 6.7 (Cu $3d$)[51] | 0.59 (0.54) | 0.95 |
| SSG | SWP | Unconventional magnons | | | MSG | Spin axis |
| 129.123.2.1 ($P_c{}^14/{}^1n{}^1m{}^1m^{\infty m}1$) | Cu 4g | F:($u$,1/2,$w$) | $F_1^SF_2^S(4)$ | QNPL | 65.489 ($C_ammm$) | [010] |
| Magnetic structure | | | Magnon dispersion | | | |

Supplementary Table LXXXVIII. Unconventional magnons in collinear Uτ AFM UGeS.

| ID | Formula | SG | $T_C$ (K) | $U$ (eV) | Moment ($\mu_B$) | $E_{gap}$ (eV) |
|---|---|---|---|---|---|---|
| 1.426 | UGeS† | 129 ($P4/nmm$) | 90 | 0[52] | 2.11 (1.26) | 0 |
| SSG | SWP | Unconventional magnons | | | MSG | Spin axis |
| 129.129.2.1 ($P_c{}^14/{}^1n{}^1m{}^1m^{\infty m}1$) | U 4c | F:($u$,1/2,$w$) | $F_1^SF_2^S(4)$ | QNPL | 130.432 ($P_c4/ncc$) | [001] |
| Magnetic structure | | | Magnon dispersion | | | |

Supplementary Table LXXXIX. Unconventional magnons in collinear Uτ AFM $BaCoF_4$.

| ID | Formula | SG | $T_C$ (K) | $U$ (eV) | Moment ($\mu_B$) | $E_{gap}$ (eV) |
|---|---|---|---|---|---|---|
| 1.438 | $BaCoF_4$ | 36 ($Cmc2_1$) | 69.6 | 2.5 (Co 3$d$)[41] | 2.73 (3.40) | 2.33 |
| SSG | SWP | Unconventional magnons | | | MSG | Spin axis |
| 4.4.2.1 ($P_a{}^12_1{}^{\infty m}1$) | Co 4a | G:($u$, 1/2, $w$) | $G_1^SG_2^S(4)$ | QNPL | 4.10 ($P_a2_1$) | [100] |
| Magnetic structure | | | Magnon dispersion | | | |

Supplementary Table XC. Unconventional magnons in collinear Uτ AFM $TbMn_2Si_2$.

| ID | Formula | SG | $T_C$ (K) | $U$ (eV) | Moment ($\mu_B$) | $E_{gap}$ (eV) |
|---|---|---|---|---|---|---|
| 1.468 | $TbMn_2Si_2$ | 139 ($I4/mmm$) | 293 | 0[53] | 1.98 (1.83) | 0 |
| SSG | SWP | Unconventional magnons | | | MSG | Spin axis |
| 129.139.2.1 ($P_I{}^14/{}^1n{}^1m{}^1m{}^{\infty m}1$) | Mn 4d | F:($u$,1/2,$w$) | $F_1^SF_2^S(4)$ | QNPL | 126.386 ($P_I4/nnc$) | [001] |
| Magnetic structure | | | Magnon dispersion | | | |

Supplementary Table XCI. Unconventional magnons in collinear Uτ AFM $YMn_2Si_2$.

| ID | Formula | SG | $T_C$ (K) | $U$ (eV) | Moment ($\mu_B$) | $E_{gap}$ (eV) |
|---|---|---|---|---|---|---|
| 1.469, 1.495 | $YMn_2Si_2$ | 139 ($I4/mmm$) | 410 | 0[54] | 1.93 (2.40) | 0 |
| SSG | SWP | Unconventional magnons | | | MSG | Spin axis |
| 129.139.2.1 ($P_I{}^14/{}^1n{}^1m{}^{\infty m}1$) | Mn 4d | F:($u$,1/2,$w$) | $F_1^S F_2^S(4)$ | QNPL | 126.386 ($P_I4/nnc$) | [001] |
| Magnetic structure | | | Magnon dispersion | | | |
| | | | | | | |

Supplementary Table XCII. Unconventional magnons in collinear Uτ AFM $CeMn_2Si_2$.

| ID | Formula | SG | $T_C$ (K) | $U$ (eV) | Moment ($\mu_B$) | $E_{gap}$ (eV) |
|---|---|---|---|---|---|---|
| 1.488, 1.489, 1.490 | $CeMn_2Si_2$ | 139 ($I4/mmm$) | 379 | 0[54] | 2.11 (1.90) | 0 |
| SSG | SWP | Unconventional magnons | | | MSG | Spin axis |
| 129.139.2.1 ($P_I{}^14/{}^1n{}^1m{}^{\infty m}1$) | Mn 4d | F:($u$,1/2,$w$) | $F_1^S F_2^S(4)$ | QNPL | 126.386 ($P_I4/nnc$) | [001] |
| Magnetic structure | | | Magnon dispersion | | | |
| | | | | | | |

Supplementary Table XCIII. Unconventional magnons in collinear Uτ AFM $PrMn_2Si_2$.

| ID | Formula | SG | $T_C$ (K) | $U$ (eV) | Moment ($\mu_B$) | $E_{gap}$ (eV) |
|---|---|---|---|---|---|---|
| 1.491, 1.492 | $PrMn_2Si_2$ | 139 ($I4/mmm$) | 348 | 0[54] | 2.16 (2.03) | 0 |
| SSG | SWP | Unconventional magnons | | | MSG | Spin axis |
| 129.139.2.1 ($P_I{}^14/{}^1n{}^1m{}^{\infty m}1$) | Mn 4d | F:($u$,1/2,$w$) | $F_1^SF_2^S(4)$ | QNPL | 126.386 ($P_I4/nnc$) | [001] |
| Magnetic structure | | Magnon dispersion | | | | |

Supplementary Table XCIV. Unconventional magnons in collinear Uτ AFM $YMn_2Ge_2$.

| ID | Formula | SG | $T_C$ (K) | $U$ (eV) | Moment ($\mu_B$) | $E_{gap}$ (eV) |
|---|---|---|---|---|---|---|
| 1.496, 1.691, 1.692 | $YMn_2Ge_2$ | 139 ($I4/mmm$) | 395 | 0[54] | 2.33 (2.23) | 0 |
| SSG | SWP | Unconventional magnons | | | MSG | Spin axis |
| 129.139.2.1 ($P_I{}^14/{}^1n{}^1m{}^{\infty m}1$) | Mn 4d | F:($u$,1/2,$w$) | $F_1^SF_2^S(4)$ | QNPL | 126.386 ($P_I4/nnc$) | [001] |
| Magnetic structure | | Magnon dispersion | | | | |

Supplementary Table XCV. Unconventional magnons in collinear Uτ AFM GdCuSn.

| ID | Formula | SG | $T_C$ (K) | $U$ (eV) | Moment ($\mu_B$) | $E_{gap}$ (eV) |
|---|---|---|---|---|---|---|
| 1.504 | GdCuSn | 186 ($P6_3mc$) | 24 | 4.0 (Gd 4$f$) | 7.12 (6.00) | 0 |
| SSG | SWP | Unconventional magnons | | | MSG | Spin axis |
| 26.36.2.1 ($P_C{}^1m^1c^12_1{}^{\infty m}1$) | Gd 4a | W:($u$,$v$,1/2) | $W_1^S W_1^S(4)$ | QNPL | 33.154 ($P_Cna2_1$) | [001] |
| Magnetic structure | | | | Magnon dispersion | | |

Supplementary Table XCVI. Unconventional magnons in collinear Uτ AFM GdAgSn.

| ID | Formula | SG | $T_C$ (K) | $U$ (eV) | Moment ($\mu_B$) | $E_{gap}$ (eV) |
|---|---|---|---|---|---|---|
| 1.505 | GdAgSn | 186 ($P6_3mc$) | 34 | 4.0 (Gd 4$f$) | 7.13 (7.20) | 0 |
| SSG | SWP | Unconventional magnons | | | MSG | Spin axis |
| 26.36.2.1 ($P_C{}^1m^1c^12_1{}^{\infty m}1$) | Gd 4a | W:($u$,$v$,1/2) | $W_1^S W_1^S(4)$ | QNPL | 33.154 ($P_Cna2_1$) | [001] |
| Magnetic structure | | | | Magnon dispersion | | |

Supplementary Table XCVII. Unconventional magnons in collinear Uτ AFM GdAuSn.

| ID | Formula | SG | $T_C$ (K) | $U$ (eV) | Moment ($\mu_B$) | $E_{gap}$ (eV) |
|---|---|---|---|---|---|---|
| 1.506 | GdAuSn | 186 ($P6_3mc$) | 35 | 4.0 (Gd 4$f$) | 7.11 (5.60) | 0 |
| SSG | SWP | Unconventional magnons | | | MSG | Spin axis |
| 26.36.2.1 ($P_C{}^1m^1c^12_1{}^{\infty m}1$) | Gd 4a | W:($u$,$v$,1/2) | $W_1^SW_1^S(4)$ | QNPL | 33.154 ($P_Cna2_1$) | [001] |
| Magnetic structure | | | Magnon dispersion | | | |

Supplementary Table XCVIII. Unconventional magnons in collinear Uτ AFM $NiSO_4$.

| ID | Formula | SG | $T_C$ (K) | $U$ (eV) | Moment ($\mu_B$) | $E_{gap}$ (eV) |
|---|---|---|---|---|---|---|
| 1.520 | $NiSO_4$ | 63 ($Cmcm$) | 37 | 5.1 (Ni 3$d$)[26] | 1.77 (2.10) | 2.57 |
| SSG | SWP | Unconventional magnons | | | MSG | Spin axis |
| 51.63.2.1 ($P_A{}^1m^1m^1a^{\infty m}1$) | Ni 4a | L:(1/2,$v$,$w$) | $L_1^SL_2^S(4)$ | QNPL | 60.431 ($P_Cbcn$) | [010] |
| Magnetic structure | | | Magnon dispersion | | | |

Supplementary Table XCIX. Unconventional magnons in collinear Uτ AFM $FeSO_4$.

| ID | Formula | SG | $T_C$ (K) | $U$ (eV) | Moment ($\mu_B$) | $E_{gap}$ (eV) |
|---|---|---|---|---|---|---|
| 1.521 | $FeSO_4$ | 63 ($Cmcm$) | 21 | 4.0 (Fe $3d$)[28] | 3.78 (4.10) | 3.44 |
| SSG | SWP | Unconventional magnons | | | MSG | Spin axis |
| 51.63.2.1 ($P_A{}^1m{}^1m{}^1a{}^{\infty m}1$) | Fe 4a | L:(1/2,$v$,$w$) | $L_1^S L_2^S$(4) | QNPL | 60.431 ($P_Cbcn$) | [010] |
| Magnetic structure | | | Magnon dispersion | | | |

Supplementary Table C. Unconventional magnons in collinear Uτ AFM $VPO_4$.

| ID | Formula | SG | $T_C$ (K) | $U$ (eV) | Moment ($\mu_B$) | $E_{gap}$ (eV) |
|---|---|---|---|---|---|---|
| 1.523 | $VPO_4$ | 63 ($Cmcm$) | 10.3 | 0 | 1.8 (0.8) | 0 |
| SSG | SWP | Unconventional magnons | | | MSG | Spin axis |
| 60.57.2.1 ($P_b{}^1b{}^1c{}^1n{}^{\infty m}1$) | V 8c | P:(1/2,$v$,1/2) | $P_1^S P_1^S$(8) | ONL | 62.452 ($P_cnma$) | [010] |
| | | T:(0,1/2,1/2) | $T_1^S T_2^S$(8) | OP | | |
| | | L:(1/2,$v$,$w$) | $L_1^S L_2^S$(4) | QNPL | | |
| | | W:($u$,$v$,1/2) | $W_1^S W_2^S$(4) | QNPL | | |
| Magnetic structure | | | Magnon dispersion | | | |

Supplementary Table CI. Unconventional magnons in collinear Uτ AFM $LiCoF_4$.

| ID | Formula | SG | $T_C$ (K) | $U$ (eV) | Moment ($\mu_B$) | $E_{gap}$ (eV) |
|---|---|---|---|---|---|---|
| 1.526 | $LiCoF_4$ | 14 ($P2_1/c$) | 150 | 5.3 (Co 3$d$)[55] | 3.36 (3.62) | 3.10 |
| SSG | SWP | Unconventional magnons | | | MSG | Spin axis |
| 14.14.2.1 ($P_a{}^{1}2_1/{}^{1}c^{\infty m}1$) | Co 4a | G:($u$, 1/2, $w$) | $G_1^S G_2^S(4)$ | QNPL | 14.80 ($P_a 2_1/c$) | [001] +90° ([100]) |
| Magnetic structure | | | | Magnon dispersion | | |
| | | | | | | |

Supplementary Table CII. Unconventional magnons in collinear Uτ AFM $CsNiF_3$.

| ID | Formula | SG | $T_C$ (K) | $U$ (eV) | Moment ($\mu_B$) | $E_{gap}$ (eV) |
|---|---|---|---|---|---|---|
| 1.527 | $CsNiF_3$ | 194 ($P6_3/mmc$) | 2.67 | 0 | 1.64 (2.25) | 1.27 |
| SSG | SWP | Unconventional magnons | | | MSG | Spin axis |
| 51.63.2.1 ($P_A{}^{1}m{}^{1}m{}^{1}a^{\infty m}1$) | Ni 4a | L:(1/2,$v$,$w$) | $L_1^S L_2^S(4)$ | QNPL | 58.402 ($P_B nnm$) | [010] |
| Magnetic structure | | | | Magnon dispersion | | |
| | | | | | | |

Supplementary Table CIII. Unconventional magnons in collinear Uτ AFM RbMnAs.

| ID | Formula | SG | $T_C$ (K) | $U$ (eV) | Moment ($\mu_B$) | $E_{gap}$ (eV) |
|---|---|---|---|---|---|---|
| 1.543, 1.544 | RbMnAs | 129 ($P4/nmm$) | 295 | 0[56] | 3.98 (4.05) | 0.80 |
| SSG | SWP | Unconventional magnons | | | MSG | Spin axis |
| 137.129.2.1 ($P_c{}^14_2/{}^1n{}^1m{}^1c{}^{\infty m}1$) | Mn 4b | F:($u$,1/2,$w$) | $F_1^SF_2^S(4)$ | QNPL | 138.528 ($P_c4_2/ncm$) | [001] |
| Magnetic structure | | | Magnon dispersion | | | |
| | | | | | | |

Supplementary Table CIV. Unconventional magnons in collinear Uτ AFM RbMnBi.

| ID | Formula | SG | $T_C$ (K) | $U$ (eV) | Moment ($\mu_B$) | $E_{gap}$ (eV) |
|---|---|---|---|---|---|---|
| 1.545 | RbMnBi | 129 ($P4/nmm$) | 295 | 0[56] | 4.00 (4.24) | 0.37 |
| SSG | SWP | Unconventional magnons | | | MSG | Spin axis |
| 137.129.2.1 ($P_c{}^14_2/{}^1n{}^1m{}^1c{}^{\infty m}1$) | Mn 4b | F:($u$,1/2,$w$) | $F_1^SF_2^S(4)$ | QNPL | 138.528 ($P_c4_2/ncm$) | [001] |
| Magnetic structure | | | Magnon dispersion | | | |
| | | | | | | |

Supplementary Table CV. Unconventional magnons in collinear Uτ AFM CsMnBi.

| ID | Formula | SG | $T_C$ (K) | $U$ (eV) | Moment ($\mu_B$) | $E_{gap}$ (eV) |
|---|---|---|---|---|---|---|
| 1.546 | CsMnBi | 129 ($P4/nmm$) | 295 | 0[56] | 4.06 (4.33) | 0.40 |
| SSG | SWP | Unconventional magnons | | | MSG | Spin axis |
| 137.129.2.1 ($P_c{}^14_2/{}^1n{}^1m{}^1c^{\infty m}1$) | Mn 4b | F:($u$,1/2,$w$) | $F_1^S F_2^S(4)$ | QNPL | 138.528 ($P_c4_2/ncm$) | [001] |
| Magnetic structure | | | Magnon dispersion | | | |

Supplementary Table CVI. Unconventional magnons in collinear Uτ AFM CsMnP.

| ID | Formula | SG | $T_C$ (K) | $U$ (eV) | Moment ($\mu_B$) | $E_{gap}$ (eV) |
|---|---|---|---|---|---|---|
| 1.547, 1.548 | CsMnP | 129 ($P4/nmm$) | 295 | 0[56] | 3.68 (3.55) | 0.59 |
| SSG | SWP | Unconventional magnons | | | MSG | Spin axis |
| 137.129.2.1 ($P_c{}^14_2/{}^1n{}^1m{}^1c^{\infty m}1$) | Mn 4b | F:($u$,1/2,$w$) | $F_1^S F_2^S(4)$ | QNPL | 138.528 ($P_c4_2/ncm$) | [001] |
| Magnetic structure | | | Magnon dispersion | | | |

Supplementary Table CVII. Unconventional magnons in collinear Uτ AFM LiMnAs.

| ID | Formula | SG | $T_C$ (K) | $U$ (eV) | Moment ($\mu_B$) | $E_{gap}$ (eV) |
|---|---|---|---|---|---|---|
| 1.550, 1.551, 1.552 | LiMnAs | 129 ($P4/nmm$) | 393 | 0[56] | 3.81 (3.75) | 0.40 |
| SSG | SWP | Unconventional magnons | | | MSG | Spin axis |
| 137.129.2.1 ($P_c{}^14_2/{}^1n{}^1m{}^1c^{\infty m}1$) | Mn 4b | F:($u$,1/2,$w$) | $F_1^SF_2^S(4)$ | QNPL | 138.528 ($P_c4_2/ncm$) | [001] |
| Magnetic structure | | | Magnon dispersion | | | |

Supplementary Table CVIII. Unconventional magnons in collinear Uτ AFM KMnAs.

| ID | Formula | SG | $T_C$ (K) | $U$ (eV) | Moment ($\mu_B$) | $E_{gap}$ (eV) |
|---|---|---|---|---|---|---|
| 1.553, 1.554 | KMnAs | 129 ($P4/nmm$) | 293 | 0[56] | 3.86 (4.03) | 0.76 |
| SSG | SWP | Unconventional magnons | | | MSG | Spin axis |
| 137.129.2.1 ($P_c{}^14_2/{}^1n{}^1m{}^1c^{\infty m}1$) | Mn 4b | F:($u$,1/2,$w$) | $F_1^SF_2^S(4)$ | QNPL | 138.528 ($P_c4_2/ncm$) | [001] |
| Magnetic structure | | | Magnon dispersion | | | |

Supplementary Table CIX. Unconventional magnons in collinear Uτ AFM BaCoSO.

| ID | Formula | SG | $T_C$ (K) | $U$ (eV) | Moment ($\mu_B$) | $E_{gap}$ (eV) |
|---|---|---|---|---|---|---|
| 1.593 | BaCoSO | 63 ($Cmcm$) | 220 | 2.1 (Co 3$d$) | 2.36 (2.75) | 1.33 |
| SSG | SWP | Unconventional magnons | | | MSG | Spin axis |
| 62.57.2.1 ($P_c{}^1n{}^1m{}^1a{}^{\infty m}1$) | Co 8d | Q:(1/2,1/2,$w$) | $Q_1^SQ_1^S(8)$ | OP | 57.386 ($P_abcm$) | [001] |
| | | L:(1/2,$v$,$w$) | $L_1^SL_2^S(4)$ | QNPL | | |
| | | N:($u$,1/2,$w$) | $N_1^SN_2^S(4)$ | QNPL | | |
| | | W:($u$,$v$,1/2) | $W_1^SW_2^S(4)$ | QNPL | | |
| Magnetic structure | | | Magnon dispersion | | | |

Supplementary Table CX. Unconventional magnons in collinear Uτ AFM $ErMn_2Si_2$.

| ID | Formula | SG | $T_C$ (K) | $U$ (eV) | Moment ($\mu_B$) | $E_{gap}$ (eV) |
|---|---|---|---|---|---|---|
| 1.636, 1.637 | $ErMn_2Si_2$ | 139 ($I4/mmm$) | 293 | 0[54] | 1.97 (2.24) | 0 |
| SSG | SWP | Unconventional magnons | | | MSG | Spin axis |
| 129.139.2.1 ($P_I{}^14/{}^1n{}^1m{}^1m{}^{\infty m}1$) | Mn 4d | F:($u$,1/2,$w$) | $F_1^SF_2^S(4)$ | QNPL | 126.386 ($P_I4/nnc$) | [001] |
| Magnetic structure | | | Magnon dispersion | | | |

Supplementary Table CXI. Unconventional magnons in collinear Uτ AFM $ErMn_2Ge_2$.

| ID | Formula | SG | $T_C$ (K) | $U$ (eV) | Moment ($\mu_B$) | $E_{gap}$ (eV) |
|---|---|---|---|---|---|---|
| 1.638, 1.639, 1.640 | $ErMn_2Ge_2$ | 139 ($I4/mmm$) | 293 | 0[54] | 2.31 (2.21) | 0 |
| SSG | SWP | Unconventional magnons | | | MSG | Spin axis |
| 129.139.2.1 ($P_I{}^14/{}^1n{}^1m{}^1m{}^{\infty m}1$) | Mn 4d | F:($u$,1/2,$w$) | $F_1^SF_2^S(4)$ | QNPL | 126.386 ($P_I4/nnc$) | [001] |
| Magnetic structure | | | Magnon dispersion | | | |

Supplementary Table CXII. Unconventional magnons in collinear Uτ AFM $Ba_2FeSi_2O_7$.

| ID | Formula | SG | $T_C$ (K) | $U$ (eV) | Moment ($\mu_B$) | $E_{gap}$ (eV) |
|---|---|---|---|---|---|---|
| 1.641 | $Ba_2FeSi_2O_7$ | 113 ($P\bar{4}2_1m$) | 5.2 | 4.0 (Fe $3d$)[28] | 3.67 (2.95) | 1.90 |
| SSG | SWP | Unconventional magnons | | | MSG | Spin axis |
| 114.113.2.1 ($P_c{}^1\bar{4}{}^12_1{}^1c{}^{\infty m}1$) | Fe 4b | F:($u$,1/2,$w$) | $F_1^SF_2^S(4)$ | QNPL | 36.177 ($C_cmc2_1$) | [110] |
| Magnetic structure | | | Magnon dispersion | | | |

Supplementary Table CXIII. Unconventional magnons in collinear Uτ AFM DyOCl.

| ID | Formula | SG | $T_C$ (K) | $U$ (eV) | Moment ($\mu_B$) | $E_{gap}$ (eV) |
| --- | --- | --- | --- | --- | --- | --- |
| 1.643 | DyOCl | 129 ($P4/nmm$) | 10 | 6.0 (Dy 4$f$)[57] | 8.71 (10.10) | 3.64 |
| SSG | SWP | Unconventional magnons | | | MSG | Spin axis |
| 129.129.2.1 ($P_c{}^14/{}^1n{}^1m{}^1m^{\infty m}1$) | Dy 4c | F:($u$,1/2,$w$) | $F_1^SF_2^S(4)$ | QNPL | 62.450 ($P_anma$) | [100] |
| Magnetic structure | | | Magnon dispersion | | | |
| | | | | | | |

Supplementary Table CXIV. Unconventional magnons in collinear Uτ AFM $Fe_{0.35}NbS_2$.

| ID | Formula | SG | $T_C$ (K) | $U$ (eV) | Moment ($\mu_B$) | $E_{gap}$ (eV) |
| --- | --- | --- | --- | --- | --- | --- |
| 1.677 | $Fe_{0.35}NbS_2$† | 182 ($P6_322$) | 45 | 0.9 (Fe 3$d$)[58] | 3.23 (2.90) | 0 |
| SSG | SWP | Unconventional magnons | | | MSG | Spin axis |
| 19.18.2.1 ($P_c{}^12_1{}^12_1{}^12_1{}^{\infty m}1$) | Fe 8c | R:(1/2,1/2,1/2) | $R_1^SR_1^S(8)$ | $C$−4 OP | 18.21 ($P_c2_12_12$) | [001] |
| | | L:(1/2,$v$,$w$) | $L_1^SL_1^S(4)$ | QNPL | | |
| | | N:($u$,1/2,$w$) | $N_1^SN_1^S(4)$ | QNPL | | |
| | | W:($u$,$v$,1/2) | $W_1^SW_1^S(4)$ | QNPL | | |
| Magnetic structure | | Magnon dispersion | | | | |
| | | | | | | |

Supplementary Table CXV. Unconventional magnons in collinear Uτ AFM $YBaCo_2O_5$.

| ID | Formula | SG | $T_C$ (K) | $U$ (eV) | Moment ($\mu_B$) | $E_{gap}$ (eV) |
|---|---|---|---|---|---|---|
| 1.703 | $YBaCo_2O_5$ | 123 ($P4/mmm$) | 220 | 6.0 (Co $3d$)[59] | 2.68 (2.70)<br>3.08 (4.20) | 1.36 |
| SSG | SWP | Unconventional magnons | | | MSG | Spin axis |
| 51.51.2.1 ($P_b{}^1m^1m^1a^{\infty m}1$) | Co 4e, 4e | L:(1/2,$v$,$w$) | $L_1^S L_2^S(4)$ | QNPL | 53.330 ($P_amna$) | [010] |
| Magnetic structure | | | Magnon dispersion | | | |

Supplementary Table CXVI. Unconventional magnons in collinear Uτ AFM $CeAuBi_2$.

| ID | Formula | SG | $T_C$ (K) | $U$ (eV) | Moment ($\mu_B$) | $E_{gap}$ (eV) |
|---|---|---|---|---|---|---|
| 1.714 | $CeAuBi_2$ | 129 ($P4/nmm$) | 19 | 6.0 (Ce $4f$)[60] | 0.89 (2.30) | 0 |
| SSG | SWP | Unconventional magnons | | | MSG | Spin axis |
| 129.129.2.1 ($P_c{}^14/{}^1n^1m^{\infty m}1$) | Ce 4c | F:($u$,1/2,$w$) | $F_1^S F_2^S(4)$ | QNPL | 130.432 ($P_c4/ncc$) | [001] |
| Magnetic structure | | | Magnon dispersion | | | |

Supplementary Table CXVII. Unconventional magnons in collinear Uτ AFM $BaFe_2O_4$.

| ID | Formula | SG | $T_C$ (K) | $U$ (eV) | Moment ($\mu_B$) | $E_{gap}$ (eV) |
|---|---|---|---|---|---|---|
| 1.754 | $BaFe_2O_4$ | 36 ($Cmc2_1$) | 890.5 | 4.0 (Fe 3$d$)[61] | 3.74 (4.02) | 3.15 |
| SSG | SWP | Unconventional magnons | | | MSG | Spin axis |
| 33.36.2.1 ($P_C{}^1n^1a^1 2_1{}^{\infty m}1$) | Fe 8b, 8b | U:(1/2,0,1/2) | $U_1^S U_1^S$(8) | OP | 29.109 ($P_Cca2_1$) | [010] |
| | | W:($u$,$v$,1/2) | $W_1^S W_1^S$(4) | QNPL | | |
| Magnetic structure | | | Magnon dispersion | | | |

*The W nodal plane is quasi-octuple due to the small interlayer exchange.

## S5.5. Chiral magnons in altermagnets

Supplementary Table CXVIII. Unconventional magnons in altermagnet $LaMnO_3$.

| ID | Formula | SG | $T_C$ (K) | $U$ (eV) | Moment ($\mu_B$) | $E_{gap}$ (eV) |
|---|---|---|---|---|---|---|
| 0.1, 0.642 | $LaMnO_3$ | 62 ($Pnma$) | 139.5 | 2.6 (Mn 3$d$)[62] | 3.77 (3.87) | 1.09 |
| SSG | SWP | Chiral magnons | | | MSG | Spin axis |
| 14.62.1.1 ($P^{\bar{1}}n^{\bar{1}}m^{1}a^{\infty m}1$) | Mn 4b | V:($u$,$v$,0) W:($u$,$v$,1/2) GP:($u$,$v$,$w$) | | | 62.448 ($Pn'ma'$) | [100] |
| Magnetic structure | | Magnon dispersion | | | | |

Supplementary Table CXIX. Unconventional magnons in altermagnet $MnF_2$.

| ID | Formula | SG | $T_C$ (K) | $U$ (eV) | Moment ($\mu_B$) | $E_{gap}$ (eV) |
|---|---|---|---|---|---|---|
| 0.15 | $MnF_2$ | 136 ($P4_2/mnm$) | 67 | 5.0 (Mn 3$d$)[63] | 4.69 (4.60) | 3.87 |
| SSG | SWP | Chiral magnons | | | MSG | Spin axis |
| 65.136.1.1 ($P^{\bar{1}}4_2/^{1}m^{\bar{1}}n^{1}m^{\infty m}1$) | Mn 2a | S:($u$,$u$,1/2) Σ:($u$,$u$,0) C:($u$,$u$,$w$) D:($u$,$v$,0) E:($u$,$v$,1/2) GP:($u$,$v$,$w$) | | | 136.499 ($P4_2'/mnm'$) | [001] |
| Magnetic structure | | Magnon dispersion | | | | |

Supplementary Table CXX. Unconventional magnons in altermagnet $PbNiO_3$.

| ID | Formula | SG | $T_C$ (K) | $U$ (eV) | Moment ($\mu_B$) | $E_{gap}$ (eV) |
|---|---|---|---|---|---|---|
| 0.21 | $PbNiO_3$ | 161 ($R3c$) | > 300 | 4.6 (Ni 3$d$)[64] | 1.70 (1.69) | 0.42 |
| **SSG** | **SWP** | **Chiral magnons** | | | **MSG** | **Spin axis** |
| 146.161.1.1 ($R^{1}3^{\bar{1}}c^{\infty m}1$) | Ni 6a | GP:($u$,$v$,$w$)<br>* Here $u$=0.3,$v$=0.4,$w$=0.35 | | | 161.69 ($R3c$) | [001] |
| **Magnetic structure** | | **Magnon dispersion** | | | | |
| | | *The band is quasi-doubly degenerate. | | | | |

Supplementary Table CXXI. Unconventional magnons in altermagnet $La_2NiO_4$.

| ID | Formula | SG | $T_C$ (K) | $U$ (eV) | Moment ($\mu_B$) | $E_{gap}$ (eV) |
|---|---|---|---|---|---|---|
| 0.45 | $La_2NiO_4$ | 138 ($P4_2/ncm$) | 80 | 4.2 (Ni 3$d$)[40] | 1.60 (1.68) | 2.05 |
| **SSG** | **SWP** | **Chiral magnons** | | | **MSG** | **Spin axis** |
| 14.56.1.1 ($P^{\bar{1}}c^{1}c^{\bar{1}}n^{\infty m}1$) | Ni 4a | M:($u$,0,$w$) N:($u$,1/2,$w$) GP:($u$,$v$,$w$) | | | 56.369 ($Pc'c'n$) | [100] |
| **Magnetic structure** | | **Magnon dispersion** | | | | |
| | | * The bands are approximately doubly degenerate in the whole Brillouin zone due to the weak out-of-plane exchange interaction. | | | | |

Supplementary Table CXXII. Unconventional magnons in altermagnet $MnTiO_3$.

| ID | Formula | SG | $T_C$ (K) | *U* (eV) | Moment ($\mu_B$) | $E_{gap}$ (eV) |
|---|---|---|---|---|---|---|
| 0.50 | $MnTiO_3$ | 161 ($R3c$) | 28 | 3.9 (Mn 3*d*)[65] | 4.57 (3.90) | 2.13 |
| SSG | SWP | Chiral magnons | | | MSG | Spin axis |
| 146.161.1.1 ($R^{1}3^{\bar{1}}c^{\infty m}1$) | Mn 6a | GP:(*u*,*v*,*w*)<br>* Here *u*=0.3,*v*=0.4,*w*=0.35 | | | 9.39 ($Cc'$) | [100] |
| Magnetic structure | | Magnon dispersion | | | | |
| | | *The band is quasi-doubly degenerate. | | | | |

Supplementary Table CXXIII. Unconventional magnons in altermagnet $Ba_2CoGe_2O_7$.

| ID | Formula | SG | $T_C$ (K) | *U* (eV) | Moment ($\mu_B$) | $E_{gap}$ (eV) |
|---|---|---|---|---|---|---|
| 0.56 | $Ba_2CoGe_2O_7$ | 113 ($P\bar{4}2_1m$) | 6.7 | 3.0 (Co 3*d*)[66] | 2.68 (2.90) | 2.52 |
| SSG | SWP | Chiral magnons | | | MSG | Spin axis |
| 81.113.1.1 ($P^{1}\bar{4}^{\bar{1}}2_1{}^{\bar{1}}m^{\infty m}1$) | Co 2a | D:(*u*,*v*,0) E:(*u*,*v*,1/2) GP:(*u*,*v*,*w*) | | | 35.167 ($Cm'm2'$) | [110] |
| Magnetic structure | | Magnon dispersion | | | | |
| | | *The band is quasi-doubly degenerate. | | | | |

Supplementary Table CXXIV. Unconventional magnons in altermagnet $Fe_2O_3$−alpha.

| ID | Formula | SG | $T_C$ (K) | $U$ (eV) | Moment ($\mu_B$) | $E_{gap}$ (eV) |
|---|---|---|---|---|---|---|
| 0.65, 0.66 | $Fe_2O_3$−alpha | 167 ($R\bar{3}c$) | 955 | 4.0 (Fe $3d$)[67] | 4.15 (4.22) | 2.15 |
| SSG | SWP | Chiral magnons | | | MSG | Spin axis |
| 148.167.1.1 ($R^{1}\bar{3}^{1}c^{\infty m}1$) | Fe 12c | GP:($u$,$v$,$w$) $u$ = 0.25, $v$ = 0.5, $w$ =-0.25 | | | 15.89 ($C2'/c'$) | [100] |
| Magnetic structure | | | Magnon dispersion | | | |

* For 0.66, $T_C$ = 259.1 K, the spin axis is "[001] +11° ([100])" and the MSG is 2.4 ($P\bar{1}$).

Supplementary Table CXXV. Unconventional magnons in altermagnet $LiFeP_2O_7$.

| ID | Formula | SG | $T_C$ (K) | $U$ (eV) | Moment ($\mu_B$) | $E_{gap}$ (eV) |
|---|---|---|---|---|---|---|
| 0.83 | $LiFeP_2O_7$ | 4 ($P2_1$) | 22 | 4.9 (Fe $3d$)[68] | 4.41 (4.91) | 2.79 |
| SSG | SWP | Chiral magnons | | | MSG | Spin axis |
| 1.4.1.1 ($P^{\bar{1}}2_1{}^{\infty m}1$) | Fe 2a | GP:($u$,$v$,$w$) | | | [100] +11° ([001]) | 4.7 ($P2_1$) |
| Magnetic structure | | | Magnon dispersion | | | |

*The band is quasi-doubly degenerate.

Supplementary Table CXXVI. Unconventional magnons in altermagnet $Ba_5Co_5ClO_{13}$.

| ID | Formula | SG | $T_C$ (K) | $U$ (eV) | Moment ($\mu_B$) | $E_{gap}$ (eV) |
|---|---|---|---|---|---|---|
| 0.118 | $Ba_5Co_5ClO_{13}$ | 194<br>$(P6_3/mmc)$ | 110 | 0 | 0.73 (0.61)<br>2.49 (2.21)<br>0.31 (0.35) | 0 |
| SSG | SWP | Chiral magnons | | | MSG | Spin axis |
| 164.194.1.1<br>$(P^{\bar{1}}6_3/^{\bar{1}}m^{1}m^{\bar{1}}c^{\infty m}1)$ | Co 2a, 4e, 4f | D:($u$,0,$w$) GP:($u$,$v$,$w$) | | | 194.268<br>$(P6_3'/m'm'c)$ | [001] |
| Magnetic structure | | | Magnon dispersion | | | |
| | | | *The band is quasi-doubly degenerate. | | | |

Supplementary Table CXXVII. Unconventional magnons in altermagnet $Cu_2V_2O_7$.

| ID | Formula | SG | $T_C$ (K) | $U$ (eV) | Moment ($\mu_B$) | $E_{gap}$ (eV) |
|---|---|---|---|---|---|---|
| 0.137 | $Cu_2V_2O_7$† | 43<br>$(Fdd2)$ | 33.4 | 6.5<br>(Cu 3$d$)[69] | 0.74 (0.93) | 2.11 |
| SSG | SWP | Chiral magnons | | | MSG | Spin axis |
| 9.43.1.1<br>$(F^{\bar{1}}d^{1}d^{\bar{1}}2^{\infty m}1)$ | Cu 8a | L:(1/2,1/2,1/2) J:($u$,0,$w$) GP:($u$,$v$,$w$) | | | 43.227<br>$(Fd'd'2)$ | [100] |
| Magnetic structure | | | Magnon dispersion | | | |

Supplementary Table CXXVIII. Unconventional magnons in altermagnet $La_2LiRuO_6$.

| ID | Formula | SG | $T_C$ (K) | $U$ (eV) | Moment ($\mu_B$) | $E_{gap}$ (eV) |
|---|---|---|---|---|---|---|
| 0.148 | $La_2LiRuO_6$ | 14 ($P2_1/c$) | 30 | 2.1 (Ru 4$d$)[70] | 2.01 (2.20) | 1.25 |
| SSG | SWP | Chiral magnons | | | MSG | Spin axis |
| 2.14.1.1 ($P^{\bar{1}}2_1/^{\bar{1}}c^{\infty m}1$) | Ru 2c | GP:($u$,$v$,$w$) | | | 14.75 ($P2_1/c$) | [100] +39° ([001]) |
| Magnetic structure | | | Magnon dispersion | | | |

Supplementary Table CXXIX. Unconventional magnons in altermagnet $CoF_2$.

| ID | Formula | SG | $T_C$ (K) | $U$ (eV) | Moment ($\mu_B$) | $E_{gap}$ (eV) |
|---|---|---|---|---|---|---|
| 0.178 | $CoF_2$ | 136 ($P4_2/mnm$) | 39 | 6.0 (Co 3$d$)[71] | 2.84 (2.60) | 4.69 |
| SSG | SWP | Chiral magnons | | | MSG | Spin axis |
| 65.136.1.1 ($P^{\bar{1}}4_2/^{1}m^{\bar{1}}n^{1}m^{\infty m}1$) | Co 2a | S:($u$,$u$,1/2) Σ:($u$,$u$,0) C:($u$,$u$,$w$) D:($u$,$v$,0) E:($u$,$v$,1/2) GP:($u$,$v$,$w$) | | | 136.499 ($P4_2'/mnm'$) | [001] |
| Magnetic structure | | | Magnon dispersion | | | |

Supplementary Table CXXX. Unconventional magnons in altermagnet $CeMn_2Ge_4O_{12}$.

| ID | Formula | SG | $T_C$ (K) | $U$ (eV) | Moment ($\mu_B$) | $E_{gap}$ (eV) |
|---|---|---|---|---|---|---|
| 0.189 | $CeMn_2Ge_4O_{12}$ | 125 ($P4/nbm$) | 8 | 5.0 (Mn 3$d$)[72] | 4.65 (4.61) | 0.93 |
| SSG | SWP | Chiral magnons | | | MSG | Spin axis |
| 67.125.1.1 ($P^{\bar{1}}4/{}^{1}n^{\bar{1}}b^{1}m^{\infty m}1$) | Mn 4f | S:($u$,$u$,1/2) Σ:($u$,$u$,0) C:($u$,$u$,$w$) D:($u$,$v$,0) E:($u$,$v$,1/2) GP:($u$,$v$,$w$) | | | 125.367 ($P4'/nbm'$) | [001] |
| Magnetic structure | | | Magnon dispersion | | | |
| 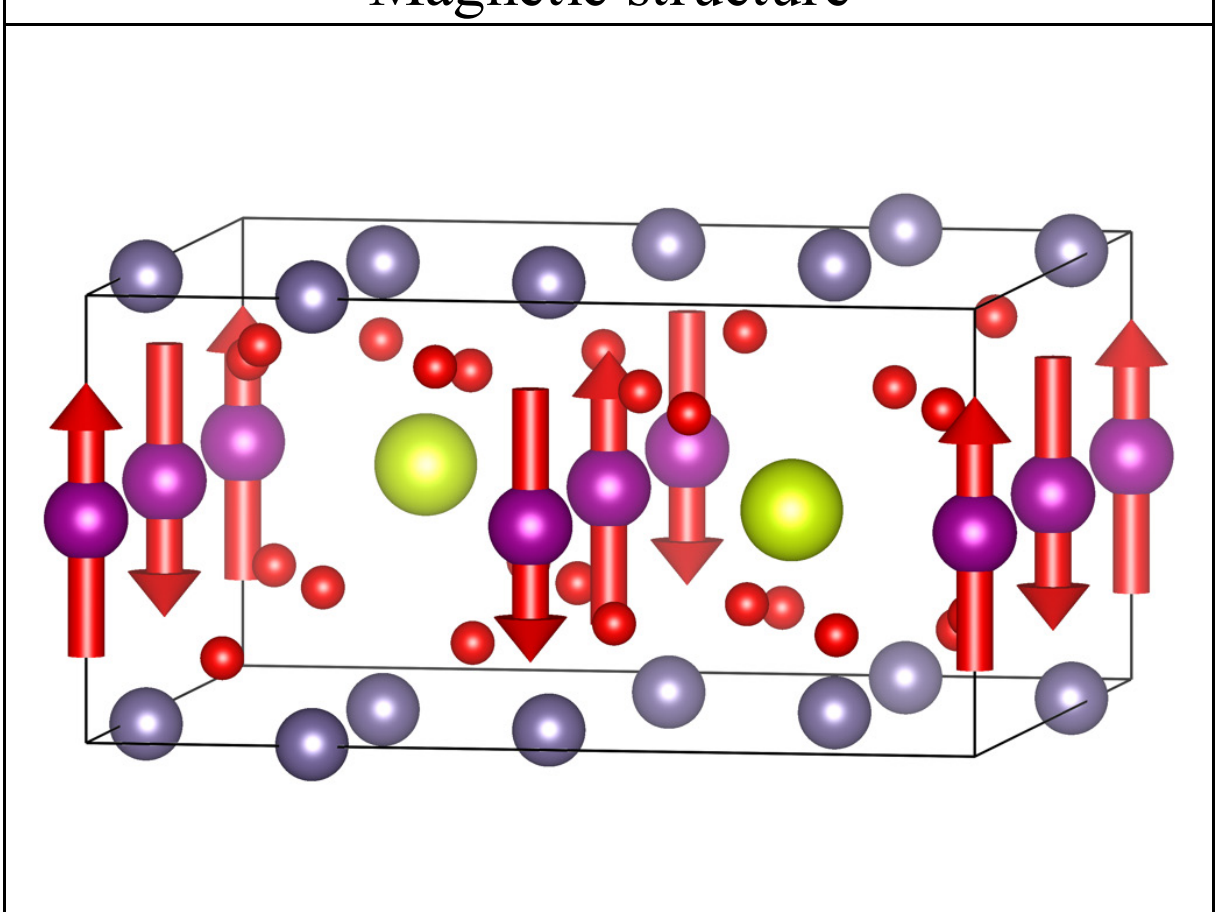 | | | 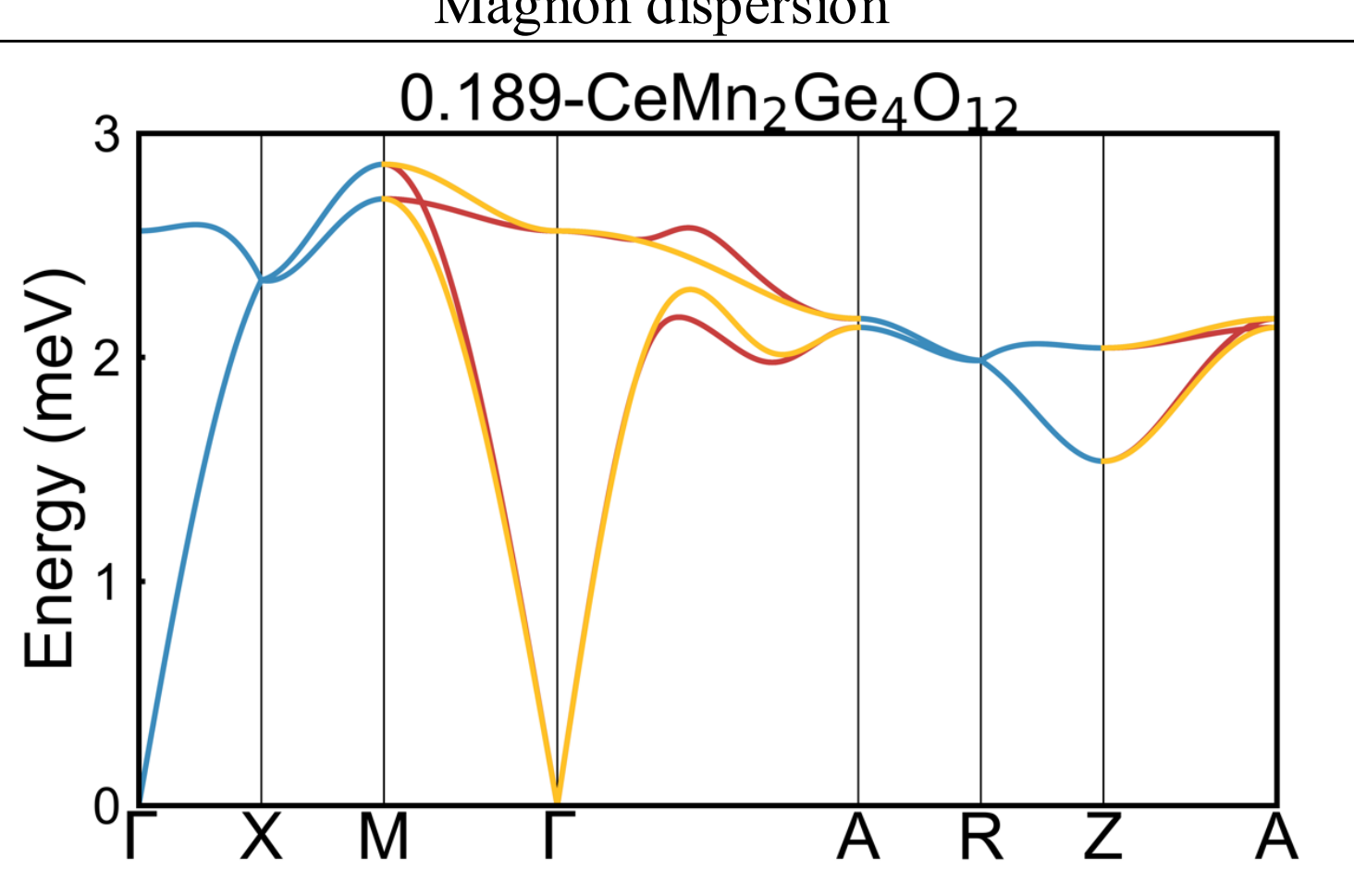 | | | |

Supplementary Table CXXXI. Unconventional magnons in altermagnet $Ba_2MnSi_2O_7$.

| ID | Formula | SG | $T_C$ (K) | $U$ (eV) | Moment ($\mu_B$) | $E_{gap}$ (eV) |
|---|---|---|---|---|---|---|
| 0.229 | $Ba_2MnSi_2O_7$ | 113 ($P\bar{4}2_1m$) | 2.95 | 4.0 (Mn 3$d$)[73] | 4.60 (4.10) | 3.35 |
| SSG | SWP | Chiral magnons | | | MSG | Spin axis |
| 81.113.1.1 ($P^{1}\bar{4}^{\bar{1}}2_1{}^{\bar{1}}m^{\infty m}1$) | Mn 2a | D:($u$,$v$,0) E:($u$,$v$,1/2) GP:($u$,$v$,$w$) | | | 113.267 ($P\bar{4}2_1m$) | [001] |
| Magnetic structure | | | Magnon dispersion | | | |
| 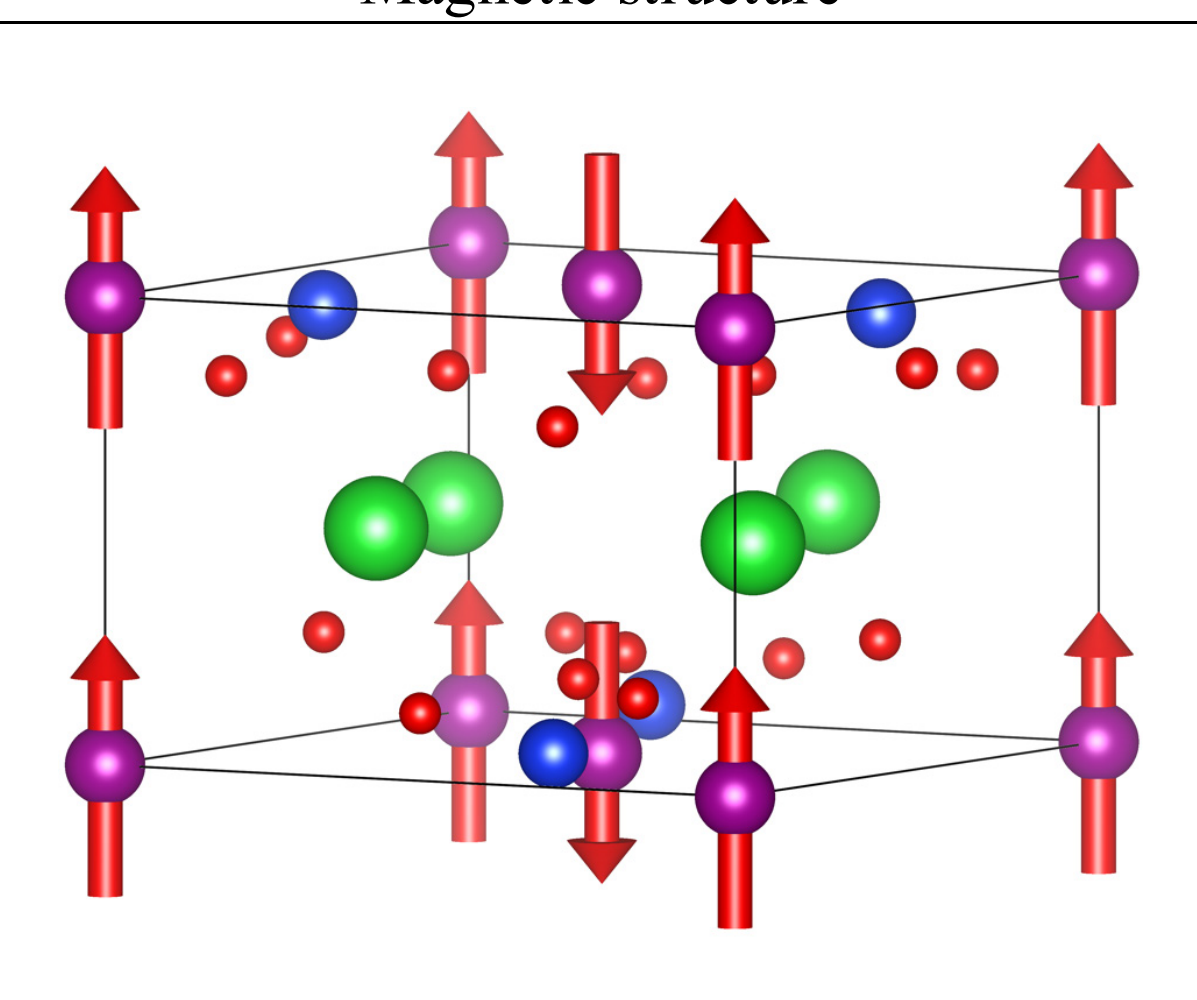 | | | 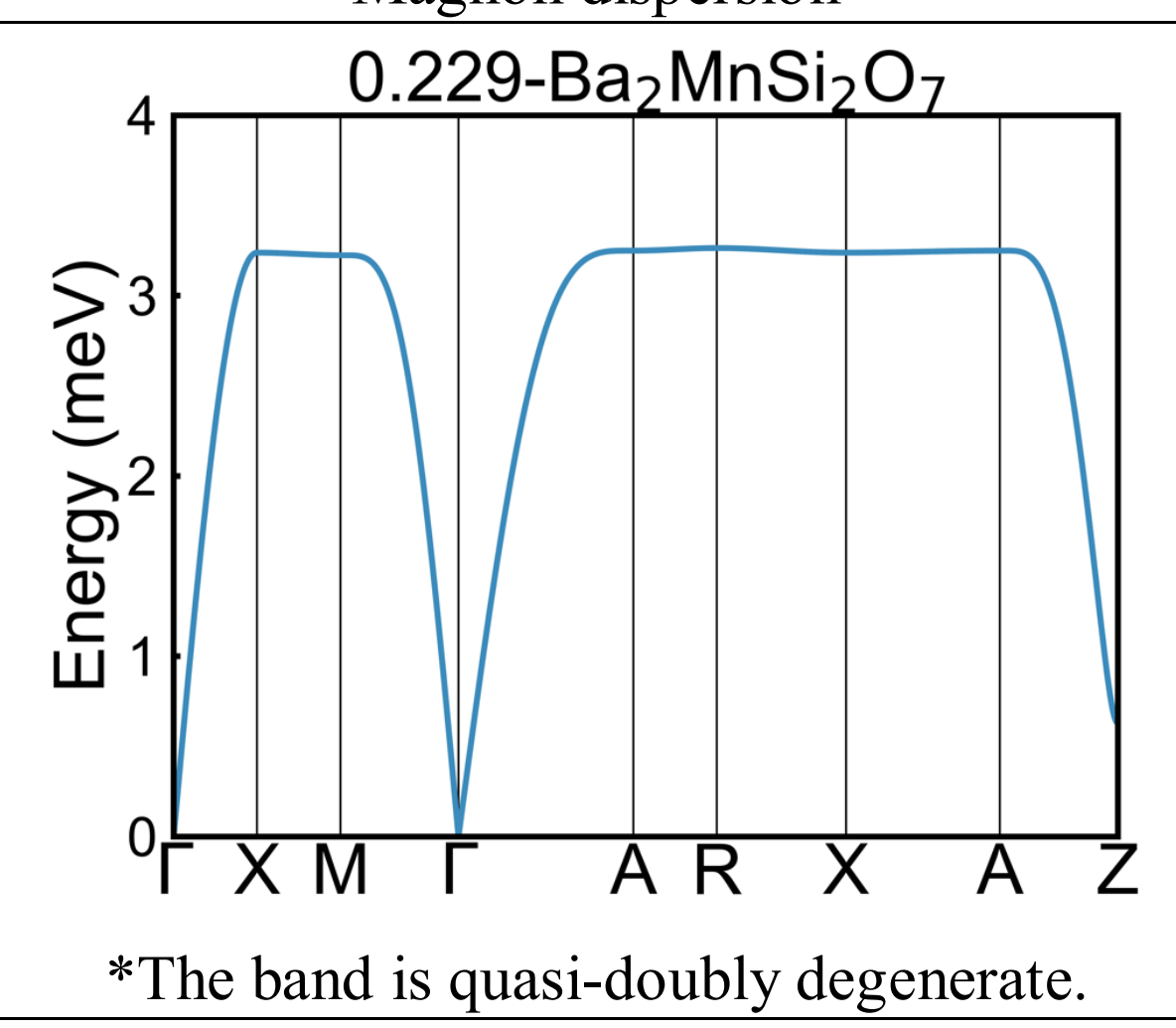 *The band is quasi-doubly degenerate. | | | |

Supplementary Table CXXXII. Unconventional magnons in altermagnet $CuFePO_5$.

| ID | Formula | SG | $T_C$ (K) | $U$ (eV) | Moment ($\mu_B$) | $E_{gap}$ (eV) |
|---|---|---|---|---|---|---|
| 0.260 | $CuFePO_5$ | 62 ($Pnma$) | 195 | 4.0 (Cu 3$d$)<br>4.9 (Fe 3$d$)[74] | 0.66 (0.95)<br>4.32 (4.28) | 1.66 |
| SSG | SWP | Chiral magnons | | | MSG | Spin axis |
| 11.62.1.1 ($P^{\bar{1}}n^{1}m^{\bar{1}}a^{\infty m}1$) | Cu 4a<br>Fe 4e | M:($u$,0,$w$) N:($u$,1/2,$w$) GP:($u$,$v$,$w$) | | | 62.441 ($Pnma$) | [010] |
| Magnetic structure | | Magnon dispersion | | | | |

Supplementary Table CXXXIII. Unconventional magnons in altermagnet $NiFePO_5$.

| ID | Formula | SG | $T_C$ (K) | $U$ (eV) | Moment ($\mu_B$) | $E_{gap}$ (eV) |
|---|---|---|---|---|---|---|
| 0.261 | $NiFePO_5$ | 62 ($Pnma$) | 178 | 4.6 (Ni 3$d$)<br>4.9 (Fe 3$d$)[74] | 1.72 (2.03)<br>4.32 (4.09) | 2.58 |
| SSG | SWP | Chiral magnons | | | MSG | Spin axis |
| 11.62.1.1 ($P^{\bar{1}}n^{1}m^{\bar{1}}a^{\infty m}1$) | Ni 4a<br>Fe 4e | M:($u$,0,$w$) N:($u$,1/2,$w$) GP:($u$,$v$,$w$) | | | 62.441 ($Pnma$) | [010] |
| Magnetic structure | | Magnon dispersion | | | | |

Supplementary Table CXXXIV. Unconventional magnons in altermagnet $Sr_2CoTeO_6$.

| ID | Formula | SG | $T_C$ (K) | $U$ (eV) | Moment ($\mu_B$) | $E_{gap}$ (eV) |
|---|---|---|---|---|---|---|
| 0.301 | $Sr_2CoTeO_6$ | 14 ($P2_1/c$) | 15 | 4.1 (Co $3d$)[75] | 2.73 (2.01) | 2.11 |
| SSG | SWP | Chiral magnons | | | MSG | Spin axis |
| 2.14.1.1 ($P^{\bar{1}}2_1/^{\bar{1}}c^{\infty m}1$) | Co 2b | GP:($u$,$v$,$w$) | | | 14.75 ($P2_1/c$) | [001] +26° ([100]) |
| Magnetic structure | | | Magnon dispersion | | | |

0.301-$Sr_2CoTeO_6$

*The band is quasi-doubly degenerate.

Supplementary Table CXXXV. Unconventional magnons in altermagnet $BaCrF_5$.

| ID | Formula | SG | $T_C$ (K) | $U$ (eV) | Moment ($\mu_B$) | $E_{gap}$ (eV) |
|---|---|---|---|---|---|---|
| 0.303 | $BaCrF_5$ | 19 ($P2_12_12_1$) | 3.4 | 3.5 (Cr $3d$)[26] | 2.88 (3.0) | 4.23 |
| SSG | SWP | Chiral magnons | | | MSG | Spin axis |
| 4.19.1.1 ($P^{\bar{1}}2_1{}^{\bar{1}}2_1{}^{1}2_1{}^{\infty m}1$) | Cr 4a | V:($u$,$v$,0) W:($u$,$v$,1/2) GP:($u$,$v$,$w$) | | | 19.27 ($P2_1'2_1'2_1$) | [001] |
| Magnetic structure | | | Magnon dispersion | | | |

0.303-$BaCrF_5$

*The band is quasi-doubly degenerate.

Supplementary Table CXXXVI. Unconventional magnons in altermagnet $ScCrO_3$.

| ID | Formula | SG | $T_C$ (K) | $U$ (eV) | Moment ($\mu_B$) | $E_{gap}$ (eV) |
|---|---|---|---|---|---|---|
| 0.307 | $ScCrO_3$ | 62 ($Pnma$) | 73 | 2.0 (Cr 3$d$)[76] | 2.78 (2.70) | 2.22 |
| **SSG** | **SWP** | **Chiral magnons** | | | **MSG** | **Spin axis** |
| 11.62.1.1 ($P^{\bar{1}}n^{1}m^{\bar{1}}a^{\infty m}1$) | Cr 4a | M:($u$,0,$w$) N:($u$,1/2,$w$) GP:($u$,$v$,$w$) | | | 62.441 ($Pnma$) | [010] |
| **Magnetic structure** | | | **Magnon dispersion** | | | |

Supplementary Table CXXXVII. Unconventional magnons in altermagnet $InCrO_3$.

| ID | Formula | SG | $T_C$ (K) | $U$ (eV) | Moment ($\mu_B$) | $E_{gap}$ (eV) |
|---|---|---|---|---|---|---|
| 0.308 | $InCrO_3$† | 62 ($Pnma$) | 93 | 2.0 (Cr 3$d$)[76] | 2.84 (2.50) | 1.44 |
| **SSG** | **SWP** | **Chiral magnons** | | | **MSG** | **Spin axis** |
| 11.62.1.1 ($P^{\bar{1}}n^{1}m^{\bar{1}}a^{\infty m}1$) | Cr 4a | M:($u$,0,$w$) N:($u$,1/2,$w$) GP:($u$,$v$,$w$) | | | 62.441 ($Pnma$) | [010] |
| **Magnetic structure** | | | **Magnon dispersion** | | | |

Supplementary Table CXXXVIII. Unconventional magnons in altermagnet $TlCrO_3$.

| ID | Formula | SG | $T_C$ (K) | $U$ (eV) | Moment ($\mu_B$) | $E_{gap}$ (eV) |
|---|---|---|---|---|---|---|
| 0.309 | $TlCrO_3$ | 62 ($Pnma$) | 89 | 2.0 (Cr 3$d$)[76] | 2.82 (2.46) | 0.26 |
| SSG | SWP | Chiral magnons | | | MSG | Spin axis |
| 11.62.1.1 ($P^{\bar{1}}n^{1}m^{\bar{1}}a^{\infty m}1$) | Cr 4a | M:($u$,0,$w$) N:($u$,1/2,$w$) GP:($u$,$v$,$w$) | | | 62.441 ($Pnma$) | [010] |
| Magnetic structure | | | Magnon dispersion | | | |
| | | | | | | |

Supplementary Table CXXXIX. Unconventional magnons in altermagnet $ZrMn_2Ge_4O_{12}$.

| ID | Formula | SG | $T_C$ (K) | $U$ (eV) | Moment ($\mu_B$) | $E_{gap}$ (eV) |
|---|---|---|---|---|---|---|
| 0.315 | $ZrMn_2Ge_4O_{12}$ | 125 ($P4/nbm$) | 8 | 5.0 (Mn 3$d$)[72] | 2.66 (2.68) | 2.54 |
| SSG | SWP | Chiral magnons | | | MSG | Spin axis |
| 67.125.1.1 ($P^{\bar{1}}4/^{1}n^{\bar{1}}b^{1}m^{\infty m}1$) | Mn 4f | S:($u$,$u$,1/2) Σ:($u$,$u$,0) C:($u$,$u$,$w$) D:($u$,$v$,0) E:($u$,$v$,1/2) GP:($u$,$v$,$w$) | | | 125.367 ($P4'/nbm'$) | [001] |
| Magnetic structure | | | Magnon dispersion | | | |
| | | | | | | |

Supplementary Table CXL. Unconventional magnons in altermagnet $RbMnF_4$.

| ID | Formula | SG | $T_C$ (K) | $U$ (eV) | Moment ($\mu_B$) | $E_{gap}$ (eV) |
|---|---|---|---|---|---|---|
| 0.329 | $RbMnF_4$ | 14 ($P2_1/c$) | 3.7 | 3.9 (Mn 3$d$)[77] | 3.78, 3.83 (2.97) | 1.91 |
| SSG | SWP | Chiral magnons | | | MSG | Spin axis |
| 2.14.1.1 ($P^{\bar{1}}2_1/^{\bar{1}}c^{\infty m}1$) | Mn 2a, 2b | GP:($u$,$v$,$w$) | | | 2.4 ($P\bar{1}$) | [110] +20° ([001]) |
| Magnetic structure | | | Magnon dispersion | | | |
| | | | *The band is quasi-doubly degenerate. | | | |

Supplementary Table CXLI. Unconventional magnons in altermagnet $Fe_2Mo_3O_8$.

| ID | Formula | SG | $T_C$ (K) | $U$ (eV) | Moment ($\mu_B$) | $E_{gap}$ (eV) |
|---|---|---|---|---|---|---|
| 0.331 | $Fe_2Mo_3O_8$[†] | 186 ($P6_3mc$) | 59 | 4.0 (Fe 3$d$)[78] | 3.69 (4.6) 3.76 (4.6) | 0.79 |
| SSG | SWP | Chiral magnons | | | MSG | Spin axis |
| 156.186.1.1 ($P^{\bar{1}}6_3{}^{1}m^{\bar{1}}c^{\infty m}1$) | Fe 2b, 2b | D:($u$,0,$w$) GP:($u$,$v$,$w$) | | | 186.205 ($P6_3'm'c$) | [001] |
| Magnetic structure | | | Magnon dispersion | | | |
| | | | *The band is quasi-doubly degenerate. | | | |

Supplementary Table CXLII. Unconventional magnons in altermagnet $CoF_3$.

| ID | Formula | SG | $T_C$ (K) | $U$ (eV) | Moment ($\mu_B$) | $E_{gap}$ (eV) |
|---|---|---|---|---|---|---|
| 0.334 | $CoF_3$ | 167 ($R\bar{3}c$) | 460 | 5.3 (Co $3d$)[55] | 3.34 (3.21) | 2.71 |
| SSG | SWP | Chiral magnons | | | MSG | Spin axis |
| 148.167.1.1 ($R^1\bar{3}^{\bar{1}}c^{\infty m}1$) | Co 6a | GP:($u$,$v$,$w$) | | | 167.103 ($R\bar{3}c$) | [001] |
| Magnetic structure | | | Magnon dispersion | | | |
| | | | *The band is quasi-doubly degenerate. | | | |

Supplementary Table CXLIII. Unconventional magnons in altermagnet $FeF_3$.

| ID | Formula | SG | $T_C$ (K) | $U$ (eV) | Moment ($\mu_B$) | $E_{gap}$ (eV) |
|---|---|---|---|---|---|---|
| 0.335, 0.581 | $FeF_3$ | 167 ($R\bar{3}c$) | 460 | 4.9 (Fe $3d$)[55] | 4.44 (4.45) | 4.18 |
| SSG | SWP | Chiral magnons | | | MSG | Spin axis |
| 148.167.1.1 ($R^1\bar{3}^{\bar{1}}c^{\infty m}1$) | Fe 6a | GP:($u$,$v$,$w$) | | | 15.89 ($C2'/c'$) | [100] |
| Magnetic structure | | | Magnon dispersion | | | |
| | | | *The band is quasi-doubly degenerate. | | | |

Supplementary Table CXLIV. Unconventional magnons in altermagnet $NdFeO_3$.

| ID | Formula | SG | $T_C$ (K) | $U$ (eV) | Moment ($\mu_B$) | $E_{gap}$ (eV) |
|---|---|---|---|---|---|---|
| 0.336 | $NdFeO_3$ | 62 ($Pnma$) | 760 | 5.0 (Fe $3d$)[79] | 4.18 (3.84) | 2.70 |
| **SSG** | **SWP** | **Chiral magnons** | | | **MSG** | **Spin axis** |
| 14.62.1.4 ($P^{1}n^{\bar{1}}m^{\bar{1}}a^{\infty m}1$) | Fe 4b | K:$(0,v,w)$ L:$(1/2,v,w)$ GP:$(u,v,w)$ | | | 62.448 ($Pn'ma'$) | [100] |
| **Magnetic structure** | **Magnon dispersion** | | | | | |

Supplementary Table CXLV. Unconventional magnons in altermagnet $TbFeO_3$.

| ID | Formula | SG | $T_C$ (K) | $U$ (eV) | Moment ($\mu_B$) | $E_{gap}$ (eV) |
|---|---|---|---|---|---|---|
| 0.351 | $TbFeO_3$ | 62 ($Pnma$) | 681 | 4.0 (Fe $3d$)[80] | 4.13 (4.80) | 2.32 |
| **SSG** | **SWP** | **Chiral magnons** | | | **MSG** | **Spin axis** |
| 14.62.1.4 ($P^{1}n^{\bar{1}}m^{\bar{1}}a^{\infty m}1$) | Fe 4b | K:$(0,v,w)$ L:$(1/2,v,w)$ GP:$(u,v,w)$ | | | 62.448 ($Pn'ma'$) | [100] |
| **Magnetic structure** | **Magnon dispersion** | | | | | |

Supplementary Table CXLVI. Unconventional magnons in altermagnet $TbCrO_3$.

| ID | Formula | SG | $T_C$ (K) | $U$ (eV) | Moment ($\mu_B$) | $E_{gap}$ (eV) |
|---|---|---|---|---|---|---|
| 0.354 | $TbCrO_3$† | 62 ($Pnma$) | 158 | 3.0 (Cr $3d$)[68] | 2.83 (2.55) | 2.74 |
| **SSG** | **SWP** | **Chiral magnons** | | | **MSG** | **Spin axis** |
| 14.62.1.4 ($P^1n^{\bar{1}}m^{\bar{1}}a^{\infty m}1$) | Fe 4b | K:$(0,v,w)$ L:$(1/2,v,w)$ GP:$(u,v,w)$ | | | 62.446 ($Pn'm'a$) | [001] |
| **Magnetic structure** | **Magnon dispersion** | | | | | |

Supplementary Table CXLVII. Unconventional magnons in altermagnet $SmFeO_3$.

| ID | Formula | SG | $T_C$ (K) | $U$ (eV) | Moment ($\mu_B$) | $E_{gap}$ (eV) |
|---|---|---|---|---|---|---|
| 0.379, 0.380 | $SmFeO_3$ | 62 ($Pnma$) | 480 | 4.0 (Fe $3d$)[80] | 4.13 (3.69) | 2.35 |
| **SSG** | **SWP** | **Chiral magnons** | | | **MSG** | **Spin axis** |
| 14.62.1.4 ($P^1n^{\bar{1}}m^{\bar{1}}a^{\infty m}1$) | Fe 4b | K:$(0,v,w)$ L:$(1/2,v,w)$ GP:$(u,v,w)$ | | | 62.446 ($Pn'm'a$) | [001] |
| **Magnetic structure** | **Magnon dispersion** | | | | | |

For 0.380, $T_C$ = 670 K, the spin axis is [100] and the MSG is 62.448 ($Pn'ma'$).

Supplementary Table CXLVIII. Unconventional magnons in altermagnet $KMnF_3$.

| ID | Formula | SG | $T_C$ (K) | $U$ (eV) | Moment ($\mu_B$) | $E_{gap}$ (eV) |
|---|---|---|---|---|---|---|
| 0.432 | $KMnF_3$ | 62 ($Pnma$) | 77 | 3.54 (Mn $3d$)[81] | 4.60 (4.80) | 2.95 |
| **SSG** | **SWP** | **Chiral magnons** | | | **MSG** | **Spin axis** |
| 14.62.1.4 ($P^1n^{\bar{1}}m^{\bar{1}}a^{\infty m}1$) | Mn 4a | K:(0,$v$,$w$) L:(1/2,$v$,$w$) GP:($u$,$v$,$w$) | | | 62.448 ($Pn'ma'$) | [001] |
| **Magnetic structure** | | **Magnon dispersion** | | | | |

Supplementary Table CXLIX. Unconventional magnons in altermagnet $KMnF_3$.

| ID | Formula | SG | $T_C$ (K) | $U$ (eV) | Moment ($\mu_B$) | $E_{gap}$ (eV) |
|---|---|---|---|---|---|---|
| 0.433 | $KMnF_3$ | 140 ($I4/mcm$) | 86.8 | 3.54 (Mn $3d$)[81] | 4.63 (4.80) | 3.85 |
| **SSG** | **SWP** | **Chiral magnons** | | | **MSG** | **Spin axis** |
| 87.140.1.1 ($I^14/^1m^{\bar{1}}c^{\bar{1}}m^{\infty m}1$) | Mn 4a | C:($u$,$v$,0) GP:($u$,$v$,$w$) * C = (0.4, 0.45, 0) | | | 140.541 ($I4/mcm$) | [001] |
| **Magnetic structure** | | **Magnon dispersion** | | | | |

*The band is quasi-doubly degenerate.

Supplementary Table CL. Unconventional magnons in altermagnet $LiFe_2F_6$.

| ID | Formula | SG | $T_C$ (K) | $U$ (eV) | Moment ($\mu_B$) | $E_{gap}$ (eV) |
|---|---|---|---|---|---|---|
| 0.501 | $LiFe_2F_6$† | 136 ($P4_2/mnm$) | 105 | 4.0 (Fe 3$d$)[82] | 4.11 (4.4) | 0.15 |
| SSG | SWP | Chiral magnons | | | MSG | Spin axis |
| 65.136.1.1 ($P^{\bar{1}}4_2/^{1}m^{\bar{1}}n^{1}m^{\infty m}1$) | Fe 4e | S:($u$,$u$,1/2) Σ:($u$,$u$,0) C:($u$,$u$,$w$) D:($u$,$v$,0) E:($u$,$v$,1/2) GP:($u$,$v$,$w$) | | | 136.499 ($P4_2'/mnm'$) | [001] |
| Magnetic structure | Magnon dispersion | | | | | |

Supplementary Table CLI. Unconventional magnons in altermagnet $YRuO_3$.

| ID | Formula | SG | $T_C$ (K) | $U$ (eV) | Moment ($\mu_B$) | $E_{gap}$ (eV) |
|---|---|---|---|---|---|---|
| 0.513 | $YRuO_3$† | 62 ($Pnma$) | 97 | 2.1 (Ru 4$d$)[83] | 0.74 (0.33) | 0.66 |
| SSG | SWP | Chiral magnons | | | MSG | Spin axis |
| 14.62.1.4 ($P^{1}n^{\bar{1}}m^{\bar{1}}a^{\infty m}1$) | Ru 4a | K:(0,$v$,$w$) L:(1/2,$v$,$w$) GP:($u$,$v$,$w$) | | | 62.448 ($Pn'ma'$) | [001] |
| Magnetic structure | Magnon dispersion | | | | | |

Supplementary Table CLII. Unconventional magnons in altermagnet CrSb.

| ID | Formula | SG | $T_C$ (K) | $U$ (eV) | Moment ($\mu_B$) | $E_{gap}$ (eV) |
|---|---|---|---|---|---|---|
| 0.528 | CrSb | 194 ($P6_3/mmc$) | > 600 | 0[84] | 2.78 (2.45) | 0 |
| SSG | SWP | Chiral magnons | | | MSG | Spin axis |
| 164.194.1.1 ($P^{\bar{1}}6_3/^{\bar{1}}m^{1}m^{\bar{1}}c^{\infty m}1$) | Cr 2a | D:($u$,0,$w$) GP:($u$,$v$,$w$) | | | 194.268 ($P6_3'/m'm'c$) | [001] |
| Magnetic structure | | Magnon dispersion | | | | |

Supplementary Table CLIII. Unconventional magnons in altermagnet $ZnFeF_5(H_2O)_2$.

| ID | Formula | SG | $T_C$ (K) | $U$ (eV) | Moment ($\mu_B$) | $E_{gap}$ (eV) |
|---|---|---|---|---|---|---|
| 0.575 | $ZnFeF_5(H_2O)_2$ | 44 ($Imm2$) | 9 | 0 | 4.12 (3.78) | 1.32 |
| SSG | SWP | Chiral magnons | | | MSG | Spin axis |
| 8.44.1.1 ($I^{\bar{1}}m^{1}m^{\bar{1}}2^{\infty m}1$) | Fe 4a | R:(1/2,0,1/2) B:($u$,0,$w$) Q:(1/2,$v$,1/2) GP:($u$,$v$,$w$) | | | 44.229 ($Imm2$) | [010] |
| Magnetic structure | | Magnon dispersion | | | | |
| | | *The band is quasi-doubly degenerate. | | | | |

Supplementary Table CLIV. Unconventional magnons in altermagnet $YCrO_3$.

| ID | Formula | SG | $T_C$ (K) | $U$ (eV) | Moment ($\mu_B$) | $E_{gap}$ (eV) |
|---|---|---|---|---|---|---|
| 0.586, 0.947 | $YCrO_3$ † | 62 ($Pnma$) | 141 | 4.0 (Cr 3$d$)[85] | 2.88 (2.96) | 3.16 |
| SSG | SWP | Chiral magnons | | | MSG | Spin axis |
| 14.62.1.4 ($P^{1}n^{\bar{1}}m^{\bar{1}}a^{\infty m}1$) | Cr 4b | K:(0,$v$,$w$) L:(1/2,$v$,$w$) GP:($u$,$v$,$w$) | | | 62.448 ($Pn'ma'$) | [100] |
| Magnetic structure | | Magnon dispersion | | | | |
| | | | | | | |

Supplementary Table CLV. Unconventional magnons in altermagnet $TmCrO_3$.

| ID | Formula | SG | $T_C$ (K) | $U$ (eV) | Moment ($\mu_B$) | $E_{gap}$ (eV) |
|---|---|---|---|---|---|---|
| 0.587 | $TmCrO_3$ † | 62 ($Pnma$) | 124 | 3.0 (Cr 3$d$)[86] | 2.86 (2.58) | 2.78 |
| SSG | SWP | Chiral magnons | | | MSG | Spin axis |
| 14.62.1.4 ($P^{1}n^{\bar{1}}m^{\bar{1}}a^{\infty m}1$) | Cr 4b | K:(0,$v$,$w$) L:(1/2,$v$,$w$) GP:($u$,$v$,$w$) | | | 62.448 ($Pn'ma'$) | [100] |
| Magnetic structure | | Magnon dispersion | | | | |
| | | | | | | |

Supplementary Table CLVI. Unconventional magnons in altermagnet $ErCrO_3$.

| ID | Formula | SG | $T_C$ (K) | $U$ (eV) | Moment ($\mu_B$) | $E_{gap}$ (eV) |
|---|---|---|---|---|---|---|
| 0.591 | $ErCrO_3$ | 62 ($Pnma$) | 133 | 4.0 (Cr $3d$)[87] | 2.88 (2.45) | 3.17 |
| SSG | SWP | Chiral magnons | | | MSG | Spin axis |
| 14.62.1.4 ($P^{1}n^{\bar{1}}m^{\bar{1}}a^{\infty m}1$) | Cr 4b | K:(0,$v$,$w$) L:(1/2,$v$,$w$) GP:($u$,$v$,$w$) | | | 62.448 ($Pn'ma'$) | [100] |
| Magnetic structure | | Magnon dispersion | | | | |

Supplementary Table CLVII. Unconventional magnons in altermagnet $DyCrO_3$.

| ID | Formula | SG | $T_C$ (K) | $U$ (eV) | Moment ($\mu_B$) | $E_{gap}$ (eV) |
|---|---|---|---|---|---|---|
| 0.592 | $DyCrO_3$ | 62 ($Pnma$) | 146 | 4.0 (Cr $3d$)[87] | 2.85 (2.60) | 3.12 |
| SSG | SWP | Chiral magnons | | | MSG | Spin axis |
| 14.62.1.4 ($P^{1}n^{\bar{1}}m^{\bar{1}}a^{\infty m}1$) | Cr 4b | K:(0,$v$,$w$) L:(1/2,$v$,$w$) GP:($u$,$v$,$w$) | | | 62.446 ($Pn'm'a$) | [001] |
| Magnetic structure | | Magnon dispersion | | | | |

Supplementary Table CLVIII. Unconventional magnons in altermagnet $RuO_2$.

| ID | Formula | SG | $T_C$ (K) | $U$ (eV) | Moment ($\mu_B$) | $E_{gap}$ (eV) |
|---|---|---|---|---|---|---|
| 0.607 | $RuO_2$ | 136 ($P4_2/mnm$) | > 300 | 1.6 (Ru 4$d$)[88] | 1.02 (0.05) | 0 |
| SSG | SWP | Chiral magnons | | | MSG | Spin axis |
| 65.136.1.1 ($P^{\bar{1}}4_2/^{1}m^{\bar{1}}n^{1}m^{\infty m}1$) | Ru 2a | S:($u$,$u$,1/2) Σ:($u$,$u$,0) C:($u$,$u$,$w$) D:($u$,$v$,0) E:($u$,$v$,1/2) GP:($u$,$v$,$w$) | | | 136.499 ($P4_2'/mnm'$) | [001] |
| Magnetic structure | | | Magnon dispersion | | | |

Supplementary Table CLIX. Unconventional magnons in altermagnet $NdMnO_3$.

| ID | Formula | SG | $T_C$ (K) | $U$ (eV) | Moment ($\mu_B$) | $E_{gap}$ (eV) |
|---|---|---|---|---|---|---|
| 0.609 | $NdMnO_3$† | 62 ($Pnma$) | 85 | 2.6 (Mn 3$d$)[62] | 3.60 (3.42) | 1.10 |
| SSG | SWP | Chiral magnons | | | MSG | Spin axis |
| 14.62.1.1 ($P^{\bar{1}}n^{\bar{1}}m^{1}a^{\infty m}1$) | Mn 4c | V:($u$,$v$,0) W:($u$,$v$,1/2) GP:($u$,$v$,$w$) | | | 11.50 ($P2_1/m$) | [010] −36° ([100]) |
| Magnetic structure | | Magnon dispersion | | | | |

Supplementary Table CLX. Unconventional magnons in altermagnet $CrNb_4S_8$.

| ID | Formula | SG | $T_C$ (K) | $U$ (eV) | Moment ($\mu_B$) | $E_{gap}$ (eV) |
|---|---|---|---|---|---|---|
| 0.708 | $CrNb_4S_8$ | 194 ($P6_3/mmc$) | | 3.0 (Cr $3d$)[89] | 3.07 (1.5) | 0 |
| SSG | SWP | Chiral magnons | | | MSG | Spin axis |
| 164.194.1.1 ($P^{\bar{1}}6_3/^{\bar{1}}m^{1}m^{\bar{1}}c^{\infty m}1$) | Cr 2a | D:($u$,0,$w$) GP:($u$,$v$,$w$) | | | 194.268 ($P6_3'/m'm'c$) | [001] |
| Magnetic structure | | Magnon dispersion | | | | |

Supplementary Table CLXI. Unconventional magnons in altermagnet $VNb_3S_6$.

| ID | Formula | SG | $T_C$ (K) | $U$ (eV) | Moment ($\mu_B$) | $E_{gap}$ (eV) |
|---|---|---|---|---|---|---|
| 0.712 | $VNb_3S_6$ | 182 ($P6_322$) | 50 | 0 | 1.76 (1.5) | 0 |
| SSG | SWP | Chiral magnons | | | MSG | Spin axis |
| 149.182.1.1 ($P^{\bar{1}}6_3{}^{\bar{1}}2^{1}2^{\infty m}1$) | V 2a | P:(1/3,1/3,$w$) C:($u$,$u$,$w$) GP:($u$,$v$,$w$) | | | 20.33 ($C2'2'2_1$) | [100] |
| Magnetic structure | | Magnon dispersion | | | | |

Supplementary Table CLXII. Unconventional magnons in altermagnet $Li_2Ni(SO_4)_2$.

| ID | Formula | SG | $T_C$ (K) | $U$ (eV) | Moment ($\mu_B$) | $E_{gap}$ (eV) |
|---|---|---|---|---|---|---|
| 0.714 | $Li_2Ni(SO_4)_2$ | 14 ($P2_1/c$) | 15 | 2.5 (Ni 3$d$)[90] | 1.72 (2.29) | 2.84 |
| SSG | SWP | Chiral magnons | | | MSG | Spin axis |
| 2.14.1.1 ($P^{\bar{1}}2_1/^{\bar{1}}c^{\infty m}1$) | Ni 2a | GP:($u$,$v$,$w$) | | | 14.75 ($P2_1/c$) | [100] |
| Magnetic structure | | | Magnon dispersion | | | |
| | | | | | | |

Supplementary Table CLXIII. Unconventional magnons in altermagnet $Mn_2SeO_3F_2$.

| ID | Formula | SG | $T_C$ (K) | $U$ (eV) | Moment ($\mu_B$) | $E_{gap}$ (eV) |
|---|---|---|---|---|---|---|
| 0.755 | $Mn_2SeO_3F_2$ | 62 ($Pnma$) | 26 | 3.6 (Mn 3$d$)[91] | 4.64 (4.34) | 3.76 |
| SSG | SWP | Chiral magnons | | | MSG | Spin axis |
| 14.62.1.4 ($P^{1}n^{\bar{1}}m^{\bar{1}}a^{\infty m}1$) | Mn 8d | K:(0,$v$,$w$) L:(1/2,$v$,$w$) GP:($u$,$v$,$w$) | | | 62.448 ($Pn'ma'$) | [001] |
| Magnetic structure | | | Magnon dispersion | | | |
| | | | | | | |

Supplementary Table CLXIV. Unconventional magnons in altermagnet $CeFeO_3$.

| ID | Formula | SG | $T_C$ (K) | $U$ (eV) | Moment ($\mu_B$) | $E_{gap}$ (eV) |
|---|---|---|---|---|---|---|
| 0.757, 0.758 | $CeFeO_3$ | 62 ($Pnma$) | 720 | 4.0 (Fe $3d$)[28] | 4.12 (4.05) | 2.34 |
| SSG | SWP | Chiral magnons | | | MSG | Spin axis |
| 14.62.1.4 ($P^{1}n^{\bar{1}}m^{\bar{1}}a^{\infty m}1$) | Fe 4b | K:(0,$v$,$w$) L:(1/2,$v$,$w$) GP:($u$,$v$,$w$) | | | 62.448 ($Pn'ma'$) | [100] |
| Magnetic structure | | Magnon dispersion | | | | |

For 0.758, $T_C$ = 220K, the spin axis is [010] and the MSG is 62.441 ($Pnma$).

Supplementary Table CLXV. Unconventional magnons in altermagnet $FeOHSO_4$.

| ID | Formula | SG | $T_C$ (K) | $U$ (eV) | Moment ($\mu_B$) | $E_{gap}$ (eV) |
|---|---|---|---|---|---|---|
| 0.760 | $FeOHSO_4$ | 15 ($C2/c$) | 125 | 4.0 (Fe $3d$)[28] | 4.29 (4.09) | 2.28 |
| SSG | SWP | Chiral magnons | | | MSG | Spin axis |
| 2.15.1.1 ($C^{\bar{1}}2/^{\bar{1}}c^{\infty m}1$) | Fe 4c | L:(1/2,1/2,1/2) V:(1/2,1/2,0) GP:($u$,$v$,$w$) | | | 15.89 ($C2'/c'$) | [010] |
| Magnetic structure | | Magnon dispersion | | | | |

Supplementary Table CLXVI. Unconventional magnons in altermagnet $NdVO_3$.

| ID | Formula | SG | $T_C$ (K) | $U$ (eV) | Moment ($\mu_B$) | $E_{gap}$ (eV) |
|---|---|---|---|---|---|---|
| 0.786 | $NdVO_3$† | 62 ($Pnma$) | 135 | 0 | 1.60 (1.06) | 0 |
| SSG | SWP | Chiral magnons | | | MSG | Spin axis |
| 11.62.1.1 ($P^{\bar{1}}n^{1}m^{\bar{1}}a^{\infty m}1$) | V 4b | M:($u$,0,$w$) N:($u$,1/2,$w$) GP:($u$,$v$,$w$) | | | 11.54 ($P2_1'/m'$) | [010] +31° ([100]) |
| Magnetic structure | | | Magnon dispersion | | | |

Supplementary Table CLXVII. Unconventional magnons in altermagnet $YVO_3$.

| ID | Formula | SG | $T_C$ (K) | $U$ (eV) | Moment ($\mu_B$) | $E_{gap}$ (eV) |
|---|---|---|---|---|---|---|
| 0.787 | $YVO_3$ | 62 ($Pnma$) | 77 | 2.85 (V $3d$)[31] | 1.79 (1.72) | 1.56 |
| SSG | SWP | Chiral magnons | | | MSG | Spin axis |
| 14.62.1.4 ($P^{1}n^{\bar{1}}m^{\bar{1}}a^{\infty m}1$) | V 4b | K:(0,$v$,$w$) L:(1/2,$v$,$w$) GP:($u$,$v$,$w$) | | | 62.446 ($Pn'm'a$) | [001] |
| Magnetic structure | | | Magnon dispersion | | | |

Supplementary Table CLXVIII. Unconventional magnons in altermagnet MnTe.

| ID | Formula | SG | $T_C$ (K) | $U$ (eV) | Moment ($\mu_B$) | $E_{gap}$ (eV) |
|---|---|---|---|---|---|---|
| 0.800 | MnTe | 194 ($P6_3/mmc$) | 323 | 4.8 (Mn $3d$)[92] | 4.53 (4.60) | 0.80 |
| SSG | SWP | Chiral magnons | | | MSG | Spin axis |
| 164.194.1.1 ($P^{\bar{1}}6_3/^{\bar{1}}m^{1}m^{\bar{1}}c^{\infty m}1$) | Mn 2a | D:($u$,0,$w$) GP:($u$,$v$,$w$) | | | 63.457 ($Cmcm$) | [110] |
| Magnetic structure | | | Magnon dispersion | | | |

Supplementary Table CLXIX. Unconventional magnons in altermagnet $CuFeS_2$.

| ID | Formula | SG | $T_C$ (K) | $U$ (eV) | Moment ($\mu_B$) | $E_{gap}$ (eV) |
|---|---|---|---|---|---|---|
| 0.802 | $CuFeS_2$ | 122 ($I\bar{4}2d$) | 823 | 4.7 (Fe $3d$)[93] | 3.74 (3.82) | 0 |
| SSG | SWP | Chiral magnons | | | MSG | Spin axis |
| 82.122.1.1 ($I^{1}\bar{4}^{\bar{1}}2^{\bar{1}}d^{\infty m}1$) | Fe 4b | C:($u$,$v$,0) GP:($u$,$v$,$w$) | | | [001] | 122.333 ($I\bar{4}2d$) |
| Magnetic structure | | Magnon dispersion | | | | |

Supplementary Table CLXX. Unconventional magnons in altermagnet $Fe_2WO_6$.

| ID | Formula | SG | $T_C$ (K) | $U$ (eV) | Moment ($\mu_B$) | $E_{gap}$ (eV) |
|---|---|---|---|---|---|---|
| 0.811, 0.813 | $Fe_2WO_6$ | 60 ($Pbcn$) | 240 | 4.0 (Fe 3$d$)<br>4.0 (W 4$d$)[94] | 4.25 (3.60)<br>4.20 (4.50) | 1.52 |
| SSG | SWP | Chiral magnons | | | MSG | Spin axis |
| 13.60.1.1 ($^{\bar{1}}P^{\bar{1}}b^{1}c^{\bar{1}}n^{\infty m}1$) | Fe 4e, 4e | M:($u$,0,$w$) N:($u$,1/2,$w$) GP:($u$,$v$,$w$) | | | 60.423 ($Pbc'n'$) | [001] |
| Magnetic structure | | Magnon dispersion | | | | |

Supplementary Table CLXXI. Unconventional magnons in altermagnet $Sr_2MnGaO_5$.

| ID | Formula | SG | $T_C$ (K) | $U$ (eV) | Moment ($\mu_B$) | $E_{gap}$ (eV) |
|---|---|---|---|---|---|---|
| 0.823 | $Sr_2MnGaO_5$ | 62 ($Pnma$) | 180 | 3.0<br>(Mn 3$d$)[95] | 3.69 (3.16) | 1.14 |
| SSG | SWP | Chiral magnons | | | MSG | Spin axis |
| 5.46.1.1 ($^{\bar{1}}I^{\bar{1}}m^{\bar{1}}a^{1}2^{\infty m}1$) | Mn 4c | T:(1/2,1/2,0) P:(1/2,1/2,$w$) GP:($u$,$v$,$w$) | | | 46.243 ($Im'a2'$) | [100] |
| Magnetic structure | | Magnon dispersion | | | | |

*The band is quasi-doubly degenerate.

Supplementary Table CLXXII. Unconventional magnons in altermagnet $Ca_2MnGaO_5$.

| ID | Formula | SG | $T_C$ (K) | $U$ (eV) | Moment ($\mu_B$) | $E_{gap}$ (eV) |
|---|---|---|---|---|---|---|
| 0.825 | $Ca_2MnGaO_5$ | 62 ($Pnma$) | 160 | 3.0 (Mn $3d$)[95] | 3.74 (3.60) | 0.30 |
| SSG | SWP | Chiral magnons | | | MSG | Spin axis |
| 14.62.1.1 ($P^{\bar{1}}n^{\bar{1}}m^{1}a^{\infty m}1$) | Mn 4a | V:$(u,v,0)$ W:$(u,v,1/2)$ GP:$(u,v,w)$ | | | 62.447 ($Pnm'a'$) | [010] |
| Magnetic structure | | | Magnon dispersion | | | |

*The band is quasi-doubly degenerate.

Supplementary Table CLXXIII. Unconventional magnons in altermagnet $DyFeO_3$.

| ID | Formula | SG | $T_C$ (K) | $U$ (eV) | Moment ($\mu_B$) | $E_{gap}$ (eV) |
|---|---|---|---|---|---|---|
| 0.836, 0.837, 0.838, 0.839, 0.840, 0.841 | $DyFeO_3$ | 62 ($Pnma$) | 650 | 3.0 (Fe $3d$)[96] | 4.06 (4.0) | 1.95 |
| SSG | SWP | Chiral magnons | | | MSG | Spin axis |
| 14.62.1.4 ($P^{1}n^{\bar{1}}m^{\bar{1}}a^{\infty m}1$) | Fe 4b | K:$(0,v,w)$ L:$(1/2,v,w)$ GP:$(u,v,w)$ | | | 62.448 ($Pn'ma'$) | [100] |
| Magnetic structure | | Magnon dispersion | | | | |

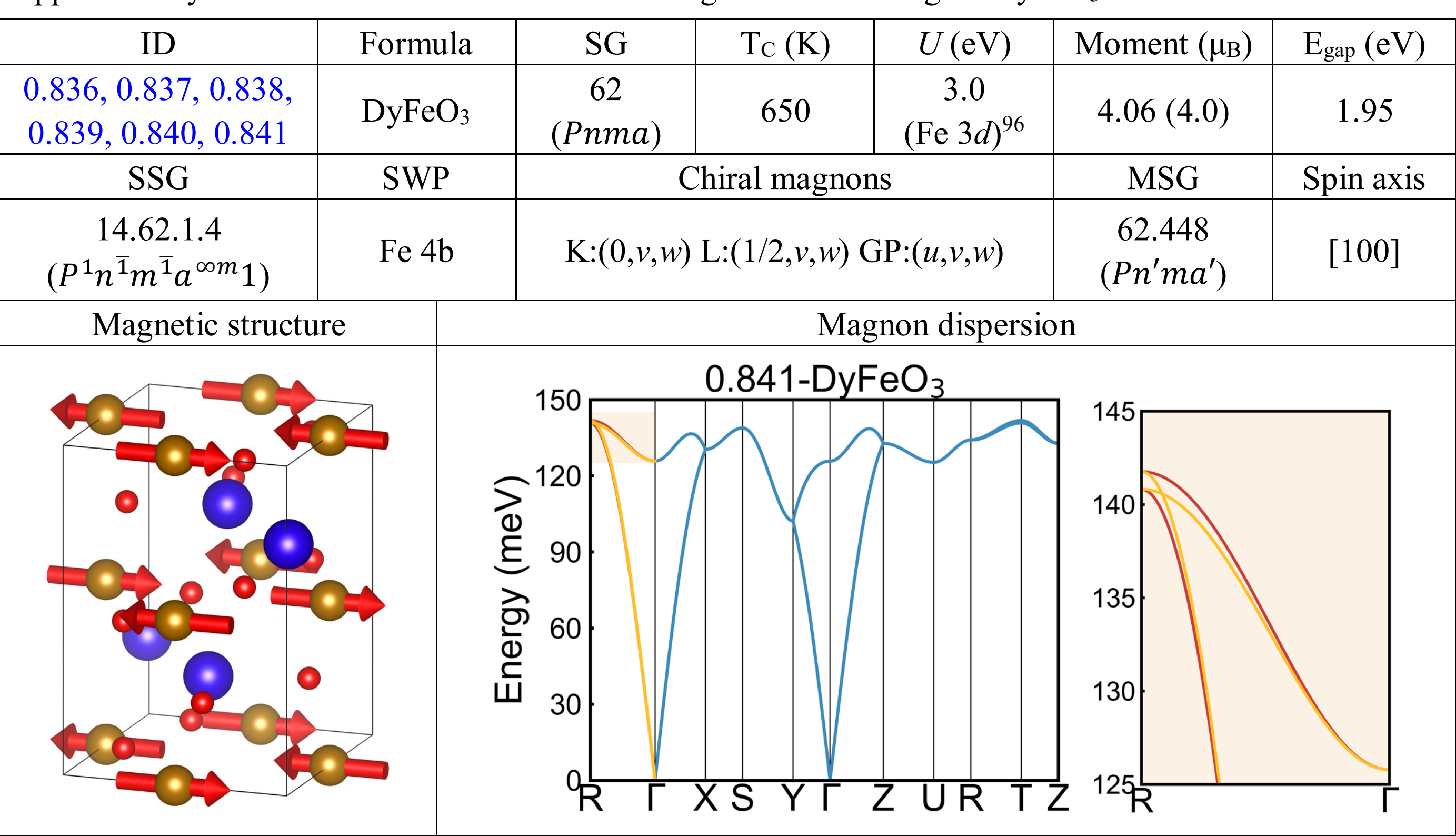

For 0.840, 0.841, the spin axis is [010] and the MSG is 62.441 ($Pnma$).

Supplementary Table CLXXIV. Unconventional magnons in altermagnet $NiCrO_4$.

| ID | Formula | SG | $T_C$ (K) | $U$ (eV) | Moment ($\mu_B$) | $E_{gap}$ (eV) |
| --- | --- | --- | --- | --- | --- | --- |
| 0.896 | $NiCrO_4$ | 63 ($Cmcm$) | | 5.1 (Ni 3$d$)[26] | 1.71 (0.64) | 1.81 |
| SSG | SWP | Chiral magnons | | | MSG | Spin axis |
| 12.63.1.1 ($C^{1}m^{\bar{1}}c^{\bar{1}}m^{\infty m}1$) | Ni 4a | K:(0,$v$,$w$) GP:($u$,$v$,$w$) | | | 63.457 ($Cmcm$) | [100] |
| Magnetic structure | | Magnon dispersion | | | | |
| | | *The band is quasi-doubly degenerate. | | | | |

Supplementary Table CLXXV. Unconventional magnons in altermagnet $LuVO_3$.

| ID | Formula | SG | $T_C$ (K) | $U$ (eV) | Moment ($\mu_B$) | $E_{gap}$ (eV) |
| --- | --- | --- | --- | --- | --- | --- |
| 0.984 | $LuVO_3$† | 62 ($Pnma$) | 82 | 2.85 (V 3$d$)[31] | 1.80 (1.20) | 1.51 |
| SSG | SWP | Chiral magnons | | | MSG | Spin axis |
| 14.62.1.4 ($P^{1}n^{\bar{1}}m^{\bar{1}}a^{\infty m}1$) | V 4b | K:(0,$v$,$w$) L:(1/2,$v$,$w$) GP:($u$,$v$,$w$) | | | 62.446 ($Pn'm'a$) | [001] |
| Magnetic structure | | Magnon dispersion | | | | |

Supplementary Table CLXXVI. Unconventional magnons in altermagnet $SrMnO_3$.

| ID | Formula | SG | $T_C$ (K) | $U$ (eV) | Moment ($\mu_B$) | $E_{gap}$ (eV) |
|---|---|---|---|---|---|---|
| 0.1018, 0.1019 | $SrMnO_3$ | 20 ($C222_1$) | 278 | 2.1 (Mn 3$d$) | 2.76 (2.27) | 1.93 |
| SSG | SWP | Chiral magnons | | | MSG | Spin axis |
| 4.20.1.1 ($C^{\bar{1}}2^{\bar{1}}2^{1}2_1{}^{\infty m}1$) | Mn 8c | R:(1/2,1/2,1/2) S:(1/2,1/2,0) D:(1/2,1/2,$w$) P:($u$,$v$,0) Q:($u$,$v$,1/2) GP:($u$,$v$,$w$) | | | 20.34 ($C22'2_1'$) | [100] |
| Magnetic structure | | | Magnon dispersion | | | |
| | | | *Altermagnetic two-fold nodal plane ($T_2$-R) | | | |

Supplementary Table CLXXVII. Unconventional magnons in altermagnet $MnSe_2$.

| ID | Formula | SG | $T_C$ (K) | $U$ (eV) | Moment ($\mu_B$) | $E_{gap}$ (eV) |
|---|---|---|---|---|---|---|
| 1.0.47 | $MnSe_2$† | 205 ($Pa\bar{3}$) | 49 | 3.0 (Mn 3$d$)[97] | 4.45 (1.0) | 0.77 |
| SSG | SWP | Chiral magnons | | | MSG | Spin axis |
| 14.61.1.1 ($P^{\bar{1}}b^{\bar{1}}c^{1}a^{\infty m}1$) | Mn 4a, 8e | V:($u$,$v$,0) W:($u$,$v$,1/2) GP:($u$,$v$,$w$) | | | 61.433 ($Pbca$) | [100] |
| Magnetic structure | | | Magnon dispersion | | | |

Supplementary Table CLXXVIII. Unconventional magnons in altermagnet $CrVO_4$.

| ID | Formula | SG | $T_C$ (K) | $U$ (eV) | Moment ($\mu_B$) | $E_{gap}$ (eV) |
|---|---|---|---|---|---|---|
| 1.522 | $CrVO_4$ | 63 ($Cmcm$) | 50 | 3.0 (Cr $3d$)[98] | 2.89 (2.11) | 2.62 |
| SSG | SWP | Chiral magnons | | | MSG | Spin axis |
| 12.63.1.1 $(C^{1}m^{\bar{1}}c^{\bar{1}}m^{\infty m}1)$ | Cr 4a | V:$(u,v,0)$ W:$(u,v,1/2)$ GP:$(u,v,w)$ | | | 2.4 ($P\bar{1}$) | [210] +9° ([001]) |
| Magnetic structure | | | Magnon dispersion | | | |

*The band is quasi-doubly degenerate.

Supplementary Table CLXXIX. Unconventional magnons in altermagnet $CuF_2$.

| ID | Formula | SG | $T_C$ (K) | $U$ (eV) | Moment ($\mu_B$) | $E_{gap}$ (eV) |
|---|---|---|---|---|---|---|
| | $CuF_2$† | 14 ($P2_1/c$) | 69 | 9.0 (Cu $3d$) | 0.77 (0.73) | 2.02 |
| SSG | SWP | Chiral magnons | | | MSG | Spin axis |
| 2.14.1.1 $(P^{\bar{1}}2_1/^{\bar{1}}c^{\infty m}1)$ | Cu 2b | GP:$(u,v,w)$ | | | 14.79 ($P2_1'/c'$) | [010] |
| Magnetic structure | | | Magnon dispersion | | | |

Supplementary Table CLXXX. Unconventional magnons in altermagnet $NiF_2$.

<table>
<tr><td>ID</td><td>Formula</td><td>SG</td><td>$T_C$ (K)</td><td>$U$ (eV)</td><td>Moment ($\mu_B$)</td><td>$E_{gap}$ (eV)</td></tr>
<tr><td></td><td>$NiF_2$†</td><td>136<br>($P4_2/mnm$)</td><td>74</td><td>5.3<br>(Ni 3$d$)[99]</td><td>1.81 (2.0)</td><td>4.51</td></tr>
<tr><td>SSG</td><td>SWP</td><td colspan="3">Chiral magnons</td><td>MSG</td><td>Spin axis</td></tr>
<tr><td>65.136.1.1<br>($P^{\bar{1}}4_2/{}^{1}m^{\bar{1}}n^{1}m^{\infty m}1$)</td><td>Ni 2a</td><td colspan="3">S:($u$,$u$,1/2) Σ:($u$,$u$,0) C:($u$,$u$,$w$)<br>D:($u$,$v$,0) E:($u$,$v$,1/2) GP:($u$,$v$,$w$)</td><td>58.398<br>($Pnn'm'$)</td><td>[100]</td></tr>
<tr><td colspan="3">Magnetic structure</td><td colspan="4">Magnon dispersion</td></tr>
<tr><td colspan="7">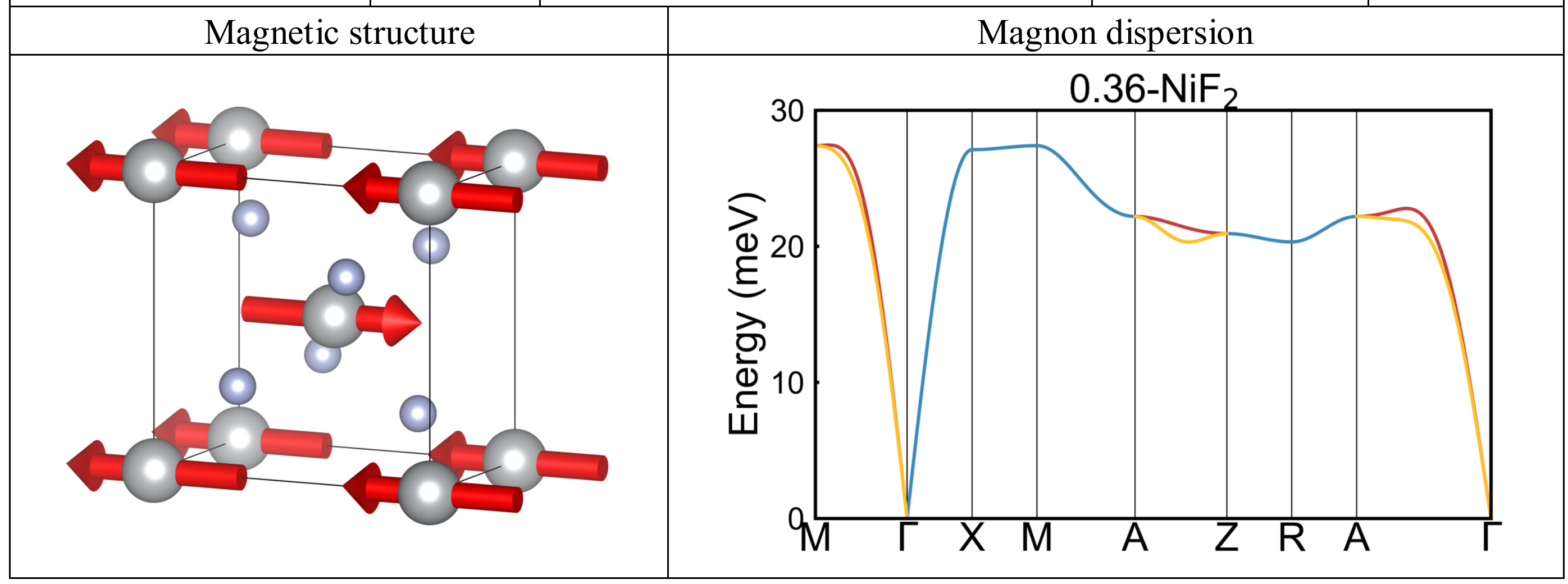
</td></tr>
</table>

## S6. Calculated bilinear exchange interactions

This section provides various bilinear exchange interactions of candidate materials obtained from high-throughput calculations, including the Heisenberg term ($J$), Dzyaloshinskii-Moriya interaction ($D$), exchange anisotropy ($J^{ani}$), and off-diagonal symmetric anisotropy ($\Gamma$). The subscripts of the exchange interactions are labeled in ascending order of bond length. For bonds of the same length but not equivalent (or symmetry-connected), additional letters are introduced to identify them. Furthermore, we also provide the magnetic atoms and the lattice translation vectors corresponding to each exchange interaction. The truncation of exchange interactions is not chosen based on a fixed distance but is performed when a certain exchange parameter and all subsequent exchange parameters are less than 1000 times smaller than the largest Heisenberg exchange parameters, or numerically less than 0.001 meV. The $U$ parameters in DFT calculations are also provided in Supplementary Table CLXXXI.

The method for evaluating the SOC-related exchange interactions is as follows: for $D$= ($D_x$, $D_y$, $D_z$), we calculate the absolute value, and for $J^{ani}$ = ($J_{xx}$, $J_{yy}$, $J_{zz}$), we calculate the difference between the maximum and minimum values of the three. For $\Gamma$ = ($J_{xy}$, $J_{xz}$, $J_{yz}$), we find the term with the largest absolute value. Then, in Supplementary Table CLXXXI, we provide the largest one for each SOC-induced exchange interaction. Finally, we compare these with the largest Heisenberg term, to determine if their values are greater than 10%. The terms that meet this criterion are bolded in the table and are marked with a † symbol after the chemical formula of material. Additionally, it should be noted that according to the TB2J method, positive J tends to ferromagnetic alignment, while negative J tends to antiferromagnetic alignment, and the spin vector S is normalized in the TB2J calculation, meaning that in the tight-binding model calculations of the magnon band structures using linear spin wave theory, S is set to 1.

Supplementary Table CLXXXI. Detailed information about the exchange interactions (in meV), where $J$, $D$, $J^{ani}$ and $\Gamma$ represent the Heisenberg, Dzyaloshinskii-Moriya, diagonal exchange anisotropy and off-diagonal symmetric anisotropic interactions, respectively.

| ID | Formula | magnetic ions | $U$ (eV) | exchange interaction (meV) |
|---|---|---|---|---|
| 0.275 | $Mn_3AlN$ | Mn1 (0.0, 0.5, 0.5)<br>Mn2 (0.5, 0.0, 0.5)<br>Mn3 (0.5, 0.5, 0.0) | 0 | $J_1$=0.118, $D_1$= (0, 0, 0.050),<br>$\Gamma_1$=(-0.002, 0, 0)<br>[Mn1-Mn2 (-1,0,0)]<br>$J_{2a}$=4.151, $J_{2a}^{ani}$= (-0.038, 0.007, 0.007)<br>[Mn1-Mn1 (-1,0,0)]<br>$J_{2b}$=2.187 [Mn1-Mn1 (0,-1,0)]<br>$J_3$=0.476 [Mn1-Mn2 (-1,0,-1)] |
| | | | | $J_{4a}$=-0.354 [Mn1-Mn1 (-1,-1,0)]<br>$J_{4b}$=0.668 [Mn1-Mn1 (0,-1,-1)]<br>$J_5$=-0.292 [Mn1-Mn2 (-1,-1,0)]<br>$J_6$=-0.059 [Mn1-Mn1 (-1,-1,-1)]<br>$J_7$=-0.102 [Mn1-Mn2 (-1,-1,-1)]<br>$J_{8a}$=0.368 [Mn1-Mn1 (-2,0,0)]<br>$J_{8b}$=-0.026 [Mn1-Mn1 (0,-2,0)]<br>$J_{9a}$=0.081 [Mn1-Mn2 (-1,0,-2)]<br>$J_{9b}$=0.525 [Mn1-Mn2 (-2,-1,0)] |
| | $Lu_2V_2O_7$ | V1 (0.625, 0.625, 0.125)<br>V2 (0.625, 0.875, 0.375)<br>V3 (0.875, 0.875,0.125)<br>V4 (0.875, 0.625, 0.375) | 3.1<br>(V 3*d*) | $J_1$=0.863, $D_1$= (0, 0.049, -0.049)<br>$J_1^{ani}$= (-0.012, -0.011, -0.011),<br>$\Gamma_1$=(0, 0, 0.001)<br>[V1-V2 (0,0,0)]<br>$J_2$=-0.013 [V1-V2 (-1/2,0,-1/2)]<br>$J_{3a}$=0.036 [V1-V1 (0,-1/2,-1/2)]<br>$J_{3b}$=0.013 [V1-V1 (0,-1/2,1/2)] |
| 0.35 | $Cu_2OSeO_3$[†] | Cu1 (0.886,0.886,0.886)<br>Cu2 (0.614,0.114,0.386)<br>Cu3 (0.386,0.614,0.114)<br>Cu4 (0.114,0.386,0.614)<br>Cu5 (0.133,0.121,0.872)<br>Cu6 (0.872,0.133,0.121)<br>Cu7 (0.121,0.872,0.133)<br>Cu8 (0.366,0.879,0.372)<br>Cu9 (0.372,0.366,0.879)<br>Cu10 (0.879,0.372,0.366)<br>Cu11 (0.867,0.621,0.628)<br>Cu12 (0.628,0.867,0.621)<br>Cu13 (0.621,0.628,0.867)<br>Cu14 (0.633,0.379,0.128)<br>Cu15 (0.128,0.633,0.379)<br>Cu16 (0.379,0.128,0.633) | 6.5<br>(Cu 3*d*) | $J_1$=-1.187, **$D_1$= (0.239, -0.354, 0.058)**<br>$J_1^{ani}$= (-0.007, -0.013, -0.005),<br>$\Gamma_1$=(0.002, -0.003, -0.002)<br>[Cu1-Cu5 (1,1,0)]<br>$J_2$=0.183 [Cu5-Cu9 (0,0,0)]<br>$J_3$=0.001 [Cu5-Cu6 (-1,0,1)]<br>$J_4$=-0.486 [Cu1-Cu11 (0,0,0)]<br>$J_5$=0.002 [Cu5-Cu8 (0,-1,0)]<br>$J_6$=-0.039 [Cu1-Cu14 (0,1,1)]<br>$J_7$=0.006 [Cu5-Cu10 (-1,0,1)]<br>$J_8$=0.072 [Cu1-Cu2 (0,1,0)]<br>$J_9$=-0.004 [Cu1-Cu14 (0,0,1)]<br>$J_{10}$=-0.013 [Cu5-Cu11 (-1,-1,0)]<br>$J_{11}$=-0.058 [Cu5-Cu14 (-1,0,1)]<br>$J_{12}$=0.094 [Cu5-Cu10 (-1,0,0)]<br>$J_{13}$=-0.040 [Cu1-Cu8 (1,0,1)]<br>$J_{14}$=0.006 [Cu5-Cu13 (0,-1,0)]<br>$J_{15}$=-0.044 [Cu5-Cu13 (0,0,0)]<br>$J_{16}$=0.002 [Cu1-Cu8 (1,0,0)]<br>$J_{17}$=-0.032 [Cu5-Cu13 (-1,-1,0)]<br>$J_{18}$=-0.580 [Cu1-Cu8 (0,0,1)] |
| 0.613,<br>0.614,<br>0.615 | $FeCr_2S_4$ | Fe1 (0.125,0.125,0.125)<br>Fe2 (0.375,0.375,0.875)<br>Fe3 (0.625,0.625,0.125)<br>Fe4 (0.375,0.875,0.375)<br>Fe5 (0.125,0.625,0.625)<br>Fe6 (0.875,0.375,0.375)<br>Fe7 (0.625,0.125,0.625) | 3.8<br>(Fe 3*d*)<br>4.0<br>(Cr 3*d*) | $J_1$=0.372 [Cr1-Cr4 (0,-1/2,1/2)]<br>$J_2$=1.303, $D_2$= (0.052, -0.038, -0.013)<br>$J_2^{ani}$= (-0.01, -0.009, -0.013),<br>$\Gamma_2$=(0.001, 0.001, 0.001)<br>[Cr1-Fe3 (1/2,1/2,0)]<br>$J_3$=-0.220 [Fe1-Fe2 (-1/2,0,-1/2)]<br>$J_4$=0.092 [Cr1-Cr7 (1/2,-1/2,0)] |
| | | Fe8 (0.875,0.875,0.875)<br>Cr1 (0.5,0.5,0.5)<br>Cr2 (0.75,0,0.25)<br>Cr3 (0,0.75,0.25)<br>Cr4 (0.25,0.25,0.5)<br>Cr5 (0,0.5,0)<br>Cr6 (0.5,0.25,0.25)<br>Cr7 (0.75,0.25,0)<br>Cr8 (0.25,0.75,0)<br>Cr9 (0.5,0,0)<br>Cr10 (0.25,0.5,0.25)<br>Cr11 (0.25,0,0.75)<br>Cr12 (0.5,0.75,0.75)<br>Cr13 (0,0,0.5)<br>Cr14 (0,0.25,0.75)<br>Cr15 (0.75,0.75,0.5)<br>Cr16 (0.75,0.5,0.75) | | $J_{5a}$=0.090 [Cr1-Fe1 (0,0,0)]<br>$J_{5b}$=-0.109 [Cr1-Fe3 (1/2,1/2,1)]<br>$J_{6a}$=-0.592 [Cr1-Cr5 (-1/2,0,-1/2)]<br>$J_{6b}$=-0.344 [Cr1-Cr5 (-1/2,0,1/2)]<br>$J_{6c}$=-0.021 [Fe1-Fe3 (-1/2,-1/2,0)]<br>$J_7$=0.090 [Cr1-Cr4 (0,1/2,1/2)]<br>$J_8$=0.007 [Cr1-Fe1 (0,0,1)]<br>$J_9$=-0.143 [Fe1-Fe2 (-3/2,0,-1/2)] |
| 0.672 | $CaCu_3Fe_2$<br>$Sb_2O_{12}$ | Cu1 (0.25,0.75,0.75)<br>Cu2 (0.75,0.25,0.25)<br>Cu3 (0.75,0.25,0.75)<br>Cu4 (0.25,0.75,0.25)<br>Cu5 (0.75,0.75,0.25)<br>Cu6 (0.25,0.25,0.75)<br>Fe1 (0,0,0)<br>Fe2 (0.5,0.5,0)<br>Fe3 (0.5,0,0.5)<br>Fe4 (0,0.5,0.5) | 7.0<br>(Cu 3*d*)<br>4.0<br>(Fe 3*d*) | $J_1$=-4.106, $D_1$= (0.067, -0.018, 0.026)<br>$J_1^{ani}$= (-0.013, -0.012, -0.019),<br>$\Gamma_1$=(0.001, 0.004, 0)<br>[Cu1-Fe1 (0,1,1)]<br>$J_2$=0.000 [Cu1-Cu4 (0,0,0)]<br>$J_3$=0.022 [Cu1-Cu3 (-1,0,0)]<br>$J_{4a}$=-1.336 [Fe1-Fe2 (-1,-1,0)]<br>$J_{4b}$=-2.337 [Fe1-Fe2 (-1,0,0)]<br>$J_{5a}$=0.015 [Cu1-Fe1 (0,0,1)]<br>$J_{5b}$=-0.343 [Cu1-Fe1 (0,1,0)]<br>$J_{5c}$=-0.781 [Cu1-Fe1 (1,1,1)]<br>$J_{6a}$=0.020 [Cu1-Cu2 (-1,0,0)]<br>$J_{6b}$=0.013 [Cu1-Cu2 (-1,0,1)]<br>$J_{7a}$=0.006 [Cu1-Cu1 (0,0,-1)]<br>$J_{7b}$=0.000 [Cu1-Cu1 (-1,0,0)]<br>$J_{7c}$=-0.000 [Cu1-Cu1 (0,-1,0)]<br>$J_{7d}$=0.124 [Fe1-Fe1 (-1,0,0)] |
| | $LiFe_5O_8$ | Fe1 (0.995,0.995,0.005)<br>Fe2 (0.505,0.005,0.505)<br>Fe3 (0.005,0.495,0.495)<br>Fe4 (0.495,0.505,0.995)<br>Fe5 (0.245,0.745,0.245)<br>Fe6 (0.255,0.255,0.745)<br>Fe7 (0.745,0.755,0.755)<br>Fe8 (0.755,0.245,0.255)<br>Fe9 (0.125,0.369,0.119)<br>Fe10 (0.375,0.631,0.619) | 4.0<br>(Fe 3*d*) | $J_1$=-1.594 [Fe9-Fe14 (0,0,0)]<br>$J_2$=-9.443, $D_2$= (0.183, 0.177, 0.019)<br>[Fe1-Fe9 (1,1,0)]<br>$J_3$=-11.975, $J_3^{ani}$= (-0.108, -0.124, -0.126),<br>$\Gamma_3$=(0.011, -0.003, 0.002)<br>[Fe1-Fe12 (0,1,-1)]<br>$J_4$=-8.811 [Fe1-Fe11 (0,0,0)]<br>$J_5$=0.284 [Fe1-Fe5 (1,0,0)]<br>$J_6$=0.150 [Fe1-Fe6 (1,1,-1)]<br>$J_7$=-0.006 [Fe9-Fe13 (-1,0,-1)] |
| | | Fe11 (0.875,0.869,0.381)<br>Fe12 (0.625,0.131,0.881)<br>Fe13 (0.881,0.125,0.631)<br>Fe14 (0.381,0.375,0.369)<br>Fe15 (0.619,0.875,0.131)<br>Fe16 (0.119,0.625,0.869)<br>Fe17 (0.369,0.881,0.875)<br>Fe18 (0.631,0.381,0.625)<br>Fe19 (0.869,0.619,0.125)<br>Fe20 (0.131,0.119,0.375) | | $J_8$=0.112 [Fe9-Fe12 (-1,0,-1)]<br>$J_{9a}$=-0.034 [Fe9-Fe10 (0,0,-1)]<br>$J_{9b}$=-0.078 [Fe9-Fe13 (-1,0,0)]<br>$J_{10}$=0.081 [Fe1-Fe11 (0,0,-1)]<br>$J_{11}$=0.023 [Fe1-Fe9 (1,0,0)]<br>$J_{12}$=-0.021 [Fe1-Fe10 (1,0,-1)]<br>$J_{13}$=0.049 [Fe1-Fe12 (1,1,-1)]<br>$J_{14}$=-0.145 [Fe9-Fe15 (0,-1,0)]<br>$J_{15}$=-0.683 [Fe1-Fe2 (0,1,-1)]<br>$J_{16}$=0.294 [Fe9-Fe15 (-1,-1,0)]<br>$J_{17}$=-0.220 [Fe1-Fe2 (1,1,-1)]<br>$J_{18}$=-0.264 [Fe9-Fe15 (-1,0,0)]<br>$J_{19}$=-0.001 [Fe9-Fe14 (-1,0,0)]<br>$J_{20}$=-0.013 [Fe9-Fe14 (0,0,-1)]<br>$J_{21}$=0.001 [Fe1-Fe10 (1,0,0)]<br>$J_{22}$=0.009 [Fe1-Fe10 (0,0,-1)]<br>$J_{23}$=0.162 [Fe1-Fe5 (0,0,0)]<br>$J_{24}$=0.230 [Fe1-Fe6 (0,1,-1)] |
| 0.24 | $LiMnPO_4$ | Mn1 (0.280,0.25,0.976)<br>Mn2 (0.780,0.25,0.524)<br>Mn3 (0.720,0.75,0.024)<br>Mn4 (0.220,0.75,0.476) | 3.92<br>(Mn 3*d*) | $J_1$=-1.187, $D_1$= (0.022, -0.016, -0.026)<br>$J_1^{ani}$= (-0.003, -0.003, -0.002),<br>$\Gamma_1$=(0, 0, 0.001)<br>[Mn1-Mn4 (0,-1,0)]<br>$J_2$=0.218 [Mn1-Mn1 (0,0,-1)]<br>$J_3$=0.475 [Mn1-Mn3 (0,-1,1)]<br>$J_4$=0.476 [Mn1-Mn2 (-1,0,0)]<br>$J_5$=0.166 [Mn1-Mn2 (-1,0,1)]<br>$J_6$=-0.079 [Mn1-Mn1 (0,-1,0)]<br>$J_7$=-0.010 [Mn1-Mn3 (-1,-1,1)]<br>$J_8$=0.021 [Mn1-Mn3 (0,-1,0)]<br>$J_9$=0.027 [Mn1-Mn3 (0,-1,2)]<br>$J_{10}$=0.018 [Mn1-Mn1 (0,-1,-1)]<br>$J_{11}$=0.007 [Mn1-Mn4 (0,-1,-1)]<br>$J_{12}$=0.017 [Mn1-Mn3 (-1,-1,0)] |
| 0.75,<br>0.144 | $Cr_2WO_6$ | Cr1 (0,0,0.334)<br>Cr2 (0.5,0.5,0.166)<br>Cr3 (0,0,0.666)<br>Cr4 (0.5,0.5,0.834) | 4.0<br>(Cr 3*d*) | $J_1$=-1.874, $J_1^{ani}$= (-0.004, -0.004, 0.001),<br>$\Gamma_1$=(-0.003, 0, 0)<br>[Cr1-Cr3 (0,0,0)]<br>$J_2$=1.365 [Cr1-Cr2 (-1,-1,0)]<br>$J_3$=-0.286, $D_3$= (0.010, 0.046, 0)<br>[Cr1-Cr1 (-1,0,0)]<br>$J_4$=-0.421 [Cr1-Cr3 (-1,0,0)]<br>$J_{5a}$=-0.161 [Cr1-Cr4 (-1,-1,0)]<br>$J_{5b}$=-0.297 [Cr1-Cr4 (-1,-1,-1)]<br>$J_6$=-0.303 [Cr1-Cr3 (0,0,-1)]<br>$J_{7a}$=-0.187 [Cr1-Cr1 (-1,-1,0)] |

| | | | | $J_{7b}$=-0.424 [Cr1-Cr1 (-1,1,0)]<br>$J_{8a}$=-0.357 [Cr1-Cr3 (-1,-1,0)]<br>$J_{8b}$=-0.208 [Cr1-Cr3 (-1,1,0)] |
|---|---|---|---|---|
| 0.87 | $NaFePO_4$ | Fe1 (0.213,0.25,0.486)<br>Fe2 (0.713,0.25,0.014)<br>Fe3 (0.787,0.75,0.514)<br>Fe4 (0.287,0.75,0.986) | 3.71<br>(Fe 3$d$) | $J_1$=-1.523, $D_1$= (0.029, 0.009, 0.021)<br>$J_1^{ani}$= (-0.005, -0.006, -0.004),<br>$\Gamma_1$=(-0.001, 0.001, 0)<br>[Fe1-Fe4 (0,-1,-1)]<br>$J_2$=-0.108 [Fe1-Fe1 (0,0,-1)]<br>$J_3$=-0.573 [Fe1-Fe3 (-1,-1,0)]<br>$J_4$=-0.001 [Fe1-Fe2 (-1,0,0)]<br>$J_5$=-0.078 [Fe1-Fe2 (-1,0,1)]<br>$J_6$=-0.561 [Fe1-Fe1 (0,-1,0)]<br>$J_7$=-0.003 [Fe1-Fe3 (0,-1,0)] |
| 0.161 | $CoSe_2O_5$ | Co1 (0,0.061,0.25)<br>Co2 (0.5,0.439,0.75)<br>Co3 (0,0.939,0.75)<br>Co4 (0.5,0.561,0.25) | 4.0<br>(Co 3$d$) | $J_1$=-1.077, $D_1$= (0.008, -0.044, -0.023)<br>$J_1^{ani}$= (-0.004, -0.005, -0.004),<br>$\Gamma_1$=(0.001, -0.002, 0.003)<br>[Co1-Co3 (0,-1,-1)]<br>$J_2$=-0.101 [Co1-Co2 (-1,0,-1)]<br>$J_3$=-0.011 [Co1-Co1 (0,0,-1)]<br>$J_4$=-0.491 [Co1-Co4 (-1,-1,0)]<br>$J_5$=-0.009 [Co1-Co1 (-1,0,0)]<br>$J_{6a}$=-0.012 [Co1-Co3 (-1,-1,-1)]<br>$J_{6b}$=-0.027 [Co1-Co3 (-1,-1,0)]<br>$J_7$=-0.004 [Co1-Co2 (-1,-1,-1)]<br>$J_8$=-0.001 [Co1-Co4 (-1,-1,-1)]<br>$J_{9a}$=-0.111 [Co1-Co1 (-1,0,-1)]<br>$J_{9b}$=0.000 [Co1-Co1 (-1,0,1)] |
| 0.193,<br>0.383,<br>0.384 | $LiCoPO_4$ | Co1 (0.279,0.25,0.979)<br>Co2 (0.779,0.25,0.521)<br>Co3 (0.721,0.75,0.021)<br>Co4 (0.221,0.75,0.479) | 5.05<br>(Co 3$d$) | $J_1$=-1.075, $D_1$= (0.015, -0.012, -0.008)<br>$J_1^{ani}$= (-0.006, -0.005, -0.006),<br>$\Gamma_1$=(0.001, 0, 0.001)<br>[Co1-Co4 (0,-1,0)]<br>$J_2$=-0.084 [Co1-Co1 (0,0,-1)]<br>$J_3$=-0.290 [Co1-Co3 (0,-1,1)]<br>$J_4$=-0.036 [Co1-Co2 (-1,0,0)]<br>$J_5$=-0.067 [Co1-Co2 (-1,0,1)]<br>$J_6$=-0.372 [Co1-Co1 (0,-1,0)]<br>$J_7$=-0.010 [Co1-Co3 (-1,-1,1)] |
| 0.217,<br>0.961,<br>0.962,<br>0.963,<br>0.964 | $LiCrGe_2O_6$ | Cr1 (0.251,0.659,0.212)<br>Cr2 (0.251,0.841,0.712)<br>Cr3 (0.749,0.159,0.288)<br>Cr4 (0.749,0.341,0.788) | 2.0<br>(Cr 3$d$) | $J_1$=0.094, $D_1$= (0.003, 0, -0.001)<br>$J_1^{ani}$= (-0.002, -0.001, -0.002),<br>$\Gamma_1$=(0.002, -0.001, 0)<br>[Cr1-Cr2 (0,0,-1)]<br>$J_2$=0.063 [Cr1-Cr1 (0,0,-1)]<br>$J_3$=-0.274 [Cr1-Cr4 (-1,0,-1)]<br>$J_4$=-0.292 [Cr1-Cr4 (0,0,0)] |
| | | | | $J_5$=0.146 [Cr1-Cr3 (0,0,0)]<br>$J_6$=0.155 [Cr1-Cr4 (0,0,-1)]<br>$J_7$=0.129 [Cr1-Cr3 (-1,0,0)]<br>$J_8$=0.022 [Cr1-Cr4 (-1,0,0)] |
| 0.399 | FeOOH | Fe1 (0.548,0.646,0.75)<br>Fe2 (0.952,0.146,0.75)<br>Fe3 (0.452,0.354,0.25)<br>Fe4 (0.048,0.854,0.25) | 5.0<br>(Fe 3*d*) | $J_1$=-12.250 [Fe1-Fe1 (0,0,-1)]<br>$J_2$=-6.814 [Fe1-Fe3 (0,0,0)]<br>$J_3$=-24.937, $D_3$= (-0.631, 0.459, 0.063)<br>$J_3^{ani}$= (-0.132, -0.136, -0.125),<br>$\Gamma_3$=(-0.019, -0.015, 0.005)<br>[Fe1-Fe4 (0,0,0)]<br>$J_4$=-0.110 [Fe1-Fe1 (-1,0,0)]<br>$J_5$=-0.443 [Fe1-Fe3 (1,0,0)]<br>$J_6$=0.011 [Fe1-Fe2 (0,0,0)]<br>$J_7$=-0.086 [Fe1-Fe3 (0,0,-1)]<br>$J_8$=0.104 [Fe1-Fe4 (0,0,-1)]<br>$J_9$=-0.038 [Fe1-Fe1 (-1,0,-1)]<br>$J_{10}$=-0.154 [Fe1-Fe2 (-1,0,0)]<br>$J_{11}$=-0.051 [Fe1-Fe3 (-1,0,0)]<br>$J_{12}$=-0.533 [Fe1-Fe1 (0,0,-2)]<br>$J_{13}$=-0.053 [Fe1-Fe2 (0,0,-1)]<br>$J_{14}$=-0.160 [Fe1-Fe2 (-1,0,-1)]<br>$J_{15}$=-0.010 [Fe1-Fe3 (1,0,-1)]<br>$J_{16}$=-0.007 [Fe1-Fe3 (0,1,0)]<br>$J_{17}$=-0.005 [Fe1-Fe4 (-1,0,0)]<br>$J_{18}$=-0.101 [Fe1-Fe3 (-1,0,-1)] |
| 0.410 | $GdAlO_3$ | Gd1 (0.466,0.75,0.483)<br>Gd2 (0.966,0.75,0.017)<br>Gd3 (0.534,0.25,0.517)<br>Gd4 (0.034,0.25,0.983) | 6.0<br>(Gd 4*f*) | $J_1$=0.009 [Gd1-Gd2 (-1,0,0)]<br>$J_2$=-0.089, $J_2^{ani}$= (-0.008, -0.005, -0.007),<br>$\Gamma_2$=(0, 0.006, 0)<br>[Gd1-Gd3 (0,0,0)]<br>$J_3$=-0.016 [Gd1-Gd2 (-1,0,1)]<br>$J_4$=0.030, $D_4$= (-0.001, -0.002, -0.002)<br>[Gd1-Gd4 (0,0,-1)]<br>$J_5$=0.016 [Gd1-Gd1 (0,0,-1)]<br>$J_6$=-0.002 [Gd1-Gd1 (-1,0,0)]<br>$J_7$=0.013 [Gd1-Gd4 (1,0,-1)]<br>$J_8$=0.002 [Gd1-Gd3 (-1,0,0)] |
| 0.455 | $RbFeO_2$ | Fe1 (0.755,0.006,0.19)<br>Fe2 (0.255,0.494,0.81)<br>Fe3 (0.245,0.506,0.31)<br>Fe4 (0.745,0.994,0.69)<br>Fe5 (0.245,0.994,0.81)<br>Fe6 (0.745,0.506,0.19)<br>Fe7 (0.755,0.494,0.69)<br>Fe8 (0.255,0.006,0.31) | 4.0<br>(Fe 3*d*) | $J_1$ =-29.978, $D_1$ = (0.020, -0.132, -0.096),<br>$J_1^{ani}$= (-0.091, -0.052, -0.077),<br>$\Gamma_1$=(0.006, 0.009, 0.017)<br>[Fe1-Fe8 (0,0,0)]<br>$J_2$=-27.144 [Fe1-Fe12 (0,-1,0)]<br>$J_3$=-30.940 [Fe9-Fe10 (0,0,0)]<br>$J_4$=-30.535 [Fe1-Fe15 (0,0,0)]<br>$J_5$=0.600 [Fe1-Fe13 (1,-1,0)] |
| | | Fe9 (0.784,0.24,0.563)<br>Fe10 (0.284,0.26,0.437)<br>Fe11 (0.216,0.74,0.937)<br>Fe12 (0.716,0.76,0.063)<br>Fe13 (0.216,0.76,0.437)<br>Fe14 (0.716,0.74,0.563)<br>Fe15 (0.784,0.26,0.063)<br>Fe16 (0.284,0.24,0.937) | | $J_6$=0.573 [Fe1-Fe16 (0,0,-1)]<br>$J_7$=0.617 [Fe1-Fe10 (0,0,0)]<br>$J_{8a}$=0.615 [Fe1-Fe1 (-1,0,0)]<br>$J_{8b}$=0.643 [Fe9-Fe9 (-1,0,0)]<br>$J_9$=0.578 [Fe1-Fe6 (0,-1,0)]<br>$J_{10}$=0.614 [Fe9-Fe14 (0,-1,0)]<br>$J_{11}$=0.661 [Fe1-Fe11 (1,-1,-1)]<br>$J_{12}$=0.560 [Fe1-Fe16 (1,0,-1)]<br>$J_{13}$=0.494 [Fe1-Fe13 (0,-1,0)]<br>$J_{14}$=0.570 [Fe1-Fe10 (1,0,0)]<br>$J_{15}$=0.660 [Fe1-Fe11 (0,-1,-1)] |
| 0.457 | $CsFeO_2$ | Fe1 (0.751,0.002,0.188)<br>Fe2 (0.251,0.498,0.812)<br>Fe3 (0.249,0.502,0.312)<br>Fe4 (0.749,0.998,0.688)<br>Fe5 (0.249,0.998,0.812)<br>Fe6 (0.749,0.502,0.188)<br>Fe7 (0.751,0.498,0.688)<br>Fe8 (0.251,0.002,0.312)<br>Fe9 (0.773,0.246,0.564)<br>Fe10 (0.273,0.254,0.436)<br>Fe11 (0.227,0.746,0.936)<br>Fe12 (0.727,0.754,0.064)<br>Fe13 (0.227,0.754,0.436)<br>Fe14 (0.727,0.746,0.564)<br>Fe15 (0.773,0.254,0.064)<br>Fe16 (0.273,0.246,0.936) | 4.0<br>(Fe 3$d$) | $J_1$=-32.759, $J_1^{ani}$= (-0.096, -0.056, -0.079),<br>$\Gamma_1$=(0.002, -0.017, -0.003)<br>[Fe1-Fe8 (0,0,0)]<br>$J_2$=-33.036, $D_2$= (0.050, -0.037, 0.058)<br>[Fe1-Fe12 (0,-1,0)]<br>$J_3$=-29.854 [Fe9-Fe10 (0,0,0)]<br>$J_4$=-34.010 [Fe1-Fe15 (0,0,0)]<br>$J_5$=0.667 [Fe1-Fe13 (1,-1,0)]<br>$J_6$=0.541 [Fe1-Fe16 (0,0,-1)]<br>$J_7$=0.670 [Fe1-Fe10 (0,0,0)]<br>$J_{8a}$=0.602 [Fe1-Fe1 (-1,0,0)]<br>$J_{8b}$=0.481 [Fe9-Fe9 (-1,0,0)]<br>$J_9$=0.613 [Fe1-Fe11 (1,-1,-1)]<br>$J_{10}$=0.659 [Fe1-Fe6 (0,-1,0)]<br>$J_{11}$=0.644 [Fe9-Fe14 (0,-1,0)]<br>$J_{12}$=0.587 [Fe1-Fe13 (0,-1,0)]<br>$J_{13}$=0.492 [Fe1-Fe16 (1,0,-1)]<br>$J_{14}$=0.607 [Fe1-Fe10 (1,0,0)]<br>$J_{15}$=0.582 [Fe1-Fe11 (0,-1,-1)] |
| 0.459,<br>0.460 | $KFeO_2$ | Fe1 (0.754,0.008,0.188)<br>Fe2 (0.254,0.492,0.812)<br>Fe3 (0.246,0.508,0.312)<br>Fe4 (0.746,0.992,0.688)<br>Fe5 (0.246,0.992,0.812)<br>Fe6 (0.746,0.508,0.188)<br>Fe7 (0.754,0.492,0.688)<br>Fe8 (0.254,0.008,0.312)<br>Fe9 (0.784,0.235,0.565)<br>Fe10 (0.284,0.265,0.435)<br>Fe11 (0.216,0.735,0.935)<br>Fe12 (0.716,0.765,0.065)<br>Fe13 (0.216,0.765,0.435)<br>Fe14 (0.716,0.735,0.565) | 4.0<br>(Fe 3$d$) | $J_1$=-40.343, $\Gamma_1$=(0.001, 0.025, 0.010)<br>[Fe1-Fe12 (0,-1,0)]<br>$J_2$=-45.019, $D_2$= (0.289, -0.091, -0.050)<br>[Fe1-Fe8 (0,0,0)]<br>$J_3$=-37.023, $J_3^{ani}$= (-0.115, -0.096, -0.096),<br>[Fe9-Fe10 (0,0,0)]<br>$J_4$=-42.401 [Fe1-Fe15 (0,0,0)]<br>$J_5$=0.759 [Fe1-Fe13 (1,-1,0)]<br>$J_6$=0.918 [Fe1-Fe16 (0,0,-1)]<br>$J_7$=0.660 [Fe1-Fe10 (0,0,0)]<br>$J_{8a}$=0.698 [Fe1-Fe1 (-1,0,0)]<br>$J_{8b}$=0.653 [Fe9-Fe9 (-1,0,0)]<br>$J_9$=0.505 [Fe1-Fe16 (1,0,-1)]<br>$J_{10}$=0.674 [Fe1-Fe6 (0,-1,0)] |
| | | Fe15 (0.784,0.265,0.065)<br>Fe16 (0.284,0.235,0.935) | | $J_{11}$=0.499 [Fe9-Fe14 (0,-1,0)]<br>$J_{12}$=0.435 [Fe1-Fe13 (0,-1,0)]<br>$J_{13}$=0.930 [Fe1-Fe11 (1,-1,-1)]<br>$J_{14}$=0.511 [Fe1-Fe10 (1,0,0)]<br>$J_{15}$=0.636 [Fe1-Fe11 (0,-1,-1)] |
| 0.505 | $Pb_2VO(PO_4)_2$ | V1 (0.256,0.499,0.062)<br>V2 (0.244,0.999,0.938)<br>V3 (0.744,0.501,0.938)<br>V4 (0.756,0.001,0.062) | 3.1<br>(V 3*d*) | $J_1$=0.133 [V1-V3 (-1,0,-1)]<br>$J_2$=0.110, $J_2^{ani}$= (-0.001, 0, -0.001),<br>[V1-V3 (0,0,-1)]<br>$J_3$=0.136 [V1-V2 (0,-1,-1)]<br>$J_4$=-0.134 [V1-V4 (-1,0,0)]<br>$J_5$=-0.205, $D_5$= (0.004, 0.004, 0.004)<br>[V1-V4 (-1,1,0)]<br>$J_6$=-0.000 [V1-V1 (-1,0,0)]<br>$\Gamma_i$=(0, 0, 0) |
| 0.798 | $MnPd_2$ | Mn1 (0.093,0.25,0.072)<br>Mn2 (0.593,0.25,0.428)<br>Mn3 (0.907,0.75,0.928)<br>Mn4 (0.407,0.75,0.572) | 0 | J1=-50.133, $J_1^{ani}$= (0.073, -0.009, -0.002),<br>$\Gamma_1$=(0.012, -0.01, -0.012)<br>[Mn1-Mn3 (-1,-1,-1)]<br>J2=3.161 [Mn1-Mn2 (-1,0,0)]<br>J3=1.121 [Mn1-Mn1 (0,-1,0)]<br>J4=-3.049, $D_4$= (-0.015, 0.125, 0.007)<br>[Mn1-Mn4 (0,-1,-1)]<br>J5=-2.765 [Mn1-Mn3 (0,-1,-1)]<br>J6=-2.105 [Mn1-Mn1 (-1,0,0)]<br>J7=-0.617 [Mn1-Mn2 (-1,-1,0)]<br>J8=-0.340 [Mn1-Mn4 (-1,-1,-1)]<br>J9=-0.478 [Mn1-Mn2 (-1,0,-1)]<br>J10=-0.334 [Mn1-Mn3 (-1,-2,-1)]<br>J11=0.184 [Mn1-Mn1 (-1,-1,0)]<br>J12=0.285 [Mn1-Mn3 (-2,-1,-1)]<br>J13=0.102 [Mn1-Mn2 (-1,-1,-1)]<br>J14=-1.106 [Mn1-Mn3 (-1,-1,0)]<br>J15=-0.477 [Mn1-Mn4 (0,-2,-1)]<br>J16=-1.183 [Mn1-Mn3 (0,-2,-1)]<br>J17=0.017 [Mn1-Mn1 (0,-2,0)]<br>J18=1.098 [Mn1-Mn1 (0,0,-1)]<br>J19=0.305 [Mn1-Mn4 (-1,-2,-1)]<br>J20=-0.134 [Mn1-Mn4 (1,-1,-1)]<br>J21=-0.012 [Mn1-Mn3 (0,-1,0)]<br>J22=0.097 [Mn1-Mn2 (-2,0,0)]<br>J23=-0.131 [Mn1-Mn3 (-2,-2,-1)] |
| 0.815 | $MnNb_2O_6$ | Mn1 (0,0.201,0.25)<br>Mn2 (0.5,0.299,0.75)<br>Mn3 (0,0.799,0.75)<br>Mn4 (0.5,0.701,0.25) | 0 | J1=-2.888, $J_1^{ani}$= (-0.003, -0.002, -0.002),<br>$\Gamma_1$=(-0.001, 0.001, -0.001)<br>[Mn1-Mn3 (0,-1,-1)]<br>J2=-1.643 [Mn1-Mn3 (0,0,-1)] |
| | | | | J3=-0.973, $D_3$= (0.001, -0.003, 0.002)<br>[Mn1-Mn1 (0,0,-1)]<br>J4=-0.183 [Mn1-Mn1 (0,-1,0)]<br>J5=-0.039 [Mn1-Mn2 (-1,0,-1)]<br>J6=-0.004 [Mn1-Mn1 (0,-1,-1)]<br>J7=-0.054 [Mn1-Mn4 (-1,-1,0)]<br>J8=-0.025 [Mn1-Mn3 (0,-1,-2)]<br>J9=-0.014 [Mn1-Mn3 (0,0,-2)]<br>J10=-0.014 [Mn1-Mn3 (0,-2,-1)]<br>J11=-0.016 [Mn1-Mn2 (-1,-1,-1)]<br>J12a=-0.029 [Mn1-Mn4 (-1,-1,-1)]<br>J12b=-0.023 [Mn1-Mn4 (-1,-1,1)]<br>J13=-0.001 [Mn1-Mn3 (0,1,-1)]<br>J14=-0.034 [Mn1-Mn2 (-1,1,-1)] |
| 0.816 | $MnTa_2O_6$ | Mn1 (0,0.174,0.25)<br>Mn2 (0.5,0.327,0.75)<br>Mn3 (0,0.827,0.75)<br>Mn4 (0.5,0.674,0.25) | 0 | J1=-4.753, $J_1^{ani}$= (-0.022, -0.005, -0.012),<br>$\Gamma_1$=(0.003, 0.006, -0.002)<br>[Mn1-Mn3 (0,-1,-1)]<br>J2=-0.589 [Mn1-Mn3 (0,0,-1)]<br>J3=-0.626, $D_3$= (0.041, 0.004, -0.017)<br>[Mn1-Mn1 (0,0,-1)]<br>J4=-0.178 [Mn1-Mn1 (0,-1,0)]<br>J5=-0.012 [Mn1-Mn1 (0,-1,-1)]<br>J6=-0.038 [Mn1-Mn2 (-1,0,-1)]<br>J7=-0.047 [Mn1-Mn4 (-1,-1,0)]<br>J8=-0.043 [Mn1-Mn3 (0,-1,-2)]<br>J9=-0.029 [Mn1-Mn3 (0,-2,-1)]<br>J10=-0.007 [Mn1-Mn3 (0,0,-2)]<br>J11=-0.033 [Mn1-Mn2 (-1,-1,-1)]<br>J12a=-0.044 [Mn1-Mn4 (-1,-1,-1)]<br>J12b=-0.023 [Mn1-Mn4 (-1,-1,1)]<br>J13=-0.001 [Mn1-Mn3 (0,1,-1)]<br>J14=-0.028 [Mn1-Mn2 (-1,1,-1)]<br>J15=-0.000 [Mn1-Mn1 (0,0,-2)]<br>J16=-0.019 [Mn1-Mn2 (-1,0,-2)]<br>J17=-0.001 [Mn1-Mn3 (0,-2,-2)]<br>J18=-0.010 [Mn1-Mn4 (-1,-2,0)] |
| 0.821 | $SrGd_2O_4$ | Gd1 (0.074,0.613,0.75)<br>Gd2 (0.574,0.886,0.75)<br>Gd3 (0.926,0.387,0.25)<br>Gd4 (0.426,0.113,0.25)<br>Gd5 (0.081,0.112,0.75)<br>Gd6 (0.581,0.388,0.75)<br>Gd7 (0.919,0.888,0.25)<br>Gd8 (0.419,0.612,0.25) | 6.0<br>(Gd 4*f*) | J1a=-0.355, $D_{1a}$= (-0.017, -0.009, 0),<br>$J_{1a}^{ani}$= (-0.013, -0.011, -0.006),<br>[Gd1-Gd1 (0,0,-1)]<br>J1b=-0.215 [Gd5-Gd5 (0,0,-1)]<br>J2=-0.331,<br>$\Gamma_2$=(0.002, 0, 0.003)<br>[Gd1-Gd3 (-1,0,0)]<br>J3=-0.231 [Gd5-Gd7 (-1,-1,0)] |
|  |  |  |  | J4=-0.044 [Gd1-Gd8 (0,0,0)]<br>J5=-0.133 [Gd1-Gd7 (-1,0,0)]<br>J6=-0.004 [Gd1-Gd6 (-1,0,0)]<br>J7=-0.000 [Gd1-Gd6 (0,0,0)]<br>J8=-0.002 [Gd1-Gd5 (0,1,0)] |
| 0.966 | $V_2WO_6$[†] | V1 (0,0,0.335)<br>V2 (0.5,0.5,0.165)<br>V3 (0,0,0.665)<br>V4 (0.5,0.5,0.835) | 2.85<br>(V 3*d*) | J1=0.172, **$D_1$= (0.108, -0.872, -1.133)**,<br>**$\Gamma_1$=(-0.003, -0.338, -0.104)**<br>[V1-V3 (0,0,0)]<br>J2=-0.169, **$J_2^{ani}$= (0.665, 0.310, -0.848)**,<br>[V1-V2 (-1,-1,0)]<br>J3=-0.038 [V1-V1 (-1,0,0)]<br>J4=0.042 [V1-V3 (-1,0,0)]<br>J5a=-0.039 [V1-V4 (-1,-1,0)]<br>J5b=-0.029 [V1-V4 (-1,-1,-1)]<br>J6=0.089 [V1-V3 (0,0,-1)]<br>J7a=-0.019 [V1-V1 (-1,-1,0)]<br>J7b=-0.075 [V1-V1 (-1,1,0)] |
| 0.969 | $CaFe_2O_4$ | Fe1 (0.067,0.11,0.75)<br>Fe2 (0.433,0.61,0.25)<br>Fe3 (0.567,0.39,0.75)<br>Fe4 (0.933,0.89,0.25)<br>Fe5 (0.08,0.608,0.75)<br>Fe6 (0.42,0.108,0.25)<br>Fe7 (0.58,0.892,0.75)<br>Fe8 (0.92,0.392,0.25) | 4.0<br>(Fe 3*d*) | J1a=-5.354, $J_{1a}^{ani}$= (-0.04, -0.046, -0.024),<br>[Fe1-Fe1 (0,0,-1)]<br>J1b=-7.601 [Fe5-Fe5 (0,0,-1)]<br>J2=-6.127 [Fe1-Fe4 (-1,-1,0)]<br>J3=-7.573 [Fe5-Fe8 (-1,0,0)]<br>J4=10.049 [Fe1-Fe6 (0,0,0)]<br>J5=15.053, $D_5$= (0.868, 0.171, 0.339),<br>$\Gamma_5$=(0.001, 0, 0.004)<br>[Fe1-Fe8 (-1,0,0)]<br>J6=-0.302 [Fe1-Fe7 (-1,-1,0)]<br>J7=-0.033 [Fe1-Fe4 (-1,-1,-1)]<br>J8=0.007 [Fe1-Fe7 (0,-1,0)]<br>J9=-0.512 [Fe5-Fe8 (-1,0,-1)]<br>J10=-0.326 [Fe1-Fe5 (0,0,0)]<br>J11=0.089 [Fe1-Fe5 (0,-1,0)]<br>J12=-0.059 [Fe1-Fe3 (-1,0,0)]<br>J13=0.060 [Fe5-Fe7 (-1,0,0)]<br>J14=-0.101 [Fe1-Fe6 (0,0,-1)]<br>J15=-0.150 [Fe1-Fe8 (-1,0,-1)]<br>J16=-0.136 [Fe1-Fe7 (-1,-1,-1)]<br>J17a=-0.549 [Fe1-Fe1 (0,0,-2)]<br>J17b=-0.418 [Fe5-Fe5 (0,0,-2)]<br>J18=-0.198 [Fe1-Fe7 (0,-1,-1)]<br>J19=-0.161 [Fe1-Fe5 (0,0,-1)]<br>J20=-0.395 [Fe1-Fe6 (-1,0,0)]<br>J21=-0.649 [Fe1-Fe5 (0,-1,-1)]<br>J22=-0.181 [Fe1-Fe3 (-1,0,-1)] |

| | | | | J23=-0.268 [Fe5-Fe7 (-1,0,-1)] |
|---|---|---|---|---|
| | $Cu_3TeO_6$ | Cu1 (0.969,0,0.25)<br>Cu2 (-0.469,0,-0.25)<br>Cu3 (0.25,0.969,0)<br>Cu4 (-0.25,-0.469,0)<br>Cu5 (0,0.25,0.969)<br>Cu6 (0,-0.25,-0.469)<br>Cu7 (-0.969,0,-0.25)<br>Cu8 (0.469,0,0.25)<br>Cu9 (-0.25,-0.969,0)<br>Cu10 (0.25,0.469,0)<br>Cu11 (0,-0.25,-0.969)<br>Cu12 (0,0.25,0.469) | | J1=4.40, $D_1$= (0.34, -0.05, -0.25)<br>[Cu1-Cu9 (1, 1, 0)]<br>J2=-0.41 [Cu1-Cu3 (1, -1, 0)]<br>J3=-1.63 [Cu1-Cu8 (0, 0, 0)]<br>J4=1.31 [Cu1-Cu2 (3/2, 1/2, 1/2)]<br>J5=1.70 [Cu1-Cu7 (2, 0, 0)]<br>J6=-0.21 [Cu1-Cu3 (1/2, -1/2, 1/2)]<br>J7=-0.14 [Cu1-Cu9 (3/2, 1/2, 1/2)]<br>J8=0.054 [Cu1-Cu3 (1/2, -3/2, 1/2)]<br>J9=4.26 [Cu1-Cu9 (3/2, 3/2,1/2)] |
| | FeS | Fe1 (0.073,0.394,0.121)<br>Fe2 (0.606,0.678,0.121)<br>Fe3 (0.322,0.927,0.121)<br>Fe4 (0.073,0.394,0.379)<br>Fe5 (0.606,0.678,0.379)<br>Fe6 (0.322,0.927,0.379)<br>Fe7 (0.394,0.073,0.879)<br>Fe8 (0.678,0.606,0.879)<br>Fe9 (0.927,0.322,0.879)<br>Fe10 (0.394,0.073,0.621)<br>Fe11 (0.678,0.606,0.621)<br>Fe12 (0.927,0.322,0.621) | 1.0<br>(Fe 3*d*) | J1=-15.657, $J_{1a}^{ani}$= (-0.187, -0.203, -0.202),<br>$\Gamma_1$=(0.004, 0.007, 0.013)<br>[Fe1-Fe2 (0,0,0)]<br>J2=16.099 [Fe1-Fe9 (-1,0,-1)]<br>J3=21.515 [Fe1-Fe4 (0,0,0)]<br>J4=0.225 [Fe1-Fe2 (-1,-1,0)]<br>J5=0.439 [Fe1-Fe2 (-1,0,0)]<br>J6=18.588, $D_6$= (0.511, 0.319, -0.390)<br>[Fe1-Fe5 (0,0,0)]<br>J7=2.479 [Fe1-Fe8 (-1,0,-1)]<br>J8=2.006 [Fe1-Fe7 (0,0,-1)]<br>J9=6.029 [Fe1-Fe7 (-1,0,-1)]<br>J10=4.182 [Fe1-Fe5 (-1,-1,0)]<br>J11=4.127 [Fe1-Fe5 (-1,0,0)]<br>J12=7.722 [Fe1-Fe8 (-1,-1,-1)]<br>J13=-0.384 [Fe1-Fe12 (-1,0,-1)] |
| 1.3 | $Sr_2IrO_4$† | Ir1 (0,0.25,0.375)<br>Ir2 (0.5,0.75,0.875)<br>Ir3 (0.5,0.25,0.625)<br>Ir4 (0,0.75,0.125)<br>Ir5 (0,0.75,0.625)<br>Ir6 (0.5,0.25,0.125)<br>Ir7 (0.5,0.75,0.375)<br>Ir8 (0,0.25,0.875) | 1.53<br>(Ir 5*d*) | J1=-0.495, $D_1$= (0, 0, 0.408),<br>$J_{1a}^{ani}$= (-0.006, -0.01, 0.68),<br>$\Gamma_1$=(0.162, 0, 0.001)<br>[Ir1-Ir7 (-1,-1,0)]<br>J2a=-0.156 [Ir1-Ir1 (0,-1,0)]<br>J2b=-0.154 [Ir1-Ir1 (-1,0,0)]<br>J3a=-0.009 [Ir1-Ir3 (-1,0,0)]<br>J3b=-0.007 [Ir1-Ir6 (-1,0,0)]<br>J3c=0.015 [Ir1-Ir4 (0,-1,0)]<br>J3d=0.017 [Ir1-Ir5 (0,-1,0)]<br>J4=-0.026 [Ir1-Ir1 (-1,-1,0)]<br>J5=-0.026 [Ir1-Ir7 (-1,-2,0)] |
| 1.18 | $MnS_2$ | Mn1 (0,0,0)<br>Mn2 (0.5,0.25,0)<br>Mn3 (0.5,0,0.5) | 5.0<br>(Mn 3*d*) | J1a=-4.776, $D_{1a}$= (-0.002, -0.020, 0.018),<br>$J_{1a}^{ani}$= (-0.007, 0.007, 0.008),<br>$\Gamma_{1a}$=(0, 0, 0.001) |
| | | Mn4 (0,0.25,0.5)<br>Mn5 (0,0.5,0)<br>Mn6 (0.5,0.75,0)<br>Mn7 (0.5,0.5,0.5)<br>Mn8 (0,0.75,0.5) | | [Mn1-Mn2 (-1,0,0)]<br>J1b=-4.977 [Mn1-Mn3 (-1,0,0)]<br>J1c=-4.884 [Mn1-Mn3 (-1,0,-1)]<br>J1d=-4.376 [Mn1-Mn8 (0,-1,-1)]<br>J1e=-3.346 [Mn1-Mn4 (0,0,-1)]<br>J1f=-3.846 [Mn1-Mn6 (-1,-1,0)]<br>J2a=-1.284 [Mn1-Mn1 (0,0,-1)]<br>J2b=-1.146 [Mn1-Mn1 (-1,0,0)]<br>J2c=-1.474 [Mn1-Mn5 (0,-1,0)]<br>J3a=-0.001 [Mn1-Mn2 (-1,0,-1)]<br>J3b=-0.219 [Mn1-Mn2 (-1,0,1)]<br>J3c=-0.067 [Mn1-Mn4 (-1,0,0)]<br>J3d=-0.237 [Mn1-Mn4 (-1,0,-1)]<br>J3e=-0.227 [Mn1-Mn6 (-1,-1,-1)]<br>J3f=-0.040 [Mn1-Mn6 (-1,-1,1)]<br>J3g=-0.271 [Mn1-Mn7 (-1,-1,-1)]<br>J3h=-0.032 [Mn1-Mn7 (-1,0,0)]<br>J3i=-0.037 [Mn1-Mn7 (-1,0,-1)]<br>J3j=-0.167 [Mn1-Mn7 (0,0,0)]<br>J3k=-0.222 [Mn1-Mn8 (-1,-1,-1)]<br>J3l=-0.109 [Mn1-Mn8 (-1,-1,0)]<br>J4a=0.039 [Mn1-Mn1 (-1,0,-1)]<br>J4b=-0.030 [Mn1-Mn1 (-1,0,1)]<br>J4c=-0.013 [Mn1-Mn5 (0,-1,1)]<br>J4d=-0.050 [Mn1-Mn5 (-1,0,0)]<br>J4e=0.015 [Mn1-Mn5 (-1,-1,0)]<br>J5a=-0.131 [Mn1-Mn2 (-1,-1,0)]<br>J5b=-0.069 [Mn1-Mn2 (-2,0,0)]<br>J5c=-0.172 [Mn1-Mn3 (-2,0,-1)]<br>J5d=-0.171 [Mn1-Mn3 (1,0,-1)]<br>J5e=-0.054 [Mn1-Mn3 (-1,0,1)]<br>J5f=-0.0578 [Mn1-Mn3 (-1,0,-2)]<br>J5g=-0.026 [Mn1-Mn4 (0,-1,-1)]<br>J5h=-0.129[Mn1-Mn4 (0,0,-2)]<br>J5i=-0.142 [Mn1-Mn6 (-1,0,0)]<br>J5j=-0.023 [Mn1-Mn6 (-2,-1,0)]<br>J5k=-0.183 [Mn1-Mn8 (0,-1,-2)]<br>J5l=-0.034 [Mn1-Mn8 (0,0,-1)] |
| 1.20 | $HoMnO_3$ | Mn1 (0,0,0.5)<br>Mn2 (0.25,0,0)<br>Mn3 (0,0.5,0.5)<br>Mn4 (0.25,0.5,0)<br>Mn5 (0.5,0,0.5)<br>Mn6 (0.75,0,0) | 3.1<br>(Mn 3$d$) | J1=5.792, $D_1$= (0.861, 0, 0.368),<br>$\Gamma_1$=(0, 0.172, 0)<br>[Mn1-Mn3 (0,-1,0)]<br>J2a=9.557, $J_{2a}^{ani}$= (0.544, -0.061, 0.369),<br>[Mn1-Mn2 (0,0,0)]<br>J2b=-10.260 [Mn1-Mn6 (-1,0,0)] |

| | | Mn7 (0.5,0.5,0.5)<br>Mn8 (0.75,0.5,0) | | J3=0.133 [Mn1-Mn1 (0,0,-1)]<br>J4a=-0.020 [Mn1-Mn4 (0,-1,0)]<br>J4b=-0.044 [Mn1-Mn4 (0,-1,1)]<br>J4c=-0.281 [Mn1-Mn8 (-1,-1,0)]<br>J4d=-0.451 [Mn1-Mn8 (-1,-1,1)]<br>J5=-0.085 [Mn1-Mn5 (-1,0,0)]<br>J6a=-0.006 [Mn1-Mn3 (0,-1,-1)]<br>J6b=-0.075 [Mn1-Mn3 (0,0,-1)]<br>J7a=0.010 [Mn1-Mn7 (-1,-1,0)]<br>J7b=0.065 [Mn1-Mn7 (-1,0,0)]<br>J8=0.403 [Mn1-Mn1 (0,-1,0)]<br>J9a=0.111 [Mn1-Mn5 (-1,0,1)]<br>J9b=-0.461 [Mn1-Mn5 (-1,0,-1)]<br>J10a=-0.092 [Mn1-Mn2 (0,-1,0)]<br>J10b=-0.107 [Mn1-Mn6 (-1,-1,0)]<br>J10c=-0.048 [Mn1-Mn6 (-1,1,0)] |
|---|---|---|---|---|
| 1.23,<br>1.405 | $La_2CuO_4$ | Cu1 (0,0,0)<br>Cu2 (0,0.5,0.5)<br>Cu3 (0.5,0,0.5)<br>Cu4 (0.5,0.5,0) | 7.2<br>(Cu 3*d*) | J1=-7.094, $D_1$= (-0.046, 0.105, 0),<br>[Cu1-Cu4 (-1,-1,0)]<br>J2=1.446 [Cu1-Cu1 (-1,0,0)]<br>J3=1.458 [Cu1-Cu1 (0,-1,0)]<br>J4=-0.000 [Cu1-Cu3 (-1,0,-1)]<br>J5=-0.000 [Cu1-Cu2 (0,-1,-1)]<br>J6=-0.054 [Cu1-Cu1 (-1,-1,0)]<br>J7=-0.076 [Cu1-Cu4 (-2,-1,0)]<br>J8=-0.073 [Cu1-Cu4 (-1,-2,0)]<br>$J_i^{ani}$=$\Gamma_i$=(0, 0, 0) |
| 1.24 | $ZnV_2O_4$ | V1 (0,0,0.5)<br>V2 (0.75,0.25,0.25)<br>V3 (0.75,0.75,0.75)<br>V4 (0.5,0.5,0)<br>V5 (0,0.5,0.5)<br>V6 (0.5,0,0)<br>V7 (0.25,0.75,0.75)<br>V8 (0.25,0.25,0.25) | 4.1<br>(V 3*d*) | J1a=-2.561, $D_1$= (0.089, -0.089, 0.125),<br>$J_{1a}^{ani}$= (-0.015, -0.015, -0.017),<br>$\Gamma_{1a}$=(0.001, 0, 0)<br>[V1-V2 (-1,0,0)]<br>J1b=-2.943 [V1-V7 (0,-1,0)]<br>J2=-8.619 [V1-V5 (0,-1,0)]<br>J3a=-0.021 [V1-V6 (-1,0,0)]<br>J3b=-0.000 [V1-V6 (0,0,0)]<br>J4a=0.005 [V1-V2 (-1,-1,0)]<br>J4b=0.008 [V1-V7 (-1,-1,0)]<br>J5a=-0.040 [V1-V4 (-1,-1,0)]<br>J5b=0.008 [V1-V4 (-1,-1,1)]<br>J6a=-0.023 [V1-V1 (-1,0,0)]<br>J6b=-0.010 [V1-V1 (0,-1,0)] |
| 1.26 | $CsFe_2Se_3$ | Fe1 (0.86,0,0.43)<br>Fe2 (0.14,0.5,0.07)<br>Fe3 (0.14,0,0.57)<br>Fe4 (0.86,0.5,0.93) | 2.0<br>(Fe 3*d*) | J1=10.883 [Fe1-Fe3 (1,0,0)]<br>J2=-36.531, $D_2$= (-1.055, 0, 0.710),<br>$J_2^{ani}$= (-0.905, -0.763, -0.64),<br>$\Gamma_2$=(-0.905, -0.763, -0.640) |
| | | Fe5 (0.86,0,0.93)<br>Fe6 (0.14,0.5,0.57)<br>Fe7 (0.14,0,0.07)<br>Fe8 (0.86,0.5,0.43) | | [Fe1-Fe8 (0,-1,0)]<br>J3=-6.684 [Fe1-Fe6 (1,-1,0)]<br>J4=-1.658 [Fe1-Fe1 (0,-1,0)]<br>J5a=-0.022 [Fe1-Fe3 (0,0,0)]<br>J5b=-0.076 [Fe1-Fe7 (1,0,0)]<br>J6a=1.009 [Fe1-Fe3 (1,-1,0)]<br>J6b=1.045 [Fe1-Fe3 (1,1,0)]<br>J7a=-0.016 [Fe1-Fe2 (1,-1,0)]<br>J7b=0.021 [Fe1-Fe6 (0,-1,0)]<br>J8=-0.182 [Fe1-Fe7 (0,0,0)] |
| 1.28,<br>1.678 | CrN | Cr1 (0.14,0.25,0.25)<br>Cr2 (0.64,0.25,0.25)<br>Cr3 (0.86,0.75,0.75)<br>Cr4 (0.36,0.75,0.75) | 0.5<br>(Cr 3*d*) | J1=-19.024 [Cr1-Cr4 (0,-1,-1)]<br>J2=-14.639 [Cr1-Cr2 (-1,0,0)]<br>J3=-4.496, $J_3^{ani}$= (-0.015, -0.015, -0.004),<br>$\Gamma_3$=(-0.002, -0.001, 0.001)<br>[Cr1-Cr1 (0,-1,0)]<br>J4=-0.080 [Cr1-Cr3 (-1,-1,-1)]<br>J5=-0.337 [Cr1-Cr1 (0,0,-1)]<br>J6=-0.533, $D_6$= (0, 0, 0.046),<br>[Cr1-Cr2 (-1,-1,0)]<br>J7=0.614 [Cr1-Cr3 (0,-1,-1)]<br>J8=0.152 [Cr1-Cr2 (-1,0,-1)]<br>J9=0.359 [Cr1-Cr4 (0,-2,-1)]<br>J10=0.322 [Cr1-Cr1 (0,-1,-1)]<br>J11=0.174 [Cr1-Cr4 (-1,-1,-1)]<br>J12=0.243 [Cr1-Cr3 (-1,-2,-1)]<br>J13=0.692 [Cr1-Cr1 (-1,0,0)]<br>J14=-0.191 [Cr1-Cr2 (-1,-1,-1)]<br>J15=-0.110 [Cr1-Cr1 (0,-2,0)]<br>J16=0.979 [Cr1-Cr3 (0,-2,-1)]<br>J17=0.947 [Cr1-Cr1 (-1,-1,0)]<br>J18=0.458 [Cr1-Cr4 (0,-1,-2)]<br>J19=0.223 [Cr1-Cr3 (-1,-1,-2)]<br>J20=0.625 [Cr1-Cr2 (-1,-2,0)]<br>J21=0.481 [Cr1-Cr4 (-1,-2,-1)] |
| 1.42 | $La_2NiO_4$ | Ni1 (0,0,0)<br>Ni2 (0.5,0,0.5)<br>Ni3 (0,0.5,0.5)<br>Ni4 (0.5,0.5,0) | 4.2<br>(Ni 3*d*) | J1a=-12.357, $D_{1a}$= (0.148, -0.19, 0.044),<br>$J_{1a}^{ani}$= (-0.005, -0.005, -0.010),<br>$\Gamma_{1a}$=(0.002, 0, 0)<br>[Ni1-Ni4 (-1,-1,0)]<br>J1b=-13.327 [Ni1-Ni4 (-1,0,0)]<br>J2=0.705 [Ni1-Ni1 (-1,0,0)]<br>J3=-0.014 [Ni1-Ni2 (-1,0,-1)]<br>J4=-0.010 [Ni1-Ni3 (0,-1,-1)]<br>J5a=0.250 [Ni1-Ni1 (-1,-1,0)]<br>J5b=0.266 [Ni1-Ni1 (-1,1,0)] |

| 1.62 | CuO | Cu1 (0.875,0.75,0)<br>Cu2 (0.125,0.25,0.5)<br>Cu3 (0.125,0.25,0)<br>Cu4 (0.875,0.75,0.5)<br>Cu5 (0.375,0.75,0)<br>Cu6 (0.625,0.25,0.5)<br>Cu7 (0.625,0.25,0)<br>Cu8 (0.375,0.75,0.5) | 7.0<br>(Cu 3*d*) | J1a=0.130 [Cu1-Cu3 (1,0,0)]<br>J1b=0.012 [Cu1-Cu3 (1,1,0)]<br>J1c=-0.291 [Cu1-Cu7 (0,0,0)]<br>J1d=0.024 [Cu1-Cu7 (0,1,0)]<br>J2a=-0.374, $D_{2a}$ = (-0.08, 0.138, 0.061),<br>[Cu1-Cu2 (1,0,0)]<br>J2b=-0.272 [Cu1-Cu6 (0,0,-1)]<br>J3=-0.216, $J_3^{ani}$= (-0.027, -0.001, -0.003),<br>$\Gamma_3$=(-0.008, 0.009, 0.004)<br>[Cu1-Cu4 (0,0,-1)]<br>J4=-0.003 [Cu1-Cu1 (0,-1,0)]<br>J5=-7.655 [Cu1-Cu8 (0,0,-1)]<br>J6a=0.138 [Cu1-Cu4 (0,1,-1)]<br>J6b=0.067 [Cu1-Cu4 (0,-1,-1)]<br>J7=-0.055 [Cu1-Cu5 (0,0,0)]<br>J8a=0.035 [Cu1-Cu8 (0,1,-1)]<br>J8b=0.004 [Cu1-Cu8 (0,-1,-1)]<br>J9=-0.171 [Cu1-Cu5 (0,0,-1)]<br>J10a=0.003 [Cu1-Cu2 (1,0,-1)]<br>J10b=0.002 [Cu1-Cu6 (0,0,0)]<br>J11a=-0.029 [Cu1-Cu3 (1,0,1)]<br>J11b=-0.027 [Cu1-Cu3 (1,1,1)]<br>J11c=-0.050 [Cu1-Cu7 (0,0,-1)]<br>J11d=-0.006 [Cu1-Cu7 (0,1,-1)]<br>J12a=-0.000 [Cu1-Cu3 (1,-1,0)]<br>J12b=-0.036 [Cu1-Cu3 (1,2,0)]<br>J12c=-0.061 [Cu1-Cu7 (0,-1,0)]<br>J12d=-0.000 [Cu1-Cu7 (0,2,0)]<br>J13a=-0.002 [Cu1-Cu2 (1,-1,0)]<br>J13b=0.003 [Cu1-Cu6 (0,-1,-1)]<br>J14a=-0.891 [Cu1-Cu5 (0,-1,0)]<br>J14b=-0.019 [Cu1-Cu5 (0,1,0)]<br>J15a=0.062 [Cu1-Cu2 (0,0,-1)]<br>J15b=0.117 [Cu1-Cu6 (1,0,0)]<br>J16a=0.002 [Cu1-Cu5 (0,-1,-1)]<br>J16b=-0.011 [Cu1-Cu5 (0,1,-1)]<br>J17a=0.026 [Cu1-Cu3 (0,0,-1)]<br>J17b=0.038 [Cu1-Cu3 (0,1,-1)]<br>J17c=-0.092 [Cu1-Cu7 (1,0,1)]<br>J17d=-0.073 [Cu1-Cu7 (1,1,1)]<br>J18=-0.001 [Cu1-Cu1 (0,0,-1)]<br>J19=0.001 [Cu1-Cu1 (0,-2,0)]<br>J20=-0.000 [Cu1-Cu8 (0,0,0)]<br>J21a=0.006 [Cu1-Cu2 (1,-1,-1)] |
|---|---|---|---|---|
| | | | | J21b=0.003 [Cu1-Cu6 (0,-1,0)]<br>J22a=-0.145 [Cu1-Cu1 (0,-1,1)]<br>J22b=-0.002 [Cu1-Cu1 (0,-1,-1)]<br>J23a=0.000 [Cu1-Cu3 (0,0,0)]<br>J23b=0.000 [Cu1-Cu3 (0,1,0)]<br>J23c=-0.003 [Cu1-Cu7 (1,0,0)]<br>J23d=0.000 [Cu1-Cu7 (1,1,0)]<br>J24a=-0.048 [Cu1-Cu3 (1,-1,1)]<br>J24b=-0.002 [Cu1-Cu3 (1,2,1)]<br>J24c=-0.004 [Cu1-Cu7 (0,-1,-1)]<br>J24d=-0.027 [Cu1-Cu7 (0,2,-1)]<br>J25=0.246 [Cu1-Cu1 (-1,0,-1)] |
| 1.64 | $BaNiF_4$ | Ni1 (0.457,0,0.828)<br>Ni2 (0.543,0.5,0.172)<br>Ni3 (0.957,0,0.828)<br>Ni4 (0.043,0.5,0.172) | 1.5<br>(Ni 3*d*) | J1=-7.448, $D_{1a}$= (0.558, 0, 0),<br>$J_{1a}^{ani}$= (-0.048, -0.01, -0.011),<br>$\Gamma_{1a}$=(0, 0, 0.004)<br>[Ni1-Ni2 (0,-1,1)]<br>J2=-8.664 [Ni1-Ni3 (-1,0,0)]<br>J3=0.121 [Ni1-Ni4 (0,-1,1)]<br>J4a=0.006 [Ni1-Ni2 (0,-1,0)]<br>J4b=0.009 [Ni1-Ni4 (0,-1,0)]<br>J5=-0.001 [Ni1-Ni3 (-1,-1,0)]<br>J6a=0.000 [Ni1-Ni1 (0,0,-1)]<br>J6b=0.000 [Ni1-Ni3 (-1,0,-1)]<br>J7=0.203 [Ni1-Ni1 (-1,0,0)] |
| 1.101 | $LuMnO_3$ | Mn1 (0.5,0,0)<br>Mn2 (0,0.25,0)<br>Mn3 (0.5,0,0.5)<br>Mn4 (0,0.25,0.5)<br>Mn5 (0.5,0.5,0)<br>Mn6 (0,0.75,0)<br>Mn7 (0.5,0.5,0.5)<br>Mn8 (0,0.75,0.5) | 3.1<br>(Mn 3*d*) | J1a=2.207 [Mn1-Mn3 (0,0,-1)]<br>J1b=2.550 [Mn1-Mn3 (0,0,0)]<br>J2a=6.081, $D_{2a}$= (0.342, 0.054, -0.423),<br>$J_{2a}^{ani}$= (-0.052, -0.046, -0.051),<br>$\Gamma_{2a}$=(0, 0, 0.004)<br>[Mn1-Mn2 (0,0,0)]<br>J2b=-5.360 [Mn1-Mn6 (1,-1,0)]<br>J3=0.011 [Mn1-Mn1 (-1,0,0)]<br>J4a=0.117 [Mn1-Mn4 (0,0,0)]<br>J4b=-0.125 [Mn1-Mn4 (0,0,-1)]<br>J4c=-0.312 [Mn1-Mn8 (0,-1,0)]<br>J4d=-0.477 [Mn1-Mn8 (0,-1,-1)]<br>J5=-0.407 [Mn1-Mn5 (0,-1,0)]<br>J6a=0.008 [Mn1-Mn3 (-1,0,-1)]<br>J6b=-0.009 [Mn1-Mn3 (-1,0,0)]<br>J7a=-0.000 [Mn1-Mn7 (0,-1,-1)]<br>J7b=-0.007 [Mn1-Mn7 (0,-1,0)]<br>J8=0.027 [Mn1-Mn1 (0,0,-1)]<br>J9a=-0.079 [Mn1-Mn5 (-1,0,0)]<br>J9b=-0.861 [Mn1-Mn5 (-1,-1,0)] |
| | | | | J10a=-0.025 [Mn1-Mn2 (0,0,-1)]<br>J10b=-0.021 [Mn1-Mn2 (0,0,1)]<br>J10c=0.004 [Mn1-Mn6 (0,-1,-1)]<br>J10d=0.006 [Mn1-Mn6 (0,-1,1)]<br>J11=0.003 [Mn1-Mn2 (-1,0,0)]<br>J12a=0.015 [Mn1-Mn7 (-1,-1,-1)]<br>J12b=-0.009 [Mn1-Mn7 (-1,0,-1)]<br>J12c=-0.029 [Mn1-Mn7 (-1,0,0)]<br>J13a=0.003 [Mn1-Mn1 (-1,0,-1)]<br>J13b=-0.001 [Mn1-Mn1 (-1,0,1)]<br>J14a=-0.003 [Mn1-Mn2 (0,-1,0)]<br>J14b=-0.015 [Mn1-Mn6 (0,0,0)] |
| 1.121 | $NaFeSO_4F$ | Fe1 (0,0,0)<br>Fe2 (0,0,0.5)<br>Fe3 (0.5,0.5,0)<br>Fe4 (0.5,0.5,0.5) | 4.0<br>(Fe 3*d*) | J1=-5.854, $D_1$= (-0.007, -0.012, -0.036),<br>$J_1^{ani}$= (-0.011, -0.016, -0.019),<br>$\Gamma_1$=(-0.002, 0.009, 0)<br>[Fe1-Fe2 (0,0,0)]<br>J2a=-0.373 [Fe1-Fe3 (-1,-1,0)]<br>J2b=-0.002 [Fe1-Fe3 (-1,0,0)]<br>J3=-0.070 [Fe1-Fe4 (-1,-1,-1)]<br>J4=-0.076 [Fe1-Fe2 (-1,0,-1)]<br>J5=-0.000 [Fe1-Fe1 (-1,0,0)]<br>J6=0.014 [Fe1-Fe1 (0,0,-1)]<br>J7=-0.002 [Fe1-Fe4 (-1,-1,0)] |
| 1.130 | $Cr_2As$ | Cr1 (0,0,0)<br>Cr2 (0,0,0.5)<br>Cr3 (0.5,0.5,0.5)<br>Cr4 (0.5,0.5,0)<br>Cr5 (0.5,0,0.338)<br>Cr6 (0.5,0,0.838)<br>Cr7 (0,0.5,0.163)<br>Cr8 (0,0.5,0.663) | 0 | J1=-7.385 [Cr1-Cr4 (-1,-1,0)]<br>J2a=-9.479 [Cr1-Cr6 (-1,0,-1)]<br>J2b=-7.011 [Cr1-Cr7 (0,-1,0)]<br>J3=-0.023, $\Gamma_3$=(-0.004, -0.001, -0.003)<br>[Cr5-Cr7 (0,-1,0)]<br>J4a=3.711 [Cr1-Cr1 (0,-1,0)]<br>J4b=3.467 [Cr1-Cr1 (-1,0,0)]<br>J4c=7.190, $J_{4c}^{ani}$= (0.008, -0.002, 0.013),<br>[Cr5-Cr5 (-1,0,0)]<br>J4d=6.736 [Cr5-Cr5 (0,-1,0)]<br>J5a=-0.400 [Cr1-Cr6 (-1,-1,-1)]<br>J5b=-0.901 [Cr1-Cr7 (-1,-1,0)]<br>J6a=1.644 [Cr1-Cr5 (-1,0,0)]<br>J6b=1.639 [Cr1-Cr8 (0,-1,-1)]<br>J7=-7.817 [Cr5-Cr8 (0,-1,0)]<br>J8a=0.972 [Cr1-Cr1 (-1,-1,0)]<br>J8b=-2.418, $D_{8b}$= (0.168, -0.127, 0),<br>[Cr5-Cr5 (-1,-1,0)]<br>J9a=0.142 [Cr1-Cr4 (-1,-2,0)]<br>J9b=-0.049 [Cr1-Cr4 (-2,-1,0)]<br>J10a=0.269 [Cr1-Cr6 (-2,0,-1)] |
| | | | | J10b=0.273 [Cr1-Cr7 (0,-2,0)]<br>J11a=0.326 [Cr1-Cr5 (-1,-1,0)]<br>J11b=0.436 [Cr1-Cr8 (-1,-1,-1)]<br>J12a=0.468 [Cr5-Cr7 (-1,-1,0)]<br>J12b=0.082 [Cr5-Cr7 (0,-2,0)]<br>J13a=-0.564 [Cr1-Cr2 (0,0,-1)]<br>J13b=-0.408 [Cr5-Cr6 (0,0,-1)]<br>J14a=-0.552 [Cr1-Cr6 (-2,-1,-1)]<br>J14b=-0.134 [Cr1-Cr7 (-1,-2,0)]<br>J15a=0.080 [Cr1-Cr3 (-1,-1,-1)]<br>J15b=0.027 [Cr1-Cr3 (-1,-1,0)]<br>J16a=-0.297 [Cr1-Cr5 (-2,0,0)]<br>J16b=-0.032 [Cr1-Cr8 (0,-2,-1)]<br>J17a=-0.530 [Cr5-Cr8 (-1,-1,0)]<br>J17b=-0.604 [Cr5-Cr8 (0,-2,0)]<br>J18a=0.354 [Cr1-Cr1 (-2,0,0)]<br>J18b=0.114 [Cr1-Cr1 (0,-2,0)]<br>J18c=0.380 [Cr5-Cr5 (0,-2,0)]<br>J18d=-0.455 [Cr5-Cr5 (-2,0,0)]<br>J19a=-0.066 [Cr1-Cr2 (-1,0,-1)]<br>J19b=-0.122 [Cr1-Cr2 (0,-1,-1)]<br>J19c=-0.398 [Cr5-Cr6 (-1,0,-1)]<br>J19d=-0.595 [Cr5-Cr6 (0,-1,-1)]<br>J20=-0.043 [Cr1-Cr4 (-2,-2,0)]<br>J21a=-0.045 [Cr1-Cr6 (-1,-2,-1)]<br>J21b=0.068 [Cr1-Cr7 (-2,-1,0)]<br>J22a=-0.227 [Cr1-Cr5 (-2,-1,0)]<br>J22b=-0.208 [Cr1-Cr8 (-1,-2,-1)]<br>J23=0.180 [Cr5-Cr7 (-1,-2,0)] |
| 1.131 | $Fe_2As$ | Fe1 (0,0,0)<br>Fe2 (0,0,0.5)<br>Fe3 (0.5,0.5,0)<br>Fe4 (0.5,0.5,0.5)<br>Fe5 (0,0.5,0.165)<br>Fe6 (0,0.5,0.665)<br>Fe7 (0.5,0,0.835)<br>Fe8 (0.5,0,0.335) | 0 | J1=6.886 [Fe1-Fe3 (-1,-1,0)]<br>J2=4.277 [Fe1-Fe5 (0,-1,0)]<br>J3=4.208 [Fe1-Fe7 (-1,0,-1)]<br>J4=-2.189 [Fe5-Fe8 (-1,0,0)]<br>J5a=-0.893 [Fe1-Fe1 (-1,0,0)]<br>J5b=-0.010, $D_{5b}$= (0, 0.239, 0),<br>[Fe5-Fe5 (-1,0,0)]<br>J6=-1.759 [Fe1-Fe8 (-1,0,0)]<br>J7=-1.737 [Fe1-Fe6 (0,-1,-1)]<br>J8=1.183 [Fe1-Fe5 (-1,-1,0)]<br>J9=1.165 [Fe1-Fe7 (-1,-1,-1)]<br>J10=5.481, $\Gamma_{10}$=(0.001, 0.007, 0.007)<br>[Fe5-Fe7 (-1,0,-1)]<br>J11a=-3.929 [Fe1-Fe1 (-1,-1,0)]<br>J11b=1.048 [Fe5-Fe5 (-1,-1,0)] |
| | | | | J12=0.178, $J_{12}^{ani}$= (0.01, 0.011, -0.014),<br>[Fe1-Fe8 (-1,-1,0)]<br>J13=0.190 [Fe1-Fe6 (-1,-1,-1)]<br>J14=-0.309 [Fe1-Fe3 (-1,-2,0)]<br>J15=0.119 [Fe1-Fe5 (0,-2,0)]<br>J16=0.126 [Fe1-Fe7 (-2,0,-1)]<br>J17a=3.061 [Fe1-Fe2 (0,0,-1)]<br>J17b=1.217 [Fe5-Fe6 (0,0,-1)]<br>J18=0.213 [Fe5-Fe8 (-1,-1,0)]<br>J19=-0.743 [Fe1-Fe4 (-1,-1,-1)]<br>J20=-0.057 [Fe1-Fe8 (-2,0,0)]<br>J21=-0.053 [Fe1-Fe6 (0,-2,-1)]<br>J22=-0.120 [Fe1-Fe5 (-1,-2,0)]<br>J23=-0.112 [Fe1-Fe7 (-2,-1,-1)]<br>J24=1.165 [Fe5-Fe7 (-1,-1,-1)]<br>J25a=0.222 [Fe1-Fe2 (-1,0,-1)]<br>J25b=0.407 [Fe5-Fe6 (-1,0,-1)]<br>J26=0.118 [Fe1-Fe1 (-2,0,0)]<br>J27a=0.048 [Fe5-Fe5 (-2,0,0)]<br>J28=0.056 [Fe1-Fe8 (-2,-1,0)]<br>J29=0.058 [Fe1-Fe6 (-1,-2,-1)]<br>J30=-0.425 [Fe1-Fe3 (-2,-2,0)] |
| 1.132 | $Mn_2As$ | Mn1 (0,0,0)<br>Mn2 (0,0,0.5)<br>Mn3 (0.5,0.5,0)<br>Mn4 (0.5,0.5,0.5)<br>Mn5 (0.5,0,0.335)<br>Mn6 (0.5,0,0.835)<br>Mn7 (0,0.5,0.665)<br>Mn8 (0,0.5,0.165) | 0 | J1=-3.321, $J_1^{ani}$= (-0.017, -0.017, -0.049),<br>[Mn1-Mn3 (-1,-1,0)]<br>J2=-11.963, $D_2$= (0.236, 0, 0),<br>[Mn1-Mn8 (0,-1,0)]<br>J3=-12.517 [Mn1-Mn6 (-1,0,-1)]<br>J4=-15.953, $\Gamma_4$=(0.017, 0.006, -0.006)<br>[Mn5-Mn8 (0,-1,0)]<br>J5a=-1.405 [Mn1-Mn1 (-1,0,0)]<br>J5b=-2.049 [Mn5-Mn5 (-1,0,0)]<br>J6=1.062 [Mn1-Mn5 (-1,0,0)]<br>J7=0.865 [Mn1-Mn7 (0,-1,-1)]<br>J8=1.158 [Mn1-Mn8 (-1,-1,0)]<br>J9=1.071 [Mn1-Mn6 (-1,-1,-1)]<br>J10=-1.743 [Mn5-Mn7 (0,-1,0)]<br>J11a=-1.013 [Mn1-Mn1 (-1,-1,0)]<br>J11b=-0.120 [Mn5-Mn5 (-1,-1,0)]<br>J12=-1.164 [Mn1-Mn5 (-1,-1,0)]<br>J13=-0.859 [Mn1-Mn7 (-1,-1,-1)]<br>J14=-0.151 [Mn1-Mn3 (-1,-2,0)]<br>J15=-1.097 [Mn1-Mn8 (0,-2,0)]<br>J16=-0.965 [Mn1-Mn6 (-2,0,-1)]<br>J17a=0.173 [Mn1-Mn2 (0,0,-1)] |
| | | | | J17b=-1.299 [Mn5-Mn6 (0,0,-1)]<br>J18=0.589 [Mn5-Mn8 (-1,-1,0)]<br>J19=-0.314 [Mn1-Mn4 (-1,-1,-1)]<br>J20=0.724 [Mn1-Mn5 (-2,0,0)]<br>J21=0.764 [Mn1-Mn7 (0,-2,-1)]<br>J22=-0.476 [Mn1-Mn8 (-1,-2,0)]<br>J23=-0.326 [Mn1-Mn6 (-2,-1,-1)]<br>J24=0.857 [Mn5-Mn7 (-1,-1,0)]<br>J25a=-0.469 [Mn1-Mn2 (-1,0,-1)]<br>J25b=1.063 [Mn5-Mn6 (-1,0,-1)]<br>J26a=-0.611 [Mn1-Mn1 (-2,0,0)]<br>J27a=-0.355 [Mn5-Mn5 (-2,0,0)]<br>J28=0.056 [Mn1-Mn5 (-2,-1,0)]<br>J29=-0.174 [Mn1-Mn7 (-1,-2,-1)]<br>J30=0.583 [Mn1-Mn3 (-2,-2,0)] |
| 1.156 | $LaMn_3Cr_4O_{12}$ | Mn1 (0,0.5,0.5)<br>Mn2 (0.5,0,0.5)<br>Mn3 (0.5,0.5,0)<br>Mn4 (0.5,0,0)<br>Mn5 (0,0.5,0)<br>Mn6 (0,0,0.5)<br>Cr1 (0.25,0.25,0.25)<br>Cr2 (0.75,0.75,0.75)<br>Cr3 (0.75,0.75,0.25)<br>Cr4 (0.25,0.75,0.75)<br>Cr5 (0.75,0.25,0.75)<br>Cr6 (0.25,0.25,0.75)<br>Cr7 (0.75,0.25,0.25)<br>Cr8 (0.25,0.75,0.25) | 4.0<br>(Mn 3*d*)<br>3.0<br>(Cr 3*d*) | J1a=3.104, $D_1$= (-0.015, 0.083, 0.061),<br>$J_1^{ani}$= (-0.011, -0.01, -0.012),<br>$\Gamma_1$=(0.001, -0.002, 0.009)<br>[Cr1-Mn1 (0,0,0)]<br>J1b=-3.540 [Cr1-Mn4 (0,0,0)]<br>J2a=-2.340 [Cr1-Cr6 (0,0,-1)]<br>J2b=-0.947 [Mn1-Mn5 (0,0,0)]<br>J3a=-0.223 [Cr1-Cr3 (-1,-1,0)]<br>J3b=-0.183 [Cr1-Cr3 (0,0,0)]<br>J3c=-0.029 [Cr1-Cr3 (0,-1,0)]<br>J3d=0.068 [Cr1-Cr3 (-1,0,0)]<br>J3e=-0.083 [Mn1-Mn2 (-1,0,0)]<br>J4a=-0.069 [Cr1-Mn1 (1,0,0)]<br>J4b=-0.043 [Cr1-Mn1 (0,0,-1)]<br>J4c=0.065 [Cr1-Mn1 (0,-1,0)]<br>J4d=0.080 [Cr1-Mn4 (0,1,0)]<br>J4e=-0.055 [Cr1-Mn4 (0,0,1)]<br>J5a=-0.103 [Cr1-Cr2 (0,-1,0)]<br>J5b=-0.012 [Cr1-Cr2 (-1,-1,-1)]<br>J5c=0.027 [Mn1-Mn4 (0,1,0)]<br>J6a=-0.096 [Mn1-Mn1 (0,0,-1)]<br>J6b=0.003 [Mn1-Mn1 (-1,0,0)]<br>J6c=0.099 [Cr1-Cr1 (-1,0,0)] |
| 1.195 | $Er_2Ni_2In$† | Er1 (0.875,0.625,0.114)<br>Er2 (0.625,0.875,0.886)<br>Er3 (0.375,0.125,0.386)<br>Er4 (0.125,0.375,0.614)<br>Er5 (0.125,0.375,0.886)<br>Er6 (0.375,0.125,0.114) | 6.0<br>(Er 4*f*) | J1=-0.138, **$J_1^{ani}$= (0.072, 0.035, 0.023)**<br>[Er1-Er6 (0,0,0)]<br>J2a=0.024 [Er1-Er2 (0,0,-1)]<br>J2b=-0.025 [Er1-Er5 (1,0,-1)]<br>J3=0.262, $D_3$= (-0.016, -0.004, 0.002)<br>[Er1-Er8 (0,0,0)] |

| | | Er7 (0.625,0.875,0.614)<br>Er8 (0.875,0.625,0.386) | | J4=-0.053 [Er1-Er6 (0,1,0)]<br>J5a=-0.051, $\Gamma_{5a}$=(0, 0.004, 0)<br>[Er1-Er2 (0,-1,-1)]<br>J5b=0.004 [Er1-Er5 (0,0,-1)]<br>J6=-0.050 [Er1-Er3 (0,0,0)]<br>J7=0.001 [Er1-Er1 (-1,0,0)]<br>J8=-0.070 [Er1-Er3 (0,1,0)]<br>J9=0.021 [Er1-Er8 (-1,0,0)] |
|---|---|---|---|---|
| 1.215 | $UP_2$† | U1 (0,0,0.14)<br>U2 (0.5,0.5,0.36)<br>U3 (0.5,0.5,0.86)<br>U4 (0,0,0.64) | 0 | J1=5.226, $D_1$= (0.142, -0.001, 0)<br>[U1-U1 (-1,0,0)]<br>J2=-0.410, **$J_2^{ani}$= (3.214, 3.216, 0.678)**,<br>$\Gamma_2$=(-0.505, 0.25, 0.249)<br>[U1-U2 (-1,-1,0)]<br>J3=-0.667 [U1-U3 (-1,-1,-1)]<br>J4=3.237 [U1-U1 (-1,-1,0)]<br>J5=0.032 [U1-U2 (-1,-2,0)]<br>J6=0.225 [U1-U3 (-1,-2,-1)]<br>J7=0.175 [U1-U1 (-2,0,0)]<br>J8=-0.733 [U1-U4 (0,0,-1)]<br>J9=-0.183 [U1-U1 (-1,-2,0)]<br>J10=-0.069 [U1-U4 (-1,0,-1)]<br>J11=-0.159 [U1-U2 (-2,-2,0)]<br>J12=0.268 [U1-U3 (-2,-2,-1)]<br>J13=0.038 [U1-U4 (-1,-1,-1)]<br>J14=0.195 [U1-U2 (-1,-3,0)]<br>J15=0.104 [U1-U3 (-1,-3,-1)]<br>J16=-0.369 [U1-U1 (-2,-2,0)]<br>J17=0.144 [U1-U4 (-2,0,-1)]<br>J18=0.149 [U1-U1 (-3,0,0)]<br>J19=-0.009 [U1-U3 (-1,-1,0)]<br>J20=0.079 [U1-U4 (-1,-2,-1)]<br>J21=-0.197 [U1-U2 (-2,-3,0)] |
| 1.219 | $CuF_2$ | Cu1 (0,0,0)<br>Cu2 (0.5,0,0)<br>Cu3 (0,0.5,0.5)<br>Cu4 (0.5,0.5,0.5) | 4.0<br>(Cu 3*d*) | J1=0.102 [Cu1-Cu2 (-1,0,0)]<br>J2=-3.954, $D_2$= (-0.031, -0.264, -0.188) ,<br>$J_2^{ani}$= (-0.008, -0.011, -0.011),<br>$\Gamma_2$=(0.001, -0.001, -0.008)<br>[Cu1-Cu3 (0,-1,-1)]<br>J3=0.109 [Cu1-Cu4 (-1,-1,-1)]<br>J4=0.129 [Cu1-Cu1 (0,-1,0)]<br>J5=-0.016 [Cu1-Cu2 (-1,0,-1)]<br>J6=-0.016 [Cu1-Cu1 (0,0,-1)]<br>J7a=-0.003 [Cu1-Cu2 (-1,1,0)]<br>J7b=-0.210 [Cu1-Cu2 (-1,-1,0)]<br>J8=-0.004 [Cu1-Cu4 (-1,-1,0)] |
| | | | | J9=0.003 [Cu1-Cu1 (-1,0,-1)]<br>J10=0.000 [Cu1-Cu3 (-1,-1,-1)]<br>J11a=0.000 [Cu1-Cu2 (-1,1,-1)]<br>J11b=-0.427 [Cu1-Cu2 (-1,-1,-1)]<br>J12=0.000 [Cu1-Cu1 (-1,0,0)]<br>J13=0.039 [Cu1-Cu1 (0,-1,-1)] |
| 1.252 | $CaCo_2P_2$ | Co1 (0,0.5,0.25)<br>Co2 (0.5,0,0.75)<br>Co3 (0.5,0,0.25)<br>Co4 (0,0.5,0.75) | 2.0<br>(Co 3*d*) | J1=3.732, $J_1^{ani}$= (0.001, 0.001, 0.019),<br>$\Gamma_1$=(0.005, 0.008, -0.008)<br>[Co1-Co3 (-1,0,0)]<br>J2=-0.408, $D_2$= (0, -0.068, 0)<br>[Co1-Co1 (-1,0,0)]<br>J3=0.170 [Co1-Co4 (0,0,-1)]<br>J4=1.307 [Co1-Co1 (-1,-1,0)]<br>J5=0.069 [Co1-Co2 (-1,0,-1)]<br>J6=-0.145 [Co1-Co3 (-1,-1,0)]<br>J7a=-0.267 [Co1-Co4 (-1,0,-1)]<br>J7b=0.086 [Co1-Co4 (-1,0,0)]<br>J8=0.026 [Co1-Co4 (-1,-1,-1)]<br>J9a=-0.027 [Co1-Co1 (-2,0,0)]<br>J9b=-0.025 [Co1-Co1 (0,-2,0)]<br>J10=-0.063 [Co1-Co2 (-1,-1,-1)]<br>J11=0.274 [Co1-Co3 (-2,-1,0)] |
| 1.253 | $CeCo_2P_2$ | Co1 (0,0.5,0.25)<br>Co2 (0.5,0,0.25)<br>Co3 (0,0.5,0.75)<br>Co4 (0.5,0,0.75) | 2.0<br>(Co 3*d*) | J1=7.617, $J_1^{ani}$= (-0.011, -0.011, -0.004),<br>$\Gamma_1$=(0.002, 0.006, -0.006)<br>[Co1-Co2 (-1,0,0)]<br>J2=0.206 [Co1-Co1 (-1,0,0)]<br>J3=-0.353 [Co1-Co3 (0,0,0)]<br>J4=1.256, $D_4$= (0.036, 0.035, 0)<br>[Co1-Co1 (-1,-1,0)]<br>J5=-0.380 [Co1-Co4 (-1,0,-1)]<br>J6=-0.045 [Co1-Co2 (-1,-1,0)]<br>J7a=0.190 [Co1-Co3 (-1,0,0)]<br>J7b=-0.025 [Co1-Co3 (-1,0,-1)]<br>J8=0.083 [Co1-Co3 (-1,-1,-1)]<br>J9=0.154 [Co1-Co1 (-2,0,0)]<br>J10=-0.002 [Co1-Co4 (-1,-1,-1)]<br>J11=0.047 [Co1-Co2 (-2,-1,0)]<br>J12=-0.012 [Co1-Co1 (-1,-2,0)]<br>J13a=0.039 [Co1-Co3 (-2,0,-1)]<br>J13b=0.035 [Co1-Co3 (2,0,0)]<br>J14=-0.065 [Co1-Co4 (-2,-1,-1)]<br>J15=-0.173 [Co1-Co1 (0,0,-1)] |
| 1.263 | $Ca_3Ru_2O_7$<br>† | Ru1 (0.253,0.751,0.401)<br>Ru2 (0.747,0.251,0.599) | 0 | J1=3.014, $\Gamma_1$=(0.082, -0.013, 0.043)<br>[Ru1-Ru7 (0,0,0)] |
| | | Ru3 (0.247,0.251,0.901)<br>Ru4 (0.753,0.751,0.099)<br>Ru5 (0.753,0.751,0.901)<br>Ru6 (0.247,0.251,0.099)<br>Ru7 (0.747,0.251,0.401)<br>Ru8 (0.253,0.751,0.599) | | J2=7.293, $D_2$= (0.314, 1.063, 0),<br>$J_2^{ani}$= (-0.551, -0.154, -0.758)<br>[Ru1-Ru8 (0,0,0)]<br>J3=1.817 [Ru1-Ru7 (-1,0,0)]<br>J4=-0.073 [Ru1-Ru1 (-1,0,0)]<br>J5=-0.150 [Ru1-Ru2 (0,0,0)]<br>J6=-0.202 [Ru1-Ru2 (-1,0,0)]<br>J7=-0.101 [Ru1-Ru1 (0,-1,0)]<br>J9=-0.019 [Ru1-Ru4 (-1,0,0)]<br>J10=-0.087 [Ru1-Ru6 (0,0,0)]<br>J11=-0.004 [Ru1-Ru8 (-1,0,0)]<br>J12=-0.110 [Ru1-Ru8 (0,-1,0)]<br>J13a=-0.321 [Ru1-Ru1 (-1,-1,0)]<br>J13b=-0.189 [Ru1-Ru1 (-1,1,0)]<br>J14=0.040 [Ru1-Ru6 (1,0,0)]<br>J15=0.016 [Ru1-Ru6 (-1,0,0)]<br>J16=0.152 [Ru1-Ru7 (1,0,0)]<br>J17a=0.018 [Ru1-Ru4 (-1,-1,0)]<br>J17b=-0.021 [Ru1-Ru4 (-1,1,0)]<br>J18=0.064 [Ru1-Ru7 (-2,0,0)]<br>J19=0.155 [Ru1-Ru8 (-1,-1,0)] |
| 1.271 | CeSbTe | Ce1 (0.25,0.25,0.137)<br>Ce2 (0.75,0.75,0.363)<br>Ce3 (0.75,0.75,0.863)<br>Ce4 (0.25,0.25,0.637) | 4.0<br>(Ce 4*f*) | J1=0.108, $D_1$= (0, -0.003, 0),<br>$J_1^{ani}$= (-0.005, -0.01, -0.004)<br>[Ce1-Ce1 (-1,0,0)]<br>J2=-0.004 [Ce1-Ce2 (-1,-1,0)]<br>J3=0.034 [Ce1-Ce3 (-1,-1,-1)]<br>J4=0.015 [Ce1-Ce1 (-1,-1,0)]<br>J5=-0.004 [Ce1-Ce2 (-1,-2,0)]<br>J6=0.019 [Ce1-Ce3 (-1,-2,-1)]<br>J7=0.000 [Ce1-Ce1 (-2,0,0)]<br>J8=-0.004 [Ce1-Ce4 (0,0,-1)]<br>$\Gamma_{\mathrm{i}}$=(0, 0, 0) |
| 1.305 | $Mn_5Si_3$ | Mn1 (0.382,0.117,0.25)<br>Mn2 (0.882,0.383,0.75)<br>Mn3 (0.618,0.117,0.25)<br>Mn4 (0.118,0.383,0.75)<br>Mn5 (0.618,0.883,0.75)<br>Mn6 (0.118,0.617,0.25)<br>Mn7 (0.382,0.883,0.75)<br>Mn8 (0.882,0.617,0.25) | 0 | J1=-6.085, $D_1$= (0, -0.089, -0.589),<br>$J_1^{ani}$= (0.034, 0.014, 0.046)<br>$\Gamma_1$=(-0.010, -0.006, 0.006)<br>[Mn13-Mn15 (0,0,0)]<br>J2=1.850 [Mn13-Mn19 (0,-1,-1)]<br>J3=4.078 [Mn13-Mn17 (0,-1,-1)]<br>J4=-1.674 [Mn13-Mn16 (0,0,-1)]<br>J5=0.553 [Mn13-Mn18 (0,-1,0)]<br>J6=2.026 [Mn13-Mn13 (0,0,-1)]<br>J7=0.337 [Mn13-Mn15 (0,0,-1)]<br>J8=-2.397 [Mn13-Mn19 (0,0,-1)]<br>J9=-0.438 [Mn13-Mn16 (0,-1,-1)] |
| | | | | J10=-1.269 [Mn13-Mn17 (0,0,-1)]<br>J11=0.165 [Mn13-Mn14 (-1,0,-1)]<br>J12=-0.013 [Mn13-Mn18 (0,-1,-1)]<br>J13a=0.587 [Mn13-Mn13 (0,-1,0)]<br>J13b=0.371 [Mn13-Mn20 (-1,0,0)]<br>J13c=0.031 [Mn13-Mn20 (-1,-1,0)]<br>J14=0.051 [Mn13-Mn19 (0,-1,-2)]<br>J15=-0.372 [Mn13-Mn15 (0,-1,0)]<br>J16=-0.033 [Mn13-Mn17 (0,-1,-2)]<br>J17=0.618 [Mn13-Mn16 (0,0,-2)]<br>J18=-0.206 [Mn13-Mn14 (-1,-1,-1)]<br>J19a=0.356 [Mn13-Mn13 (0,-1,-1)]<br>J19b=0.334 [Mn13-Mn20 (-1,-1,-1)]<br>J19c=-0.056 [Mn13-Mn20 (-1,0,-1)]<br>J20=-0.077 [Mn13-Mn19 (0,-2,-1)]<br>J21=0.032 [Mn13-Mn15 (0,-1,-1)]<br>J22=0.255 [Mn13-Mn19 (0,0,-2)]<br>J23=0.328 [Mn13-Mn15 (-1,0,0)]<br>J24=-0.189 [Mn13-Mn17 (0,-2,-1)]<br>J25=0.432 [Mn13-Mn16 (1,0,-1)]<br>J26=-0.179 [Mn13-Mn16 (0,-1,-2)]<br>J27=0.210 [Mn13-Mn17 (0,0,-2)]<br>J28=0.099 [Mn13-Mn18 (1,-1,0)]<br>J29=0.091 [Mn13-Mn14 (-1,0,-2)]<br>J30=0.409 [Mn13-Mn17 (-1,-1,-1)]<br>J31=0.054 [Mn13-Mn16 (0,1,-1)]<br>J32=-0.790 [Mn13-Mn13 (0,0,-2)]<br>J33=-0.218 [Mn13-Mn15 (0,0,-2)]<br>J34=0.025 [Mn13-Mn15 (-1,0,-1)]<br>J35=0.049 [Mn13-Mn16 (1,-1,-1)]<br>J36=0.284 [Mn13-Mn18 (1,-1,-1)]<br>J37=0.283 [Mn16-Mn19 (-1,-1,-1)]<br>J38=-0.204 [Mn13-Mn14 (-1,-1,-2)]<br>J39=-0.083 [Mn13-Mn18 (0,-1,-2)]<br>J40=0.098 [Mn13-Mn17 (-1,0,-1)]<br>J41=0.255 [Mn13-Mn18 (0,-2,0)]<br>J42=0.091 [Mn13-Mn14 (-1,1,-1)]<br>J43=0.154 [Mn13-Mn19 (0,-2,-2)]<br>J44=-0.032 [Mn13-Mn15 (-1,-1,0)]<br>J45=0.177 [Mn13-Mn17 (0,-2,-2)]<br>J46=0.149 [Mn13-Mn16 (1,0,-2)]<br>J47=0.195 [Mn13-Mn17 (-1,-1,-2)]<br>J48a=-0.133 [Mn13-Mn13 (0,-1,-2)]<br>J48b=-0.234 [Mn13-Mn20 (-1,-1,-2)] |
| | | | | J48c=0.165 [Mn13-Mn20 (-1,0,-2)]<br>J49=0.011 [Mn13-Mn18 (0,-2,-1)]<br>J50a=0.336 [Mn13-Mn20 (-1,1,0)]<br>J50b=0.018 [Mn13-Mn20 (-1,-2,0)]<br>J51=-0.076 [Mn13-Mn13 (-1,0,0)]<br>J52=-0.348 [Mn13-Mn19 (0,-1,-3)]<br>J53=0.025 [Mn13-Mn15 (0,-1,-2)] |
| 1.341 | $TmMnO_3$ | Mn1 (0,0,0.5)<br>Mn2 (0.25,0,0)<br>Mn3 (0,0.5,0.5)<br>Mn4 (0.25,0.5,0)<br>Mn5 (0.5,0,0.5)<br>Mn6 (0.75,0,0)<br>Mn7 (0.5,0.5,0.5)<br>Mn8 (0.75,0.5,0) | 3.1<br>(Mn 3*d*) | J1a=2.019 [Mn1-Mn3 (0,-1,0)]<br>J1b=2.287 [Mn1-Mn3 (0,0,0)]<br>J2a=7.293, $D_{2a}$= (0.110, -0.135, 0.087),<br>$J_{2a}^{ani}$=(-0.04, -0.041, -0.041)<br>[Mn1-Mn2 (0,0,0)]<br>J2b=5.971 [Mn1-Mn6 (-1,0,0)]<br>J3=0.029 [Mn1-Mn1 (0,0,-1)]<br>J4a=0.045 [Mn1-Mn4 (0,-1,1)]<br>J4b=-0.083 [Mn1-Mn4 (0,-1,0)]<br>J4c=-0.387 [Mn1-Mn8 (-1,-1,1)]<br>J4d=-0.474 [Mn1-Mn8 (-1,-1,0)]<br>J5=-0.347 [Mn1-Mn5 (-1,0,0)]<br>J6a=0.001 [Mn1-Mn3 (0,-1,-1)]<br>J6b=-0.020 [Mn1-Mn3 (0,0,-1)]<br>J7a=0.001 [Mn1-Mn7 (-1,-1,0)]<br>J7b=0.006 [Mn1-Mn7 (-1,0,0)]<br>J8=0.114 [Mn1-Mn1 (0,-1,0)]<br>J9a=-0.886 [Mn1-Mn5 (-1,0,-1)]<br>J9b=-0.055 [Mn1-Mn5 (-1,0,1)]<br>$J_{i}^{ani}$=$\Gamma_{i}$=(0, 0, 0) |
| 1.349 | $CoNb_3S_6$ | Co1 (0.167,0.5,0.25)<br>Co2 (0.333,0,0.75)<br>Co3 (0.667,0,0.25)<br>Co4 (0.833,0.5,0.75) | 1.0<br>(Co 3*d*) | J1a=-2.267, $J_{1a}^{ani}$= (0.015, 0.007, 0.052)<br>[Co1-Co1 (0,-1,0)]<br>J1b=-0.868 [Co1-Co3 (-1,0,0)]<br>J2a=-1.477, $D_{2a}$= (0.080, 0.026, -0.097)<br>[Co1-Co2 (0,0,-1)]<br>J2b=-1.144 [Co1-Co4 (-1,0,-1)]<br>J3a=0.824, $\Gamma_{3a}$=(0, 0.012, 0)<br>[Co1-Co4 (-1,-1,-1)]<br>J3b=0.226 [Co1-Co4 (0,0,-1)]<br>J4a=-0.423 [Co1-Co1 (-1,0,0)]<br>J4b=-0.049 [Co1-Co3 (-1,-1,0)]<br>J5a=0.086 [Co1-Co2 (-1,0,-1)]<br>J5b=0.562 [Co1-Co2 (0,-1,-1)]<br>J5c=0.369 [Co1-Co4 (0,-1,-1)]<br>J6a=0.050 [Co1-Co1 (-1,-1,0)]<br>J6b=-0.283 [Co1-Co1 (0,-2,0)]<br>J7=0.230 [Co1-Co1 (0,0,-1)] |
| | | | | J8a=-0.478 [Co1-Co1 (0,-1,-1)]<br>J8b=-0.433 [Co1-Co1 (0,-1,1)]<br>J8c=-0.058 [Co1-Co3 (-1,0,-1)]<br>J8d=0.039 [Co1-Co3 (-1,0,1)]<br>J9a=-0.038 [Co1-Co2 (-1,-1,-1)]<br>J9b=-0.142 [Co1-Co2 (1,0,-1)]<br>J9c=0.036 [Co1-Co4 (-1,-2,-1)]<br>J10a=0.026 [Co1-Co4 (-2,0,-1)]<br>J10b=-0.005 [Co1-Co4 (0,-2,-1)]<br>J11a=-0.110 [Co1-Co1 (-1,-2,0)]<br>J11b=0.084 [Co1-Co3 (-1,-2,0)]<br>J11c=0.020 [Co1-Co3 (-2,0,0)]<br>J12a=-0.041 [Co1-Co1 (-1,0,-1)]<br>J12b=-0.051 [Co1-Co3 (-1,-1,-1)]<br>J12c=-0.033 [Co1-Co3 (-1,-1,1)]<br>J13a=0.080 [Co1-Co2 (0,-2,-1)]<br>J13b=0.026 [Co1-Co2 (1,-1,-1)]<br>J13c=-0.009 [Co1-Co4 (-2,-1,-1)]<br>J14a=0.001 [Co1-Co1 (-1,-1,-1)]<br>J14b=-0.063 [Co1-Co1 (-1,-1,1)]<br>J14c=0.353 [Co1-Co1 (0,-2,1)]<br>J14d=0.090 [Co1-Co1 (0,-2,-1)]<br>J15a=0.211 [Co1-Co1 (0,-3,0)]<br>J15b=-0.068 [Co1-Co3 (-2,-1,0)] |
| 1.351 | $Ba_2Co_2F_7Cl$ | Co1 (0.413,0.25,0.734)<br>Co2 (0.587,0.75,0.266)<br>Co3 (0.913,0.25,0.734)<br>Co4 (0.087,0.75,0.266)<br>Co5 (0.409,0.25,0.175)<br>Co6 (0.591,0.75,0.825)<br>Co7 (0.909,0.25,0.175)<br>Co8 (0.091,0.75,0.825) | 3.3<br>(Co 3*d*) | J1=-4.755, $D_1$= (0.066, 0.180, 0.029),<br>$J_1^{ani}$= (-0.021, -0.039, -0.017)<br>$\Gamma_1$=(0.004, 0.001, 0.004)<br>[Co1-Co6 (0,-1,0)]<br>J2=-4.075 [Co1-Co5 (0,0,1)]<br>J3=-3.818 [Co1-Co5 (0,0,0)]<br>J4=0.013 [Co1-Co2 (0,-1,1)]<br>J5=-0.006 [Co5-Co8 (0,-1,-1)]<br>J6=-0.058 [Co1-Co1 (0,-1,0)]<br>J7=-0.011 [Co1-Co8 (0,-1,0)]<br>J8=-0.002 [Co1-Co4 (0,-1,0)] |
| 1.371 | $Nd_2NiO_4$ | Ni1 (0, 0, 0)<br>Ni2 (0.5, 0.5, 0)<br>Ni3 (0.5, 0, 0.5)<br>Ni4 (0, 0.5, 0.5) | 4.2<br>(Ni 3*d*) | J1=-11.846, $D_1$= (-0.142, -0.112, -0.002),<br>$J_1^{ani}$= (-0.006, -0.005, -0.008)<br>$\Gamma_1$=(0.002, 0, 0)<br>[Ni1-Ni2 (-1,-1,0)]<br>J2=0.535 [Ni1-Ni1 (-1,0,0)]<br>J3=0.614 [Ni1-Ni1 (0,-1,0)]<br>J4=-0.002 [Ni1-Ni3 (-1,0,-1)]<br>J5=-0.001 [Ni1-Ni4 (0,-1,-1)]<br>J6=0.256 [Ni1-Ni1 (-1,-1,0)] |

| | | | | J7=-0.019 [Ni1-Ni2 (-2,-1,0)] |
|---|---|---|---|---|
| 1.384 | $USb_2$† | U1 (0.25,0.25,0.14)<br>U2 (0.75,0.75,0.36)<br>U3 (0.75,0.75,0.86)<br>U4 (0.25,0.25,0.64) | 0 | J1=8.782, $\boldsymbol{D_1}$= **(0, -3.654, -0.006)**,<br>$J_1^{ani}$= (-4.681, -4.743, -5.1)<br>$\Gamma_1$=(-0.582, 0.072, 0.072)<br>[U1-U1 (-1,0,0)]<br>J2=5.950 [U1-U2 (-1,-1,0)]<br>J3=-0.233 [U1-U3 (-1,-1,-1)]<br>J4=4.106 [U1-U1 (-1,-1,0)]<br>J5=-1.633 [U1-U2 (-1,-2,0)]<br>J6=-0.002 [U1-U3 (-1,1,-1)]<br>J7=0.106 [U1-U1 (-2,0,0)]<br>J8=-0.434 [U1-U4 (0,0,-1)]<br>J9=0.182 [U1-U1 (-1,-2,0)]<br>J10=0.166 [U1-U4 (-1,0,-1)]<br>J11=0.446 [U1-U2 (-2,-2,0)]<br>J12=0.155 [U1-U3 (-2,-2,-1)]<br>J13=0.277 [U1-U4 (-1,-1,-1)]<br>J14=-0.093 [U1-U2 (-1,-3,0)]<br>J15=-0.085 [U1-U3 (-1,-3,-1)]<br>J16=-0.475 [U1-U1 (-2,-2,0)]<br>J17=0.180 [U1-U4 (-2,0,-1)]<br>J18=0.046 [U1-U1 (-3,0,0)]<br>J19=0.063 [U1-U3 (-1,-1,0)]<br>J20=0.308 [U1-U4 (-1,-2,-1)]<br>J21=-0.424 [U1-U2 (-2,-3,0)] |
| 1.388 | $La_2NiO_3F_2$ | Ni1 (0.25,0.25,0)<br>Ni2 (0.25,0.75,0.5)<br>Ni3 (0.75,0.75,0)<br>Ni4 (0.75,0.25,0.5) | 4.2<br>(Ni 3*d*) | J1=-10.446, $D_1$= (0, 0.033, 0.067),<br>$J_1^{ani}$= (-0.012, -0.008, -0.009)<br>[Ni1-Ni2 (0,-1,-1)]<br>J2=0.475, $\Gamma_2$=(0, 0.001, -0.032)<br>[Ni1-Ni1 (0,0,-1)]<br>J3=0.745 [Ni1-Ni1 (0,-1,0)]<br>J4=0.000 [Ni1-Ni4 (-1,0,-1)]<br>J5a=-0.029 [Ni1-Ni3 (-1,-1,0)]<br>J5b=-0.007 [Ni1-Ni3 (-1,0,0)]<br>J6=0.056 [Ni1-Ni1 (0,-1,-1)]<br>J7=-0.029 [Ni1-Ni2 (0,-1,-2)]<br>J8a=-0.000 [Ni1-Ni3 (-1,-1,-1)]<br>J8b=-0.000 [Ni1-Ni3 (-1,0,-1)]<br>J9=-0.001 [Ni1-Ni4 (-1,-1,-1)]<br>J10=-0.035 [Ni1-Ni2 (0,-2,-1)] |
| 1.389 | $Sr_2CoO_3Cl$† | Co1 (0,0,0.795)<br>Co2 (0.5,0,0.205)<br>Co3 (0.5,0.5,0.795)<br>Co4 (0,0.5,0.205) | 3.3<br>(Co 3*d*) | J1=-20.925, $\boldsymbol{D_1}$= **(1.908, -1.907, -0.002)**,<br>$J_1^{ani}$= (-0.301, -0.301, -0.234)<br>$\Gamma_1$=(-0.049, 0, 0)<br>[Co1-Co3 (-1,-1,0)] |
| | | | | J2=0.819 [Co1-Co1 (-1,0,0)]<br>J3a=-0.179 [Co1-Co2 (-1,0,1)]<br>J3b=-0.217 [Co1-Co4 (0,-1,1)]<br>J4=0.294 [Co1-Co1 (-1,-1,0)] |
| 1.414 | $CeNiGe_3$† | Ce1 (0.168,0,0.5)<br>Ce2 (0.668,0.5,0.5)<br>Ce3 (0.332,0.5,0.5)<br>Ce4 (0.832,0,0.5) | 6.0<br>(Ce 4*f*) | J1=-0.005 [Ce1-Ce3 (0,-1,0)]<br>J2=0.044, **$D_2$= (0, 0, 0.012)**,<br>$J_2^{ani}$= (0.002, 0, 0.004)<br>[Ce1-Ce1 (0,-1,0)]<br>J3=0.076 [Ce1-Ce1 (0,0,-1)]<br>J5=0.034 [Ce1-Ce3 (0,-1,-1)]<br>J6=0.041 [Ce1-Ce1 (0,-1,-1)]<br>J7=0.006 [Ce1-Ce3 (0,-2,0)]<br>J8=0.015 [Ce1-Ce4 (-1,0,0)]<br>J9=-0.001 [Ce1-Ce1 (0,-2,0)]<br>J10=-0.010 [Ce1-Ce3 (0,-2,-1)]<br>J11=-0.007 [Ce1-Ce1 (0,0,-2)]<br>J12=0.001 [Ce1-Ce4 (-1,-1,0)]<br>J13=-0.010 [Ce1-Ce4 (-1,0,-1)]<br>J14=0.004 [Ce1-Ce1 (0,-2,-1)]<br>J15=-0.001 [Ce1-Ce3 (0,-1,-2)]<br>J16=-0.002 [Ce1-Ce1 (0,-1,-2)]<br>J17=0.001 [Ce1-Ce4 (-1,-1,-1)]<br>J18=-0.001 [Ce1-Ce3 (0,-3,0)]<br>J19=0.009 [Ce1-Ce3 (0,-2,-2)]<br>$\Gamma_1$=(0, 0, 0) |
| 1.420 | $YBa_2Cu_3O_6$ | Cu1 (0.75,0.75,0.639)<br>Cu2 (0.75,0.75,0.361)<br>Cu3 (0.25,0.25,0.361)<br>Cu4 (0.25,0.25,0.639) | 6.7<br>(Cu 3*d*) | J1=-0.345 [Cu1-Cu2 (0,0,0)]<br>J2=-7.158, $D_2$= (0.255, 0.255, 0),<br>$\Gamma_2$=(0.002, 0, 0)<br>[Cu1-Cu4 (0,0,0)]<br>J3=0.011 [Cu1-Cu3 (0,0,0)]<br>J4=1.608, $J_4^{ani}$= (-0.006, -0.005, -0.008)<br>[Cu1-Cu1 (-1,0,0)]<br>J5=-0.062 [Cu1-Cu2 (-1,0,0)]<br>J6=-0.605 [Cu1-Cu1 (-1,-1,0)]<br>J7=0.003 [Cu1-Cu2 (-1,-1,0)]<br>J8=-0.000 [Cu1-Cu2 (0,0,1)]<br>J9=-0.159 [Cu1-Cu4 (-1,0,0)]<br>J10=0.007 [Cu1-Cu3 (-1,0,0)]<br>J11=-0.000 [Cu1-Cu2 (-1,0,1)]<br>J12=0.087 [Cu1-Cu1 (-2,0,0)]<br>J13=-0.005 [Cu1-Cu2 (-2,0,0)]<br>J14=0.015 [Cu1-Cu4 (-1,-1,0)] |
| 1.425 | UGeTe† | U1 (0,0,0.124)<br>U2 (0,0,0.876) | 0 | J1=9.354, **$D_1$= (0, 1.551, 0)**,<br>[U1-U1 (-1,0,0)] |
| | | U3 (0.5,0.5,0.376)<br>U4 (0.5,0.5,0.624) | | J2=0.026 [U1-U2 (0,0,-1)]<br>J3=-2.114, **$J_3^{ani}$= (1.824, 1.825, 0.826)**,<br>$\Gamma_3$=(0.719, 0.155, 0.154)<br>[U1-U3 (-1,-1,0)]<br>J4=0.745 [U1-U1 (-1,-1,0)]<br>J5=2.724 [U1-U2 (-1,0,-1)]<br>J6=1.092 [U1-U2 (-1,-1,-1)]<br>J7=-0.540 [U1-U3 (-1,-2,0)]<br>J8=-0.494 [U1-U1 (-2,0,0)]<br>J9=-0.221 [U1-U1 (-1,-2,0)]<br>J10=-0.274 [U1-U4 (-1,-1,-1)]<br>J11=0.621 [U1-U2 (-2,0,-1)]<br>J12=-0.698 [U1-U3 (-2,-2,0)]<br>J13=0.180 [U1-U2 (-1,-2,-1)]<br>J14=-0.232 [U1-U4 (-1,-2,-1)]<br>J15=-0.064 [U1-U3 (-1,-3,0)]<br>J16=-0.545 [U1-U1 (-2,-2,0)]<br>J17=-0.183 [U1-U1 (-3,0,0)]<br>J18=-0.027 [U1-U4 (-2,-2,-1)]<br>J19=0.480 [U1-U2 (-2,-2,-1)] |
| 1.426 | UGeS† | U1 (0,0.5,0.144)<br>U2 (0.5,0,0.856)<br>U3 (0.5,0,0.356)<br>U4 (0,0.5,0.644) | 0 | J1=6.960, **$D_1$= (1.834, -0.003, 0)**,<br>**$J_1^{ani}$= (5.736, -0.093, 0.369)**<br>[U1-U1 (-1,0,0)]<br>J2=0.753, **$\Gamma_2$=(-0.946, -0.231, -0.231)**<br>[U1-U3 (-1,0,0)]<br>J3=2.842 [U1-U1 (-1,-1,0)]<br>J4=-0.579 [U1-U2 (-1,0,-1)]<br>J5=-0.984 [U1-U3 (-1,-1,0)]<br>J6=1.949 [U1-U1 (-2,0,0)]<br>J7=0.157 [U1-U2 (-1,-1,-1)]<br>J8=-0.012 [U1-U4 (0,0,-1)]<br>J9=0.405 [U1-U1 (-1,-2,0)]<br>J10=-0.321 [U1-U3 (-2,-1,0)]<br>J11=-0.216 [U1-U4 (-1,0,-1)]<br>J12=0.143 [U1-U2 (-2,-1,-1)]<br>J13=-0.045 [U1-U4 (-1,-1,-1)]<br>J14=0.561 [U1-U3 (-1,-2,0)]<br>J15=-0.270 [U1-U1 (-2,-2,0)]<br>J16=-0.078 [U1-U2 (-1,-2,-1)]<br>J17=-0.130 [U1-U4 (-2,0,-1)]<br>J18=0.915 [U1-U1 (-3,0,0)] |
| 1.438 | $BaCoF_4$ | Co1 (0.294,0,0.175)<br>Co2 (0.706,0.5,0.825)<br>Co3 (0.794,0,0.175) | 2.5<br>(Co 3*d*) | J1=-2.737, $J_1^{ani}$= (-0.024, -0.022, -0.029)<br>[Co1-Co4 (0,-1,-1)]<br>J2=-4.662, $D_2$= (0, 0.107, 0.039), |
| | | Co4 (0.206,0.5,0.825) | | $\Gamma_2$=(0, 0, -0.003)<br>[Co1-Co3 (-1,0,0)]<br>J3=-0.004 [Co1-Co2 (-1,-1,-1)]<br>J4=-0.035 [Co1-Co1 (0,-1,0)]<br>J5a=0.002 [Co1-Co2 (0,-1,0)]<br>J5b=0.003 [Co1-Co4 (0,-1,0)]<br>J6=0.001 [Co1-Co3 (-1,-1,0)]<br>J7a=0.001 [Co1-Co1 (0,0,-1)]<br>J7b=-0.000 [Co1-Co3 (-1,0,-1)]<br>J8=0.035 [Co1-Co1 (-1,0,0)] |
| 1.468 | $TbMn_2Si_2$ | Mn1 (0,0.5,0.25)<br>Mn2 (0.5,0,0.25)<br>Mn3 (0,0.5,0.75)<br>Mn4 (0.5,0,0.75) | 0 | J1=1.878, $J_1^{ani}$= (-0.018, -0.018, 0.046)<br>$\Gamma_1$=(-0.004, -0.019, 0.019)<br>[Mn1-Mn2 (-1,0,0)]<br>J2=5.495 [Mn1-Mn1 (-1,0,0)]<br>J3=0.610 [Mn1-Mn3 (0,0,-1)]<br>J4=-0.149 [Mn1-Mn1 (-1,-1,0)]<br>J5=1.753, $D_5$= (-0.046, -0.046, -0.007)<br>[Mn1-Mn4 (-1,0,-1)]<br>J6=1.006 [Mn1-Mn2 (-1,-1,0)]<br>J7a=-0.348 [Mn1-Mn3 (-1,0,0)]<br>J7b=-2.543 [Mn1-Mn3 (-1,0,-1)]<br>J8=-0.234 [Mn1-Mn3 (-1,-1,-1)]<br>J9=0.143 [Mn1-Mn1 (-2,0,0)]<br>J10=0.040 [Mn1-Mn4 (-1,-1,-1)]<br>J11=0.987 [Mn1-Mn2 (-2,-1,0)]<br>J12=-0.061 [Mn1-Mn1 (-1,-2,0)]<br>J13a=-0.040 [Mn1-Mn3 (-2,0,0)]<br>J13b=-0.525 [Mn1-Mn3 (-2,0,-1)]<br>J14=0.193 [Mn1-Mn4 (-2,-1,-1)]<br>J15=-0.017 [Mn1-Mn2 (-1,-2,0)]<br>J16a=0.093 [Mn1-Mn3 (-1,-2,0)]<br>J16b=-0.113 [Mn1-Mn3 (-1,-2,-1)]<br>J17=0.749 [Mn1-Mn1 (0,0,-1)]<br>J18=0.087 [Mn1-Mn2 (-1,0,-1)]<br>J19=0.130 [Mn1-Mn1 (-2,-2,0)]<br>J20=-0.000 [Mn1-Mn1 (-1,0,-1)]<br>J21=-0.044 [Mn1-Mn4 (-1,-2,-1)]<br>J22=-0.136 [Mn1-Mn2 (-2,-2,0)]<br>J23=-0.034 [Mn1-Mn1 (-3,0,0)]<br>J24=-0.236 [Mn1-Mn1 (-1,-1,-1)] |
| 1.469,<br>1.495 | $YMn_2Si_2$ | Mn1 (0,0.5,0.25)<br>Mn2 (0.5,0,0.25)<br>Mn3 (0,0.5,0.75)<br>Mn4 (0.5,0,0.75) | 0 | J1=3.305, $\Gamma_1$=(-0.025, 0.002, -0.002)<br>[Mn1-Mn2 (-1,0,0)]<br>J2=5.334, $D_2$= (0, -0.085, 0),<br>$J_2^{ani}$= (-0.017, 0.010, 0.006) |
| | | | | [Mn1-Mn1 (-1,0,0)]<br>J3=0.243 [Mn1-Mn3 (0,0,-1)]<br>J4=-0.116 [Mn1-Mn1 (-1,-1,0)]<br>J5=1.562 [Mn1-Mn4 (-1,0,-1)]<br>J6=0.909 [Mn1-Mn2 (-1,-1,0)]<br>J7a=-0.588 [Mn1-Mn3 (-1,0,0)]<br>J7b=-2.099 [Mn1-Mn3 (-1,0,-1)]<br>J8=-0.181 [Mn1-Mn3 (-1,-1,-1)]<br>J9=0.056 [Mn1-Mn1 (-2,0,0)]<br>J10=0.056 [Mn1-Mn4 (-1,-1,-1)]<br>J11=0.768 [Mn1-Mn2 (-2,-1,0)]<br>J12=-0.171 [Mn1-Mn1 (-1,-2,0)]<br>J13a=-0.125 [Mn1-Mn3 (-2,0,0)]<br>J13b=-0.327 [Mn1-Mn3 (-2,0,-1)]<br>J14=0.208 [Mn1-Mn4 (-2,-1,-1)]<br>J15=0.041 [Mn1-Mn2 (-1,-2,0)]<br>J16a=0.002 [Mn1-Mn3 (-1,-2,0)]<br>J16b=-0.118 [Mn1-Mn3 (-1,-2,-1)]<br>J17=0.657 [Mn1-Mn1 (0,0,-1)]<br>J18=-0.097 [Mn1-Mn2 (-1,0,-1)]<br>J19=0.172 [Mn1-Mn1 (-2,-2,0)]<br>J20=-0.020 [Mn1-Mn1 (-1,0,-1)]<br>J21=-0.025 [Mn1-Mn4 (-1,-2,-1)]<br>J22=-0.152 [Mn1-Mn2 (-2,-2,0)]<br>J23=-0.162 [Mn1-Mn1 (-3,0,0)]<br>J24=-0.151 [Mn1-Mn1 (-1,-1,-1)] |
| 1.472 | CaOFeS | Fe1 (0.75,0.25,0)<br>Fe2 (0.25,0.75,0.5)<br>Fe3 (0.75,0.25,0.5)<br>Fe4 (0.25,0.75,0) | 2.0<br>(Fe 3*d*) | J1a=-3.936, $D_1$= (0.276, 0, -0.325),<br>$J_1^{ani}$= (-0.023, -0.03, -0.043)<br>$\Gamma_1$=(0.011, 0.005, 0.007)<br>[Fe1-Fe1 (0,-1,0)]<br>J1b=-6.511 [Fe1-Fe4 (0,-1,0)]<br>J2=-0.357 [Fe1-Fe3 (0,0,-1)]<br>J3a=0.001 [Fe1-Fe1 (-1,0,0)]<br>J3b=-0.129 [Fe1-Fe4 (0,-2,0)]<br>J4a=-0.237 [Fe1-Fe3 (0,-1,-1)]<br>J4b=-0.229 [Fe1-Fe2 (0,-1,-1)]<br>J4c=-0.104 [Fe1-Fe2 (1,-1,-1)]<br>J5a=-0.286 [Fe1-Fe1 (-1,-1,0)]<br>J5b=-0.018 [Fe1-Fe1 (0,-2,0)]<br>J6a=-0.002 [Fe1-Fe2 (0,-2,-1)]<br>J6b=-0.029 [Fe1-Fe2 (1,-2,-1)]<br>J6c=-0.039 [Fe1-Fe3 (-1,0,-1)]<br>J6d=0.000 [Fe1-Fe3 (1,0,-1)]<br>J7a=-0.004 [Fe1-Fe3 (1,-1,-1)] |
| | | | | J7b=-0.004 [Fe1-Fe3 (-1,-1,-1)]<br>J7c=0.001 [Fe1-Fe3 (0,-2,-1)]<br>J8a=-0.019 [Fe1-Fe1 (-1,-2,0)]<br>J8b=-0.004 [Fe1-Fe4 (-1,-1,0)]<br>J8c=-0.002 [Fe1-Fe4 (0,-3,0)]<br>J9a=-0.006 [Fe1-Fe1 (0,-3,0)]<br>J9b=-0.019 [Fe1-Fe4 (-1,-2,0)] |
| 1.488,<br>1.489,<br>1.490 | $CeMn_2Si_2$ | Mn1 (0,0.5,0.25)<br>Mn2 (0.5,0,0.25)<br>Mn3 (0,0.5,0.75)<br>Mn4 (0.5,0,0.75) | 0 | J1=4.726, $\Gamma_1$=(0.002, -0.018, 0.018)<br>[Mn1-Mn2 (-1,0,0)]<br>J2=5.974, $D_2$= (-0.087, 0, 0),<br>$J_2^{ani}$= (-0.001, -0.003, 0.086)<br>[Mn1-Mn1 (-1,0,0)]<br>J3=-1.358 [Mn1-Mn3 (0,0,-1)]<br>J4=-0.302 [Mn1-Mn1 (-1,-1,0)]<br>J5=-2.022 [Mn1-Mn4 (-1,0,-1)]<br>J6=0.853 [Mn1-Mn2 (-1,-1,0)]<br>J7a=0.821 [Mn1-Mn3 (-1,0,0)]<br>J7b=-2.837 [Mn1-Mn3 (-1,0,-1)]<br>J8=-0.091 [Mn1-Mn3 (-1,-1,-1)]<br>J9=0.017 [Mn1-Mn1 (-2,0,0)]<br>J10=-0.123 [Mn1-Mn4 (-1,-1,-1)]<br>J11=0.982 [Mn1-Mn2 (-2,-1,0)]<br>J12=-0.184 [Mn1-Mn1 (-1,-2,0)]<br>J13a=0.028 [Mn1-Mn3 (-2,0,0)]<br>J13b=-0.324 [Mn1-Mn3 (-2,0,-1)]<br>J14=-0.027 [Mn1-Mn4 (-2,-1,-1)]<br>J15=-0.016 [Mn1-Mn2 (-1,-2,0)]<br>J16a=0.194 [Mn1-Mn3 (-1,-2,0)]<br>J16b=-0.245 [Mn1-Mn3 (-1,-2,-1)]<br>J17=0.953 [Mn1-Mn1 (0,0,-1)]<br>J18=0.401 [Mn1-Mn2 (-1,0,-1)]<br>J19=-0.096 [Mn1-Mn1 (-1,0,-1)]<br>J20=-0.083 [Mn1-Mn1 (-2,-2,0)] |
| 1.491,<br>1.492 | $PrMn_2Si_2$ | Mn1 (0,0.5,0.25)<br>Mn2 (0.5,0,0.25)<br>Mn3 (0,0.5,0.75)<br>Mn4 (0.5,0,0.75) | 0 | J1=6.663, $J_1^{ani}$ = (-0.018, -0.018, 0.075),<br>$\Gamma_1$= (0.003, -0.018, 0.018)<br>[Mn1-Mn2 (-1,0,0)]<br>J2=5.797, $D_2$= (0, -0.049, 0)<br>[Mn1-Mn1 (-1,0,0)]<br>J3=0.860 [Mn1-Mn3 (0,0,0)]<br>J4=-0.247 [Mn1-Mn1 (-1,-1,0)]<br>J5=1.789 [Mn1-Mn4 (-1,0,-1)]<br>J6=0.796 [Mn1-Mn2 (-1,-1,0)]<br>J7a=0.051 [Mn1-Mn3 (-1,0,0)]<br>J7b=-3.292 [Mn1-Mn3 (-1,0,-1)] |
| | | | | J8=-0.198 [Mn1-Mn3 (-1,-1,-1)]<br>J9=-0.007 [Mn1-Mn1 (-2,0,0)]<br>J10=-0.047 [Mn1-Mn4 (-1,-1,-1)]<br>J11=1.180 [Mn1-Mn2 (-2,-1,0)]<br>J12=0.088 [Mn1-Mn1 (-1,-2,0)]<br>J13a=-0.105 [Mn1-Mn3 (-2,0,0)]<br>J13b=-0.613 [Mn1-Mn3 (-2,0,-1)]<br>J14=0.039 [Mn1-Mn4 (-2,-1,-1)]<br>J15=-0.100 [Mn1-Mn2 (-1,-2,0)]<br>J16a=0.205 [Mn1-Mn3 (-1,-2,0)]<br>J16b=-0.355 [Mn1-Mn3 (-1,-2,-1)]<br>J17=0.792 [Mn1-Mn1 (0,0,-1)]<br>J18=0.159 [Mn1-Mn2 (-1,0,-1)]<br>J19=-0.117 [Mn1-Mn1 (-1,0,-1)]<br>J20=-0.036 [Mn1-Mn1 (-2,-2,0)]<br>J21=-0.052 [Mn1-Mn4 (-1,-2,-1)]<br>J22=-0.052 [Mn1-Mn2 (-2,-2,0)]<br>J23=-0.552 [Mn1-Mn1 (-1,-1,-1)]<br>J24=-0.122 [Mn1-Mn1 (-3,0,0)]<br>J25a=0.204 [Mn1-Mn2 (-1,-1,-1)]<br>J25b=0.081 [Mn1-Mn2 (-1,-1,1)] |
| 1.493, 1.494 | $NdMn_2Si_2$ | Mn1 (0,0.5,0.25)<br>Mn2 (0.5,0,0.25)<br>Mn3 (0,0.5,0.75)<br>Mn4 (0.5,0,0.75) | 0 | J1=-1.161, $J_1^{ani}$= (-0.019, -0.019, 0.071), $\Gamma_1$= (-0.005, -0.02, 0.02)<br>[Mn1-Mn2 (-1,0,0)]<br>J2=6.005 [Mn1-Mn1 (-1,0,0)]<br>J3=-0.085 [Mn1-Mn3 (0,0,-1)]<br>J4=-0.927 [Mn1-Mn1 (-1,-1,0)]<br>J5=1.995, $D_5$= (-0.059, -0.059, -0.029)<br>[Mn1-Mn4 (-1,0,-1)]<br>J6=0.728 [Mn1-Mn2 (-1,-1,0)]<br>J7a=-0.322 [Mn1-Mn3 (-1,0,0)]<br>J7b=-2.438 [Mn1-Mn3 (-1,0,-1)]<br>J8=-0.125 [Mn1-Mn3 (-1,-1,-1)]<br>J9=-0.142 [Mn1-Mn1 (-2,0,0)]<br>J10=-0.002 [Mn1-Mn4 (-1,-1,-1)]<br>J11=0.838 [Mn1-Mn2 (-2,-1,0)]<br>J12=-0.265 [Mn1-Mn1 (-1,-2,0)]<br>J13a=-0.080 [Mn1-Mn3 (-2,0,0)]<br>J13b=-0.461 [Mn1-Mn3 (-2,0,-1)]<br>J14=0.124 [Mn1-Mn4 (-2,-1,-1)]<br>J15=0.077 [Mn1-Mn2 (-1,-2,0)]<br>J16a=0.000 [Mn1-Mn3 (-1,-2,0)]<br>J16b=-0.243 [Mn1-Mn3 (-1,-2,-1)]<br>J17=0.500 [Mn1-Mn1 (0,0,-1)] |
| 1.496, 1.691, 1.692 | $YMn_2Ge_2$ | Mn1 (0,0.5,0.25)<br>Mn2 (0.5,0,0.25)<br>Mn3 (0,0.5,0.75)<br>Mn4 (0.5,0,0.75) | 0 | J1=9.870, $J_1^{ani}$= (0.009, 0.009, 0.055), $\Gamma_1$= (-0.002, -0.009, 0.009) [Mn1-Mn2 (-1,0,0)]<br>J2=6.041 [Mn1-Mn1 (-1,0,0)]<br>J3=1.157 [Mn1-Mn3 (0,0,-1)]<br>J4=0.631 [Mn1-Mn1 (-1,-1,0)]<br>J5=1.134, $D_5$= (0.08, 0.08, 0.004) [Mn1-Mn4 (-1,0,-1)]<br>J6=1.302 [Mn1-Mn2 (-1,-1,0)]<br>J7a=0.109 [Mn1-Mn3 (-1,0,0)]<br>J7b=-3.708 [Mn1-Mn3 (-1,0,-1)]<br>J8=-0.026 [Mn1-Mn3 (-1,-1,-1)]<br>J9=-0.174 [Mn1-Mn1 (-2,0,0)]<br>J10=-0.169 [Mn1-Mn4 (-1,-1,-1)]<br>J11=2.110 [Mn1-Mn2 (-2,-1,0)]<br>J12=0.176 [Mn1-Mn1 (-1,-2,0)]<br>J13a=0.180 [Mn1-Mn3 (-2,0,0)]<br>J13b=-0.792 [Mn1-Mn3 (-2,0,-1)]<br>J14=0.111 [Mn1-Mn4 (-2,-1,-1)]<br>J15=-0.126 [Mn1-Mn2 (-1,-2,0)]<br>J16a=0.227 [Mn1-Mn3 (-1,-2,0)]<br>J16b=-0.185 [Mn1-Mn3 (-1,-2,-1)]<br>J17=0.240 [Mn1-Mn1 (0,0,-1)]<br>J18=0.033 [Mn1-Mn2 (-1,0,-1)]<br>J19=-0.026 [Mn1-Mn1 (-2,-2,0)]<br>J20=0.012 [Mn1-Mn4 (-1,-2,-1)]<br>J21=-0.167 [Mn1-Mn1 (-1,0,-1)]<br>J22=-0.106 [Mn1-Mn2 (-2,-2,0)]<br>J23=-0.058 [Mn1-Mn1 (-3,0,0)]<br>J24=-0.288 [Mn1-Mn1 (-1,-1,-1)]<br>J25=-0.038 [Mn1-Mn3 (-2,-2,-1)]<br>J26a=0.304 [Mn1-Mn2 (-1,-1,-1)]<br>J26b=0.085 [Mn1-Mn2 (-1,-1,1)]<br>J27=-0.103 [Mn1-Mn1 (-1,-3,0)] |
| 1.504 | GdCuSn | Gd1 (0,0,0.5)<br>Gd2 (0,0,0)<br>Gd3 (0.5,0.5,0)<br>Gd4 (0.5,0.5,0.5) | 4.0 (Gd 4*f*) | J1=1.961, $D_1$= (0.004, -0.014, -0.031) [Gd1-Gd2 (0,0,0)]<br>J2=0.226, $J_2^{ani}$= (-0.016, -0.017, -0.016), $\Gamma_2$= (-0.001, 0.001, -0.001) [Gd1-Gd1 (0,-1,0)]<br>J3=0.228 [Gd1-Gd4 (-1,-1,0)]<br>J4a=-0.199 [Gd1-Gd2 (0,-1,0)]<br>J4b=-0.192 [Gd1-Gd3 (-1,-1,0)]<br>J4c=-0.159 [Gd1-Gd3 (0,-1,0)]<br>J5=0.233 [Gd1-Gd1 (0,0,-1)] |
| | | | | J6a=-0.011 [Gd1-Gd1 (-1,0,0)]<br>J6b=-0.028 [Gd1-Gd4 (-1,-2,0)]<br>J7=-0.034 [Gd1-Gd1 (0,-1,-1)]<br>J8a=-0.054 [Gd1-Gd4 (-1,-1,-1)]<br>J8b=-0.028 [Gd1-Gd4 (-1,-1,1)]<br>J9a=0.018 [Gd1-Gd2 (-1,0,0)]<br>J9b=0.005 [Gd1-Gd3 (-1,-2,0)]<br>J9c=0.008 [Gd1-Gd3 (0,-2,0)]<br>J10a=0.054 [Gd1-Gd1 (-1,-1,0)]<br>J10b=0.060 [Gd1-Gd1 (0,-2,0)]<br>J11a=0.043 [Gd1-Gd2 (0,-2,0)]<br>J11b=0.026 [Gd1-Gd2 (-1,-1,0)]<br>J11c=0.025 [Gd1-Gd2 (1,-1,0)]<br>J12a=-0.004 [Gd1-Gd1 (-1,0,1)]<br>J12b=-0.011 [Gd1-Gd4 (-1,-2,-1)]<br>J12c=-0.040 [Gd1-Gd1 (-1,0,-1)]<br>J12d=-0.045 [Gd1-Gd4 (-1,-2,1)]<br>J13=0.036 [Gd1-Gd2 (0,0,-1)] |
| 1.505 | GdAgSn | Gd1 (0,0,0.5)<br>Gd2 (0,0,0)<br>Gd3 (0.5,0.5,0)<br>Gd4 (0.5,0.5,0.5) | 4.0<br>(Gd 4*f*) | J1=2.231, $J_1^{ani}$= (0.134, 0.151, 0.057)<br>[Gd1-Gd2 (0,0,0)]<br>J2a=0.215, $D_{2a}$= (-0.032, 0.001, 0.043)<br>[Gd1-Gd1 (0,-1,0)]<br>J2b=0.171 [Gd1-Gd4 (-1,-1,0)]<br>J3a=-0.312 [Gd1-Gd2 (0,-1,0)]<br>J3b=-0.302 [Gd1-Gd3 (-1,-1,0)]<br>J3c=-0.287 [Gd1-Gd3 (0,-1,0)]<br>J4=0.270 [Gd1-Gd1 (0,0,-1)]<br>J5a=-0.002 [Gd1-Gd1 (-1,0,0)]<br>J5b=-0.026 [Gd1-Gd4 (-1,-2,0)]<br>J6a=-0.024 [Gd1-Gd1 (0,-1,-1)]<br>J6b=-0.031 [Gd1-Gd4 (-1,-1,-1)]<br>J6c=-0.015 [Gd1-Gd4 (-1,-1,1)]<br>J7a=0.028 [Gd1-Gd2 (-1,0,0)]<br>J7b=0.005 [Gd1-Gd2 (1,0,0)]<br>J7c=-0.003 [Gd1-Gd3 (-1,-2,0)]<br>J7d=0.004 [Gd1-Gd3 (0,-2,0)]<br>J8a=0.052 [Gd1-Gd1 (-1,-1,0)]<br>J8b=0.047 [Gd1-Gd1 (0,-2,0)]<br>J9a=0.037 [Gd1-Gd2 (0,-2,0)]<br>J9b=0.026 [Gd1-Gd2 (1,-1,0)]<br>J9c=0.021 [Gd1-Gd2 (-1,-1,0)]<br>J10a=-0.050 [Gd1-Gd1 (-1,0,-1)]<br>J10b=-0.005 [Gd1-Gd1 (-1,0,1)]<br>J10c=-0.060 [Gd1-Gd4 (-1,-2,1)] |
| | | | | J10d=-0.008 [Gd1-Gd4 (-1,-2,-1)]<br>J11=0.038 [Gd1-Gd2 (0,0,-1)]<br>$\Gamma_{i}$= (0, 0, 0) |
| 1.506 | GdAuSn | Gd1 (0,0,0.5)<br>Gd2 (0,0,0)<br>Gd3 (0.5,0.5,0)<br>Gd4 (0.5,0.5,0.5) | 4.0<br>(Gd 4*f*) | J1=2.569, $J_1^{ani}$= (-0.008, 0.034, 0.007),<br>$\Gamma_1$= (0, 0.001, -0.003)<br>[Gd1-Gd2 (0,0,0)]<br>J2a=0.130, $D_{2a}$= (-0.092, 0.006, 0.044)<br>[Gd1-Gd1 (0,-1,0)]<br>J2b=0.076 [Gd1-Gd4 (-1,0,0)]<br>J3a=-0.279 [Gd1-Gd2 (0,-1,0)]<br>J3b=-0.263 [Gd1-Gd3 (-1,-1,0)]<br>J3c=-0.206 [Gd1-Gd3 (0,-1,0)]<br>J4=0.295 [Gd1-Gd1 (0,0,-1)]<br>J5a=0.016 [Gd1-Gd1 (-1,0,0)]<br>J5b=-0.017 [Gd1-Gd4 (-1,-2,0)]<br>J6a=-0.026 [Gd1-Gd1 (0,-1,1)]<br>J6b=-0.034 [Gd1-Gd1 (0,-1,-1)]<br>J6c=-0.052 [Gd1-Gd4 (-1,-1,-1)]<br>J7a=0.040 [Gd1-Gd2 (-1,0,0)]<br>J7b=-0.011 [Gd1-Gd2 (1,0,0)]<br>J7c=-0.021 [Gd1-Gd3 (-1,-2,0)]<br>J7d=0.019 [Gd1-Gd3 (0,-2,0)]<br>J8a=0.009 [Gd1-Gd1 (-1,-1,0)]<br>J8b=0.010 [Gd1-Gd1 (-1,1,0)]<br>J8c=0.018 [Gd1-Gd1 (0,-2,0)]<br>J9a=-0.008 [Gd1-Gd2 (-1,-1,1)]<br>J9b=0.012 [Gd1-Gd2 (0,-2,0)]<br>J9c=-0.006 [Gd1-Gd2 (1,-1,0)]<br>J10a=-0.032 [Gd1-Gd1 (-1,0,-1)]<br>J10b=-0.003 [Gd1-Gd1 (-1,0,1)]<br>J10c=-0.005 [Gd1-Gd4 (-1,-2,-1)]<br>J10d=-0.050 [Gd1-Gd4 (-1,-2,1)]<br>J11=0.073 [Gd1-Gd2 (0,0,-1)] |
| 1.520 | $NiSO_4$ | Ni1 (0,0,0)<br>Ni2 (0.5,0.5,0)<br>Ni3 (0,0,0.5)<br>Ni4 (0.5,0.5,0.5) | 5.1<br>(Ni 3*d*) | J1=-0.455, $D_1$= (-0.03, 0.001, -0.001),<br>$J_1^{ani}$= (-0.003, -0.006, -0.002)<br>[Ni1-Ni3 (0,0,-1)]<br>J2=-0.254 [Ni1-Ni2 (-1,-1,0)]<br>J3=-0.797 [Ni1-Ni1 (-1,0,0)]<br>J4=-0.579 [Ni1-Ni4 (-1,-1,-1)]<br>J5=0.060 [Ni1-Ni3 (-1,0,-1)]<br>J6=-0.010 [Ni1-Ni1 (0,0,-1)]<br>J7=0.004 [Ni1-Ni1 (0,-1,0)]<br>$\Gamma_{i}$= (0, 0, 0) |
| 1.521 | $FeSO_4$ | Fe1 (0,0,0) | 4.0 | J1=4.962, $D_1$= (0.055, 0, 0), |
| | | Fe2 (0.5,0.5,0)<br>Fe3 (0,0,0.5)<br>Fe4 (0.5,0.5,0.5) | (Fe 3$d$) | $J_1^{ani}$= (0.001, 0, 0.002)<br>[Fe1-Fe3 (0,0,-1)]<br>J2=-1.705 [Fe1-Fe2 (-1,-1,0)]<br>J3=0.112 [Fe1-Fe1 (-1,0,0)]<br>J4=-0.077 [Fe1-Fe4 (-1,-1,-1)]<br>J5=0.335 [Fe1-Fe3 (-1,0,-1)]<br>J6=0.021 [Fe1-Fe1 (0,0,-1)]<br>J7=0.005 [Fe1-Fe1 (0,-1,0)]<br>J8a=0.015 [Fe1-Fe2 (-1,-1,-1)]<br>J8b=0.002 [Fe1-Fe2 (-1,-1,1)]<br>J9=0.014 [Fe1-Fe1 (-1,0,-1)]<br>J10=0.011 [Fe1-Fe3 (0,-1,-1)]<br>J11=0.019 [Fe1-Fe2 (-2,-1,0)]<br>J12=0.010 [Fe1-Fe4 (-2,-1,-1)]<br>J13=0.003 [Fe1-Fe1 (-1,-1,0)]<br>J14=0.001 [Fe1-Fe3 (0,0,-2)]<br>J15=0.003 [Fe1-Fe3 (-1,-1,-1)]<br>J16a=0.076 [Fe1-Fe1 (0,-1,-1)]<br>J16b=-0.000 [Fe1-Fe1 (0,-1,1)]<br>$\Gamma_i$= (0, 0, 0) |
| 1.523 | $VPO_4$ | V1 (0,0,0)<br>V2 (0.25,0.5,0.5)<br>V3 (0.75,0.5,0.5)<br>V4 (0.5,0,0)<br>V5 (0.75,0.5,0)<br>V6 (0.5,0,0.5)<br>V7 (0,0,0.5)<br>V8 (0.25,0.5,0) | 0 | J1=-10.656, $D_1$= (0.196, 0.028, 0),<br>$J_1^{ani}$= (0.004, -0.01, -0.024)<br>[V1-V7 (0,0,-1)]<br>J2a=-0.251, $\Gamma_{2a}$= (0, 0, 0.001)<br>[V1-V5 (-1,-1,0)]<br>J2b=-0.033 [V1-V8 (0,-1,0)]<br>J3=0.162 [V1-V4 (-1,0,0)]<br>J4a=1.384 [V1-V2 (0,-1,-1)]<br>J4b=-0.945 [V1-V3 (-1,-1,-1)]<br>J5=0.315 [V1-V6 (-1,0,-1)]<br>J6=0.402 [V1-V1 (0,0,-1)]<br>J7=0.086 [V1-V1 (0,-1,0)]<br>J8a=0.071 [V1-V5 (-1,-1,-1)]<br>J8b=-0.009 [V1-V5 (-1,-1,1)]<br>J8c=0.287 [V1-V8 (0,-1,-1)]<br>J8d=0.227 [V1-V8 (0,-1,1)] |
| 1.526 | $LiCoF_4$ | Co1 (0,0,0)<br>Co2 (0,0.5,0.5)<br>Co3 (0.5,0,0)<br>Co4 (0.5,0.5,0.5) | 5.3<br>(Co 3$d$) | J1=5.435, $D_1$= (0.128, 0.158, -0.482),<br>$J_1^{ani}$= (0.026, 0.008, -0.004),<br>$\Gamma_1$= (0.002, 0.007, 0.007)<br>[Co1-Co2 (0,-1,-1)]<br>J2=0.440 [Co1-Co1 (0,-1,0)]<br>J3=0.101 [Co1-Co3 (-1,0,0)]<br>J4=0.021 [Co1-Co4 (-1,-1,-1)]<br>J5=0.327 [Co1-Co1 (0,0,-1)] |
| | | | | J6=-0.082 [Co1-Co3 (-1,0,-1)]<br>J7a=0.005 [Co1-Co3 (-1,-1,0)]<br>J7b=0.000 [Co1-Co3 (-1,1,0)]<br>J8a=-0.142 [Co1-Co1 (0,-1,-1)]<br>J8b=-0.109 [Co1-Co1 (0,-1,1)] |
| 1.527 | $CsNiF_3$ | Ni1 (0,0,0)<br>Ni2 (0.5,0.5,0.5)<br>Ni3 (0.5,0,0.5)<br>Ni4 (0,0.5,0) | 0 | J1=0.583, $J_1^{ani}$= (0.015, 0.046, 0.015)<br>[Ni1-Ni4 (0,-1,0)]<br>J2=-0.030 [Ni1-Ni1 (0,-1,0)]<br>J3a=-0.014 [Ni1-Ni1 (0,0,-1)]<br>J3b=-0.014 [Ni1-Ni3 (-1,0,-1)]<br>J4a=-0.003 [Ni1-Ni2 (-1,-1,-1)]<br>J4b=-0.003 [Ni1-Ni4 (0,-1,-1)]<br>J5=-0.001 [Ni1-Ni4 (0,-2,0)]<br>J6a=-0.000 [Ni1-Ni1 (0,-1,-1)]<br>J6b=-0.000 [Ni1-Ni3 (-1,-1,-1)]<br>J7a=-0.003 [Ni1-Ni2 (-1,-2,-1)]<br>J7b=-0.003 [Ni1-Ni4 (0,-2,-1)]<br>$D_i$=$\Gamma_i$= (0, 0, 0) |
| 1.543, 1.544 | RbMnAs | Mn1 (0,0,0)<br>Mn2 (0.5,0.5,0.5)<br>Mn3 (0.5,0.5,0)<br>Mn4 (0,0,0.5) | 0 | J1=-42.548, $D_1$= (0.035, 0.017, 1.165),<br>$J_1^{ani}$= (-0.249, -0.271, -0.178),<br>$\Gamma_1$= (0.028, 0.014, 0.005)<br>[Mn1-Mn3 (-1,-1,0)]<br>J2=-12.707 [Mn1-Mn1 (-1,0,0)]<br>J3=0.865 [Mn1-Mn1 (-1,-1,0)]<br>J4=0.399 [Mn1-Mn3 (-1,-2,0)]<br>J5=-0.207 [Mn1-Mn4 (0,0,-1)]<br>J6=0.006 [Mn1-Mn2 (-1,-1,-1)]<br>J7=-0.077 [Mn1-Mn1 (-2,0,0)]<br>J8=-0.041 [Mn1-Mn4 (-1,0,-1)]<br>J9=-0.033 [Mn1-Mn3 (-2,-2,0)]<br>J10=-0.002 [Mn1-Mn1 (-1,-2,0)]<br>J11=-0.040 [Mn1-Mn4 (-1,-1,-1)]<br>J12a=0.008 [Mn1-Mn2 (-1,-2,-1)]<br>J12b=-0.043 [Mn1-Mn2 (-1,-2,0)]<br>J13=0.043 [Mn1-Mn3 (-1,-3,0)] |
| 1.545 | RbMnBi | Mn1 (0,0,0)<br>Mn2 (0.5,0.5,0.5)<br>Mn3 (0.5,0.5,0)<br>Mn4 (0,0,0.5) | 0 | J1=-31.563, $J_1^{ani}$= (-3.131, -3.327, -3.678),<br>$\Gamma_1$= (-0.048, 0.06, -0.057)<br>[Mn1-Mn3 (-1,-1,0)]<br>J2=-10.235, $D_2$= (2.441, -0.026, -0.105)<br>[Mn1-Mn1 (-1,0,0)]<br>J3=0.422 [Mn1-Mn1 (-1,-1,0)]<br>J4=0.607 [Mn1-Mn3 (-1,-2,0)]<br>J5=-0.008 [Mn1-Mn4 (0,0,-1)]<br>J6=0.013 [Mn1-Mn2 (-1,-1,-1)] |
| | | | | J7=-0.151 [Mn1-Mn1 (-2,0,0)] |
| 1.546 | CsMnBi | Mn1 (0,0,0)<br>Mn2 (0.5,0.5,0.5)<br>Mn3 (0.5,0.5,0)<br>Mn4 (0,0,0.5) | 0 | J1=-28.834, $J_1^{ani}$= (-3.372, -3.37, -3.461),<br>$\Gamma_1$= (-0.036, 0.063, -0.063)<br>[Mn1-Mn3 (-1,-1,0)]<br>J2=-9.637, $D_2$= (-0.009, 1.13, -0.003)<br>[Mn1-Mn1 (-1,0,0)]<br>J3=0.108 [Mn1-Mn1 (-1,-1,0)]<br>J4=0.440 [Mn1-Mn3 (-1,-2,0)]<br>J5=-0.205 [Mn1-Mn4 (0,0,-1)]<br>J6=-0.042 [Mn1-Mn2 (-1,-1,-1)]<br>J7=-0.225 [Mn1-Mn1 (-2,0,0)] |
| 1.547,<br>1.548 | CsMnP | Mn1 (0,0,0)<br>Mn2 (0.5,0.5,0.5)<br>Mn3 (0.5,0.5,0)<br>Mn4 (0,0,0.5) | 0 | J1=-59.305, $\Gamma_1$= (-0.016, -0.027, 0.026)<br>[Mn1-Mn3 (-1,-1,0)]<br>J2=-21.071, $D_2$= (-0.833, 0.009, 0.041)<br>$J_2^{ani}$= (-0.064, -0.021, -0.052),<br>[Mn1-Mn1 (-1,0,0)]<br>J3=2.075 [Mn1-Mn1 (-1,-1,0)]<br>J4=0.580 [Mn1-Mn3 (-1,-2,0)]<br>J5=-0.156 [Mn1-Mn1 (-2,0,0)]<br>J6=-0.289 [Mn1-Mn4 (0,0,-1)]<br>J7=-0.071 [Mn1-Mn3 (-2,-2,0)]<br>J8=-0.018 [Mn1-Mn2 (-1,-1,-1)] |
| 1.550,<br>1.551,<br>1.552 | LiMnAs | Mn1 (0,0,0)<br>Mn2 (0.5,0.5,0.5)<br>Mn3 (0.5,0.5,0)<br>Mn4 (0,0,0.5) | 0 | J1=-51.892, $D_1$= (0.144, 0.147, -0.127),<br>$J_1^{ani}$= (-0.077, -0.07, -0.074),<br>$\Gamma_1$= (0.025, 0, 0)<br>[Mn1-Mn3 (-1,-1,0)]<br>J2=-19.968 [Mn1-Mn1 (-1,0,0)]<br>J3=1.784 [Mn1-Mn1 (-1,-1,0)]<br>J4=0.245 [Mn1-Mn4 (0,0,-1)]<br>J5=0.254 [Mn1-Mn3 (-1,-2,0)]<br>J6=-0.006 [Mn1-Mn2 (-1,-1,-1)]<br>J7=-0.248 [Mn1-Mn4 (-1,0,-1)]<br>J8=0.008 [Mn1-Mn1 (-2,0,0)]<br>J9=-0.127 [Mn1-Mn4 (-1,-1,-1)]<br>J10=0.022 [Mn1-Mn3 (-2,-2,0)]<br>J11a=0.131 [Mn1-Mn2 (-1,-2,0)]<br>J11b=-0.248 [Mn1-Mn2 (-1,-2,-1)]<br>J12=-0.137 [Mn1-Mn1 (-1,-2,0)] |
| 1.553,<br>1.554 | KMnAs | Mn1 (0,0,0)<br>Mn2 (0.5,0.5,0.5)<br>Mn3 (0.5,0.5,0)<br>Mn4 (0,0,0.5) | 0 | J1=-48.740, $D_1$= (0.508, -0.029, 0.301),<br>$J_1^{ani}$= (-0.055, -0.06, -0.111),<br>$\Gamma_1$= (-0.025, -0.009, 0.01)<br>[Mn1-Mn3 (-1,-1,0)]<br>J2=-15.573 [Mn1-Mn1 (-1,0,0)]<br>J3=1.199 [Mn1-Mn1 (-1,-1,0)] |
| | | | | J4=0.529 [Mn1-Mn3 (-1,-2,0)]<br>J5=-0.024 [Mn1-Mn4 (0,0,-1)]<br>J6=0.013 [Mn1-Mn2 (-1,-1,-1)]<br>J7=-0.060 [Mn1-Mn1 (-2,0,0)]<br>J8=-0.040 [Mn1-Mn4 (-1,0,-1)]<br>J9=-0.078 [Mn1-Mn3 (-2,-2,0)]<br>J10=-0.068 [Mn1-Mn1 (-1,-2,0)]<br>J11=-0.090 [Mn1-Mn4 (-1,-1,-1)]<br>J12a=0.004 [Mn1-Mn2 (-1,-2,-1)]<br>J12b=-0.075 [Mn1-Mn2 (-1,-2,0)]<br>J13=0.031 [Mn1-Mn3 (-1,-3,0)] |
| 1.593 | BaCoSO | Co1 (0,0.095,0.25)<br>Co2 (0,0.905,0.75)<br>Co3 (0.25,0.595,0.25)<br>Co4 (0.25,0.405,0.75)<br>Co5 (0.5,0.095,0.25)<br>Co6 (0.5,0.905,0.75)<br>Co7 (0.75,0.595,0.25)<br>Co8 (0.75,0.405,0.75) | 2.1<br>(Co 3*d*) | J1=-25.015, $D_1$= (0.001, 0.001, 0.603),<br>$J_1^{ani}$= (-0.055, -0.092, -0.091)<br>[Co1-Co2 (0,-1,-1)]<br>J2=-7.525 [Co1-Co5 (-1,0,0)]<br>J3a=-0.470 [Co1-Co4 (0,0,-1)]<br>J3b=-0.565 [Co1-Co8 (-1,0,-1)]<br>J4=0.154, $\Gamma_4$= (0.003, 0.002, 0)<br>[Co1-Co6 (-1,-1,-1)]<br>J5=0.407 [Co1-Co1 (0,0,-1)]<br>J6a=-0.098 [Co1-Co3 (0,-1,0)]<br>J6b=-0.100 [Co1-Co7 (-1,-1,0)] |
| 1.628 | $PrMnSi_2$ | Mn1 (0,0.75,0.25)<br>Mn2 (0,0.25,0.75)<br>Mn3 (0.5,0.75,0.75)<br>Mn4 (0.5,0.25,0.25) | 0 | J1=-1.182, $D_1$= (-0.206, -0.224, -0.008),<br>$J_1^{ani}$= (-0.139, -0.015, -0.043),<br>$\Gamma_1$= (0.015, -0.007, 0.01)<br>[Mn1-Mn3 (-1,0,-1)]<br>J2a=-0.729 [Mn1-Pr2 (0,0,-1)]<br>J2b=-0.975 [Mn1-Pr2 (0,0,0)]<br>J3=-1.510 [Mn1-Pr4 (-1,0,0)]<br>J4a=4.194 [Mn1-Mn1 (0,0,-1)]<br>J4b=0.062 [Pr1-Pr1 (0,0,-1)]<br>J5=-0.009 [Pr1-Pr2 (0,-1,-1)]<br>J6a=7.323 [Mn1-Mn1 (-1,0,0)]<br>J6b=0.090 [Pr1-Pr1 (-1,0,0)]<br>J7=-0.044 [Mn1-Pr4 (-1,0,-1)]<br>J8=-0.075 [Mn1-Pr2 (-1,0,-1)]<br>J9a=-1.917 [Mn1-Mn1 (-1,0,-1)]<br>J9b=-0.044 [Pr1-Pr1 (-1,0,-1)]<br>J10=0.031 [Pr1-Pr2 (-1,-1,-1)]<br>J11=-0.002 [Pr1-Pr3 (-1,0,-1)]<br>J12=-0.016 [Mn1-Pr1 (0,1,0)]<br>J13=0.670 [Mn1-Mn3 (-1,0,-2)]<br>J14=0.875 [Mn1-Mn3 (-2,0,-1)] |
| 1.636, | $ErMn_2Si_2$ | Mn1 (0,0.5,0.25) | 0 | J1=2.705, $J_1^{ani}$= (-0.014, -0.014, 0.051), |
| 1.637 | | Mn2 (0.5,0,0.25)<br>Mn3 (0,0.5,0.75)<br>Mn4 (0.5,0,0.75) | | $\Gamma_1$= (-0.006, -0.019, 0.019)<br>[Mn1-Mn2 (-1,0,0)]<br>J2=4.795 [Mn1-Mn1 (-1,0,0)]<br>J3=0.428 [Mn1-Mn3 (0,0,-1)]<br>J4=-0.009 [Mn1-Mn1 (-1,-1,0)]<br>J5=1.655, $D_5$= (-0.03, -0.031, -0.007)<br>[Mn1-Mn4 (-1,0,-1)]<br>J6=0.979 [Mn1-Mn2 (-1,-1,0)]<br>J7a=-0.615 [Mn1-Mn3 (-1,0,0)]<br>J7b=-2.362 [Mn1-Mn3 (-1,0,-1)]<br>J8=-0.194 [Mn1-Mn3 (-1,-1,-1)]<br>J9=0.188 [Mn1-Mn1 (-2,0,0)]<br>J10=0.045 [Mn1-Mn4 (-1,-1,-1)]<br>J11=0.783 [Mn1-Mn2 (-2,-1,0)]<br>J12=-0.068 [Mn1-Mn1 (-1,-2,0)]<br>J13a=-0.036 [Mn1-Mn3 (-2,0,0)]<br>J13b=-0.494 [Mn1-Mn3 (-2,0,-1)]<br>J14=0.173 [Mn1-Mn4 (-2,-1,-1)]<br>J15=-0.007 [Mn1-Mn2 (-1,-2,0)]<br>J16a=0.055 [Mn1-Mn3 (-1,-2,0)]<br>J16b=-0.089 [Mn1-Mn3 (-1,-2,-1)]<br>J17=0.718 [Mn1-Mn1 (0,0,-1)]<br>J18=0.026 [Mn1-Mn2 (-1,0,-1)]<br>J19=0.093 [Mn1-Mn1 (-2,-2,0)]<br>J20=-0.015 [Mn1-Mn1 (-1,0,-1)]<br>J21=-0.053 [Mn1-Mn4 (-1,-2,-1)]<br>J22=-0.157 [Mn1-Mn2 (-2,-2,0)]<br>J23=-0.111 [Mn1-Mn1 (-3,0,0)]<br>J24=-0.199 [Mn1-Mn1 (-1,-1,-1)]<br>J25a=0.046 [Mn1-Mn2 (-1,-1,1)]<br>J25b=-0.051 [Mn1-Mn2 (-1,-1,-1)]<br>J26=-0.115 [Mn1-Mn3 (-2,-2,-1)] |
| 1.638,<br>1.639,<br>1.640 | $ErMn_2Ge_2$ | Mn1 (0,0.5,0.25)<br>Mn2 (0.5,0,0.25)<br>Mn3 (0,0.5,0.75)<br>Mn4 (0.5,0,0.75) | 0 | J1=7.833, $J_1^{ani}$= (0, 0, 0.07),<br>$\Gamma_1$= (0.003, -0.013, 0.014)<br>[Mn1-Mn2 (-1,0,0)]<br>J2=5.718 [Mn1-Mn1 (-1,0,0)]<br>J3=1.126 [Mn1-Mn3 (0,0,-1)]<br>J4=0.769 [Mn1-Mn1 (-1,-1,0)]<br>J5=1.104, $D_5$= (0.068, 0.068, -0.037)<br>[Mn1-Mn4 (-1,0,-1)]<br>J6=1.260 [Mn1-Mn2 (-1,-1,0)]<br>J7a=0.188 [Mn1-Mn3 (-1,0,0)]<br>J7b=-3.919 [Mn1-Mn3 (-1,0,-1)]<br>J8=-0.057 [Mn1-Mn3 (-1,-1,-1)] |

| | | | | J9=-0.015 [Mn1-Mn1 (-2,0,0)]<br>J10=-0.163 [Mn1-Mn4 (-1,-1,-1)]<br>J11=1.781 [Mn1-Mn2 (-2,-1,0)]<br>J12=0.317 [Mn1-Mn1 (-1,-2,0)]<br>J13a=0.222 [Mn1-Mn3 (-2,0,0)]<br>J13b=-0.728 [Mn1-Mn3 (-2,0,-1)]<br>J14=0.155 [Mn1-Mn4 (-2,-1,-1)]<br>J15=-0.123 [Mn1-Mn2 (-1,-2,0)]<br>J16=-0.123 [Mn3-Mn4 (-1,-2,0)]<br>J17a=0.222 [Mn1-Mn3 (-1,-2,0)]<br>J17b=-0.182 [Mn1-Mn3 (-1,-2,-1)]<br>J18=0.236 [Mn1-Mn1 (0,0,-1)]<br>J19=0.000 [Mn1-Mn2 (-1,0,-1)]<br>J20=-0.137 [Mn1-Mn1 (-2,-2,0)]<br>J21=0.018 [Mn1-Mn4 (-1,-2,-1)]<br>J22=-0.130 [Mn1-Mn1 (-1,0,-1)]<br>J23=-0.151 [Mn1-Mn2 (-2,-2,0)]<br>J24=-0.062 [Mn1-Mn1 (-3,0,0)]<br>J25=-0.222 [Mn1-Mn1 (-1,-1,-1)]<br>J26=-0.077 [Mn1-Mn3 (-2,-2,-1)]<br>J27a=0.307 [Mn1-Mn2 (-1,-1,-1)]<br>J27b=0.070 [Mn1-Mn2 (-1,-1,1)] |
|---|---|---|---|---|
| 1.641 | $Ba_2FeSi_2O_7$ | Fe1 (0,0,0)<br>Fe2 (0,0,0.5)<br>Fe3 (0.5,0.5,0)<br>Fe4 (0.5,0.5,0.5) | 4.0<br>(Fe 3*d*) | J1=-0.028 [Fe1-Fe2 (0,0,-1)]<br>J2=-0.247, $D_2$= (0.008, 0.006, -0.008)<br>[Fe1-Fe3 (-1,-1,0)]<br>J3=0.001 [Fe1-Fe4 (-1,-1,-1)]<br>J4=0.080 [Fe1-Fe1 (-1,0,0)]<br>$J_i^{ani}$=$\Gamma_i$= (0, 0, 0) |
| 1.643 | DyOCl | Dy1 (0.25,0.25,0.085)<br>Dy2 (0.25,0.25,0.585)<br>Dy3 (0.75,0.75,0.415)<br>Dy4 (0.75,0.75,0.915) | 6.0<br>(Dy 4*f*) | J1=-0.105, $J_1^{ani}$= (-0.026, -0.026, -0.023),<br>$\Gamma_1$= (0.001, 0.004, -0.004)<br>[Dy1-Dy4 (-1,-1,-1)]<br>J2=-0.049, $D_2$= (0.007, 0, 0),<br>[Dy1-Dy1 (-1,0,0)]<br>J3=0.000 [Dy1-Dy3 (-1,-1,0)]<br>J4=-0.001 [Dy1-Dy1 (-1,-1,0)] |
| 1.677 | $Fe_{0.35}NbS_2$† | Fe1 (0,0.667,0.25)<br>Fe2 (0.5,0.667,0.25)<br>Fe3 (0,0.333,0.75)<br>Fe4 (0.5,0.333,0.75)<br>Fe5 (0.25,0.167,0.25)<br>Fe6 (0.75,0.167,0.25)<br>Fe7 (0.25,0.833,0.75)<br>Fe8 (0.75,0.833,0.75) | 0.9<br>(Fe 3*d*) | J1a=-1.915, **$D_1$= (0.021, -0.26, -0.834)**,<br>**$J_1^{ani}$= (0.696, 0.951, 0.266)**,<br>$\Gamma_1$= (0.002, -0.004, 0.002)<br>[Fe1-Fe2 (-1,0,0)]<br>J1b=1.375 [Fe1-Fe5 (0,0,0)]<br>J1c=-1.644 [Fe1-Fe6 (-1,0,0)]<br>J2a=-0.228 [Fe1-Fe3 (0,0,-1)]<br>J2b=-1.221 [Fe1-Fe3 (0,0,0)]<br>J2c=-0.925 [Fe1-Fe7 (0,0,-1)] |
| | | | | J2d=0.677 [Fe1-Fe8 (-1,0,-1)]<br>J3a=0.982 [Fe1-Fe3 (0,1,-1)]<br>J3b=0.801 [Fe1-Fe3 (0,1,0)]<br>J3c=0.087 [Fe1-Fe4 (-1,0,-1)]<br>J3d=0.262 [Fe1-Fe4 (-1,0,0)]<br>J4a=0.202 [Fe1-Fe1 (0,-1,0)]<br>J4b=-0.120 [Fe1-Fe5 (-1,0,0)]<br>J4c=-0.214 [Fe1-Fe6 (0,0,0)] |
| 1.689 | $LuMn_2Ge_2$ | Mn1 (0,0.5,0.25)<br>Mn2 (0.5,0,0.25)<br>Mn3 (0,0.5,0.75)<br>Mn4 (0.5,0,0.75) | 0 | J1=-5.703, $J_1^{ani}$= (-0.01, -0.01, 0.061),<br>$\Gamma_1$= (0, -0.016, 0.015)<br>[Mn1-Mn2 (-1,0,0)]<br>J2=5.613 [Mn1-Mn1 (-1,0,0)]<br>J3=1.356 [Mn1-Mn3 (0,0,-1)]<br>J4=0.522 [Mn1-Mn1 (-1,-1,0)]<br>J5=1.067, $D_5$= (0.044, 0.044, -0.04)<br>[Mn1-Mn4 (-1,0,-1)]<br>J6=1.227 [Mn1-Mn2 (-1,-1,0)]<br>J7a=0.003 [Mn1-Mn3 (-1,0,0)]<br>J7b=-3.568 [Mn1-Mn3 (-1,0,-1)]<br>J8=-0.105 [Mn1-Mn3 (-1,-1,-1)]<br>J9=0.236 [Mn1-Mn1 (-2,0,0)]<br>J10=-0.104 [Mn1-Mn4 (-1,-1,-1)]<br>J11=1.773 [Mn1-Mn2 (-2,-1,0)]<br>J12=0.338 [Mn1-Mn1 (-1,-2,0)]<br>J13a=0.330 [Mn1-Mn3 (-2,0,0)]<br>J13b=-0.689 [Mn1-Mn3 (-2,0,-1)]<br>J14=0.205 [Mn1-Mn4 (-2,-1,-1)]<br>J15=-0.065 [Mn1-Mn2 (-1,-2,0)]<br>J16a=0.144 [Mn1-Mn3 (-1,-2,0)]<br>J16b=-0.141 [Mn1-Mn3 (-1,-2,-1)]<br>J17=0.171 [Mn1-Mn1 (0,0,-1)]<br>J18=-0.132 [Mn1-Mn1 (-2,-2,0)]<br>J19=0.098 [Mn1-Mn2 (-1,0,-1)]<br>J20=0.013 [Mn1-Mn4 (-1,-2,-1)]<br>J21=-0.195 [Mn1-Mn2 (-2,-2,0)]<br>J22=-0.234 [Mn1-Mn1 (-1,0,-1)]<br>J23=-0.002 [Mn1-Mn1 (-3,0,0)]<br>J24=-0.201 [Mn1-Mn1 (-1,-1,-1)]<br>J25=-0.031 [Mn1-Mn3 (-2,-2,-1)]<br>J26=-0.108 [Mn1-Mn1 (-1,-3,0)]<br>J27a=0.262 [Mn1-Mn2 (-1,-1,-1)]<br>J27b=0.029 [Mn1-Mn2 (-1,-1,1)] |
| 1.690 | $TmMn_2Ge_2$ | Mn1 (0,0.5,0.25)<br>Mn2 (0.5,0,0.25) | 0 | J1=-5.425, $J_1^{ani}$= (-0.023, -0.023, 0.072),<br>$\Gamma_1$= (0.005, -0.014, 0.014) |
| | | Mn3 (0,0.5,0.75)<br>Mn4 (0.5,0,0.75) | | [Mn1-Mn2 (-1,0,0)]<br>J2=6.002 [Mn1-Mn1 (-1,0,0)]<br>J3=1.061 [Mn1-Mn3 (0,0,-1)]<br>J4=0.749 [Mn1-Mn1 (-1,-1,0)]<br>J5=1.010, $D_5$= (0.046, 0.047, -0.039)<br>[Mn1-Mn4 (-1,0,-1)]<br>J6=1.293 [Mn1-Mn2 (-1,-1,0)]<br>J7a=0.119 [Mn1-Mn3 (-1,0,0)]<br>J7b=-3.460 [Mn1-Mn3 (-1,0,-1)]<br>J8=-0.061 [Mn1-Mn3 (-1,-1,-1)]<br>J9=0.175 [Mn1-Mn1 (-2,0,0)]<br>J10=-0.119 [Mn1-Mn4 (-1,-1,-1)]<br>J11=1.527 [Mn1-Mn2 (-2,-1,0)]<br>J12=0.313 [Mn1-Mn1 (-1,-2,0)]<br>J13a=0.279 [Mn1-Mn3 (-2,0,0)]<br>J13b=-0.728 [Mn1-Mn3 (-2,0,-1)]<br>J14=0.142 [Mn1-Mn4 (-2,-1,-1)]<br>J15=-0.101 [Mn1-Mn2 (-1,-2,0)]<br>J16a=0.206 [Mn1-Mn3 (-1,-2,0)]<br>J16b=-0.124 [Mn1-Mn3 (-1,-2,-1)]<br>J17=0.460 [Mn1-Mn1 (0,0,-1)]<br>J18=-0.010 [Mn1-Mn1 (-2,-2,0)]<br>J19=-0.022 [Mn1-Mn2 (-1,0,-1)]<br>J20=0.028 [Mn1-Mn4 (-1,-2,-1)]<br>J21=-0.138 [Mn1-Mn2 (-2,-2,0)]<br>J22=-0.101 [Mn1-Mn1 (-1,0,-1)]<br>J23=-0.025 [Mn1-Mn1 (-3,0,0)]<br>J24=-0.253 [Mn1-Mn1 (-1,-1,-1)]<br>J25=-0.098 [Mn1-Mn3 (-2,-2,-1)]<br>J26=-0.068 [Mn1-Mn1 (-1,-3,0)]<br>J27a=0.284 [Mn1-Mn2 (-1,-1,-1)]<br>J27b=0.017 [Mn1-Mn2 (-1,-1,1)] |
| 1.694 | $TbMn_2Ge_2$ | Mn1 (0,0.5,0.25)<br>Mn2 (0.5,0,0.25)<br>Mn3 (0,0.5,0.75)<br>Mn4 (0.5,0,0.75) | 0 | J1=-5.526, $J_1^{ani}$= (-0.031, -0.031, 0.061),<br>$\Gamma_1$= (0, -0.02, 0.02)<br>[Mn1-Mn2 (-1,0,0)]<br>J2=8.475, $D_2$= (0, 0.099, 0)<br>[Mn1-Mn1 (-1,0,0)]<br>J3=0.829 [Mn1-Mn3 (0,0,-1)]<br>J4=-0.963 [Mn1-Mn1 (-1,-1,0)]<br>J5=1.219 [Mn1-Mn4 (-1,0,-1)]<br>J6=1.141 [Mn1-Mn2 (-1,-1,0)]<br>J7a=0.118 [Mn1-Mn3 (-1,0,0)]<br>J7b=-2.969 [Mn1-Mn3 (-1,0,-1)]<br>J8=-0.385 [Mn1-Mn3 (-1,-1,-1)] |
| | | | | J9=0.071 [Mn1-Mn1 (-2,0,0)]<br>J10=-0.017 [Mn1-Mn4 (-1,-1,-1)]<br>J11=1.692 [Mn1-Mn2 (-2,-1,0)]<br>J12=-0.020 [Mn1-Mn1 (-1,-2,0)]<br>J13a=0.117 [Mn1-Mn3 (-2,0,0)]<br>J13b=-0.264 [Mn1-Mn3 (-2,0,-1)]<br>J14=0.118 [Mn1-Mn4 (-2,-1,-1)]<br>J15=-0.000 [Mn1-Mn2 (-1,-2,0)]<br>J16a=0.038 [Mn1-Mn3 (-1,-2,0)]<br>J16b=-0.171 [Mn1-Mn3 (-1,-2,-1)]<br>J17=0.424 [Mn1-Mn1 (0,0,-1)]<br>J18=0.044 [Mn1-Mn2 (-1,0,-1)]<br>J19=0.428 [Mn1-Mn1 (-2,-2,0)]<br>J20=-0.030 [Mn1-Mn4 (-1,-2,-1)]<br>J21=-0.067 [Mn1-Mn1 (-1,0,-1)]<br>J22=-0.011 [Mn1-Mn2 (-2,-2,0)]<br>J23=-0.133 [Mn1-Mn1 (-3,0,0)]<br>J24=-0.266 [Mn1-Mn1 (-1,-1,-1)]<br>J25=-0.062 [Mn1-Mn3 (-2,-2,-1)]<br>J26a=0.137 [Mn1-Mn2 (-1,-1,-1)]<br>J26b=-0.003 [Mn1-Mn2 (-1,-1,1)] |
| 1.703 | $YBaCo_2O_5$ | Co1 (0.25,0.25,0.271)<br>Co2 (0.75,0.25,0.729)<br>Co3 (0.25,0.75,0.271)<br>Co4 (0.75,0.75,0.729)<br>Co5 (0.25,0.25,0.755)<br>Co6 (0.75,0.25,0.245)<br>Co7 (0.25,0.75,0.755)<br>Co8 (0.75,0.75,0.245) | 6.0<br>(Co 3*d*) | J1=-1.502 [Co1-Co5 (0,0,0)]<br>J2=-20.570, $\Gamma_2$= (0.004, 0.002, 0)<br>[Co1-Co5 (0,0,-1)]<br>J3a=-3.108 [Co1-Co3 (0,-1,0)]<br>J3b=-16.966, $D_{3b}$= (1.326, 0.25, 0.008),<br>$J_{3b}^{ani}$= (-0.164, -0.224, -0.153)<br>[Co5-Co7 (0,-1,0)]<br>J4=-9.320 [Co1-Co6 (0,0,0)]<br>J5=-0.245 [Co1-Co2 (-1,0,0)]<br>J6=-0.177 [Co1-Co7 (0,-1,0)]<br>J7=-0.944 [Co5-Co6 (-1,0,1)]<br>J8=-0.092 [Co5-Co6 (-1,0,0)]<br>J9=0.081 [Co1-Co7 (0,-1,-1)]<br>J10=0.153 [Co1-Co8 (-1,-1,0)]<br>J11=0.029 [Co3-Co4 (-1,0,-1)]<br>J12=-0.031 [Co1-Co4 (-1,-1,0)]<br>J13=-0.139 [Co5-Co8 (-1,-1,1)]<br>J14=-0.015 [Co5-Co8 (-1,-1,0)]<br>J15=0.006 [Co1-Co4 (-1,-1,-1)]<br>J16a=0.191 [Co1-Co1 (0,0,-1)]<br>J16b=-0.092 [Co5-Co5 (0,0,-1)]<br>J17a=-0.060 [Co1-Co1 (0,-1,0)]<br>J17b=0.025 [Co5-Co5 (0,-1,0)] |
| | | | | J18a=0.005 [Co1-Co1 (-1,0,0)]<br>J18b=0.102 [Co5-Co5 (-1,0,0)] |
| 1.714 | CeAu$Bi_2$ | Ce1 (0.25,0.25,0.119)<br>Ce2 (0.75,0.75,0.881)<br>Ce3 (0.75,0.75,0.381)<br>Ce4 (0.25,0.25,0.619) | 6.0<br>(Ce 4*f*) | J1=-0.077, $D_1$= (0.007, 0, 0),<br>$J_1^{ani}$= (0.002, -0.002, 0.001),<br>$\Gamma_1$= (0, 0.003, 0.003)<br>[Ce1-Ce1 (-1,0,0)]<br>J2=-0.022 [Ce1-Ce2 (-1,-1,-1)]<br>J3=-0.023 [Ce1-Ce3 (-1,-1,0)]<br>J4=-0.070 [Ce1-Ce1 (-1,-1,0)]<br>J5=0.012 [Ce1-Ce2 (-1,-2,-1)]<br>J6=-0.005 [Ce1-Ce3 (-1,-2,0)] |
| 1.746 | $YMn_2$ | Mn1 (0.75,0.5,0)<br>Mn2 (0.25,0,0)<br>Mn3 (0.5,0.25,0)<br>Mn4 (0.5,0.75,0)<br>Mn5 (0,0.75,0)<br>Mn6 (0,0.25,0)<br>Mn7 (0.75,0.5,0.5)<br>Mn8 (0.25,0,0.5) | 0 | J1a=-16.616, $D_{1a}$= (-0.077, -0.5, 0.569),<br>$\Gamma_{1a}$= (-0.064, -0.001, 0.008)<br>[Mn1-Mn3 (0,0,0)]<br>J1b=-19.429 [Mn1-Mn4 (0,0,0)]<br>J1c=-22.904 [Mn1-Mn7 (0,0,-1)]<br>J2a=-3.227, $J_{2a}^{ani}$= (-0.123, -0.259, -0.224),<br>[Mn1-Mn3 (0,0,-1)]<br>J2b=-6.444 [Mn1-Mn4 (0,-1,0)]<br>J2c=-10.612 [Mn1-Mn8 (0,0,-1)]<br>J2d=-4.131 [Mn1-Mn8 (0,1,-1)]<br>J3a=-2.094 [Mn1-Mn1 (0,-1,0)]<br>J3b=-5.915 [Mn1-Mn1 (0,0,-1)]<br>J3c=-0.947 [Mn1-Mn2 (0,0,-1)]<br>J3d=-1.893 [Mn1-Mn2 (0,0,0)]<br>J4a=1.841 [Mn1-Mn3 (0,1,-1)]<br>J4b=2.168 [Mn1-Mn3 (1,0,1)]<br>J4c=2.125 [Mn1-Mn7 (0,1,-1)]<br>J4d=3.789 [Mn1-Mn4 (0,-1,-1)]<br>J4e=-0.365 [Mn1-Mn4 (1,0,1)]<br>J4f=4.244 [Mn3-Mn7 (0,-1,-1)]<br>J4g=4.541 [Mn1-Mn7 (0,-1,-1)] |
| 1.754 | $BaFe_2O_4$ | Fe1 (0.794,0.03,0.528)<br>Fe2 (0.206,0.97,0.028)<br>Fe3 (0.706,0.03,0.028)<br>Fe4 (0.294,0.97,0.528)<br>Fe5 (0.294,0.53,0.528)<br>Fe6 (0.706,0.47,0.028)<br>Fe7 (0.206,0.53,0.028)<br>Fe8 (0.794,0.47,0.528)<br>Fe9 (0.959,0.039,0.51)<br>Fe10 (0.041,0.961,0.01)<br>Fe11 (0.541,0.039,0.01)<br>Fe12 (0.459,0.961,0.51) | 4.0<br>(Fe 3*d*) | J1=-35.413, $D_1$= (-0.011, 0.03, -0.203),<br>$J_1^{ani}$= (-0.087, -0.043, -0.052),<br>$\Gamma_1$= (-0.002, 0.001, -0.055)<br>[Fe1-Fe9 (0,0,0)]<br>J2=-33.329 [Fe1-Fe3 (0,0,0)]<br>J3=-31.602 [Fe9-Fe10 (1,-1,0)]<br>J4=-47.410 [Fe9-Fe16 (0,0,0)]<br>J5=-50.383 [Fe1-Fe8 (0,0,0)]<br>J6=-0.008 [Fe1-Fe8 (0,-1,0)]<br>J7=0.514 [Fe1-Fe16 (0,0,0)]<br>J8=-0.089 [Fe9-Fe16 (0,-1,0)]<br>J9=0.726 [Fe1-Fe6 (0,0,0)] |
| | | Fe13 (0.459,0.539,0.51)<br>Fe14 (0.541,0.461,0.01)<br>Fe15 (0.041,0.539,0.01)<br>Fe16 (0.959,0.461,0.51) | | J10a=0.447 [Fe9-Fe15 (1,-1,1)]<br>J10b=0.690 [Fe9-Fe15 (1,-1,0)]<br>J11a=0.539 [Fe1-Fe1 (0,0,-1)]<br>J11b=0.272 [Fe9-Fe9 (0,0,-1)]<br>J12=0.244 [Fe1-Fe10 (1,-1,1)]<br>J13=0.693 [Fe1-Fe11 (0,0,1)]<br>J14=0.736 [Fe1-Fe10 (1,-1,0)]<br>J15=0.714 [Fe1-Fe11 (0,0,0)] |
| 0.1, 0.642 | $LaMnO_3$ | Mn1 (0,0,0)<br>Mn2 (0.5,0,0.5)<br>Mn3 (0,0.5,0)<br>Mn4 (0.5,0.5,0.5) | 2.6<br>(Mn 3*d*) | J1=1.126 [Mn1-Mn3 (0,-1,0)]<br>J2=6.894, $D_2$= (0.285, -0.582, -0.174)<br>[Mn1-Mn2 (-1,0,-1)]<br>J3a=-0.778, $J_{3a}^{ani}$= (-0.124, -0.138, -0.121),<br>$\Gamma_{3a}$= (0.001, 0.007, -0.004)<br>[Mn1-Mn4 (-1,-1,-1)]<br>J3b=-0.707 [Mn1-Mn4 (-1,-1,0)]<br>J4=-0.060 [Mn1-Mn1 (0,0,-1)]<br>J5=-0.389 [Mn1-Mn1 (-1,0,0)]<br>J6=-0.004 [Mn1-Mn3 (0,-1,-1)]<br>J7=-0.040 [Mn1-Mn3 (-1,-1,0)]<br>J8=0.019 [Mn1-Mn1 (0,-1,0)]<br>J9a=-2.479 [Mn1-Mn1 (-1,0,-1)]<br>J9b=-0.113 [Mn1-Mn1 (-1,0,1)]<br>J10a=0.021 [Mn1-Mn2 (-1,-1,-1)]<br>J10b=0.009 [Mn1-Mn2 (-1,1,-1)]<br>J11=-0.003 [Mn1-Mn2 (-1,0,-2)]<br>J12a=-0.130 [Mn1-Mn3 (-1,-1,-1)]<br>J12b=-0.070 [Mn1-Mn3 (-1,-1,1)] |
| 0.15 | $MnF_2$ | Mn1 (0,0,0)<br>Mn2 (0.5,0.5,0.5) | 5.0<br>(Mn 3*d*) | J1=-1.584 [Mn1-Mn1 (0,0,-1)]<br>J2=-2.239, $D_2$= (-0.008, 0.018, -0.01)<br>[Mn1-Mn2 (-1,-1,-1)]<br>J3=-0.010 [Mn1-Mn1 (-1,0,0)]<br>J4=-0.024 [Mn1-Mn1 (-1,0,-1)]<br>J5=0.020 [Mn1-Mn2 (-1,-1,-2)]<br>J6=-0.059 [Mn1-Mn1 (0,0,-2)]<br>J7a=-0.080 [Mn1-Mn1 (-1,-1,0)]<br>J7b=-0.001 [Mn1-Mn1 (-1,1,0)]<br>J8a=0.004 [Mn1-Mn1 (-1,-1,-1)]<br>J8b=0.006 [Mn1-Mn1 (-1,1,-1)]<br>$J_i^{ani}$=$\Gamma_i$= (0, 0, 0) |
| 0.21 | $PbNiO_3$ | Ni1 (0,0,0)<br>Ni2 (0.5,0.5,0.5) | 4.6<br>(Ni 3*d*) | J1a=-3.364, $D_{1a}$= (0.192, 0.168, 0),<br>$J_{1a}^{ani}$= (-0.006, -0.005, -0.002),<br>$\Gamma_{1a}$= (-0.001, -0.001, 0)<br>[Ni1-Ni2 (0,0,1)]<br>J1b=-3.373 [Ni1-Ni2 (0,1,0)] |
| | | | | J2a=-0.011 [Ni1-Ni1 (-1,1,0)]<br>J2b=-0.008 [Ni1-Ni1 (-1,0,1)]<br>J2c=-0.006 [Ni1-Ni1 (0,-1,1)]<br>J3a=-0.033 [Ni1-Ni1 (-1,0,0)]<br>J3b=-0.029 [Ni1-Ni1 (0,-1,0)]<br>J3c=-0.027 [Ni1-Ni1 (0,0,-1)]<br>J4=-0.022 [Ni1-Ni2 (-1,1,1)]<br>J5=-0.021 [Ni1-Ni2 (0,0,0)]<br>J6=0.037 [Ni1-Ni1 (-1,-1,1)]<br>J7=-0.006 [Ni1-Ni2 (-1,0,2)]<br>J8a=-0.003 [Ni1-Ni2 (-1,0,1)]<br>J8b=0.002 [Ni1-Ni2 (-1,1,0)]<br>J9=-0.029 [Ni1-Ni1 (-1,-1,2)]<br>J10a=-0.003 [Ni1-Ni1 (-1,2,0)]<br>J10b=-0.001 [Ni1-Ni1 (-1,0,2)] |
| 0.23 | $Ca_3Mn_2O_7$ | Mn1 (0.097,0.753,0.75)<br>Mn2 (0.903,0.247,0.25)<br>Mn3 (0.903,0.753,0.75)<br>Mn4 (0.097,0.247,0.25) | 4.5<br>(Mn 3*d*) | J1=-0.058, $D_1$= (0.103, 0, -0.075),<br>$J_1^{ani}$= (-0.017, -0.012, -0.015),<br>$\Gamma_1$= (0, 0, 0.002)<br>[Mn1-Mn4 (0,1,0)]<br>J2=-2.265 [Mn1-Mn4 (0,0,0)]<br>J3=0.206 [Mn1-Mn3 (-1,0,0)]<br>J4=-0.060 [Mn1-Mn1 (0,0,-1)]<br>J5=-0.004 [Mn1-Mn1 (0,-1,0)]<br>J6=-0.155 [Mn1-Mn2 (-1,1,0)]<br>J7=-0.041 [Mn1-Mn2 (-1,0,0)]<br>J8=-0.008 [Mn1-Mn3 (-1,0,-1)]<br>J9=0.002 [Mn1-Mn3 (-1,-1,0)]<br>J10=0.051 [Mn1-Mn2 (-1/2,1/2,0)]<br>J11=0.070 [Mn1-Mn3 (-1/2,-1/2,0)]<br>J12a=0.122 [Mn1-Mn1 (0,-1,-1)]<br>J12b=0.172 [Mn1-Mn1 (0,-1,1)]<br>J13=0.022 [Mn1-Mn4 (0,2,0)]<br>J14=0.028 [Mn1-Mn4 (0,1,-1)]<br>J15=0.036 [Mn1-Mn4 (0,0,-1)]<br>J16a=0.027 [Mn1-Mn4 (0,-1,0)]<br>J16b=0.001 [Mn1-Mn2 (-1/2,3/2,0)]<br>J17a=0.053 [Mn1-Mn3 (-1,-1,-1)]<br>J17b=0.045 [Mn1-Mn3 (-1,-1,1)] |
| 0.25 | $NaOsO_3$[†] | Os1 (0,0,0.5)<br>Os2 (0.5,0.5,0)<br>Os3 (0,0.5,0.5)<br>Os4 (0.5,0,0) | 1.58<br>(Os 5*d*) | J1=-7.996, **$D_1$= (0.626, -0.746, -0.008)**,<br>$\Gamma_1$= (-0.006, 0.463, -0.154)<br>[Os1-Os4 (-1,0,0)]<br>J2=-6.311, **$J_2^{ani}$= (-0.902, 0.261, -0.451)**,<br>[Os1-Os3 (0,-1,0)]<br>J3=0.731 [Os1-Os1 (0,0,-1)] |
| | | | | J4a=0.771 [Os1-Os2 (-1,-1,0)]<br>J4b=0.750 [Os1-Os2 (-1,-1,1)]<br>J5=0.552 [Os1-Os1 (-1,0,0)]<br>J6=-0.010 [Os1-Os3 (0,-1,-1)]<br>J7=-0.019 [Os1-Os3 (-1,-1,0)]<br>J8a=0.682 [Os1-Os1 (-1,0,-1)]<br>J8b=0.595 [Os1-Os1 (-1,0,1)]<br>J9=0.493 [Os1-Os1 (0,-1,0)]<br>J10=-0.036 [Os1-Os4 (-1,0,-1)]<br>J11a=-0.023 [Os1-Os3 (-1,-1,-1)]<br>J11b=-0.103 [Os1-Os3 (-1,-1,1)]<br>J12a=-0.027 [Os1-Os4 (-1,-1,0)]<br>J12b=-0.042 [Os1-Os4 (-1,1,0)] |
| 0.45 | $La_2NiO_4$ | Ni1 (0,0,0)<br>Ni2 (0.5,0,0.5)<br>Ni3 (0,0.5,0.5)<br>Ni4 (0.5,0.5,0) | 4.2<br>(Ni 3*d*) | J1a=-12.357, $D_{1a}$= (0.148, -0.19, 0.044),<br>$J_{1a}^{ani}$= (-0.006, -0.006, -0.011),<br>$\Gamma_{1a}$= (0.002, 0, 0)<br>[Ni1-Ni4 (-1,-1,0)]<br>J1b=-13.327 [Ni1-Ni4 (-1,0,0)]<br>J2=0.705 [Ni1-Ni1 (-1,0,0)]<br>J3=-0.014 [Ni1-Ni2 (-1,0,-1)]<br>J4=-0.010 [Ni1-Ni3 (0,-1,-1)]<br>J5a=0.250 [Ni1-Ni1 (-1,-1,0)]<br>J5b=0.266 [Ni1-Ni1 (-1,1,0)] |
| 0.50 | $MnTiO_3$ | Mn1 (0.284,0.284,0.284)<br>Mn2 (0.784,0.784,0.784) | 3.9<br>(Mn 3*d*) | J1a=-1.506, $D_{1a}$= (0.148, -0.19, 0.044)<br>[Mn1-Mn2 (-1,0,-1)]<br>J1b=-1.514 [Mn1-Mn2 (-1,-1,0)]<br>J2=-0.478 [Mn1-Mn1 (-1,0,1)]<br>J3=-0.373 [Mn1-Mn1 (-1,0,0)]<br>J4=-0.211 [Mn1-Mn2 (-1,-1,1)]<br>J5=-0.007 [Mn1-Mn2 (-1,-1,-1)]<br>J6=-0.022 [Mn1-Mn1 (-1,-1,1)]<br>J7a=-0.033 [Mn1-Mn2 (-1,-2,1)]<br>J7b=-0.008 [Mn1-Mn2 (-1,1,-2)]<br>J8a=-0.012 [Mn1-Mn2 (-1,-2,0)]<br>J8b=-0.007 [Mn1-Mn2 (-1,0,-2)]<br>J9=-0.005 [Mn1-Mn1 (-1,-1,2)]<br>J10a=-0.002 [Mn1-Mn1 (-1,0,2)]<br>J10b=-0.000 [Mn1-Mn1 (-1,2,0)]<br>$J_i^{ani}$=$\Gamma_i$= (0, 0, 0) |
| 0.56 | $Ba_2CoGe_2O_7$ | Co1 (0,0,0)<br>Co2 (0.5,0.5,0) | 3.0<br>(Co 3*d*) | J1=0.002, $D_1$= (-0.006, -0.034, 0)<br>[Co1-Co1 (0,0,-1)]<br>J2=-0.487 [Co1-Co2 (-1,-1,0)]<br>J3=-0.002 [Co1-Co2 (-1,-1,-1)]<br>J4=0.284 [Co1-Co1 (-1,0,0)] |
| | | | | J5=0.001 [Co1-Co1 (-1,0,-1)]<br>$J_i^{ani}$=$\Gamma_i$= (0, 0, 0) |
| 0.65, 0.66 | $Fe_2O_3$−alpha | Fe1 (0.355,0.355,0.355)<br>Fe2 (0.645,0.645,0.645)<br>Fe3 (0.145,0.145,0.145)<br>Fe4 (0.855,0.855,0.855) | 4.0<br>(Fe 3*d*) | J1=-3.528, $D_1$= (0.197, -0.116, 0.469),<br>$J_1^{ani}$= (-0.089, -0.089, -0.084),<br>$\Gamma_1$= (0.003, -0.003, -0.007)<br>[Fe1-Fe3 (0,0,0)]<br>J2=-4.153 [Fe1-Fe2 (-1,0,0)]<br>J3=-15.645 [Fe1-Fe3 (0,0,1)]<br>J4=-9.986 [Fe1-Fe4 (-1,-1,0)]<br>J5=0.377 [Fe1-Fe2 (0,0,0)]<br>J6a=0.231 [Fe1-Fe1 (-1,0,1)]<br>J6b=0.225 [Fe1-Fe1 (0,-1,1)]<br>J6c=0.166 [Fe1-Fe1 (-1,1,0)]<br>J7=0.095 [Fe1-Fe1 (-1,0,0)]<br>J8a=-0.136 [Fe1-Fe3 (0,1,-1)]<br>J8b=-0.132 [Fe1-Fe3 (-1,0,1)]<br>J8c=-0.127 [Fe1-Fe3 (-1,1,0)]<br>J9=0.092 [Fe1-Fe2 (-1,-1,1)]<br>J10=-0.264 [Fe1-Fe2 (-1,-1,0)]<br>J11=-0.449 [Fe1-Fe3 (-1,1,1)]<br>J12=-0.048 [Fe1-Fe4 (-1,-1,1)]<br>J13a=0.308 [Fe1-Fe2 (-1,0,1)]<br>J13b=-0.498 [Fe1-Fe2 (-1,1,0)]<br>J14=0.004 [Fe1-Fe4 (-1,-1,-1)]<br>J15=0.007 [Fe1-Fe3 (0,1,1)]<br>J16=0.078 [Fe1-Fe1 (-1,-1,1)] |
| 0.79 | $CaIrO_3^{\dagger}$ | Ir1 (0, 0, 0)<br>Ir2 (0, 0, 0.5) | 2.0<br>(Ir 5*d*) | J1=0.719 [Ir1-Ir1 (-1,0,0)]<br>J2=-4.373, $\boldsymbol{D_2}$**= (1.569, 0, 0)**,<br>$\boldsymbol{J_2^{ani}}$**= (-0.503, 0.42, 1.038)**,<br>[Ir1-Ir2 (0,0,-1)]<br>J3=0.044 [Ir1-Ir2 (-1,0,-1)]<br>J4=-0.013 [Ir1-Ir1 (-1/2,-1/2,0)]<br>J5=-0.072 [Ir1-Ir1 (-2,0,0)]<br>J6=0.002 [Ir1-Ir2 (-1/2,-1/2,-1)]<br>J7=-0.004 [Ir1-Ir1 (-3/2,-1/2,0)]<br>J8=-0.001 [Ir1-Ir2 (-2,0,-1)]<br>J9=0.070 [Ir1-Ir1 (0,0,-1)]<br>$\Gamma_i$= (0, 0, 0) |
| 0.83 | $LiFeP_2O_7$ | Fe1 (0.22,0.25,0.234)<br>Fe2 (0.78,0.75,0.766) | 4.9<br>(Fe 3*d*) | J1=-0.647 [Fe1-Fe1 (-1,0,0)]<br>J2=-0.263 [Fe1-Fe2 (-1,-1,-1)]<br>J3=-0.975, $D_3$= (-0.009, -0.039, -0.03)<br>[Fe1-Fe2 (0,-1,0)]<br>J4=-0.932 [Fe1-Fe2 (-1,-1,0)]<br>J5=-0.785 [Fe1-Fe2 (0,-1,-1)] |
| | | | | J6=0.015 [Fe1-Fe1 (0,0,-1)]<br>J7=0.043 [Fe1-Fe1 (-1,0,-1)]<br>J8=-0.072 [Fe1-Fe2 (-2,-1,-1)]<br>J9=-0.002 [Fe1-Fe1 (0,-1,0)]<br>J10=-0.001 [Fe1-Fe2 (1,-1,0)]<br>J11a=-0.000 [Fe1-Fe1 (-1,-1,0)]<br>J11b=-0.002 [Fe1-Fe1 (-1,1,0)]<br>$J_i^{ani}$=$\Gamma_i$= (0, 0, 0) |
| 0.115 | $MnCO_3$ | Mn1 (0,0,0)<br>Mn2 (0.5,0.5,0.5) | 3.5<br>(Mn 3*d*) | J1a=-0.902 [Mn1-Mn2 (-1,-1,0)]<br>J1b=-1.005, $D_{1b}$= (0.047, -0.027, -0.061)<br>[Mn1-Mn2 (-1,0,-1)]<br>J2=1.267 [Mn1-Mn1 (-1,1,0)]<br>J3=0.137 [Mn1-Mn1 (-1,0,0)]<br>J4=0.696 [Mn1-Mn2 (-1,-1,1)]<br>J5=-0.005 [Mn1-Mn1 (-1,-1,1)]<br>J6=0.065 [Mn1-Mn2 (-1,-2,1)]<br>J7=-0.005 [Mn1-Mn2 (-1,-1,-1)]<br>J8a=0.081 [Mn1-Mn1 (-2,1,1)]<br>J8b=0.070 [Mn1-Mn1 (-1,2,-1)]<br>J8c=0.069 [Mn1-Mn1 (-1,-1,2)]<br>J9a=0.023 [Mn1-Mn1 (-1,0,2)]<br>J9b=-0.000 [Mn1-Mn1 (-1,2,0)]<br>$J_i^{ani}$=$\Gamma_i$= (0, 0, 0) |
| 0.118 | $Ba_5Co_5ClO_{13}$ | Co1 (0,0,0.603)<br>Co2 (0,0,0.103)<br>Co3 (0,0,0.397)<br>Co4 (0,0,0.897)<br>Co5 (0.333,0.667,0.823)<br>Co6 (0.667,0.333,0.323)<br>Co7 (0.667,0.333,0.177)<br>Co8 (0.333,0.667,0.677)<br>Co9 (0,0,0)<br>Co10 (0,0,0.5) | 0 | J1=0.936 [Co1-Co10 (0,0,0)]<br>J2=-43.566, $J_2^{ani}$= (0.418, 0.417, 0.537)<br>[Co5-Co8 (0,0,0)]<br>J3=8.034, $\Gamma_3$= (0.019, 0.02, -0.012)<br>[Co1-Co8 (-1,-1,0)]<br>J4=-0.030 [Co1-Co3 (0,0,0)]<br>J5=0.476, $D_5$= (-0.138, 0, 0)<br>[Co5-Co9 (0,0,1)]<br>J6a=0.313 [Co1-Co1 (-1,-1,0)]<br>J6b=0.163 [Co5-Co5 (-1,-1,0)]<br>J6c=0.034 [Co9-Co9 (-1,-1,0)]<br>J7=-0.003 [Co1-Co10 (-1,-1,0)]<br>J8=-0.559 [Co1-Co5 (-1,-1,0)] |
| 0.137 | $Cu_2V_2O_7$[†] | Cu1 (0.166,0.365,0.752)<br>Cu2 (0.834,0.635,0.752)<br>Cu3 (0.084,0.615,0.002)<br>Cu4 (0.416,0.885,0.002) | 6.5<br>(Cu 3*d*) | J1=-0.994, **$D_1$= (-0.76, -0.673, 0.088)**,<br>**$J_1^{ani}$= (-0.009, 0.089, -0.079)**,<br>$\Gamma_1$= (0.057, 0.018, -0.01)<br>[Cu1-Cu3 (0,0,1)]<br>J2=-0.101 [Cu1-Cu2 (-1/2,-1/2,0)]<br>J3=-1.451 [Cu1-Cu2 (-1/2,0,-1/2)]<br>J4a=-0.109 [Cu1-Cu1 (0,-1/2,1/2)]<br>J4b=0.015 [Cu1-Cu1 (0,-1/2,-1/2)] |
| | | | | J5=-0.010 [Cu1-Cu4 (-1/2,-1/2,1)]<br>J6=-0.165 [Cu1-Cu3 (0,0,0)]<br>J7=-0.007 [Cu1-Cu1 (0,0,-1)]<br>J8=-0.011 [Cu1-Cu3 (0,-1,1)]<br>J9=-0.020 [Cu1-Cu4 (-1/2,-1,1/2)]<br>J10=-0.003 [Cu1-Cu4 (-1/2,0,1/2)]<br>J11=-0.004 [Cu1-Cu4 (-1/2,-1/2,0)]<br>J12=-0.007 [Cu1-Cu2 (-1,0,0)]<br>J13=-0.499 [Cu1-Cu2 (-1/2,1/2,0)] |
| 0.148 | $La_2LiRuO_6$ | Ru1 (0, 0.5, 0)<br>Ru2 (0.5, 0, 0.5) | 2.1<br>(Ru 4*d*) | J1=0.693, $\Gamma_1$= (0.002, -0.013, -0.001)<br>[Ru1-Ru1 (-1,0,0)]<br>J2=-0.704, $D_2$= (-0.018, 0.021, -0.016)<br>[Ru1-Ru2 (-1,0,-1)]<br>J3=-1.135 [Ru1-Ru2 (-1,0,0)]<br>J4=-1.361, $J_4^{ani}$= (-0.022, -0.013, -0.029)<br>[Ru1-Ru1 (0,-1,0)]<br>J5=0.018 [Ru1-Ru1 (0,0,-1)]<br>J6a=0.009 [Ru1-Ru1 (-1,-1,0)]<br>J6b=0.004 [Ru1-Ru1 (-1,1,0)]<br>J7=-0.001 [Ru1-Ru1 (-1,0,-1)]<br>J8=-0.006 [Ru1-Ru1 (-1,0,1)]<br>J9=-0.001 [Ru1-Ru2 (-2,0,-1)]<br>J10=0.003 [Ru1-Ru2 (-2,0,0)]<br>J11a=0.000 [Ru1-Ru1 (0,-1,-1)]<br>J11b=-0.003 [Ru1-Ru1 (0,-1,1)]<br>J12=-0.002 [Ru1-Ru2 (-1,-1,-1)]<br>J13=0.004 [Ru1-Ru2 (-1,-1,0)]<br>J15=-0.000 [Ru1-Ru1 (-2,0,0)]<br>J16a=0.002 [Ru1-Ru1 (-1,1,-1)]<br>J16b=-0.004 [Ru1-Ru1 (-1,-1,-1)]<br>J17a=0.003 [Ru1-Ru1 (-1,-1,1)]<br>J17b=0.002 [Ru1-Ru1 (-1,1,1)] |
| 0.178 | $CoF_2$ | Co1 (0,0,0)<br>Co2 (0.5,0.5,0.5) | 6.0<br>(Co 3*d*) | J1=-0.715 [Co1-Co1 (0,0,-1)]<br>J2=-1.439, $D_2$= (0.077, -0.078, 0.023),<br>$J_2^{ani}$= (-0.007, -0.007, -0.016)<br>[Co1-Co2 (-1,-1,-1)]<br>J3=0.007 [Co1-Co1 (-1,0,0)]<br>J4=-0.026 [Co1-Co1 (-1,0,-1)]<br>J5=0.007 [Co1-Co2 (-1,-1,-2)]<br>J6=-0.034 [Co1-Co1 (0,0,-2)]<br>J7a=-0.096 [Co1-Co1 (-1,-1,0)]<br>J7b=-0.001 [Co1-Co1 (-1,1,0)]<br>J8a=0.003 [Co1-Co1 (-1,-1,-1)]<br>J8b=0.004 [Co1-Co1 (-1,1,-1)] |

| | | | | $\Gamma_i$= (0, 0, 0) |
|---|---|---|---|---|
| 0.189 | $CeMn_2Ge_4O_{12}$ | Mn1 (0,0,0.5)<br>Mn2 (0,0.5,0.5)<br>Mn3 (0.5,0,0.5)<br>Mn4 (0.5,0.5,0.5) | 5.0<br>(Mn 3*d*) | J1=-0.079, $D_2$= (0, 0.006, -0.002)<br>[Mn1-Mn1 (0,0,-1)]<br>J2=-0.084 [Mn1-Mn2 (0,-1,0)]<br>J3a=-0.233 [Mn1-Mn2 (0,-1,-1)]<br>J3b=-0.059 [Mn1-Mn2 (0,-1,1)]<br>J4a=-0.027 [Mn1-Mn4 (-1,-1,0)]<br>J4b=-0.015 [Mn1-Mn4 (-1,0,0)]<br>J5a=-0.005 [Mn1-Mn4 (-1,-1,-1)]<br>J5b=-0.002 [Mn1-Mn4 (-1,0,-1)]<br>$J_i^{ani}$=$\Gamma_i$= (0, 0, 0) |
| 0.229 | $Ba_2MnSi_2O_7$ | Mn1 (0,0,0)<br>Mn2 (0.5,0.5,0) | 4.0<br>(Mn 3*d*) | J1=0.003 [Mn1-Mn1 (0,0,-1)]<br>J2=-0.404 [Mn1-Mn2 (-1,-1,0)]<br>J3=-0.001 [Mn1-Mn2 (-1,-1,-1)]<br>J4=-0.002 [Mn1-Mn1 (-1,0,0)]<br>$D_i$=$J_i^{ani}$=$\Gamma_i$= (0, 0, 0) |
| 0.260 | $CuFePO_5$ | Cu1 (0,0,0)<br>Cu2 (0.5,0.5,0.5)<br>Cu3 (0,0.5,0)<br>Cu4 (0.5,0,0.5)<br>Fe1 (0.173,0.25,0.712)<br>Fe2 (0.673,0.25,0.788)<br>Fe3 (0.827,0.75,0.288)<br>Fe4 (0.327,0.75,0.212) | 4.0<br>(Cu 3*d*)<br>4.9<br>(Fe 3*d*) | J1=-0.417, $D_1$= (0.338, -0.085, -0.334),<br>$J_1^{ani}$= (-0.045, -0.024, -0.025),<br>$\Gamma_1$= (0.011, 0.011, 0.003)<br>[Cu1-Fe1 (0,0,-1)]<br>J2=-0.977 [Cu1-Cu3 (0,-1,0)]<br>J3=-8.238 [Cu1-Fe2 (-1,0,-1)]<br>J4=-7.474 [Fe1-Fe2 (-1,0,0)]<br>J5=-0.106 [Fe1-Fe4 (0,-1,0)]<br>J6=1.385 [Fe1-Fe3 (-1,-1,0)]<br>J7=0.005 [Cu1-Cu4 (-1,0,-1)]<br>J8=-0.008 [Cu1-Fe1 (0,-1,-1)]<br>J9=-0.037 [Cu1-Fe1 (0,0,0)]<br>J10=0.008 [Cu1-Fe2 (0,0,-1)]<br>J11=0.009 [Cu1-Fe2 (-1,-1,-1)]<br>J12=-0.046 [Fe1-Fe3 (-1,-1,1)]<br>J13=-0.014 [Cu1-Cu2 (-1,-1,-1)]<br>J14=-0.506 [Cu1-Fe2 (-1,0,0)]<br>J15=-0.011 [Cu1-Cu1 (0,-1,0)]<br>J16=-1.654 [Fe1-Fe1 (0,-1,0)]<br>J17=0.295 [Fe1-Fe3 (0,-1,0)] |
| 0.261 | $NiFePO_5$ | Ni1 (0,0,0)<br>Ni2 (0.5,0.5,0.5)<br>Ni3 (0,0.5,0)<br>Ni4 (0.5,0,0.5)<br>Fe1 (0.144,0.25,0.707)<br>Fe2 (0.644,0.25,0.793)<br>Fe3 (0.856,0.75,0.293)<br>Fe4 (0.356,0.75,0.207) | 4.6<br>(Ni 3*d*)<br>4.9<br>(Fe 3*d*) | J1=-0.341 [Fe1-Ni1 (0,0,1)]<br>J2=-0.448, $D_2$= (0.086, 0, -0.015)<br>[Ni1-Ni3 (0,-1,0)]<br>J3=-6.689 [Fe1-Ni2 (0,0,0)]<br>J4=-13.902, $J_4^{ani}$= (-0.098, -0.082, -0.104)<br>[Fe1-Fe2 (-1,0,0)]<br>J5=-1.271 [Fe1-Fe3 (-1,-1,0)]<br>J6=-0.170 [Fe1-Ni2 (-1,0,0)] |
| | | | | J7=-0.099 [Fe1-Fe4 (0,-1,0)]<br>J8=-0.031 [Ni1-Ni4 (-1,0,-1)]<br>J9=-0.016 [Fe1-Ni1 (0,1,1)]<br>J10=-0.084 [Fe1-Ni1 (0,0,0)]<br>J11=-0.052 [Fe1-Ni2 (0,-1,0)]<br>J12=-0.055 [Fe1-Fe3 (-1,-1,1)]<br>J13a=-0.484 [Ni1-Ni2 (-1,-1,-1)]<br>J13b=-0.001 [Ni1-Ni2 (-1,-1,0)]<br>J14=-1.750 [Fe1-Fe1 (0,-1,0)]<br>$\Gamma_i$= (0, 0, 0) |
| 0.263 | $Fe_2PO_5$ | Fe1 (0,0,0)<br>Fe2 (0.5,0.5,0.5)<br>Fe3 (0,0.5,0)<br>Fe4 (0.5,0,0.5)<br>Fe5 (0.341,0.75,0.213)<br>Fe6 (0.841,0.75,0.287)<br>Fe7 (0.659,0.25,0.787)<br>Fe8 (0.159,0.25,0.713) | 4.0<br>(Fe 3*d*) | J1=-0.956, $D_1$= (0.007, -0.265, 0.255)<br>[Fe1-Fe6 (-1,-1,0)]<br>J2=-2.599 [Fe1-Fe3 (0,-1,0)]<br>J3=-13.409, $J_3^{ani}$= (0.117, 0.099, 0.096),<br>$\Gamma_3$= (0.002, 0.014, 0.019)<br>[Fe1-Fe5 (0,-1,0)]<br>J4=-11.864 [Fe5-Fe6 (-1,0,0)]<br>J5=-1.486 [Fe5-Fe7 (0,0,-1)]<br>J6=-0.222 [Fe5-Fe8 (0,0,-1)]<br>J7=-0.062 [Fe1-Fe4 (-1,0,-1)]<br>J8=-0.224 [Fe1-Fe5 (-1,-1,0)]<br>J9=-0.051 [Fe1-Fe6 (-1,0,0)]<br>J10=-0.119 [Fe1-Fe6 (-1,-1,-1)]<br>J11=-0.039 [Fe1-Fe5 (0,0,0)]<br>J12=-0.035 [Fe5-Fe7 (0,0,0)]<br>J13a=0.038 [Fe1-Fe2 (-1,-1,0)]<br>J13b=-0.248 [Fe1-Fe2 (-1,-1,-1)]<br>J14a=-0.002 [Fe1-Fe1 (0,-1,0)]<br>J14b=-2.129 [Fe5-Fe5 (0,-1,0)]<br>J15=-0.460 [Fe1-Fe5 (0,-1,-1)]<br>J16=0.170 [Fe5-Fe7 (-1,0,-1)]<br>J17=0.051 [Fe1-Fe6 (0,-1,0)]<br>J18=-0.016 [Fe1-Fe5 (-1,0,0)]<br>J19=-0.018 [Fe1-Fe6 (-1,0,-1)]<br>J20=-0.044 [Fe5-Fe7 (-1,0,0)]<br>J21a=-0.001 [Fe1-Fe1 (-1,0,0)]<br>J21b=0.171 [Fe5-Fe5 (-1,0,0)] |
| 0.301 | $Sr_2CoTeO_6$ | Co1 (0, 0, 0.5)<br>Co2 (0.5, 0.5, 0) | 4.1<br>(Co 3*d*) | J1=-1.489 [Co1-Co1 (0,-1,0)]<br>J2=-1.031, $\Gamma_2$= (0.002, 0.002, -0.001)<br>[Co1-Co2 (-1,-1,1)]<br>J3=-0.850, $D_3$= (0.048, 0.008, -0.016)<br>$J_3^{ani}$= (-0.006, -0.003, -0.004)<br>[Co1-Co2 (-1,-1,0)]<br>J4=1.207 [Co1-Co1 (-1,0,0)] |
| | | | | J5=0.015 [Co1-Co1 (0,0,-1)]<br>J6a=0.074 [Co1-Co1 (-1,1,0)]<br>J6b=0.062 [Co1-Co1 (-1,-1,0)] |
| 0.303 | $BaCrF_5$ | Cr1 (0.165,0.603,0.425)<br>Cr2 (0.665,0.897,0.575)<br>Cr3 (0.335,0.397,0.925)<br>Cr4 (0.835,0.103,0.075) | 3.5<br>(Cr 3*d*) | J1=-0.580, $D_1$= (-0.01, -0.004, -0.014)<br>[Cr1-Cr3 (0,0,-1)]<br>J2=-0.000 [Cr1-Cr1 (0,0,-1)]<br>J3=-0.051 [Cr1-Cr4 (-1,0,0)]<br>J4=0.003 [Cr1-Cr3 (0,1,-1)]<br>J5=-0.003 [Cr1-Cr1 (0,-1,0)]<br>J6=0.013 [Cr1-Cr4 (-1,0,1)]<br>J7=0.001 [Cr1-Cr2 (-1,0,0)]<br>J8a=0.001 [Cr1-Cr1 (0,-1,-1)]<br>J8b=0.001 [Cr1-Cr1 (0,-1,1)]<br>$J_i^{ani}$=$\Gamma_i$= (0, 0, 0) |
| 0.307 | $ScCrO_3$ | Cr1 (0,0,0.5)<br>Cr2 (0.5,0.5,0)<br>Cr3 (0,0.5,0.5)<br>Cr4 (0.5,0,0) | 2.0<br>(Cr 3*d*) | J1=-1.227, $D_1$= (-0.008, 0.004, 0.054)<br>[Cr1-Cr4 (-1,0,0)]<br>J2=1.268 [Cr1-Cr3 (0,-1,0)]<br>J3=0.075 [Cr1-Cr1 (0,0,-1)]<br>J4a=-0.075 [Cr1-Cr2 (-1,-1,0)]<br>J4b=0.023 [Cr1-Cr2 (-1,-1,1)]<br>J5=0.108 [Cr1-Cr1 (-1,0,0)]<br>J6=0.021 [Cr1-Cr3 (0,-1,-1)]<br>J7=-0.000 [Cr1-Cr3 (-1,-1,0)]<br>J8a=0.005 [Cr1-Cr1 (-1,0,-1)]<br>J8b=-0.071 [Cr1-Cr1 (-1,0,1)]<br>J9=0.101 [Cr1-Cr1 (0,-1,0)]<br>$J_i^{ani}$=$\Gamma_i$= (0, 0, 0) |
| 0.308 | $InCrO_3$† | Cr1 (0,0,0.5)<br>Cr2 (0.5,0.5,0)<br>Cr3 (0,0.5,0.5)<br>Cr4 (0.5,0,0) | 2.0<br>(Cr 3*d*) | J1=-1.150, **$D_1$= (0.092, -0.134, 0.03)**<br>[Cr1-Cr4 (-1,0,0)]<br>J2=1.385 [Cr1-Cr3 (0,-1,0)]<br>J3=-0.188 [Cr1-Cr1 (0,0,-1)]<br>J4a=-0.210 [Cr1-Cr2 (-1,-1,0)]<br>J4b=-0.061 [Cr1-Cr2 (-1,-1,1)]<br>J5=0.195 [Cr1-Cr1 (-1,0,0)]<br>J6=0.002 [Cr1-Cr3 (0,-1,-1)]<br>J7=-0.007 [Cr1-Cr3 (-1,-1,0)]<br>J8a=0.179 [Cr1-Cr1 (-1,0,-1)]<br>J8b=0.091 [Cr1-Cr1 (-1,0,1)]<br>J9=0.184 [Cr1-Cr1 (0,-1,0)]<br>$J_i^{ani}$=$\Gamma_i$= (0, 0, 0) |
| 0.309 | $TlCrO_3$ | Cr1 (0,0,0.5)<br>Cr2 (0.5,0.5,0)<br>Cr3 (0,0.5,0.5)<br>Cr4 (0.5,0,0) | 2.0<br>(Cr 3*d*) | J1=-1.417, $D_1$= (0.021, -0.084, 0.02)<br>[Cr1-Cr4 (-1,0,0)]<br>J2=0.897 [Cr1-Cr3 (0,-1,0)]<br>J3=-0.522 [Cr1-Cr1 (0,0,-1)] |
| | | | | J4a=-0.517 [Cr1-Cr2 (-1,-1,0)]<br>J4b=-0.187 [Cr1-Cr2 (-1,-1,1)]<br>J5=0.145 [Cr1-Cr1 (-1,0,0)]<br>J6=-0.064 [Cr1-Cr3 (0,-1,-1)]<br>J7=-0.046 [Cr1-Cr3 (-1,-1,0)]<br>J8a=0.196 [Cr1-Cr1 (-1,0,-1)]<br>J8b=0.086 [Cr1-Cr1 (-1,0,1)]<br>J9=0.215 [Cr1-Cr1 (0,-1,0)]<br>J10=0.012 [Cr1-Cr4 (-1,0,-1)]<br>J11a=-0.014 [Cr1-Cr3 (-1,-1,-1)]<br>J11b=-0.000 [Cr1-Cr3 (-1,-1,1)]<br>J12=0.005 [Cr1-Cr4 (-2,0,0)]<br>J13a=0.017 [Cr1-Cr4 (-1,-1,0)]<br>J13b=0.005 [Cr1-Cr4 (-1,1,0)]<br>J14a=-0.012 [Cr1-Cr2 (-1,-1,-1)]<br>J14b=-0.005 [Cr1-Cr2 (-1,-1,2)]<br>$J_i^{ani}$=$\Gamma_i$= (0, 0, 0) |
| 0.315 | $ZrMn_2Ge_4O_{12}$ | Mn1 (0,0,0.5)<br>Mn2 (0,0.5,0.5)<br>Mn3 (0.5,0,0.5)<br>Mn4 (0.5,0.5,0.5) | 5.0<br>(Mn 3*d*) | J1=-0.151, $D_1$= (-0.004, -0.014, -0.001)<br>[Mn1-Mn2 (0,-1,0)]<br>J2=-0.059 [Mn1-Mn1 (0,0,-1)]<br>J3a=-0.019 [Mn1-Mn4 (-1,0,0)]<br>J3b=-0.009 [Mn1-Mn4 (-1,-1,0)]<br>J4a=-0.391 [Mn1-Mn2 (0,-1,-1)]<br>J4b=-0.115 [Mn1-Mn2 (0,-1,1)]<br>J5a=0.001 [Mn1-Mn4 (-1,-1,-1)]<br>J5b=-0.005 [Mn1-Mn4 (-1,0,-1)]<br>$J_i^{ani}$=$\Gamma_i$= (0, 0, 0) |
| 0.329 | $RbMnF_4$ | Mn1 (0,0,0)<br>Mn2 (0.5,0.5,0)<br>Mn3 (0.5,0,0)<br>Mn4 (0,0.5,0) | 3.9<br>(Mn 3*d*) | J1=-2.911, $D_1$= (-0.042, -0.022, 0.011),<br>$J_1^{ani}$= (-0.014, -0.008, -0.009),<br>$\Gamma_1$= (-0.001, 0.001, -0.001)<br>[Mn1-Mn4 (0,-1,0)]<br>J2=2.646 [Mn1-Mn3 (-1,0,0)]<br>J3a=-0.147 [Mn1-Mn2 (-1,-1,0)]<br>J3b=0.003 [Mn3-Mn4 (0,-1,0)]<br>J4a=0.057 [Mn1-Mn1 (0,0,-1)]<br>J4b=0.066 [Mn3-Mn3 (0,0,-1)]<br>J5=0.002 [Mn1-Mn3 (-1,0,-1)]<br>J6a=-0.000 [Mn1-Mn4 (0,-1,-1)]<br>J6b=-0.001 [Mn1-Mn4 (0,-1,1)]<br>J7=0.005 [Mn1-Mn3 (-1,0,1)]<br>J8a=-0.040 [Mn1-Mn1 (0,-1,0)]<br>J8b=-0.307 [Mn3-Mn3 (0,-1,0)]<br>J9a=-0.170 [Mn1-Mn1 (-1,0,0)]<br>J9b=-0.033 [Mn3-Mn3 (-1,0,0)] |
| 0.331 | $Fe_2Mo_3O_8$ † | Fe1 (0.333,0.667,0.957)<br>Fe2 (0.667,0.333,0.457)<br>Fe3 (0.333,0.667,0.512)<br>Fe4 (0.667,0.333,0.012) | 4.0<br>(Fe 3*d*) | J1=-1.515, $\boldsymbol{D}_1$**= (-0.226, -0.053, -0.006)**,<br>$J_1^{ani}$= (-0.021, -0.018, -0.008),<br>$\Gamma_1$= (0.003, 0.001, 0)<br>[Fe1-Fe4 (-1,0,1)]<br>J2=-0.261 [Fe1-Fe3 (0,0,0)]<br>J3=-0.053 [Fe1-Fe3 (0,0,1)]<br>J4a=-0.185 [Fe1-Fe1 (-1,0,0)]<br>J4b=0.049 [Fe3-Fe3 (-1,0,0)]<br>J5a=-0.176 [Fe1-Fe2 (0,1,1)]<br>J5b=-0.059 [Fe3-Fe4 (0,1,1)] |
| 0.334 | $CoF_3$ | Co1 (0, 0, 0)<br>Co2 (0, 0, 0.5) | 5.3<br>(Co 3*d*) | J1=-6.552, $D_1$= (0.049, 0.176, 0.24),<br>$J_1^{ani}$= (-0.022, -0.079, -0.027)<br>[Co1-Co2 (-1/3,-2/3,-2/3)]<br>J2=0.505, $\Gamma_2$= (0.004, 0.004, 0)<br>[Co1-Co1 (0,-1,0)]<br>J3=0.481 [Co1-Co1 (-1/3,-2/3,1/3)]<br>J4=0.009 [Co1-Co2 (-4/3,-2/3,-2/3)]<br>J5=0.008 [Co1-Co2 (0,0,-1)]<br>J6a=0.051 [Co1-Co1 (2/3,4/3,1/3)]<br>J6b=0.026 [Co1-Co1 (2/3,-2/3,1/3)]<br>J6c=0.023 [Co1-Co1 (-4/3,-2/3,1/3)] |
| 0.335,<br>0.581 | $FeF_3$ | Fe1 (0, 0, 0)<br>Fe2 (0.5, 0.5, 0.5) | 4.9<br>(Fe 3*d*) | J1a=-11.837, $D_1$= (-0.001, -0.154, 0.152),<br>$J_1^{ani}$= (-0.032, -0.039, -0.034),<br>$\Gamma_1$= (0.004, -0.007, 0.004)<br>[Fe1-Fe2 (-1,-1,1)]<br>J1b=-11.855 [Fe1-Fe2 (-1,0,0)]<br>J2a=0.173 [Fe1-Fe1 (-1,0,1)]<br>J2b=0.169 [Fe1-Fe1 (0,-1,1)]<br>J2c=0.162 [Fe1-Fe1 (-1,1,0)]<br>J3a=0.187 [Fe1-Fe1 (0,0,-1)]<br>J3b=0.179 [Fe1-Fe1 (0,-1,0)]<br>J3c=0.174 [Fe1-Fe1 (1,0,0)]<br>J4=0.001 [Fe1-Fe2 (-1,-1,2)]<br>J5=-0.001 [Fe1-Fe2 (-1,-1,0)]<br>J6a=0.114 [Fe1-Fe1 (-1,1,-1)]<br>J6b=0.114 [Fe1-Fe1 (-1,1,1)]<br>J6c=0.111 [Fe1-Fe1 (-1,-1,1)] |
| 0.336 | $NdFeO_3$ | Fe1 (0.5,0,0)<br>Fe2 (0,0.5,0.5)<br>Fe3 (0.5,0,0.5)<br>Fe4 (0,0.5,0) | 5.0<br>(Fe 3*d*) | J1=-8.738 [Fe1-Fe3 (0,0,-1)]<br>J2=-8.887, $D_2$= (0.218, -0.138, -0.212),<br>$J_2^{ani}$= (-0.088, -0.087, -0.085),<br>$\Gamma_2$= (-0.011, -0.004, 0.004)<br>[Fe1-Fe4 (0,-1,0)]<br>J3=0.263 [Fe1-Fe1 (-1,0,0)]<br>J4a=0.269 [Fe1-Fe2 (0,-1,0)] |
| | | | | J4b=0.234 [Fe1-Fe2 (0,-1,-1)]<br>J5=0.133 [Fe1-Fe1 (0,-1,0)]<br>J6=-0.014 [Fe1-Fe3 (-1,0,-1)]<br>J7=-0.015 [Fe1-Fe3 (0,-1,-1)]<br>J8=0.154 [Fe1-Fe1 (0,0,-1)]<br>J9a=0.218 [Fe1-Fe1 (-1,-1,0)]<br>J9b=0.194 [Fe1-Fe1 (-1,1,0)]<br>J10=-0.009 [Fe1-Fe4 (-1,-1,0)]<br>J11a=-0.011 [Fe1-Fe4 (0,-1,-1)]<br>J11b=-0.022 [Fe1-Fe4 (0,-1,1)]<br>J12=-0.009 [Fe1-Fe3 (-1,-1,-1)]<br>J13=-0.018 [Fe1-Fe4 (0,-2,0)] |
| 0.351 | $TbFeO_3$ | Fe1 (0.5,0,0)<br>Fe2 (0,0.5,0.5)<br>Fe3 (0.5,0,0.5)<br>Fe4 (0,0.5,0) | 4.0<br>(Fe 3*d*) | J1=-8.444, $D_1$= (0.224, -0.33, 0),<br>$J_1^{ani}$= (-0.096, -0.097, -0.095),<br>$\Gamma_1$= (-0.004, 0, 0)<br>[Fe1-Fe3 (0,0,-1)]<br>J2=-9.223 [Fe1-Fe4 (0,-1,0)]<br>J3=0.151 [Fe1-Fe1 (-1,0,0)]<br>J4a=0.150 [Fe1-Fe2 (0,-1,0)]<br>J4b=0.133 [Fe1-Fe2 (0,-1,-1)]<br>J5=-0.090 [Fe1-Fe1 (0,-1,0)]<br>J6=-0.005 [Fe1-Fe3 (-1,0,-1)]<br>J7=-0.025 [Fe1-Fe3 (0,-1,-1)]<br>J8=0.187 [Fe1-Fe1 (0,0,-1)]<br>J9a=0.241 [Fe1-Fe1 (-1,1,0)]<br>J9b=0.285 [Fe1-Fe1 (-1,-1,0)]<br>J10=-0.008 [Fe1-Fe4 (-1,-1,0)]<br>J11a=-0.012 [Fe1-Fe4 (0,-1,-1)]<br>J11b=-0.031 [Fe1-Fe4 (0,-1,1)]<br>J12a=-0.012 [Fe1-Fe3 (-1,1,-1)]<br>J12b=-0.006 [Fe1-Fe3 (-1,-1,-1)]<br>J13=-0.023 [Fe1-Fe4 (0,-2,0)] |
| 0.354 | $TbCrO_3$† | Cr1 (0.5,0,0)<br>Cr2 (0,0.5,0.5)<br>Cr3 (0.5,0,0.5)<br>Cr4 (0,0.5,0) | 3.0<br>(Cr 3*d*) | J1=-1.875, **$D_1$= (-0.011, 0.199, 0)**,<br>$J_1^{ani}$= (-0.007, -0.014, -0.005),<br>$\Gamma_1$= (-0.001, 0, 0)<br>[Cr1-Cr3 (0,0,-1)]<br>J2=-1.842 [Cr1-Cr4 (0,-1,0)]<br>J3=0.013 [Cr1-Cr1 (-1,0,0)]<br>J4a=0.014 [Cr1-Cr2 (0,-1,-1)]<br>J4b=-0.042 [Cr1-Cr2 (0,-1,0)]<br>J5=-0.068 [Cr1-Cr1 (0,-1,0)]<br>J6=0.002 [Cr1-Cr3 (-1,0,-1)]<br>J7=-0.005 [Cr1-Cr3 (0,-1,-1)]<br>J8=0.139 [Cr1-Cr1 (0,0,-1)] |
| | | | | J9a=0.062 [Cr1-Cr1 (-1,-1,0)]<br>J9b=0.078 [Cr1-Cr1 (-1,1,0)] |
| 0.358 | $CaFe_5O_7$† | Fe1 (0.274,0.425,0.548)<br>Fe2 (0.726,0.925,0.452)<br>Fe3 (0.726,0.575,0.452)<br>Fe4 (0.274,0.075,0.548)<br>Fe5 (0.47,0.654,0.926)<br>Fe6 (0.53,0.154,0.074)<br>Fe7 (0.53,0.346,0.074)<br>Fe8 (0.47,0.846,0.926)<br>Fe9 (0,0.5,0)<br>Fe10 (0,0,0) | 3.0<br>(Fe 3*d*) | J1=3.596 [Fe1-Fe7 (0,0,1)]<br>J2a=6.821 [Fe1-Fe1 (-1,0,0)]<br>J2b=4.386, $\Gamma_{2b}$= (-0.031, 0.004, -0.007)<br>[Fe5-Fe5 (-1,0,0)]<br>J2c=-3.173 [Fe9-Fe9 (-1,0,0)]<br>J2d=1.529 [Fe1-Fe9 (1,0,1)]<br>J3=-2.164 [Fe1-Fe9 (0,0,1)]<br>J4=-0.698 [Fe1-Fe9 (0,0,0)]<br>J5a=-0.978 [Fe1-Fe3 (-1,0,0)]<br>J5b=-0.393 [Fe1-Fe3 (0,0,0)]<br>J6=-10.569 [Fe1-Fe7 (0,0,0)]<br>J7a=-6.837 [Fe1-Fe7 (-1,0,0)]<br>J7b=-0.159 [Fe5-Fe9 (1,0,1)]<br>J8=-1.281 [Fe5-Fe9 (0,0,1)]<br>J9=-16.058 [Fe5-Fe8 (0,0,0)]<br>J10=-12.392, **$D_{10}$= (0.384, -2.045, -0.279)**<br>$J_{10}^{ani}$= (0.559, 0.881, 0.581)<br>[Fe1-Fe7 (1,0,1)]<br>J11=-13.000 [Fe1-Fe7 (-1,0,1)]<br>J12=-8.815 [Fe1-Fe9 (-1,0,0)]<br>J13=-10.201 [Fe1-Fe9 (1,0,0)]<br>J14=-7.588 [Fe1-Fe5 (0,0,0)]<br>J15=0.150 [Fe5-Fe8 (-1,0,0)] |
| 0.379,<br>0.380 | $SmFeO_3$ | Fe1 (0,0.5,0)<br>Fe2 (0.5,0,0.5)<br>Fe3 (0,0.5,0.5)<br>Fe4 (0.5,0,0) | 4.0<br>(Fe 3*d*) | J1=-9.673, $D_1$= (0.167, 0.316, 0),<br>$J_1^{ani}$= (-0.097, -0.096, -0.1),<br>$\Gamma_1$= (0.003, 0, 0)<br>[Fe1-Fe3 (0,0,-1)]<br>J2=-9.876 [Fe1-Fe4 (-1,0,0)]<br>J3=0.229 [Fe1-Fe1 (-1,0,0)]<br>J4a=0.234 [Fe1-Fe2 (-1,0,-1)]<br>J4b=0.184 [Fe1-Fe2 (-1,0,0)]<br>J5=0.036 [Fe1-Fe1 (0,-1,0)]<br>J6=-0.012 [Fe1-Fe3 (-1,0,-1)]<br>J7=-0.021 [Fe1-Fe3 (0,-1,-1)]<br>J8=0.216 [Fe1-Fe1 (0,0,-1)]<br>J9a=0.293 [Fe1-Fe1 (-1,1,0)]<br>J9b=0.260 [Fe1-Fe1 (-1,-1,0)]<br>J10=-0.011 [Fe1-Fe4 (-2,0,0)]<br>J11a=-0.015 [Fe1-Fe4 (-1,0,-1)]<br>J11b=-0.030 [Fe1-Fe4 (-1,0,1)]<br>J12a=-0.015 [Fe1-Fe3 (-1,-1,-1)]<br>J12b=-0.012 [Fe1-Fe3 (-1,1,-1)] |

| | | | | J13=-0.023 [Fe1-Fe4 (-1,-1,0)] |
|---|---|---|---|---|
| 0.420 | $Sr_2LuRuO_6$ | Ru1 (0.5, 0, 0.5)<br>Ru2 (0, 0.5, 0) | 2.1<br>(Ru 4*d*) | J1=-1.752, $D_1$= (0.018, 0.014, 0.001)<br>[Ru1-Ru2 (0,-1,0)]<br>J2=-1.792 [Ru1-Ru1 (-1,0,0)]<br>J3=-2.021, $J_3^{ani}$= (-0.018, -0.03, -0.042),<br>$\Gamma_3$= (-0.004, 0.012, 0.002)<br>[Ru1-Ru1 (0,-1,0)]<br>J4=-1.553 [Ru1-Ru2 (0,-1,1)]<br>J5=0.010 [Ru1-Ru1 (0,0,-1)]<br>J6a=0.049 [Ru1-Ru1 (-1,-1,0)]<br>J6b=0.049 [Ru1-Ru1 (-1,1,0)]<br>J7=-0.002 [Ru1-Ru1 (-1,0,-1)]<br>J8=0.002 [Ru1-Ru2 (-1,-1,0)]<br>J9a=0.002 [Ru1-Ru1 (0,-1,1)]<br>J9b=0.001 [Ru1-Ru1 (0,-1,-1)]<br>J10=0.005 [Ru1-Ru2 (0,-2,0)]<br>J11=0.001 [Ru1-Ru2 (0,-2,1)]<br>J12=0.002 [Ru1-Ru2 (-1,-1,1)]<br>J13=0.003 [Ru1-Ru1 (-1,0,1)]<br>J14a=0.009 [Ru1-Ru1 (-1,1,-1)]<br>J14b=0.007 [Ru1-Ru1 (-1,-1,-1)]<br>J15=0.008 [Ru1-Ru1 (-2,0,0)]<br>J16=0.001 [Ru1-Ru1 (0,-2,0)]<br>J17a=0.007 [Ru1-Ru1 (-1,1,1)]<br>J17b=0.005 [Ru1-Ru1 (-1,-1,1)]<br>J18=0.000 [Ru1-Ru2 (0,-1,-1)]<br>J19=-0.001 [Ru1-Ru2 (-1,-2,0)]<br>J20a=0.004 [Ru1-Ru1 (-2,-1,0)]<br>J20b=-0.000 [Ru1-Ru1 (-2,1,0)] |
| 0.432 | $KMnF_3$ | Mn1 (0.5,0,0)<br>Mn2 (0,0.5,0.5)<br>Mn3 (0.5,0.5,0)<br>Mn4 (0,0,0.5) | 3.54<br>(Mn 3*d*) | J1=-3.917, $D_1$= (-0.011, 0.001, 0.013),<br>$\Gamma_1$= (0, 0.004, 0)<br>[Mn1-Mn4 (0,0,-1)]<br>J2=-4.651, $J_2^{ani}$= (-0.016, -0.026, -0.016)<br>[Mn1-Mn3 (0,-1,0)]<br>J3=0.003 [Mn1-Mn1 (-1,0,0)]<br>J4a=0.017 [Mn1-Mn2 (0,-1,0)]<br>J4b=0.005 [Mn1-Mn2 (0,-1,-1)]<br>J5=0.003 [Mn1-Mn1 (0,0,-1)]<br>J6=-0.001 [Mn1-Mn3 (-1,-1,0)]<br>J7=-0.001 [Mn1-Mn3 (0,-1,-1)]<br>J8a=0.018 [Mn1-Mn1 (-1,0,-1)]<br>J8b=0.005 [Mn1-Mn1 (-1,0,1)]<br>J9=0.025 [Mn1-Mn1 (0,-1,0)] |
| 0.433 | $KMnF_3$ | Mn1 (0, 0, 0) | 3.54 | J1=-4.206, $D_1$= (0, 0, 0.025), |
| | | Mn2 (0, 0, 0.5) | (Mn 3*d*) | $J_1^{ani}$= (-0.012, -0.012, -0.009),<br>$\Gamma_1$= (0.003, 0, 0)<br>[Mn1-Mn2 (-1/2,-1/2,-1/2)]<br>J2=-4.337 [Mn1-Mn2 (0,0,-1)]<br>J3=0.014 [Mn1-Mn1 (-1,0,0)]<br>J4=0.014 [Mn1-Mn1 (-1/2,-1/2,-1/2)]<br>J5=0.000 [Mn1-Mn2 (-1,0,-1)]<br>J6=-0.001 [Mn1-Mn1 (-1,-1,0)]<br>J7=-0.002 [Mn1-Mn1 (0,0,-1)]<br>J8=-0.001 [Mn1-Mn2 (-1/2,-3/2,-1/2)] |
| 0.434,<br>0.553 | $K_2ReI_6$† | Re1 (0, 0, 0)<br>Re2 (0.5, 0.5, 0.5) | 2.0<br>(Re 5*d*) | J1=-0.073, **$\boldsymbol{J_1^{ani}}$= (-0.023, -0.047, -0.055)**,<br>**$\boldsymbol{\Gamma_1}$= (-0.014, -0.002, 0.006)**<br>[Re1-Re1 (-1,0,0)]<br>J2=-0.096 [Re1-Re1 (0,-1,0)]<br>J3=-0.029, **$\boldsymbol{D_3}$= (-0.006, -0.004, -0.013)**<br>[Re1-Re2 (-1,-1,-1)]<br>J4=-0.028 [Re1-Re2 (-1,-1,0)]<br>J5a=0.005 [Re1-Re1 (-1,-1,0)]<br>J5b=0.005 [Re1-Re1 (-1,1,0)]<br>J6=0.004 [Re1-Re1 (0,0,-1)] |
| 0.501 | $LiFe_2F_6$† | Fe1 (0,0,0.333)<br>Fe2 (0.5,0.5,0.833)<br>Fe3 (0.5,0.5,0.167)<br>Fe4 (0,0,0.667) | 4.0<br>(Fe 3*d*) | J1=9.593, **$\boldsymbol{J_1^{ani}}$= (-9.464, -9.458, -7.498)**,<br>$\Gamma_1$= (0.048, 0, 0)<br>[Fe1-Fe4 (0,0,0)]<br>J2=-6.255, $D_2$= (0.141, -0.141, 0)<br>[Fe1-Fe3 (-1,-1,0)]<br>J3=0.020 [Fe1-Fe1 (-1,0,0)]<br>J4=-0.068 [Fe1-Fe4 (-1,0,0)]<br>J5a=0.113 [Fe1-Fe2 (-1,-1,-1)]<br>J5b=-0.112 [Fe1-Fe2 (-1,-1,0)]<br>J6=-0.036 [Fe1-Fe4 (0,0,-1)]<br>J7a=-0.265 [Fe1-Fe1 (-1,-1,0)]<br>J7b=0.014 [Fe1-Fe1 (-1,1,0)]<br>J8a=-0.018 [Fe1-Fe4 (-1,-1,0)]<br>J8b=-0.006 [Fe1-Fe4 (-1,1,0)]<br>J9=-0.000 [Fe1-Fe3 (-1,-2,0)]<br>J10=0.003 [Fe1-Fe4 (-1,0,-1)]<br>J11=0.114 [Fe1-Fe3 (-1,-1,1)] |
| 0.513 | $YRuO_3$† | Ru1 (0, 0, 0.5)<br>Ru2 (0.5, 0.5, 0)<br>Ru3 (0, 0.5, 0.5)<br>Ru4 (0.5, 0, 0) | 2.1<br>(Ru 4*d*) | J1=-0.810, **$\boldsymbol{D_1}$= (-0.284, 0, -0.184)**,<br>[Ru1-Ru3 (0,-1,0)]<br>J2=-0.953, **$\boldsymbol{J_2^{ani}}$= (-1.353, -0.489, -1.225)**,<br>**$\boldsymbol{\Gamma_2}$= (-0.189, -0.064, -0.006)**<br>[Ru1-Ru4 (-1,0,0)]<br>J3=0.110 [Ru1-Ru1 (0,0,-1)]<br>J4a=0.032 [Ru1-Ru2 (-1,-1,0)] |
| | | | | J4b=0.023 [Ru1-Ru2 (0,-1,1)]<br>J5=0.029 [Ru1-Ru1 (-1,0,0)]<br>J6=0.007 [Ru1-Ru3 (0,-1,-1)]<br>J7=-0.005 [Ru1-Ru3 (-1,-1,0)]<br>J8=0.024 [Ru1-Ru1 (0,-1,0)]<br>J9a=0.172 [Ru1-Ru1 (-1,0,-1)]<br>J9b=-0.014 [Ru1-Ru1 (-1,0,1)]<br>J10=0.009 [Ru1-Ru4 (-1,0,-1)]<br>J11a=0.011 [Ru1-Ru4 (-1,-1,0)]<br>J11b=0.005 [Ru1-Ru4 (-1,1,0)]<br>J12a=-0.007 [Ru1-Ru3 (-1,-1,-1)]<br>J12b=0.001 [Ru1-Ru3 (-1,-1,1)] |
| 0.528 | CrSb | Cr1 (0, 0, 0)<br>Cr2 (0, 0, 0.5) | 0 | J1=-19.509, $J_1^{ani}$= (0.088, 0.091, 0.421),<br>[Cr1-Cr2 (0,0,-1)]<br>J2=5.606, $\Gamma_2$= (0.014, -0.006, 0.002)<br>[Cr1-Cr1 (-1,-1,0)]<br>J3=-3.023, $D_3$= (-0.107, 0, 0.173)<br>[Cr1-Cr2 (-1,-1,-1)]<br>J4=1.313 [Cr1-Cr1 (0,0,-1)]<br>J5=0.587 [Cr1-Cr1 (-1,-1,-1)]<br>J6=-0.469 [Cr1-Cr1 (-1,-2,0)]<br>J7=0.781 [Cr1-Cr2 (-1,-2,-1)]<br>J8=0.669 [Cr1-Cr2 (0,0,-2)]<br>J9=0.246 [Cr1-Cr1 (-2,-2,0)]<br>J10=0.012 [Cr1-Cr2 (-2,-2,-1)]<br>J11a=-1.921 [Cr1-Cr1 (-1,-2,-1)]<br>J11b=-0.373 [Cr1-Cr1 (-1,-2,1)] |
| 0.575 | $ZnFeF_5(H_2O)_2$ | Fe1 (0.25,0,0)<br>Fe2 (0.75,0,0) | 0 | J1a=-21.784, $D_1$= (0, 0.392, 0),<br>$J_1^{ani}$= (-0.111, -0.071, 0.087)<br>[Fe1-Fe2 (-1,0,0)]<br>J1b=-11.787 [Fe1-Fe2 (0,0,0)]<br>J2=-0.337 [Fe1-Fe2 (-1/2,-1/2,-1/2)]<br>J3=-0.024 [Fe1-Fe1 (0,0,-1)]<br>J4a=-0.010 [Fe1-Fe1 (-1/2,-1/2,-1/2)]<br>J4b=-0.008 [Fe1-Fe1 (-1/2,-1/2,1/2)]<br>J5=0.692 [Fe1-Fe1 (-1,0,0)]<br>J6a=-0.047 [Fe1-Fe2 (-1,0,-1)]<br>J6b=-0.007 [Fe1-Fe2 (0,0,-1)]<br>J7a=-0.001 [Fe1-Fe2 (-3/2,-1/2,-1/2)]<br>J7b=0.000 [Fe1-Fe2 (1/2,-1/2,-1/2)]<br>J8a=-0.001 [Fe1-Fe1 (-1,0,1)]<br>J8b=-0.001 [Fe1-Fe1 (-1,0,-1)]<br>J9=-0.001 [Fe1-Fe1 (0,-1,0)]<br>J10=-0.000 [Fe1-Fe2 (-1/2,-1/2,-3/2)] |
| | | | | J11a=-0.046 [Fe1-Fe2 (-2,0,0)]<br>J11b=-0.022 [Fe1-Fe2 (1,0,0)]<br>$\Gamma_1$= (0, 0, 0) |
| 0.586,<br>0.947 | $YCrO_3$† | Cr1 (0.5,0,0)<br>Cr2 (0,0.5,0.5)<br>Cr3 (0.5,0,0.5)<br>Cr4 (0,0.5,0) | 4.0<br>(Cr 3*d*) | J1=-1.099, **$D_1$= (0.014, 0.163, 0.087)**,<br>$J_1^{ani}$= (-0.007, -0.011, -0.009),<br>[Cr1-Cr3 (0,0,-1)]<br>J2=-1.059 [Cr1-Cr4 (0,-1,0)]<br>J3=-0.002 [Cr1-Cr1 (-1,0,0)]<br>J4a=0.012 [Cr1-Cr2 (0,-1,-1)]<br>J4b=0.009 [Cr1-Cr2 (0,-1,0)]<br>J5=-0.017 [Cr1-Cr1 (0,-1,0)]<br>J6=-0.008 [Cr1-Cr3 (-1,0,-1)]<br>J7=-0.001 [Cr1-Cr3 (0,-1,-1)]<br>J8=0.102 [Cr1-Cr1 (0,0,-1)]<br>J9a=0.042 [Cr1-Cr1 (-1,1,0)]<br>J9b=0.034 [Cr1-Cr1 (-1,-1,0)]<br>$\Gamma_1$= (0, 0, 0) |
| 0.587 | $TmCrO_3$† | Cr1 (0.5,0,0)<br>Cr2 (0,0.5,0.5)<br>Cr3 (0.5,0,0.5)<br>Cr4 (0,0.5,0) | 3.0<br>(Cr 3*d*) | J1=-0.271, **$D_1$= (-0.147, 0.042, 0.299)**,<br>**$J_1^{ani}$= (-0.123, 0.362, -0.598)**,<br>**$\Gamma_1$= (-0.001, -0.298, 0.067)**<br>[Cr1-Cr3 (0,0,-1)]<br>J2=-0.303 [Cr1-Cr4 (0,-1,0)]<br>J3=-0.002 [Cr1-Cr1 (-1,0,0)]<br>J4a=0.000 [Cr1-Cr2 (0,-1,-1)]<br>J4b=-0.004 [Cr1-Cr2 (0,-1,0)]<br>J5=-0.008 [Cr1-Cr1 (0,-1,0)]<br>J6=0.000 [Cr1-Cr3 (-1,0,-1)]<br>J7=0.000 [Cr1-Cr3 (0,-1,-1)]<br>J8=0.024 [Cr1-Cr1 (0,0,-1)]<br>J9a=-0.035 [Cr1-Cr1 (-1,1,0)]<br>J9b=-0.020 [Cr1-Cr1 (-1,-1,0)] |
| 0.591 | $ErCrO_3$ | Cr1 (0.5,0,0)<br>Cr2 (0,0.5,0.5)<br>Cr3 (0.5,0,0.5)<br>Cr4 (0,0.5,0) | 4.0<br>(Cr 3*d*) | J1=-0.789, $D_1$= (0.018, 0.028, 0),<br>$J_1^{ani}$= (-0.004, -0.007, -0.002),<br>$\Gamma_1$= (-0.001, 0, 0)<br>[Cr1-Cr3 (0,0,-1)]<br>J2=-1.685 [Cr1-Cr4 (0,-1,0)]<br>J3=0.008 [Cr1-Cr1 (-1,0,0)]<br>J4a=0.037 [Cr1-Cr2 (0,-1,-1)]<br>J4b=0.012 [Cr1-Cr2 (0,0,0)]<br>J5=-0.025 [Cr1-Cr1 (0,-1,0)]<br>J6=-0.008 [Cr1-Cr3 (-1,0,-1)]<br>J7=-0.001 [Cr1-Cr3 (0,-1,-1)]<br>J8=0.101 [Cr1-Cr1 (0,0,-1)]<br>J9a=0.048 [Cr1-Cr1 (-1,1,0)] |

| | | | | J9b=0.046 [Cr1-Cr1 (-1,-1,0)] |
|---|---|---|---|---|
| 0.592 | $DyCrO_3$ | Cr1 (0.5,0,0)<br>Cr2 (0,0.5,0.5)<br>Cr3 (0.5,0,0.5)<br>Cr4 (0,0.5,0) | 4.0<br>(Cr 3*d*) | J1=-2.123, $D_1$= (0.026, 0.189, 0.084),<br>$J_1^{ani}$= (-0.006, -0.01, -0.01),<br>$\Gamma_1$= (0, 0, -0.001)<br>[Cr1-Cr4 (0,-1,0)]<br>J2=-1.304 [Cr1-Cr3 (0,0,-1)]<br>J3=-0.019 [Cr1-Cr1 (0,-1,0)]<br>J4=-0.049 [Cr1-Cr1 (-1,0,0)]<br>J5a=0.062 [Cr1-Cr2 (0,-1,-1)]<br>J5b=-0.004 [Cr1-Cr2 (0,-1,0)]<br>J6=-0.015 [Cr1-Cr3 (0,-1,-1)]<br>J7=-0.000 [Cr1-Cr3 (-1,0,-1)]<br>J8a=0.056 [Cr1-Cr1 (-1,-1,0)]<br>J8b=0.044 [Cr1-Cr1 (-1,1,0)]<br>J9=0.103 [Cr1-Cr1 (0,0,-1)] |
| 0.607 | $RuO_2$ | Ru1 (0, 0, 0)<br>Ru2 (0.5, 0.5, 0.5) | 1.6<br>(Ru 4*d*) | J1=-9.662, $\Gamma_1$= (0.039, 0, 0)<br>[Ru1-Ru1 (0,0,-1)]<br>J2=-7.231 $D_2$= (-0.203, 0.203, 0)<br>[Ru1-Ru2 (-1,-1,-1)]<br>J3=-0.184, $J_3^{ani}$= (-0.019, -0.243, -0.071)<br>[Ru1-Ru1 (-1,0,0)]<br>J4=0.123 [Ru1-Ru1 (-1,0,-1)]<br>J5=0.632 [Ru1-Ru2 (-1,-1,-2)]<br>J6=1.668 [Ru1-Ru1 (0,0,-2)]<br>J7a= 1.922 [Ru1-Ru1 (-1,1,0)]<br>J7b=1.683 [Ru1-Ru1 (-1,-1,0)]<br>J8a=1.084 [Ru1-Ru1 (-1,-1,-1)]<br>J8b=0.313 [Ru1-Ru1 (-1,1,-1)] |
| 0.608 | $PrMnO_3$ | Mn1 (0.5,0,0)<br>Mn2 (0,0.5,0.5)<br>Mn3 (0.5,0,0.5)<br>Mn4 (0,0.5,0) | 2.6<br>(Mn 3*d*) | J1=7.519 [Mn1-Mn3 (0,0,-1)]<br>J2=10.424, $D_2$= (-0.119, 0.053, 0.582),<br>$J_2^{ani}$= (-0.134, -0.13, -0.146),<br>$\Gamma_2$= (0.005, 0.003, -0.002)<br>[Mn1-Mn4 (0,-1,0)]<br>J3a=-1.058 [Mn1-Mn2 (0,-1,0)]<br>J3b=-1.283 [Mn1-Mn2 (0,-1,-1)]<br>J4=0.005 [Mn1-Mn1 (-1,0,0)]<br>J5=-0.360 [Mn1-Mn1 (0,-1,0)]<br>J6=-0.008 [Mn1-Mn3 (-1,0,-1)]<br>J7=-0.050 [Mn1-Mn3 (0,-1,-1)]<br>J8=0.013 [Mn1-Mn1 (0,0,-1)]<br>J9a=-1.958 [Mn1-Mn1 (-1,-1,0)]<br>J9b=-0.333 [Mn1-Mn1 (-1,1,0)]<br>J10a=0.020 [Mn1-Mn4 (0,-1,1)]<br>J10b=0.015 [Mn1-Mn4 (0,-1,-1)] |
| | | | | J11=-0.002 [Mn1-Mn4 (-1,-1,0)]<br>J12a=-0.149 [Mn1-Mn3 (-1,-1,-1)]<br>J12b=-0.198 [Mn1-Mn3 (-1,1,-1)] |
| 0.609 | $NdMnO_3$† | Mn1 (0.5,0,0)<br>Mn2 (0.5,0,0.5)<br>Mn3 (0,0.5,0)<br>Mn4 (0,0.5,0.5) | 2.6<br>(Mn 3*d*) | J1=-0.532 [Mn1-Mn2 (0,0,-1)]<br>J2=1.185, **$D_2$= (0.11, -0.161, -0.281)**,<br>$J_2^{ani}$= (-0.041, -0.048, -0.047),<br>$\Gamma_2$= (0.001, 0.002, 0)<br>[Mn1-Mn3 (0,-1,0)]<br>J3=-0.160 [Mn1-Mn1 (-1,0,0)]<br>J4a=-0.367 [Mn1-Mn4 (0,-1,0)]<br>J4b=-0.456 [Mn1-Mn4 (0,-1,-1)]<br>J5=-0.352 [Mn1-Mn1 (0,-1,0)]<br>J6=-0.066 [Mn1-Mn2 (-1,0,-1)]<br>J7=-0.035 [Mn1-Mn2 (0,-1,-1)]<br>J8=0.242 [Mn1-Mn1 (0,0,-1)]<br>J9a=-0.116 [Mn1-Mn1 (-1,-1,0)]<br>J9b=-0.592 [Mn1-Mn1 (-1,1,0)]<br>J10a=-0.023 [Mn1-Mn3 (0,-1,-1)]<br>J10b=-0.005 [Mn1-Mn3 (0,-1,1)]<br>J11=0.006 [Mn1-Mn3 (-1,-1,0)]<br>J12a=-0.042 [Mn1-Mn2 (-1,-1,-1)]<br>J12b=-0.015 [Mn1-Mn2 (-1,1,-1)]<br>J13=-0.011 [Mn1-Mn3 (0,-2,0)] |
| 0.708 | $CrNb_4S_8$ | Cr1 (0, 0, 0)<br>Cr2 (0, 0, 0.5) | 3.0<br>(Cr 3*d*) | J1=5.298, $D_1$= (-0.014, 0.024, -0.003),<br>$J_1^{ani}$= (0.025, 0.024, 0.032),<br>$\Gamma_1$= (0, 0.001, 0.002)<br>[Cr1-Cr2 (0,0,-1)]<br>J2a=-0.172 [Cr1-Cr1 (-1,-1,0)]<br>J2b=-0.171 [Cr1-Cr1 (0,-1,0)]<br>J2c=-0.153 [Cr1-Cr1 (-1,0,0)]<br>J3=0.038 [Cr1-Cr2 (-1,-1,-1)]<br>J4=0.143 [Cr1-Cr1 (-1,-2,0)]<br>J5=-1.025 [Cr1-Cr1 (0,0,-1)]<br>J6=0.133 [Cr1-Cr2 (-1,-2,-1)]<br>J7a=0.023 [Cr1-Cr1 (0,-2,0)]<br>J7b=0.023 [Cr1-Cr1 (-2,-2,0)]<br>J7c=0.024 [Cr1-Cr1 (-2,0,0)]<br>J8=-0.341 [Cr1-Cr1 (-1,-1,-1)]<br>J9=0.047 [Cr1-Cr2 (-2,-2,-1)]<br>J10a=0.024 [Cr1-Cr1 (-1,-2,-1)]<br>J10b=0.019 [Cr1-Cr1 (-1,-2,1)]<br>J11=-0.146 [Cr1-Cr1 (-1,-3,0)]<br>J12=0.037 [Cr1-Cr1 (-2,-2,-1)]<br>J13=-1.047 [Cr1-Cr2 (0,0,1)] |

| | | | | J14=-0.104 [Cr1-Cr2 (-1,-3,-1)] |
|---|---|---|---|---|
| 0.712 | $VNb_3S_6$ | V1 (0.333, 0.667, 0.25)<br>V2 (0.667, 0.333, 0.75) | 0 | J1a=0.291 [V1-V1 (-1,-1,0)]<br>J1b=0.306 [V1-V1 (-1,0,0)]<br>J2a=-0.518 [V1-V2 (-1,0,-1)]<br>J2b=-0.520, $D_{2b}$= (-0.002, 0, 0.024)<br>[V1-V2 (0,1,-1)]<br>J3a=-0.296 [V1-V2 (-1,-1,-1)]<br>J3b=-0.297 [V1-V2 (-1,1,-1)]<br>J4a=0.118 [V1-V1 (-1,-2,0)]<br>J4b=0.118 [V1-V1 (-1,1,0)]<br>J5a=-0.001 [V1-V2 (-2,-1,-1)]<br>J5b=-0.001 [V1-V2 (-2,0,-1)]<br>J6a=0.013 [V1-V1 (-2,-2,0)]<br>J6b=0.013 [V1-V1 (-2,0,0)]<br>J7=-0.010 [V1-V1 (0,0,-1)]<br>J8=-0.005 [V1-V2 (-1,-2,-1)]<br>J9a=0.018 [V1-V1 (-1,-1,1)]<br>J9b=-0.017 [V1-V1 (-1,0,1)]<br>$J_i^{ani}=\Gamma_i$= (0, 0, 0) |
| 0.714 | $Li_2Ni(SO_4)_2$ | Ni1 (0, 0, 0)<br>Ni2 (0, 0.5, 0.5) | 2.5<br>(Ni 3*d*) | J1=0.524, $D_1$= (0.001, -0.019, 0.027)<br>[Ni1-Ni1 (-1,0,0)]<br>J2=-0.363 [Ni1-Ni2 (0,-1,-1)]<br>J3=0.053 [Ni1-Ni2 (-1,-1,-1)]<br>J4=0.009 [Ni1-Ni1 (-1,0,-1)]<br>J5=0.003 [Ni1-Ni1 (0,-1,0)]<br>J6=-0.001 [Ni1-Ni1 (0,0,-1)]<br>J7=0.022 [Ni1-Ni2 (-1,-1,0)]<br>J8=0.001 [Ni1-Ni1 (-2,0,-1)]<br>J9=0.009 [Ni1-Ni2 (-2,-1,-1)]<br>J10a=0.006 [Ni1-Ni1 (-1,1,0)]<br>J10b=0.001 [Ni1-Ni1 (-1,-1,0)]<br>J11=0.001 [Ni1-Ni1 (-2,0,0)]<br>J12a=0.044 [Ni1-Ni1 (-1,-1,-1)]<br>J12b=0.001 [Ni1-Ni1 (-1,1,-1)]<br>$J_i^{ani}=\Gamma_i$= (0, 0, 0) |
| 0.748 | $Ba_3NiRu_2O_9$ | Ni1 (0,0,0)<br>Ni2 (0,0,0.5)<br>Ru1 (0.333,0.667,0.155)<br>Ru2 (0.667,0.333,0.655)<br>Ru3 (0.667,0.333,0.845)<br>Ru4 (0.333,0.667,0.345) | 6.0<br>(Ni 3*d*)<br>2.1<br>(Ru 4*d*) | J1=-12.079, $D_1$= (0.08, 0.138, 0.001),<br>$J_1^{ani}$= (-0.195, -0.195, -0.217),<br>$\Gamma_1$= (0.001, -0.006, -0.004)<br>[Ru1-Ru4 (0,0,0)]<br>J2=2.812 [Ni1-Ru1 (-1,-1,0)]<br>J3=-3.219 [Ru1-Ru3 (-1,0,-1)]<br>J4a=-0.332 [Ni1-Ni1 (-1,-1,0)]<br>J4b=-0.332 [Ni1-Ni1 (-1,0,0)]<br>J4c=-0.570 [Ru1-Ru1 (-1,-1,0)] |
| | | | | J4d=-0.555 [Ru1-Ru1 (-1,0,0)]<br>J5=-0.134 [Ni1-Ru2 (-1,-1,-1)]<br>J6a=-0.070 [Ru1-Ru4 (-1,-1,0)]<br>J6b=-0.065 [Ru1-Ru4 (-1,0,0)]<br>J7a=-0.005 [Ni1-Ru1 (-1,-2,0)]<br>J7b=-0.004 [Ni1-Ru1 (-1,0,0)]<br>J8=0.002 [Ni1-Ni2 (0,0,-1)]<br>J9=0.137 [Ru1-Ru2 (-1,0,-1)]<br>J10a=0.026 [Ru1-Ru3 (-1,-1,-1)]<br>J10b=0.029 [Ru1-Ru3 (-1,1,-1)] |
| 0.755 | $Mn_2SeO_3F_2$ | Mn1 (0.096,0.067,0.249)<br>Mn2 (0.904,0.567,0.751)<br>Mn3 (0.904,0.933,0.751)<br>Mn4 (0.096,0.433,0.249)<br>Mn5 (0.596,0.433,0.251)<br>Mn6 (0.404,0.933,0.749)<br>Mn7 (0.404,0.567,0.749)<br>Mn8 (0.596,0.067,0.251) | 3.6<br>(Mn 3*d*) | J1=1.994, $D_1$= (0.017, -0.171, 0),<br>$J_1^{ani}$= (-0.028, -0.023, -0.002),<br>$\Gamma_1$= (-0.006, 0.001, 0)<br>[Mn1-Mn3 (-1,-1,-1)]<br>J2=-1.405 [Mn1-Mn3 (-1,-1,0)]<br>J3=-1.108 [Mn1-Mn8 (-1,0,0)]<br>J4=-2.513 [Mn1-Mn4 (0,0,0)]<br>J5a=-1.685 [Mn1-Mn6 (0,-1,-1)]<br>J5b=-1.489 [Mn2-Mn5 (0,0,0)]<br>J6=-0.174 [Mn1-Mn5 (-1,0,0)]<br>J7=-0.011 [Mn1-Mn1 (0,0,-1)]<br>J8=-0.011 [Mn1-Mn2 (-1,-1,-1)]<br>J9=0.034 [Mn1-Mn6 (-1,-1,-1)]<br>J10=-0.022 [Mn1-Mn2 (-1,-1,0)]<br>J11a=-0.141 [Mn1-Mn7 (0,-1,0)]<br>J11b=-0.015 [Mn1-Mn7 (0,-1,-1)]<br>J12=-0.674 [Mn1-Mn4 (0,-1,0)]<br>J13=0.005 [Mn1-Mn8 (-1,0,-1)]<br>J14=-0.026 [Mn1-Mn8 (-1,0,1)]<br>J15=-0.012 [Mn1-Mn4 (0,0,-1)]<br>J16=0.009 [Mn1-Mn3 (0,-1,-1)]<br>J17=-0.126 [Mn1-Mn3 (0,-1,0)] |
| 0.757,<br>0.758 | $CeFeO_3$ | Fe1 (0.5,0,0)<br>Fe2 (0,0.5,0.5)<br>Fe3 (0.5,0,0.5)<br>Fe4 (0,0.5,0) | 4.0<br>(Fe 3*d*) | J1=-10.591 [Fe1-Fe3 (0,0,-1)]<br>J2=-11.077, $D_2$= (0.256, -0.137, -0.191),<br>$J_2^{ani}$= (-0.101, -0.099, -0.095),<br>$\Gamma_2$= (-0.015, -0.004, 0.004)<br>[Fe1-Fe4 (0,-1,0)]<br>J3=0.297 [Fe1-Fe1 (-1,0,0)]<br>J4a=0.280 [Fe1-Fe2 (0,-1,0)]<br>J4b=0.247 [Fe1-Fe2 (0,-1,-1)]<br>J5=0.175 [Fe1-Fe1 (0,-1,0)]<br>J6=-0.026 [Fe1-Fe3 (-1,0,-1)]<br>J7=-0.028 [Fe1-Fe3 (0,-1,-1)]<br>J8=0.228 [Fe1-Fe1 (0,0,-1)] |
| | | | | J9a=0.277 [Fe1-Fe1 (-1,1,0)]<br>J9b=0.311 [Fe1-Fe1 (-1,-1,0)]<br>J10=-0.017 [Fe1-Fe4 (-1,-1,0)]<br>J11a=-0.017 [Fe1-Fe4 (0,-1,-1)]<br>J11b=-0.029 [Fe1-Fe4 (0,-1,1)]<br>J12a=-0.018 [Fe1-Fe3 (-1,-1,-1)]<br>J12b=-0.017 [Fe1-Fe3 (-1,1,-1)]<br>J13=-0.027 [Fe1-Fe4 (0,-2,0)] |
| 0.760 | $FeOHSO_4$ | Fe1 (0, 0.5, 0)<br>Fe2 (0.5, 0.5, 0.5) | 4.0<br>(Fe 3*d*) | J1=-14.221, $D_1$= (-0.523, 0.51, 0.283),<br>$J_1^{ani}$= (-0.11, -0.109, -0.117),<br>$\Gamma_1$= (-0.006, -0.022, 0.021)<br>[Fe1-Fe2 (-1,0,-1)]<br>J2a=-0.302 [Fe1-Fe1 (0,-1,0)]<br>J2b=-0.091 [Fe1-Fe1 (-1,0,0)]<br>J2c=-0.195 [Fe1-Fe2 (-1,-1,-1)]<br>J3=-1.443 [Fe1-Fe2 (-1,1,0)]<br>J4=-0.008 [Fe1-Fe1 (-1,-1,0)]<br>J5a=0.037 [Fe1-Fe1 (-1,-1,-1)]<br>J5b=-0.030 [Fe1-Fe1 (0,0,-1)]<br>J5c=0.003 [Fe1-Fe2 (-1,1,-1)]<br>J6=0.413 [Fe1-Fe1 (-1,0,-1)]<br>J7a=0.074 [Fe1-Fe1 (0,-1,-1)]<br>J7b=0.042 [Fe1-Fe1 (-1,1,0)]<br>J8=-0.016 [Fe1-Fe2 (-1,-1,0)] |
| 0.786 | $NdVO_3$† | V1 (0.5, 0, 0)<br>V2 (0.5, 0, 0.5)<br>V3 (0, 0.5, 0)<br>V4 (0, 0.5, 0.5) | 0 | J1=-9.468 [V1-V2 (0,0,-1)]<br>J2=4.072, **$D_2$= (0.779, -1.359, -0.163)**,<br>$J_2^{ani}$= (-0.271, -0.133, -0.296),<br>$\Gamma_2$= (-0.011, -0.006, 0.014)<br>[V1-V3 (0,-1,0)]<br>J3=-2.258 [V1-V1 (-1,0,0)]<br>J4a=-1.483 [V1-V4 (0,-1,0)]<br>J4b=-1.938 [V1-V4 (0,-1,-1)]<br>J5=-3.584 [V1-V1 (0,-1,0)]<br>J6=-0.188 [V1-V2 (-1,0,-1)]<br>J7=-0.537 [V1-V2 (0,-1,-1)]<br>J8=0.709 [V1-V1 (0,0,-1)]<br>J9a=-0.268 [V1-V1 (-1,-1,0)]<br>J9b=0.110 [V1-V1 (-1,1,0)]<br>J10=0.309 [V1-V3 (-1,-1,0)]<br>J11=0.057 [V1-V3 (0,-1,-1)]<br>J12=0.242 [V1-V2 (-1,-1,-1)]<br>J13=0.746 [V1-V3 (0,-2,0)]<br>J14a=0.105 [V1-V4 (-1,-1,-1)]<br>J14b=0.096 [V1-V4 (-1,-1,0)] |
| | | | | J15a=-0.011 [V1-V1 (-1,0,-1)]<br>J15b=-0.268 [V1-V1 (-1,0,1)]<br>J16a=-0.253 [V1-V1 (0,-1,-1)]<br>J16b=-0.169 [V1-V1 (0,-1,1)]<br>J17=0.173 [V1-V4 (0,-2,-1)]<br>J18=-0.075 [V1-V1 (-2,0,0)]<br>J19a=-0.105 [V1-V1 (-1,-1,-1)]<br>J19b=-0.457 [V1-V1 (-1,1,-1)]<br>J20=-0.708 [V1-V1 (0,-2,0)]<br>J21=0.023 [V1-V2 (-2,0,-1)]<br>J22a=-0.123 [V1-V3 (-1,-1,-1)]<br>J22b=-0.089 [V1-V3 (-1,-1,1)]<br>J23=1.347 [V1-V2 (0,0,-2)]<br>J24=0.042 [V1-V3 (-1,-2,0)]<br>J25a=0.004 [V1-V3 (0,-2,-1)]<br>J25b=-0.015 [V1-V3 (0,-2,1)]<br>J26=0.038 [V1-V2 (0,-2,-1)]<br>J27a=0.100 [V1-V1 (-2,1,0)]<br>J27b=0.048 [V1-V1 (-2,-1,0)]<br>J28a=0.483 [V1-V4 (0,-1,-2)]<br>J28b=0.425 [V1-V4 (0,-1,1)]<br>J29a=0.037 [V1-V4 (-1,-2,0)]<br>J29b=-0.019 [V2-V3 (-1,-2,1)]<br>J30a=0.162 [V1-V1 (-1,2,0)]<br>J30b=-0.414 [V1-V1 (-1,-2,0)] |
| 0.787 | $YVO_3$ | V1 (0.5, 0, 0)<br>V2 (0, 0.5, 0.5)<br>V3 (0.5, 0, 0.5)<br>V4 (0, 0.5, 0) | 2.85<br>(V 3*d*) | J1=-2.056 [V1-V3 (0,0,-1)]<br>J2=-2.257, $D_2$= (-0.036, 0.049, 0.023),<br>[V1-V4 (0,-1,0)]<br>J3=-0.004 [V1-V1 (-1,0,0)]<br>J4a=0.009 [V1-V2 (0,-1,0)]<br>J4b=-0.023 [V1-V2 (0,-1,-1)]<br>J5=-0.239 [V1-V1 (0,-1,0)]<br>J6=-0.002 [V1-V3 (-1,0,-1)]<br>J7=-0.006 [V1-V3 (0,-1,-1)]<br>J8=0.074 [V1-V1 (0,0,-1)]<br>J9a=-0.135 [V1-V1 (-1,1,0)]<br>J9b=0.003 [V1-V1 (-1,-1,0)]<br>$J_i^{ani}$=$\Gamma_i$= (0, 0, 0) |
| 0.795 | $Sr_2YRuO_6$ | Ru1 (0, 0, 0)<br>Ru2 (0.5, 0.5, 0.5) | 2.0<br>(Ru 3*d*) | J1=-1.687 [Ru1-Ru2 (-1,-1,-1)]<br>J2=-1.498 [Ru1-Ru1 (-1,0,0)]<br>J3=-1.304, $D_3$= (-0.005, 0.001, 0.021),<br>$\Gamma_3$= (-0.023, -0.004, 0.003)<br>[Ru1-Ru2 (-1,-1,0)]<br>J4=-1.960, $J_4^{ani}$= (-0.014, -0.043, -0.047) |

| | | | | [Ru1-Ru1 (0,-1,0)]<br>J5=-0.040 [Ru1-Ru1 (0,0,-1)]<br>J6a=0.007 [Ru1-Ru1 (-1,-1,0)]<br>J6b=0.018 [Ru1-Ru1 (-1,1,0)] |
|---|---|---|---|---|
| 0.800 | MnTe | Mn1 (0, 0, 0)<br>Mn2 (0, 0, 0.5) | 4.8<br>(Mn 3*d*) | J1=-12.740 [Mn1-Mn2 (0,0,-1)]<br>J2=-0.554, $\Gamma_2$= (0, -0.016, 0.014)<br>[Mn1-Mn1 (-1,-1,0)]<br>J3=-1.694, $D_3$= (0.058, 0.152, -0.206),<br>$J_3^{ani}$= (-0.111, -0.081, -0.133)<br>[Mn1-Mn2 (-1,-1,-1)]<br>J4=-0.080 [Mn1-Mn1 (0,0,-1)]<br>J5=-0.071 [Mn1-Mn1 (-1,-2,0)]<br>J6=-0.109 [Mn1-Mn1 (-1,-1,-1)]<br>J7=-0.023 [Mn1-Mn2 (-1,-2,-1)]<br>J8=-0.084 [Mn1-Mn1 (-2,-2,0)]<br>J9=-0.075 [Mn1-Mn2 (-2,-2,-1)]<br>J10a=-0.237 [Mn1-Mn1 (-1,1,-1)]<br>J10b=0.007 [Mn1-Mn1 (-1,-2,1)]<br>J11=-0.001 [Mn1-Mn2 (0,0,-2)]<br>J12=-0.029 [Mn1-Mn1 (-2,-2,-1)]<br>J13=-0.025 [Mn1-Mn2 (-1,-1,-2)] |
| 0.802 | $CuFeS_2$ | Fe1 (0,0,0.5)<br>Fe2 (0,0.5,0.75)<br>Fe3 (0.5,0.5,0)<br>Fe4 (0.5,0,0.25) | 4.7<br>(Fe 3*d*) | J1=-31.658, $D_1$= (0.444, 0, 0.154),<br>$J_1^{ani}$= (-0.063, -0.09, -0.076),<br>$\Gamma_1$= (0, -0.027, 0)<br>[Fe1-Fe4 (-1/2,-1/2,1/2)]<br>J2=0.496 [Fe1-Fe1 (-1,0,0)]<br>J3=1.086 [Fe1-Fe1 (-1/2,-1/2,-1/2)]<br>J4=-0.097 [Fe1-Fe4 (-3/2,-1/2,1/2)]<br>J5=-1.165 [Fe1-Fe1 (-1,-1,0)]<br>J6=-0.020 [Fe1-Fe4 (-1/2,-1/2,-1/2)]<br>J7=0.064 [Fe1-Fe4 (-1/2,-3/2,1/2)]<br>J8=-0.201 [Fe1-Fe4 (-3/2,-1/2,-1/2)]<br>J9a=-0.067 [Fe1-Fe1 (-1/2,-3/2,-1/2)]<br>J9b=-0.118 [Fe1-Fe1 (-1/2,3/2,-1/2)]<br>J10=0.155 [Fe1-Fe4 (-3/2,-3/2,1/2)]<br>J11=0.011 [Fe1-Fe1 (0,0,-1)]<br>J12=0.084 [Fe1-Fe1 (-2,0,0)]<br>J13=-0.134 [Fe1-Fe4 (-1/2,-3/2,-1/2)]<br>J14=-0.007 [Fe1-Fe4 (-5/2,-1/2,1/2)]<br>J15=0.022 [Fe1-Fe1 (-1,0,-1)]<br>J16a=-0.087 [Fe1-Fe1 (-1,-2,0)]<br>J16b=0.001 [Fe1-Fe1 (-1,2,0)] |
| 0.811,<br>0.813 | $Fe_2WO_6$ | Fe1 (0,0.056,0.25)<br>Fe2 (0.5,0.444,0.75) | 4.0<br>(Fe 3*d*) | J1=-2.146 [Fe1-Fe3 (0,-1,-1)]<br>J2=14.455, $D_2$= (0.529, -0.178, 0.235), |
| | | Fe3 (0,0.944,0.75)<br>Fe4 (0.5,0.556,0.25)<br>Fe5 (0,0.724,0.25)<br>Fe6 (0.5,0.776,0.75)<br>Fe7 (0,0.276,0.75)<br>Fe8 (0.5,0.224,0.25) | 4.0<br>(W 4*d*) | $J_2^{ani}$= (-0.103, -0.105, -0.099),<br>$\Gamma_2$= (0.021, -0.03, -0.007)<br>[Fe5-Fe6 (-1,0,-1)]<br>J3=8.500 [Fe1-Fe8 (-1,0,0)]<br>J4=-0.597 [Fe1-Fe7 (0,0,-1)]<br>J5a=-0.802 [Fe1-Fe1 (-1,0,0)]<br>J5b=-0.302 [Fe5-Fe5 (-1,0,0)]<br>J6a=-1.183 [Fe1-Fe1 (0,0,-1)]<br>J6b=0.054 [Fe5-Fe5 (0,0,-1)]<br>J7a=-0.827 [Fe1-Fe3 (-1,-1,-1)]<br>J7b=-1.141 [Fe1-Fe3 (-1,-1,0)]<br>J8=-0.484 [Fe1-Fe5 (0,-1,0)]<br>J9a=-0.456 [Fe1-Fe6 (-1,-1,-1)]<br>J9b=-0.248 [Fe1-Fe6 (-1,-1,0)]<br>J10a=-0.441 [Fe1-Fe8 (-1,0,-1)]<br>J10b=-0.854 [Fe1-Fe8 (-1,0,1)]<br>J11a=-2.722 [Fe1-Fe7 (-1,0,0)]<br>J11b=-0.195 [Fe1-Fe7 (-1,0,-1)]<br>J12a=-0.676 [Fe1-Fe1 (-1,0,1)]<br>J12b=-0.530 [Fe1-Fe1 (-1,0,-1)]<br>J12c=-0.158 [Fe5-Fe5 (-1,0,-1)]<br>J12d=0.028 [Fe5-Fe5 (-1,0,1)]<br>J13=-0.709 [Fe1-Fe5 (-1,-1,0)] |
| 0.823 | $Sr_2MnGaO_5$ | Mn1 (0, 0, 0)<br>Mn2 (0.5, 0.5, 0) | 3.0<br>(Mn 3*d*) | J1=-3.624, $D_1$= (-0.009, -0.038, 0.015),<br>$J_1^{ani}$= (-0.019, -0.019, -0.03),<br>$\Gamma_1$= (-0.007, 0, 0)<br>[Mn1-Mn2 (-1,-1,0)]<br>J2a=-0.313 [Mn1-Mn1 (-1,0,0)]<br>J2b=-0.367 [Mn1-Mn1 (0,-1,0)]<br>J3=0.293 [Mn1-Mn1 (-1,-1,0)]<br>J4=-1.111 [Mn1-Mn2 (-1,-1,-1)]<br>J5=-0.035 [Mn1-Mn2 (-2,-1,0)]<br>J6=-0.067 [Mn1-Mn2 (-1,-2,0)]<br>J7a=0.010 [Mn1-Mn1 (-1,-1,-1)]<br>J7b=0.001 [Mn1-Mn1 (-1,0,-1)]<br>J8=-0.002 [Mn1-Mn2 (-1,0,1)]<br>J9=-0.004 [Mn1-Mn2 (-1,-2,-1)]<br>J10=-0.004 [Mn1-Mn2 (-1,0,-1)]<br>J11=-0.004 [Mn1-Mn1 (-2,0,0)]<br>J12=-0.010 [Mn1-Mn1 (0,-2,0)]<br>J13=0.005 [Mn1-Mn2 (-1,1,1)]<br>J14=0.004 [Mn1-Mn2 (-1,-1,1)]<br>J15=0.025 [Mn1-Mn2 (-2,-2,0)] |
| 0.825 | $Ca_2MnGa$ | Mn1 (0,0,0) | 3.0 | J1=3.965, $D_1$= (-0.046, 0.002, 0.054), |
| | $O_5$ | Mn2 (0.5,0.5,0.5)<br>Mn3 (0,0.5,0)<br>Mn4 (0.5,0,0.5) | (Mn 3$d$) | $J_1^{ani}$= (-0.027, -0.031, -0.026),<br>$\Gamma_1$= (0.001, -0.014, -0.001)<br>[Mn1-Mn4 (-1,0,-1)]<br>J2=-0.435 [Mn1-Mn1 (-1,0,0)]<br>J3=-0.192 [Mn1-Mn1 (0,0,-1)]<br>J4a=0.685 [Mn1-Mn1 (-1,0,1)]<br>J4b=0.520 [Mn1-Mn1 (-1,0,-1)]<br>J5=-0.390 [Mn1-Mn3 (0,-1,0)]<br>J6=0.023 [Mn1-Mn4 (-2,0,-1)]<br>J7a=0.122 [Mn1-Mn2 (-1,-1,0)]<br>J7b=0.061 [Mn1-Mn2 (-1,-1,-1)]<br>J8=-0.016 [Mn1-Mn4 (-1,0,-2)]<br>J9=0.018 [Mn1-Mn3 (-1,-1,0)]<br>J10=0.028 [Mn1-Mn3 (0,-1,-1)]<br>J11=-0.026 [Mn1-Mn1 (-2,0,0)]<br>J12a=-0.015 [Mn1-Mn3 (-1,-1,-1)]<br>J12b=-0.137 [Mn1-Mn3 (-1,-1,1)]<br>J13=-0.024 [Mn1-Mn1 (0,0,-2)]<br>J14=-0.029 [Mn1-Mn2 (-2,-1,-1)]<br>J15=0.083 [Mn1-Mn4 (-2,0,-2)]<br>J16=-0.036 [Mn1-Mn2 (-1,-1,-2)]<br>J17a=0.005 [Mn1-Mn1 (-2,0,-1)]<br>J17b=-0.030 [Mn1-Mn1 (-2,0,1)]<br>J18a=0.002 [Mn1-Mn1 (-1,0,-2)]<br>J18b=-0.039 [Mn1-Mn1 (-1,0,2)] |
| 0.836,<br>0.837,<br>0.838,<br>0.839,<br>0.840,<br>0.841 | $DyFeO_3$ | Fe1 (0, 0.5, 0)<br>Fe2 (0.5, 0, 0.5)<br>Fe3 (0, 0.5, 0.5)<br>Fe4 (0.5, 0, 0) | 3.0<br>(Fe 3$d$) | J1=-10.262 [Fe1-Fe3 (0,0,-1)]<br>J2=-10.915, $D_2$= (-0.329, 0.172, -0.23),<br>$J_2^{ani}$= (-0.121, -0.118, -0.115),<br>$\Gamma_2$= (-0.014, 0.007, -0.007)<br>[Fe1-Fe4 (-1,0,0)]<br>J3=0.141 [Fe1-Fe1 (-1,0,0)]<br>J4a=0.161 [Fe1-Fe2 (-1,0,-1)]<br>J4b=0.101 [Fe1-Fe2 (-1,0,0)]<br>J5=-0.165 [Fe1-Fe1 (0,-1,0)]<br>J6=-0.010 [Fe1-Fe3 (-1,0,-1)]<br>J7=-0.037 [Fe1-Fe3 (0,-1,-1)]<br>J8=0.288 [Fe1-Fe1 (0,0,-1)]<br>J9a=0.392 [Fe1-Fe1 (-1,1,0)]<br>J9b=0.337 [Fe1-Fe1 (-1,-1,0)]<br>J10=-0.014 [Fe1-Fe4 (-2,0,0)]<br>J11a=-0.021 [Fe1-Fe4 (-1,0,-1)]<br>J11b=-0.044 [Fe1-Fe4 (-1,0,1)]<br>J12a=-0.022 [Fe1-Fe3 (-1,-1,-1)]<br>J12b=-0.014 [Fe1-Fe3 (-1,1,-1)] |
| | | | | J13=-0.032 [Fe1-Fe4 (-1,-1,0)] |
| 0.896 | $NiCrO_4$ | Ni1 (0, 0, 0)<br>Ni2 (0, 0, 0.5) | 5.1<br>(Ni 3*d*) | J1=-3.096, $D_1$= (-0.051, 0.002, -0.003)<br>[Ni1-Ni2 (0,0,-1)]<br>J2=-0.312 [Ni1-Ni1 (-1/2,-1/2,0)]<br>J3=-0.252 [Ni1-Ni1 (-1,0,0)]<br>J4=-0.049 [Ni1-Ni2 (-1/2,-1/2,-1)]<br>J5=0.025 [Ni1-Ni1 (0,0,-1)]<br>J6=0.003 [Ni1-Ni2 (-1,0,-1)]<br>J7a=0.004 [Ni1-Ni1 (-1/2,-1/2,-1)]<br>J7b=0.004 [Ni1-Ni1 (-1/2,-1/2,1)]<br>J8=-0.002 [Ni1-Ni1 (0,-1,0)]<br>J9=-0.014 [Ni1-Ni1 (-1,0,-1)]<br>$J_i^{ani}$=$\Gamma_i$= (0, 0, 0) |
| 0.917 | $Sr_2ScOsO_6$† | Os1 (0.5, 0, 0.5)<br>Os2 (0, 0.5, 0) | 3.0<br>(Os 5*d*) | J1=-1.651, $D_1$= (-0.013, -0.02, -0.005),<br>$\Gamma_1$= (0.109, 0.01, 0.092)<br>[Os1-Os2 (0,-1,0)]<br>J2=-2.070, $\boldsymbol{J_2^{ani}}$**= (-0.154, -0.229, -0.371)**<br>[Os1-Os1 (0,-1,0)]<br>J3=-1.352 [Os1-Os1 (-1,0,0)]<br>J4=-1.476 [Os1-Os2 (0,-1,1)]<br>J5a=0.113 [Os1-Os1 (-1,1,0)]<br>J5b=0.117 [Os2-Os2 (-1,1,0)]<br>J6=0.040 [Os1-Os1 (0,0,-1)]<br>J7=0.002 [Os1-Os2 (-1,-1,0)]<br>J8=0.003 [Os1-Os1 (-1,0,-1)]<br>J9=0.006 [Os1-Os2 (0,-2,0)]<br>J10=0.003 [Os1-Os2 (0,-2,1)]<br>J11a=0.003 [Os1-Os1 (0,-1,-1)]<br>J11b=0.002 [Os1-Os1 (0,-1,1)]<br>J12=0.002 [Os1-Os2 (-1,-1,1)]<br>J13=0.003 [Os1-Os1 (-1,0,1)]<br>J14a=0.015 [Os1-Os1 (-1,-1,-1)]<br>J14b=0.015 [Os1-Os1 (-1,1,-1)]<br>J15=0.011 [Os1-Os1 (0,-2,0)]<br>J16=0.007 [Os1-Os1 (-2,0,0)]<br>J17a=0.012 [Os1-Os1 (-1,-1,1)]<br>J17b=0.012 [Os1-Os1 (-1,1,1)] |
| 0.984 | $LuVO_3$† | V1 (0.5,0,0)<br>V2 (0,0.5,0.5)<br>V3 (0.5,0,0.5)<br>V4 (0,0.5,0) | 2.85<br>(V 3*d*) | J1=0.010, $\boldsymbol{\Gamma_1}$**= (0.002, -0.219, 0.049)**<br>[V1-V3 (0,0,-1)]<br>J2=0.124, $\boldsymbol{D_2}$**= (0.145, -0.274, 0.667)**,<br>$\boldsymbol{J_2^{ani}}$**= (-0.705, -0.005, -1.05)**,<br>[V1-V4 (0,-1,0)]<br>J3=0.003 [V1-V1 (-1,0,0)]<br>J4a=-0.010 [V1-V2 (0,-1,0)] |

| | | | | J4b=-0.022 [V1-V2 (0,-1,-1)]<br>J5=-0.050 [V1-V1 (0,-1,0)]<br>J6=0.000 [V1-V3 (-1,0,-1)]<br>J7=-0.001 [V1-V3 (0,-1,-1)]<br>J8=0.012 [V1-V1 (0,0,-1)]<br>J9a=-0.044 [V1-V1 (-1,1,0)]<br>J9b=-0.005 [V1-V1 (-1,-1,0)] |
|---|---|---|---|---|
| 0.1018, 0.1019 | $SrMnO_3$ † | Mn1 (0.988,0.333,0.613)<br>Mn2 (0.012,0.333,0.887)<br>Mn3 (0.988,0.667,0.387)<br>Mn4 (0.012,0.667,0.113) | 2.1<br>(Mn 3*d*) | J1=-0.606, $\boldsymbol{D_1}$**= (-0.024, 0.305, -1.208)**<br>[Mn1-Mn2 (1,0,0)]<br>J2=-1.048, $\boldsymbol{\Gamma_2}$**= (-0.002, -0.137, 0.013)**<br>[Mn1-Mn3 (0,0,0)]<br>J3=-1.022, $\boldsymbol{J_3^{ani}}$**= (2.322, -0.292, -2.031)**<br>[Mn1-Mn7 (-1/2,-1/2,0)]<br>J4a=-0.033 [Mn1-Mn5 (-1/2,1/2,0)]<br>J4b=0.013 [Mn1-Mn5 (-1/2,-1/2,0)]<br>J5=-0.005 [Mn1-Mn1 (-1,0,0)] |
| 1.0.9, 1.0.38 | $CsCoCl_3$ | Co1 (0,0,0)<br>Co2 (0,0,0.5)<br>Co3 (0.333,0.667,0)<br>Co4 (0.667,0.333,0.5)<br>Co5 (0.667,0.333,0)<br>Co6 (0.333,0.667,0.5) | 3.3<br>(Co 3*d*) | J1a=-3.693, $D_1$= (0.005, 0.093, 0.003),<br>$J_1^{ani}$= (-0.022, -0.025, -0.01)<br>[Co1-Co2 (0,0,-1)]<br>J1b=-3.494 [Co3-Co6 (0,0,-1)]<br>J1c=-3.689 [Co4-Co5 (0,0,0)]<br>J2a=-0.039 [Co1-Co1 (0,0,-1)]<br>J2b=-0.034 [Co3-Co3 (0,0,-1)]<br>J2c=-0.038 [Co4-Co4 (0,0,-1)]<br>J3a=-0.043 [Co1-Co3 (-1,-1,0)]<br>J3b=-0.036 [Co3-Co5 (-1,0,0)] |
| 1.0.26 | $RbCoCl_3$ | Co1 (0,0,0)<br>Co2 (0,0,0.5)<br>Co3 (0.333,0.667,0)<br>Co4 (0.667,0.333,0.5)<br>Co5 (0.667,0.333,0)<br>Co6 (0.333,0.667,0.5) | 3.3<br>(Co 3*d*) | J1a=-4.436, $D_1$= (0.061, 0.104, 0.008),<br>$J_1^{ani}$= (-0.145, -0.154, -0.146)<br>[Co1-Co2 (0,0,-1)]<br>J1b=-4.449 [Co3-Co6 (0,0,-1)]<br>J2a=-0.019 [Co1-Co1 (0,0,-1)]<br>J2b=-0.015 [Co3-Co3 (0,0,-1)]<br>J3a=-0.142 [Co1-Co3 (-1,-1,0)]<br>J3b=-0.123 [Co3-Co5 (-1,0,0)]<br>$\Gamma_i$= (0, 0, 0) |
| 1.0.39 | $BaMnO_3$ | Mn1 (0,0,0)<br>Mn2 (0,0,0.5)<br>Mn3 (0.333,0.667,0)<br>Mn4 (0.667,0.333,0.5)<br>Mn5 (0.667,0.333,0)<br>Mn6 (0.333,0.667,0.5) | 2.7<br>(Mn 3*d*) | J1a=-18.574, $D_1$= (-0.002, -0.004, -0.098)<br>[Mn1-Mn2 (0,0,-1)]<br>J1b=-18.582 [Mn3-Mn6 (0,0,-1)]<br>J2a=0.654 [Mn1-Mn1 (0,0,-1)]<br>J2b=0.663 [Mn3-Mn3 (0,0,-1)]<br>J3a=-0.165 [Mn1-Mn3 (-1,-1,0)]<br>J3b=-0.150 [Mn3-Mn5 (-1,0,0)]<br>J4=-0.004 [Mn1-Mn4 (-1,-1,-1)]<br>J5=-0.051 [Mn1-Mn2 (0,0,-2)] |

| | | | | $J_i^{ani}$=$\Gamma_i$= (0, 0, 0) |
|---|---|---|---|---|
| 1.0.47 | $MnSe_2$† | Mn1 (0,0,0)<br>Mn2 (0.5,0.5,0)<br>Mn3 (0,0.5,0.5)<br>Mn4 (0.5,0,0.5)<br>Mn5 (0.167,0,0.5)<br>Mn6 (0.667,0.5,0.5)<br>Mn7 (0.833,0.5,0)<br>Mn8 (0.333,0,0)<br>Mn9 (0.833,0,0.5)<br>Mn10 (0.333,0.5,0.5)<br>Mn11 (0.167,0.5,0)<br>Mn12 (0.667,0,0) | 3.0<br>(Mn 3*d*) | J1a=-0.725, **$D_1$= (-0.125, 0.079, -0.067)**,<br>$J_1^{ani}$= (-0.017, -0.022, -0.018),<br>$\Gamma_1$= (0.005, -0.001, -0.001)<br>[Mn1-Mn3 (0,-1,-1)]<br>J1b=-0.813 [Mn5-Mn8 (0,0,0)]<br>J1c=1.357 [Mn1-Mn5 (0,0,0)]<br>J1d=1.331 [Mn1-Mn5 (0,0,-1)]<br>J1e=-0.782 [Mn1-Mn7 (-1,-1,0)]<br>J1f=-0.715 [Mn1-Mn7 (-1,0,0)]<br>J1g=1.270 [Mn5-Mn10 (0,-1,0)]<br>J1h=-0.748 [Mn5-Mn11 (0,-1,0)]<br>J1i=-0.758 [Mn5-Mn11 (0,0,0)]<br>J2a=0.141 [Mn1-Mn8 (0,0,0)]<br>J2b=-0.132 [Mn1-Mn1 (0,0,-1)]<br>J2c=-0.208 [Mn5-Mn5 (0,0,1)]<br>J2d=-0.226 [Mn5-Mn9 (-1,0,0)]<br>J2e=-0.253 [Mn1-Mn1 (0,-1,0)]<br>J2f=-0.259 [Mn5-Mn5 (0,-1,0)]<br>J3a=-0.149 [Mn5-Mn10 (0,-1,1)]<br>J3b=-0.146 [Mn1-Mn5 (0,1,-1)]<br>J3c=-0.141 [Mn1-Mn10 (0,-1,0)]<br>J3d=-0.136 [Mn1-Mn6 (-1,-1,-1)]<br>J3e=-0.135 [Mn1-Mn5 (0,1,0)]<br>J3f=-0.095 [Mn5-Mn8 (0,1,0)]<br>J3g=-0.092 [Mn10-Mn12 (0,0,1)]<br>J3h=-0.085 [Mn1-Mn11 (0,0,1)]<br>J3i=-0.077 [Mn1-Mn11 (0,-1,-1)]<br>J3j=-0.029 [Mn1-Mn5 (0,-1,-1)]<br>J3k=-0.028 [Mn1-Mn10 (0,-1,-1)]<br>J3l=-0.024 [Mn1-Mn10 (0,0,-1)]<br>J3m=-0.021 [Mn5-Mn8 (0,-1,0)]<br>J3n=-0.020 [Mn1-Mn5 (0,-1,0)]<br>J3o=-0.019 [Mn5-Mn10 (0,-1,-1)]<br>J3p=-0.011 [Mn5-Mn7 (-1,-1,0)]<br>J3q=-0.006 [Mn1-Mn11 (0,-1,1)]<br>J3r=0.005 [Mn1-Mn7 (-1,-1,1)]<br>J4a=-0.087 [Mn1-Mn1 (0,-1,1)]<br>J4b=-0.085 [Mn5-Mn5 (0,-1,-1)]<br>J4c=-0.080 [Mn1-Mn12 (-1,0,1)]<br>J4d=-0.076 [Mn5-Mn9 (-1,1,0)]<br>J4e=-0.072 [Mn1-Mn8 (0,-1,0)]<br>J4f=-0.072 [Mn5-Mn9 (-1,0,-1)]<br>J4g=0.018 [Mn5-Mn9 (-1,0,1)] |
| | | | | J4h=0.018 [Mn1-Mn8 (0,1,0)]<br>J4i=0.020 [Mn1-Mn1 (0,-1,-1)]<br>J4j=0.021 [Mn5-Mn5 (0,-1,1)]<br>J4k=0.021 [Mn5-Mn9 (-1,-1,0)]<br>J4l=0.025 [Mn1-Mn12 (-1,0,-1)]<br>J5a=-0.111 [Mn5-Mn12 (-1,0,0)]<br>J5b=-0.102 [Mn5-Mn12 (-1,0,1)]<br>J5c=-0.097 [Mn1-Mn11 (0,-2,0)]<br>J5d=-0.096 [Mn1-Mn11 (0,1,0)]<br>J5e=-0.092 [Mn5-Mn10 (0,-2,0)]<br>J5f=-0.091 [Mn5-Mn11 (0,0,-1)]<br>J5g=-0.091 [Mn1-Mn3 (0,-1,-2)]<br>J5h=-0.090 [Mn1-Mn4 (-1,0,-1)]<br>J5i=-0.085 [Mn5-Mn11 (0,-1,-1)]<br>J5j=-0.015 [Mn5-Mn8 (0,0,-1)]<br>J5k=-0.014 [Mn5-Mn6 (-1,0,0)]<br>J5l=-0.011 [Mn1-Mn5 (0,0,-2)]<br>J5m=-0.011 [Mn1-Mn2 (-1,-1,0)]<br>J5n=-0.010 [Mn5-Mn11 (0,1,0)]<br>J5o=-0.009 [Mn5-Mn6 (-1,-1,0)]<br>J5p=-0.009 [Mn1-Mn5 (0,0,1)]<br>J5q=-0.006 [Mn1-Mn3 (0,-2,-1)]<br>J5r=-0.005 [Mn5-Mn11 (0,-2,0)] |
| 1.522 | $CrVO_4$ | Cr1 (0, 0, 0)<br>Cr2 (0, 0, 0.5) | 3.0<br>(Cr 3*d*) | J1=-4.097, $D_1$= (-0.012, 0.002, -0.003),<br>[Cr1-Cr2 (0,0,-1)]<br>J2=0.442 [Cr1-Cr1 (-1,0,0)]<br>J3=-0.371 [Cr1-Cr1 (-1,-1,0)]<br>J4=-0.133 [Cr1-Cr2 (-1,0,-1)]<br>J5=-0.052 [Cr1-Cr1 (0,0,-1)]<br>J6=-0.217 [Cr1-Cr2 (-1,-1,-1)]<br>J7a=0.004 [Cr1-Cr1 (-1,0,-1)]<br>J7b=0.003 [Cr1-Cr1 (-1,0,1)]<br>$J_i^{ani}$=$\Gamma_i$= (0, 0, 0) |
| | $CuF_2$† | Cu1 (0, 0, 0)<br>Cu2 (0, 0.5, 0.5) | 9.0<br>(Cu 3*d*) | J1=0.104 [Cu1-Cu1 (-1,0,0)]<br>J2=-3.974, **$D_2$= (-0.056, -0.289, -0.336),**<br>$J_2^{ani}$= (-0.008, -0.011, -0.024),<br>$\Gamma_2$= (0.001, -0.002, -0.013)<br>[Cu1-Cu2 (0,-1,-1)]<br>J3=0.103 [Cu1-Cu2 (-1,-1,-1)]<br>J4=0.144 [Cu1-Cu1 (0,-1,0)]<br>J5=-0.014 [Cu1-Cu1 (-1,0,-1)]<br>J6=0.025 [Cu1-Cu1 (0,0,-1)]<br>J7a=-0.209 [Cu1-Cu1 (-1,-1,0)]<br>J7b=-0.005 [Cu1-Cu1 (-1,1,0)] |
| | | | | J8=-0.003 [Cu1-Cu2 (-1,-1,0)]<br>J9=0.003 [Cu1-Cu1 (-2,0,-1)]<br>J10=0.000 [Cu1-Cu2 (-2,-1,-1)]<br>J11a=-0.435 [Cu1-Cu1 (-1,-1,-1)]<br>J11b=0.001 [Cu1-Cu1 (-1,1,-1)]<br>J12=0.000 [Cu1-Cu1 (-2,0,0)]<br>J13a=0.037 [Cu1-Cu1 (0,-1,-1)]<br>J13b=0.031 [Cu1-Cu1 (0,-1,1)] |
| | $NiF_2$[†] | Ni1 (0, 0, 0)<br>Ni2 (0.5, 0.5, 0.5) | 5.3<br>(Ni 3*d*) | J1=-0.825, $D_2$= (0.124, -0.124, 0),<br>$\boldsymbol{J_2^{ani}}$**= (-0.006, -0.006, -0.351)**,<br>[Ni1-Ni1 (0,0,-1)]<br>J2=-1.790 [Ni1-Ni2 (-1,-1,-1)]<br>J3=0.003 [Ni1-Ni1 (-1,0,0)]<br>J4=-0.038 [Ni1-Ni1 (-1,0,-1)]<br>J5=0.004 [Ni1-Ni2 (-1,-1,-2)]<br>J6=-0.045 [Ni1-Ni1 (0,0,-2)]<br>J7a=-0.135 [Ni1-Ni1 (-1,-1,0)]<br>J7b=0.002 [Ni1-Ni1 (-1,1,0)]<br>J8a=0.005 [Ni1-Ni1 (-1,-1,-1)]<br>J8b=0.006 [Ni1-Ni1 (-1,1,-1)]<br>$\Gamma_i$= (0, 0, 0) |

# S7. Computational methods and examples

## S7.1. DFT parameters and benchmark

We first conduct SOC-free self-consistent calculations and construct the Wannier Hamiltonians for two spin channels, followed by obtaining the exchange interactions using TB2J. To assess the influence of SOC, we also perform self-consistent calculations and construct Wannier Hamiltonians under SOC, and then derive the exchange parameters under SOC using TB2J. Since the calculations of spin-axis-related exchange terms in TB2J are unphysical, three calculations with spin axis along the $x$, $y$, and $z$ directions are performed. We find that when the exchange interactions related to SOC are small, the differences in the Heisenberg coefficients derived from SOC-free and SOC calculations are not significant. For the high-throughput calculations for magnon band structures, we use the Heisenberg exchange parameters under SOC.

A brief comparison of the CPU hour statistics between electronic and magnon band structures in our work is shown in Supplementary Table CLXXXII. Notice that the computational cost of electronic structures in our calculations is similar to the high-throughput calculations of topological magnetic materials, where 41.8 and 1163.8 CPU hours are used in 10-atom and 66-atom systems for SOC-included electronic structure calculations in Y. Xu' work[44]. However, the calculation of magnon dispersion for a single magnetic material costs 20-50 times more CPU hours than the electronic structure (with SOC).

Comparing with electronic structure calculations, the calculations of exchange interactions are more susceptible to numerical convergence and nonlocalized Wannier functions due to the three SOC calculations with spin axis along $x$, $y$, $z$. Additionally, even if all exchange coefficients can be successfully computed, the magnon Hamiltonian may still be negative-definite (corresponding to the instability of the current magnetic configuration), thus making it impossible to obtain a magnon dispersion with real eigenvalues. Nonetheless, we have performed calculations on 348 collinear magnets, ultimately evaluating the bilinear magnetic exchange interactions for 223 systems and deducing the corresponding magnon dispersion for 203 systems.

Supplementary Table CLXXXII. A brief comparison of the CPU hour statistics between the electronic band structures and the magnon band structures.

| CPU hour statistics | Electronic structure | | Magnon band structure | |
|---|---|---|---|---|
| | w/o SOC | w/ SOC | w/o SOC | w/ SOC |
| 10-atom system | 5.4 | 58.1 | 67.1 | 1230.5 |

| ($CaCo_2P_2$) | | | | |
|---|---|---|---|---|
| 20-atom system ($LaMnO_3$) | 21.6 | 215.8 | 335.7 | 1587.6 |
| 48-atom system ($VPO_4$) | 83.9 | 258.0 | 699.2 | 4565.8 |
| 80-atom system ($Fe_{0.35}NbS_2$) | 402.7 | 1297.4 | 8490.3 | 70703.7 |

## S7.2. Linear spin wave theory and beyond

To calculate the magnon band structures of materials, we use the standard Holstein-Primakoff transformation:

$$S^{+} = \sqrt{2S - a^{\dagger}a} \cdot a \approx \sqrt{2S} \cdot a \tag{61}$$

$$S^{-} = a^{\dagger} \cdot \sqrt{2S - a^{\dagger}a} \approx \sqrt{2S} \cdot a^{\dagger} \tag{62}$$

$$S^{z} = S - a^{\dagger}a \tag{63}$$

where the boson operators $a^{\dagger}, a$ satisfy the commutation relations $[a, a^{\dagger}] = 1$.

At low temperatures or weak excitation, we can disregard all but the zeroth order in $a^{\dagger}a/2S$ in the series expansion of the square root.

By performing the Holstein-Primakoff transformation and Fourier transformation, the ground state Heisenberg spin Hamiltonian can be changed into quadratic Hamiltonian:

$$H = \sum_{k} \psi^{\dagger}(k) \begin{pmatrix} h(k) & g(k) \\ g(k)^{\dagger} & h(-k)^{T} \end{pmatrix} \psi(k) \tag{64}$$

where $\psi^{\dagger}(k) = \left(a_{k1}^{\dagger} \dots a_{km}^{\dagger} a_{-k1} \dots a_{-km}\right)^{T}$, and

$$h(k)_{ab} = S\left[\sum_{R_{ij}} \left(\alpha_{ab} J_{\tau_a, \tau_b + R_{ij}}\right) \cdot e^{ikR_{ij}} + \delta_{ab} \sum_{R_{ij},\, c} \left(\gamma_{ac} J_{\tau_a, \tau_c + R_{ij}}\right)\right] \tag{65}$$

$$g(k)_{ab} = S \sum_{R_{ij}} \left(\lambda_{ab} J_{\tau_a, \tau_b + R_{ij}}\right) \cdot e^{ikR_{ij}} \tag{66}$$

$$\alpha_{ab} = \frac{1}{2}\{[\sin\theta_a \sin\theta_b + (\cos\theta_a \cos\theta_b + 1)\cos(\varphi_a - \varphi_b)] + i[(\cos\theta_a + \cos\theta_b)\sin(\varphi_b - \varphi_a)]\} \tag{67}$$

$$\lambda_{ab} = \frac{1}{2}\{[\sin\theta_a \sin\theta_b + (\cos\theta_a \cos\theta_b - 1)\cos(\varphi_a - \varphi_b)] - i[(\cos\theta_a - \cos\theta_b)\sin(\varphi_b - \varphi_a)]\} \tag{68}$$

$$\gamma_{ab} = -[\cos\theta_a \cos\theta_b + \sin\theta_a \sin\theta_b \cos(\varphi_a - \varphi_b)] \quad (69)$$

where the spin vector $S_i = S(\sin\theta\cos\varphi, \sin\theta\sin\varphi, \cos\theta)$, $J_{\tau_a,\tau_b+R_{ij}}$ represent the Heisenberg exchange interaction, $\tau_a$ and $R_{ij}$ stand for the position of magnetic ion in lattice basis and the lattice translation vector. For collinear magnetic configuration: $\alpha_{ab}$=1, $\lambda_{ab}$=0, $\gamma_{ab}$=-1 when $S_a$ is parallel to $S_b$, while $\alpha_{ab}$=0, $\lambda_{ab}$=-1, $\gamma_{ab}$=1 when $S_a$ is antiparallel to $S_b$.

In the framework of collinear SSGs and Heisenberg limit, the magnon Hamiltonian is $\mathrm{SO}(2)$ symmetric, the magnon carry a good spin quantum number S = ±1. In other words, even if we introduce the nonlinear terms, they will carry the irreps of $SO(2)$ in spin space and the identity irrep in real space. As a result, the nonlinear terms will not mix the magnon spin and the magnon band degeneracy stays robust. However, once the $SO(2)$ symmetry is broken by either noncollinear magnetic configuration or SOC-related exchange interaction, the nonlinear term may bring different topology due to the alternation of irreps of a certain SSG or MSG. For example, in the prototypical honeycomb ferromagnets, the Dirac points at K point are gapped as the effective time reversal $\{TU_{\boldsymbol{n}}(\pi)||E\}$ symmetry is broken by DM, leading to the existence of chiral edge magnons. When incorporating the nonlinear magnon terms, the chiral edge magnons undergoes another topological phase transition, where the Chern number reverses, i.e., topological magnon phase transitions from left- to right-handed chirality[100,101].

## S7.3. Material realizations and band representations

Below, we present the construction of the magnon Hamiltonian for the candidate materials in the main text, along with a comparison of the band representations under collinear SSGs and MSGs, as well as some relevant discussions. Here we use the material name (unconventional magnons) to denote every section.

### S7.3.1. $Gd_4Sb_3$ (sextuple magnon, triple magnon)

Rare earth intermetallic compound $Gd_4Sb_3$ crystallizes in body-centered cubic system of space group $I\bar{4}3d$ (220) with lattice constants a = 9.220 Å. The magnetic ground state of $Gd_4Sb_3$ is a collinear FM with easy axis along [001] direction, which belongs to the MSG $I\bar{4}2'd'$ (122.337). The SSG of collinear FM $Gd_4Sb_3$ is $I^1\bar{4}^13^1d^{\infty m}1$ (220.220.1.1) and the magnetic ions Gd locate at the spin Wyckoff position 16*c* (*a*, *a*, *a* | 0, 0, z). The lattice basis for constructing magnon Hamiltonian is $\mathrm{a}_1 = (\mathrm{a}, 0, 0), \mathrm{a}_2 = (0, \mathrm{a}, 0), \mathrm{a}_3 = (0, 0, \mathrm{a})$. Here we adopt the exchange parameters using energy mapping method, where $J_1 = 2.369$meV, $J_2 = 0.699$meV, $J_3 = 0.663$meV, *S* = 7/2. The 8 equivalent magnetic ions in magnetic primitive cell are given in Supplementary Table CLXXXIII, and the detailed information of bonds are shown in Supplementary Tables CLXXXIV~ CLXXXVI. Finally, a comparison of describing magnon dispersion of $Gd_4Sb_3$ using band representations of collinear SSGs and MSGs are shown in Supplementary Table CLXXXVII.

Supplementary Table CLXXXIII. The coordinates of the 8 Gd ions in the conventional cell basis vectors (*a* = 0.08333). All the magnetic moments are set to [001] direction.

| $n$ | $r_n$ |
|---|---|
| 1 | (*a*, *a*, *a*) |
| 2 | (-*a*+1/2, -*a*, *a*+1/2) |
| 3 | (-*a*, *a*+1/2, -*a*+1/2) |
| 4 | (*a*+1/2, -*a*+1/2, -*a*) |
| 5 | (*a*+1/4, *a*+1/4, *a*+1/4) |
| 6 | (-*a*+1/4, -*a*+3/4, *a*+3/4) |
| 7 | (*a*+3/4, -*a*+1/4, -*a*+3/4) |
| 8 | (-*a*+3/4, *a*+3/4, -*a*+1/4) |

Supplementary Table CLXXXIV. The 12 bonds for the 1st-neighbor exchange parameter $J_1$.

| $n$ | $n'$ | $R_{ij}$ | $R_l$ |
|---|---|---|---|
| 1 | 2 | (1/2-2*a*, -2*a*, 1/2) | (-1/2, 1/2, -1/2) |
| 1 | 3 | (-2*a*, 1/2, 1/2-2*a*) | (1/2, -1/2, -1/2) |
| 1 | 4 | (1/2, 1/2-2*a*, -2*a*) | (-1/2, -1/2, 1/2) |
| 2 | 3 | (-1/2, 2*a*+1/2, -2*a*) | (1/2, -1/2, 1/2) |
| 2 | 4 | (2*a*, 1/2, -2*a*-1/2) | (-1/2, -1/2, 1/2) |
| 3 | 4 | (2*a*+1/2, -2*a*, -1/2) | (-1/2, 1/2, 1/2) |
| 5 | 6 | (-2*a*, 1/2-2*a*, 1/2) | (1/2, -1/2, -1/2) |
| 5 | 7 | (1/2, -2*a*, 1/2-2*a*) | (-1/2, 1/2, -1/2) |
| 5 | 8 | (1/2-2*a*, 1/2, -2*a*) | (-1/2, -1/2, 1/2) |
| 6 | 7 | (1/2+2*a*, -1/2, -2*a*) | (-1/2, 1/2, 1/2) |
| 6 | 8 | (1/2, 2*a*, -1/2-2*a*) | (-1/2, -1/2, 1/2) |

| 7 | 8 | (-2$a$, 2$a$+1/2, -1/2) | (1/2, -1/2, 1/2) |
|---|---|---|---|

Supplementary Table CLXXXV. The 8 bonds for the 2nd-neighbor exchange parameter $J_2$.

| $n$ | $n'$ | $R_{ij}$ | $R_l$ |
|---|---|---|---|
| 1 | 5 | (1/4, 1/4, 1/4) | (0, 0, 0) |
| | | | (-1/2, -1/2, -1/2) |
| 2 | 6 | (-1/4, 3/4, 1/4) | (0, -1, 0) |
| | | | (1/2, -1/2, -1/2) |
| 3 | 8 | (3/4, 1/4, -1/4) | (-1, 0, 0) |
| | | | (-1/2, -1/2, 1/2) |
| 4 | 7 | (1/4, -1/4, 3/4) | (0, 0, -1) |
| | | | (-1/2, 1/2, -1/2) |

Supplementary Table CLXXXVI. The 24 bonds for the 3rd-neighbor exchange parameter $J_3$.

| $n$ | $n'$ | $R_{ij}$ | $R_l$ |
|---|---|---|---|
| 1 | 6 | (1/4-2$a$, 3/4-2$a$, 3/4) | (0, -1, -1) |
| | | | (-1/2, -1/2, -1/2) |
| | 7 | (3/4, 1/4-2$a$, 3/4-2$a$) | (-1, 0, -1) |
| | | | (-1/2, -1/2, -1/2) |
| | 8 | (3/4-2$a$, 3/4, 1/4-2$a$) | (-1, -1, 0) |
| | | | (-1/2, -1/2, -1/2) |
| 2 | 5 | (-1/4+2$a$, 1/4+2$a$, -1/4) | (0, 0, 0) |
| | | | (1/2, -1/2, 1/2) |
| | 7 | (1/4+2$a$, 1/4, 1/4-2$a$) | (0, 0, 0) |
| | | | (-1/2, -1/2, -1/2) |
| | 8 | (1/4, 3/4+2$a$, -1/4-2$a$) | (0, -1, 0) |
| | | | (-1/2, -1/2, 1/2) |
| 3 | 5 | (2$a$+1/4, -1/4, -1/4+2$a$) | (0, 0, 0) |
| | | | (-1/2, 1/2, 1/2) |
| | 6 | (1/4, 1/4-2$a$, 1/4+2$a$) | (0, 0, 0) |
| | | | (-1/2, -1/2, -1/2) |
| | 7 | (3/4+2$a$, -1/4-2$a$, 1/4) | (-1, 0, 0) |
| | | | (-1/2, 1/2, -1/2) |
| 4 | 5 | (-1/4, -1/4+2$a$, 1/4+2$a$) | (0, 0, 0) |
| | | | (1/2, 1/2, -1/2) |
| | 6 | (-1/4-2$a$, 1/4, 3/4+2$a$) | (0, 0, -1) |
| | | | (1/2, -1/2, -1/2) |
| | 8 | (1/4-2$a$, 1/4+2$a$, 1/4) | (0, 0, 0) |
| | | | (-1/2, -1/2, -1/2) |

By using Eqs. (65)~(69), the magnon dispersion of $Gd_4Sb_3$ can be obtained, as shown in Supplementary Table XXXI. We also provide the magnon band representations of the corresponding SSG and MSG. We find that the SSG effectively describes the band degeneracy.

Supplementary Table CLXXXVII. Band representation of the MSG $I\bar{4}2'd'$ (122.337) and the collinear SSG $I^1\bar{4}^13^1d^{\infty m}1$ (220.220.1.1) induced from the local representation of the Wyckoff position 16*c*.

| | Magnetic space group<br>$I\bar{4}2'd'$ (122.337) | Spin space group<br>$I^1\bar{4}^13^1d^{\infty m}1$ (220.220.1.1) |
|---|---|---|
| **Band rep** | $\mathbf{A} \uparrow \mathbf{G}(\mathbf{8})$ | $\mathbf{A^S} \uparrow \mathbf{G}(\mathbf{8})$ |
| Γ (0, 0, 0) | $2\Gamma_1(1) + 2\Gamma_2(1) + 2\Gamma_3(1) + 2\Gamma_4(1)$ | $\Gamma_1^S(1) + \Gamma_2^S(1) + \Gamma_4^S(3)+\Gamma_5^S(3)$ |
| H (0, 1, 0) | $2H_1H_2(2) + 2H_3H_4(2)$ | $H_1^SH_2^S(2) + H_4^SH_5^S(6)$ |
| N (1/2, 1/2, 0) | $8N_1(1)$ | $4N_1^S(2)$ |
| P (1/2, 1/2, 1/2) | $2P_1(1) + 2P_2(1) + 2P_3(1) + 2P_4(1)$ | $2P_3^S(4)$ |
| Λ (u, u, u) | $4\Lambda_1(1) + 4\Lambda_2(1)$ | $2\Lambda_1^S(1) + 2\Lambda_2^S(1) + 2\Lambda_3^S(2)$ |
| D (1/2, 1/2, w) | $4D_1(1) + 4D_2(1)$ | $4D_1^S(2)$ |
| Δ (0, v, 0) | $8\Delta_1(1)$ | $2\Delta_1^S(1) + 2\Delta_2^S(1)$<br>$+2\Delta_3^S\Delta_4^S(2)$ |
| A (u, v, 0) | $8A_1(1)$ | $8A_1^S(1)$ |
| Σ (u, u, 0) | $8\Sigma_1(1)$ | $4\Sigma_1^S(1) + 4\Sigma_2^S(1)$ |
| GP (u, v, w) | $8GP_1(1)$ | $8GP_1^S(1)$ |

### S7.3.2. $Cu_3TeO_6$ (sextuple magnon)

Spin-web compound $Cu_3TeO_6$ crystallizes in body-centered cubic system of space group $Ia\bar{3}$ (206) with lattice constants a = 9.537 Å. The magnetic ground state of $Cu_3TeO_6$ is a collinear AFM with Néel order along [111] direction, which belongs to the MSG $R\bar{3}'$ (148.19). The SSG of collinear AFM $Cu_3TeO_6$ is $I^{\bar{1}}a^{1}\bar{3}^{\infty m}1$ (199.206.1.1) and the magnetic ions Cu locate at the spin Wyckoff position 24$d$ ($a$, 0, 1/4 | 0, 0, z). The lattice basis for constructing magnon Hamiltonian is $\mathrm{a}_1 = (\mathrm{a}, 0, 0), \mathrm{a}_2 = (0, \mathrm{a}, 0), \mathrm{a}_3 = (0, 0, \mathrm{a})$. Here we adopt the reported exchange parameters[102,103], where $J_1 = 4.49\,\mathrm{meV}$, $J_2 = -0.22\,\mathrm{meV}$, $J_3 = -1.49\,\mathrm{meV}$, $J_4 = 1.33\,\mathrm{meV}$, $J_5 = 1.79\,\mathrm{meV}$, $J_6 = -0.21\,\mathrm{meV}$, $J_7 = -0.14\,\mathrm{meV}$, $J_8 = 0.11\,\mathrm{meV}$, $J_9 = 4.51\mathrm{meV}$, $S$ = 1/2. The 12 equivalent magnetic ions in magnetic primitive cell are given in Supplementary Table CLXXXVIII, and the detailed information of bonds are shown in Supplementary Tables CLXXXIX ~ CXCVII. Finally, a comparison of describing magnon dispersion of $Cu_3TeO_6$ using band representations of collinear SSGs and MSGs are shown in Supplementary Table CXCVIII.

Supplementary Table CLXXXVIII. The coordinates of the 12 Cu ions in the conventional basis vectors ($a$ = 0.96907). Here "up" and "down" represent the [111] and [$\bar{1}\bar{1}\bar{1}$] direction, respectively.

| $n$ | $r_n$ | magnetic moment |
|---|---|---|
| 1 | ($a$, 0, 1/4) | up |
| 2 | (1/2-$a$, 0, -1/4) | up |
| 3 | (1/4, $a$, 0) | up |
| 4 | (-1/4, 1/2-$a$, 0) | up |
| 5 | (0, 1/4, $a$) | up |
| 6 | (0, -1/4, 1/2-$a$) | up |
| 7 | (-$a$, 0, -1/4) | down |
| 8 | (-1/2+$a$, 0, 1/4) | down |
| 9 | (-1/4, -$a$, 0) | down |
| 10 | (1/4, -1/2+$a$, 0) | down |
| 11 | (0, -1/4, -$a$) | down |
| 12 | (0, 1/4, -1/2+$a$) | down |

TABLE CLXXXIX. The 24 bonds for the 1st-NN exchange parameter $J_1$.

| $n$ | $n'$ | $R_{ij}$ | $R_l$ |
|---|---|---|---|
| 1 | 9 | (-1/4-$a$, -$a$, -1/4) | (1, 1, 0) |
|  | 10 | (1/4-$a$, $a$-1/2, -1/4) | (1/2, -1/2, 1/2) |
|  | 11 | (-$a$, -1/4, -$a$-1/4) | (1, 0, 1) |
|  | 12 | (-$a$, 1/4, $a$-3/4) | (1, 0, 0) |
| 2 | 9 | ($a$-3/4, -$a$, 1/4) | (0, 1, 0) |
| | 10 | ($a$-1/4, $a$-1/2, 1/4) | (-1/2, -1/2, -1/2) |
| | 11 | ($a$-1/2, -1/4, 1/4-$a$) | (-1/2, 1/2, 1/2) |
| | 12 | ($a$-1/2, 1/4, $a$-1/4) | (-1/2, -1/2, -1/2) |
| 3 | 7 | (-1/4-$a$, -$a$, -1/4) | (1, 1, 0) |
| | 8 | ($a$-3/4, -$a$, 1/4) | (0, 1, 0) |
| | 11 | (-1/4, -1/4-a, -$a$) | (0, 1, 1) |
| | 12 | (-1/4, 1/4-$a$, $a$-1/2) | (1/2, 1/2, -1/2) |
| 4 | 7 | (1/4-$a$, $a$-1/2, -1/4) | (1/2, -1/2, 1/2) |
| | 8 | ($a$-1/4, $a$-1/2, 1/4) | (-1/2, -1/2, -1/2) |
| | 11 | (1/4, $a$-3/4, -$a$) | (0, 0, 1) |
| | 12 | (1/4, $a$-1/4, $a$-1/2) | (-1/2, -1/2, -1/2) |
| 5 | 7 | (-$a$, -1/4, -$a$-1/4) | (1, 0, 1) |
| | 8 | ($a$-1/2, -1/4, 1/4-$a$) | (-1/2, 1/2, 1/2) |
| | 9 | (-1/4, -1/4-a, -$a$) | (0, 1, 1) |
| | 10 | (1/4, $a$-3/4, -$a$) | (0, 0, 1) |
| 6 | 7 | (-$a$, 1/4, $a$-3/4) | (1, 0, 0) |
| | 8 | ($a$-1/2, 1/4, $a$-1/4) | (-1/2, -1/2, -1/2) |
| | 9 | (-1/4, 1/4-$a$, $a$-1/2) | (1/2, 1/2, -1/2) |
| | 10 | (1/4, $a$-1/4, $a$-1/2) | (-1/2, -1/2, -1/2) |

TABLE CXC. The 24 bonds for the 2nd-NN exchange parameter $J_2$.

| $n$ | $n'$ | $R_{ij}$ | $R_l$ |
|---|---|---|---|
| 1 | 3 | (1/4-$a$, $a$, -1/4) | (1, -1, 0) |
| | 4 | (-1/4-$a$, 1/2-$a$, -1/4) | (3/2, 1/2, 1/2) |
| | 5 | (-$a$, 1/4, $a$-1/4) | (1, 0, -1) |
| | 6 | (-$a$, -1/4, 1/4-$a$) | (1, 0, 1) |
| 2 | 3 | ($a$-1/4, $a$, 1/4) | (-1, -1, 0) |
| | 4 | ($a$-3/4, 1/2-$a$, 1/4) | (-1/2, 1/2, -1/2) |
| | 5 | ($a$-1/2, 1/4, $a$+1/4) | (-1/2, -1/2, -3/2) |
| | 6 | ($a$-1/2, -1/4, 3/4-$a$) | (-1/2, 1/2, 1/2) |
| 3 | 5 | (-1/4, 1/4-$a$, $a$) | (0, 1, -1) |
| | 6 | (-1/4, -1/4-$a$, 1/2-$a$) | (1/2, 3/2, 1/2) |
| 4 | 5 | (1/4, $a$-1/4, $a$) | (0, -1, -1) |
| | 6 | (1/4, $a$-3/4, 1/2-$a$) | (-1/2, -1/2, 1/2) |
| 7 | 9 | (-1/4+$a$, -$a$, 1/4) | (-1, 1, 0) |
| | 10 | (1/4+$a$, -1/2+$a$, 1/4) | (-3/2, -1/2, -1/2) |
| | 11 | ($a$, -1/4, -$a$+1/4) | (-1, 0, 1) |
| | 12 | ($a$, 1/4, -1/4+$a$) | (-1, 0, -1) |
| 8 | 9 | (-$a$+1/4, -$a$, -1/4) | (1, 1, 0) |
| | 10 | (-$a$+3/4, -1/2+$a$, -1/4) | (1/2, -1/2, 1/2) |
| | 11 | (-$a$+1/2, -1/4, -$a$-1/4) | (1/2, 1/2, 3/2) |
| | 12 | (-$a$+1/2, 1/4, -3/4$a$) | (1/2, -1/2, -1/2) |
| 9 | 11 | (1/4, -1/4+$a$, -$a$) | (0, -1, 1) |
| | 12 | (1/4, 1/4+$a$, -1/2+$a$) | (-1/2, -3/2, -1/2) |
| 10 | 11 | (-1/4, -$a$+1/4, -$a$) | (0, 1, 1) |
| | 12 | (-1/4, -$a$+3/4, -1/2+$a$) | (1/2, 1/2, -1/2) |

TABLE CXCI. The 12 bonds for the 3rd-NN exchange parameter $J_3$.

| $n$ | $n'$ | $R_{ij}$ | $R_l$ |
|---|---|---|---|
| 1 | 8 | (-1/2, 0, 0) | (0, 0, 0) |
| | | | (1, 0, 0) |
| 2 | 7 | (-1/2, 0, 0) | (0, 0, 0) |
| | | | (1, 0, 0) |
| 3 | 10 | (0, -1/2, 0) | (0, 0, 0) |
| | | | (0, 1, 0) |
| 4 | 9 | (0, -1/2, 0) | (0, 0, 0) |
| | | | (0, 1, 0) |
| 5 | 12 | (0, 0, -1/2) | (0, 0, 0) |
| | | | (0, 0, 1) |
| 6 | 11 | (0, 0, -1/2) | (0, 0, 0) |
| | | | (0, 0, 1) |

TABLE CXCII. The 12 bonds for the 4th-NN exchange parameter $J_4$.

| $n$ | $n'$ | $R_{ij}$ | $R_l$ |
|---|---|---|---|
| 1 | 2 | (1/2-2$a$, 0, -1/2) | (3/2, 1/2, 1/2) |
| | | | (3/2, -1/2, 1/2) |
| 3 | 4 | (-1/2, 1/2-2$a$, 0) | (1/2, 3/2, 1/2) |
| | | | (1/2, 3/2, -1/2) |
| 5 | 6 | (0, -1/2, 1/2-2$a$) | (1/2, 1/2, 3/2) |
| | | | (-1/2, 1/2, 3/2) |
| 7 | 8 | (-1/2+2$a$, 0, 1/2) | (-3/2, -1/2, -1/2) |
| | | | (-3/2, 1/2, -1/2) |
| 9 | 10 | (1/2, -1/2+2$a$, 0) | (-1/2, -3/2, -1/2) |
| | | | (-1/2, -3/2, 1/2) |
| 11 | 12 | (0, 1/2, -1/2+2$a$) | (-1/2, -1/2, -3/2) |
| | | | (1/2, -1/2, -3/2) |

TABLE CXCIII. The 12 bonds for the 5th-NN exchange parameter $J_5$.

| $n$ | $n'$ | $R_{ij}$ | $R_l$ |
|---|---|---|---|
| 1 | 7 | (-2$a$, 0, -1/2) | (2, 0, 0) |
| | | | (2, 0, 1) |
| 2 | 8 | (2$a$-1, 0, 1/2) | (-1, 0, 0) |
| | | | (-1, 0, -1) |
| 3 | 9 | (-1/2, -2$a$, 0) | (0, 2, 0) |
|  |  |  | (1, 2, 0) |
| 4 | 10 | (1/2, 2a-1, 0) | (0, -1, 0) |
|  |  |  | (-1, -1, 0) |
| 5 | 11 | (0, -1/2, -2a) | (0, 0, 2) |
|  |  |  | (0, 1, 2) |
| 6 | 12 | (0, 1/2, 2a-1) | (0, 0, -1) |
|  |  |  | (0, -1, -1) |

TABLE CXCIV. The 24 bonds for the 6th-NN exchange parameter $J_6$.

| $n$ | $n'$ | $R_{ij}$ | $R_l$ |
|---|---|---|---|
| 1 | 3 | (1/4-a, a, -1/4) | (1/2, -1/2, 1/2) |
|  | 4 | (-1/4-a, 1/2-a, -1/4) | (1, 0, 0) |
|  | 5 | (-a, 1/4, a-1/4) | (1/2, -1/2, -1/2) |
|  | 6 | (-a, -1/4, 1/4-a) | (1/2, 1/2, 1/2) |
| 2 | 3 | (a-1/4, a, 1/4) | (-1/2, -1/2, -1/2) |
|  | 4 | (a-3/4, 1/2-a, 1/4) | (0, 0, 0) |
|  | 5 | (a-1/2, 1/4, a+1/4) | (0, 0, -1) |
|  | 6 | (a-1/2, -1/4, 3/4-a) | (0, 0, 0) |
| 3 | 5 | (-1/4, 1/4-a, a) | (1/2, 1/2, -1/2) |
|  | 6 | (-1/4, -1/4-a, 1/2-a) | (0, 1, 0) |
| 4 | 5 | (1/4, a-1/4, a) | (-1/2, -1/2, -1/2) |
|  | 6 | (1/4, a-3/4, 1/2-a) | (0, 0, 0) |
| 7 | 9 | (-1/4+a, -a, 1/4) | (-1/2, 1/2, -1/2) |
|  | 10 | (1/4+a, -1/2+a, 1/4) | (-1, 0, 0) |
|  | 11 | (a, -1/4, -a+1/4) | (-1/2, 1/2, 1/2) |
|  | 12 | (a, 1/4, -1/4+a) | (-1/2, -1/2, -1/2) |
| 8 | 9 | (-a+1/4, -a, -1/4) | (1/2, 1/2, 1/2) |
|  | 10 | (-a+3/4, -1/2+a, -1/4) | (0, 0, 0) |
|  | 11 | (-a+1/2, -1/4, -a-1/4) | (0, 0, 1) |
|  | 12 | (-a+1/2, 1/4, -3/4a) | (0, 0, 0) |
| 9 | 11 | (1/4, -1/4+a, -a) | (-1/2, -1/2, 1/2) |
|  | 12 | (1/4, 1/4+a, -1/2+a) | (0, -1, 0) |
| 10 | 11 | (-1/4, -a+1/4, -a) | (1/2, 1/2, 1/2) |
|  | 12 | (-1/4, -a+3/4, -1/2+a) | (0, 0, 0) |

Supplementary Table CXCV. The 24 bonds for the 7th-NN exchange parameter $J_7$.

| $n$ | $n'$ | $R_{ij}$ | $R_l$ |
|---|---|---|---|
| 1 | 9 | (-1/4-a, -a, -1/4) | (3/2, 1/2, 1/2) |
|  | 10 | (1/4-a, a-1/2, -1/4) | (1, 0, 0) |
|  | 11 | (-a, -1/4, -a-1/4) | (1/2, 1/2, 3/2) |
|  | 12 | (-a, 1/4, a-3/4) | (1/2, -1/2, -1/2) |
| 2 | 9 | (a-3/4, -a, 1/4) | (-1/2, 1/2, -1/2) |
| | 10 | ($a$-1/4, $a$-1/2, 1/4) | (-1, 0, 0) |
| | 11 | ($a$-1/2, -1/4, 1/4-$a$) | (0, 0, 1) |
| | 12 | ($a$-1/2, 1/4, $a$-1/4) | (0, 0, -1) |
| 3 | 7 | (-1/4-$a$, -$a$, -1/4) | (3/2, 1/2, 1/2) |
| | 8 | ($a$-3/4, -$a$, 1/4) | (-1/2, 1/2, -1/2) |
| | 11 | (-1/4, -1/4-a, -$a$) | (1/2, 3/2, 1/2) |
| | 12 | (-1/4, 1/4-$a$, $a$-1/2) | (0, 1, 0) |
| 4 | 7 | (1/4-$a$, $a$-1/2, -1/4) | (1, 0, 0) |
| | 8 | ($a$-1/4, $a$-1/2, 1/4) | (-1, 0, 0) |
| | 11 | (1/4, $a$-3/4, -$a$) | (-1/2, -1/2, 1/2) |
| | 12 | (1/4, $a$-1/4, $a$-1/2) | (0, -1, 0) |
| 5 | 7 | (-$a$, -1/4, -$a$-1/4) | (1/2, 1/2, 3/2) |
| | 8 | ($a$-1/2, -1/4, 1/4-$a$) | (0, 0, 1) |
| | 9 | (-1/4, -1/4-a, -$a$) | (1/2, 3/2, 1/2) |
| | 10 | (1/4, $a$-3/4, -$a$) | (-1/2, -1/2, 1/2) |
| 6 | 7 | (-$a$, 1/4, $a$-3/4) | (1/2, -1/2, -1/2) |
| | 8 | ($a$-1/2, 1/4, $a$-1/4) | (0, 0, -1) |
| | 9 | (-1/4, 1/4-$a$, $a$-1/2) | (0, 1, 0) |
| | 10 | (1/4, $a$-1/4, $a$-1/2) | (0, -1, 0) |

Supplementary Table CXCVI. The 24 bonds for the 8th-NN exchange parameter $J_8$.

| $n$ | $n'$ | $R_{ij}$ | $R_l$ |
|---|---|---|---|
| 1 | 3 | (1/4-$a$, $a$, -1/4) | (1/2, -3/2, 1/2) |
| | 4 | (-1/4-$a$, 1/2-$a$, -1/4) | (1, 1, 0) |
| | 5 | (-$a$, 1/4, $a$-1/4) | (3/2, -1/2, -1/2) |
| | 6 | (-$a$, -1/4, 1/4-$a$) | (3/2, 1/2, 1/2) |
| 2 | 3 | ($a$-1/4, $a$, 1/4) | (-1/2, -3/2, -1/2) |
| | 4 | ($a$-3/4, 1/2-$a$, 1/4) | (0, 1, 0) |
| | 5 | ($a$-1/2, 1/4, $a$+1/4) | (-1, 0, -1) |
| | 6 | ($a$-1/2, -1/4, 3/4-$a$) | (-1, 0, 0) |
| 3 | 5 | (-1/4, 1/4-$a$, $a$) | (1/2, 1/2, -3/2) |
| | 6 | (-1/4, -1/4-$a$, 1/2-$a$) | (0, 1, 1) |
| 4 | 5 | (1/4, $a$-1/4, $a$) | (-1/2, -1/2, -3/2) |
| | 6 | (1/4, $a$-3/4, 1/2-$a$) | (0, 0, 1) |
| 7 | 9 | (-1/4+$a$, -$a$, 1/4) | (-1/2, 3/2, -1/2) |
| | 10 | (1/4+$a$, -1/2+$a$, 1/4) | (-1, -1, 0) |
| | 11 | ($a$, -1/4, -$a$+1/4) | (-3/2, 1/2, 1/2) |
| | 12 | ($a$, 1/4, -1/4+$a$) | (-3/2, -1/2, -1/2) |
| 8 | 9 | (-$a$+1/4, -$a$, -1/4) | (1/2, 3/2, 1/2) |
| | 10 | (-$a$+3/4, -1/2+$a$, -1/4) | (0, -1, 0) |
| | 11 | (-$a$+1/2, -1/4, -$a$-1/4) | (1, 0, 1) |
| | 12 | (-$a$+1/2, 1/4, -3/4$a$) | (1, 0, 0) |

| 9 | 11 | (1/4, -1/4+$a$, -$a$) | (-1/2, -1/2, 3/2) |
|---|---|---|---|
|  | 12 | (1/4, 1/4+$a$, -1/2+$a$) | (0, -1, -1) |
| 10 | 11 | (-1/4, -$a$+1/4, -$a$) | (1/2, 1/2, 3/2) |
|  | 12 | (-1/4, -$a$+3/4, -1/2+$a$) | (0, 0, -1) |

Supplementary Table CXCVII. The 24 bonds for the 9th-NN exchange parameter $J_9$.

| $n$ | $n'$ | $R_{ij}$ | $R_l$ |
|---|---|---|---|
| 1 | 9 | (-1/4-$a$, -$a$, -1/4) | (3/2, 3/2, 1/2) |
|  | 10 | (1/4-$a$, $a$-1/2, -1/4) | (1, -1, 0) |
|  | 11 | (-$a$, -1/4, -$a$-1/4) | (3/2, 1/2, 3/2) |
|  | 12 | (-$a$, 1/4, $a$-3/4) | (3/2, -1/2, -1/2) |
| 2 | 9 | ($a$-3/4, -$a$, 1/4) | (-1/2, 3/2, -1/2) |
|  | 10 | ($a$-1/4, $a$-1/2, 1/4) | (-1, -1, 0) |
|  | 11 | ($a$-1/2, -1/4, 1/4-$a$) | (-1, 0, 1) |
|  | 12 | ($a$-1/2, 1/4, $a$-1/4) | (-1, 0, -1) |
| 3 | 7 | (-1/4-$a$, -$a$, -1/4) | (3/2, 3/2, 1/2) |
|  | 8 | ($a$-3/4, -$a$, 1/4) | (-1/2, 3/2, -1/2) |
|  | 11 | (-1/4, -1/4-a, -$a$) | (1/2, 3/2, 3/2) |
|  | 12 | (-1/4, 1/4-$a$, $a$-1/2) | (0, 1, -1) |
| 4 | 7 | (1/4-$a$, $a$-1/2, -1/4) | (1, -1, 0) |
|  | 8 | ($a$-1/4, $a$-1/2, 1/4) | (-1, -1, 0) |
|  | 11 | (1/4, $a$-3/4, -$a$) | (-1/2, -1/2, 3/2) |
|  | 12 | (1/4, $a$-1/4, $a$-1/2) | (0, -1, -1) |
| 5 | 7 | (-$a$, -1/4, -$a$-1/4) | (3/2, 1/2, 3/2) |
|  | 8 | ($a$-1/2, -1/4, 1/4-$a$) | (-1, 0, 1) |
|  | 9 | (-1/4, -1/4-a, -$a$) | (1/2, 3/2, 3/2) |
|  | 10 | (1/4, $a$-3/4, -$a$) | (-1/2, -1/2, 3/2) |
| 6 | 7 | (-$a$, 1/4, $a$-3/4) | (3/2, -1/2, -1/2) |
|  | 8 | ($a$-1/2, 1/4, $a$-1/4) | (-1, 0, -1) |
|  | 9 | (-1/4, 1/4-$a$, $a$-1/2) | (0, 1, -1) |
|  | 10 | (1/4, $a$-1/4, $a$-1/2) | (0, -1, -1) |

By using Eqs. (65)~(69), the magnon dispersion of $Cu_3TeO_6$ can be obtained, as shown in Supplementary Table LV. We also provide the magnon band representations of the corresponding SSG and MSG. We find that the SSG effectively describes the band degeneracy.

Supplementary Table CXCVIII. Band representation of the MSG $R\bar{3}'$ (148.19) and the collinear SSG $I^{\bar{1}}a^{\bar{1}}\bar{3}^{\infty m}1$ (199.206.1.1) induced from the local representation of the Wyckoff position 24$d$.

|  | Magnetic space group<br>$R\bar{3}'$ (148.19) | Collinear spin space group |
|---|---|---|

| | | $I^{\bar{1}}a^{\bar{1}}\bar{3}^{\infty m}1$ (199.206.1.1) |
|---|---|---|
| **Band rep** | $\boldsymbol{2A \uparrow G\ (12)}$ | $\boldsymbol{A^S \uparrow G\ (12)}$ |
| $\Gamma$ (0, 0, 0) | $4\Gamma_1(1) + 4\Gamma_2\Gamma_3(2)$ | $\Gamma_1^S(2) + \Gamma_2^S\Gamma_3^S(4)$+$\Gamma_4^S(6)$ |
| $H$ (0, 1, 0) | $4\mathrm{H}_1(1) + 4\mathrm{H}_2\mathrm{H}_3(2)$ | $2\mathrm{H}_4^S(6)$ |
| $N$ (1/2, 1/2, 0) | $12\mathrm{N}_1(1)$ | $3\mathrm{N}_1^S(2) + 3\mathrm{N}_2^S(2)$ |
| $P$ (1/2, 1/2, 1/2) | $4\mathrm{P}_1(1) + 4\mathrm{P}_2\mathrm{P}_3(2)$ | $\mathrm{P}_1^S(4) + \mathrm{P}_2^S(4) + \mathrm{P}_3^S(4)$ |
| $\Lambda$ ($u$, $u$, $u$) | $4\Lambda_1(1) + 4\Lambda_2\Lambda_3(2)$ | $2\Lambda_1^S(2) + 2\Lambda_2^S(2) + 2\Lambda_3^S(2)$ |
| $D$ (1/2, 1/2, $w$) | $12\mathrm{D}_1(1)$ | $3\mathrm{D}_1^S(2) + 3\mathrm{D}_2^S(2)$ |
| $\Delta$ (0, $v$, 0) | $4\Delta_1(1) + 4\Delta_2\Delta_3(2)$ | $4\Delta_1^S(2) + 2\Delta_2^S(2)$ |
| A (u, v, 0) | $12\mathrm{A}_1(1)$ | $6\mathrm{A}_1^S(2)$ |
| Σ (u, u, 0) | $12\Sigma_1(1)$ | $6\Sigma_1^S(2)$ |
| GP (u, v, w) | $12\mathrm{GP}_1(1)$ | $6\mathrm{GP}_1^S(2)$ |

The magnon dispersion of $Cu_3TeO_6$ with single-ion anisotropy included is shown in Supplementary Fig. 9, where the magnon band is overall shifted. Besides, the linear dispersion around Γ point is softed. However, the single-ion anisotropy does not affect the symmetry of the magnon Hamiltonian, the band representations and the nodal topology.

Furthermore, we noticed that the role of DM interaction in the magnon dispersion in a collinear AFM $Cu_3TeO_6$ has been discussed recently[104]. According to Fig. S1 in that article, it can be observed that the magnon dispersion is almost unchanged when considering $D = 0.1J$. Even when $D = 0.25J$, the energy scale for band degeneracy lifting is still very small, at the magnitude of 0.1meV. These results fully support our SOC benchmark results that SSG serves as an ideal framework for describing magnon band nodes in most 3*d*, 4*d* and 4*f* collinear magnets, where the SOC-induced interactions at least an order of magnitude smaller than the Heisenberg interaction.

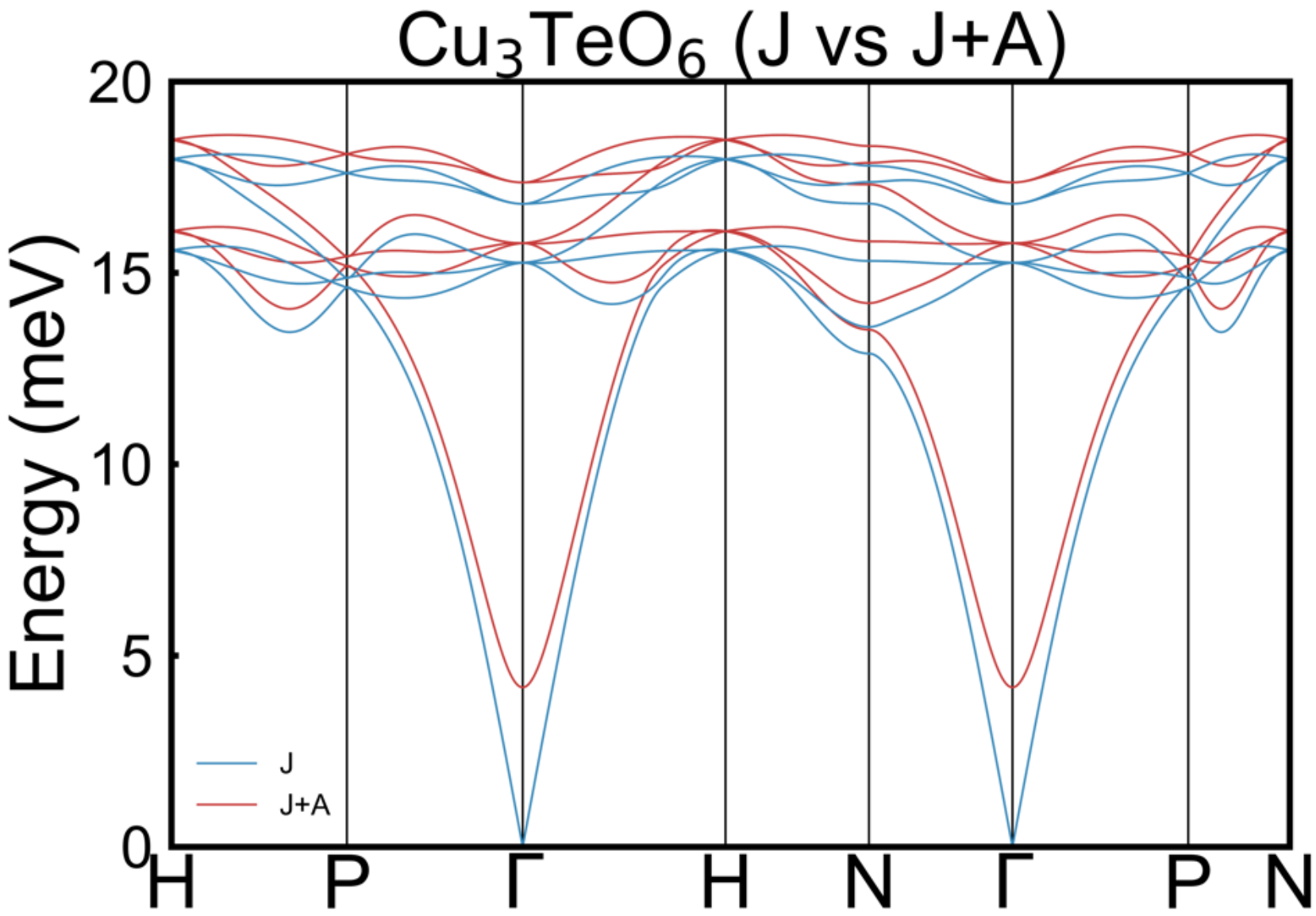

Supplementary Fig. 9. Magnon band structure of $Cu_3TeO_6$ when considering the absence and presence of single-ion anisotropy A, depicting with blue and red lines, respectively.

### S7.3.2.1. Topological charge (PT)

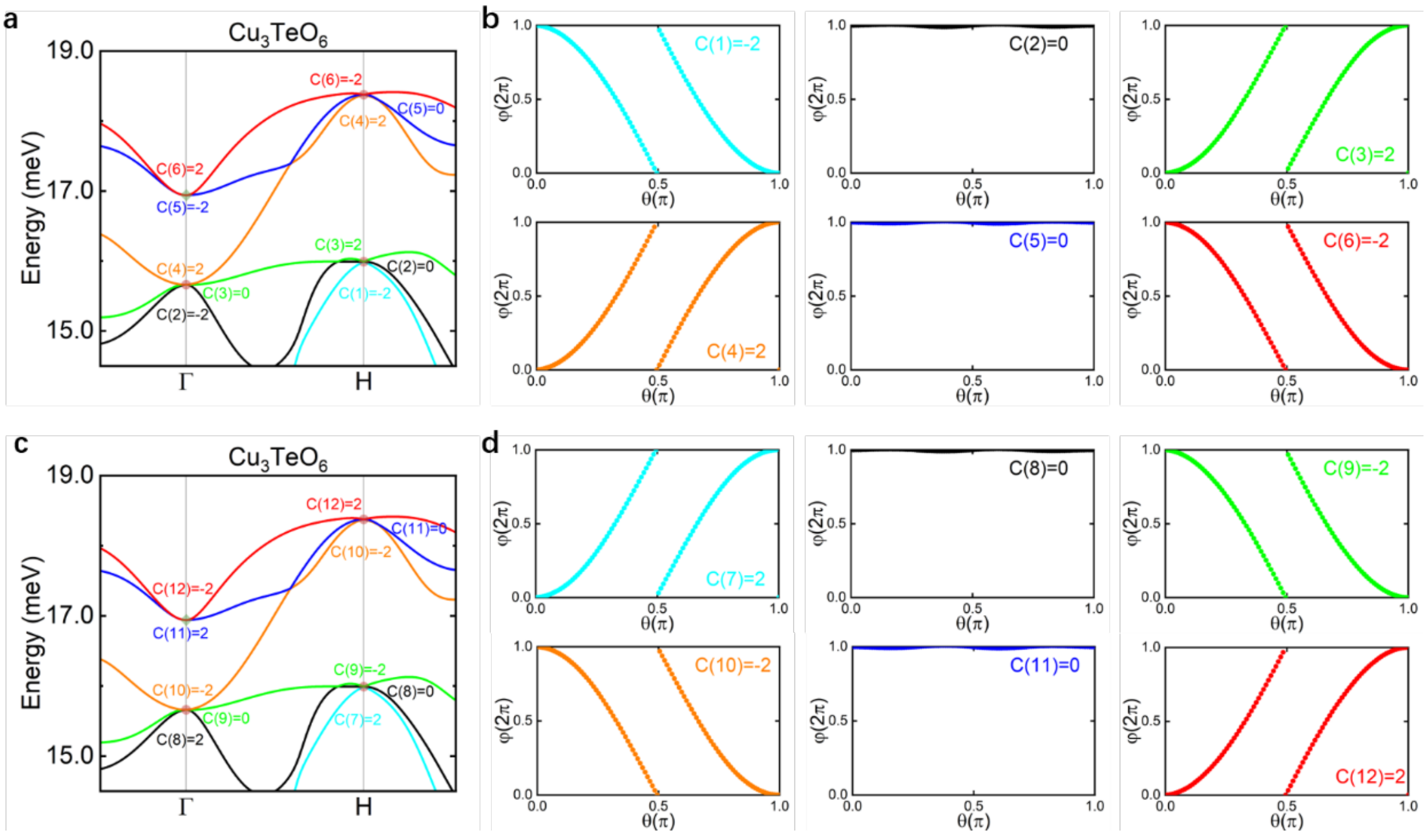

Supplementary Fig. 10. Magnon band structure of collinear AFM $Cu_3TeO_6$ for the **a**. S=1 (up) channel and the **c**. S=-1 (down) channel. Here S denote the spin angular momentum of magnon. Wilson loop of **b**. magnon

bands 1-6 in spin up channel and **d**. magnon bands 7-12 in spin down channel at H point.

As shown in Supplementary Fig. 10, we can conclude that, in type-II collinear SSGs, PT-connected two opposite-spin channel hosts opposite topological charges as stated in the manuscript. This is because that PT reverses the topological charges of nodal points of two degenerate magnon modes.

### S7.3.3. FeS (octuple magnon, chiral magnon)

Troilite FeS crystallizes in hexagonal system of space group $P\bar{6}2c$ (190) with lattice constants a = 5.896 Å and c = 11.421 Å. The magnetic ground state of FeS is a collinear AFM with Néel order along [001] direction, which belongs to the MSG $P\bar{6}'2c'$ (190.230). The SSG of collinear AFM FeS is $P^{\bar{1}}\bar{6}^{\bar{1}}2^{1}c^{\infty m}1$ (159.190.1.1) and the magnetic ions Fe locate at the spin Wyckoff position 12$i$ ($a$, $b$, $c$ | 0, 0, z). The lattice basis for constructing magnon Hamiltonian is $\mathrm{a}_1 = (\mathrm{a}, 0, 0), \mathrm{a}_2 = \left(-\frac{1}{2}\mathrm{a}, \frac{\sqrt{3}}{2}\mathrm{a}, 0\right), \mathrm{a}_3 = (0, 0, \mathrm{c})$. The exchange parameters are given in Supplementary Table CLXXXI. The 12 equivalent magnetic ions in magnetic primitive cell are given in Supplementary Table CXCIX, and the detailed information of bonds are shown in Supplementary Tables CC ~ CCXII. Finally, a comparison of describing magnon dispersion of FeS using band representations of collinear SSGs and MSGs are shown in Supplementary Table CCXIII.

Supplementary Table CXCIX. The coordinates of the 12 Fe ions in the conventional cell basis vectors ($a$ = 0.0725, $b$ = 0.3942, $c$ = 0.1210). Here "up" and "down" represent the [001] and [00$\bar{1}$] direction, respectively.

| $n$ | $r_n$ | magnetic moment |
|---|---|---|
| 1 | ($a$, $b$, $c$) | up |
| 2 | (-$b$, $a$-$b$, $c$) | up |
| 3 | ($b$-$a$, -$a$, $c$) | up |
| 4 | ($a$, $b$, 1/2-$c$) | down |
| 5 | (-$b$, $a$-$b$, 1/2-$c$) | down |
| 6 | ($b$-$a$, -$a$, 1/2-$c$) | down |
| 7 | ($b$, $a$, -$c$) | down |
| 8 | ($a$-$b$, -$b$, -$c$) | down |
| 9 | (-$a$, $b$-$a$, -$c$) | down |
| 10 | ($b$, $a$, 1/2+$c$) | up |
| 11 | ($a$-$b$, -$b$, 1/2+$c$) | up |
| 12 | (-$a$, $b$-$a$, 1/2+$c$) | up |

Supplementary Table CC. The 12 bonds for the 1st-NN exchange parameter $J_1$.

| $n$ | $n'$ | $R_{ij}$ | $R_l$ |
|---|---|---|---|
| 1 | 2 | (-$a$-$b$, $a$-2$b$, 0) | (1, 1, 0) |
|  | 3 | (-2$a$+$b$, -$a$-$b$, 0) | (0, 1, 0) |
| 2 | 3 | (2$b$-$a$, $b$-2$a$, 0) | (-1, 0, 0) |
| 4 | 5 | (-$a$-$b$, $a$-2$b$, 0) | (1, 1, 0) |
|  | 6 | (-2$a$+$b$, -$a$-$b$, 0) | (0, 1, 0) |
| 5 | 6 | (2$b$-$a$, $b$-2$a$, 0) | (-1, 0, 0) |

| 7 | 8 | ($a$-2$b$, -$a$-$b$, 0) | (1, 1, 0) |
|---|---|---|---|
| | 9 | (-$a$-$b$, $b$-2$a$, 0) | (1, 0, 0) |
| 8 | 9 | ($b$-2$a$, 2$b$-$a$, 0) | (0, -1, 0) |
| 10 | 11 | ($a$-2$b$, -$a$-$b$, 0) | (1, 1, 0) |
| | 12 | (-$a$-$b$, $b$-2$a$, 0) | (1, 0, 0) |
| 11 | 12 | ($b$-2$a$, 2$b$-$a$, 0) | (0, -1, 0) |

Supplementary Table CCI. The 6 bonds for the 2nd-NN exchange parameter $J_2$.

| $n$ | $n'$ | $R_{ij}$ | $R_l$ |
|---|---|---|---|
| 1 | 9 | (-2$a$, -$a$, -2$c$) | (0, 0, 0) |
| 2 | 8 | ($a$, -$a$, -2$c$) | (0, 0, 0) |
| 3 | 7 | ($a$, 2$a$, -2$c$) | (0, 0, 0) |
| 4 | 12 | (-2$a$, -$a$, 2$c$) | (0, 0, 0) |
| 5 | 11 | ($a$, -$a$, 2$c$) | (0, 0, 0) |
| 6 | 10 | ($a$, 2$a$, 2$c$) | (0, 0, 0) |

Supplementary Table CCII. The 6 bonds for the 3rd-NN exchange parameter $J_3$.

| $n$ | $n'$ | $R_{ij}$ | $R_l$ |
|---|---|---|---|
| 1 | 4 | (0, 0, 1/2-2$c$) | (0, 0, 0) |
| 2 | 5 | (0, 0, 1/2-2$c$) | (0, 0, 0) |
| 3 | 6 | (0, 0, 1/2-2$c$) | (0, 0, 0) |
| 7 | 10 | (0, 0, 1/2+2$c$) | (0, 0, -1) |
| 8 | 11 | (0, 0, 1/2+2$c$) | (0, 0, -1) |
| 9 | 12 | (0, 0, 1/2+2$c$) | (0, 0, -1) |

Supplementary Table CCIII. The 12 bonds for the 4th-NN exchange parameter $J_4$.

| $n$ | $n'$ | $R_{ij}$ | $R_l$ |
|---|---|---|---|
| 1 | 2 | (-$a$-$b$, $a$-2$b$, 0) | (0, 0, 0) |
| | 3 | (-2$a$+$b$, -$a$-$b$, 0) | (0, 0, 0) |
| 2 | 3 | (2$b$-$a$, $b$-2$a$, 0) | (0, 0, 0) |
| 4 | 5 | (-$a$-$b$, $a$-2$b$, 0) | (0, 0, 0) |
| | 6 | (-2$a$+$b$, -$a$-$b$, 0) | (0, 0, 0) |
| 5 | 6 | (2$b$-$a$, $b$-2$a$, 0) | (0, 0, 0) |
| 7 | 8 | ($a$-2$b$, -$a$-$b$, 0) | (0, 0, 0) |
| | 9 | (-$a$-$b$, $b$-2$a$, 0) | (0, 0, 0) |
| 8 | 9 | ($b$-2$a$, 2$b$-$a$, 0) | (0, 0, 0) |
| 10 | 11 | ($a$-2$b$, -$a$-$b$, 0) | (0, 0, 0) |
| | 12 | (-$a$-$b$, $b$-2$a$, 0) | (0, 0, 0) |
| 11 | 12 | ($b$-2$a$, 2$b$-$a$, 0) | (0, 0, 0) |

Supplementary Table CCIV. The 12 bonds for the 5th-NN exchange parameter $J_5$.

| $n$ | $n'$ | $R_{ij}$ | $R_l$ |
|---|---|---|---|
| 1 | 2 | (-*a*-*b*, *a*-2*b*, 0) | (0, 1, 0) |
| | 3 | (-2*a*+*b*, -*a*-*b*, 0) | (-1, 0, 0) |
| 2 | 3 | (2*b*-*a*, *b*-2*a*, 0) | (-1, -1, 0) |
| 4 | 5 | (-*a*-*b*, *a*-2*b*, 0) | (0, 1, 0) |
| | 6 | (-2*a*+*b*, -*a*-*b*, 0) | (-1, 0, 0) |
| 5 | 6 | (2*b*-*a*, *b*-2*a*, 0) | (-1, -1, 0) |
| 7 | 8 | (*a*-2*b*, -*a*-*b*, 0) | (1, 0, 0) |
| | 9 | (-*a*-*b*, *b*-2*a*, 0) | (0, -1, 0) |
| 8 | 9 | (*b*-2*a*, 2*b*-*a*, 0) | (-1, -1, 0) |
| 10 | 11 | (*a*-2*b*, -*a*-*b*, 0) | (1, 0, 0) |
| | 12 | (-*a*-*b*, *b*-2*a*, 0) | (0, -1, 0) |
| 11 | 12 | (*b*-2*a*, 2*b*-*a*, 0) | (-1, -1, 0) |

Supplementary Table CCV. The 12 bonds for the 6th-NN exchange parameter $J_6$.

| $n$ | $n'$ | $R_{ij}$ | $R_l$ |
|---|---|---|---|
| 1 | 5 | (-*a*-*b*, *a*-2*b*, 1/2-2*c*) | (1, 1, 0) |
| | 6 | (*b*-2*a*, -*a*-*b*, 1/2-2*c*) | (0, 1, 0) |
| 2 | 4 | (*a*+*b*, 2*b*-*a*, 1/2-2*c*) | (-1, -1, 0) |
| | 6 | (2*b*-*a*, *b*-2*a*, 1/2-2*c*) | (-1, 0, 0) |
| 3 | 4 | (2*a*-*b*, *a*+*b*, 1/2-2*c*) | (0, -1, 0) |
| | 5 | (*a*-2*b*, 2*a*-*b*, 1/2-2*c*) | (1, 0, 0) |
| 7 | 11 | (*a*-2*b*, -*a*-*b*, 1/2+2*c*) | (1, 1, -1) |
| | 12 | (-*a*-*b*, 2*a*-*b*, 1/2+2*c*) | (1, 0, -1) |
| 8 | 10 | (2*b*-*a*, *a*+*b*, 1/2+2*c*) | (-1, -1, -1) |
| | 12 | (*b*-2*a*, 2*b*-*a*, 1/2+2*c*) | (0, -1, -1) |
| 9 | 10 | (*a*+*b*, 2*a*-*b*, 1/2+2*c*) | (-1, 0, -1) |
| | 11 | (2*a*-*b*, *a*-2*b*, 1/2+2*c*) | (0, 1, -1) |

Supplementary Table CCVI. The 12 bonds for the 7th-NN exchange parameter $J_7$.

| $n$ | $n'$ | $R_{ij}$ | $R_l$ |
|---|---|---|---|
| 1 | 8 | (-*b*, -2*b*, -2*c*) | (0, 1, 0) |
| | | | (1, 1, 0) |
| 2 | 7 | (2*b*, *b*, -2*c*) | (-1, 0, 0) |
| | | | (-1, -1, 0) |
| 3 | 9 | (-*b*, *b*, -2*c*) | (1, 0, 0) |
| | | | (0, -1, 0) |
| 4 | 11 | (-*b*, -2*b*, 2*c*) | (0, 1, 0) |
| | | | (1, 1, 0) |
| 5 | 10 | (2*b*, *b*, 2*c*) | (-1, 0, 0) |

| | | | (-1, -1, 0) |
|---|---|---|---|
| 6 | 12 | (-*b*, *b*, 2*c*) | (1, 0, 0) |
| | | | (0, -1, 0) |

Supplementary Table CCVII. The 6 bonds for the 8th-NN exchange parameter $J_8$.

| $n$ | $n'$ | $R_{ij}$ | $R_l$ |
|---|---|---|---|
| 1 | 7 | (*b*-*a*, *a*-*b*, -2*c*) | (0, 0, 0) |
| 2 | 9 | (*b*-*a*, 2*b*-2*a*, -2*c*) | (0, 0, 0) |
| 3 | 8 | (2*a*-2*b*, *a*-*b*, -2*c*) | (0, 0, 0) |
| 4 | 10 | (*b*-*a*, *a*-*b*, 2*c*) | (0, 0, 0) |
| 5 | 12 | (*b*-*a*, 2*b*-2*a*, 2*c*) | (0, 0, 0) |
| 6 | 11 | (2*a*-2*b*, *a*-*b*, 2*c*) | (0, 0, 0) |

Supplementary Table CCVIII. The 12 bonds for the 9th-NN exchange parameter $J_9$

| $n$ | $n'$ | $R_{ij}$ | $R_l$ |
|---|---|---|---|
| 1 | 7 | (*b*-*a*, *a*-*b*, -2*c*) | (-1, 0, 0) |
| | | | (0, 1, 0) |
| 2 | 9 | (*b*-*a*, 2*b*-2*a*, -2*c*) | (0, -1, 0) |
| | | | (-1, -1, 0) |
| 3 | 8 | (2*a*-2*b*, *a*-*b*, -2*c*) | (1, 0, 0) |
| | | | (1, 1, 0) |
| 4 | 10 | (*b*-*a*, *a*-*b*, 2*c*) | (-1, 0, 0) |
| | | | (0, 1, 0) |
| 5 | 12 | (*b*-*a*, 2*b*-2*a*, 2*c*) | (0, -1, 0) |
| | | | (-1, -1, 0) |
| 6 | 11 | (2*a*-2*b*, *a*-*b*, 2*c*) | (1, 0, 0) |
| | | | (1, 1, 0) |

Supplementary Table CCIX. The 12 bonds for the 10th-NN exchange parameter $J_{10}$

| $n$ | $n'$ | $R_{ij}$ | $R_l$ |
|---|---|---|---|
| 1 | 5 | (-*a*-*b*, *a*-2*b*, 1/2-2*c*) | (0, 0, 0) |
| | 6 | (*b*-2*a*, -*a*-*b*, 1/2-2*c*) | (0, 0, 0) |
| 2 | 4 | (*a*+*b*, 2*b*-*a*, 1/2-2*c*) | (0, 0, 0) |
| | 6 | (2*b*-*a*, *b*-2*a*, 1/2-2*c*) | (0, 0, 0) |
| 3 | 4 | (2*a*-*b*, *a*+*b*, 1/2-2*c*) | (0, 0, 0) |
| | 5 | (*a*-2*b*, 2*a*-*b*, 1/2-2*c*) | (0, 0, 0) |
| 7 | 11 | (*a*-2*b*, -*a*-*b*, 1/2+2*c*) | (0, 0, -1) |
| | 12 | (-*a*-*b*, 2*a*-*b*, 1/2+2*c*) | (0, 0, -1) |
| 8 | 10 | (2*b*-*a*, *a*+*b*, 1/2+2*c*) | (0, 0, -1) |
| | 12 | (*b*-2*a*, 2*b*-*a*, 1/2+2*c*) | (0, 0, -1) |
| 9 | 10 | (*a*+*b*, 2*a*-*b*, 1/2+2*c*) | (0, 0, -1) |
| | 11 | (2*a*-*b*, *a*-2*b*, 1/2+2*c*) | (0, 0, -1) |

Supplementary Table CCX. The 12 bonds for the 11th-NN exchange parameter $J_{11}$

| $n$ | $n'$ | $R_{ij}$ | $R_l$ |
|---|---|---|---|
| 1 | 5 | (-*a*-*b*, *a*-2*b*, 1/2-2*c*) | (0, 1, 0) |
| | 6 | (*b*-2*a*, -*a*-*b*, 1/2-2*c*) | (-1, 0, 0) |
| 2 | 4 | (*a*+*b*, 2*b*-*a*, 1/2-2*c*) | (0, -1, 0) |
| | 6 | (2*b*-*a*, *b*-2*a*, 1/2-2*c*) | (-1, -1, 0) |
| 3 | 4 | (2*a*-*b*, *a*+*b*, 1/2-2*c*) | (1, 0, 0) |
| | 5 | (*a*-2*b*, 2*a*-*b*, 1/2-2*c*) | (1, 1, 0) |
| 7 | 11 | (*a*-2*b*, -*a*-*b*, 1/2+2*c*) | (1, 0, -1) |
| | 12 | (-*a*-*b*, 2*a*-*b*, 1/2+2*c*) | (0, -1, -1) |
| 8 | 10 | (2*b*-*a*, *a*+*b*, 1/2+2*c*) | (-1, 0, -1) |
| | 12 | (*b*-2*a*, 2*b*-*a*, 1/2+2*c*) | (-1, -1, -1) |
| 9 | 10 | (*a*+*b*, 2*a*-*b*, 1/2+2*c*) | (0, 1, -1) |
| | 11 | (2*a*-*b*, *a*-2*b*, 1/2+2*c*) | (1, 1, -1) |

Supplementary Table CCXI. The 6 bonds for the 12th-NN exchange parameter $J_{12}$

| $n$ | $n'$ | $R_{ij}$ | $R_l$ |
|---|---|---|---|
| 1 | 8 | (-*b*, -2*b*, -2*c*) | (0, 0, 0) |
| 2 | 7 | (2*b*, *b*, -2*c*) | (0, 0, 0) |
| 3 | 9 | (-*b*, *b*, -2*c*) | (0, 0, 0) |
| 4 | 11 | (-*b*, -2*b*, 2*c*) | (0, 0, 0) |
| 5 | 10 | (2*b*, *b*, 2*c*) | (0, 0, 0) |
| 6 | 12 | (-*b*, *b*, 2*c*) | (0, 0, 0) |

Supplementary Table CCXII. The 12 bonds for the 13th-NN exchange parameter $J_{13}$

| $n$ | $n'$ | $R_{ij}$ | $R_l$ |
|---|---|---|---|
| 1 | 12 | (-2*a*, -*a*, 1/2) | (0, 0, 0) |
| | | | (0, 0, -1) |
| 2 | 11 | (*a*, -*a*, 1/2) | (0, 0, 0) |
| | | | (0, 0, -1) |
| 3 | 10 | (*a*, 2*a*, 1/2) | (0, 0, 0) |
| | | | (0, 0, -1) |
| 4 | 9 | (-2*a*, -*a*, -1/2) | (0, 0, 0) |
| | | | (0, 0, 1) |
| 5 | 8 | (*a*, -*a*, -1/2) | (0, 0, 0) |
| | | | (0, 0, 1) |
| 6 | 7 | (*a*, 2*a*, -1/2) | (0, 0, 0) |
| | | | (0, 0, 1) |

By using Eqs. (65)~(69), the magnon dispersion of FeS can be obtained, as shown in Supplementary

Table LVI. We also provide the magnon band representations of the corresponding SSG and MSG. We find that the SSG effectively describes the band degeneracy.

Supplementary Table CCXIII. Band representation of the MSG $P\bar{6}'2c'$ (190.230) and the SSG $P^{\bar{1}}\bar{6}^{\bar{1}}2^{1}c^{\infty m}1$ (159.190.1.1) and the local representation of the Wyckoff position 12$i$.

| | Magnetic space group<br>$P\bar{6}'2c'$ (190.230) | Spin space group<br>$P^{\bar{1}}\bar{6}^{\bar{1}}2^{1}c^{\infty m}1$ (159.190.1.1) |
|---|---|---|
| **Band rep** | $\boldsymbol{A\uparrow G\ (12)}$ | $\boldsymbol{A^S\uparrow G\ (12)}$ |
| Γ (0, 0, 0) | $2\Gamma_1(1)+2\Gamma_2(1)+4\Gamma_3(2)$ | $\Gamma_1^S(2)+\Gamma_2^S(2)+2\Gamma_3^S(4)$ |
| M (1/2, 0, 0) | $6\mathrm{M}_1(1)+6\mathrm{M}_2(1)$ | $3\mathrm{M}_1^S(2)+3\mathrm{M}_2^S(2)$ |
| K (1/3, 1/3, 0) | $2\mathrm{K}_1(1)+2\mathrm{K}_2(1)+4\mathrm{K}_3(2)$ | $\mathrm{K}_1^S(2)+\mathrm{K}_2^S(2)+2\mathrm{K}_3^S(4)$ |
| H (1/3, 1/3, 1/2) | $2\mathrm{H}_1(1)+2\mathrm{H}_2(1)+4\mathrm{H}_3(2)$ | $\mathrm{H}_1^S(2)+\mathrm{H}_2^S(2)+2\mathrm{H}_3^S(4)$ |
| L (1/2, 0, 1/2) | $6L_1L_2(2)$ | $3L_1^SL_2^S(4)$ |
| A (0, 0, 1/2) | $2A_1A_2(2)+2A_3A_3(4)$ | $A_1^SA_2^S(4)+A_3^SA_3^S(8)$ |
| Λ (u, u, 0) | $6\Lambda_1(1)+6\Lambda_2(1)$ | $3\Lambda_1^S(2)+3\Lambda_2^S(2)$ |
| Σ (u, 0, 0) | $12\Sigma_1(1)$ | $6\Sigma_1^S(2)$ |
| B (u, v, 0) | $12B_1(1)$ | $6\mathrm{B}_1^S(2)$ |
| E (u, v, 1/2) | $12E_1(1)$ | $6E_1^S(2)$ |
| R (u, 0, 1/2) | $6R_1R_1(2)$ | $3R_1^SR_1^S(4)$ |
| Q (u, u, 1/2) | $6Q_1(1)+6Q_2(1)$ | $3Q_1^S(2)+3Q_2^S(2)$ |
| Δ (0, 0, w) | $4\Delta_1(1)+4\Delta_2\Delta_3(2)$ | $2\Delta_1^S(1)+2\Delta_2^S(1)+4\Delta_3^S(2)$ |
| P (1/3, 1/3, w) | $4P_1(1)+4P_2(1)+4P_3(1)$ | $2\mathrm{P}_1^S(1)+2\mathrm{P}_2^S(1)+4\mathrm{P}_3^S(2)$ |
| GP (u, v, w) | $12GP_1(1)$ | $12GP_1^S(1)$ |

### S7.3.4. $Fe_{0.35}NbS_2$ (C-4 octuple magnon, quadruple nodal plane magnon)

Intercalated transition-metal dichalcogenide $Fe_{0.35}NbS_2$ crystallizes in hexagonal system of space group $P6_322$ (182) with lattice constants a = 11.519 Å, b = 9.976 Å, c = 12.191 Å. The magnetic ground state of $Fe_{0.35}NbS_2$ is a collinear zigzag AFM with Néel order along [001] direction, which belongs to the MSG $P_c2_12_12$ (18.21). The SSG of collinear AFM $Fe_{0.35}NbS_2$ is $P_a{}^12_1{}^12_1{}^12_1{}^{\infty m}1$ (19.18.2.1) and the magnetic ions Fe locate at the spin Wyckoff position 8$c$ ($a$, $b$, $c$ | 0, 0, z). The lattice basis for constructing magnon Hamiltonian is $\mathrm{a}_1 = (\mathrm{a}, 0, 0), \mathrm{a}_2 = (0, \mathrm{b}, 0), \mathrm{a}_3 = (0, 0, \mathrm{c})$. In DFT calculations, we use AFM zigzag $Fe_{1/3}NbS_2$ phase to model the $Fe_{0.35}NbS_2$. The exchange parameters are given in Supplementary Table CLXXXI. The 8 equivalent magnetic ions in magnetic primitive cell are given in Supplementary Table CCXIV, and the detailed information of bonds are shown in Supplementary Tables CCXV~ CCXXVIII. Finally, a comparison of describing magnon dispersion of $Fe_{0.35}NbS_2$ using band representations of collinear SSGs and MSGs are shown in Supplementary Table CCXXIX.

Supplementary Table CCXIV. The coordinates of the 8 Fe ions in the non-standard conventional cell basis vectors. Here "up" and "down" represent the [001] and [00$\bar{1}$] direction, respectively.

| $n$ | $r_n$ | magnetic moment |
|---|---|---|
| 1 | (0, 2/3, 1/4) | up |
| 2 | (1/2, 2/3, 1/4) | down |
| 3 | (0, 1/3, 3/4) | down |
| 4 | (1/2, 1/3, 3/4) | up |
| 5 | (1/4, 1/6, 1/4) | up |
| 6 | (3/4, 1/6, 1/4) | down |
| 7 | (1/4, 5/6, 3/4) | down |
| 8 | (3/4, 5/6, 3/4) | up |

Supplementary Table CCXV. The 8 bonds for the 1st-NN exchange parameter $J_{1a}$.

| $n$ | $n'$ | $R_{ij}$ | $R_l$ |
|---|---|---|---|
| 1 | 2 | (1/2, 0, 0) | (-1, 0, 0) |
|  |  |  | (0, 0, 0) |
| 3 | 4 | (1/2, 0, 0) | (-1, 0, 0) |
|  |  |  | (0, 0, 0) |
| 5 | 6 | (1/2, 0, 0) | (-1, 0, 0) |
|  |  |  | (0, 0, 0) |
| 7 | 8 | (1/2, 0, 0) | (-1, 0, 0) |
|  |  |  | (0, 0, 0) |

Supplementary Table CCXVI. The 8 bonds for the 1st-NN exchange parameter $J_{1b}$.

| $n$ | $n'$ | $R_{ij}$ | $R_l$ |
|---|---|---|---|
| 1 | 5 | (1/4, -1/2, 0) | (0, 0, 0) |
| | | | (0, 1, 0) |
| 2 | 6 | (1/4, -1/2, 0) | (0, 0, 0) |
| | | | (0, 1, 0) |
| 3 | 7 | (1/4, 1/2, 0) | (0, 0, 0) |
| | | | (0, -1, 0) |
| 4 | 8 | (1/4, 1/2, 0) | (0, 0, 0) |
| | | | (0, -1, 0) |

Supplementary Table CCXVII. The 8 bonds for the 1st-NN exchange parameter $J_{1c}$.

| $n$ | $n'$ | $R_{ij}$ | $R_l$ |
|---|---|---|---|
| 1 | 6 | (3/4, -1/2, 0) | (-1, 0, 0) |
| | | | (-1, 1, 0) |
| 2 | 5 | (-1/4, -1/2, 0) | (0, 0, 0) |
| | | | (0, 1, 0) |
| 3 | 8 | (3/4, 1/2, 0) | (-1, 0, 0) |
| | | | (-1, -1, 0) |
| 4 | 7 | (-1/4, 1/2, 0) | (0, 0, 0) |
| | | | (0, -1, 0) |

Supplementary Table CCXVIII. The 4 bonds for the 2nd-NN exchange parameter $J_{2a}$.

| $n$ | $n'$ | $R_{ij}$ | $R_l$ |
|---|---|---|---|
| 1 | 3 | (0, -1/3, 1/2) | (0, 0, -1) |
| 2 | 4 | (0, -1/3, 1/2) | (0, 0, -1) |
| 5 | 7 | (0, 2/3, 1/2) | (0, -1, 0) |
| 6 | 8 | (0, 2/3, 1/2) | (0, -1, 0) |

Supplementary Table CCXIX. The 4 bonds for the 2nd-NN exchange parameter $J_{2b}$.

| $n$ | $n'$ | $R_{ij}$ | $R_l$ |
|---|---|---|---|
| 1 | 3 | (0, -1/3, 1/2) | (0, 0, 0) |
| 2 | 4 | (0, -1/3, 1/2) | (0, 0, 0) |
| 5 | 7 | (0, 2/3, 1/2) | (0, -1, -1) |
| 6 | 8 | (0, 2/3, 1/2) | (0, -1, -1) |

Supplementary Table CCXX. The 8 bonds for the 2nd-NN exchange parameter $J_{2c}$.

| $n$ | $n'$ | $R_{ij}$ | $R_l$ |
|---|---|---|---|
| 1 | 7 | (1/4, 1/6, 1/2) | (0, 0, 0) |
| | | | (0, 0, -1) |

| 2 | 8 | (1/4, 1/6, 1/2) | (0, 0, 0) |
|---|---|---|---|
|  |  |  | (0, 0, -1) |
| 3 | 5 | (1/4, -1/6, -1/2) | (0, 0, 0) |
|  |  |  | (0, 0, 1) |
| 4 | 6 | (1/4, -1/6, -1/2) | (0, 0, 0) |
|  |  |  | (0, 0, 1) |

Supplementary Table CCXXI. The 8 bonds for the 2nd-NN exchange parameter $J_{2d}$.

| $n$ | $n'$ | $R_{ij}$ | $R_l$ |
|---|---|---|---|
| 1 | 8 | (3/4, 1/6, 1/2) | (-1, 0, 0) |
|  |  |  | (-1, 0, -1) |
| 2 | 7 | (-1/4, 1/6, 1/2) | (0, 0, 0) |
|  |  |  | (0, 0, -1) |
| 3 | 6 | (3/4, -1/6, -1/2) | (-1, 0, 0) |
|  |  |  | (-1, 0, 1) |
| 4 | 5 | (-1/4, -1/6, -1/2) | (0, 0, 0) |
|  |  |  | (0, 0, 1) |

Supplementary Table CCXXII. The 4 bonds for the 3rd-NN exchange parameter $J_{3a}$.

| $n$ | $n'$ | $R_{ij}$ | $R_l$ |
|---|---|---|---|
| 1 | 3 | (0, -1/3, 1/2) | (0, 1, -1) |
| 2 | 4 | (0, -1/3, 1/2) | (0, 1, -1) |
| 5 | 7 | (0, 2/3, 1/2) | (0, 0, 0) |
| 6 | 8 | (0, 2/3, 1/2) | (0, 0, 0) |

Supplementary Table CCXXIII. The 4 bonds for the 3rd-NN exchange parameter $J_{3b}$.

| $n$ | $n'$ | $R_{ij}$ | $R_l$ |
|---|---|---|---|
| 1 | 3 | (0, -1/3, 1/2) | (0, 1, 0) |
| 2 | 4 | (0, -1/3, 1/2) | (0, 1, 0) |
| 5 | 7 | (0, 2/3, 1/2) | (0, 0, -1) |
| 6 | 8 | (0, 2/3, 1/2) | (0, 0, -1) |

Supplementary Table CCXXIV. The 8 bonds for the 3rd-NN exchange parameter $J_{3c}$.

| $n$ | $n'$ | $R_{ij}$ | $R_l$ |
|---|---|---|---|
| 1 | 4 | (1/2, -1/3, 1/2) | (-1, 0, -1) |
|  |  |  | (0, 0, -1) |
| 2 | 3 | (-1/2, -1/3, 1/2) | (1, 0, -1) |
|  |  |  | (0, 0, -1) |
| 5 | 8 | (1/2, 2/3, 1/2) | (-1, -1, 0) |

| | | | (0, -1, 0) |
|---|---|---|---|
| 6 | 7 | (-1/2, 2/3, 1/2) | (0, -1, 0) |
| | | | (1, -1, 0) |

Supplementary Table CCXXV. The 8 bonds for the 3rd-NN exchange parameter $J_{3d}$.

| $n$ | $n'$ | $R_{ij}$ | $R_l$ |
|---|---|---|---|
| 1 | 4 | (1/2, -1/3, 1/2) | (-1, 0, 0) |
| | | | (0, 0, 0) |
| 2 | 3 | (-1/2, -1/3, 1/2) | (1, 0, 0) |
| | | | (0, 0, 0) |
| 5 | 8 | (1/2, 2/3, 1/2) | (-1, -1, -1) |
| | | | (0, -1, -1) |
| 6 | 7 | (-1/2, 2/3, 1/2) | (0, -1, -1) |
| | | | (1, -1, -1) |

Supplementary Table CCXXVI. The 16 bonds for the 4th-NN exchange parameter $J_{4a}$.

| $n$ | $n'$ | $R_{ij}$ | $R_l$ |
|---|---|---|---|
| $i$ | $i$ | (0, 0, 0) | (0, 1, 0) |
| | | | (0, -1, 0) |

Supplementary Table CCXXVII. The 8 bonds for the 4th-NN exchange parameter $J_{4b}$.

| $n$ | $n'$ | $R_{ij}$ | $R_l$ |
|---|---|---|---|
| 1 | 5 | (1/4, -1/2, 0) | (-1, 0, 0) |
| | | | (-1, 1, 0) |
| 2 | 6 | (1/4, -1/2, 0) | (-1, 0, 0) |
| | | | (-1, 1, 0) |
| 3 | 7 | (1/4, 1/2, 0) | (-1, 0, 0) |
| | | | (-1, -1, 0) |
| 4 | 8 | (1/4, 1/2, 0) | (-1, 0, 0) |
| | | | (-1, -1, 0) |

Supplementary Table CCXXVIII. The 8 bonds for the 4th-NN exchange parameter $J_{4c}$.

| $n$ | $n'$ | $R_{ij}$ | $R_l$ |
|---|---|---|---|
| 1 | 6 | (3/4, -1/2, 0) | (0, 0, 0) |
| | | | (0, 1, 0) |
| 2 | 5 | (-1/4, -1/2, 0) | (1, 0, 0) |
| | | | (1, 1, 0) |
| 3 | 8 | (3/4, 1/2, 0) | (0, 0, 0) |
| | | | (0, -1, 0) |
| 4 | 7 | (-1/4, 1/2, 0) | (1, 0, 0) |

| | | | (1, -1, 0) |
|---|---|---|---|

By using Eqs. (65)~(69), the magnon dispersion of $Fe_{0.35}NbS_2$ can be obtained, as shown in Supplementary Fig. 11. Apart from the C-4 octuple magnon at the R point, there are quadrupole nodal magnons existing on the at three perpendicular k-planes ($k_x$, $k_y$, $k_z = \pi$), thus forming quadrupole nodal magnon net. We also provide the magnon band representations of the corresponding SSG and MSG. We find that the SSG effectively describes the band degeneracy.

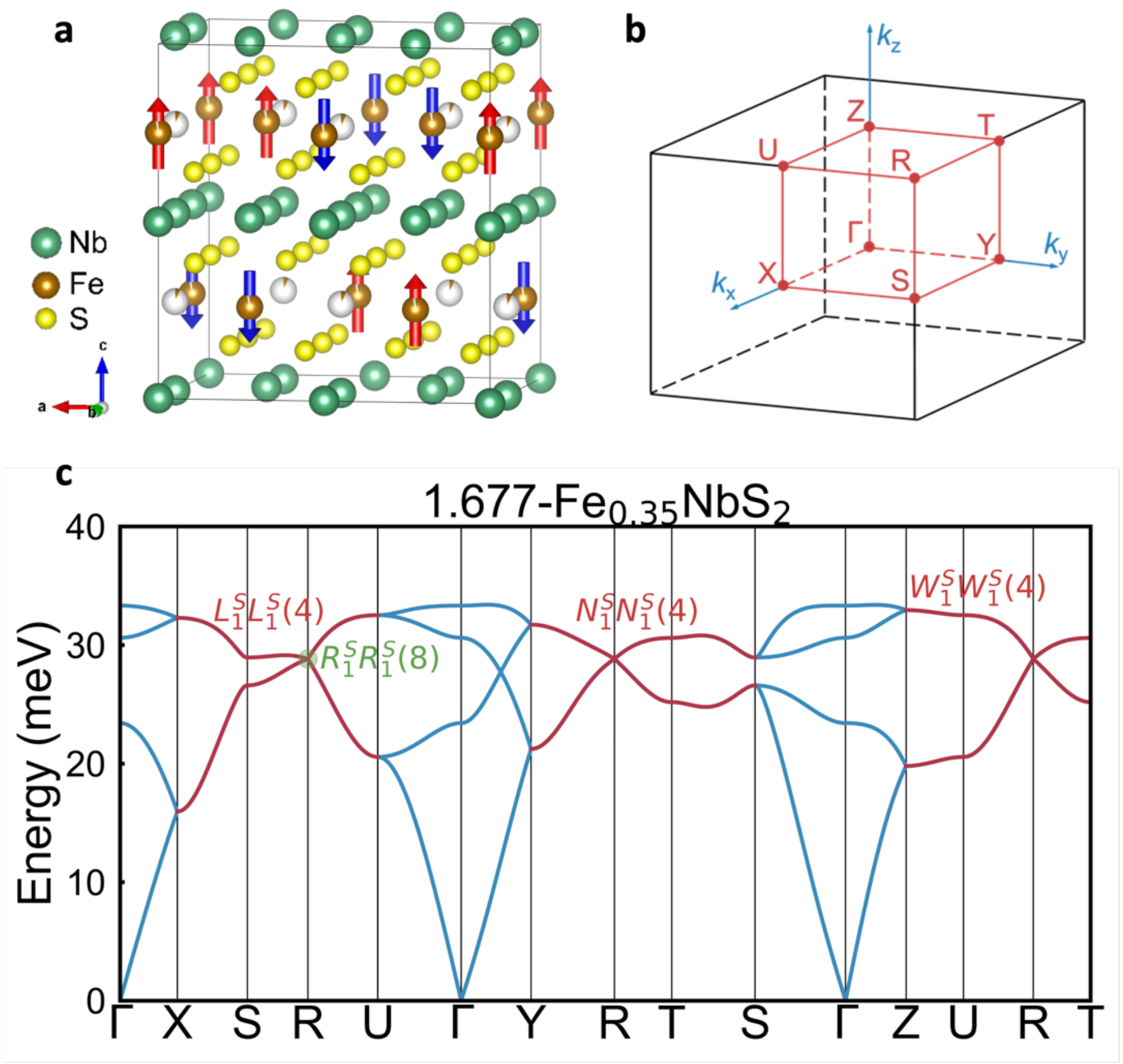

Supplementary Fig. 11. **a**, Magnetic structure, **b**, spin Brillouin zone and **c**, magnon dispersion of $Fe_{0.35}NbS_2$.

Supplementary Table CCXXIX. Band representation of the MSG $P_c2_12_12$ (18.21) and the SSG $P_a{}^12_1{}^12_1{}^12_1{}^{\infty m}1$ (19.18.2.1) induced from the local representation of the Wyckoff position (MSG: 8c, SSG: 8a).

| | Magnetic space group<br>$P_c2_12_12$ (18.21) | Spin space group<br>$P_a{}^12_1{}^12_1{}^12_1{}^{\infty m}1$ (19.18.2.1) |
|---|---|---|
| **Band rep** | $\boldsymbol{A \uparrow G}$ **(8)** | $\boldsymbol{A^S \uparrow G}$ **(8)** |
| Γ (0, 0, 0) | $2\Gamma_1(1) + 2\Gamma_2(1)$ | $\Gamma_1^S(2) + \Gamma_2^S(2)$ |

| | $+2\Gamma_3(1) + 2\Gamma_4(1)$ | $+\Gamma_3^S(2) + \Gamma_4^S(2)$ |
|---|---|---|
| X (1/2, 0, 0) | $2X_1X_2(2) + 2X_3X_4(2)$ | $2X_1^S(4)$ |
| S (1/2, 1/2, 0) | $2S_1S_1(4)$ | $S_1^SS_2^S(4) + S_3^SS_4^S(4)$ |
| Y (0, 1/2, 0) | $4Y_1(2)$ | $2Y_1^S(4)$ |
| Z (0, 0, 1/2) | $4Z_1(2)$ | $2Z_1^S(4)$ |
| U (1/2, 0, 1/2) | $2U_1U_1(4)$ | $U_1^SU_4^S(4) + U_2^SU_3^S(4)$ |
| R (1/2, 1/2, 1/2) | $R_1R_1(2) + R_2R_2(2)$<br>$+R_3R_3(2) + R_4R_4(2)$ | $R_1^SR_1^S(8)$ |
| T (0, 1/2, 1/2) | $2T_1T_2(2) + 2T_3T_4(2)$ | $T_1^ST_3^S(4) + T_2^ST_4^S(4)$ |
| Σ (u, 0, 0) | $4\Sigma_1(1) + 4\Sigma_2(1)$ | $2\Sigma_1^S(2) + 2\Sigma_2^S(2)$ |
| D (1/2, v, 0) | $4D_1D_2(2)$ | $2D_1^SD_2^S(4)$ |
| C (u, 1/2, 0) | $4C_1C_2(2)$ | $2C_1^SC_2^S(4)$ |
| Δ (0, v, 0) | $4\Delta_1(1) + 4\Delta_2(1)$ | $2\Delta_1^S(2) + 2\Delta_2^S(2)$ |
| Λ (0, 0, w) | $4\Lambda_1(1) + 4\Lambda_2(1)$ | $2\Lambda_1^S(2) + 2\Lambda_2^S(2)$ |
| A (u, 0, 1/2) | $4A_1A_2(2)$ | $2A_1^SA_2^S(4)$ |
| P (1/2, v, 1/2) | $2P_1P_1(2) + 2P_2P_2(2)$ | $P_1^SP_1^S(4) + P_2^SP_2^S(4)$ |
| E (u, 1/2, 1/2) | $2E_1E_1(2) + 2E_2E_2(2)$ | $E_1^SE_1^S(4) + E_2^SE_2^S(4)$ |
| B (0, v, 1/2) | $4B_1B_2(2)$ | $2B_1^SB_2^S(4)$ |
| L (1/2, v, w) | $4L_1L_1(2)$ | $2L_1^SL_1^S(4)$ |
| N (u, 1/2, w) | $4N_1N_1(2)$ | $2N_1^SN_1^S(4)$ |
| W (u, v, 1/2) | $4W_1W_1(2)$ | $2W_1^SW_1^S(4)$ |
| GP (u, v, w) | $8GP_1(1)$ | $4GP_1^S(2)$ |

### S7.3.4.1. Topological charge (Uτ)

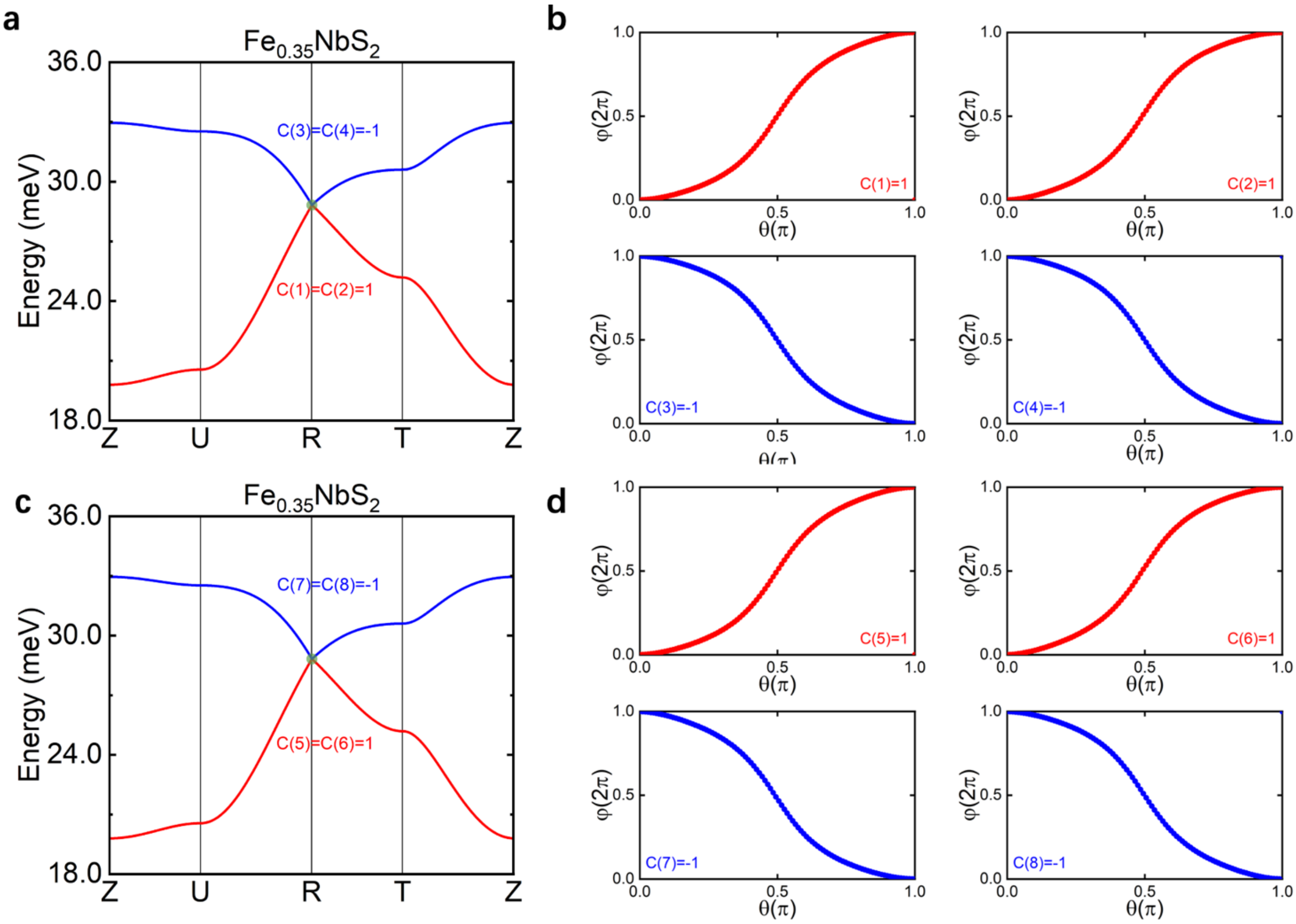

Supplementary Fig. 12. Magnon band structure of collinear AFM $Fe_{0.35}NbS_2$ (stripe AFM phase) for the **a**. S=1 (up) channel and the **c**. S=-1 (down) channel. Here S denote the spin angular momentum of magnon. Wilson loop of **b**. magnon bands 1-4 of up channel and **d**. magnon bands 5-8 of down channel at R point.

As shown in Supplementary Fig. 12, we can conclude that, in type-IV collinear SSGs, Uτ-connected two opposite-spin channel hosts identical topological charges as stated in the manuscript, which resulting in the double-valued topological charges. Here the octuple nodal point R host topological charges $C$=+4. This is because that Uτ keeps the same topological charges between nodal points of two degenerate magnon modes.

In addition, we found two Dirac points with topological charge $C$ = -2 at two k-points, $\Delta_1(0, 0.283, 0)$ and $\Delta_2(0, -0.283, 0)$, connected by $T\tau$ symmetry (shown in Supplementary Fig. 13). Therefore, the total topological charge of the entire system is 0, obeying the Nielsen-Ninomiya theorem[105].

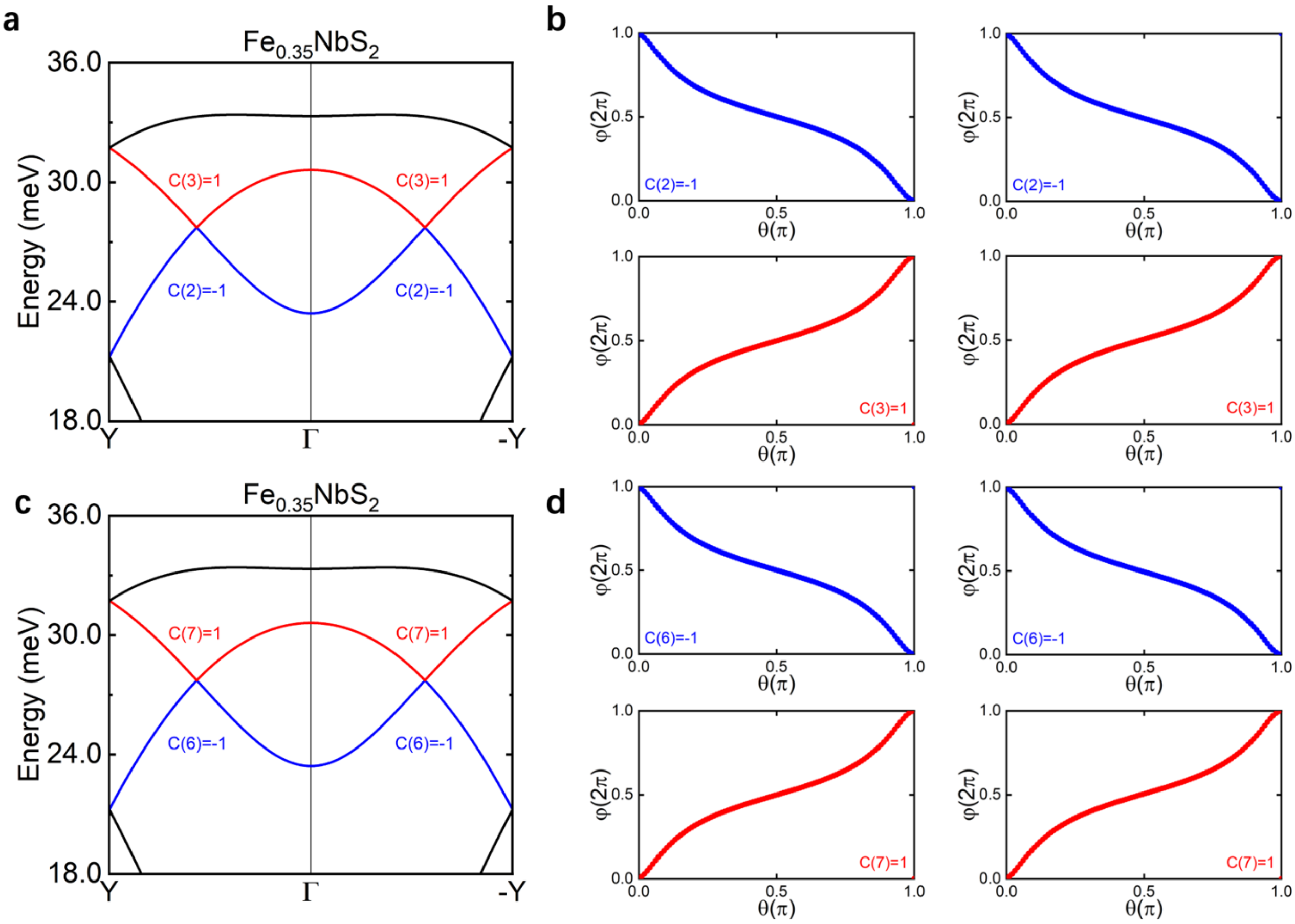

Supplementary Fig. 13. Magnon band structure of collinear AFM $Fe_{0.35}NbS_2$ (stripe AFM phase) for the **a**. S=1 (up) channel and the **c**. S=-1 (down) channel along the (Y)-(Γ)-(-Y) path. Here S denote the spin angular momentum of magnon. Wilson loop of **b**. magnon bands 2-3 of up channel and **d**. magnon bands 6-7 of down channel at $\Delta_1$ point (left panel) and $\Delta_2$ point (right panel), respectively.

### S7.3.5. $Pr_5Mo_3O_{16}$ (duodecuple magnon, quadruple nodal line magnon)

Praseodymium molybdate $Pr_5Mo_3O_{16}$ crystallizes in cubic system of space group $Pn\bar{3}n$ (222) with lattice constants a = 11.21 Å. Although the material has been experimentally synthesized[106], its magnetic structure is not yet understood. In order to demonstrate the possibility of a duodecuple magnon, we assumed a collinear AFM ground state with Néel order along [111] direction for the material, which belongs to the MSG $R\bar{3}'c'$ (167.106). The SSG of collinear AFM $Pr_5Mo_3O_{16}$ is $P^{\bar{1}}n^{\bar{1}}\bar{3}^{1}n^{\infty m}1$ (218.222.1.1) and the magnetic ions Mo locate at the spin Wyckoff position 12*d* (0, 3/4, 1/4 | 0, 0, z), as shown in Supplementary Fig. 14**a**. The lattice basis for constructing magnon Hamiltonian is $\mathrm{a}_1 = (\mathrm{a}, 0, 0), \mathrm{a}_2 = (0, \mathrm{a}, 0), \mathrm{a}_3 = (0, 0, \mathrm{a})$. Here we adopt the artificial exchange parameters, where $J_2 = -0.12J_1$, $J_3 = -0.01J_1$, $J_4 = 0.05J_1$, $S = 1$. The 12 equivalent magnetic ions in magnetic primitive cell are given in Supplementary Table CCXXX, and the detailed information of bonds are shown in Supplementary Tables CCXXXI ~ CCXXXIII.

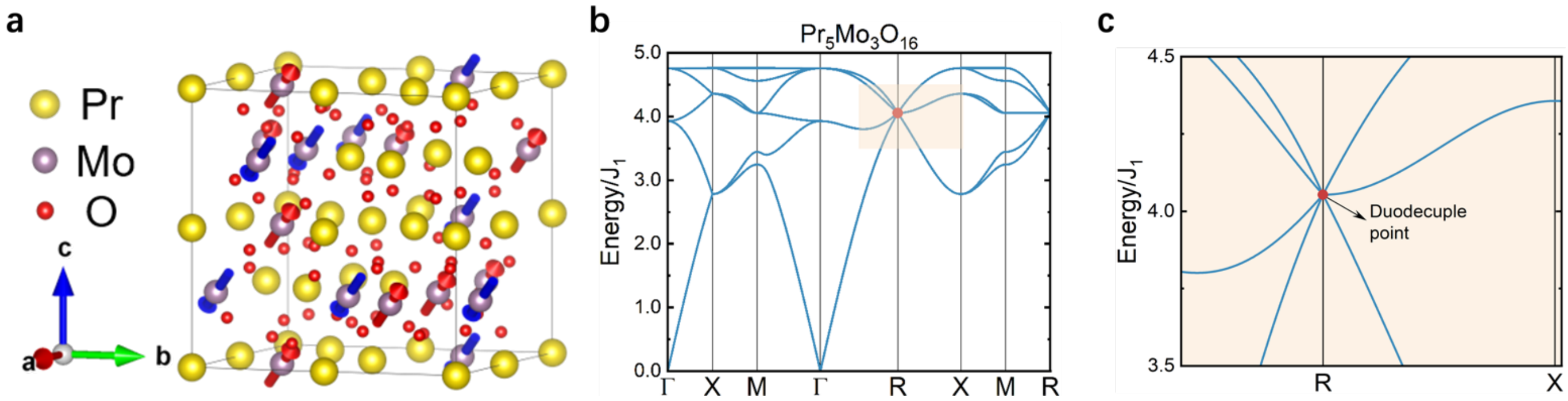

Supplementary Fig. 14. **a**, Magnetic structure and **b,** magnon dispersion of collinear AFM $Pr_5Mo_3O_{16}$. **c**, the duodecuple magnon at R point.

By using Eqs. (65)~(69), the magnon dispersion of $Pr_5Mo_3O_{16}$ can be obtained, as shown in Supplementary Fig. 14**b**. We also provide the magnon band representations of the corresponding SSG and MSG in Supplementary Table CCXXXIV. We find that the magnon band is overall doubly degenerate due to the combination effect of $PT$ and $SO(2)$. Specially, the coexistence of the ASS and ARS at R point results in both the doubled degeneracy in spin space and real space, contributing to the twelve-dimensional $R_4^S R_5^S(12)$ co-irreps of the little group $^{\bar{1}}m^{\bar{1}}\bar{3}^{1}m^{\infty m}1$.

Supplementary Table CCXXX. The coordinates of the 12 Mo ions in the conventional cell basis vectors.

| $n$ | $r_n$ | magnetic moment |
|---|---|---|
| 1 | (0, 3/4, 1/4) | up |
| 2 | (1/2, 3/4, 1/4) | up |
| 3 | (1/4, 0, 3/4) | up |

| 4 | (3/4, 1/4, 0) | up |
|---|---|---|
| 5 | (3/4, 1/4, 1/2) | up |
| 6 | (1/4, 1/2, 3/4) | up |
| 7 | (0, 1/4, 3/4) | down |
| 8 | (1/2, 1/4, 3/4) | down |
| 9 | (3/4, 0, 1/4) | down |
| 10 | (1/4, 3/4, 0) | down |
| 11 | (1/4, 3/4, 1/2) | down |
| 12 | (3/4, 1/2, 1/4) | down |

Supplementary Table CCXXXI. The 24 bonds for the 1st-NN exchange parameter $J_1$.

| $n$ | $n'$ | $R_{ij}$ | $R_l$ |
|---|---|---|---|
| 1 | 9 | (3/4, -3/4, 0) | (-1, 1, 0) |
| | 10 | (1/4, 0, -1/4) | (0, 0, 0) |
| | 11 | (1/4, 0, 1/4) | (0, 0, 0) |
| | 12 | (3/4, -1/4, 0) | (-1, 0, 0) |
| 2 | 9 | (1/4, -3/4, 0) | (0, 1, 0) |
| | 10 | (-1/4, 0, -1/4) | (0, 0, 0) |
| | 11 | (-1/4, 0, 1/4) | (0, 0, 0) |
| | 12 | (1/4, -1/4, 0) | (0, 0, 0) |
| 3 | 7 | (-1/4, 1/4, 0) | (0, 0, 0) |
| | 8 | (1/4, 1/4, 0) | (0, 0, 0) |
| | 10 | (0, 3/4, -3/4) | (0, -1, 1) |
| | 11 | (0, 3/4, -1/4) | (0, -1, 0) |
| 4 | 7 | (-3/4, 0, 3/4) | (1, 0, -1) |
| | 8 | (-1/4, 0, 3/4) | (0, 0, -1) |
| | 9 | (0, -1/4, 1/4) | (0, 0, 0) |
| | 12 | (0, 1/4, 1/4) | (0, 0, 0) |
| 5 | 7 | (-3/4, 0, 1/4) | (1, 0, 0) |
| | 8 | (-1/4, 0, 1/4) | (0, 0, 0) |
| | 9 | (0, -1/4, -1/4) | (0, 0, 0) |
| | 12 | (0, 1/4, -1/4) | (0, 0, 0) |
| 6 | 7 | (-1/4, -1/4, 0) | (0, 0, 0) |
| | 8 | (1/4, -1/4, 0) | (0, 0, 0) |
| | 10 | (0, 1/4, -3/4) | (0, 0, 1) |
| | 11 | (0, 1/4, -1/4) | (0, 0, 0) |

Supplementary Table CCXXXII. The 12 bonds for the 2nd-NN exchange parameter $J_2$.

| $n$ | $n'$ | $R_{ij}$ | $R_l$ |
|---|---|---|---|
| 1 | 2 | (1/2, 0, 0) | (0, 0, 0) |
| | | | (-1, 0, 0) |
| 3 | 6 | (0, 1/2, 0) | (0, 0, 0) |

| | | | (0, -1, 0) |
|---|---|---|---|
| 4 | 5 | (0, 0, 1/2) | (0, 0, 0) |
| | | | (0, 0, -1) |
| 7 | 8 | (1/2, 0, 0) | (0, 0, 0) |
| | | | (-1, 0, 0) |
| 9 | 12 | (0, 1/2, 0) | (0, 0, 0) |
| | | | (0, -1, 0) |
| 10 | 11 | (0, 0, 1/2) | (0, 0, 0) |
| | | | (0, 0, -1) |

Supplementary Table CCXXXIII. The 48 bonds for the 3rd-NN exchange parameter $J_3$

| $n$ | $n'$ | $R_{ij}$ | $R_l$ |
|---|---|---|---|
| 1 | 3 | (1/4, -3/4, 1/2) | (0, 1, 0) |
| | | | (0, 1, -1) |
| | 4 | (3/4, -1/2, -1/4) | (-1, 0, 0) |
| | | | (-1, 1, 0) |
| | 5 | (3/4, -1/2, 1/4) | (-1, 0, 0) |
| | | | (-1, 1, 0) |
| | 6 | (1/4, -1/4, 1/2) | (0, 0, 0) |
| | | | (0, 0, -1) |
| 2 | 3 | (-1/4, -3/4, 1/2) | (0, 1, 0) |
| | | | (0, 1, -1) |
| | 4 | (1/4, -1/2, -1/4) | (0, 0, 0) |
| | | | (0, 1, 0) |
| | 5 | (1/4, -1/2, 1/4) | (0, 0, 0) |
| | | | (0, 1, 0) |
| | 6 | (-1/4, -1/4, 1/2) | (0, 0, 0) |
| | | | (0, 0, -1) |
| 3 | 4 | (1/2, 1/4, -3/4) | (0, 0, 1) |
| | | | (-1, 0, 1) |
| | 5 | (1/2, 1/4, -1/4) | (0, 0, 0) |
| | | | (-1, 0, 0) |
| 4 | 6 | (-1/2, 1/4, 3/4) | (0, 0, -1) |
| | | | (1, 0, -1) |
| 5 | 6 | (-1/2, 1/4, 1/4) | (0, 0, 0) |
| | | | (1, 0, 0) |
| 7 | 9 | (3/4, -1/4, -1/2) | (-1, 0, 0) |
| | | | (-1, 0, 1) |
| | 10 | (1/4, 1/2, -3/4) | (0, 0, 1) |
| | | | (0, -1, 1) |
| | 11 | (1/4, 1/2, -1/4) | (0, 0, 0) |
| | | | (0, -1, 0) |

| | 12 | (3/4, 1/4, -1/2) | (-1, 0, 0) |
|---|---|---|---|
| | | | (-1, 0, 1) |
| 8 | 9 | (1/4, -1/4, -1/2) | (0, 0, 0) |
| | | | (0, 0, 1) |
| | 10 | (-1/4, 1/2, -3/4) | (0, 0, 1) |
| | | | (0, -1, 1) |
| | 11 | (-1/4, 1/2, -1/4) | (0, 0, 0) |
| | | | (0, -1, 0) |
| | 12 | (1/4, 1/4, -1/2) | (0, 0, 0) |
| | | | (0, 0, 1) |
| 9 | 10 | (-1/2, 3/4, -1/4) | (0, -1, 0) |
| | | | (1, -1, 0) |
| | 11 | (-1/2, 3/4, 1/4) | (0, -1, 0) |
| | | | (1, -1, 0) |
| 10 | 12 | (1/2, -1/4, 1/4) | (0, 0, 0) |
| | | | (-1, 0, 0) |
| 11 | 12 | (1/2, -1/4, -1/4) | (0, 0, 0) |
| | | | (-1, 0, 0) |

Supplementary Table CCXXXIV. Band representation of the SSG $R\bar{3}'c'$ (167.106) and the collinear SSG $P^{\bar{1}}n^{\bar{1}}\bar{3}^{1}n^{\infty m}1$ (218.222.1.1) induced from the local representation of the spin Wyckoff position 12$d$.

| | Magnetic space group $R\bar{3}'c'$ (167.106) | Spin space group $P^{\bar{1}}n^{\bar{1}}\bar{3}^{1}n^{\infty m}1$ (218.222.1.1) |
|---|---|---|
| **Band rep** | $\boldsymbol{A \uparrow G}$ $\mathbf{(12)}$ | $\boldsymbol{A^S \uparrow G}$ $\mathbf{(12)}$ |
| Γ (0, 0, 0) | $2\Gamma_1(1) + 2\Gamma_2(1) + 4\Gamma_3(2)$ | $\Gamma_1^S(2) + \Gamma_3^S(4) + \Gamma_5^S(6)$ |
| Δ (0, v, 0) | $12\Delta_1(1)$ | $2\Delta_1^S(2) + 2\Delta_2^S(2) + \Delta_3^S\Delta_4^S(4)$ |
| X (0, 1/2, 0) | $6X_1X_2(2)$ | $X_1^SX_2^S(4) + 2X_5^S(4)$ |
| Z (u, 1/2, 0) | $12Z_1(1)$ | $2Z_1^S(2) + 4Z_2^S(2)$ |
| M (1/2, 1/2, 0) | $6M_1(1) + 6M_2(1)$ | $M_2^S(2) + 2M_3^S(2) + M_4^S(2) + M_5^S(4)$ |
| Σ (u, u, 0) | $6\Sigma_1(1) + 6\Sigma_2(1)$ | $3\Sigma_1^S(2) + 3\Sigma_2^S(2)$ |
| Λ (u, u, u) | $4\Lambda_1(2) + 4\Lambda_2\Lambda_3(2)$ | $\Lambda_1^S(2) + \Lambda_2^S(2) + 2\Lambda_3^S(4)$ |
| R (1/2, 1/2, 1/2) | $2R_1R_2(2) + 2R_3R_3(4)$ | $R_4^SR_5^S(12)$ |
| S (u, 1/2, u) | $12S_1(1)$ | $3S_1^SS_2^S(4)$ |

| T (1/2, 1/2, w) | $12T_1(1)$ | $T_1^S T_2^S(4) + 2T_3^S(2) + 2T_4^S(2)$ |
| --- | --- | --- |
| GP (u, v, w) | $12GP_1(1)$ | $6GP_1^S(2)$ |

### S7.3.6. $RuO_2$ (chiral magnon)

Rutile $RuO_2$ crystallizes in tetragonal system of space group $P4_2/mnm$ (136) with lattice constants a = 4.533 Å, c = 3.124 Å. As the first system predicted to be a candidate material for altermagnets, there is still debate about its magnetic ground state. Taken the reported collinear AFM ground state with Néel order along [001] direction as an example, we demonstrate the description of its chiral magnons by collinear SSGs and the importance of U values in describing the magnon band structure calculations. For this magnetic structure, the MSG is $P4_2'/mnm'$ (136.499). It belongs to the collinear SSG is $P^{\bar{1}}4_2/^{1}m^{\bar{1}}n^{1}m^{\infty m}1$ (65.136.1.1) and the magnetic ions Ru locate at the spin Wyckoff position $2a$ (0, 0, 0 | 0, 0, z). The lattice basis for constructing magnon Hamiltonian is $\mathrm{a}_1 = (\mathrm{a}, 0, 0), \mathrm{a}_2 = (0, \mathrm{a}, 0), \mathrm{a}_3 = (0, 0, \mathrm{c})$. The 2 equivalent magnetic ions in magnetic primitive cell are given in Supplementary Table CCXXXV, and the detailed information of bonds are shown in Supplementary Tables CCXXXVI ~ CCXLIII.

One thing worth noting about altermagnet is that the chiral splitting of its magnon dispersion does not necessarily require a set of the same bond length, but due to the different exchange parameters brought about by unequivalent coordination environments. This situation often arises when the symmetry group of magnetic ions is higher than the system's symmetry group, such as $RuO_2$. In cases where considering only the magnetic ions does not bring about higher symmetry, there is no need for these bond configurations, such as FeS.

Supplementary Table CCXXXV. The coordinates of the 2 Ru ions in the conventional cell basis vectors

| $n$ | $r_n$ | magnetic moment |
|---|---|---|
| 1 | (0, 0, 0) | up |
| 2 | (1/2, 1/2, 1/2) | down |

Supplementary Table CCXXXVI. The 4 bonds for the 1st-NN exchange parameter $J_1$

| $n$ | $n'$ | $R_{ij}$ | $R_l$ |
|---|---|---|---|
| 1 | 1 | (0, 0, 0) | (0, 0, 1) |
| | | | (0, 0, -1) |
| 2 | 2 | (0, 0, 0) | (0, 0, 1) |
| | | | (0, 0, -1) |

Supplementary Table CCXXXVII. The 8 bonds for the 2rd-NN exchange parameter $J_2$

| $n$ | $n'$ | $R_{ij}$ | $R_l$ |
|---|---|---|---|
| 1 | 2 | (1/2, 1/2, 1/2) | (0, 0, 0) |
| | | | (-1, 0, 0) |
| | | | (0, -1, 0) |

| | | | (0, 0, -1) |
|---|---|---|---|
| | | | (-1, -1, 0) |
| | | | (-1, 0, -1) |
| | | | (0, -1, -1) |
| | | | (-1, -1, -1) |

Supplementary Table CCXXXVIII. The 8 bonds for the 3rd-NN exchange parameter $J_3$

| $n$ | $n'$ | $R_{ij}$ | $R_l$ |
|---|---|---|---|
| 1 | 1 | (0, 0, 0) | (1, 0, 0) |
| | | | (-1, 0, 0) |
| | | | (0, 1, 0) |
| | | | (0, -1, 0) |
| 2 | 2 | (0, 0, 0) | (1, 0, 0) |
| | | | (-1, 0, 0) |
| | | | (0, 1, 0) |
| | | | (0, -1, 0) |

Supplementary Table CCXXXIX. The 16 bonds for the 4th-NN exchange parameter $J_4$

| $n$ | $n'$ | $R_{ij}$ | $R_l$ |
|---|---|---|---|
| 1 | 1 | (0, 0, 0) | (1, 0, 1) |
| | | | (-1, 0, 1) |
| | | | (1, 0, -1) |
| | | | (-1, 0, -1) |
| | | | (0, 1, 1) |
| | | | (0, -1, 1) |
| | | | (0, 1, -1) |
| | | | (0, -1, -1) |
| 2 | 2 | (0, 0, 0) | (1, 0, 1) |
| | | | (-1, 0, 1) |
| | | | (1, 0, -1) |
| | | | (-1, 0, -1) |
| | | | (0, 1, 1) |
| | | | (0, -1, 1) |
| | | | (0, 1, -1) |
| | | | (0, -1, -1) |

Supplementary Table CCXL. The 8 bonds for the 5th-NN exchange parameter $J_5$

| $n$ | $n'$ | $R_{ij}$ | $R_l$ |
|---|---|---|---|
| 1 | 2 | (1/2, 1/2, 1/2) | (0, 0, 1) |
| | | | (0, 0, -2) |

| | | | (0, -1, 1) |
|---|---|---|---|
| | | | (0, -1, -2) |
| | | | (-1, 0, 1) |
| | | | (-1, 0, -2) |
| | | | (-1, -1, 1) |
| | | | (-1, -1, -2) |

Supplementary Table CCXLI. The 4 bonds for the 6th-NN exchange parameter $J_6$

| $n$ | $n'$ | $R_{ij}$ | $R_l$ |
|---|---|---|---|
| 1 | 1 | (0, 0, 0) | (0, 0, 2) |
| | | | (0, 0, -2) |
| 2 | 2 | (0, 0, 0) | (0, 0, 2) |
| | | | (0, 0, -2) |

Supplementary Table CCXLII. The 4 bonds for the 7th-NN exchange parameter $J_7^a$

| $n$ | $n'$ | $R_{ij}$ | $R_l$ |
|---|---|---|---|
| 1 | 1 | (0, 0, 0) | (1, 1, 0) |
| | | | (-1, -1, 0) |
| 2 | 2 | (0, 0, 0) | (1, -1, 0) |
| | | | (-1, 1, 0) |

Supplementary Table CCXLIII. The 4 bonds for the 7th-NN exchange parameter $J_7^b$

| $n$ | $n'$ | $R_{ij}$ | $R_l$ |
|---|---|---|---|
| 1 | 1 | (0, 0, 0) | (1, -1, 0) |
| | | | (-1, 1, 0) |
| 2 | 2 | (0, 0, 0) | (1, 1, 0) |
| | | | (-1, -1, 0) |

By using Eqs. (65)~(69), the magnon dispersion of $RuO_2$ can be obtained, as shown in Supplementary Table CLVIII. We also provide the magnon band representations of the corresponding SSG and MSG. We find that the SSG effectively describes the band degeneracy.

Supplementary Table CCXLIV. Band representation of the magnetic space group $P4_2'/mnm'$ (136.499) and the spin space group $P^{\bar{1}}4_2/^{1}m^{\bar{1}}n^{1}m^{\infty m}1$ (65.136.1.1) and the Wyckoff position 2a.

| | Magnetic space group $P4_2'/mnm'$ (136.499) | Spin space group $P^{\bar{1}}4_2/^{1}m^{\bar{1}}n^{1}m^{\infty m}1$ (65.136.1.1) |
|---|---|---|

| Band rep | $\boldsymbol{B_g} \uparrow \boldsymbol{G}$ (**2**) | $\boldsymbol{A_g^S} \uparrow \boldsymbol{G}$ (**2**) |
| --- | --- | --- |
| Γ (0, 0, 0) | $\Gamma_3^+\Gamma_4^+(2)$ | $\Gamma_1^{S,+}(2)$ |
| Δ (0, v, 0) | $\Delta_2(1) + \Delta_3(1)$ | $\Delta_1^S(2)$ |
| X (0, 1/2, 0) | $X_1(2)$ | $X_1^{S,+}(2)$ |
| Y (u, 1/2, 0) | $Y_2(1) + Y_3(1)$ | $Y_1^S(2)$ |
| M (1/2, 1/2, 0) | $M_3^+(1) + M_4^+(1)$ | $M_1^{S,+}(2)$ |
| Σ (u, u, 0) | $2\Sigma_2(1)$ | $2\Sigma_1^S(1)$ |
| Λ (0, 0, w) | $\Lambda_3\Lambda_4(2)$ | $\Lambda_1^S(2)$ |
| Z (0, 0, 1/2) | $Z_2(2)$ | $Z_1^{S,+}(2)$ |
| U (0, v, 1/2) | $U_1(2)$ | $U_1^S(2)$ |
| R (0, 1/2, 1/2) | $R_1^+(2)$ | $R_1^{S,+}(2)$ |
| T (u, 1/2, 1/2) | $T_1(2)$ | $T_1^S(2)$ |
| A (1/2, 1/2, 1/2) | $A_2(2)$ | $A_1^{S,+}(2)$ |
| S (u, u, 1/2) | $S_1(1) + S_2(1)$ | $2S_1^S(1)$ |
| GP (u, v, w) | $2GP_1(1)$ | $2GP_1^S(1)$ |

### S7.3.6.1. U effect

We illustrate the potential effects of $U_{eff}$ values using $RuO_2$ as an example, offering a semi-quantitative perspective. Our calculations on nonmagnetic (NM) and antiferromagnetic (AFM) phases of $RuO_2$ with varying $U_{eff}$ values (Supplementary Fig. 15**a**) reveal that the NM phase is favored when $U_{eff} \leq 1.2$ eV, while the AFM phase becomes preferred for larger $U_{eff}$ values. This underscores the significance of selecting an appropriate $U_{eff}$ value for accurately portraying the magnetic ground state and subsequent magnon dispersion.

To compare the magnon band structures with different $U_{eff}$ values, the $U_{eff}$-dependent bilinear exchange interactions are given in Supplementary Table CCXLV, where the maximal Heisenberg exchange parameter $J_1$ increases gradually with $U_{eff}$. In addition, the three SOC-related interactions $D_2$, $J_3^{ani}$, and $\Gamma_1$ are also augmented with increasing $U_{eff}$. Using linear spin wave theory, the $U_{eff}$-dependent SOC-free magnon band structures are depicted in Supplementary Fig. 15**b-d**. When the magnetic structure is not affected by the $U_{eff}$ values, for example, as selected in Supplementary Fig. 15 with $U_{eff}$ = 1.6, 2.0, and 3.0 eV. Qualitatively, four types of interaction J, DM, $J^{ani}$ and $\Gamma$ increase gradually with the increased $U_{eff}$ values. However, varied $U_{eff}$ values keep the dimension of the band representation, the magnon band degeneracy and the momentum position of the magnon chirality splitting invariant. More specifically, the variation in $U_{eff}$ value leads to the different splitting pattern of chiral magnons, including changes in the energy levels and energy differences between left and right-handed chiral magnons.

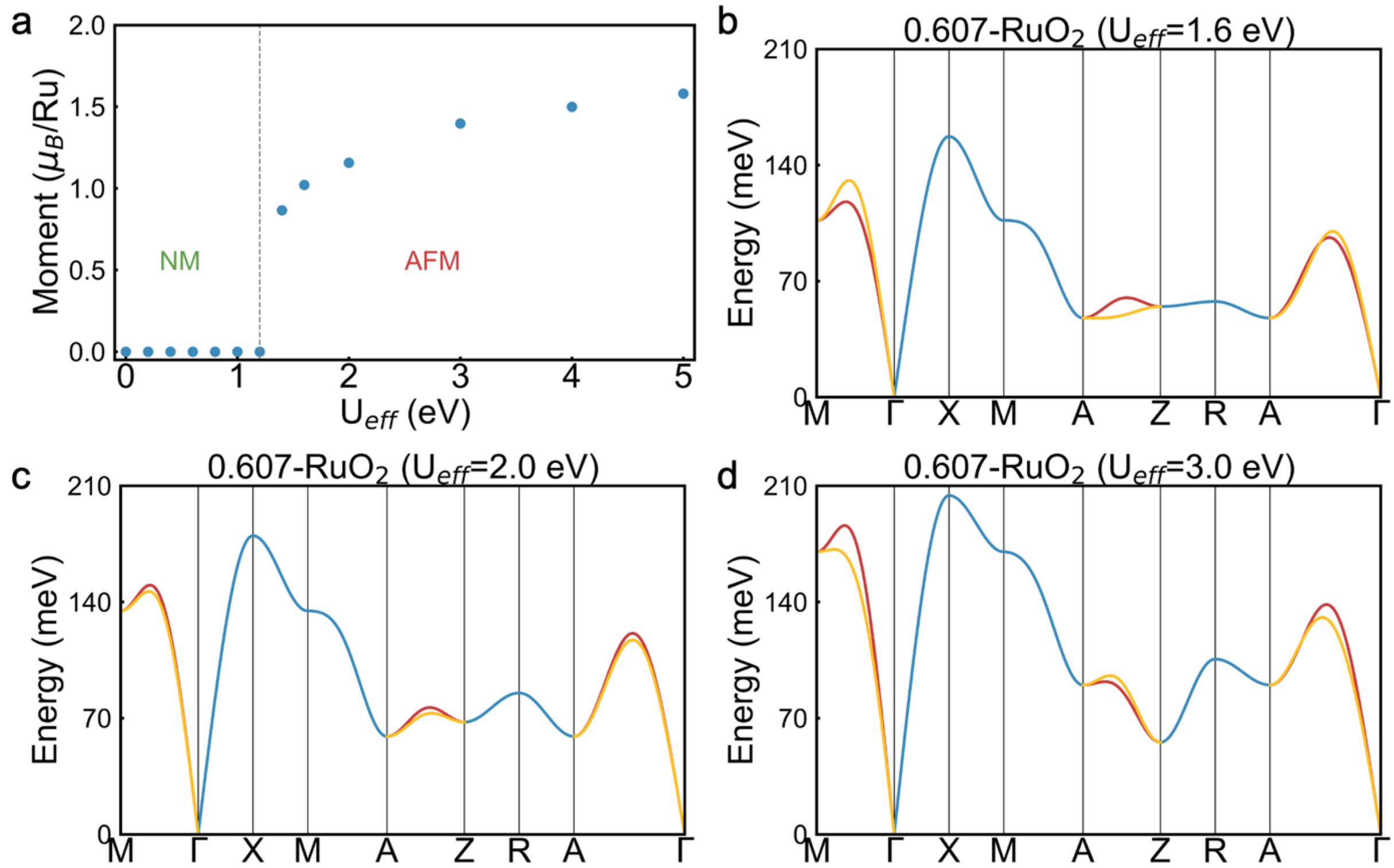

Supplementary Fig. 15. **a**, Local magnetization of Ru ion in $RuO_2$ as a function of $U_{eff}$. **b-d**, Magnon dispersions as a function of different $U_{eff}$ values. Here the red and yellow lines represent the magnon spin up (S = 1) and spin down (S = -1) channels, while the blue lines are magnon spin degenerate.

In our magnon database, we adopt the reported $U_{eff}$ value of 1.6 eV for $RuO_2$ as reported[88]. The selection of an appropriate $U_{eff}$ is crucial for ensuring a positive-definite magnon Hamiltonian. Therefore, we utilize the $U_{eff}$ values reported in the literature for all the high-throughput calculations. Finally, we remark that due to the considerable computational cost as mentioned earlier, attempting to iterate through U values for the calculation of all materials is quite challenging.

Supplementary Table CCXLV. All bilinear exchange interactions (meV) as a function of $U_{eff}$ (eV). The maximal Heisenberg exchange terms is highlighted.

| $U_{eff}$ | 1.6 | 2.0 | 3.0 |
|---|---|---|---|
| $\boldsymbol{J_1}$ | **-9.662** | **-10.338** | **-11.763** |
| $J_2$ | -7.231 | -9.194 | -10.446 |
| $J_3$ | -0.184 | -0.230 | 1.568 |
| $J_4$ | 0.123 | 0.156 | -0.295 |
| $J_5$ | 0.632 | 0.861 | 0.788 |
| $J_6$ | 1.668 | 1.985 | 1.551 |
| $J_{7a}$ | 1.922 | 1.867 | 1.826 |
| $J_{7b}$ | 1.683 | 2.369 | 2.838 |
| $J_{8a}$ | 1.084 | 0.363 | -0.273 |
| $J_{8b}$ | 0.313 | 0.391 | 0.551 |
| $D_2$ | (-0.2027,0.2027,0) | (-0.3703,0.3703,0) | (-0.7855,0.7855,0) |
| $J_3^{ani}$ | (-0.019,-0.243,-0.071) | (-0.021,-0.349,-0.163) | (-0.569,-0.887,-0.702) |
| $\Gamma_1$ | (0.039, 0, 0) | (0.070, 0, 0) | (0.201, 0, 0) |
| $\|D_2/J_{max}\|$ | 3.0% | 5.1% | 9.4% |
| $\|J_3^{ani}/J_{max}\|$ | 2.3% | 3.2% | 2.7% |
| $\|\Gamma_1/J_{max}\|$ | 0.4% | 0.7% | 1.7% |

## Supplementary References

1. Landau, L. D. & Lífshíts, E. M. *Electrodynamics of continuous media*. (Pergamon Press Oxford, 1946).

2. Liu, P. et al. Spin-Group Symmetry in Magnetic Materials with Negligible Spin-Orbit Coupling. *Phys. Rev. X* **12**, 021016 (2022).

3. https://findspingroup.com/.

4. Wintgen, G. Zur Darstellungstheorie der Raumgruppen. *Math. Ann.* **118**, 195-215 (1941).

5. Angelova, M. N. & Boyle, L. On the classification and enumeration of the irreducible co-representations of magnetic space groups. *J. Phys. A: Math. Gen.* **29**, 993 (1996).

6. Wilson, A. 1.4. Arithmetic crystal classes and symmorphic space groups. (2006).

7. Dimmock, J. O. & Wheeler, R. G. Symmetry Properties of Wave Functions in Magnetic Crystals. *Phys. Rev.* **127**, 391 (1962).

8. Chen, X. et al. Enumeration and representation theory of spin space groups. *Phys. Rev. X* **14**, 031038 (2024).

9. Ritter, C. et al. Magnetic ordering of the $R_4Sb_3$ compounds (R= Pr, Nd, Sm) and of $Pr_2Nd_2Sb_3$. *J. Alloys Compd.* **494**, 28-33 (2010).

10. Shamoto, S.-i., Nakano, T., Nozue, Y. & Kajitani, T. Substitution effects on ferromagnetic Mott insulator $Lu_2V_2O_7$. *J. Phys. Chem. Solids* **63**, 1047-1050 (2002).

11. Kim, S.-Y. et al. First-principles study on structural, electronic, magnetic and thermodynamic properties of lithium ferrite $LiFe_5O_8$. *RSC Adv.* **12**, 15973-15979 (2022).

12. Herak, M. et al. Novel spin lattice in $Cu_3TeO_6$: an antiferromagnetic order and domain dynamics. *J. Phys.: Condens. Matter* **17**, 7667 (2005).

13. Ricci, F. & Bousquet, E. Unveiling the room-temperature magnetoelectricity of troilite FeS. *Phys. Rev. Lett.* **116**, 227601 (2016).

14. Qiang, B. et al. Magnetic and transport properties of chiral antiferromagnetic $Co_{2-x}Pd_xMo_3N$ thin films. *AIP Adv.* **13**(2023).

15. Joenk, R. & Bozorth, R. Magnetic properties of $CuF_2$. *J. Appl. Phys.* **36**, 1167-1168 (1965).

16. Brown, P. & Forsyth, J. A neutron diffraction study of weak ferromagnetism in nickel fluoride. *J. Phys. C: Solid State Phys.* **14**, 5171 (1981).

17. Xiang, H. et al. Single-ion anisotropy, Dzyaloshinskii-Moriya interaction, and negative magnetoresistance of the spin-1/2 pyrochlore $R_2V_2O_7$. *Phys. Rev. B* **83**, 174402 (2011).

18. Yang, J.-H. et al. Strong Dzyaloshinskii-Moriya interaction and origin of ferroelectricity in $Cu_2OSeO_3$. *Phys. Rev. Lett.* **109**, 107203 (2012).

19. Santos-Carballal, D., Roldan, A., Grau-Crespo, R. & de Leeuw, N. H. First-principles study of the

inversion thermodynamics and electronic structure of Fe $M_2X_4$ (thio) spinels (M= Cr, Mn, Co, Ni; X= O, S). *Phys. Rev. B* **91**, 195106 (2015).
20. Wang, D., Shaikh, M., Ghosh, S. & Sanyal, B. Prediction of half-metallic ferrimagnetic quadruple perovskites $ACu_3Fe_2Re_2O_{12}$ (A= Ca, Sr, Ba, Pb, Sc, Y, La) with high Curie temperatures. *Phys. Rev. Mater.* **5**, 054405 (2021).
21. Zhou, F. et al. First-principles prediction of redox potentials in transition-metal compounds with LDA+ U. *Phys. Rev. B* **70**, 235121 (2004).
22. Zhu, M. et al. Ferromagnetic superexchange in insulating $Cr_2MoO_6$ by controlling orbital hybridization. *Phys. Rev. B* **92**, 094419 (2015).
23. Tai, B. et al. Two-dimensional CoSe structures: Intrinsic magnetism, strain-tunable anisotropic valleys, magnetic Weyl point, and antiferromagnetic metal state. *Phys. Rev. B* **102**, 224422 (2020).
24. Janson, O. et al. Magnetic pyroxenes $LiCrGe_2O_6$ and $LiCrSi_2O_6$: Dimensionality crossover in a nonfrustrated S= 3/2 Heisenberg model. *Phys. Rev. B* **90**, 214424 (2014).
25. Otte, K., Schmahl, W. W. & Pentcheva, R. DFT+ U study of arsenate adsorption on FeOOH surfaces: evidence for competing binding mechanisms. *J. Phys. Chem. C* **117**, 15571-15582 (2013).
26. Calderon, C. E. et al. The AFLOW standard for high-throughput materials science calculations. *Comput. Mater. Sci.* **108**, 233-238 (2015).
27. Atanelov, J. & Mohn, P. Electronic and magnetic properties of $GaFeO_3$: ab initio calculations for varying Fe/Ga ratio, inner cationic site disorder, and epitaxial strain. *Phys. Rev. B* **92**, 104408 (2015).
28. Balachandran, P. V. et al. Predictions of new $ABO_3$ perovskite compounds by combining machine learning and density functional theory. *Phys. Rev. Mater.* **2**, 043802 (2018).
29. Shao, D.-F., Gurung, G., Zhang, S.-H. & Tsymbal, E. Y. Dirac nodal line metal for topological antiferromagnetic spintronics. *Phys. Rev. Lett.* **122**, 077203 (2019).
30. Johnson, R. D. et al. $MnSb_2O_6$: a polar magnet with a chiral crystal structure. *Phys. Rev. Lett.* **111**, 017202 (2013).
31. Jmal, S. A., Balli, M., Bouhani, H. & Mounkachi, O. First principles investigations of the electronic and magnetic properties of $RVO_3$ (R= Er, Ho, Y, Lu) vanadates. *arXiv preprint arXiv:2211.07554*(2022).
32. Das, R. et al. Ferrimagnetic cluster formation due to oxygen vacancies in $CaFe_2O_{4-\delta}$. *Phys. Rev. B* **98**, 144404 (2018).
33. Liu, P. et al. Anisotropic magnetic couplings and structure-driven canted to collinear transitions in $Sr_2IrO_4$ by magnetically constrained noncollinear DFT. *Phys. Rev. B* **92**, 054428 (2015).
34. Yuan, L.-D., Wang, Z., Luo, J.-W. & Zunger, A. Prediction of low-Z collinear and noncollinear antiferromagnetic compounds having momentum-dependent spin splitting even without spin-orbit coupling. *Phys. Rev. Mater.* **5**, 014409 (2021).

35. Picozzi, S. & Ederer, C. First principles studies of multiferroic materials. *J. Phys.: Condens. Matter* **21**, 303201 (2009).
36. Pesant, S. & Côté, M. DFT+ U study of magnetic order in doped $La_2CuO_4$ crystals. *Phys. Rev. B* **84**, 085104 (2011).
37. Maitra, T. & Valenti, R. Orbital order in $ZnV_2O_4$. *Phys. Rev. Lett.* **99**, 126401 (2007).
38. Zhang, Y. et al. Block antiferromagnetism and possible ferroelectricity in $KFe_2Se_2$. *Phys. Status Solidi* **10**, 757-761 (2016).
39. Gui, Z. et al. Improper multiferroiclike transition in a metal. *Phys. Rev. B* **105**, L180101 (2022).
40. Chen, X. et al. Asymmetric interfaces sandwiched between infinite-layer oxides and perovskite oxides. *Phys. Rev. B* **104**, 045310 (2021).
41. Zhou, S. et al. $BaMF_4$ (M= Mn, Co, Ni): New electrode materials for hybrid supercapacitor with layered polar structure. *J. Power Sources* **359**, 585-591 (2017).
42. Zhang, Y., Brgoch, J. & Miller, G. J. Magnetic ordering in tetragonal 3*d* metal arsenides $M_2As$ (M= Cr, Mn, Fe): an ab initio investigation. *Inorg. Chem.* **52**, 3013-3021 (2013).
43. Wang, X. et al. Observation of magnetoelectric multiferroicity in a cubic perovskite system: $LaMn_3Cr_4O_{12}$. *Phys. Rev. Lett.* **115**, 087601 (2015).
44. Xu, Y. et al. High-throughput calculations of magnetic topological materials. *Nature* **586**, 702-707 (2020).
45. Zheng, Y. et al. First-principles studies on the structural and electronic properties of Li-ion battery cathode material $CuF_2$. *Solid State Commun.* **152**, 1703-1706 (2012).
46. Marković, I. et al. Electronically driven spin-reorientation transition of the correlated polar metal $Ca_3Ru_2O_7$. *Proc. Natl. Acad. Sci. USA* **117**, 15524-15529 (2020).
47. Schoop, L. M. et al. Tunable Weyl and Dirac states in the nonsymmorphic compound CeSbTe. *Sci. Adv.* **4**, eaar2317 (2018).
48. Biniskos, N. et al. Complex magnetic structure and spin waves of the noncollinear antiferromagnet $Mn_5Si_3$. *Phys. Rev. B* **105**, 104404 (2022).
49. Miyahara, Y., Miyazaki, K., Fukutsuka, T. & Abe, T. Strontium cobalt oxychlorides: enhanced electrocatalysts for oxygen reduction and evolution reactions. *Chem. Comm.* **53**, 2713-2716 (2017).
50. Winiarski, M. & Samsel-Czekała, M. The electronic structure of $CeNiGe_3$ and $YNiGe_3$ superconductors by ab initio calculations. *Solid State Commun.* **179**, 6-10 (2014).
51. Anisimov, V. I., Zaanen, J. & Andersen, O. K. Band theory and Mott insulators: Hubbard U instead of Stoner I. *Phys. Rev. B* **44**, 943 (1991).
52. Chen, Q. et al. Orbital-Selective Kondo Entanglement and Antiferromagnetic Order in $USb_2$. *Phys. Rev. Lett.* **123**, 106402 (2019).
53. Gerasimov, E., Mushnikov, N. & Gaviko, V. Magnetic Phase Transitions in $La_{1-x}R_xMn_2Si_2$ (R= Gd, Tb,

Dy) Compounds. *Solid State Phenomena* **190**, 171-174 (2012).

54. Korotin, D. M. et al. Electronic structure of $RMn_2Si_2$ (R= Y, La) intermetallics: DFT and XPS studies. *J. Alloys Compd.* **695**, 1663-1671 (2017).

55. Mattsson, S. & Paulus, B. Density Functional Theory Calculations of Structural, Electronic, and Magnetic Properties of the 3*d* Metal Trifluorides $MF_3$ (M= Ti-Ni) in the Solid State. *J. Comput. Chem.* **40**, 1190-1197 (2019).

56. Beleanu, A. et al. Large resistivity change and phase transition in the antiferromagnetic semiconductors LiMnAs and LaOMnAs. *Phys. Rev. B* **88**, 184429 (2013).

57. Tian, C. et al. DyOCl: A rare-earth based two-dimensional van der Waals material with strong magnetic anisotropy. *Phys. Rev. B* **104**, 214410 (2021).

58. Weber, S. F. & Neaton, J. B. Origins of anisotropic transport in the electrically switchable antiferromagnet $Fe_{1/3}NbS_2$. *Phys. Rev. B* **103**, 214439 (2021).

59. Han, F. et al. Perpendicular magnetic anisotropy induced by $La_{2/3}Sr_{1/3}MnO3$–$YBaCo_2O_{5+\delta}$ interlayer coupling. *J. Phys. D: Appl. Phys.* **54**, 185302 (2021).

60. Yu, X. et al. First-principles study on structural, mechanical, and electronic properties of $REAuBi_2$ (RE= La–Pr, Sm) intermetallic compounds. *AIP Adv.* **12**(2022).

61. Orlandi, F. et al. γ-$BaFe_2O_4$: a fresh playground for room temperature multiferroicity. *Nat. Commun.* **13**, 7968 (2022).

62. Park, J.-H. et al. Electronic aspects of the ferromagnetic transition in manganese perovskites. *Phys. Rev. Lett.* **76**, 4215 (1996).

63. Yuan, L.-D. et al. Giant momentum-dependent spin splitting in centrosymmetric low-Z antiferromagnets. *Phys. Rev. B* **102**, 014422 (2020).

64. Hao, X. et al. Exceptionally large room-temperature ferroelectric polarization in the $PbNiO_3$ multiferroic nickelate: First-principles study. *Phys. Rev. B* **86**, 014116 (2012).

65. Kang, S. G. First-principles analysis of ferroelectric transition in $MnSnO_3$ and $MnTiO_3$ perovskites. *J. Solid State Chem.* **262**, 251-255 (2018).

66. Yamauchi, K., Barone, P. & Picozzi, S. Theoretical investigation of magnetoelectric effects in $Ba_2$ $CoGe_2O_7$. *Phys. Rev. B* **84**, 165137 (2011).

67. Feng, Y., Wang, N., Guo, X. & Zhang, S. Dopant screening of modified $Fe_2O_3$ oxygen carriers in chemical looping hydrogen production. *Fuel* **262**, 116489 (2020).

68. Mali, B. et al. Re-entrant spin reorientation transition and Griffiths-like phase in antiferromagnetic $TbFe_{0.5}Cr_{0.5}O_3$. *Phys. Rev. B* **102**, 014418 (2020).

69. Turnbull, R. et al. Pressure-induced phase transition and band gap decrease in semiconducting β-

$Cu_2V_2O_7$. *Inorg. Chem.* **61**, 3697-3707 (2022).
70. Chen, X. et al. Tuning magnetic anisotropy and Dzyaloshinskii-Moriya interaction via interface engineering in nonisostructural $SrCuO_2/SrRuO_3$ heterostructures. *Phys. Rev. B* **105**, 214428 (2022).
71. Barreda-Argüeso, J. A. et al. Pressure-induced phase-transition sequence in $CoF_2$: An experimental and first-principles study on the crystal, vibrational, and electronic properties. *Phys. Rev. B* **88**, 214108 (2013).
72. Xu, D. et al. Experimental and computational study of the magnetic properties of $ZrMn_{2-x}Co_xGe_4O_{12}$. *J. Chem. Soc., Dalton Trans.* **46**, 6921-6933 (2017).
73. Yamauchi, K., Barone, P. & Picozzi, S. Theoretical investigation of magnetoelectric effects in $Ba_2CoGe_2O_7$. *Phys. Rev. B* **84**, 165137 (2011).
74. Guo, Y. et al. Spin-split collinear antiferromagnets: A large-scale ab-initio study. *Mater. Today Phys.* **32**, 100991 (2023).
75. Morrow, R. et al. Independent ordering of two interpenetrating magnetic sublattices in the double perovskite $Sr_2CoOsO_6$. *J. Am. Chem. Soc.* **135**, 18824-18830 (2013).
76. Ding, L. et al. Unusual magnetic structure of the high-pressure synthesized perovskites $ACrO_3$ (A= Sc, In, Tl). *Phys. Rev. B* **95**, 054432 (2017).
77. Baskurt, M., Nair, R. R., Peeters, F. M. & Sahin, H. Ultra-thin structures of manganese fluorides: conversion from manganese dichalcogenides by fluorination. *Phys. Chem. Chem. Phys.* **23**, 10218-10224 (2021).
78. Wu, F. et al. Fluctuation-enhanced phonon magnetic moments in a polar antiferromagnet. *Nat. Phys.* **19**, 1868-1875 (2023).
79. Wang, Y. et al. Improved conductivity of $NdFeO_3$ through partial substitution of Nd by Ca: a theoretical study. *Phys. Chem. Chem. Phys.* **17**, 29097-29102 (2015).
80. Weber, M. C. et al. Raman spectroscopy of rare-earth orthoferrites $RFeO_3$ (R= La, Sm, Eu, Gd, Tb, Dy). *Phys. Rev. B* **94**, 214103 (2016).
81. Wang, Y.-C. & Jiang, H. Local screened Coulomb correction approach to strongly correlated d-electron systems. *J. Chem. Phys.* **150**(2019).
82. Lin, L.-F. et al. Ferroelectric ferrimagnetic $LiFe_2F_6$: Charge-ordering-mediated magnetoelectricity. *Phys. Rev. Mater.* **1**, 071401 (2017).
83. Gupta, K., Mandal, B. & Mahadevan, P. Strain-induced metal-insulator transition in ultrathin films of $SrRuO_3$. *Phys. Rev. B* **90**, 125109 (2014).
84. Šmejkal, L., Sinova, J. & Jungwirth, T. Beyond conventional ferromagnetism and antiferromagnetism: A phase with nonrelativistic spin and crystal rotation symmetry. *Phys. Rev. X* **12**, 031042 (2022).
85. Bhowmik, T. K. & Sinha, T. P. Al-dependent electronic and magnetic properties of $YCrO_3$ with magnetocaloric application: an ab-initio and Monte Carlo approach. *Physica B Condens. Matter* **606**, 412659

(2021).
86. Shi, J. et al. Structural and electronic properties of rare-earth chromites: A computational and experimental study. *Phys. Rev. B* **106**, 165117 (2022).
87. Rajput, S. et al. Coexisting magnetic structures and spin reorientation in $Er_{0.5}Dy_{0.5}FeO_3$: Bulk magnetization, neutron scattering, specific heat, and density functional theory studies. *Phys. Rev. B* **105**, 214436 (2022).
88. Feng, Z. et al. An anomalous Hall effect in altermagnetic ruthenium dioxide. *Nat. Electron.* **5**, 735-743 (2022).
89. Hatanaka, T., Nomoto, T. & Arita, R. Magnetic interactions in intercalated transition metal dichalcogenides: A study based on abinitio model construction. *Phys. Rev. B* **107**, 184429 (2023).
90. Jha, P. K. et al. First principles investigation of anionic redox in bisulfate lithium battery cathodes. *Phys. Chem. Chem. Phys.* **24**, 22756-22767 (2022).
91. Zhu, T. et al. Cycloidal Magnetic Order Promoted by Labile Mixed Anionic Paths in $M_2(SeO_3)F_2$ (M= $Mn^{2+}$, $Ni^{2+}$). *Inorg. Chem.* **60**, 12001-12008 (2021).
92. Mu, S. et al. Phonons, magnons, and lattice thermal transport in antiferromagnetic semiconductor MnTe. *Phys. Rev. Mater.* **3**, 025403 (2019).
93. Brekke, B. et al. Low-energy properties of electrons and holes in $CuFeS_2$. *Phys. Rev. B* **106**, 224421 (2022).
94. Schuler, R., Bianchini, F., Norby, T. & Fjellvag, H. Near-Broken-Gap Alignment between $FeWO_4$ and $Fe_2WO_6$ for Ohmic Direct p–n Junction Thermoelectrics. *ACS Appl. Mater. Interfaces* **13**, 7416-7422 (2021).
95. Shin, Y. & Rondinelli, J. M. Pressure effects on magnetism in $Ca_2Mn_2O_5$-type ferrites and manganites. *Phys. Rev. B* **102**, 104426 (2020).
96. Stroppa, A., Marsman, M., Kresse, G. & Picozzi, S. The multiferroic phase of $DyFeO_3$: an ab initio study. *New J. Phys.* **12**, 093026 (2010).
97. Hu, L. et al. Two-dimensional magneto-photoconductivity in non-van der Waals manganese selenide. *Mater. Horiz.* **8**, 1286-1296 (2021).
98. Wisesa, P., Li, C., Wang, C. & Mueller, T. Materials with the $CrVO_4$ structure type as candidate superprotonic conductors. *RSC Adv.* **9**, 31999-32009 (2019).
99. Kim, Y., Kim, S. & Choi, S. First-principles study of the change in the electronic properties of $NiF_2$ by the introduction of oxygen. *J. Mol. Struct.* **1054**, 293-296 (2013).
100. Mook, A., Plekhanov, K., Klinovaja, J. & Loss, D. Interaction-stabilized topological magnon insulator in ferromagnets. *Phys. Rev. X* **11**, 021061 (2021).
101. Lu, Y.-S., Li, J.-L. & Wu, C.-T. Topological phase transitions of Dirac magnons in honeycomb ferromagnets. *Phys. Rev. Lett.* **127**, 217202 (2021).

102. Yao, W. et al. Topological spin excitations in a three-dimensional antiferromagnet. *Nat. Phys.* **14**, 1011 (2018).

103. Wang, D., Bo, X., Tang, F. & Wan, X. Calculated magnetic exchange interactions in the Dirac magnon material $Cu_3TeO_6$. *Phys. Rev. B* **99**, 035160 (2019).

104. Liu, Y. et al. Spin Nernst effect and spatiotemporal dynamic simulation of topological magnons in the antiferromagnet $Cu_3TeO_6$. *Phys. Rev. B* **108**, 094427 (2023).

105. Nielsen, H. B. & Ninomiya, M. Absence of neutrinos on a lattice:(II). Intuitive topological proof. *Nucl. Phys. B* **193**, 173-194 (1981).

106. Martinez-Lope, M., Alonso, J., Sheptyakov, D. & Pomjakushin, V. Preparation and structural study from neutron diffraction data of $Pr_5Mo_3O_{16}$. *J. Solid State Chem.* **183**, 2974-2978 (2010).